\documentclass[withindex,glossary]{cam-thesis}

\usepackage{subfiles}

\usepackage{amsmath}
\usepackage{amsthm}
\usepackage{amssymb}
\usepackage{epigraph}
\usepackage{mathtools} 
\renewcommand{\vec}[1]{\textbf{#1}} 
\usepackage{braket} 
\usepackage{tensor} 

\usepackage{framed}

\usepackage{slashed}

\usepackage[numbers]{natbib}

 \usepackage[inline]{enumitem}
 
 \usepackage{tabularx}
 \newcolumntype{C}[1]{>{\centering\arraybackslash}p{#1}}
 \usepackage{booktabs}
 
\usepackage{graphicx}
\usepackage{float}
\usepackage{caption}
\usepackage[linesnumbered,lined,ruled,commentsnumbered]{algorithm2e}

\usepackage{tikz}
 \usetikzlibrary{decorations.pathmorphing}
  \usetikzlibrary{decorations.markings}
\tikzset{snake it/.style={decorate, decoration=snake}}
\usetikzlibrary{shapes.geometric, arrows}
\tikzstyle{process} = [rectangle, minimum width=3cm, minimum height=1cm, text centered, draw=black, fill=orange!30]
\tikzstyle{arrow} = [thick,->,>=stealth]
\usetikzlibrary{arrows.meta,bending}
\usepackage{pgf}
\usepackage{subcaption}
\usetikzlibrary{arrows}
\usetikzlibrary{patterns}

\usepackage{multirow}
\usepackage{neuralnetwork}
\usepackage{xstring}
\usepackage{xpatch}
\usepackage{array}

\newcommand{\simunet}{\textsc{SIMUnet}}
\newcommand{\nnpdf}{\textsc{NNPDF4.0}}
\newcommand{\fitm}{\textsc{fitmaker}}
\newcommand{\smefit}{\textsc{SMEFiT}}

\def\gsim{\mathrel{\rlap{\lower4pt\hbox{\hskip1pt$\sim$}}
    \raise1pt\hbox{$>$}}}

\makeatletter
\def\thickhline{%
             \noalign{\ifnum0 =`}\fi\hrule \@height \thickarrayrulewidth \futurelet
             \reserved@a\@xthickhline}
\def\@xthickhline{\ifx\reserved@a\thickhline
                \vskip\doublerulesep
                \vskip -\thickarrayrulewidth
                \fi
                \ifnum0 =`{\fi}}
\xpatchcmd{\linklayers}{\nn@lastnode}{\lastnode}{}{}
\xpatchcmd{\linklayers}{\nn@thisnode}{\thisnode}{}{}
\makeatother
\newlength{\thickarrayrulewidth}
\newcommand*{\tmp}[4]{\ensuremath{%
    {#4%
    \ifx\empty#3\empty\ifx\empty#1\empty\else^{#1}\fi\else^{#1(#3)}\fi%
    \ifx\empty#2\empty\else_{#2}\fi}%
}}

\newcommand{\madgraph}{\texttt{MadGraph5\_aMC@NLO}{}}

\newcommand{\la}{\left\langle}
\newcommand{\ra}{\right\rangle}

\newcommand{\be}{\begin{equation}}
\newcommand{\ee}{\end{equation}}
\newcommand{\bea}{\begin{eqnarray}}
\newcommand{\eea}{\end{eqnarray}}
\newcommand{\bi}{\begin{itemize}}
\newcommand{\ei}{\end{itemize}}
\newcommand{\ben}{\begin{enumerate}}
\newcommand{\een}{\end{enumerate}}
\newcommand{\lc}{\left[}
\newcommand{\rc}{\right]}
\newcommand{\lp}{\left(}
\newcommand{\rp}{\right)}
\newcommand{\as}{\alpha_s}

\def\frac#1#2{{{#1}\over {#2}}}
\def\gsim{\mathrel{\rlap{\lower4pt\hbox{\hskip1pt$\sim$}}
    \raise1pt\hbox{$>$}}}       
\def\lsim{\mathrel{\rlap{\lower4pt\hbox{\hskip1pt$\sim$}}
    \raise1pt\hbox{$<$}}}

\newcommand{\draft}[1]{}

\def\beq{\begin{equation}}
\def\eeq{\end{equation}}

\numberwithin{equation}{chapter}
\numberwithin{figure}{chapter}
\numberwithin{table}{chapter}

\newcommand{\nnpdfnotop}{\texttt{NNPDF4.0-notop} }

\usepackage{tabularx}

\definecolor{darkblue}{rgb}{0.0,0,0.5}
\definecolor{darkgreen}{rgb}{0.0,0.3,0.0}
\definecolor{redish}{rgb}{0.675,0,0.2}
\definecolor{red}{rgb}{0.8,0,0}
\definecolor{green}{rgb}{0,0.6,0}
\definecolor{bluish}{rgb}{0.2,0.2,0.675}
\definecolor{mygrey}{rgb}{0.6,0.6,0.6}

\definecolor{commentsColor}{rgb}{0.497495, 0.497587, 0.497464}
\definecolor{keywordsColor}{rgb}{0.000000, 0.000000, 0.635294}
\definecolor{stringColor}{rgb}{0.558215, 0.000000, 0.135316}
\definecolor{backgroundColor}{rgb}{0.9, 0.9, 0.9}

\usepackage{pythonhighlight}

\newcommand{\web}{\href{https://hep-pbsp.github.io/SIMUnet/}{\url{https://hep-pbsp.github.io/SIMUnet/}}}
\newcommand{\nnpdfcode}{\texttt{NNPDF}}
\newcommand{\evolventhreefit}{\texttt{evolven3fit}}

\newcommand{\fitmaker}{\texttt{fitmaker}}
\newcommand{\apfel}{\texttt{apfel}}

\title{Precision phenomenology \\
of the PDF-BSM interplay}

\author{Manuel Morales Alvarado}

\college{Girton College}

\collegeshield{CollegeShields/Girton}

\submissiondate{Wednesday 31st July, 2024}

\submissionnotice{
  \centering
  This thesis is submitted on Wednesday 31st July, 2024, \\ for the degree of Doctor of Philosophy.
  }

\date{July, 2024}

\subjectline{High Energy Physics}

\abstract{%
The Standard Model (SM) stands as one of the most successful theories ever conceived, representing a triumph of
human intellect and collaboration. The SM provides a very good description of visible matter, with the proton being a
central part of this understanding. The structure of the proton, in terms of its elementary constituents, can be parametrised
in terms of parton distribution functions (PDFs). PDFs, essential ingredients in theoretical predictions, cannot be
calculated from first principles and must be extracted from fits to experimental data.

Despite the immense success of the SM, we know that it cannot be the most complete theory of nature as it
leaves important questions unanswered. In this context, the high energy physics community looks for physics
beyond the SM (BSM). One way to conduct these searches is by studying subtle deviations of our theoretical
predictions from the observed experimental values. In this context, as it reaches unprecedented levels of precision,
the Large Hadron Collider (LHC) is carrying out one of the most rigorous tests of the SM. However, despite the large amounts
of data collected so far, no direct evidence of BSM physics has been found.

PDFs are crucial inputs in LHC physics. However, they are typically fitted under the assumption of the SM, which can introduce potential inconsistencies when they are used to generate predictions in BSM searches. In this way, a simultaneous determination of PDFs and BSM parameters is crucial to avoid biases in the interpretation of LHC data and to maximise the potential of the LHC to discover new physics. Additionally, BSM physics can be absorbed by the PDFs, rendering the interpretation of the data very challenging. The interplay between PDFs and BMS physics is an issue that is often overlooked in the literature and, in this thesis, we aim to explore it further.

In Chapter \ref{ch:introduction}, we begin by providing a review of the SM and, in particular, quantum chromodynamics and PDFs. Additionally, we introduce aspects of BSM physics and effective field theories (EFTs). We also formulate the problem of the PDF-BSM interplay and discuss the potential implications of this interplay in the context of LHC physics.
After that, in Chapter \ref{ch:simunet}, we discuss aspects
of modern machine learning (ML) techniques that are useful for the analyses carried out in this thesis and we introduce \simunet{}: an open-source deep learning methodology to perform simultaneous fits of PDFs and BSM parameters, and to assess the potential absorption of new physics effects in PDF fits.
In Chapter \ref{ch:fits} we apply the \simunet{} methodology to perform fits of PDFs and EFT coefficients in the top sector and in global datasets. We compare the results of these fits with those obtained from standard SM PDF-only and EFT-only fits, and we discuss the implications of these results in terms of a simultaneous determination.
Then, in Chapter \ref{ch:contamination}, we apply the \simunet{} methodology to address the possible contamination of BSM physics in PDF fits. We show how this absorption can lead to apparent but spurious tensions between theoretical predictions and experimental measurements, and how to disentangle the effects of this contamination in the observables.
Afterwards, in Chapter \ref{ch:smeft_dglap}, we discuss whether BSM physics, formulated in terms of EFTs, can modify the renormalisation group evolution of the PDFs in the DGLAP equations, which describes how PDFs evolve with the energy scale.
In Chapter \ref{ch:sr}, we shift our focus from PDFs to ways in which novel ML techniques can be used for precision calculations in LHC physics in general. To pave the way for the use of interpretable ML methodologies in collider physics, we explore how symbolic regression can be used for precision calculations in collider observables. 
In Chapter \ref{ch:conclusions} we summarise the main results of this thesis and discuss potential future directions.

Quotes are used to introduce each chapter. If the original quote is not in English, the original and the translated version are provided. If an official or well-known translation is available, it is used. Otherwise, a translation by the author, indicated by `trad. M. M. A.', is provided.
}

\acknowledgements{%

This thesis, and the work that has led to it, would not have been possible without the support and guidance of many people. I would like to express my deepest gratitude to all of them.

First and foremost, I would like to thank Prof. Maria Ubiali, my PhD supervisor, for her guidance and support throughout these years. Her enthusiasm for research and her dedication to PBSP, her group, have been a constant source of inspiration that will remain with me. I am grateful for the many opportunities she has given me to grow as a physicist and for the trust she has placed in me. She is not only an exceptional scientist but also a wonderful mentor that, since the day that I arrived in Cambridge, and in the middle of an ongoing global pandemic, has been there for us.

I would like to thank Prof. Ben Allanach and Prof. Luca Silvestrini for agreeing to be my examiners and participating in the viva. 

Thank you to the members of the PBSP group, past and present, for their support and friendship. My deepest gratitude to you, James, for being such an amazing friend and collaborator. I am grateful not only for the work we have done, but also for the many laughs, trips, and good memories that we have shared since that day at The Eagle. Thank you, Luca, for the many discussions and memorable conferences and workshops that we have shared. Thanks to Mark, Elie, and Francesco, our younger students, for your enthusiasm that permeates the whole group. My gratitude to you, Shayan, Maeve, Zahari, and Cameron, for all your help and support when the PhD journey was just starting. Many thanks to Ben, Bernat, Nico, Hannah, and the extended HEP Phenomenology Group, for the great atmosphere that you have created. These past three and a half years would not have been the same without all of you.

I would like to thank Fernando Quevedo and Verónica Sanz for being great mentors beyond my research group, and helping me navigate the world of academia. 

To you, my dear friends outside the department, I am also truly grateful. Michelle, Joaquín, Maddy, Cindy, Bob, Mario, Bernardo, Daniel, Caro, Jaco, Fabian, and many others, thank you for your friendship. Countless gatherings, outings to the pub, dinners, music concerts, salsa nights, intense thesis writings days, and coffees with you have made my time in Cambridge unforgettable.

Many people, away from Cambridge, have been with me in this journey. Heartfelt thanks to my friends from undergrad days, Nicolás Bitar, Giovanni Pais, and Nicolás Cárdenas, for your invaluable and constant friendship that crosses borders.

I could not be more grateful to my lifelong friends who have become family. Thank you Sebastián, Agustín, Joaquín, and Ignacio, for being the best friends anyone could ask for. Thank you for the most honest friendship that can come from the naive and innocent hearts of childhood.

Thank you to my extended family, both in Santiago and in Patagonia, for your unconditional love and support.

Finally, words cannot express how thankful I am to my parents, Nancy and Kenneth, for \textit{simply everything}. Everything good that I am, I am because of you. Thank you for your kindness, wisdom, courage, but above all, for showing me what the purest form of love is. This thesis is for you. 

\vspace{1em} 
\noindent\rule{\textwidth}{0.4pt}
\vspace{1em} 

I would like to thank the support of the European Research Council under the European Union's Horizon 2020 Research and Innovation Programme (Grant Agreement N. 950246).

}

\begin{document}

\frontmatter{}

\chapter{Introduction}
\label{ch:introduction}

\epigraph{When you find yourself in the thick of it \\ Help yourself to a bit of what is all around you}{\textit{\\ from Martha My Dear, \\ by The Beatles.}}

In this chapter we provide a brief introduction to the main topics that will be discussed in the thesis. We start by discussing the Standard Model (SM) of particle physics. We introduce the concepts of deep inelastic scattering, and the parton model as a way of understanding hadrons in terms of elementary constituents. We then introduce the concept of parton distribution functions (PDFs) and their evolution. We finish this chapter by presenting some aspects of physics beyond the SM, effective field theories, and their interplay with the PDFs, which will be the main focus of the thesis.

\section{The Standard Model}
\label{sec:intro_sm}

'\textit{Nothing exists except atoms and empty space; everything else is merely thought to exist}', said Democritus in the writings of Diogenes Laertius \cite{laertius2020lives}. Human curiosity moves us to understand and describe the world around us at the most fundamental level. This curiosity was the driving force behind the ideas of ancient philosophers like Democritus himself, who famously proposed the existence of \textit{atoms}. Over time, our understanding of the fundamental constituents of matter has evolved, leading to the development of modern frameworks.

Up to the early 1800s, we had theories that accounted for two interactions: gravity and electromagnetism. After that, the 20th century saw rapid progress in the field of physics ranging from the first experiments in radioactivity, to the discovery of the electron, the first fundamental particle to be discovered, and to the development of quantum mechanics and relativity. These developments culminated in the creation of the \textit{Standard Model} (SM), one of the most successful theories ever devised. It accounts for three of the four fundamental forces of nature: electromagnetism, the weak nuclear force, and the strong nuclear force. The SM does not describe gravity. The SM is a theory that describes particles and their interactions in terms of relativistic quantum field theories. 

To Democritus of Abdera's contentment, we will first briefly formulate some aspects of the SM in terms of its spacetime and matter content, and then we will delve deeper into its \textit{proper} formulation in terms of quantum fields and symmetries.

To talk about the SM we first need a notion of \textit{spacetime}. Throughout this thesis we will work in the framework of special relativity, which describes the geometry of spacetime in the absence of gravity. In this context, we work in a flat $(3+1)$-dimensional Minkowski spacetime, with a metric $g$ given by
\begin{equation}
    \label{eq:flat_metric}
    g = \begin{pmatrix}
    1 & 0 & 0 & 0 \\
    0 & -1 & 0 & 0 \\
    0 & 0 & -1 & 0 \\
    0 & 0 & 0 & -1 \\
    \end{pmatrix}.
\end{equation}
Its elements are denoted by $g_{\mu\nu}$, with $\mu, \nu = 0, 1, 2, 3$, going over the time and space components of the metric\footnote{Note that Eq. (\ref{eq:flat_metric}) describes the metric of a flat spacetime. In the presence of gravity, the metric is no longer flat and it becomes a dynamical field, which is described by Einstein's field equations in general relativity.}

We also need the notion of \textit{fields}, which represent particles. In the SM, we have gauge fields, which are associated with the fundamental forces, and matter fields, which are associated with the matter particles. Gauge fields mediate the interactions between the matter fields, so they can be though of as the 'messenger' particles of the forces. 

The matter fields can be split into two categories: \textit{leptons} and \textit{quarks}. In the SM there are three generations of leptons
\begin{equation*}
    \begin{array}{c}
    \begin{pmatrix} e \\ \nu_e \end{pmatrix}, \quad
    \begin{pmatrix} \mu \\ \nu_{\mu} \end{pmatrix}, \quad
    \begin{pmatrix} \tau \\ \nu_{\tau} \end{pmatrix},
    \end{array}
\end{equation*}
where $e$ is the electron, $\mu$ is the muon, and $\tau$ is the tau lepton. Each one of these charged leptons has an associated neutrino respectively denoted as $\nu_e$, $\nu_{\mu}$, and $\nu_{\tau}$.
Neutrinos are neutral particles, and they are much lighter than the charged leptons\footnote{The SM predicts massless neutrinos.}. 

In the SM there are also quarks. Unlike leptons, quarks have colour charge, and therefore they interact through the strong nuclear force. There are 3 generations of quarks
\begin{equation*}
    \begin{array}{c}
    \left( \begin{array}{c} u \\ d \end{array} \right), \quad \left( \begin{array}{c} c \\ s \end{array} \right), \quad \left( \begin{array}{c} t \\ b \end{array} \right)
    \end{array},
\end{equation*}
where $u$, $c$, and $t$ are the up, charm, and top quarks, and $d$, $s$, and $b$ are the down, strange, and bottom quarks. 
%
%
Leptons and quarks interact with each other through gauge bosons, the carriers of the fundamental interactions.

The electromagnetic interaction, which describes light, heat, chemical bonds, and many other things, is mediated by the \textit{photon}, usually denoted by
\begin{equation*}
    \gamma.
\end{equation*}
The weak interaction, which is responsible for the radioactive decay of atoms, is mediated by the
\begin{equation*}
    \begin{array}{c}
    W^+, \quad W^-, \quad \text{and} \quad Z
    \end{array}
\end{equation*}
bosons. The $W^+$ and $W^-$ bosons are charged, and the $Z$ boson is neutral. Finally, the strong interaction is mediated by the \textit{gluons}
\begin{equation*}
    g.
\end{equation*}

The last ingredient of the SM is the \textit{Higgs boson}: a scalar particle that is responsible for the generation of mass of the fundamental particles, and it is usually denoted in the literature by
\begin{equation*}
    H.
\end{equation*}

After this lightning review of the 'atoms' and 'empty space' of Democritus, we will now delve deeper into the formulation of the SM in terms of relativistic quantum fields and symmetries. The SM is a \textit{gauge} theory, which means that the Lagrangian is invariant under local gauge transformations. The gauge group of the SM is
\begin{equation}
    \label{eq:sm_gauge_group}
    G_{\text{SM}} = \text{SU}(3)_C \times \text{SU}(2)_L \times \text{U}(1)_Y,
\end{equation}
where $\text{SU}(3)_C$ is the gauge group of the strong interaction, and $\text{SU}(2)_L \times \text{U}(1)_Y$ is the gauge group of the electroweak interaction. In a gauge theory, associated to every generator of a group there is a gauge boson that acts as a mediator of the interaction. In this way, there are 8 gluons associated with the $\text{SU}(3)_C$ group, and 4 gauge bosons associated with the $\text{SU}(2)_L \times \text{U}(1)_Y$ group which, after a rotation via the weak mixing angle, become the $W^+$, $W^-$, and $Z$ bosons, and the photon $\gamma$.

In addition to the matter particles and the gauge bosons, the SM Lagrangian contains the Higgs boson, which is a scalar field that is responsible for the generation of mass for the fundamental particles. The Higgs boson is a spin-$0$ particle, and it is the only scalar particle in the SM. The Higgs boson is a key ingredient of the SM, and its discovery in 2012 by the ATLAS and CMS collaborations at the LHC was a major milestone in the history of particle physics. The gauge group of Eq. (\ref{eq:sm_gauge_group}) is broken by the Higgs field acquiring a non-zero vacuum expectation value, which gives mass to the $W^+$, $W^-$, and $Z$ bosons, and the fermions. The photon remains massless, as it is the gauge boson associated with an unbroken $\text{U}(1)_\text{EM}$ subgroup.

The SM Lagrangian, before spontaneous symmetry breaking, can be written as
\begin{equation}
    \label{eq:sm_lagrangian}
    \begin{aligned}
    \mathcal{L}_{\text{SM}} = & -\frac{1}{4} G_{\mu\nu}^A G^{A\mu\nu} 
    - \frac{1}{4} W_{\mu\nu}^I W^{I\mu\nu} 
    - \frac{1}{4} B_{\mu\nu} B^{\mu\nu} \\
    & + \sum_{\psi} \overline{\psi} i \slashed{D} \psi 
    + (D_{\mu} H)^{\dagger} (D^{\mu} H) \\
    & - \lambda \left( H^{\dagger} H - \frac{1}{2} v^2 \right)^2 \\
    & - \left[ Y_d \overline{q} H d 
    + Y_u \overline{q} \tilde{H} u 
    + Y_e \overline{l} H e 
    + \text{H.c.} \right].
    \end{aligned}
\end{equation}
Let us summarise the terms in Eq. (\ref{eq:sm_lagrangian}). To do so, it will first be useful to define the covariant derivative
\begin{equation}
    \label{eq:covariant_derivative}
    D_{\mu} = \partial_{\mu} + i g^{\prime} Y B_{\mu} + i g S^{I} W_{\mu}^I + i g_s T^A G_{\mu}^A,
\end{equation}
where $g^{\prime}$, $g$, and $g_s$ are the gauge couplings of the $\text{U}(1)_Y$, $\text{SU}(2)_L$, and $\text{SU}(3)_C$ groups, respectively, $Y$ is the hypercharge, $S^I$ are generators of $\text{SU}(2)_{L}$, and $T^A$ are generators of $\text{SU}(3)_C$. The generators of $\text{SU}(2)_L$ can be chosen to be
\begin{equation}
    \label{eq:su2_generators}
    S^I = \frac{\tau^I}{2}, \quad I = 1, 2, 3,
\end{equation}
with the Pauli matrices defined in Eq. (\ref{eq:tau_matrices}) of App. \ref{app:appendix_int}. The generators of $\text{SU}(3)_C$ can be chosen to be
\begin{equation}
    \label{eq:su3_generators}
    T^A = \frac{\lambda^A}{2}, \quad A = 1, \dots, 8,
\end{equation}
where $\lambda^A$ are the Gell-Mann matrices defined in Eq. (\ref{eq:gell_mann_matrices}) of App. \ref{app:appendix_int}.

Now, let us go back to the SM Lagrangian in Eq. (\ref{eq:sm_lagrangian}). The first line contains the kinetic terms for the gauge bosons, with $G_{\mu\nu}^A$, $W_{\mu\nu}^I$, and $B_{\mu\nu}$ being the field strength tensors for the gluons, $W$ bosons, and the hypercharge boson, respectively, and it also contains the self-interactions among the bosons. The field strengths are defined as
\begin{equation}
    \begin{aligned}
        G_{\mu\nu}^A &= \partial_{\mu} G_{\nu}^A - \partial_{\nu} G_{\mu}^A - g_s f^{ABC} G_{\mu}^B G_{\nu}^C, \\
        W_{\mu\nu}^I &= \partial_{\mu} W_{\nu}^I - \partial_{\nu} W_{\mu}^I - g \epsilon^{IJK} W_{\mu}^J W_{\nu}^K, \\
        B_{\mu\nu} &= \partial_{\mu} B_{\nu} - \partial_{\nu} B_{\mu}.
    \end{aligned}
\end{equation}
where $f^{ABC}$ and $\epsilon^{IJK}$ are the structure constants of $\text{SU}(3)_C$ and $\text{SU}(2)_L$, respectively, and characterise the algebra of the groups as in Eq. (\ref{eq:commutation_relations}) of App. \ref{app:appendix_int}. The second line in Eq. (\ref{eq:sm_lagrangian}) contains the kinetic terms for the fermions, and the interaction of the fermions with the gauge bosons via the covariant derivative. The sum over $\psi$ runs over all the fermions in the SM. The second line also contains the kinetic term of the Higgs $H$ and its interaction with the gauge fields. The third line encodes the Higgs potential which, after spontaneous symmetry breaking, gives the Higgs a mass $m_H = 2 \lambda v^2$. The last line is the Yukawa sector, and it contains the interactions between the Higgs and the fermions. $d$, $u$, and $e$ are the right-handed down, up, and electron singlets, and $q$ and $l$ are the left-handed quark and lepton doublets. We also define $\tilde{H}_{i} = \epsilon_{ij} H^{*j}$. In Eq. (\ref{eq:sm_lagrangian}) we have neglected generation indices, but the Yukawa couplings $Y_u$, $Y_d$, and $Y_e$ are actually $3 \times 3$ matrices that encode the interactions between the Higgs and the generations of fermions. The Yukawa sector, after spontaneous symmetry breaking, generates the masses of the fermions.

The SM Lagrangian in Eq. (\ref{eq:sm_lagrangian}) can be used to calculate hard scattering reactions involving elementary particles, but we will see that this is not the whole story when dealing with quarks and gluons. To do this, we will take a closer look to the QCD part of the SM Lagrangian. Eq. (\ref{eq:sm_lagrangian}) provides an almost complete description of perturbative QCD, but it has to be supplemented in two aspects. First, to properly define the gluon propagator we have to add a gauge-fixing term to the Lagrangian, and second, in non-Abelian gauge theories like QCD, ghosts have to be introduced to cancel the unphysical degrees of freedom that arise in the theory \cite{Bjorken:1965zz,Peskin:257493}.
%

In nature, quarks and gluons are never found in isolation. This is a consequence of \textit{confinement}, which can be understood in terms of the running of the strong coupling. Let us consider a physical observable $R$ that depends on the strong coupling
\begin{equation}
    \label{eq:as}
    \alpha_s \equiv \frac{g_s^2}{4\pi}.
\end{equation}
When we calculate $R$ in perturbation theory including loop contributions, we find UV divergences that have to be removed by means of a renormalisation procedure at a scale $\mu$. Naturally, $R$ cannot depend on the arbitrary scale $\mu$ that we have introduced in the calculation so
\begin{equation}
    \label{eq:R_eq}
    \frac{d R}{d t} = 0,
\end{equation}
where $t \equiv \text{log} (\mu^2)$. It can be shown from Eq. (\ref{eq:R_eq}) that the scale dependence on $R$ enters through the \textit{running} of the strong coupling $\as(\mu^2)$, which is described by its $\beta$ function
\begin{equation}
    \label{eq:beta1}
    \frac{d \as}{dt} = \beta(\as), 
\end{equation}
with
\begin{equation}
    \label{eq:beta}
    \beta(\alpha_S) = -b \alpha_S^2 (1 + b^{\prime}\alpha_S + b^{\prime \prime} \alpha_S^2 + \mathcal{O}(\alpha_S^3)), 
\end{equation}
and the coefficients above are given by 
\begin{equation}
\label{eq:beta3}
\begin{aligned}
    b &= \frac{33-2 n_f}{12 \pi}, \\
    b^{\prime} &= \frac{153 - 19 n_f}{2 \pi (33 -2 n_f)}, \\
    b^{\prime\prime} &= \frac{77139 - 15099 n_f + 325 n_f^2}{288 \pi^2 (33 - 2 n_f)},   
\end{aligned}
\end{equation}
where $n_f$ is the number of active quark flavours at the energy scale $\mu$, and the first, second, and third terms on the right-hand side of Eq. (\ref{eq:beta}) represent respectively the $1$-loop, $2$-loop, and $3$-loop corrections to the running. In the SM the running is currently known up to 5-loop corrections \cite{Caswell:1974gg, Jones:1974mm, Tarasov:1980au, Larin:1993tp, vanRitbergen:1997va, Baikov:2016tgj}. Neglecting 2-loop and higher order corrections, the $\beta$ function can be integrated analytically to obtain the running of the strong coupling
\begin{equation}
    \label{eq:as_running}
    \as(\mu^2) = \frac{\as(\mu_0^2)}{1 + \as(\mu_0^2) \ b \ \text{log}(\mu^2/\mu_0^2)},
\end{equation}
where $\mu_0$ is a reference scale (where the coupling remains in the perturbative regime). From Eq. (\ref{eq:as_running}) we see that at very high energies, where the logarithm becomes very large, the strong coupling approaches zero, and the quarks and gluons become \textit{asymptotically free}. On the other hand, at low energies, the strong coupling becomes very large, and the quarks and gluons are confined into hadrons. This is the phenomenon of confinement, which is one of the most important features of QCD. In this energy regime, the strong coupling is not small any more and, therefore, perturbation theory no longer converges.

Notice that Eq. (\ref{eq:as_running}) does not describe absolute values of $\as$ but just its evolution. In practice, we extract the value of the strong coupling at a given scale from experimental data, using for example $\mu_0 = m_Z$, and then use Eq. (\ref{eq:as_running}) to predict the value of $\as$ at other scales. Another possibility is to introduce a dimensionful parameter $\Lambda_{\text{QCD}}$ to describe the evolution. $\Lambda_{\text{QCD}}$ represents the energy scale at which the strong coupling becomes non-perturbative. This value is usually take to be of around $200$ MeV, although its exact value depends on the order of the calculation and the renormalisation scheme used. In this way, the 1-loop running of Eq. (\ref{eq:as_running}) can be written as
\begin{equation*}
    \label{eq:as_running_lambda}
    \as(\mu^2) = \frac{1}{b \ \text{log}(\mu^2/\Lambda_{\text{QCD}}^2)},
\end{equation*}
and the dependence on the reference scale $\mu_0$ has been removed in favour of $\Lambda_{\text{QCD}}$. 

Asymptotic freedom and confinement are key features that will allows us to describe the dynamics of the strong interaction, and how factorisation of short distance and long distance effects can be used to understand that we observe at hadron colliders.

\section{Deep inelastic scattering}
\label{sec:intro_dis}

\begin{figure}[H]
    \centering
    \includegraphics[width=0.5\linewidth]{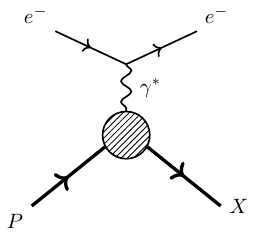}
    \caption{Deep inelastic scattering of an electron off a proton.}
    \label{fig:dis}
\end{figure}

\textit{Deep inelastic scattering} (DIS) is a process that has been instrumental in the development of our understanding of the structure of the proton (and hadrons in general). In DIS, an electron is scattered off a proton
\begin{equation*}
    e^-(k) + P(p) \to e^-(k^\prime) + X,
\end{equation*}
where $X$ is a hadronic final state. The process is characterised by the exchange of a virtual photon, as shown in Fig. \ref{fig:dis}.
We define
\begin{equation}
s = (p + k)^2, \quad q = k - k',
\end{equation}
where $s$ is the square of the center-of-mass energy, and adopt the standard notation for the kinematic variables
\begin{equation}
    \label{eq:kinematics_dis}
    Q^2 = -q^2, \quad x = \frac{Q^2}{2 p \cdot q}, \quad y = \frac{p \cdot q}{p \cdot k^\prime}.
\end{equation}
At leading order in QED the differential cross section is given by
\begin{equation}
    \label{eq:dsigma2}
    d\sigma = \frac{e^4}{2s} \frac{4\pi}{Q^4} L_{\mu\nu}(k, k') W^{\mu\nu}(p, q) \frac{d^3k'}{(2\pi)^3 2E'},
\end{equation}
where we have made explicit the phase space integration over the final state electron, and the integration over the final state hadronic system is implicit in the hadronic tensor $W^{\mu\nu}(p, q)$. Using the kinematics in Eq. (\ref{eq:kinematics_dis}), Eq. (\ref{eq:dsigma2}) can be written as
\begin{equation}
    \label{eq:dsigma}
    d\sigma = \frac{4\pi \alpha^2}{Q^4} L_{\mu\nu}(k, k') W^{\mu\nu}(p, q) \frac{y^2}{2Q^2} dQ^2 dx,
\end{equation}
where we have defined $\alpha = e^2 / (4\pi)$ and used the leptonic tensor defined as
\begin{equation}
\begin{aligned}
    L_{\mu\nu} &= \frac{1}{2} \sum_{\text{spins}} \bar{u}(k') \gamma_\mu u(k) \bar{u}(k) \gamma_\nu u(k') \\
    &= 2 \left( k_\mu k'_\nu + k_\nu k'_\mu - g_{\mu\nu} (k \cdot k') \right).
\end{aligned}
\end{equation}
By imposing conservation $q_\mu W^{\mu \nu} = q_{\nu} W^{\mu \nu} = 0$, it can be shown that the hadronic tensor $W^{\mu\nu}(p, q)$ in Eq. (\ref{eq:dsigma}) can be written in terms of Lorentz structures and scalars as
\begin{equation}
    \label{eq:w_munu}
    W_{\mu \nu}(p, q) = F_1(x, Q^2) \left( -g_{\mu \nu} + \frac{q_{\mu} q_{\nu}}{q^2} \right) + F_2(x, Q^2) \frac{1}{p \cdot q} \left( p_{\mu} - \frac{p \cdot q}{q^2} q_{\mu} \right) \left( p_{\nu} - \frac{p \cdot q}{q^2} q_{\nu} \right),
\end{equation}
where $F_1(x, Q^2)$ and $F_2(x, Q^2)$ are \textit{structure functions}. Having expressions for both the leptonic and hadronic tensors in Eq. (\ref{eq:dsigma}), we can write
\begin{equation}
    \label{eq:LW}
    L_{\mu\nu} W^{\mu\nu} = \frac{2 Q^2}{y^2}  \left( y^2 F_1(x, Q^2) + \frac{1-y}{x} F_2(x, Q^2) \right) \equiv 2 \frac{Q^2}{y^2} \mathcal{F}(x, y, Q^2),
\end{equation}
and obtain the simplified differential cross section
\begin{equation}
    \label{eq:dsigma_final}
    d \sigma = \frac{4 \pi \alpha^2}{Q^4} \mathcal{F}(x, y, Q^2) d Q^2 dx.
\end{equation}
The structure functions $F_1(x, Q^2)$ and $F_2(x, Q^2)$ can be related to the structure of the proton in terms of its elementary constituents, as we shall see in the following section.

\section{The parton model}
\label{sec:intro_pdfs}

The \textit{parton model} \cite{Bjorken:1969ja,feynmanparton} assumes that hadrons are QCD bound states of elementary particles called partons, and that interactions involving hadrons are interactions of point-like partons convoluted with functions that parametrise the structure of the hadron: \textit{parton distribution functions} (PDFs). In this way, any DIS differential cross section $d \sigma$ can be written as 
\begin{equation}
    \label{eq:dis_conv}
    d \sigma(p) = \sum_{i} \int_{0}^{1} dz f_{i}(z) d \hat{\sigma}_{i} (z p),
\end{equation}
where, for a parton of type $i$, $\hat{\sigma}_{i}$ is the \textit{partonic cross section}, which is calculated in perturbation theory, and $f_i(z)$ is its PDF, which is associated to the probability of finding a parton of type $i$ with a fraction $z$ of the total momentum of the hadron $p$. The integral accounts for the process to happen with any momentum fraction weighted by the PDF, and the sum runs over all parton types $i$. 

In the parton model, certain sum rules must be met by the PDFs. For example, the \textit{momentum sum rule} states that the total momentum of the partons in the hadron must be equal to the total momentum of the hadron
\begin{equation}
    \label{eq:momentum_sum_rule}
    \sum_{i} \int_{0}^{1} dx x f_i(x) = 1.
\end{equation}
where the sum runs over all partons $i$. Additionally there are also \textit{number sum rules} which are determined by the number of valence quarks in the hadron
\begin{equation}
    \label{eq:n_sum_rule}
    \int_{0}^{1} \left( f_{q}(x) - f_{\bar{q}(x)} \right) = N_q,
\end{equation}
where $N_q$ is the number of valence quarks of type $q$ in the hadron: $N_u = 2$, $N_d = 1$, and $N_q = 0$ for other $q$ flavours in the proton. 

Using the parton model in Eq. (\ref{eq:dis_conv}), it can be shown that the hadronic tensor can be written in terms of the PDFs as
\begin{equation}
    \label{eq:w_munu_pdf}
    W_{\mu \nu}(p, q) = \sum_i e_i^2 f_i(x) \left[ \frac{1}{2} \left( -g_{\mu \nu} + \frac{q_{\mu} q_{\nu}}{q^2} \right) + \frac{x}{p \cdot q} \left( p_{\mu} - \frac{p \cdot q}{q^2} q_{\mu} \right) \left( p_{\nu} - \frac{p \cdot q}{q^2} q_{\nu} \right) \right].
\end{equation}
Now, we have two expressions for the hadronic tensor in Eqs. (\ref{eq:w_munu}) and (\ref{eq:w_munu_pdf}), and we can identify the structure functions $F_1(x, Q^2)$ and $F_2(x, Q^2)$ with the PDFs as
\begin{equation}
    \label{eq:f1_f2}
    F_1(x, Q^2) = \frac{1}{2} \sum_i e_i^2 f_i(x), \quad F_2(x, Q^2) = x \sum_i e_i^2 f_i(x).
\end{equation}
Eq. (\ref{eq:f1_f2}) shows that the structure functions $F_1(x, Q^2)$ and $F_2(x, Q^2)$, in fact, do not depend on $Q^2$ but simply on the momentum fraction $x$ of the partons. This property is called \textit{Bjorken scaling}, and it is a consequence of the parton model. Furthermore, the relation
\begin{equation}
    2 x F_1(x) = F_2(x) \quad \implies R(x) = \frac{2 x F_1(x)}{F_2(x)} = 1,
\end{equation}
known as Callan-Gross equation \cite{callangross.22.156}, confirms with the fact that partons are spin-$1/2$ fermions. If partons were spin-$0$ particles we would have $F_1(x) = 0$, but experimentally it has been found that $R(x) \approx 1$ \cite{Perkins:1982xb}.

The Bjorken scaling holds to a good approximation in QED but, as we shall see in the next section, QCD corrections will modify this behaviour.

\section{QCD corrections to DIS}
\label{sec:intro_qcd_corrections}

\begin{figure}[H]
    \centering
    \includegraphics[width=0.4\linewidth]{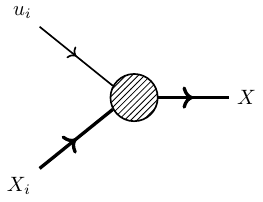}
    \caption{Incoming quark of colour $i$ interacting with some initial state $X_i$ to produce a final state $X$.}
    \label{fig:inc_quark}
\end{figure}

\begin{figure}[H]
    \centering
    \includegraphics[width=0.5\linewidth]{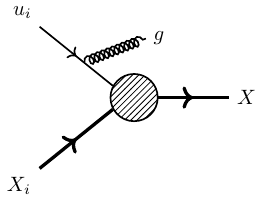}
    \caption{Incoming quark of colour $i$ interacting with some initial state $X_i$ to produce a final state $X$, with the emission of a gluon.}
    \label{fig:inc_quark_gluon}
\end{figure}

To consider QCD corrections to DIS, we will first calculate the relevant partonic cross section with no QCD emissions, and then we will compute the first-order radiative corrections. Consider an incoming quark of colour $i$ with momentum $p$ in the initial state, interacting with some initial state $X_i$ to produce a final state $X$, as shown in Fig. \ref{fig:inc_quark}.
The amplitude of this process is
\begin{equation}
    \mathcal{M}^{(0)}_q(p) = A_i(p) u_i(p),
\end{equation}
where $i$ is a color index, summed over the range $i = 1, \ldots, N_c = 3$. The parton cross section is proportional to
\begin{equation}
    \label{eq:lo_amp}
    \hat{\sigma}^{(0)}_q(p) = \frac{1}{p \cdot p'} \frac{1}{N_c} \left| \mathcal{M}^{(0)}_q(p) \right|^2 = \frac{1}{p \cdot p'} \frac{1}{N_c} A_i(p) \bar{p} \!\!\!/ \, \gamma^0 A_i^{\dagger}(p),
\end{equation}
where $p'$ is the collective momentum of $X_i$. We want to compute the first-order radiative corrections to the cross section in QCD so, now, we will consider the emission of a gluon from the quark, as shown in Fig. \ref{fig:inc_quark_gluon}.
In what follows, we will be interested in the treatment of the soft and collinear singularities that arise in the theory. For this, it will be useful to introduce the \textit{Sudakov decomposition} of the momenta of $k$, the momentum of the gluon, as
\begin{equation}
    k = (1 - z)p + k_T + \xi \eta,
\end{equation}
where the lightlike four-vector $\eta$ is chosen so that
\begin{equation}
    p \cdot k_T = \eta k_T = 0; \quad \eta^2 = 0; \quad 2 p \cdot \eta = 1.
\end{equation}
%
With these ingredients, the phase space measure for the gluon can be written as
\begin{equation}
    \frac{d^3k}{(2 \pi)^3 2k^0} = \frac{1}{16 \pi^2} \frac{d |k_T^2| d z}{1-z}.
\end{equation}
The amplitude with the emission of a gluon can be decomposed in terms of a singular and a regular part in the limit $k_T \to 0$ as
\begin{equation}
    \mathcal{M}_q^{(1)}(p, k) = \mathcal{M}_{q, \text{sing}}^{(1)} (p, k) + \mathcal{M}_{q, \text{reg}}^{(1)} (p, k).
\end{equation}
After some algebra, it can be shown that the singular part of the amplitude is
\begin{equation}
    \left| \mathcal{M}_{q, \text{sing}}^{(1)} (p, k) \right|^2 = - \frac{2 g_s^2 C_F}{k_T^2} \frac{1 + z^2}{z} \left| \mathcal{M}^{(0)}(zp) \right|^2,
\end{equation}
where we have used the leading order amplitude in Eq. (\ref{eq:lo_amp}), before the gluon emission. From the squared matrix element we can get the cross section with gluon emission as
\begin{equation}
    \hat{\sigma}_q^{(1)}(p) = \frac{\alpha_s C_F}{2\pi} \int_0^1 \frac{dz}{1 - z} \int_0^{|k_T^2|_{\text{max}}} d|k_T^2| \left[ \frac{1 + z^2}{|k_T^2|} \hat{\sigma}_q^{(0)}(zp) + R_q(k_T, z) \right],
    \label{eq:example}
\end{equation}
%
where again $R_q(k_T, z)$ is the regular part of the amplitude. The integral over $|k_T^2|$ in Eq. (\ref{eq:example}) is divergent in the collinear limit when $k_T^2 \to 0$. Additionally, the cross section also exhibits a soft singularity when $z \to 1$. To deal with them, we can introduce two cut-off parameters: $1-\epsilon$ in the upper bound of the integration over $z$, and $\lambda$ in the lower bound of the integration over $|k_T^2|$. Then, we obtain the integrals by taking the limits $\epsilon \to 0$ and $\lambda \to 0$ at the end of the calculation. It can be shown that the inclusion of virtual corrections cancel the soft singularity in the real emissions, so we are left with the collinear divergence. To do this, we introduce a momentum scale $\mu$ in charge of defining the regions of phase space with small and high transverse momentum. From Eq. (\ref{eq:example}) we obtain
\begin{equation}
    \begin{aligned}
        \hat{\sigma}_q^{(1)}(p) &= \frac{\alpha_s C_F}{2\pi} \int_0^1 \frac{dz}{1 - z} \int_{\lambda^2}^{\mu^2} \frac{d|k_T^2|}{|k_T^2|} (1 + z^2) \left[ \hat{\sigma}_q^{(0)}(zp) - \hat{\sigma}_q^{(0)}(p) \right] \\
        &\quad + \frac{\alpha_s C_F}{2\pi} \int_0^1 \frac{dz}{1 - z} \int_{\mu^2}^{|k_T^2|_{\text{max}}} \frac{d|k_T^2|}{|k_T^2|} (1 + z^2) \left[ \hat{\sigma}_q^{(0)}(zp) - \hat{\sigma}_q^{(0)}(p) \right] \\
        &\quad + \frac{\alpha_s C_F}{2\pi} \int_0^1 \frac{dz}{1 - z} \int_0^{|k_T^2|_{\text{max}}} d|k_T^2| R_q(k_T, z),
    \end{aligned}
\end{equation}
where we have separated the integral over $|k_T^2|$ into the two regions, the $\hat{\sigma}_{q}^{0}(p)$ contributions have come from the virtual corrections, and on the last line we have set $\lambda^2 = 0$ because the integrand is regular. By defining a so-called \textit{splitting function}
\begin{equation}
P_{qq}(z) = C_F \left( \frac{1 + z^2}{1 - z} \right)_+,
\end{equation}
where the $+$ subscript means that $P_{qq}(z)$ is a distribution whose action over a regular test function $f(z)$ is defined by subtracting the value of the function at $f(z = 1)$ as
\begin{equation}
\int_0^1 dz \, P_{qq}(z) f(z) = \int_0^1 dz \, C_F \frac{1 + z^2}{1 - z} [f(z) - f(1)],
\end{equation}
we get
\begin{equation}
\hat{\sigma}_q^{(1)}(p) = \frac{\alpha_s C_F}{2\pi} \int_0^1 dz \, P_{qq}(z) \hat{\sigma}_q^{(0)}(zp) \log \frac{\mu^2}{\lambda^2} + \hat{\sigma}_{q, \text{reg}}^{(1)} (p, \mu^2),
\end{equation}
where the regular part of the cross section is
\begin{equation}
    \hat{\sigma}^{(1)}_{q,\text{reg}}(p, \mu^2) = \frac{\alpha_s}{2\pi} \int_0^1 dz \left\{ P_{qq}(z) \hat{\sigma}^{(0)}_q(zp) \log \frac{|k_T^2|_{\text{max}}}{\mu^2} + \frac{C_F}{1 - z} \int_0^{|k_T^2|_{\text{max}}} d|k_T^2| R_q(k_T, z) \right\}.
\end{equation}

An additional contribution to the total cross section of this process comes from gluon splitting. A similar procedure, after a long calculation, leads to the following expression for its associated cross section
\begin{equation}
    \hat{\sigma}^{(1)}_g(p) = \frac{\alpha_s}{2\pi} \int_0^1 dz \, P_{qg}(z) \hat{\sigma}^{(0)}_q(zp) \log \frac{\mu^2}{\lambda^2} + \hat{\sigma}^{(1)}_{g,\text{reg}}(p, \mu^2),
\end{equation}
where we have defined the splitting function
\begin{equation}
    P_{qg}(z) = T_F \left[ z^2 + (1 - z)^2 \right],
\end{equation}
and the regular part of the cross section is given by
\begin{equation}
    \hat{\sigma}_{g, \text{reg}}^{(1)} (p, \mu^2) = \frac{\alpha_s}{2\pi} \int_0^1 dz \left\{ P_{qg}(z) \hat{\sigma}_q^{(0)}(zp) \log \frac{|k_T^2|_{\text{max}}}{\mu^2} + \frac{T_F}{1 - z} \int_0^{|k_T^2|_{\text{max}}} d|k_T^2| R_g(k_T, z) \right\}.
\end{equation}

So far, by including leading order QCD corrections, we have found the cross sections
\begin{align}
    \label{eq:hola1}
    \hat{\sigma}_q(p) &= \hat{\sigma}_q^{(0)}(p) + \hat{\sigma}_q^{(1)}(p) \notag{} \\
    &= \hat{\sigma}_q^{(0)}(p) + \frac{\alpha_s}{2\pi} \int_0^1 dz \, P_{qq}(z) \hat{\sigma}_q^{(0)}(zp) \log \frac{\mu^2}{\lambda^2} + \hat{\sigma}_{q, \text{reg}}^{(1)}  (p, \mu^2),
\end{align}
and
\begin{align}
    \label{eq:hola2}
    \hat{\sigma}_g(p) &= \hat{\sigma}_g^{(1)}(p) \notag{} \\
    &= \frac{\alpha_s}{2\pi} \int_0^1 dz \, P_{qg}(z) \hat{\sigma}_q^{(0)}(zp) \log \frac{\mu^2}{\lambda^2} + \hat{\sigma}_{g, \text{reg}}^{(1)}  (p, \mu^2).
\end{align}

So far, we have only dealt with partonic cross sections. Now, we will calculate the complete hadronic cross section in the parton model via Eq. (\ref{eq:dis_conv}) as
\begin{equation}
    \sigma(p) = \int_{0}^{1} dy \left[ f_q(y) \hat{\sigma}_q(yp) + f_g(y) \hat{\sigma}_g(yp) \right],
\end{equation}

In the next section, we will see how PDFs provide a mechanism to deal with the collinear singularities that arise in the theory.

\section{PDF evolution}

We now observe that the two terms proportional to $\log \lambda^2$ in Eqs. (\ref{eq:hola1}) and (\ref{eq:hola2}) can be \textit{absorbed} in a redefinition of the quark distribution function $f_q(y)$. Indeed, the first three terms in (\ref{eq:hola1}) can be written
\begin{equation}
    \label{eq:trick}
    \int_{0}^{1} dy \int_{0}^{1} dz \hat{\sigma}^{(0)}_q(zyp) \left\{ f_q(y) \left[ \delta(1-z) + \frac{\alpha_s}{2\pi} P_{qq}(z) \log \frac{\mu^2}{\lambda^2} \right] + f_g(y) \frac{\alpha_s}{2\pi} P_{qg}(z) \log \frac{\mu^2}{\lambda^2} \right\},
\end{equation}
where we have grouped the terms in the cross section by introducing an integral over $z$ to be able to factor the leading order contribution
\begin{equation}
    \int_{0}^{1} dy \left( f_q (y) \hat{\sigma}_{q}(yp)  \right) = \int_{0}^{1} dy \int_{0}^{1} dz \hat{\sigma}^{(0)}_q(zyp) f_q(y) \delta(1-z).
\end{equation}
Eq. (\ref{eq:trick}) shows that we can redefine the quark distribution function to absorb the collinear divergence. By using the change of variables
\begin{equation}
zy = x; \quad dz = \frac{dx}{y}; \quad 0 \leq x \leq y,
\end{equation}
this becomes
\begin{equation}
    \resizebox{\textwidth}{!}{$
    \begin{aligned}
    \int_{0}^{1} \frac{dy}{y} \int_{0}^{y} dx \hat{\sigma}^{(0)}_q(xp) \left\{ f_q(y) \left[ \delta \left(1 - \frac{x}{y} \right) + \frac{\alpha_s}{2\pi} P_{qq} \left( \frac{x}{y} \right) \log \frac{\mu^2}{\lambda^2} \right] + f_g(y) \frac{\alpha_s}{2\pi} P_{qg} \left( \frac{x}{y} \right) \log \frac{\mu^2}{\lambda^2} \right\} \\
    = \int_{0}^{1} dx \hat{\sigma}^{(0)}_q(xp) \int_{x}^{1} \frac{dy}{y} \left\{ f_q(y) \left[ \delta \left(1 - \frac{x}{y} \right) + \frac{\alpha_s}{2\pi} P_{qq} \left( \frac{x}{y} \right) \log \frac{\mu^2}{\lambda^2} \right] + f_g(y) \frac{\alpha_s}{2\pi} P_{qg} \left( \frac{x}{y} \right) \log \frac{\mu^2}{\lambda^2} \right\}.
    \end{aligned}
    $}
\end{equation}
We then define
\begin{equation}
\begin{aligned}
f_q(x, \mu^2) = \int_{x}^{1} \frac{dy}{y} \left\{ f_q(y) \left[ \delta \left(1 - \frac{x}{y} \right) + \frac{\alpha_s}{2\pi} P_{qq} \left( \frac{x}{y} \right) \log \frac{\mu^2}{\lambda^2} \right] + f_g(y) \frac{\alpha_s}{2\pi} P_{qg} \left( \frac{x}{y} \right) \log \frac{\mu^2}{\lambda^2} \right\},
\end{aligned}
\end{equation}
which allows us to write the hadronic cross section with scale dependent PDFs as
\begin{equation}
    \sigma(p) = \int_{0}^{1} dx \left[ f_q(x, \mu^2) \hat{\sigma}_q(xp, \mu^2) + f_g(x) \hat{\sigma}_g(xp, \mu^2) \right].
\end{equation}
We have reabsorbed the collinear singularities in the definitions of the quark PDF as
\begin{equation}
    \label{eq:quark_pdf_evol}
    f_q (x, \mu^2) = \int_{x}^{1} \frac{dy}{y} \left\{ f_q(y) \left[ \delta \left(1 - \frac{x}{y} \right) + \frac{\alpha_s}{2\pi} P_{qq} \left( \frac{x}{y} \right) \log \frac{\mu^2}{\lambda^2} \right] + f_g(y) \frac{\alpha_s}{2\pi} P_{qg} \left( \frac{x}{y} \right) \log \frac{\mu^2}{\lambda^2} \right\}.
\end{equation}
Now, if a gluon goes on to participate in the hard reaction we can repeat the process to obtain
\begin{equation}
    \label{eq:gluon_pdf_evol}
    f_{g}(x, \mu^2) = \int_{x}^{1} \frac{dy}{y} \left\{ \left[ \delta \left(1 - \frac{x}{y} \right) + \frac{\alpha_s}{2\pi} P_{gg} \left( \frac{x}{y} \right) \log \frac{\mu^2}{\lambda^2} \right] f_g(y) + \frac{\alpha_s}{2\pi} P_{gq} \left( \frac{x}{y} \right) \log \frac{\mu^2}{\lambda^2} f_q(y) \right\},
\end{equation}
where the additional splitting functions are given by 
\begin{equation}
    P_{gg}(z) = 2 C_A \left[ \frac{z}{1 - z}_{+} + \frac{1 - z}{z} + z(1 - z) \right] + \frac{33 - 2 n_f}{6} \delta(1 - z),
\end{equation}
where $n_f$ is the number of active flavours, and
\begin{equation}
    P_{gq}(z) = C_F \left[ z^2 + (1 - z)^2 \right].
\end{equation}
The evolution of the PDFs in the scale $\mu^2$ is given then by
\begin{equation}
    \label{eq:dglap}
    \begin{aligned}
    \mu^2 \frac{\partial f_q(x, \mu^2)}{\partial \mu^2} &= \frac{\alpha_s}{2\pi} \int_x^1 \frac{dy}{y} \left\{ P_{qq} \left( \frac{x}{y} \right) f_q(y, \mu^2) + P_{qg} \left( \frac{x}{y} \right) f_g(y, \mu^2) \right\}, \\
    \mu^2 \frac{\partial f_g(x, \mu^2)}{\partial \mu^2} &= \frac{\alpha_s}{2\pi} \int_x^1 \frac{dy}{y} \left\{ P_{gg} \left( \frac{x}{y} \right) f_g(y, \mu^2) + P_{gq} \left( \frac{x}{y} \right) f_q(y, \mu^2) \right\},
    \end{aligned}
\end{equation}
which are the well known \textit{Dokshitzer-Gribov-Lipatov-Altarelli-Parisi (DGLAP) evolution equations} \cite{Dokshitzer:1977sg,Gribov:1972ri,Altarelli:1977zs}. 

The procedure we have described so far considers only first order QCD corrections. In general, at higher orders the splitting functions of the DGLAP equations can be written as an expansion in $\as$ as
\begin{equation}
\label{eq:split_all_ord}
\begin{aligned}
    P_{q_i q_j}(z, \as) &= \delta_{ij} P_{qq}^{(0)}(z) + \mathcal{O}(\as), \\
    P_{q_i g}(z, \as) &= P_{q_i g}^{(0)}(z) + \mathcal{O}(\as), \\
    P_{g q_i}(z, \as) &= P_{g q_i}^{(0)}(z) + \mathcal{O}(\as), \\
    P_{g g}(z, \as) &= P_{g g}^{(0)}(z) + \mathcal{O}(\as), \\
\end{aligned}
\end{equation}
where, we consider $n_f$ active flavours, $q_i$ runs over quark and antiquarks, $i = 1, \dots, 2 n_f$, and $g$ represent the gluons.

State of the art splitting functions in the SM have been computed up to NNLO~\cite{Moch:1999eb,Vogt:2004mw,Moch:2004pa,Vermaseren:2005qc}, and partially to N$^3$LO~\cite{Davies:2016jie,Moch:2017uml,Davies:2022ofz,Henn:2019swt,Duhr:2022cob,Moch:2021qrk,Soar:2009yh,Falcioni:2023luc,Falcioni:2023vqq,Moch:2023tdj,Falcioni:2023tzp}.

It is worth mentioning that, at leading order, the splitting functions can be more simply calculated as the collinear limit of matrix elements as
\begin{equation} \label{eq:def_split}
    P_{BA}(z) = \lim_{k_T  \to  0} \left( \frac{1}{2} z (1-z) \overline{\sum_{\operatorname{pols.}}} \sum_{\operatorname{cols.}}\frac{|\mathcal{M}_{A \rightarrow B + C}|^2}{k_T^2} \right),
\end{equation}
where $A$ denotes the incoming particle, $B$ denotes the outgoing particle that participates in the hard reaction, $C$ denotes the on-shell outgoing particle, and we average over the polarisation states and the colours. The factors on the right hand side of the equation arise from the phase space integration in the Sudakov decomposition. 

Later in this thesis we will be interested in corrections coming from higher dimensional operators to the leading order splitting functions that we have described here.

Naturally, the splitting functions are essential in describing the evolution of the PDFs, but they are not the only ingredient. The DGLAP equations also depend on the strong coupling $\as$. As we described earlier in this chapter, $\as$ is not a constant, but it runs with the energy scale according to its beta function. Later in the thesis, we will also explore how the action of higher dimensional operators can modify the running of $\as$.

A final comment is in order regarding the factorisation in Eq. (\ref{eq:dis_conv}) in the DIS case. The factorisation theorems state that the full cross section of a process can be factorised into a convolution of the PDFs and the partonic cross section, separating long- and short-distance effects. In a similar way, we can also decompose hadronic cross sections, with two hadrons in the initial state, in terms of its luminosity for the $i, j$ parton channel
\begin{equation}
    \mathcal{L}_{ij}(\tau, m) = \int_{\tau}^{1} \frac{dx}{x} \, f_i \left( x, m^2 \right) f_j \left( \frac{\tau}{x}, m^2 \right) \equiv f_i \otimes f_j,
\end{equation}
where we have introduced the $\otimes$ convolution symbol. In this way, the hadronic cross section can be written as
\begin{equation}
    \sigma = \sum_{i,j} \mathcal{L}_{ij} \otimes \hat{\sigma}_{ij},
\end{equation}
where $\hat{\sigma}_{ij}$ are the partonic cross sections.

PDFs will be fundamental objects in this thesis. So far, however, we have only considered the PDFs in the context of the SM. In the next section, we will discuss some aspects of physics beyond the SM, and how it can be related to the PDFs.

\section{Beyond the Standard Model physics}
\label{sec:intro_bsm}

What we have described so far has been formulated in the context of the SM, one of the most successful theories ever devised. To excellent agreement, the SM describes experiments that had been carried out before its formulation while, at the same time, made predictions that were spectacularly confirmed by experiments later (e.g. the electroweak bosons and the Higgs). However, we also know that the SM \textit{cannot} be the full story. There are several reasons for this, with notable examples being the origin of neutrino masses, the origin of matter in the universe, the nature of dark matter and dark energy, and a quantum theory of gravity. For these reasons, the SM is considered incomplete, and we look for physics \textit{beyond the SM} (BSM).

There are several framework that account for BSM physics. One of them is by means of \textit{effective field theory} (EFT) frameworks. In this framework, we consider the SM as an effective theory that describes the interactions of particles at energies below a certain \textit{high energy scale}, $\Lambda$, while potential heavy new particles lie above this scale. At energies below $\Lambda$, the SM fields are retained as dynamical degrees of freedom, and the heavy fields are integrated out of the theory. The EFT is then constructed by expanding the Lagrangian in terms of \textit{higher dimensional operators} of increasing mass dimension, suppressed by powers of $\Lambda$. The coefficients of these operators are called \textit{Wilson coefficients}, and they encode the effects of the heavy fields that have been integrated out. The EFT can be used to study the effects of BSM physics in a fairly model-independent way, and we can indirectly study heavy new physics by constraining the parameter space of the EFT using experimental data.

In this section, we will delve deeper into the EFT frameworks, how the SM can be thought as an EFT, and how we can use EFTs to study BSM physics and their interplay with the PDFs.

\subsection{Effective field theories}

\begin{figure}[htbp]
    \centering
    \includegraphics[width=0.80\textwidth]{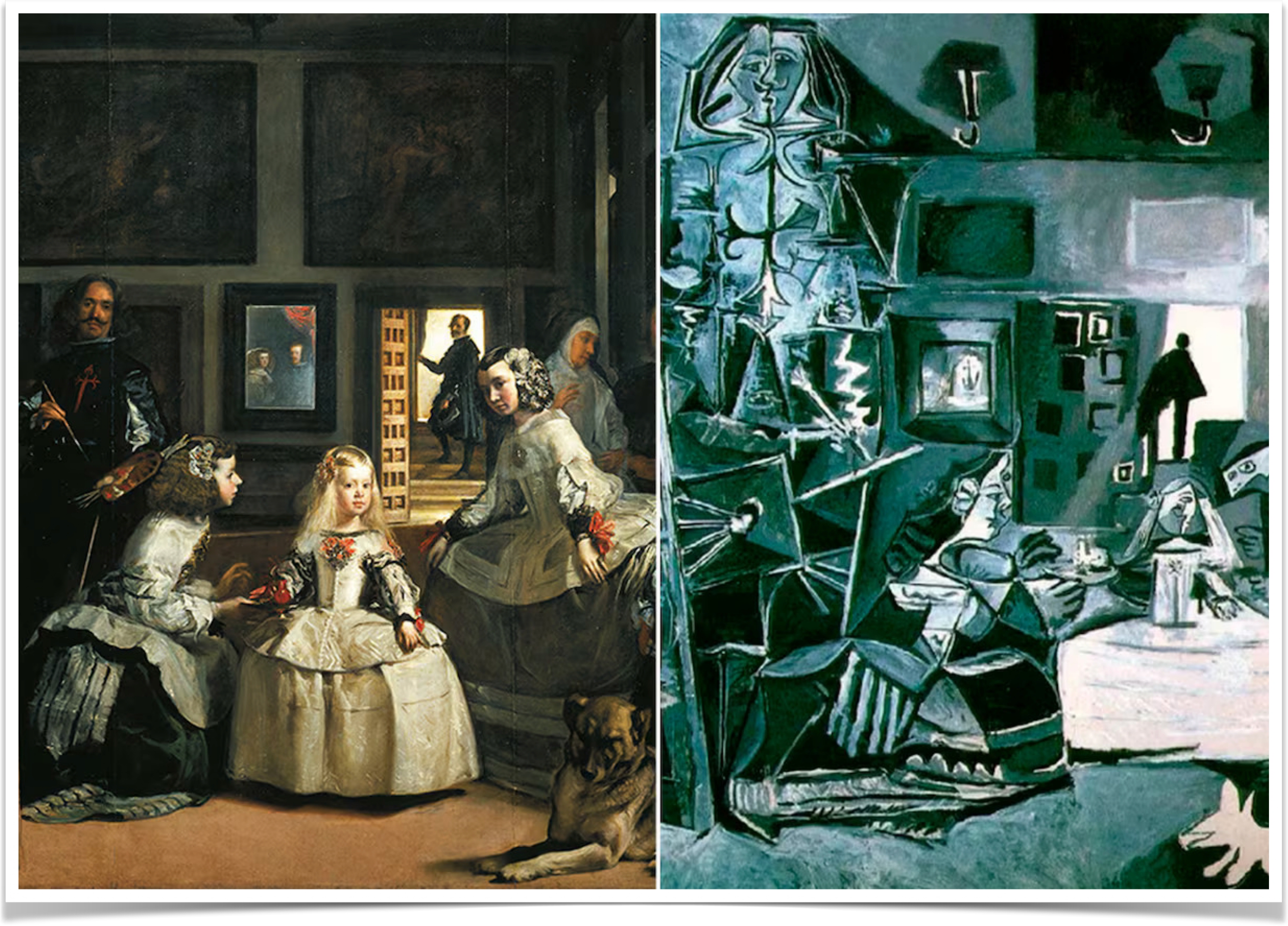}
    \caption{To the left, the original \textit{Las Meninas} by Diego Velázquez, 1656. To the right, an iteration of \textit{Las Meninas} by Pablo Picasso, 1957.}
    \label{fig:paint}
\end{figure}

In general, our description of physical phenomena is scale dependent. To describe an apple falling from a tree we can simply use Newton's laws of motion and neglect the effects of Einstein's general relativity. In this way, we do not need a full description of the laws of nature to describe, to \textit{excellent} agreement, a physical system. We could formulate this discussion, for example, in terms of a painting: we do not need to know every single brush stroke and a complete colour palette to understand a painting as a consistent message. Consider the two cases in Fig. \ref{fig:paint}. 

In the figure we see the original \textit{Las Meninas} by Diego Velázquez, and an iteration\footnote{There are many versions.} of \textit{Las Meninas} by Pablo Picasso. The original painting, famous example of the Spanish Baroque period, is a detailed and realistic (in the technical sense) representation of a room in the Royal Alcázar of Madrid. 

The painting by Velázquez is characterised by its dramatic use of light and shadow, rich colours, and sophisticated technique. In this way, the composition of the painting is complex and multi-layered. Velázquez uses tenebrism, a technique that emphasizes stark contrasts between light and dark, to create a sense of depth and volume, giving the figures a real sense of presence.

Picasso, one of the icons of the Cubist movement, reimagined the painting in his own style. The painting is still clearly recognizable as \textit{Las Meninas}, but the style is completely different as the color palette, the shapes, and the geometry of the scene are simplified. Picasso's approach is minimalistic, breaking down the intricate details and forms of Velázquez's original into basic geometric shapes and colours. Despite this reduction, Picasso manages to capture the essence of the original painting. The characters and the overall scene remain clearly identifiable, demonstrating that the fundamental components of Velázquez’s composition, such as the arrangement of figures and the interaction of light and shadow, are powerful enough to transcend the drastic stylistic changes.

Picasso’s minimalist interpretation emphasises the core elements of \textit{Las Meninas}, neglecting some of the details while retaining the essential narrative. This minimalism does not diminish the painting's impact. Rather, it highlights the underlying structure and the interplay of forms and space that define Velázquez's masterpiece. Picasso’s work exemplifies how an artist can convey the same idea through a different visual language, focusing on the effective components that define the scene.

This idea, that we can describe a system in terms of its effective components, is the basis of effective field theories (EFTs). Just as Picasso's \textit{Las Meninas} captures the essence of Velázquez's work through a minimalist approach, EFTs can capture the main features of new physics, to which we do not have direct access, by focusing on their most relevant features, allowing for a deeper understanding of their fundamental properties.

An EFT, in the context of high energy physics, describes elementary interactions at energies below an energy threshold $\Lambda$ when there is a scale separation between theories. We define the UV theory to be the complete theory at higher energies, and the EFT to be the theory that describes the low-energy dynamics. We can move between the UV and EFT domains in two directions:
\begin{itemize}
\item \textbf{Top-bottom approach:} We know the UV theory, but we find it useful (or necessary) to integrate the heavy dynamics into effective operators and couplings in the EFT. We retain only the light fields as dynamical degrees of freedom and we are interested on how would the UV theory manifest at low energies.

\item \textbf{Bottom-up approach:} We know the EFT and we are interested in mapping the constraints of the effective operators onto the parameters of the UV theory. Observables constrain the parameter space of the EFT and possible tension with an underlying theory (the SM, for example), so this allows us to study possible UV completions of the model that account for what we see.
\end{itemize}
To construct an EFT approximation of a UV theory we have to consider some steps:
\begin{itemize}
\item \textbf{Identify the separation of scales: } We must define the energy scale that separates the UV from the EFT. This scale, usually denoted as $\Lambda$, defines the upper bound of new physics and the power-counting of higher dimensional operators in the EFT. Both the UV and the EFT have to match at low energies, and the EFT has to be valid up to the scale $\Lambda$. To have a dimensionally consistent Lagrangian, possible higher dimensional operators in the EFT are weighted by inverse powers of $\Lambda$ and, the higher the dimension of the operator, the more suppressed by the high energy scale it is. 

\item \textbf{Determine the symmetries of the theory:} Symmetries are associated to conserved quantities and constrain the interactions that the model can describe. Once we define them, we expand the EFT in an operator basis that satisfies these symmetries.

\item \textbf{Determine the relevant degrees of freedom:} The separation of scales defines the light fields that remain as dynamical degrees of freedom in the EFT, and the heavy fields that are integrated out of the theory. Below the matching scale, only the light fields appear in the operator content of the effective Lagrangian, and the integrated fields will
give rise to new, higher dimensional operators and their couplings, known as Wilson coefficients.
\end{itemize}

To illustrate the points above, in the following we will discuss how to construct an EFT in a simplified model, and how to relate the EFT and the UV theories. Let us consider a simple UV model described by a heavy and a light real scalar field, denoted respectively as $\phi$ and $H$ (the idea would be fairly analogous for a set of fields). The mass of the light field is $m$ and the mass of the heavy field is $M$ with $M \gg m$.

To obtain an EFT from the UV, the heavy field $H$ must be integrated out, and only the light field $\phi$ remains as a dynamical degree of freedom. We can write the generating functional of the UV theory as
\begin{equation}
    \label{eq:uv_path}
    Z_{\text{UV}} [J_\phi, J_H] = \int \mathcal{D} \phi \mathcal{D} H \operatorname{exp} \Bigl(i \int d^4 x \left( \mathcal{L}_{\text{UV}} (\phi, H) + J_\phi \phi +  J_H H \right) \Bigr),
\end{equation}
where $J_\phi$ and $J_H$ are currents. The functional in Eq. (\ref{eq:uv_path}) characterises the UV theory as all the $n$-point correlators of $\phi$ and $H$ can be obtained by differentiating with respect to the currents. In the effective theory, we only need the correlators for the light fields $\phi$, so its generating functional can be written as
\begin{equation}
    \label{eq:eft_path}
    Z_{\text{EFT}} [J_\phi] = Z_{\text{UV}} [J_\phi, J_H = 0],
\end{equation}
which, in terms of the effective Lagrangian, can be expressed as
\begin{equation}
    \label{eq:eft_path_L}
    Z_{\text{EFT}} [J_\phi] =   \int \mathcal{D} \phi \operatorname{exp} \Bigl(i \int d^4 x \left( \mathcal{L}_{\text{EFT}} (\phi) + J_\phi \phi \right) \Bigr).
\end{equation}
In terms of the UV and EFT actions can be related by
\begin{equation}
    \label{eq:S_eff}
    \text{exp} \left( i S_{\text{EFT}} [\phi] (\Lambda) \right) = \int \mathcal{D} H \text{exp} \left( i S_{\text{UV}} [\phi, H] (\Lambda) \right),
\end{equation}
where the above defines a matching condition between the UV and the EFT at a scale $\Lambda \approx M \gg  m$. From now on, we will drop the explicit $\Lambda$ scale dependence of the action and understand that the matching is computed at that scale. The effective action can be obtained by approximating the heavy field close to its classical solution $H_c$. If we expand $H = H_c + \eta$, we can approximate the action by
\begin{equation}
    \label{eq:eff_s_2}
    S_{\text{EFT}} [\phi] = S_{\text{UV}} [\phi, H_c] + \frac{1}{2} \eta \frac{\delta^2 S_{\text{UV}}}{\delta H^2}  \Bigg|_{H_c} \eta + \dots .
\end{equation}
where the dots represent higher order derivatives in $H$, and the first functional derivatives vanish because we are expanding around the classical solution
\begin{equation*}
    \frac{\delta S_{\text{UV} [\phi, H]}}{\delta H} = 0 \quad \text{at} \quad H = H_c.
\end{equation*}
In this way, we can write the effective action in Eq. (\ref{eq:S_eff}) as 
\begin{equation}
    \text{exp} \left( i S_{\text{EFT}} [\phi] \right) = \int \mathcal{D} \eta \ \text{exp} \left( i S_{\text{UV}} [\phi, H_c + \eta]  \right) \approx \text{exp} \left( i S [\phi, H_c] \right) \left[ \text{det} \left( - \frac{\delta^2 S}{\delta H^2} \Bigg|_{H_c} \right) \right]^{-1/2},
\end{equation}
and obtain
\begin{equation}
    \label{eq:eft_final}
    S_{\text{EFT}} [\phi] = S_{\text{UV}} [\phi, H_c] + \frac{i}{2} \text{Tr} \ \text{log} \left( - \frac{\delta^2 S}{\delta H^2} \Bigg|_{H_c} \right).
\end{equation}
Eq. (\ref{eq:eft_final}) is the effective action of the EFT, which naturally contains the effective Lagrangian. It tells us that the effective action can be approximated by the UV action evaluated at the classical solution of the heavy field, which constitutes a leading order tree-level matching, plus a term that accounts for the fluctuations of the heavy field around its classical solution, which constitutes a 1-loop correction. 

By removing the heavy field from the UV theory, we generate new operators in the EFT that are built from the light fields and, in this way, the effective Lagrangian can be written as an expansion 
\begin{equation} \label{eq:pc}
\mathcal{L}_{\text{EFT}} = \mathcal{L}_{\text{light}} +  \sum_{k}^{\infty} \frac{1}{\Lambda^{k-4}} \sum_{i}^{n_k} \mathcal{C}_{(i,k)} \mathcal{O}_{(i,k)},
\end{equation}
where $\mathcal{L}_{\text{light}}$ accounts for the part of the UV Lagrangian that only contains the light fields (and therefore remains unchanged when we integrate out the heavy fields), the sum over $k$ runs over all dimensions $k \geq 5$, the sum over $i$ runs over the $n_k$ operators at dimensions $k$, $\mathcal{C}_{(i,k)}$ are couplings called Wilson coefficients, and $\mathcal{O}_{(i,k)}$ are operators that are built from the light fields. The statement above is formalised by the Appelquist–Carazzone theorem \cite{Appelquist:1974tg}, which stastes that the low energy dynamics can be effectively described using the light degrees of freedom alone, while the decoupling of the heavy fields is encoded and reflected in the coefficients of the higher dimensional operators.

To illustrate the ideas above, in the following pages we will go over a concrete example of a matching procedure between a UV theory and an EFT using the Fermi theory of weak interactions.

\textit{Matching} is the process that allows us to relate the UV theory to the EFT. The idea is to impose that the UV theory and the EFT match at low energies, and in this way we can determine the Wilson coefficients of the new operators in the EFT. The matching is done by calculating the matrix elements of the UV theory and the EFT, and equating them. Notice that EFT and the UV theories have to match at low energies, but they do not have to match at high energies since the EFT is only valid up to the scale $\Lambda$.

In the literature, matching is usually done at tree level or at the 1-loop level. Currently, the procedure to match a UV theory to an EFT can be done in a systematic and mostly automated way using a variety of tools, as described in Ref. \cite{Aebischer:2023nnv}.

\begin{figure}[ht]
    \centering
    \includegraphics[width=0.35\linewidth]{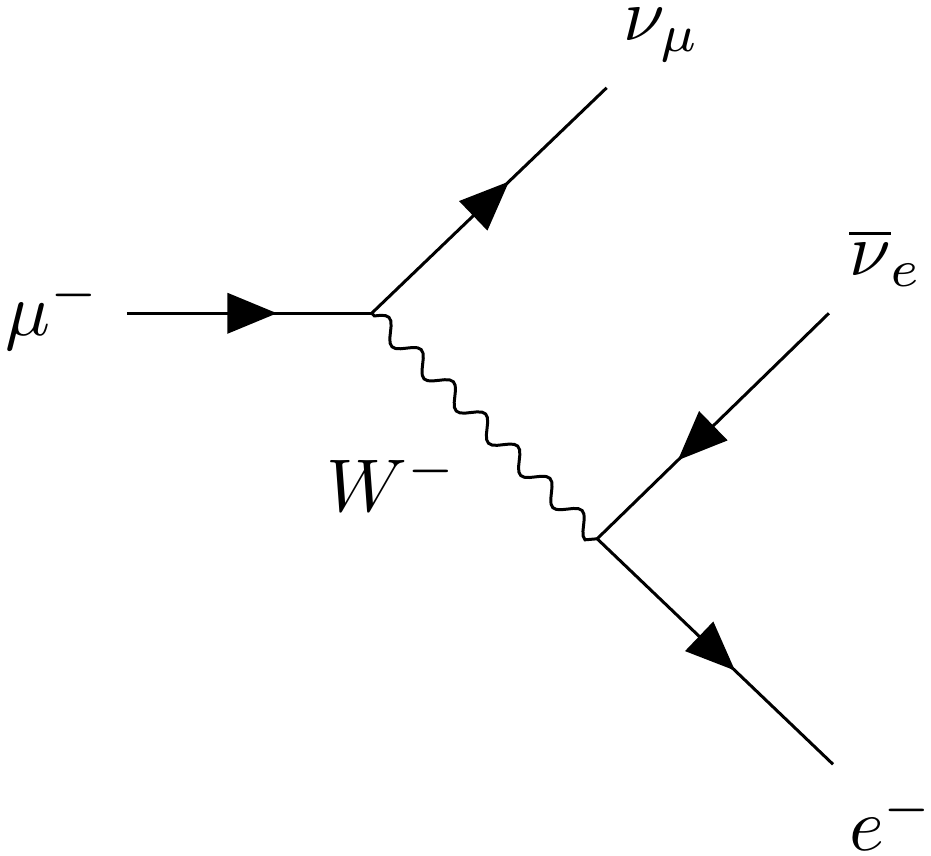}
    \caption{Diagram in the UV theory.}
    \label{fig:intro_1}
\end{figure}

\begin{figure}[ht]
    \centering
    \includegraphics[width=0.35\linewidth]{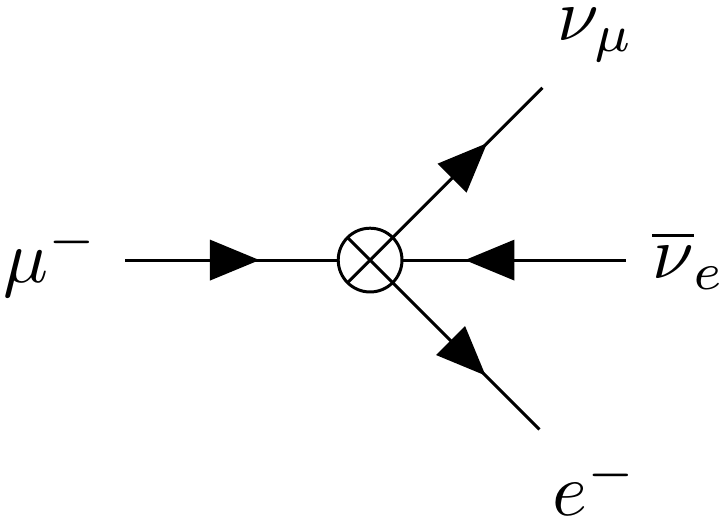}
    \caption{Diagram in the EFT.}
    \label{fig:intro_2}
\end{figure}

To illustrate the idea behind matching, let us consider an example of matching at tree level: \textit{Fermi theory}. Beta decay, $\mu^{-} \rightarrow \nu_{\mu} + e^{-} + \overline{\nu}_{e}$, is mediated by a $W$ boson. Its mass $m_W \approx 80 \ \text{GeV}$ is much larger than the masses of the other particles involved in this process. In this way, we can think of the mass difference of the particles as a clear separation of scales. We can obtain a low energy approximation of this process, an EFT, by integrating out of the theory the $W$ boson. This will generate a new, higher dimensional operator in the EFT, altogether with its coupling, the Wilson coefficient. The diagram in the UV theory (which contains the $W$ boson) is given in Fig. \ref{fig:intro_1}.
The amplitude of the UV process is given by
\begin{equation} \label{eq:fermi}
\begin{split}
i \mathcal{M}_{UV} &= \overline{\nu_\mu} \Big(\frac{ig}{\sqrt{2}} \gamma^\alpha P_L \Big) \mu \cdot \frac{-i g_{\alpha \beta}}{(p_\mu - p_{\nu_\mu})^2 - m_W^2} \cdot \overline{e} \Big(\frac{ig}{\sqrt{2}} \gamma^\beta P_L \Big) \nu_e, \\
&= \frac{ig^2}{2(q^2 - m_W^2)} (\overline{\nu_\mu} \gamma^\alpha P_L \mu )\cdot (\overline{e} \gamma_\alpha P_L \nu_e),
\end{split}
\end{equation}
where $g/\sqrt{2}$ is the $W$ coupling, $P_L$ are left-handed projectors, and we have defined $q = p_\mu - p_{\nu_\mu}$. If $m_W^2 \gg q^2$, we can expand the denominator and obtain
\begin{equation}
    \label{eq:fermi_final}
    i \mathcal{M}_{UV} = -\frac{ig^2}{2m_W^2} (\overline{\nu_\mu} \gamma^\alpha P_L \mu) \cdot (\overline{e} \gamma_\alpha P_L \nu_e) + \mathcal{O} \Biggl( \frac{1}{{m_W}^4} \Biggr).
\end{equation}
If we retain only the first term, the interaction above can be written as a four-fermion interaction coming from an effective Lagrangian
%
\begin{equation} 
    \label{eq:fermi_2}
\mathcal{L}_{\operatorname{EFT}} = \frac{- g^2}{2 m_W^2} \big(\overline{\nu}_\mu \gamma^\alpha P_L \mu \big) \cdot \big( \overline{e} \gamma_\alpha P_L \nu_e \big) + \mathcal{O} \Biggl( \frac{1}{{m_W}^4} \Biggr).
\end{equation}
From Eq. (\ref{eq:fermi_2}) we already see a higher dimensional operator that comes from the action of integrating out the $W$ boson in the UV theory, and we see a Wilson coefficient $\mathcal{C} = - g^2/(2 m_W^2)$.
Diagrammatically, the four-fermion operator can be represented as in Fig. \ref{fig:intro_2}.
In terms of the coupling, we can write it in terms of the Fermi constant $G_F$ as
\begin{equation}
    \label{eq:fermi_c}
    \frac{G_F}{\sqrt{2}} = \frac{g^2}{8 m_W^2},
\end{equation}
and obtain
\begin{equation}
    \label{eq:fermi_4}
    \mathcal{L}_{\operatorname{EFT}} = - \frac{4 G_F}{\sqrt{2}} \big(\overline{\nu}_\mu \gamma^\alpha P_L \mu \big) \cdot \big( \overline{e} \gamma_\alpha P_L \nu_e \big),
\end{equation}
where we have neglected subdominant contributions in inverse powers of $m_W$. 

We have matched a UV theory to an EFT by integrating out the heavy field, retained the leading contributions, and we have obtained a new operator in the EFT with a Wilson coefficient that is related to the coupling of the UV theory. The EFT can be used to study the effects of the heavy field at low energies, and to constrain the parameter space of the UV theory.

We can use an EFT to study observables at different energy scales (below the UV). Wilson coefficients are not constant but evolve with the energy scale at which they are measured according to renormalisation group equations. 

In many cases in the literature of global studies and fits of EFT coefficients, the running of Wilson coefficients is not usually not taken into account. However, with increasing level of precision in experiments, the full running of Wilson coefficients will have to be considered to map observables to EFT constraints in the future. In the global results of this thesis, the running of Wilson coefficients will not be taken into account (although we will discuss how they can shift the running of the strong coupling), but it is important to keep in mind that this can be a relevant aspect of the EFT framework.

When we are matching a UV theory to an EFT, the Wilson coefficients are found at the matching scale $\Lambda$. This choice ensures that we avoid the need to resum large logarithms that would otherwise appear at lower energies. However, if we want to use the EFT to calculate an observable at lower energies than the matching scale, we have to make them evolve using the \textit{renormalisation group} (RG) equations. They describe how the physical couplings of the theory evolve with the renormalisation scale. The RG equation for the $c_i$ Wilson coefficient can be written, in a simplified way, as
\begin{equation}
    \label{eq:rg_running}
    \frac{d c_i }{d \operatorname{log} \mu } = \sum_{j} \frac{1}{16 \pi^2} \gamma_{ij} c_j, 
\end{equation} 
where $\mu$ is the renormalisation scale, $\gamma_{ij}$ is the anomalous dimension matrix, and the index $j$ runs over all independent Wilson coefficients of the EFT at a given order in the expansion in $\Lambda$. At leading order, the solution to Eq. (\ref{eq:rg_running}) is given by  
\begin{equation}
    \label{eq:wc_lo}
    c_i (\mu) = c_i (\Lambda) - \sum_{j} \frac{1}{16 \pi^2} \gamma_{ij} c_j(\Lambda)\operatorname{log} \left(\frac{\Lambda}{\mu} \right).
\end{equation}
Eq. (\ref{eq:wc_lo}) is valid only in the approximation where one neglects the $\mu$ dependence of the right-hand side of Eq. (\ref{eq:rg_running}) (more in general, the dependence on the renormalisation scale $\mu$ can be accounted for by expanding the anomalous dimension matrix in a power series in the scale dependent couplings and solving the RGE numerically). Additionally, if the scales $\Lambda$ and $\mu$ are very different, the logarithm can become large and the perturbative expansion on the Wilson coefficients can break down. In this case, one has to resum the large logarithms. 

Computing the anomalous dimensions in the SMEFT is no easy task, but the running at one-loop has been computed \cite{Jenkins:2013zja,Jenkins:2013wua,Alonso:2013hga,Alonso:2014zka,Elias-Miro:2013mua,Elias-Miro:2013eta}, and useful tools to automate the process have been implemented, as discussed in Ref. \cite{Aebischer:2023nnv}, and other references therein.

In this thesis, when dealing with explicit UV completions, the matching to the EFT will be done at tree level, and the running of the Wilson coefficients will not be taken into account, thus following the rationale presented in Ref. \cite{murayama} and conforming to the most standard practice in the literature. However, it is important to keep in mind that the running of the Wilson coefficients is a relevant aspect of the EFT framework, and it will have to be considered in future studies.

In this chapter we have introduced the concept of EFTs, and discussed some aspects of its connection with a UV theory, its construction, and matching procedure, among other things. In the next section we will discuss how the SM can be thought as an EFT, and how we can use EFTs to study BSM physics and their interplay with the PDFs.

\subsection{The Standard Model Effective Field Theory}

As we said at the beginning of this section, the SM is a very powerful theory but we know that it cannot be the most complete description of nature. Still, its spectacular success and precision in describing what we see at energies currently within reach at colliders make it a good candidate to be thought as an EFT, as it encodes most of the effects that we see at current energies, while small deviations can also be included in the formalism. More specifically, the SM can be thought as an EFT that describes the interactions of particles at energies below a certain energy scale $\Lambda$, while potential heavy new particles lie above this scale. The \textit{Standard Model Effective Field Theory} (SMEFT) is a way to describe their potential effects at low energies. 

The SMEFT is constructed by supplementing the regular SM Lagrangian with towers of higher dimensional operators that are suppressed by powers of $\Lambda$. The coefficients of these operators are Wilson coefficients, and they encode the effects of potential heavy fields that have been integrated out in the EFT. The SMEFT can be used to study the effects of BSM physics in a fairly model-independent way by constraining the parameter space of the SMEFT using experimental measurements. With this information, we can indirectly study the effects of heavy new physics.

The SMEFT consists of the SM Lagrangian supplemented by higher dimensional operators
\begin{equation}
    \label{eq:smeft}
    \mathcal{L}_{\text{SMEFT}} = \mathcal{L}_{\text{SM}} +  \sum_{k = 5}^{\infty} \sum_{i=1}^{n_k} \frac{C_{k, i}}{\Lambda^{k-4}} \mathcal{O}_i^{(k)},
\end{equation}
where $\Lambda$ is the scale of new physics, the sum over $k$ spans all dimensions greater or equal to 5, the sum over $i$ goes over all $n_k$ independent operators of dimension $k$, $C_{n,i}$ are the Wilson coefficients, and $\mathcal{O}_{i}^{(k)}$ is a dimension-$k$ operator of type $i$ built from SM fields that respects the gauge group of $\text{SU}(3)_{C} \times \text{SU}(2)_{L} \times \text{U}(1)_{Y} $.

In this thesis we will mostly discuss only the leading SMEFT contributions in Eq. (\ref{eq:smeft}), up to powers of $\Lambda^{-2}$ with dimension-$6$ operators, and neglect the effects of higher dimensional operators as they are more heavily suppressed. We will also not consider the effects of dimension-$5$ $B-L$ violating operators as they give Majorana neutrino mass terms which are extremely small (in general, these operators are assumed to be generated only at very high energies close to the GUT scale).

There are many ways to parametrise the basis of dimension-$6$ operators of Eq. (\ref{eq:smeft}), but in this thesis we will choose to work in the so-called \textit{Warsaw basis} \cite{Buchmuller:1985jz,Grzadkowski:2010es}. This basis is complete and minimal, as only independent SMEFT operators have been retained in the expansion, and redundancies have been removed by means of equations of motion and integration by parts. The Warsaw basis contains 59 dimension-$6$ operators neglecting flavour indices, and 2499 operators in full generality.

The power of the SMEFT lies in that it allows us to study the effects at low energies of physics beyond the energy threshold of current experiments. This theoretical framework provides a bridge between potential UV-complete theories and observable quantities measured in experiments. By employing the SMEFT, we can translate the effect of potential BSM physics into measurable low energy effects, thus making it possible to test theories of new physics using experimental data. In particular, the coefficients of the SMEFT can be constrained by comparing SMEFT predictions with precise measurements from various experiments, and in this way we can indirectly study the effects of BSM physics that lies at higher energies.

The SMEFT is a powerful framework to constrain, identify, and parametrise potential deviations with respect to SM predictions, as discussed in Ref.~\cite{Brivio:2017vri} and other studies therein. It allows for the interpretation of experimental measurements in the context of BSM scenarios featuring heavy new particles while minimising assumptions on the nature of the underlying UV-complete theory. 

The concepts that we have discussed in this section have been related to BSM physics, EFTs, and the SMEFT, both from a theoretical point of view and in the context of fits to experimental measurements. Before, we discussed some aspects of the SM and, in particular, PDFs. In what follows, we will show that another crucial factor has to be taken into account when performing global fits in the context of SM and BSM physics studies: the interplay between the PDFs, and potential BSM physics.

\section{The PDF-BSM interplay}
\label{sec:intro_pdf_bsm}


We have seen how the determination of PDFs and potential new physics effects is essential to study physics in the SM and beyond. The problem is, however, that PDF and BSM/EFT/SMEFT studies do \textit{not} usually talk to each other and thus PDFs and, for example, Wilson coefficients are determined separately, neglecting possible cross-talk. In this thesis we will address the problem of the PDF-BSM interplay, and we will show that it is a critical issue in searches for BSM physics at the LHC.

Let us first provide a general scope of what happens in the PDF studies. As shown in Eq. (\ref{eq:dis_conv}), theoretical predictions of cross sections $\sigma$ are found by convolving PDFs $f(x, Q^2)$ and partonic cross sections $\hat{\sigma}$. In PDF fits, $\hat{\sigma}$ is assumed to be the SM. Then, after convolving the partonic cross section with the PDFs, we obtain a theoretical prediction $\sigma$ that is compared with the data, and the PDFs are determined by minimising a loss function that quantifies the disagreement between the predictions and the data. In this way, we obtain PDFs that can be used to make predictions for other processes at the LHC.

Now let us discuss what happen in BSM searches. The BSM models that we test at, for example, the LHC, are implemented at the level of the partonic cross section $\hat{\sigma}$, which receives contributions that are calculated with a BSM Lagrangian. Then, to generate theoretical predictions, we take the BSM $\hat{\sigma}$ partonic cross section and convolve it with PDFs that have been obtained from a previous fits. We compare these BSM predictions with the data, and we use the results to constrain, for example, Wilson coefficients $\{ c_{i}\}$ by also minimising a loss function.

In this way, PDF studies tend to neglect potential effects of BSM physics in their fits because it is assumed that the SM is the correct theory at short distances, and the predictions in BSM studies have been obtained with PDF sets from SM fits, potentially biasing the predictions. This is the problem of the \textit{PDF-BSM interplay}. As we have found in recent years, the PDF-BSM interplay is a critical issue in searches for BSM physics at the LHC, and is one of the main motivations of this thesis.

The determination of SMEFT Wilson coefficients from a fit of LHC data, like the determination of SM precision parameters from LHC data, might display a non-negligible interplay with the input set of PDFs used to compute theory predictions. This was shown in several recent studies~\cite{Carrazza:2019sec,Greljo:2021kvv,Gao:2022srd,Iranipour:2022iak,Kassabov:2023hbm,Hammou:2023heg}, in which such interplay between PDFs and SMEFT Wilson coefficients was quantified for the first time in relevant phenomenological scenarios. For example, in~\cite{Carrazza:2019sec} it was shown that, while the effect of four-fermion operators on DIS data can be non-negligible, if DIS data were fitted while taking the effect of such operators into account, the fit quality would deteriorate proportionally to the energy scale $Q^2$ of the data included in the determination. On the other hand, in the context of high-mass Drell-Yan, especially in a High Luminosity (HL)-LHC scenario, neglecting the cross-talk between the large-$x$ PDFs and the SMEFT effects in the tails could potentially miss new physics manifestations or misinterpret them~\cite{Greljo:2021kvv,Madigan:2021uho,Hammou:2023heg}, as the bounds on SMEFT operators are significantly broader if the PDFs are fitted by including the effect of higher dimensional operators in the high energy tails of the data. 

For this purpose, we have developed \simunet{} \cite{Iranipour:2022iak,Kassabov:2023hbm,Costantini:2024xae}, a new framework that allows for a simultaneous determination of PDFs and EFT Wilson coefficients. It is a complete open-source framework, fully documented, that can be used to study the PDF-BSM interplay in a global scenario. The \simunet{} methodology, built upon the NNPDF4.0 methodology \cite{NNPDF:2021njg,NNPDF:2021uiq} and first presented in Ref.~\cite{Iranipour:2022iak}, was used in the context of high-mass Drell-Yan distributions and the study of PDFs with a couple of BSM coefficients. Later, in Ref.~\cite{Kassabov:2023hbm}, we applied the \simunet{} methodology to analyse the broadest top quark dataset used to date (March 2023) in either PDF or SMEFT interpretations, which in particular contained all available measurements in the top sector from the ATLAS and CMS experiments based on the full Run II luminosity. In \cite{Costantini:2024xae}, we have extended the \simunet{} methodology to its full potential, and analysed the PDF-BSM interplay in a fully global footing.

\simunet{} also allows to study the potential absorption of BSM effects in PDF fits. This is a critical issue in the context of BSM searches, as the absorption of BSM effects by the PDFs can hide the effects of new physics. At the same time, the absorption of BSM effects in the PDFs can lead to spurious comparisons between experimental observables and theoretical predictions, showing apparent tension where there is none. 

In this chapter we have introduced the fundamental concepts that we will explore in this thesis, and the subsequent chapters will delve deeper into these topics. Chapters \ref{ch:simunet}, \ref{ch:fits}, and \ref{ch:contamination} are based on finalised publications. Chapter \ref{ch:simunet} will describe the \simunet{} methodology, a novel ML framework that enables the simultaneous fitting of PDFs and BSM physics parameters. We will discuss its specific features and code structure, aspects of its usage, and applications to contaminated fits. Chapter \ref{ch:fits} will present the results of \simunet{} applications in PDF and EFT fits in the top quark sector and in a global fit. Chapter \ref{ch:contamination} delves deeper into the potential absorption of BSM physics by the PDFs, exploring how they can shift, and how this absorption can potentially bias BSM searches. Chapters \ref{ch:smeft_dglap} and \ref{ch:sr}  are based on ongoing work. The former will discuss aspects of potential corrections of SMEFT operators to the DGLAP evolution equations of the PDFs, and the latter will explore the use of symbolic regression in collider physics in general. Finally, Chapter \ref{ch:conclusions} will summarise the findings and outline directions for future research.



 
\chapter{\simunet{}: an open-source tool for simultaneous global fits of EFT Wilson coefficients and PDFs}
\label{ch:simunet}
  \epigraph{Todo lenguaje es un alfabeto de símbolos cuyo ejercicio presupone
un pasado que los interlocutores comparten; ¿cómo transmitir a los otros el
infinito Aleph, que mi temerosa memoria apenas abarca? \\ \vspace{1em}  All language is a set of symbols whose use among its speakers assumes a shared past. How, then, can I translate into words the limitless Aleph, which my floundering mind can scarcely encompass?}{\textit{\\ from El Aleph, \\ by J. L. Borges}}

\textit{This chapter is based on Ref.~\cite{Costantini:2024xae}, and was written in collaboration with M. N. Costantini, E. Hammou, Z. Kassabov, M. Madigan, L. Mantani, J. Moore, and M. Ubiali. In this chapter we describe the public release of the} \simunet{} \textit{code, a new tool for the simultaneous determination of PDFs and BSM coefficients. Chronologically, the work presented in this chapter was carried out after the work presented in Chapter \ref{ch:fits} and Chapter \ref{ch:contamination}, and it is based on the same methodology. We choose to have this chapter here to provide a complete and consistent picture of the methodology that will be used later in the text. My contributions to this work include the new implementation of the} \simunet{} \textit{methodology in} \texttt{n3fit} \textit{(together with J. Moore), the implementation of the EFT and EFT-PDF analysis routines in the} \texttt{validphys} \textit{module (all the code but the principal component analysis). Additionally, I implemented of a substantial part of the open-source documentation to help users understand the code and its functionalities by enhancing its readability and usability, and created tutorials to perform simultaneous PDF-BSM fits.}

\vspace{1em} 
\noindent\rule{\textwidth}{0.4pt}
\vspace{1em} 

We present the open-source \simunet{} code, designed to fit SMEFT Wilson coefficient alongside the PDFs of the proton.
\simunet{} can perform SMEFT global fits, as well as simultaneous fits
of the PDFs and of an arbitrarily large number of SMEFT degrees of
freedom, by including both PDF-dependent and PDF-independent
observables. \simunet{} can also be used to determine whether the effects of any New Physics 
models can be fitted away in a global fit of PDFs. 
\simunet{} is built upon the open-source \nnpdf{} code and is
released together with documentation, and tutorials.
In the following chapters, we will use the \simunet{} methodology to
study the PDF-BSM interplay in the context of the top quark sector and in global fits.

\section{Introduction}
\label{sec:introduction}

In this chapter we present the \simunet{} methodology. It is an open-source code so that any user can assess the interplay between PDFs and New Physics, either by performing simultaneous fits of SMEFT operators (or operators in any other EFT expansion) alongside the PDFs of the proton, or by injecting any New Physics model in the data and checking whether a global fit of PDFs can absorb the effects induced by such a model in the data.

\simunet{} is based on the first tagged version of the public NNPDF code~\cite{nnpdfcode,NNPDF:2021uiq} associated with the \nnpdf{} PDF set release~\cite{NNPDF:2021njg,NNPDF:2024dpb,NNPDF:2024djq}, which it augments with a number of new features that allow the exploration of the correlation between a fit of PDFs and BSM degrees of freedom. First of all, it allows for the simultaneous determination of PDFs alongside any Wilson coefficients of an EFT that enters in the theory predictions. The user can specify any subset of operators that are of phenomenological relevance, compute the corresponding corrections to the SM predictions, and derive bounds on the operators from all observables entering the PDF fit and an arbitrary number of PDF-independent observables that can be added on top of the PDF-dependent ones. Histograms of the bounds on Wilson coefficients, correlation coefficients between PDF and Wilson coefficients, as well as Fisher information matrices and bound comparisons are automatically produced by the {\tt validphys} analysis framework, a package implemented in \texttt{reportengine}~\cite{zahari_kassabov_2019_2571601} that allows for the analysis and plotting of data related to the \simunet{} fit structures and input/output capabilities to other elements of the code base. Users can also inject the effects of any New Physics scenario in the data, and assess whether PDFs might absorb them and fit them away, in the context of a closure test based on artificial data, as carried out in Ref.~\cite{Hammou:2023heg}.

The structure of this chapter is the following. In Section~\ref{sec:intro_ml} we discuss the essential aspects of machine learning methods and, in particular, deep learning. In Section~\ref{sec:nnpdf_new} we provide a review of the \nnpdf{} methodology, upon which the \simunet{} code is built. Then, in Section~\ref{sec:code} we review the \simunet{} methodology as such, and describe in detail how it is implemented in the code. This section is especially relevant for those who wish to use the public code to perform simultaneous fits of PDFs and Wilson coefficients of an EFT. In Section~\ref{sec:contamination} we describe the use of the code in assessing whether New Physics signals can be absorbed by PDFs. Then, in Section~\ref{sec:conclusions} we summarise the main points of the chapter.

\section{Machine learning methods}
\label{sec:intro_ml}

PDFs can be fitted to experimental data using a variety of methodologies. Some of the most used and recent PDF fits include MSHT20 \cite{Bailey:2020ooq}, CT18 \cite{Hou:2019efy}, and \nnpdf{} \cite{NNPDF:2021njg}. While the first two approaches use fixed functional forms to perform the fits, the approach of \nnpdf{} uses a neural network (NN). The NN is trained to minimise a loss function that quantifies the agreement between the data and the predictions of the PDFs without overfitting the noise.

\begin{figure}[ht!]
    \centering
    \includegraphics[width=0.5\linewidth]{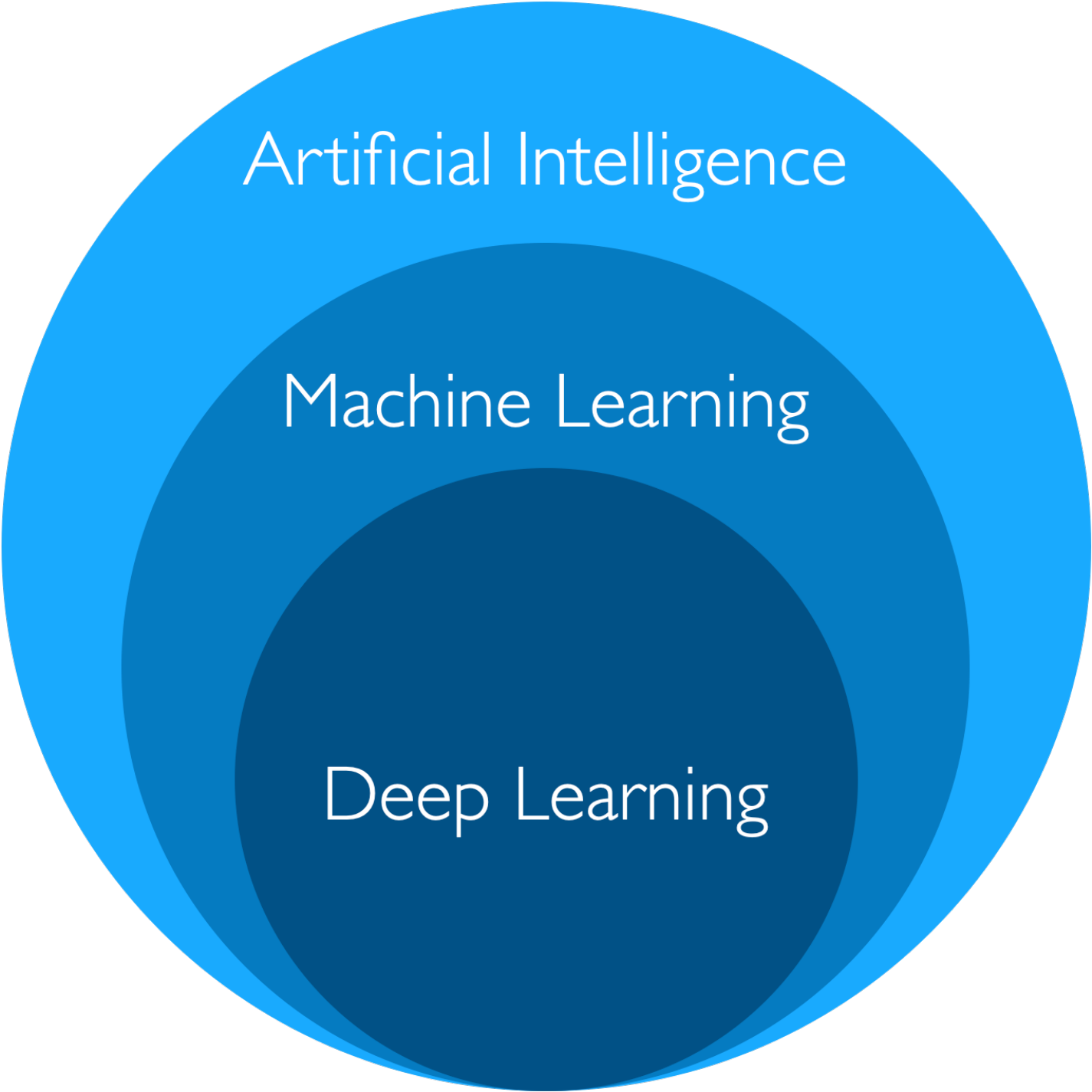}
    \caption{Venn diagram of AI, ML, and DL.}
    \label{fig:ai_ml_dl}
\end{figure}

NNs have become a powerful tool in many scientific fields, and their application in fundamental physics is not the exception. As the quest for precision and accuracy in experimental measurements and theoretical predictions advances, NNs can offer flexible tools for interpreting subtle deviations from established models in the presence of high volumes of complex data. NNs have to be understood in the context of machine learning (ML) and artificial intelligence (AI), as shown in Fig. \ref{fig:ai_ml_dl}.
There is no precise definition of AI and ML, but it is still useful to give a brief overview of these concepts. AI is a broad field of computer science focused on creating systems capable of performing tasks that normally require human intelligence. These tasks include reasoning, learning, problem-solving, perception, and language understanding.

ML is a subset of AI that involves the development of algorithms that allow computers to learn from and make predictions based on data without being explicitly programmed. Instead of being explicitly programmed to perform a task, ML algorithms use statistical techniques to improve their performance as more data becomes available.

\begin{figure}[ht!]
    \centering
    \includegraphics[width=0.8\linewidth]{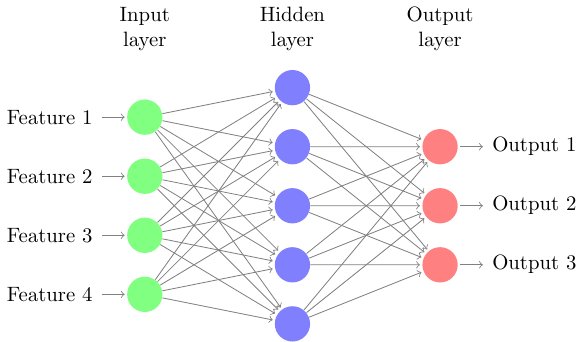}
    \caption{Neural network with a single hidden layer.}
    \label{fig:nn_tikz}
\end{figure}

DL is a further subset of ML, characterised by the use of NNs with many so-called layers (hence 'deep'). These deep NNs can model complex patterns in data and are particularly powerful for tasks such as image and speech recognition. NNs are a specific tool within DL that consist of interconnected layers of nodes (or neurons). Each connection can transmit a signal from one node to another, and the network can learn to transform input data into desired outputs through training. We show an example of a NN with a single hidden layer in Fig. \ref{fig:nn_tikz}.   
This NN consists of three layers. The first layer is the input layer, which has four neurons. Each neuron represents a different input feature of an element of the dataset. Next is the hidden layer, containing 5 neurons. These neurons are responsible for processing the input data and extracting relevant patterns or features after training which, after training, optimises the strength of the connections (also called weights) between neurons, represented by the gray arrows in the diagram. Finally, there is the output later, which consists of three neurons. These neurons produce the final output of the NN, which could be a prediction, classification, or other task depending on the specific application.

The NN processes the input features through the interconnected layers of neurons. Each connection can transmit a signal from one neuron to another. During training, the network learns to adjust the weights of these connections to transform the input data into the desired outputs by providing predictions, known as forward pass, and adjusting the weights to minimise the difference between the predicted and actual outputs, known as backpropagation. This process is repeated iteratively until the network converges to a set of weights that produce accurate predictions.

The forward pass is the process of calculating the output of a NN given an input. It involves computing the activations of each layer from the input layer to the output layer. Consider a simple neural network with $L$ layers. The input to the network is denoted as $\mathbf{x}$, and the weights and biases of the $l$-th layer are denoted as $\mathbf{W}^{(l)}$ and $\mathbf{b}^{(l)}$, respectively. For the $l$-th layer, the pre-activation $\mathbf{z}^{(l)}$ and activation $\mathbf{a}^{(l)}$ are computed as follows:
\begin{equation}
    \mathbf{z}^{(l)} = \mathbf{W}^{(l)} \mathbf{a}^{(l-1)} + \mathbf{b}^{(l)},
\end{equation}
and
\begin{equation}
    \label{eq:activation}
    \mathbf{a}^{(l)} = \sigma(\mathbf{z}^{(l)}),
\end{equation}
where $\sigma$ is the activation function (e.g., sigmoid, ReLU). The output of the network $\hat{\mathbf{y}}$ is given by:
\begin{equation*}
    \hat{\mathbf{y}} = \mathbf{a}^{(L)}.
\end{equation*}
Backpropagation is the process of calculating the gradient of the loss function with respect to each weight in the network. It involves propagating the error backward through the network. The loss function $\mathcal{L}$ measures the discrepancy between the predicted output $\hat{\mathbf{y}}$ and the true output $\mathbf{y}$. For example, for mean squared error (MSE):
\begin{equation*}
    \mathcal{L} = \frac{1}{2} \|\hat{\mathbf{y}} - \mathbf{y}\|^2.
\end{equation*}
The MSE is one of the most used loss functions for easy tasks but, in the context of global PDF or SMEFT fits, we will use more sophisticated loss functions that account for uncertainties and correlations in the data. In backpropagation, the goal is to compute the gradients $\frac{\partial \mathcal{L}}{\partial \mathbf{W}^{(l)}}$ and $\frac{\partial \mathcal{L}}{\partial \mathbf{b}^{(l)}}$ for each layer $l$. To do this, we compute the gradient of the loss $\mathcal{L}$ with respect to the output of the network and then we propagate the error backward through the layers. In this way, for the $l$-th layer we update the weights and biases using gradient descent as
\begin{equation}
\begin{aligned}
\mathbf{W}^{(l)} &\to \mathbf{W}^{(l)} - \eta \frac{\partial \mathcal{L}}{\partial \mathbf{W}^{(l)}}, \\
\mathbf{b}^{(l)} &\to \mathbf{b}^{(l)} - \eta \frac{\partial \mathcal{L}}{\partial \mathbf{b}^{(l)}},
\end{aligned}
\end{equation}
where $\eta$ is the learning rate. The learning rate controls the size of the step taken in the direction of the gradient during training. A small learning rate can lead to slow convergence, while a large learning rate can cause the network to overshoot the minimum of the loss function. Thus, the learning rate is a hyperparameter and must be tuned during training.

The forward pass and backpropagation are fundamental processes in training neural networks. The forward pass computes the output of the network, while backpropagation computes the gradients needed to update the network's parameters, minimising the loss function.

\begin{figure}[ht!]
    \centering
    \includegraphics[width=0.8\linewidth]{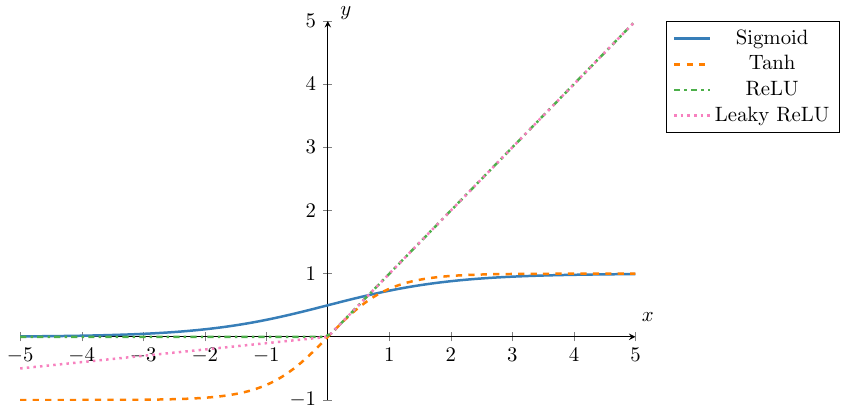}
    \caption{Some activation functions used in NN architectures.}
    \label{fig:act_func}
\end{figure}

The activation functions of Eq. (\ref{eq:activation}) are crucial for the operation of NNs. They introduce non-linearity into the network, allowing it to model complex data (otherwise the composition of linear functions remains linear). Some common activation functions used in neural networks are shown in Table \ref{tab:activation_functions}, along with their ranges.
\begin{table}[h]
    \begin{center}
    \resizebox{\textwidth}{!}{%
    \begin{tabular}{@{}c>{\centering\arraybackslash}m{0.3\textwidth}>{\centering\arraybackslash}m{0.3\textwidth}@{}}
    \toprule
    Activation Function & Formula & Range \\
    \midrule
    Sigmoid & 
    $\sigma(x) = \frac{1}{1 + e^{-x}}$ & 
    $(0, 1)$ \\
    \noalign{\smallskip} \hline \noalign{\smallskip}
    Tanh & 
    $\tanh(x) = \frac{e^x - e^{-x}}{e^x + e^{-x}}$ & 
    $(-1, 1)$ \\
    \noalign{\smallskip} \hline \noalign{\smallskip}
    ReLU & 
    $\text{ReLU}(x) = \max(0, x)$ & 
    $[0, \infty)$ \\
    \noalign{\smallskip} \hline \noalign{\smallskip}
    Leaky ReLU & 
    $\text{Leaky ReLU}(x) = \begin{cases} 
    x & \text{if } x > 0 \\ 
    \alpha x & \text{if } x \leq 0 
    \end{cases}$ & 
    $(-\infty, \infty)$ \\
    \bottomrule
    \end{tabular}
    }
    \caption{Some common activation functions used in NNs.}
    \label{tab:activation_functions}
    \end{center}
\end{table}
Figure \ref{fig:act_func} shows the graphical representation of these activation functions. The sigmoid function is commonly used in the output layer of a binary classification problem, where the output is a probability between 0 and 1. The tanh function is similar to the sigmoid function but has a range between -1 and 1. The ReLU function is the most commonly used activation function in hidden layers of NNs. It is computationally efficient and has been shown to perform well in practice. The Leaky ReLU function is a variant of the ReLU function that allows a small gradient when the input is negative, preventing the dying ReLU problem where neurons stop learning because the gradient is zero.

In ML, finding the correct amount of complexity and flexibility in the algorithm is essential to achieve good performance. This is known as the bias-variance tradeoff. The bias-variance tradeoff is a fundamental concept that affects model performance, and it involves balancing two sources of error that affect the predictive capability of ML models: bias and variance. Bias refers to the error introduced by approximating a problem by a model that is too simplistic. High bias can cause a model to miss relevant relations between features and target outputs, leading to underfitting. Variance, on the other hand, refers to the model's sensitivity to small fluctuations in the training set. If the model is too complex or flexible, high variance can cause a model to capture noise in the data, leading to overfitting.

Effective model training aims to find a balance where both bias and variance are minimised. This balance ensures that the model generalizes well to new, unseen data. In practice, techniques such as cross-validation, regularisation, and ensemble methods are employed to optimise this balance. Cross-validation helps in selecting a model that performs well on different subsets of the data (training and validation, for example), regularization adds a penalty for complexity to the loss function to reduce overfitting, and ensemble methods combine multiple models to stabilize predictions and reduce variance. 

\begin{figure}[ht!]
    \centering
    \includegraphics[width=0.8\linewidth]{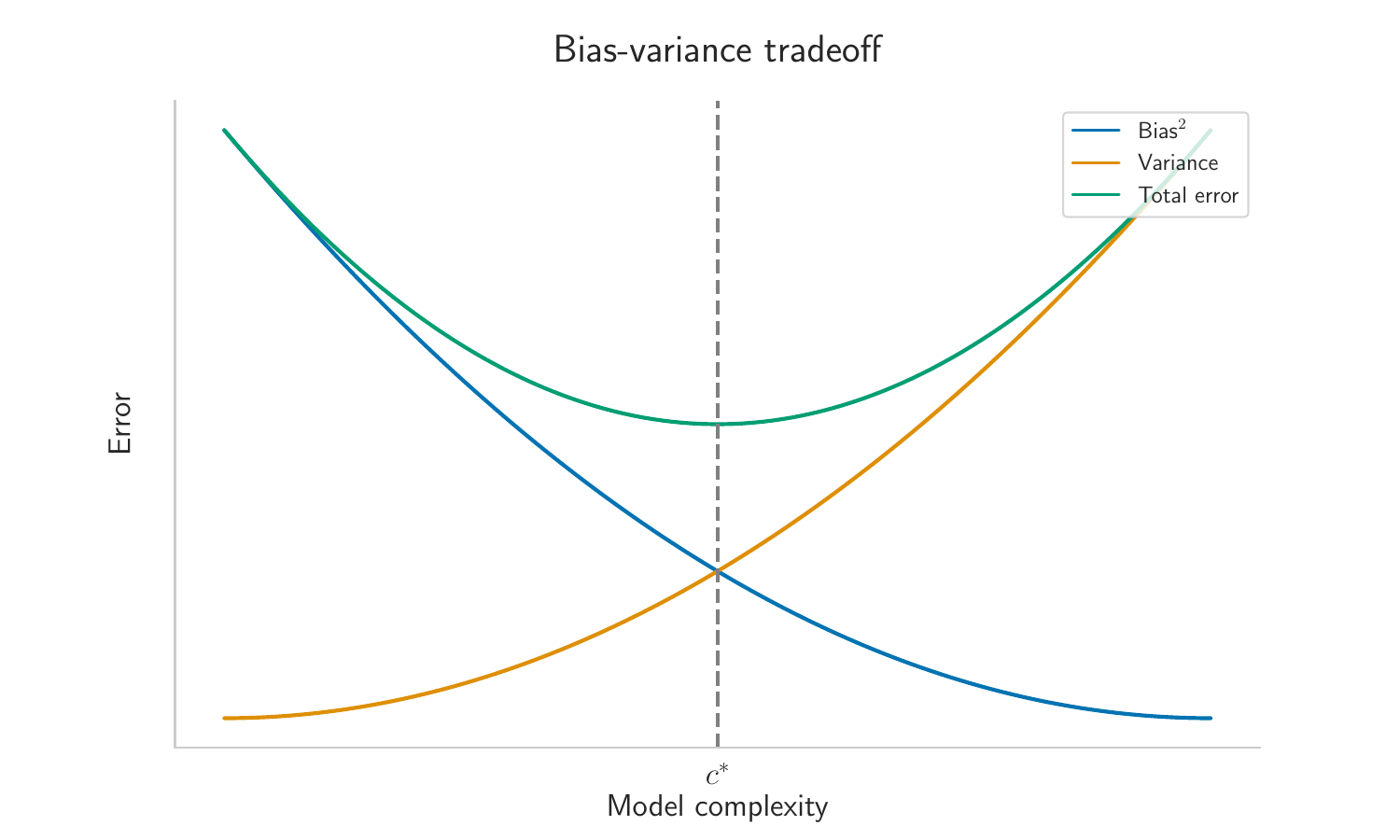}
    \caption{Schematic representation of the bias-variance tradeoff.}
    \label{fig:bias_variance_tradeoff}
\end{figure}

Figure \ref{fig:bias_variance_tradeoff} presents a schematic representation of this tradeoff, highlighting the ideal model complexity $c^*$ that minimises the total error, defines as the sum of the bias and variance errors. 
Models with low complexity can exhibit high bias and low variance, as they are unable to capture the underlying patterns in the data. On the other hand, models with high complexity can exhibit low bias and high variance, as they can capture the underlying patterns but also the noise in the data. The ideal model complexity is reached at $c^*$.

ML, and NNs in particular, provide powerful methodologies for advancing our understanding of potentially complex physical systems. As the field of ML continues to evolve, its integration with fundamental physics can provide significant advancements and insights, particularly in areas requiring comprehensive data manipulation, like collider physics. Still, the integration of NNs with physical sciences is not free of challenges. 

NNs are often considered as black boxes, meaning that the internal mechanisms of the model are not easily interpretable. This lack of interpretability can have consequences on the trustworthiness of the model and the scientific insights it provides. In the context of high energy physics the interpretability of ML models can become crucial. As physicists, we are not only interested in the performance of our frameworks, but also in the physical meaning and the causes of the results. If we continue to use ML models as black boxes in different stages of our analyses (from data acquisition, preprocessing, to theoretical predictions and inference), we risk losing the physical understanding of the phenomena we are studying and, when we claim discovery, we may not be able to provide a clear explanation of our results.

This does not mean that we should stop using deep ML models, but it is important to develop methods to interpret the results of these models in parallel. For this reason, in Chapter \ref{ch:sr} we will introduce the concept of symbolic regression, a ML algorithm that can provide accurate, easily interpretable answers to describe a dataset.

After this introduction to ML and NNs, in what follows we will focus on the application of these tools to the determination of PDFs in the context of fits to data. Now, we discuss the \nnpdf{} methodology, which uses NNs to determine the PDFs of the proton, and which serves as the basis for the \simunet{} methodology.

\section{The \nnpdf{} methodology}
\label{sec:nnpdf_new}

The central premise of the \nnpdf{} methodology~\cite{NNPDF:2021uiq,NNPDF:2021njg} is the NN parametrisation of the PDFs of the proton. The \nnpdf{} methodology assumes that one can write the eight fitted PDF flavours at the parametrisation scale $Q_0^2$, $\vec{f}(x, Q_0^2) \in \mathbb{R}^8$, in the form:\footnote{Technically, the form $f_i(x,Q_0^2) = x^{\alpha_i} (1-x)^{\beta_i} NN_{i}(x,\pmb{\omega})$ is used for each of the fitted PDF flavours, i.e. a power law scaling is used as a pre-factor; we will ignore this to streamline the presentation.  Indeed, it was recently shown in Ref.~\cite{Carrazza:2021yrg} that this pre-factor can be removed completely, and recovered using only the neural network parametrisation as presented in this text.}
\begin{equation}
    \vec{f}(x,Q_0^2) = \vec{NN}(x, \pmb{\omega}),
\end{equation}
where $\vec{NN}(\cdot,\pmb{\omega}) : [0,1] \rightarrow \mathbb{R}^8$ denotes a suitable neural network, and $\pmb{\omega}$ are the parameters of the network. Given an $N_{\text{dat}}$-dimensional dataset $\vec{D} \in \mathbb{R}^{N_{\text{dat}}}$, the corresponding theory predictions $\vec{T}(\pmb{\omega},\vec{c}) \in \mathbb{R}^{N_{\text{dat}}}$ are constructed from this neural network parametrisation via the following discretisation of the standard collinear factorisation formula:
\begin{equation}
    \label{eq:convolution_disc}
    T_i(\pmb{\omega},\vec{c}) = \begin{cases} \displaystyle \sum_{a = 1}^{N_{\text{flav}}} \sum_{\alpha=1}^{N_{\text{grid},i}}  \text{FK}_{i,a\alpha}(\vec{c}) \text{NN}_a(x_\alpha,\pmb{\omega}), & \text{if $D_i$ is DIS data;} \\[3ex] \displaystyle\sum_{a,b=1}^{N_{\text{flav}}} \sum_{\alpha,\beta=1}^{N_{\text{grid},i}} \text{FK}_{i,a\alpha b\beta}(\vec{c}) \text{NN}_a(x_\alpha,\pmb{\omega}) \text{NN}_b(x_\beta,\pmb{\omega}), & \text{if $D_i$ is hadronic data.}\end{cases}
\end{equation}
Here, $\text{FK}_{i,a\alpha}(\vec{c})$ (or $\text{FK}_{i,a\alpha b \beta}(\vec{c})$) is called the \textit{fast-kernel (FK) table} for the $i$th datapoint, which encompasses both the partonic cross section for the process associated to the $i$th datapoint and the coupled evolution of the PDFs from the parametrisation scale $Q_0^2$ to the scale $Q^2_i$ associated to the datapoint $i$, and $(x_1,...,x_{N_{\text{grid},i}})$ is a discrete $x$-grid for the $i$th datapoint. Note importantly that the theory predictions carry a dependence on the parameters of the neural network, $\pmb{\omega}$, but additionally on a vector of physical constants $\vec{c} \in \mathbb{R}^{N_{\text{param}}}$ through the FK-tables, such as the strong coupling $\alpha_s(M_z)$, the electroweak parameters in a given electroweak input scheme, the CKM matrix elements, or the masses of the heavy quarks, which are fixed to some reference values for any given \nnpdf{} fit. A choice of constants $\vec{c}$ determines a corresponding set of FK-tables, which is referred to as a `theory' in the \nnpdf{} parlance. Corrections to the FK-tables by means of \textit{cfactors} can account for theoretical uncertainties and/or higher order QCD and EW corrections (when available) in the theory predictions. In \nnpdf{}, FK-tables are generated only with SM contributions.

A PDF fit in the \nnpdf{} framework comprises the determination of the neural network parameters $\pmb{\omega}$, given a choice of theory $\vec{c} = \vec{c}^*$, together with an uncertainty estimate for these parameters. This is achieved via the \textit{Monte Carlo replica method}, described as follows. Suppose that we are given experimental data $\vec{D}$, together with a covariance matrix $\Sigma$. We generate $N_{\text{rep}}$ `pseudodata' vectors, $\vec{D}_1, ..., \vec{D}_{N_{\text{rep}}}$, as samples from the multivariate normal distribution:
\begin{equation}
	\vec{D}_i \sim \mathcal{N}(\vec{D}, \Sigma).
\end{equation}
The default choice for the covariance matrix $\Sigma$ is $\Sigma = \Sigma_{\rm exp}$, i.e. the experimental covariance matrix, which includes all information on experimental uncertainties and correlations. The latter can also be augmented by a theory covariance matrix $\Sigma_{\rm th}$, hence setting $\Sigma = \Sigma_{\rm exp} + \Sigma_{\rm th}$, where $\Sigma_{\rm th}$ includes the effects of nuclear correction uncertainties~\cite{Ball:2018twp,Ball:2020xqw} and missing higher order uncertainties~\cite{AbdulKhalek:2019bux,AbdulKhalek:2019ihb,NNPDF:2024dpb} in the theory predictions used in a PDF fit. An alternative approach to keep account of theory uncertainties in the fit is to expand the Monte Carlo sampling to the space of factorisation and renormalisation scale parameters; this approach was presented in Ref.~\cite{Kassabov:2022orn}. Both options can be implemented in the \simunet{} framework, the former in a straightforward way, the latter with some modifications that we leave to future releases.  

For each pseudodata $\vec{D}_k$, we find the corresponding best-fit PDF parameters $\pmb{\omega}_k$ by minimising the $\chi_k^2$ loss function, defined as a function of $\pmb{\omega}$ as
\begin{equation}
    \label{eq:montecarlo_chi2}
    \chi^2_k(\pmb{\omega}, \vec{c}^*) = (\vec{D}_k - \vec{T}(\pmb{\omega}, \vec{c}^*))^T \Sigma_{t_0} (\vec{D}_k - \vec{T}(\pmb{\omega}, \vec{c}^*)) \, ,
\end{equation}
in which the $t_0$ covariance matrix (rather than the experimental covariance matrix $\Sigma_{\rm exp}$) is used. The use of the $t_0$ covariance matrix is designed to avoid the so-called D'Agostini bias~\cite{DAgostini:1993arp} and ensure a faithful propagation of multiplicative uncertainties, as described in Ref.~\cite{Ball:2009qv}. The latter is defined as
\begin{equation}
  \label{eq:covt0}
  \left(\Sigma_{t_0 }\right)_{i j}=\delta_{i j}\sigma_i^{\text {(uncorr) }}\sigma_j^{\text {(uncorr) }}+\sum_{m=1}^{N_{\rm norm}} \sigma_{i, m}^{(\rm norm)} \sigma_{j, m}^{(\rm norm)}T_i^{(0)} T_j^{(0)}+\sum_{l=1}^{N_{\rm corr}} \sigma_{i, l}^{(\rm corr)} \sigma_{j, l}^{(\rm corr)}D_i D_j,
\end{equation}
where $T^{(0)}_i$  is a theoretical prediction for the $i$-th data point evaluated using a $t_0$ input PDF. This leads to a set of PDFs that is then used to iteratively compute the theoretical predictions $T_i^{(0)}$ that are needed for the construction of a new $t_0$ covariance matrix, which is then used for a new PDF determination. This procedure is iterated until convergence. The minimisation is achieved using \textit{stochastic gradient descent}, which can be applied here because the analytic dependence of $\chi^2_k(\pmb{\omega}, \vec{c}^*)$ on $\pmb{\omega}$ is known, since the neural network parametrisation is constructed from basic analytic functions as building blocks. Furthermore, the $\chi^2$-loss is additionally supplemented by \textit{positivity} and \textit{integrability} penalty terms; these ensure the positivity of observables and the integrability of the PDFs in the small-$x$ region.

Given the best-fit parameters $\pmb{\omega}_1, ..., \pmb{\omega}_{N_{\text{rep}}}$ to each of the pseudodata, we now have an ensemble of neural networks which determine an ensemble of PDFs, $\vec{f}_1, ..., \vec{f}_{N_{\text{rep}}}$, from which statistical estimators can be evaluated, such as the \textit{mean} or \textit{variance} of the initial-scale PDFs. 

The methodology sketched above is at the basis of the \nnpdfcode{} public code~\cite{nnpdfcode}. The code comprises several packages. To transform the original measurements provided by the experimental collaborations, e.g. via {\sc\small HepData}~\cite{Maguire:2017ypu}, into a standard format that is tailored for fitting, the \nnpdfcode{} code uses the {\tt buildmaster} {\tt C++} experimental data formatter. Physical observables are evaluated as a tensor sum as in Eq.~\eqref{eq:convolution_disc} via the {\tt APFELcomb}~\cite{Bertone:2016lga} package that takes hard-scattering partonic matrix element interpolators from {\tt APPLgrid}~\cite{Carli:2010rw} and {\tt FastNLO}~\cite{Wobisch:2011ij} (for hadronic processes) and \apfel~\cite{Bertone:2013vaa} (for DIS structure functions) and combines them with the QCD evolution kernels that evolve the initial-scale PDFs. The package also handles NNLO QCD and/or NLO electroweak $K$-factors when needed. The actual fitting code is implemented in the {\tt TensorFlow} framework~\cite{abadi2016tensorflow} via the \texttt{n3fit} library. The latter allows for a flexible specification of the neural network model adopted to parametrise the PDFs, whose settings can be selected automatically via the built-in hyperoptimisation tooling ~\cite{2015CS&D....8a4008B}, such as the neural network architecture, the activation functions, and the initialisation strategy; the choice of optimiser and of the hyperparameters related to the implementation in the fit of theoretical constraints such as PDF positivity~\cite{Candido:2020yat} and integrability. Finally the code comprises the {\tt validphys} analysis framework, which analyses and plots data related to the NNPDF fit structures, and governs input/output capabilities of other elements of the code base.\footnote{The full \nnpdfcode code documentation is provided at {\tt \url{https://docs.nnpdf.science}}.}

An important feature of the \nnpdf{} methodology is the \textit{closure test}, first described in Ref.~\cite{NNPDF:2014otw} and described in much more detail in Ref.~\cite{DelDebbio:2021whr}. Closure tests are regularly deployed by the NNPDF collaboration to ensure that their methodology is working reliably.

The NNPDF closure test methodology begins by supposing that we are given Nature's true PDFs at the initial scale $\vec{f}_{\text{true}}(x,Q_0^2) = \vec{NN}(x, \pmb{\omega}_{\text{true}})$,\footnote{Assuming that the PDFs can be parametrised by a neural network, and then subsequently using a neural network to perform the fit, ignores modelling error; we do not consider this here for simplicity, but both the \nnpdf{} and \simunet{} frameworks can account for it.} and true physical parameters $\vec{c}_{\text{true}}$. Under this assumption, experimental data $\vec{D}$ should appear to be a sample from the multivariate normal distribution:
\begin{equation}
    \vec{D} \sim \mathcal{N}(\vec{T}(\pmb{\omega}_{\text{true}}, \vec{c}_{\text{true}}), \Sigma_{\text{exp}}),
\end{equation}
where $\Sigma_{\text{exp}}$ is the experimental covariance matrix, and $\vec{T}$ is the theory prediction from the full discrete convolution formula Eq.~\eqref{eq:convolution_disc} (\textit{not} the \simunet{} prediction Eq.~\eqref{eq:simunet_predictions}). A sample from this distribution is called a set of \textit{level 1 pseudodata} in the NNPDF closure test language. Given a set of level 1 pseudodata, we can attempt to recover the theory of Nature using the \nnpdf{} methodology. Instead of fitting PDFs to experimental data, we can perform the fit replacing the experimental data with the level 1 pseudodata. If the resulting PDF fit has a spread covering the true PDF law $\vec{f}_{\text{true}}$, we say that the methodology has passed this `closure test'.\footnote{In practice, an ensemble of fits is produced, and if their 68\% confidence bands cover the true law in 68\% of the fits, we say that the methodology has passed the closure test. Note that using 68\% is only a prescription; however, it is valuable as it provides a stricter constraint on the reliability of the resulting PDF fit when compared to the broader 95\% confidence interval.}

Importantly, closure tests are typically performed \textit{only} fitting the true PDF law; in particular, the closure test usually assumes that the true physical parameters $\vec{c}_{\text{true}}$ are known exactly and perfectly, that is, theory predictions for the fit take the form $\vec{T}(\pmb{\omega}, \vec{c}_{\text{true}})$ where the PDF parameters $\pmb{\omega}$ are variable and fitted to the pseudodata, but the physical parameters $\vec{c}_{\text{true}}$ are fixed. With \simunet{}, more options become available, as described below.

\section{\simunet{}: methodology for simultaneous fits and code structure}
\label{sec:code}

In this section, we provide a broad overview of the general methodology and code structure of \simunet{}, in particular describing its functionality regarding simultaneous PDF-SMEFT fits. In Sect.~\ref{sec:simunet} we review the specifics of the \simunet{} methodology and its code implementation. In Sect.~\ref{sec:codedes} we give some details of usage (in particular the user-facing \texttt{yaml} runcard structure), but the user/reader is encouraged to check the supporting webpage, \web{}, for a more comprehensive description. 

\subsection{The \simunet{} framework}
\label{sec:simunet}

An important aspect of the \nnpdf{} methodology, as well as of most global PDF analyses, is that the physical parameters $\vec{c}$ must be chosen and fixed before a PDF fit, so that the resulting PDFs are produced \textit{under the assumption} of a given theory $\vec{c} = \vec{c}^*$. It is desirable (and in fact necessary in some scenarios; see for example Sect. 5.3 of Ref.~\cite{Greljo:2021kvv}) to relax this requirement, and instead to fit both the PDFs \textit{and} the physical parameters $\vec{c}$ \textit{simultaneously}.

On regarding Eq.~\eqref{eq:montecarlo_chi2}, one may assume that this problem can immediately  be solved by applying stochastic gradient descent not only to the parameters of the neural network $\pmb{\omega}$, but to the tuple $(\pmb{\omega}, \vec{c})$, without first fixing a reference value $\vec{c} = \vec{c}^*$. Unfortunately, this is not a solution in practice; the FK-tables typically have an extremely complex, non-analytical, dependence on the physical parameters $\vec{c}$, and evaluation of a complete collection of FK-tables at even one point $\vec{c} = \vec{c}^*$ requires hundreds of computational hours and an extensive suite of codes. Hence, since the dependence of the FK-tables as an analytic function of $\vec{c}$ is unavailable, it follows that $\vec{T}(\pmb{\omega},\vec{c})$ is not a known analytic function of $\vec{c}$, and stochastic gradient descent \textit{cannot} be applied.

Thus, if we would like to continue with our programme of simultaneous extraction of PDFs and physical parameters, we will need to approximate. The central conceit of the \simunet{} methodology is the observation that for many phenomenologically-interesting parameters, particularly the Wilson coefficients of the SMEFT expansion, the dependence of the theory predictions $\vec{T}(\pmb{\omega}, \vec{c})$ on the physical parameters $\vec{c}$ can be well-approximated using the linear ansatz:\footnote{Note that in Sect.~5.3 of Ref.~\cite{Iranipour:2022iak}, it was initially proposed that \simunet{} would operate in \textit{non-linear} regimes too. However, due to the findings regarding uncertainty propagation using the Monte Carlo replica method presented in App. E of Ref.~\cite{Kassabov:2023hbm} and formalised in~\cite{Costantini:2024wby} the non-linear feature has been deprecated in this public release of the code.}
\begin{equation}
  \label{eq:simunet_predictions}
  \vec{T}(\pmb{\omega}, \vec{c}) \approx \vec{T}(\pmb{\omega}, \vec{c}^*) \odot \left[ \vec{1} + \vec{K}_{\text{fac}}(\pmb{\omega}^*) (\vec{c} - \vec{c}^*) \right],
\end{equation}
where $\odot$ denotes the element-wise Hadamard product of vectors,\footnote{Recall that if $\vec{v} = \vec{w} \odot \vec{z}$, the elements of $\vec{v}$ are defined by $v_i = w_i z_i$.} $\vec{1} \in \mathbb{R}^{N_{\text{dat}}}$ is a vector of ones, $\vec{c}^*$ is some reference value of the physical parameters, and $\vec{K}_{\text{fac}}(\pmb{\omega}^*) \in \mathbb{R}^{N_{\text{dat}} \times N_{\text{param}}}$ is a matrix of pre-computed `$K$-factors':
\begin{equation}
  K_{\text{fac}}(\pmb{\omega}^*)_{ij} = \frac{T_i(\pmb{\omega}^*, c_1^*, ..., c_{j-1}^*,c_j^* + 1,c_{j+1}^*, ..., c_{N_{\text{param}}}^*) - T_i(\pmb{\omega}^*,\vec{c}^*)}{T_i(\pmb{\omega}^*, \vec{c}^*)},
\end{equation}
which are designed to approximate the appropriate (normalised) gradients:
\begin{equation}
  \label{eq:gradient_approximation}
  K_{\text{fac}}(\pmb{\omega}^*)_{ij} \approx \frac{1}{T_i(\pmb{\omega}^*, \vec{c}^*)} \frac{\partial T_i}{\partial c_j}(\pmb{\omega}^*, \vec{c}^*).
\end{equation}
Observe that the $K$-factors are determined with a fixed choice of reference PDF, $\pmb{\omega} = \pmb{\omega}^*$, where they should technically depend on PDFs freely. In practice however, this approximation is often justified, and the reliability of the approximation can always be checked post-fit; see Appendix C of Ref.~\cite{Greljo:2021kvv} for an example of this validation.

Note that the dependence of the theory in Eq.~\eqref{eq:simunet_predictions} on $\pmb{\omega}$ \textit{and} $\vec{c}$ is now known as an analytic function. The \simunet{} methodology, at its heart, now simply extends the \nnpdf{} methodology by replacing the theory predictions in Eq.~\eqref{eq:montecarlo_chi2} with those specified by Eq.~\eqref{eq:simunet_predictions}, and then running stochastic gradient descent on a series of Monte Carlo pseudodata as in the \nnpdf{} framework. The result of the fit is a series of best-fit tuples $(\pmb{\omega}_1, \vec{c}_1)$, ..., $(\pmb{\omega}_{N_{\text{rep}}}, \vec{c}_{N_{\text{rep}}})$ to the various pseudodata, from which statistical estimators can be calculated as above.

\subsection{Structure and usage of the \simunet{} code}
\label{sec:codedes} 

\begin{figure}[t]
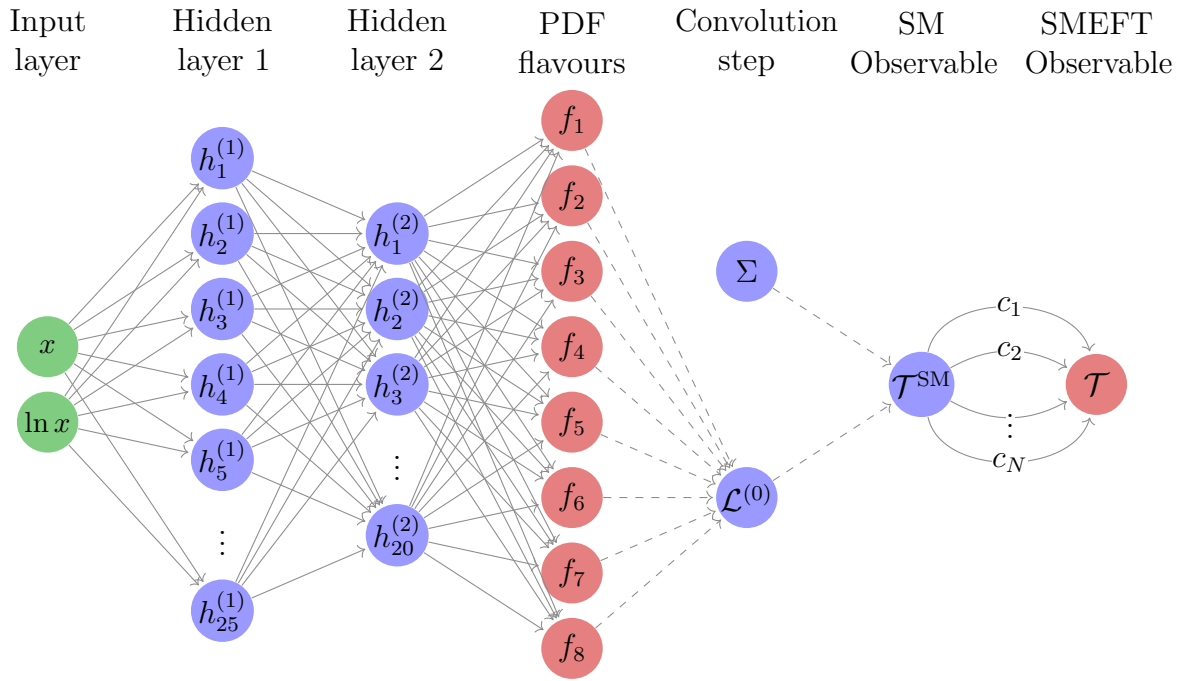

  \centering
  \begin{neuralnetwork}[height=8, layerspacing=23mm, nodesize=23pt]
    \newcommand{\x}[2]{\IfEqCase{#2}{{1}{\small $x$}{2}{\small $\ln{x}$}}}
    \newcommand{\yy}[2]{$f_#2$}
    \newcommand{\basis}[2]{\ifnum #2=4 \(\mathcal{L}^{(0)}\) \else $\Sigma$ \fi}
    \newcommand{\hfirst}[2]{\ifnum #2=7 $h^{(1)}_{25}$ \else $h^{(1)}_#2$ \fi}
    \newcommand{\hsecond}[2]{\ifnum #2=5 $h^{(2)}_{20}$ \else $h^{(2)}_#2$ \fi}
    \newcommand{\standardmodel}[2] {$\mathcal{T}^\text{SM}$}
    \newcommand{\eft}[2] {$\mathcal{T}$}
    \newcommand{\co}[4]{$c_1$}
    \newcommand{\ct}[4]{$c_2$}
    \newcommand{\cnn}[4]{$c_{N-1}$}
    \newcommand{\cn}[4]{$c_{N}$}
    \newcommand{\vd}[4]{$\vdots$}
    \newcommand{\FK}[2]{$\sigma$}
    \newcommand{\convolve}[4]{$\otimes$}
    \newcommand{\ci}[4]{$c_1$}
    \newcommand{\cj}[4]{$c_2$}
    \newcommand{\ck}[4]{$c_N$}
    \inputlayer[count=2, bias=false, title=Input\\layer, text=\x]
    \hiddenlayer[count=7, bias=false, title=Hidden\\layer 1, text=\hfirst, exclude={6}]\linklayers[not to={6}]
    \hiddenlayer[count=5, bias=false, title=Hidden\\layer 2, text=\hsecond, exclude={4}]\linklayers[not from={6}, not to={4}]
    \outputlayer[count=8, title=PDF\\flavours, text=\yy] \linklayers[not from={4}]
    \hiddenlayer[count=4, bias=false, text=\basis, title=Convolution\\step, exclude={2,3}]\linklayers[not to={1,2,3}, style={dashed}]
    \hiddenlayer[count=1, bias=false, title=SM\\Observable, text=\standardmodel]\linklayers[not from={2,3}, style={dashed}]
    \outputlayer[count=1, bias=false, text=\eft, title=SMEFT \\ Observable]
    \link[from layer = 5, to layer = 6, from node = 1, to node = 1, style={bend left=79}, label={\ci}]
    \link[from layer = 5, to layer = 6, from node = 1, to node = 1, style={bend left=30}, label={\cj}]
    \link[from layer = 5, to layer = 6, from node = 1, to node = 1, style={bend right=30}, label={\vd}]
    \link[from layer = 5, to layer = 6, from node = 1, to node = 1, style={bend right=79}, label={\ck}]
    \path (L1-5) -- node{$\vdots$} (L1-7);
    \path (L2-3) -- node{$\vdots$} (L2-5);
  \end{neuralnetwork}
\caption{\label{fig:simunet_network} A schematic representation of the neural network parametrisation of $\vec{T}(\pmb{\omega}, \vec{c})$ used by the \simunet{} code. The usual NNPDF network is represented by the layers from the `Input layer' to the `SM Observable' layer. The final `SMEFT observable' layer is related to the `SM observable' layer by edges which carry the theory parameters $\vec{c}$ as weights.}
\end{figure}

The \simunet{} code is a fork of the \nnpdf{} public code~\cite{NNPDF:2021uiq}, where the key modification is the replacement of the standard theory predictions used in the $\chi^2$-loss, Eq.~\eqref{eq:montecarlo_chi2}, by \nnpdf{} with the approximate formula for the theory predictions given in Eq.~\eqref{eq:simunet_predictions}. This replacement is effected by the inclusion of a new post-observable \textit{combination layer} in the network, specified by the \texttt{CombineCfac.py} layer added to the \texttt{n3fit/layers} directory; see Fig.~\ref{fig:simunet_network} for a schematic representation.
Beyond the inclusion of this layer, the main modification to the \nnpdf{} code comprises reading of the $K$-factors required in the approximation shown in Eq.~\eqref{eq:simunet_predictions}. The $K$-factors are implemented in a new file format, the \texttt{simu\_fac} format. One \texttt{simu\_fac} file is made available for each dataset. The \texttt{simu\_fac} files are packaged inside the relevant \nnpdf{} theory folders, inside the \texttt{simu\_factors} directory; an example of the correct structure is given inside \texttt{theory\_270}, which is a theory available as a separate download to the \simunet{} release. As an example of the files, consider the file \texttt{SIMU\_ATLAS\_CMS\_WHEL\_8TEV.yaml} from the directory \texttt{theory\_270/simu\_factors}:
\begin{python}
metadata:
  ref: arXiv:2005.03799
  author: Luca Mantani
  ufomodel: SMEFTatNLO
  flavour: dim6top
  date: 15/09/2023
  pdf: None
  info_SM_fixed: SMEFiT
  observable: F0, FL
  bins: []

SM_fixed: [0.6896346, 0.3121927]

EFT_LO:
  SM: [0.697539, 0.30189]
  OtW: [-0.0619917432825, 0.0620585212469]
  OtG: [0.0, 0.0]

EFT_NLO:
  SM: [0.689466, 0.308842]
  OtW: [-0.0609861823272, 0.061150870174]
  OtG: [0.00064474891454, -0.000673970936802]
\end{python}
The structure of the \texttt{simu\_fac} file is as follows:
\begin{itemize}
  \item There is a \texttt{metadata} namespace at the beginning of the file, containing information on the file. The metadata is never read by the code, and is only included as a convenience to the user. 
  \item The \texttt{SM\_fixed} namespace provides the best available SM predictions; these may be calculated at next-to-leading order or next-to-next-to-leading order in QCD, including or excluding EW corrections, depending on the process.
  \item The remaining parts of the file contain the information used to construct $K$-factors for different models. In this file, two models are included: the SMEFT at leading order in QCD, and the SMEFT at next-to-leading order in QCD.
\end{itemize}
In order to perform a simultaneous fit of PDFs and the parameters specified in a particular model of a \texttt{simu\_fac} file, we must create a runcard.  The runcard follows the standard \nnpdf{} format, details and examples of which are given on the website \web{}.  We provide here a discussion of the modifications necessary for a \simunet{} fit.

First, we must modify the \texttt{dataset\_inputs} configuration key. Here is an example of an appropriate \texttt{dataset\_inputs} for a simultaneous fit of PDFs and SMEFT Wilson coefficients at next-to-leading order in QCD:
\begin{python}
dataset_inputs:
- {dataset: NMC, frac: 0.75}
- {dataset: ATLASTTBARTOT7TEV, cfac: [QCD], simu_fac: "EFT_NLO"}
- {dataset: CMS_SINGLETOPW_8TEV_TOTAL, simu_fac: "EFT_NLO", use_fixed_predictions: True}
\end{python}
Here, we are fitting three datasets: the fixed-target DIS data from the New Muon Collaboration~\cite{NewMuon:1996fwh,NewMuon:1996uwk} (\texttt{NMC}), the total $t\bar{t}$ cross section measured by ATLAS at $\sqrt{s}=7$~TeV~\cite{ATLAS:2014nxi} (\texttt{ATLASTTBARTOT7TEV}) and the total associated single top and $W$ boson cross section measure by CMS at $\sqrt{s}=8$~TeV~\cite{CMS:2014fut} (\texttt{CMS\_SINGLETOPW\_8TEV\_TOTAL}), which we discuss in turn:
\begin{enumerate}
\item First, note that the dataset \texttt{NMC} is entered exactly as it would appear in an \nnpdf{} runcard; 
it is therefore treated by \simunet{} as a standard dataset for which there is no $K$-factor modification, 
i.e. it depends purely on the PDFs and no additional physics parameters. 
\item On the other hand, the dataset \texttt{ATLASTTBARTOT7TEV} has an additional key included, namely \texttt{simu\_fac}, 
set to the value \texttt{EFT\_NLO}; this tells \simunet{} to treat theory predictions for this dataset using the approximation 
Eq.~\eqref{eq:simunet_predictions}, with the $K$-factors constructed from the \texttt{EFT\_NLO} model specified in the relevant 
\texttt{simu\_fac} file. Precisely \textit{which} parameters from the model are fitted is determined by the \texttt{simu\_parameters} 
configuration key, which we describe below.
\item Finally, observe that the dataset \texttt{CMS\_SINGLETOPW\_8TEV\_TOTAL} includes two new keys: \texttt{simu\_fac} and 
\texttt{use\_fixed\_predictions}. The first key, \texttt{simu\_fac} has exactly the same interpretation as with the previous dataset; 
on the other hand, the fact that the key \texttt{use\_fixed\_predictions} is set to \texttt{True} instead removes the PDF-dependence of this dataset. 
That is, the predictions for this dataset are simplified even from Eq.~\eqref{eq:simunet_predictions}, to:
\begin{equation}
\label{eq:simunet_predictions_fixed}
\vec{T}(\pmb{\omega}, \vec{c}) \approx \vec{T}_{\text{fixed}} \odot \left[ \vec{1} + \vec{K}_{\text{fac}}(\pmb{\omega}^*) (\vec{c} - \vec{c}^*) \right],
\end{equation}
where $\vec{T}_{\text{fixed}}$ is the vector of predictions taken from the \texttt{SM\_fixed} namespace of the \texttt{simu\_fac} file. Note that the right 
hand side no longer has a dependence on $\pmb{\omega}$, so this dataset effectively becomes PDF-independent. This is appropriate for observables which do not 
depend on the PDFs (e.g. the electroweak coupling, or $W$-helicities in top decays), but also observables which depend only weakly on the PDFs and for which 
the full computation of FK-tables would be computationally expensive.
\end{enumerate}

Outside of the \texttt{dataset\_inputs} configuration key, we must also include a new key, called \texttt{simu\_parameters}. This 
specifies \textit{which} of the parameters in the model read from the \texttt{simu\_fac} files will be fitted, and additionally specifies hyperparameters 
relevant to their fit. An example is:
\begin{python}
simu_parameters:
- {name: 'OtG', scale: 0.01, initialisation: {type: uniform, minval: -1, maxval: 1}}
- {name: 'Opt', scale: 0.1, initialisation: {type: gaussian, mean: 0, std_dev: 1}}
\end{python}
This specification tells \simunet{} that for each of the datasets which have set a model value for \texttt{simu\_fac}, the relevant model in the \texttt{simu\_fac} file should be checked for the existence of each of the parameters, and their contribution should be included if they appear in the model in the file. Observe the following:

\begin{enumerate}
\item The \texttt{scale} can be used to modify the learning rate in the direction of each of the specified parameters. In detail, suppose that the 
scale $\lambda$ is chosen for the parameter $c$. Then, \simunet{} multiplies the relevant $K$-factor contribution by $1/\lambda$ so that we are effectively 
fitting the parameter $\lambda c$ instead of $c$ itself. When $K$-factor contributions from parameters are particularly large, setting a large scale can improve 
training of the network, avoiding exploring parts of the parameter space which have extremely poor $\chi^2$s; see Ref.~\cite{Iranipour:2022iak} for 
further discussion. On the other hand, setting a small scale can speed up training of the network by increasing the effective step size in a particular 
direction in parameter space; the user must tune these scales by hand to obtain optimal results. An automatic scale choice feature may be included in a future release.
\item The \texttt{initialisation} key informs \simunet{} how to initialise the parameters when training commences; this initialisation is random and there 
are three basic types available. If \texttt{uniform} is selected, the initial value of the parameter is drawn from a random uniform distribution between the 
keys \texttt{minval} and \texttt{maxval}, which must be additionally specified. If \texttt{gaussian} is selected, the initial value of the parameter is drawn 
from a random Gaussian distribution with mean and standard deviation given by the keys \texttt{mean} and \texttt{std\_dev} respectively. Finally, 
if \texttt{constant} is chosen, the key \texttt{value} is supplied and the initial value of the parameter is always set to this key (no random selection 
is made in this instance).
\end{enumerate}

It is also possible to specify \textit{linear combinations} of the model parameters to fit; this feature is useful because the supplied \texttt{simu\_fac} 
files only contain SMEFT models with Wilson coefficients in the \textit{Warsaw basis}~\cite{Grzadkowski:2010es}. For example, the user may want to fit in a 
different basis to the 
Warsaw basis, or may wish to fit a UV model which has been matched to the SMEFT at dimension $6$, with parameters given by combinations of the SMEFT 
parameters. An example is the following:
\begin{python}
simu_parameters:
  - name: 'W'
    linear_combination: 
      'Olq3': -15.94
    scale: 1000
    initialisation: {type: uniform, minval: -1, maxval: 1}

  - name: 'Y'
    linear_combination: 
      'Olq1': 1.51606
      'Oed': -6.0606
      'Oeu': 12.1394
      'Olu': 6.0606
      'Old': -3.0394
      'Oqe': 3.0394
    scale: 1000
    initialisation: {type: uniform, minval: -1, maxval: 1}
\end{python}
using the $\hat{W}$ and $\hat{Y}$ parameters studied in Ref.~\cite{Greljo:2021kvv}. Here, we fit the user-defined operators:
\begin{align*}
  \hat{W} &= -15.94 \;\mathcal{O}_{lq}^{(3)}, \\
  \hat{Y} &= 1.51606 \; \mathcal{O}_{lq}^{(1)} - 6.0606 \; \mathcal{O}_{ed} + 12.1394 \; \mathcal{O}_{eu} + 6.0606 \; \mathcal{O}_{lu} - 3.0394 \; \mathcal{O}_{ld} + 3.0394 \; \mathcal{O}_{qe},
\end{align*}
which are specified as linear combinations of SMEFT operators, the contributions of which are supplied in the \texttt{simu\_fac} files.

Once a runcard is prepared, the user simply follows the standard \nnpdf{} pipeline to perform a fit. In particular, they should run \texttt{vp-setupfit}, 
\texttt{n3fit}, \evolventhreefit{} and \texttt{postfit} in sequence in order to obtain results. Analysis of the results can be achieved with a range of tools supplied with the release; further details are given below and on the website: \web{}.

\paragraph{The fixed PDF feature of \simunet{}.} The \simunet{} release also provides functionality for fits of the physical parameters $\vec{c}$ alone, with the PDFs kept fixed (similar to tools such as \smefit{} or \fitmaker{} in the case of the SMEFT Wilson coefficients). This is achieved simply by loading the weights of a previous neural network PDF fit~\footnote{See the documentation in \web{} for a list of available fits to preload.}, then keeping these weights fixed as the remaining parameters are fitted. To specify this at the level of a \simunet{} runcard, use the syntax:
\begin{python}
  fixed_pdf_fit: True
  load_weights_from_fit: 221103-jmm-no_top_1000_iterated
\end{python}
for example. Setting \texttt{fixed\_pdf\_fit} to \texttt{True} instructs \simunet{} to perform the fit keeping the weights of the PDF part of the network constant, and the namespace \texttt{load\_weights\_from\_fit} tells \simunet{} which previous PDF fit to obtain the frozen weights from. The pipeline for a fit then proceeds exactly as in the case of a normal \simunet{} fit, beginning by running \texttt{vp-setupfit} and then \texttt{n3fit}. \textit{However}, the user need not use \evolventhreefit{} at the subsequent stage, since the PDF grid is already stored and does not need to be recomputed; it does still need to be copied into the correct fit result directory though, which should be achieved using the \texttt{vp-fakeevolve} script via:
\begin{python}
vp-fakeevolve fixed_simunet_fit num_reps
\end{python}
where \texttt{fixed\_simunet\_fit} is the name of the simultaneous fit that the user has just run, and \texttt{num\_reps} is the number of replicas in the fit. The user \textit{must} still run \texttt{postfit} after running the \texttt{vp-fakeevolve} script.

The ability to load weights from a previous fit can also be used to aid a simultaneous fit, by starting training from a good PDF solution already. For example, if we use the syntax:
\begin{python}
fixed_pdf_fit: False
load_weights_from_fit: 221103-jmm-no_top_1000_iterated
\end{python}
in a \simunet{} runcard, then the weights of the PDF part of the neural network will be initialised to the weights of the appropriate replica from the previous PDF fit \texttt{221103-jmm-no\_top\_1000\_iterated}. \textit{However}, since the namespace \texttt{fixed\_pdf\_fit} is set to \texttt{False} here, the fit will \textit{still} be simultaneous, fitting both PDFs and physical parameters. Assuming that the resulting simultaneously-determined PDF fit is reasonably close to the previous PDF fit, this can significantly decrease training time.

\paragraph{\simunet{} analysis tools.} The \simunet{} code is released with a full suite of analysis tools, 
which build on the tools already available in the \nnpdfcode{} public code. 
These tools rely on the \texttt{validphys} analysis framework implemented in 
\texttt{reportengine}, and address exclusively the PDF aspect of the fits.
They allow the user, for example, to generate PDF plots, luminosities, compare fits, and evaluate fit quality metrics, 
among many other things.

The \simunet{} code adds an extensive set of tools to study results in the SMEFT and assess the PDF-SMEFT interplay.
These analysis tools are exclusively allocated in the \texttt{simunet\_analysis.py} file, where the user can find 
documented functions to perform different tasks. 
In the context of EFT coefficients, \simunet{} computes their posterior distributions, confidence intervals (at $68\%$ and $95\%$, corresponding to the mean of the replicas $\pm 1\sigma$ and $\pm 2\sigma$, respectively), correlations, and deviations (pulls) from the SM, along with other relevant metrics.
In the next chapters we will see some examples of the results that \simunet{} can produce.

In this section, we have discussed the code structure and usage of the \simunet{} code in the context of simultaneous PDF-SMEFT fits. In what follows, we will delve deeper into one of its other features: the ability to perform contaminated PDF fits to determine if the effects of new physics can be absorbed by the fits.

\section{\simunet{} for contaminated fits}
\label{sec:contamination}

The methodology we use throughout this study is based on the NNPDF \textit{closure test} framework, first introduced in Ref.~\cite{NNPDF:2014otw}, and explained in more detail in Ref.~\cite{DelDebbio:2021whr}. This method was developed in order to assess the quality and the robustness of the NNPDF fitting methodology; in brief it follows three basic steps: (i) assume that Nature's PDFs are given by some fixed reference set; (ii) generate artificial MC data based on this assumption, which we term \textit{pseudodata}; (iii) fit PDFs to the pseudodata using the NNPDF methodology. Various statistical estimators, described in Ref.~\cite{DelDebbio:2021whr}, can then be applied to check the quality of the fit (in broad terms, assessing its difference from the true PDFs), hence verifying the accuracy of the fitting methodology. In this study, the closure test methodology is adapted to account for the fact that the true theory of Nature may \textit{not} be the SM.

The structure of the forthcoming sections is as follows. In Sect.~\ref{subsec:contamination_definition}, we define the terms \textit{baseline fit} and \textit{contaminated fit}, which shall be used throughout this chapter. In Sect.~\ref{subsec:pseudodata_generation}, we provide more details on how the MC data are generated in the presence of new physics (NP). In Sect~\ref{subsec:postfit} we give an overview of the types of analysis we can perform on the fits we obtain. Then, in Sect.~\ref{sec:ctusage} we discuss in practical terms how BSM-contaminated fits can be performed with \simunet{}.

\subsection{Basic definitions and fitting methodology}
\label{subsec:contamination_definition}
Let us suppose that the true theory of Nature is given by the SM, plus some NP contribution. 
Under this assumption, observables $T \equiv T(\theta_{\text{SM}}, \theta_{\text{NP}})$ have a dependence on both the SM 
parameters $\theta_{\text{SM}}$ (in our work, exclusively the PDFs), and the NP parameters $\theta_{\text{NP}}$ 
(for example, masses and couplings of new particles). Let us further fix notation by writing the true values of the 
parameters $\theta_{\text{SM}}, \theta_{\text{NP}}$ as $\theta_{\text{SM}}^*, \theta_{\text{NP}}^*$ respectively; 
for convenience, we shall also write the true value of the observable $T$ as $T^* \equiv T(\theta_{\text{SM}}^*, \theta_{\text{NP}}^*)$.

Suppose that we wish to perform a fit of some of the theory parameters using experimental measurements of $N_{\rm obs}$ observables, which we package as a single vector $T = (T_1, T_2,..., T_{N_{\rm obs}})$. All measurements are subject to random observational noise. For additive Gaussian uncertainties, the observed data is distributed according to:
\begin{equation}
\label{eq:observed_data}
D_0 = T^* + \eta,
\end{equation}
where $\eta$ is drawn from the multivariate Gaussian distribution $\mathcal{N}(0,\Sigma)$, with $\Sigma$ the experimental covariance matrix describing the uncertainties of the measurements and the correlations between them.
The general procedure, which also accounts for multiplicative uncertainties and positivity effects, is implemented in the NNPDF code~\cite{NNPDF:2021uiq}.

In the context of the NNPDF closure test methodology~\cite{NNPDF:2014otw}, the true values of the observables $T^*$ are referred to as \textit{level 0 pseudodata} (L0), whilst the fluctuated values $D_0$ are referred to as \textit{level 1 pseudodata} (L1).

Once we have generated a sample $D_0$ of L1 pseudodata, we may perform a fit of some of the theory parameters to this pseudodata. In this work, we shall perform various types of fits with different choices of $\theta_{\text{SM}}^*, \theta_{\text{NP}}^*$, and different choices of which parameters we are fitting. We define the types of fits as follows:
\begin{enumerate}[label = (\arabic*)]
\item \textbf{Baseline fit.} If there is no new physics, $\theta_{\text{NP}}^* \equiv 0$, then the SM is the true theory of Nature. We generate L1 pseudodata $D_0$ according to the SM. If we subsequently fit the parameters $\theta_{\text{SM}}$, we say that we are performing a \textit{baseline fit}. This is precisely equivalent to performing a standard NNPDF closure test.
\item \textbf{Contaminated fit.} If new physics exists, $\theta_{\text{NP}}^* \neq 0$, then the SM is \textit{not} the true theory of Nature. We generate L1 pseudodata $D_0$ according to the SM plus the NP contribution. If we subsequently \textit{only} fit the parameters $\theta_{\text{SM}}$ whilst ignoring the NP parameters $\theta_{\text{NP}}$, we say that we are performing a \textit{contaminated fit}.
\item \textbf{Simultaneous fit.} If new physics exists, $\theta_{\text{NP}}^* \neq 0$, we again generate L1 pseudodata $D_0$ according to the SM plus the NP contribution. If we subsequently fit \textit{both} the parameters $\theta_{\text{SM}}$ \textit{and} $\theta_{\text{NP}}$, we say that we are performing a \textit{simultaneous fit}. A closure test of this type is performed in Ref.~\cite{Iranipour:2022iak} in order to benchmark the \simunet{} methodology. However, we do not perform such fits in this work, with our main goal being to assess the possible deficiencies associated with performing \textit{contaminated} fits.
\end{enumerate}
\begin{table}
  \centering
  \begin{tabular}{l|l|l}
    \toprule
    {\bf Fit name} & {\bf Nature}  & {\bf Fitted parameters}\\
    \midrule
    Baseline & Standard Model: $\theta_{\text{NP}}^* \equiv 0$ & Standard Model only: $\theta_{\text{SM}}$ \\
    Contaminated & SM + new physics: $\theta_{\text{NP}}^* \neq 0$ & Standard Model only: $\theta_{\text{SM}}$\\
    \bottomrule
    \end{tabular}
   \caption{\label{table:definition} A summary of the definitions of \textit{baseline} and \textit{contaminated} fits used throughout this work.}
  \end{table}
We have discussed \textit{simultaneous} fits in \simunet{}. In what follows, we shall emphasise \textit{baseline} and \textit{contaminated} fits; that is, we shall only fit SM 
parameters, but we shall fit them to pseudodata generated either assuming the law of Nature is given by the SM only, 
or that it is given by the SM plus some NP contribution. A summary of the relevant definitions is given for convenient reference in Table~\ref{table:definition}.

The NNPDF methodology makes use of the Monte-Carlo (MC) replica method to propagate errors to the PDFs. This involves the generation of an additional layer of pseudodata, referred to as \textit{level 2 pseudodata} (L2). Given an L1 pseudodata sample $D_0$, we generate L2 pseudodata by augmenting $D_0$ with further random noise $\epsilon$:
\begin{equation}
\label{eq:mc_pseudodata}
D = D_0 + \epsilon = T^* + \eta + \epsilon,
\end{equation}
where $\epsilon$ is an independent multivariate Gaussian variable, distributed according to $\epsilon \sim \mathcal{N}(0,\Sigma)$, with $\Sigma$ the experimental covariance matrix. Whilst the L1 pseudodata is sampled only once, the L2 pseudodata $D$ is sampled $N_{\text{rep}}$ times, and the best-fit PDFs are obtained for each of the L2 pseudodata samples. This provides an ensemble of PDFs from which statistical estimators, in particular uncertainty bands, can be constructed.

\subsection{Pseudodata generation}
\label{subsec:pseudodata_generation}

As described above, in contaminated fits we assume that the true theory of Nature is the SM plus some new physics. The true PDF set which shall be used in the latter chapters of this thesis is the \nnpdf{} set~\cite{Ball:2021leu} (in principle, we are of course allowed to choose any PDF set).

To generate pseudodata we inject NP signals. NP scenarios can be drawn from specific UV-complete models in the case of heavy NP, for example. Furthermore, we can choose NP scenarios which are characterised by scales much higher than the energy scales probed by the data, which allows us to justify matching the UV-complete models to a SMEFT parametrisation. The advantage of this approach is that theory predictions become polynomial in the SMEFT Wilson coefficients, which is not necessarily the case in UV-complete models; this allows us maximum flexibility to trial many different values for the "true" NP parameters. In this way, we can make a $K$-factor approximation to avoid expensive computation of fast interpolation grids for the PDFs which, as in \nnpdf{}, are generated with purely SM inputs. As a result, the formula for the "true" value of an observable takes the schematic form:
\begin{equation}
\label{eq:observable_cont}
T \equiv \left(1 + cK_{\text{lin}} + c^2K_{\text{quad}}\right) \hat{\sigma}^{\text{SM}} \otimes \mathcal{L},
\end{equation}
where $\mathcal{L}$ denotes either the PDFs or PDF luminosities for \nnpdf{} (depending on whether the observable is a deep inelastic scattering or hadronic observable), $c$ denotes the SMEFT Wilson coefficient(s) under consideration, $\hat{\sigma}^{\text{SM}}$ is the SM partonic cross-section computed at NNLO in QCD perturbation theory, and $K_{\text{lin}}, K_{\text{quad}}$ are the SMEFT $K$-factors.

Note that, unlike Eq. (\ref{eq:simunet_predictions}) which describes the forward-pass of the combination layer in \simunet{} to perform a BSM fit (that can be performed only with linear contributions because of the reasons described in the previous section), the value of the observable in Eq. (\ref{eq:observable_cont}) accounts for both \textit{linear} and \textit{quadratic} BSM contributions, giving it more flexibility to describe the NP effects. In this way, \simunet{} can be used to perform simultaneous fits of PDFs and BSM coefficient and to assess the potential absorption of BSM effects by the PDFs to the best of its ability in each case.

\subsection{Post-fit analysis}
\label{subsec:postfit}

Once we have produced a contaminated fit, where PDFs have been fitted using SM theory to data produced with the SM plus some NP contribution, several natural questions arise.

\paragraph{Detection of contamination.} Is it possible to detect contamination of the PDF fit by the NP effects? If there are many datasets entering the fit that are \textit{not} affected by NP, it might be the case that datasets that \textit{are} affected by NP could appear inconsistent, and might be poorly described by the resulting fit.

In order to address this point, we use the NNPDF dataset selection criteria, discussed in detail in Ref.~\cite{Ball:2021leu}. We consider both the $\chi^2$-statistic of the resulting contaminated PDF fit to each dataset entering the fit, and also consider the number of standard deviations 
\begin{equation}
\label{eq:nsigma}
n_{\sigma} = (\chi^2 - n_{\rm dat}) / \sqrt{2n_{\rm dat}}
\end{equation} 
of the $\chi^2$-statistic from the expected $\chi^2$ for each dataset. 
If $\chi^2/n_{\rm dat} > 1.5$ and $n_{\sigma} > 2$ for a particular dataset, the dataset would be flagged by the NNPDF selection criteria, indicating an inconsistency with the other data entering the fit.

There are two possible outcomes of performing such a dataset selection analysis on a contaminated fit. In the first instance, the datasets affected by NP are flagged by the dataset selection criterion. 
If a dataset is flagged according to this condition, then a weighted fit is performed, 
i.e. a fit in which a dataset is given a
larger weight inversely proportional to the number of data points. In more detail, if the $j$th dataset has been flagged, then the $\chi^2$-loss used in the subsequent weighted fit is modified to:
\begin{equation}
\chi^2_w=\frac{1}{n_{\rm dat}-n_{\rm
    dat}^{(j)}}\,\sum_{i=1,i\neq j}^{n_{\rm exp}} \,n_{\rm
  dat}^{(i)}\,\chi^2_i \, + \,w^{(j)}\chi^2_j,
  \end{equation}
where $\chi^2_i$ denotes the usual $\chi^2$-loss for the $i$th dataset, and where the \textit{weight} is defined by:
  \begin{equation}
    \label{eq:wgts}
    w^{(j)} = n_{\rm dat}/n_{\rm dat}^{(j)}.
    \end{equation}
If the data-theory agreement improved for the set under investigation, to the extent that it now satisfies the selection criteria, \textit{and further} the
data-theory agreement of the other datasets does not deteriorate in
any statistically significant way, then the dataset is kept; otherwise, the dataset is discarded, 
on the basis of inconsistency with the remaining datasets.

In the second instance, the ``contaminated'' datasets are \textit{not} flagged, and are consistent enough that the contaminated fit 
would pass undetected as a \textit{bona fide} SM PDF fit. We introduce the following terms to describe each of these cases: 
in the former case, we say that the PDF was \textit{unable to absorb} the NP; in the latter case, we say that the PDF has \textit{absorbed} the NP.

\paragraph{Distortion of NP bounds.} Can using a contaminated fit in a subsequent fit of NP effects lead to misleading bounds? In more detail, suppose that we construct a contaminated fit which has absorbed NP - that is, the contamination would go undetected by the NNPDF dataset selection criterion. In this case, we would trust that our contaminated fit was a perfectly consistent SM PDF fit, and might try to use it to subsequently fit the underlying parameters in the NP scenario. 

There are two possible outcomes of such a fit. The contamination of the PDFs may be weak enough for the NP bounds that we obtain to be perfectly sensible, containing the true values of the NP parameters. On the other hand, the absorption of the NP may be strong enough for the NP bounds to be distorted, no longer capturing the true underlying values of the NP parameters. The second case is particularly concerning, and if it can occur, points to a clear need to disentangle PDFs and possible NP effects.

\paragraph{Distortion of SM predictions.} Finally, we ask: can using a contaminated fit lead to poor agreement on new datasets that are not affected by NP? In particular, suppose that we are again in the case where NP has been absorbed by a contaminated fit, so that the NP signal has gone undetected. If we were to use this contaminated fit to make predictions for an observable that is \textit{not} affected by the NP, it is interesting to see whether the data is well-described or not; if the contamination is sufficiently strong, it may appear that the dataset is inconsistent with the SM. This could provide a route for disentangling PDFs and NP; we shall discuss this point later in the thesis.

\paragraph{}. We have discussed the \simunet{} methodology for contaminated fits. In the next section, we shall discuss how to use the \simunet{} code to perform contaminated fits with specific examples, and how to analyse the results of such fits in order to address the questions raised above

\subsection{BSM-contaminated fits with \simunet{}} 
\label{sec:ctusage}

In terms of usage, a contaminated fit using the \simunet{} code should be based on an \nnpdf{} closure test runcard; in particular, the \texttt{closures} namespace must be specified. The required \simunet{} additions to such a runcard are twofold and described as follows. First, the \texttt{dataset\_inputs} must be modified with a new key, for example:
\begin{python}
dataset_inputs:
- {dataset: LHCB_Z_13TEV_DIELECTRON, frac: 0.75, cfac: ['QCD']}
- {dataset: CMSDY1D12, frac: 0.75, cfac: ['QCD', 'EWK'], contamination: 'EFT_LO'}
\end{python}
Here, two datasets are included: (i) the \texttt{LHCB\_Z\_13TEV\_DIELECTRON} dataset has no additional keys compared to a standard \nnpdf{} closure test runcard, and hence the level 1 pseudodata for this set is generated normally; (ii) the \texttt{CMSDY1D12} dataset has an additional key, \texttt{contamination}, set to the value \texttt{'EFT\_LO'}, which tells \simunet{} to base the level 1 pseudodata for this set, using parameters drawn from the model \texttt{EFT\_LO} in the relevant \texttt{simu\_fac} file. The precise values of the true physical parameters to generate the pseudodata are specified as part of the \texttt{closuretest} namespace, which is the second modification required to a standard closure test runcard:
\begin{python}
closuretest:
  ...
  contamination_parameters:
    - name: 'W'
      value:  -0.0012752
      linear_combination:
        'Olq3': -15.94

    - name: 'Y'
      value: 0.00015
      linear_combination:
        'Olq1': 1.51606
        'Oed': -6.0606
        'Oeu': 12.1394
        'Olu': 6.0606
        'Old': -3.0394
        'Oqe': 3.0394
\end{python}
In the example above we have specified that we inject a NP model that belongs to the so-called {\it universal theories} in which the parameters $\hat{W}$ and $\hat{Y}$, which are a combination of four-fermion operators defined for example in Refs.~\cite{Greljo:2021kvv,Iranipour:2022iak,Hammou:2023heg}, are set the values given in this example. The precise definitions of the linear combinations in this example are given by:
\begin{eqnarray*}
  \hat{W}\,\mathcal{O}_W &=& (-0.0012752) \times \left(-15.94\, \mathcal{O}_{lq3}\right),\\
  \hat{Y}\,\mathcal{O}_Y &=& (0.00015) \times \left(1.51606\, \mathcal{O}_{lq1} -6.0606\,\mathcal{O}_{ed} 
  +12.1394\,\mathcal{O}_{eu}+6.0606\,\mathcal{O}_{lu}\right.\\
  && \left.-3.0394\,\mathcal{O}_{ld}+3.0394\,\mathcal{O}_{qe}\right).
\end{eqnarray*}
The `contamination' feature of the \simunet{} code allows the user not only to study whether a theory bias can be reabsorbed into the PDFs, but can be used in conjunction with the simultaneous fit feature to perform closure tests on the simultaneous fits of PDFs and SMEFT Wilson coefficients. The closure test validates how much \simunet{} is able to replicate, not only fixed Wilson coefficients, but also a known underlying PDF. In the context of a simultaneous fitting methodology a `level 2' closure test amounts to generating SM predictions using a known PDF set (henceforth referred to as the underlying law) and mapping these to SMEFT observables by multiplying the SM theory predictions with SMEFT $K$-factors scaled by a previously determined choice of Wilson coefficients. These SMEFT observables, generated by the underlying law, replace the usual MC pseudodata replicas in the SM, and are used to train the neural network in the usual way.
 
Through this approach we are not only able to assess the degree to which the parametrisation 
is able to capture an underlying choice of Wilson coefficients, but also ensure it is
sufficiently flexible to adequately replicate a known PDF. 

\paragraph{Closure test settings.} 
In the rest of the section, we present the results of a simultaneous closure test the validates the \simunet{} methodology in
its ability to produce an underlying law comprising both PDFs and Wilson coefficients. In this example we set:
\begin{align*}
  \omega_{\rm true} &= {\tt NNPDF40\_nnlo\_as\_0118\_1000},\\
  \vec{c}_{\rm true} &= \begin{pmatrix} c_{tG} \\ c_{lq}^{(3)} \end{pmatrix} = \begin{pmatrix} 1 \\ 0.08 \end{pmatrix}.
\end{align*}
That is, \nnpdf{} serves as the true underlying PDF law, and all Wilson coefficients are set to zero except $c_{tG}$ and $c_{lq}^{(3)}$.
Note that the {\tt true} values of the Wilson coefficients that we choose in this test are 
outside the 95\% C.L. bounds on the Wilson coefficients that were found in previous 
analyses~\cite{Greljo:2021kvv,Kassabov:2023hbm} corresponding to $c_{tG}\in (-0.18,0.17)$ and 
$c_{lq}^{(3)}\in (-0.07,0.02)$ respectively, hence the success of the 
closure test is non-trivial. 

\begin{figure}[htbp]
  \centering
  \includegraphics[width=0.49\textwidth]{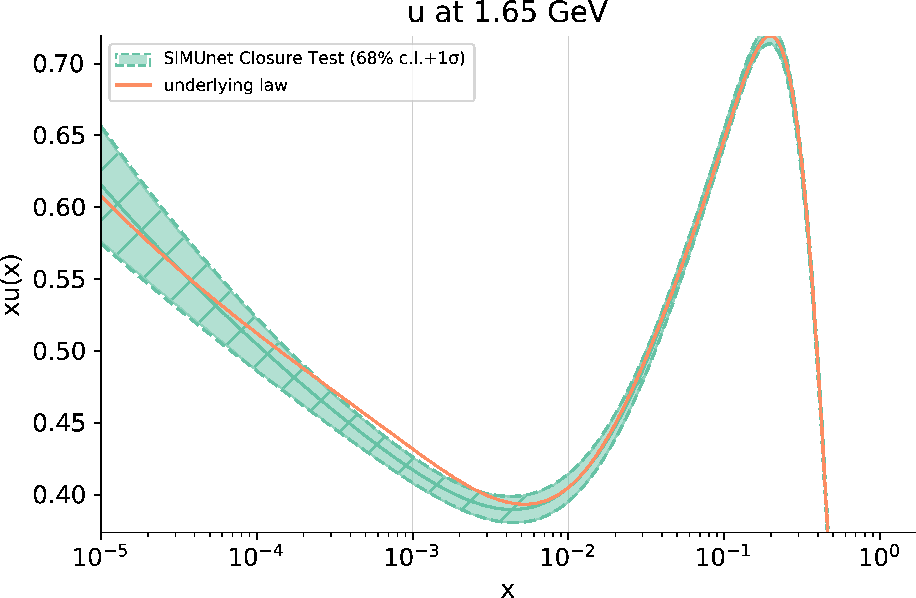}
  \includegraphics[width=0.49\textwidth]{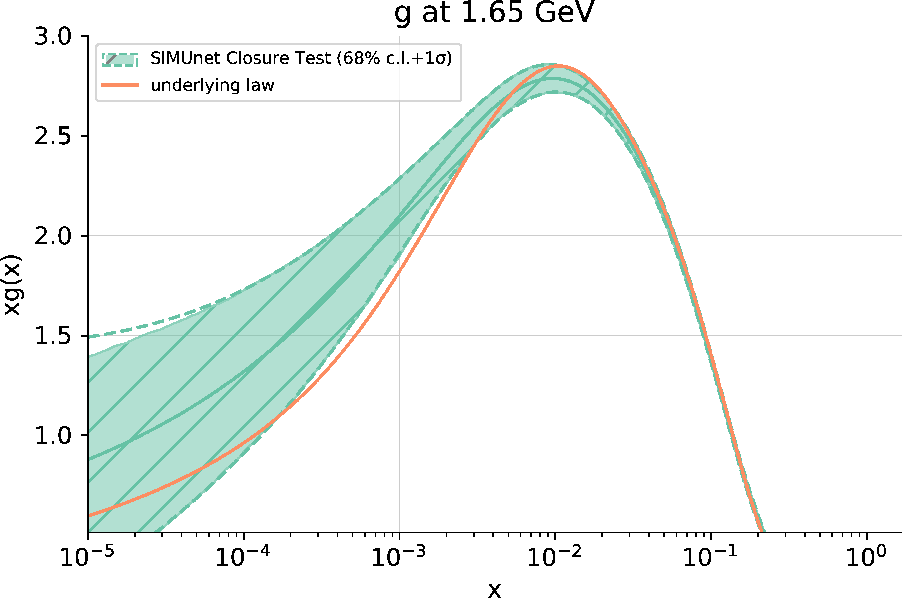}
\caption{
  The up quark (left) and gluon (right) PDFs obtained from the closure test 
framework in which both the underlying PDF set and Wilson coefficients are known. 
Shown in orange is the PDF replica used as the underlying law which generates the fake data used to train our model. 
The resulting PDFs are shown in green along with their 68\% confidence level bands. 
The fake data generated by the underlying law is subsequently modified so as to encode the ($c_{tG}$, $c_{lq}^{(3)}$) = (1, 0.08) condition.
}
\label{fig:pdf_closuretest}
\end{figure}
The results of the closure test for the gluon and up quark PDFs at the parametrisation scale 
are presented in Fig. (\ref{fig:pdf_closuretest}). Here we can see that the combination layer is
capturing the data’s dependence on the Wilson coefficients, whilst the complementary PDF
sector of the network architecture captures the data’s dependence on the underlying PDF. 
The combination layer, in effect, subtracts off the EFT
dependence, leaving behind the pure SM contribution for the PDF sector to parametrise.
\begin{figure}[htbp]
    \centering
    \includegraphics[width=0.5\textwidth]{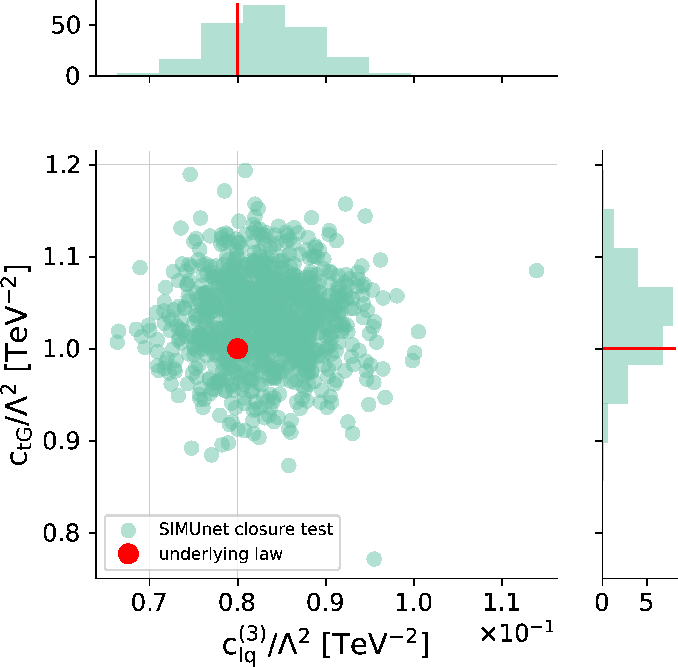}
\caption{
    Result of the closure testing framework for our methodology. 
    The distribution of ($c_{tG}$,$c_{lq}^{(3)}$) when fitting to data that has been modified 
    by setting ($c_{tG}$,$c_{lq}^{(3)}$) = ($1$, $0.08$). 
    The upper and right panels show the histograms for the distribution of the best fit values in their respective directions.
    }
\label{fig:wc_closuretest}
\end{figure}
The corresponding results for the Wilson coefficients $c_{tG}$ and $c_{lq}^{(3)}$ are 
displayed in Fig. (\ref{fig:wc_closuretest}). 
Both figures demonstrate that the parametrisation successfully captures the underlying 
law for both the PDFs and Wilson coefficients.
To verify the robustness of these findings and rule out random fluctuations, we conducted 
the closure test 25 times, 
each time using different level 1 pseudodata; the results consistently remained 
stable and aligned with those presented above.

\section{Summary}
\label{sec:conclusions}

In this chapter we have presented the public release, as an open-source
code, of the software framework underlying the \simunet{} 
global simultaneous determination of PDFs and SMEFT coefficients. 
Based on the first tagged version of the public {\tt NNPDF} code~\cite{nnpdfcode,NNPDF:2021uiq}, 
\simunet{} adds a number of new
features that allow the exploration of the interplay between PDFs and 
BSM degrees of freedom.

The code is fully documented and open-source. The phenomenological studies that can be performed with it include (i) a PDF-only fit {\it \`{a} la} \nnpdf{} with extra datasets included; (ii) a purely global EFT analysis; (iii) a simultaneous PDF-SMEFT fit in which the user can decide which observable to treat as PDF-dependent or independent by either fitting PDFs alongside the Wilson coefficients or freezing them to a baseline PDF set.

We have demonstrated that the user can use the code to inject any new physics model 
in the data and check whether the model can be fitted away by the PDF parametrisation. We have also shown that the methodology successfully passes the closure test with respect to an underlying PDF law and UV 'true' model that is injected in the data. 

The public release of \simunet{} represents a first crucial step towards the interpretation of indirect searches under a unified framework. We can assess the impact of new datasets not only on the PDFs, but now on the couplings of an EFT expansion, or on any other physical parameter. Indeed, the methodology presented here can be extended to simultaneous fitting precision SM parameters, such as the strong coupling or electroweak parameters. Indeed, our framework extends naturally, not only to BSM studies, but to any parameter which may modify the SM prediction through the use of K-factors or other kind of interpolation, see Appendix A of Ref.~\cite{Iranipour:2022iak} for the application of \simunet{} to the simultaneous fit of PDFs and $\alpha_s(M_z)$~\cite{Forte:2020pyp}.

In the next chapter, we will see the \simunet{} methodology in action. We will apply it to study the PDF-BSM interplay in the context of fits to data in the top quark sector, and in a global dataset.

\chapter{PDF and SMEFT analyses in the top sector and in global datasets}
\label{ch:fits}
  \epigraph{Hay que conocer la piedra \\ Que corona el ventisquero \\ Hay que recorrer callando \\ Los atajos del silencio \\ \vspace{1em} You have to know the stone \\
That crowns the hanging glacier \\ You have to walk quietly \\ Through the shortcuts of silence}{\textit{\\ from Arriba en la Cordillera, \\ by P. Manns (trad. M. M. A.).}}

\textit{This chapter is based on Refs.~\cite{Kassabov:2023hbm} and~\cite{Costantini:2024xae}, and was written in collaboration with Z. Kassabov, M. Madigan, L. Mantani, J. Moore, J. Rojo,  and M. Ubiali. My contributions to this work include the new implementation of the} \simunet{} \textit{methodology in} \texttt{n3fit} \textit{(together with J. Moore), the implementation of the EFT and EFT-PDF analysis routines in the} \texttt{validphys} \textit{module (all functions but the principal component analysis), and running different fits to assess the impact of top quark data on PDFs and SMEFT Wilson coefficients.}

\vspace{1em} 
\noindent\rule{\textwidth}{0.4pt}
\vspace{1em} 

In this chapter we use the \simunet{} methodology to study the PDF-BSM interplay in the context of top quark measurements at the LHC, and in a global dataset. 

First, we discuss the top-specific scenario. We assess the impact of  top quark production at the LHC on global analyses of PDFs and of Wilson coefficients in the SMEFT, both separately and in the framework of a joint interpretation. We considered the broadest top quark dataset to date (up to March 2023) containing all available measurements based on the full Run II luminosity. We determine the constraints that this dataset provides on the large-$x$ gluon PDF and study its consistency with other gluon-sensitive measurements. Afterwards, we carry out a SMEFT interpretation of the same dataset using state-of-the-art SM and EFT theory calculations, resulting in bounds on 25 Wilson coefficients modifying top quark interactions. Subsequently, we integrate the two analyses within the \simunet{} methodology to realise a simultaneous determination of the PDFs and the EFT coefficients and identify regions in the parameter space where their interplay is most phenomenologically relevant.

After that, we extend the analysis to a global dataset. We do this by including, in addition to the previous datasets, electroweak precision observables (EWPO), Higgs, and diboson measurements. We perform a global purely SMEFT fit to the combined dataset, and compare the results to existing methodologies. Then, we perform a global simultaneous fit of PDFs and Wilson coefficients, finding constraints on 40 SMEFT operators, which makes this study one of the \textit{most comprehensive} SMEFT analysis to date, and the \textit{first} to include PDFs alongside SMEFT coefficient in a global fit.

\section{Introduction}

In this chapter we will study the PDF-BSM interplay in two scenarios: in the top sector, and in a global dataset.

The top quark is one of the most remarkable particles within the SM.
Being the heaviest elementary particle known to date, with a mass around 185 times heavier than a proton,
and the only fermion with an $\mathcal{O}(1)$
Yukawa coupling to the Higgs boson, the top quark has long been suspected to play a privileged role
in potential BSM physics extensions~\cite{Isidori:2001bm,Buttazzo:2013uya,DiLuzio:2015iua,Schwienhorst:2022yqu}.
For these reasons, since its discovery at the Tevatron in 1995~\cite{CDF:1995wbb,D0:1995jca}
the properties of the top quark have been scrutinised with utmost attention and
a large number of BSM searches involving top quarks as final states have been carried out.
The focus on the top quark has further intensified since the start of operations
at the LHC, which has realised an unprecedented top factory producing more than 200 million
top quark pairs so far, for example.

In addition to this excellent potential for BSM studies, top quark production at hadron
colliders also provides unique information on a variety of SM parameters such
as the strong coupling constant $\alpha_s(m_t)$~\cite{CMS:2014rml,Cooper-Sarkar:2020twv}, 
the CKM matrix element $V_{tb}$~\cite{CMS:2020vac}, and the top quark
mass $m_t$~\cite{ATLAS:2022jbw,CMS:2023ebf},
among several others.
Furthermore, top quark production at the LHC constrains the PDFs of the proton~\cite{Gao:2017yyd,Amoroso:2022eow},
in particular the large-$x$ gluon PDF from inclusive top quark
pair production~\cite{Czakon:2013tha,Czakon:2016olj,Czakon:2019yrx}
and the quark PDF flavour separation
from inclusive single top production~\cite{Nocera:2019wyk,Campbell:2021qgd}.
Indeed, fiducial and differential measurements of top quark pair
production are part of the majority of recent global PDF determinations.
Reliably extracting SM parameters, including those parametrising the
subnuclear structure of the proton in the PDFs, from LHC top quark
production data has been made possible thanks to
recent progress in higher order QCD and electroweak calculations
of top quark production.
Inclusive top quark pair production is now known at NNLO in the QCD
expansion both for single-
and double-differential
distributions~\cite{Czakon:2016dgf,Czakon:2020qbd,Catani:2019iny,Catani:2019hip,Grazzini:2017mhc}, 
eventually complemented
with electroweak corrections~\cite{Czakon:2017wor}, threshold
resummation~\cite{Czakon:2018nun}, and matching to parton showers~\cite{Mazzitelli:2021mmm}.
NNLO QCD corrections are also known for single top quark production
at the LHC, both in the
$t$-channel~\cite{Berger:2017zof,Berger:2016oht} and in the $s$-channel~\cite{Liu:2018gxa}.

As we discussed in Chapter \ref{ch:introduction}, by using an EFT approximation, sufficiently heavy new particles whose direct production lies beyond the reach of the LHC can still provide BSM sensitivity through low-energy signatures.
These are typically revealed in the modification
of SM particle properties, such as their interactions and coupling strengths.
This is where the SMEFT, as defined in the first chapter of this thesis, comes into play.
Several groups, both from the theory community and within the experimental collaborations,
have presented interpretations of LHC top quark 
measurements in the SMEFT framework~\cite{Aguilar-Saavedra:2018ksv,
  Buckley:2015lku, Brivio:2019ius, Bissmann:2019gfc, Hartland:2019bjb,
  Durieux:2019rbz, vanBeek:2019evb, Yates:2021udl, Aoude:2022imd, Maltoni:2019aot, Aoude:2022deh, Degrande:2018fog, Faham:2021zet, Severi:2022qjy} to derive bounds
on  higher-dimensional EFT operators that distort the interactions of top quarks.
Studies have also been performed in the framework of anomalous couplings~\cite{Cao:2020npb,Cao:2015doa}.
A key feature of these analyses is that the unprecedented energy reach of the LHC
data increases the sensitivity to SMEFT operators via energy-growing
effects entering the partonic cross sections.

We have explained how the PDF-BSM interplay, in general, is important for the interpretation of LHC data in SM and BSM studies. In this chapter, our \textit{first} goal is to extend the initial explorations of~\cite{Carrazza:2019sec,Greljo:2021kvv} to a global determination of the PDFs in the SMEFT from top quark production measurements.
To this purpose, we consider the broadest top quark dataset used to date in either
PDF or EFT interpretations, which in particular
contains all available measurements from ATLAS and CMS based on the full Run II luminosity.
By combining this wide dataset with the \simunet{} methodology, 
we derive bounds on  25 independent Wilson coefficients modifying top quark
interactions, identify regions in the parameter space where the interplay between
PDFs and SMEFT signatures
is most phenomenologically relevant, and
demonstrate how to separate eventual BSM signals from QCD effects
in the interpretation of top quark measurements.
As a non-trivial by-product, we also revisit the SM-PDF and fixed-PDF analyses by quantifying the information
that our comprehensive top quark dataset provides. On the one hand, we assess the impact on the large-$x$ gluon (SM-PDF),
and on the other, we study the impact on the EFT coefficients (fixed-PDF),
and compare our findings with related studies in the literature.

Furthermore, our \textit{second} goal is to present a new global simultaneous analysis of SMEFT
Wilson coefficients and PDFs based on a broad range of 
observables, including for the first time  (i) the Higgs production and decay
rates data from the LHC, (ii) precision electroweak and diboson measurements 
from LEP and the LHC, and (iii) Drell-Yan measurements, in a global analysis of 40 dimension-6 SMEFT operators. The Wilson coefficients are determined both with fixed PDFs and simultaneously with the PDFs themselves.

Early SMEFT analyses constrained subsets of SMEFT operators relevant to sectors of observables, for example the electroweak, diboson and Higgs sectors~\cite{Han:2004az,daSilvaAlmeida:2018iqo,Ellis:2014jta,Almeida:2021asy,Biekotter:2018ohn,Kraml:2019sis,Ellis:2018gqa,Corbett:2012ja,Ethier:2021ydt}, the top sector~\cite{Buckley:2015lku,Brivio:2019ius,Hartland:2019bjb,Bissmann:2019gfc,Kassabov:2023hbm,Elmer:2023wtr} as well as Drell-Yan~\cite{Allwicher:2022gkm,Boughezal:2022nof}. More recently, global fits constraining a larger set of SMEFT operators have been performed: the Higgs, top, diboson and electroweak sectors have been combined~\cite{Ellis:2020unq,Ethier:2021bye,Giani:2023gfq}, as well as subsets of those~\cite{Cirigliano:2016nyn} along with low-energy charged-current observables~\cite{Cirigliano:2022qdm,Cirigliano:2023nol}. Also, a combinations of all those sectors alongside Drell-Yan and flavour observables has been recently published~\cite{Bartocci:2023nvp}, and its connection with UV-complete models and higher energy colliders has been studied~\cite{terHoeve:2023pvs,Celada:2024mcf}. These studies were developed assuming fixed PDFs, and the SMEFT operators were determined by fitting the data to the SM predictions modified by the SMEFT operators. In this chapter, we will extend these analyses by performing a global simultaneous fit of the SMEFT operators \textit{and} the PDFs, which will allow us to determine the impact of the SMEFT operators on the PDFs and vice versa.

The structure of this chapter is as follows.
In Sect.~\ref{sec:exp} we describe the data
inputs and the theory calculations used in our studies.
Subsequently, in Sect.~\ref{sec:baseline_sm_fits} we present the results of SM-PDF fits with the full top quark dataset, and in particular we quantify the impact on the large-$x$ gluon of recent high-statistics Run II measurements.
In Sect.~\ref{sec:res_smeft} we consider the fixed-PDF analyses in the top sector, and present the most extensive SMEFT interpretation of top quark operators from the LHC to date, including comparisons with previous results in the literature.
Then, in  Sect.~\ref{sec:joint_pdf_smeft}, we present the simultaneous determinations of the PDFs and EFT coefficients and the comparison of these  with both the fixed-PDF and SM-PDF cases in the top scenario.
In Sect.~\ref{sec:global} we extend the analysis to a \textit{global} dataset including, in addition to the top quark dataset, the Higgs, electroweak, diboson, and Drell-Yan measurements. First, we perform a global SMEFT fit with fixed PDFs, and then we extend the analysis to the first global simultaneous determination of the SMEFT operators and the PDFs in the presence of the full dataset.
We summarise our results and outline some possible future developments
in Sect.~\ref{sec:summary_top}.

\section{Experimental data and theory calculations}
\label{sec:exp}

We begin by  describing the experimental data and theoretical predictions,
both in the SM and in the SMEFT, used as input for the present analysis.
We start in Sect. \ref{sec:baseline_data} by describing the datasets and theoretical calculations that we consider in the top specific scenario. Then, to generalise our analyses to include the global dataset, we describe in Sect. \ref{subsec:data} the datasets and theoretical calculations that we consider in the global scenario.


\subsection{Top dataset and theory}
\label{sec:baseline_data}

With the exception of the top quark measurements, the dataset used in
this work for fitting the PDFs both in the SM-PDF and SMEFT-PDF cases
overlaps with that of the \nnpdf{} determination presented in Ref.~\cite{NNPDF:2021njg}.
In particular,
the no-top variant of the \nnpdf{} dataset consists of 4535 data points
corresponding to a wide variety of processes in
deep-inelastic lepton-proton scattering~\cite{Arneodo:1996kd,Arneodo:1996qe,Whitlow:1991uw,Benvenuti:1989rh,Onengut:2005kv,Goncharov:2001qe,MasonPhD,Abramowicz:2015mha,H1:2018flt}
and in hadronic proton-proton collisions~\cite{Moreno:1990sf,Webb:2003ps,Towell:2001nh,Aaltonen:2010zza,Abazov:2007jy,Abazov:2013rja,D0:2014kma,Abulencia:2007ez,Aad:2011dm,Aaboud:2016btc,Aad:2014qja,Aad:2013iua,Chatrchyan:2012xt,Chatrchyan:2013mza,Chatrchyan:2013tia,Khachatryan:2016pev, Aaij:2012mda,Aaij:2015gna,Aaij:2015vua,Aaij:2015zlq,Aad:2016zzw,Aaboud:2017ffb,Aad:2019rou,Aaij:2016qqz,Aad:2016naf,Aaij:2016mgv,Aad:2015auj,Khachatryan:2015oaa, Aaboud:2017soa, Sirunyan:2017wgx,Aad:2011fc,Aad:2013lpa,Aad:2014vwa,Chatrchyan:2012bja,Khachatryan:2015luy,Aaboud:2017dvo,Khachatryan:2016mlc,Aad:2016xcr,ATLAS:2017nah}; see~\cite{NNPDF:2021njg}
for more details.

Concerning the LHC top quark measurements considered in the present
analysis,
they partially overlap, but significantly extend, the top datasets included in
global PDF fits such as \nnpdf{}~\cite{NNPDF:2021njg} as well as in SMEFT analyses of the top quark sector~\cite{Ellis:2020unq,Ethier:2021bye}.
Here we discuss in turn the different types of measurements to be included: inclusive $t\bar{t}$ cross sections and differential distributions; $t\bar{t}$ production asymmetries; the $W$-helicity fractions;
 associated top pair production with vector bosons and heavy quarks, including $\bar{t}t Z$, $\bar{t}t W$, $\bar{t}t \gamma$, $\bar{t}t\bar{t}t$, $\bar{t}t\bar{b}b$;
 $t-$ and $s-$channel single top production;
 and associated single top and vector boson production.

 \paragraph{Choice of kinematic distribution.}
 Many of these measurements, in particular those targeting
 top quark pair production, are available
  differentially in several kinematic variables,
 as well as either absolute distributions, or distributions normalised
 to the fiducial cross-section.
We must decide which of the
 available kinematic distributions associated to
 a given measurement should be included in the fit,
 and whether it is more advantageous to consider
 absolute or normalised distributions.
 
Regarding the former, we note that correlations between
 kinematic distributions are in general not available, and only one
 distribution at a time can be included without double-counting 
 (one exception is the ATLAS $t\bar{t}$ lepton+jet measurement
 at $\sqrt{s}=8$ TeV~\cite{Aad:2015mbv}
 where the full correlation matrix is provided). Therefore, wherever
 possible we include the top-pair invariant mass $m_{t\bar{t}}$ distributions
 with the rationale that these have enhanced sensitivity to SMEFT
 operators via energy-growing effects; they also provide
 direct information on the large-$x$ PDFs.
 Otherwise, we consider the top or top-pair rapidity distributions,
 $y_{t}$ and $y_{t\bar{t}}$ respectively, which also provide
 the sought-for information on the large-$x$ PDFs;
 furthermore they benefit from moderate
higher-order QCD and electroweak corrections~\cite{Czakon:2016olj}.

Regarding the choice of absolute versus normalised distributions, we elect to use normalised
distributions together with corresponding fiducial cross-sections throughout.
Normalised distributions are typically more precise that their absolute counterparts, since
experimental and theoretical errors partially cancel out when normalising.
In addition, normalisation does not affect the PDF and EFT sensitivity of the measurement,
provided the fiducial cross section measurements used for normalising are also accounted for.
%
%
From the implementation point of view, since in a normalised measurement one bin is dependent on the others,
we choose to exclude the bin with lowest $m_{t\bar{t}}$ value (the production threshold) to avoid
losing sensitivity arising from the high-energy tails.

\paragraph{Inclusive $t\bar{t}$ production.}

\begin{table}[h!]
  \begin{center}
{\fontsize{8pt}{8pt}\selectfont
  \centering
   \renewcommand{\arraystretch}{1.5}
   \setlength{\tabcolsep}{2pt}
   \begin{tabular}{lcccccc|c|c}
     \toprule \textbf{Exp.}   & $\bf{\sqrt{s}}$ \textbf{(TeV)}    &   \textbf{Channel}
    &  \textbf{Observable} & $\mathcal{L}$ (fb${}^{-1}$) & $\mathbf{n_{\rm dat}}$ & \textbf{Ref.}
     &\textbf{New (PDF fits)}
    &  \textbf{New (SMEFT fits)}\\
    \toprule
    \multirow{1}{*}{
      \bf ATLAS}
      & 7
      & dilepton
      & $\sigma(t\bar{t})$
      & $4.6$
      & 1
      & \cite{ATLAS:2014nxi}
      &
      & ($\checkmark$)\\\midrule
      & 8
      & dilepton
      & $\sigma(t\bar{t})$
      & $20.3$
      & 1
      & \cite{ATLAS:2014nxi}
      &
      & ($\checkmark$)\\
      & 
      & 
      & $1/\sigma d\sigma/dm_{t\bar{t}}$
      & $20.2$
      & 5
      & \cite{Aaboud:2016iot}
      & ($y_{t\bar{t}} \rightarrow m_{t\bar{t}}$)
      & (absolute $\rightarrow$ ratio)\\
      & 
      & $\ell+$jets
      & $\sigma(t\bar{t})$
      & $20.2$
      & 1
      & \cite{ATLAS:2017wvi}
      & $\checkmark$
      & ($\checkmark$)\\
      & 
      & 
      & $1/\sigma d\sigma/d|y_{t}|$
      & $20.3$
      & 4
      & \cite{Aad:2015mbv}
      &
      & ($m_{t\bar{t}}, p_t^T \rightarrow |y_t|, |y_{t\bar{t}}|$)\\
      & 
      &
      & $1/\sigma d\sigma/d|y_{t\bar{t}}|$
      & $20.3$
      & 4
      & \cite{Aad:2015mbv}
      &
      & ($m_{t\bar{t}}, p_t^T \rightarrow |y_t|, |y_{t\bar{t}}|$)\\\midrule
      & 13
      & dilepton
      & $\sigma(t \bar{t})$
      & $36.1$
      & 1
      & \cite{ATLAS:2019hau}
      & $\checkmark$
      & $\checkmark$\\
      &
      & hadronic
      & $\sigma(t \bar{t})$
      & $36.1$
      & 1
      & \cite{ATLAS:2020ccu}
      & $\checkmark$
      & $\checkmark$\\
      &
      &
      & $1/\sigma d^2\sigma/d|y_{t\bar{t}}|dm_{t\bar{t}}$
      & $36.1$
      & 10
      & \cite{ATLAS:2020ccu}
      & $\checkmark$
      & $\checkmark$\\
      & 
      & $\ell+$jets
      & $\sigma(t \bar{t})$
      & $139$
      & 1
      & \cite{ATLAS:2020aln}
      & 
      & ($\checkmark$) \\
      &
      & 
      & $1/\sigma d\sigma/dm_{t\bar{t}}$
      & $36$
      & 8
      & \cite{Aad:2019ntk}
      & $\checkmark$
      & (absolute $\rightarrow$ ratio)\\
     \midrule \midrule
       \multirow{1}{*}{\bf CMS}      & 5
      & combination
      & $\sigma(t \bar{t})$
      & 0.027
      & 1
      & \cite{CMS:2017zpm}
      & 
      & $\checkmark$\\ \midrule
      & 7
      & combination
      & $\sigma(t \bar{t})$
      & $5.0$
      & 1
      & \cite{Spannagel:2016cqt}
      & 
      & $\checkmark$\\\midrule
      & 8
      & combination
      & $\sigma(t \bar{t})$
      & $19.7$
      & 1
      & \cite{Spannagel:2016cqt}
      & 
      & $\checkmark$\\     
      &
      & dilepton
      & $1/\sigma d^2\sigma/dy_{t\bar{t}}dm_{t\bar{t}}$
      & $19.7$
      & 16
      & \cite{Sirunyan:2017azo}
      & ($m_{t\bar{t}}, y_t \rightarrow m_{t\bar{t}}, y_{t\bar{t}}$)
      & \\
      &
      & $\ell+$jets
      & $1/\sigma d\sigma/dy_{t\bar{t}}$
      & $19.7$
      & 9
      & \cite{Khachatryan:2015oqa}
      & 
      & \\\midrule
	& 13
	& dilepton
	& $\sigma(t \bar{t})$
	& $43$
	  & 1
	  & \cite{CMS:2015yky}
      & 
      & ($\checkmark$)\\
      &
      &
       & $1/\sigma d\sigma/dm_{t\bar{t}}$
       & $35.9$
      & 5
      & \cite{Sirunyan:2018ucr}
      & 
      & (absolute $\rightarrow$ ratio)\\   
	& 
	& $\ell+$jets
	& $\sigma(t \bar{t})$
	& $137$
        & 1
        & \cite{CMS:2021vhb}
        & $\checkmark$
       & $\checkmark$\\
      &
      & 
      & $1/\sigma  d\sigma/dm_{t\bar{t}}$
      & $137$
      & 14
      & \cite{CMS:2021vhb}
        & $\checkmark$
       & $\checkmark$\\     
\bottomrule
   \end{tabular}
   \vspace{0.3cm}
   \caption{\small
    The inclusive cross-sections and differential distributions
    for top quark pair production from ATLAS
    and CMS that we consider in this analysis.
    For each dataset, we indicate the experiment,
    the centre of mass energy $\sqrt{s}$, the final-state
    channel, the observable(s) used in the fit, the integrated luminosity $\mathcal{L}$ in inverse femtobarns, and the
    number of data points $n_{\rm dat}$, together
    with the corresponding publication reference.
    In the last two columns, we indicate
    with a $\checkmark$ the datasets
    that are included for the first time here in a global PDF fit
    and in a SMEFT interpretation, respectively.
    The sets marked with brackets have already been included in
    previous studies but here we account for their constraints
    in different manner (e.g. by changing spectra or normalisation), as indicated in the table and in the
    text description.
    \label{tab:input_datasets_toppair}
}
}
\end{center}
\end{table}


A summary of the inclusive $t\bar{t}$ fiducial cross sections
and differential distributions considered in this work is provided
in Table~\ref{tab:input_datasets_toppair}.
We indicate in each case
 the centre of mass energy $\sqrt{s}$, the final-state
 channel, the observable(s) used in the fit, the luminosity, and the
 number of data points $n_{\rm dat}$, together
 with the corresponding publication reference.
 In the last two columns, we indicate
 with a $\checkmark$ the datasets
 that are included for the first time here in a global PDF fit (specifically, those which
 are new with respect to \nnpdf{})
 and in a SMEFT interpretation (specifically, in comparison
 with the global fits of~\cite{Ellis:2020unq,Ethier:2021bye}).
 The sets marked with brackets have already been included in
 previous studies, but are implemented here in a different manner
 (e.g. by changing spectra or normalisation), as indicated in the
 table; more details are given in each paragraph of the section. 

 The ATLAS dataset comprises six total cross section measurements
 and five differential normalised cross section measurements.
 Concerning the latter, at $8$ TeV we include three distributions
 from the dilepton and $\ell+$jets channels.
 In the $\ell+$jets channel, several kinematic distributions are available
 together with their correlations.
 Following the dataset selection analysis carried out in~\cite{NNPDF:2021njg},
 we select to fit the $y_t$ and $y_{t\bar{t}}$ distributions as done in the
 \nnpdf{} baseline.
 At $13$ TeV, we include the normalised cross sections
 differential in $m_{t\bar{t}}$ from the $\ell+$jets 
 and  hadronic channels, with both measurements being considered
 for the first time here in the context of a PDF analysis.

 Moving to CMS, in the inclusive $t\bar{t}$ category
 we consider five total cross section  and four normalised differential cross section
 measurements.
 At $\sqrt{s}=8$ TeV we include differential distributions in the 
 $\ell+$jets and dilepton channels, the latter being doubly differential
 in $y_{t\bar{t}}$ and $m_{t\bar{t}}$.
 The double-differential 8 TeV measurement is part of \nnpdf{}, but there
 the $(y_{t},m_{t\bar{t}})$ distribution was fitted instead.
 At $13$ TeV, we include the $m_{t\bar{t}}$  distributions
 in the dilepton and $\ell+$jets channels.
 In the latter case we  include the single $m_{t\bar{t}}$ distribution
 rather than the double-differential one in $(m_{t\bar{t}},
 y_{t\bar{t}})$, which is also available, since we find that the
 latter cannot be reproduced by the NNLO SM predictions. We present a
 dedicated analysis of the double-differential distribution in Sect.~5.3.
 As mentioned above, we will study the impact of our dataset selection choices
 by presenting variations of the baseline SM-PDF, fixed-PDF, and SMEFT-PDF analyses
 in the following sections.


 \paragraph{$t\bar{t}$ asymmetry measurements.} The $t\bar{t}$ production asymmetry
 at the LHC is defined as:
 \begin{equation}
   \label{eq:ac}
A_C = \frac{N(\Delta |y| > 0) - N(\Delta |y| < 0)}{N(\Delta |y| > 0) + N(\Delta |y| < 0)} \, ,
\end{equation}
with $N(P)$ being the number of events satisfying the kinematic condition $P$, and $\Delta |y| = |y_t| - |y_{\bar{t}}|$ is the difference between the absolute values of the top quark and anti-top quark rapidities.
The asymmetry $A_C$ can be measured either integrating over the fiducial phase space
or differentially, for example binning in the invariant mass $m_{t\bar{t}}$.
Measurements of $A_C$ are particularly important in constraining certain SMEFT directions,
in particular those associated to the two-light-two-heavy operators.
However, they are unlikely to have an impact on PDF fitting due to their large
experimental uncertainties; nevertheless, with the underlying 
motivation of a comprehensive SMEFT-PDF interpretation
of top quark data, we consider here the $A_C$ measurement as part of our baseline dataset,
and hence study whether or not they also provide relevant PDF information. 
A summary of the asymmetry measurements included in this work is given in Table~\ref{tab:input_datasets_topasymmetries}.

\begin{table}[h!]
  \begin{center}
{\fontsize{8pt}{8pt}\selectfont
  \centering
   \renewcommand{\arraystretch}{1.5}
   \setlength{\tabcolsep}{2pt}
   \begin{tabular}{lcccccc|c|c}
     \toprule \textbf{Experiment}     & $\bf{\sqrt{s}}$\textbf{(TeV)}
     & \textbf{Channel}  &  \textbf{Observable} & $\mathcal{L}$ (fb${}^{-1}$) & $\mathbf{n_{\rm dat}}$ & \textbf{Ref.}   &\textbf{New (PDF fits)}
    &  \textbf{New (SMEFT fits)} \\
    \toprule
      \multirow{2}{*}{ATLAS}
      & 8
      & dilepton
      & $A_C$
      & $20.3$
      & 1
      & \cite{Aad:2016ove}
      & $\checkmark$                                                                     
      &  \\
      & 13
      & $\ell+$jets
      & $A_C$
      & $139$
      & 5
      & \cite{ATLAS:2022waa}
      & $\checkmark$                                                                     
      & $\checkmark$   \\
     \midrule
      \multirow{2}{*}{CMS}
      & 8
      & dilepton
      & $A_C$
      & $19.5$
      & 3
      & \cite{Khachatryan:2016ysn}
      & $\checkmark$                                                                     
      &  \\
      & 13
      & $\ell+$jets
      & $A_C$
      & $138$
      & 3
      & \cite{CMS-PAS-TOP-21-014}
      & $\checkmark$                                                                     
      &  \\
\midrule
      ATLAS/CMS comb.
      & 8
      & $\ell+$jets
      & $A_C$
      & $20$
      & 6
      & \cite{Sirunyan:2017lvd}
      & $\checkmark$                                                                     
      &  \\
\bottomrule
   \end{tabular}
   \vspace{0.3cm}
   \caption{\small Same as Table~\ref{tab:input_datasets_toppair} for the $t\bar{t}$
     asymmetry datasets.
   \label{tab:input_datasets_topasymmetries} \label{tab:asymmetries}
}
}
\end{center}
\end{table}


\paragraph{$W$-helicity fractions.}
The $W$-helicity fractions $F_0, F_L$ and $F_R$ are PDF-independent observables
sensitive to SMEFT corrections, and the dependence of the theory predictions
with respect to the Wilson coefficients can be computed
analytically.
Since these $W$-helicity fractions are PDF-independent observables, to include them
in the joint SMEFT-PDF analysis one has to extend the methodology presented in~\cite{Iranipour:2022iak}
to include in the fit datasets that either lack, or have negligible, PDF sensitivity
and depend only on the EFT coefficients.

In Table~\ref{tab:input_datasets_helicityfractions} we list
the LHC measurements of the $W$-helicity fractions considered in the current analysis.
At $\sqrt{s}=8$ TeV we include the combined ATLAS and CMS measurement from~\cite{Aad:2020jvx},
while at $13$ TeV we consider the ATLAS measurement of the $W$-helicities from~\cite{ATLAS:2022bdg},
for the first time in a SMEFT fit.

\begin{table}[h!]
  \begin{center}
{\fontsize{7pt}{7pt}\selectfont
  \centering
   \renewcommand{\arraystretch}{2}
   \setlength{\tabcolsep}{5pt}
   \begin{tabular}{lccccc|c}
     \toprule \textbf{Experiment}     & $\bf{\sqrt{s}}$\textbf{(TeV)}
     &   \textbf{Observable} & $\mathcal{L}$ (fb${}^{-1}$) & $\mathbf{n_{\rm dat}}$ & \textbf{Ref.}   
    &  \textbf{New (SMEFT fits)} \\
     \toprule
           ATLAS \& CMS combination
      & 8
      & $F_0, F_L$
      & $20$
      & 2
      & \cite{Aad:2020jvx}
      &  \\
 \midrule
       ATLAS
      & 13
      & $F_0, F_L$
      & $139$
      & 2
      & \cite{ATLAS:2022bdg}
      & $\checkmark$     \\
\bottomrule
   \end{tabular}
   \vspace{0.3cm}
  \caption{\small Same as Table~\ref{tab:input_datasets_toppair} for
    the $W$-helicity fraction measurements.
    These helicity fractions are PDF-independent and hence
    are only relevant in constraining the EFT coefficients.
    \label{tab:whelicities}
   \label{tab:input_datasets_helicityfractions}
}
}
\end{center}
\end{table}


\paragraph{Associated top quark pair production.}
The next class of observables that we discuss is associated $t\bar{t}$ production with a $Z$- or a $W$-boson (Table~\ref{tab:input_datasets2}), a photon $\gamma$ (Table~\ref{tab:input_datasets2b}), or a heavy quark pair ($t\bar{t}b\bar{b}$ or $t\bar{t}t\bar{t}$, Table~\ref{tab:input_datasets2c}).
While measurements of $t\bar{t}V$ have been considered for SMEFT interpretations,
we use them for the first time here in the context of a PDF determination.
The rare processes $t\bar{t}\gamma$, $t\bar{t}b\bar{b}$, and $t\bar{t}t\bar{t}$ exhibit
a very weak PDF sensitivity and hence in the present analysis their theory predictions
are obtained using a fixed PDF, in the same manner as the $W$-helicity fractions
in Table~\ref{tab:input_datasets_helicityfractions}.

\begin{table}[h!]
  \begin{center}
{\fontsize{8pt}{8pt}\selectfont
  \centering
   \renewcommand{\arraystretch}{2}
   \setlength{\tabcolsep}{5pt}
   \begin{tabular}{lccccc|c|c}
     \toprule \textbf{Exp.}   & $\bf{\sqrt{s}}$ \textbf{(TeV)}    
    &  \textbf{Observable} & $\mathcal{L}$ (fb${}^{-1}$) & $\mathbf{n_{\rm dat}}$ & \textbf{Ref.}
     &\textbf{New (PDF fits)}
    &  \textbf{New (SMEFT fits)}\\
    \toprule
    {\bf ATLAS}
    & 8
    & $\sigma(t\bar{t}Z)$
    & $20.3$
    & 1
    &  \cite{Aad:2015eua}
      & $\checkmark$                                                                     
      &  \\
    & 
    & $\sigma(t\bar{t}W)$
    & $20.3$
    & 1
    & \cite{Aad:2015eua}
      & $\checkmark$                                                                     
      &  \\\midrule
    & 13
    & $\sigma(t\bar{t}Z)$
    & $36.1$
    & 1
    &\cite{Aaboud:2019njj}
       & $\checkmark$                                                                     
      &  \\
    & 
    & $1/\sigma d\sigma(t\bar{t}Z)/dp_T^Z$
    & $139$
    & 6
    &  \cite{ATLAS:2021fzm}
      & $\checkmark$                                                                     
      & $\checkmark$ \\
    & 
    & $\sigma(t\bar{t}W)$
    & $36.1$
    & 1
    &  \cite{Aaboud:2019njj}
      & $\checkmark$                                                                     
      &  \\
  \midrule
     {\bf  CMS}
    & 8
    & $\sigma(t\bar{t}Z)$
    & $19.5$
    & 1
    & \cite{Khachatryan:2015sha}
       & $\checkmark$                                                                     
      &  \\
    & 
    & $\sigma(t\bar{t}W)$
    & $19.5$
    & 1
    & \cite{Khachatryan:2015sha}
      & $\checkmark$                                                                     
      &  \\ \midrule
    & 13
    & $\sigma(t\bar{t}Z)$
    & $35.9$
    & 1
    & \cite{Sirunyan:2017uzs}
        & $\checkmark$                                                                     
      &  \\
    & 
    & $1/\sigma d\sigma(t\bar{t}Z)/dp_T(Z)$
    & $77.5$
    & 3
    & \cite{CMS:2019too}
      & $\checkmark$                                                                     
      & (absolute $\rightarrow$ ratio) \\
    & 
    & $\sigma(t\bar{t}W)$
    & $35.9$
    & 1
    &  \cite{Sirunyan:2017uzs}
       & $\checkmark$                                                                     
      &  \\
\bottomrule
   \end{tabular}
   \vspace{0.3cm}
   \caption{\small Same as Table~\ref{tab:input_datasets_toppair} for the measurements
     of top quark production
     in association with a vector boson.
     \label{tab:input_datasets2}
   }
}
\end{center}
\end{table}


\begin{table}[h!]
{\fontsize{8pt}{8pt}\selectfont
  \centering
   \renewcommand{\arraystretch}{2}
   \setlength{\tabcolsep}{5pt}
   \begin{tabular}{lccccc|c}
     \toprule \textbf{Experiment}     & $\bf{\sqrt{s}}$\textbf{(TeV)}
     &   \textbf{Observable} & $\mathcal{L}$ (fb${}^{-1}$) & $\mathbf{n_{\rm dat}}$ & \textbf{Ref.}   
    &  \textbf{New (SMEFT fits)} \\
    \toprule
    ATLAS
    & 8
    & $\sigma(t\bar{t}\gamma)$
    & $20.2$
    & 1
    &\cite{Aaboud:2017era}
    & \\
  \midrule
      CMS
    & 8
    & $\sigma(t\bar{t}\gamma)$
    & $19.7$
    & 1
    & \cite{Sirunyan:2017iyh}
    & \\
\bottomrule
   \end{tabular}
   \vspace{0.3cm}
  \caption{\small Same as Table~\ref{tab:input_datasets_toppair} for
     $t\bar{t}$ production in association with a photon.
    \label{tab:input_datasets2b}
    Theory predictions for these observables adopt a fixed PDF.
   }
   }
\end{table}


\begin{table}[h!]
  \begin{center}
{\fontsize{8pt}{8pt}\selectfont
  \centering
   \renewcommand{\arraystretch}{2.3}
   \setlength{\tabcolsep}{5pt}
   \begin{tabular}{lcccccc|c}
     \toprule \textbf{Experiment}     & $\bf{\sqrt{s}}$\textbf{(TeV)}
     &   \textbf{Channel} & \textbf{Observable} & $\mathcal{L}$ (fb${}^{-1}$) & $\mathbf{n_{\rm dat}}$ & \textbf{Ref.}   
    &  \textbf{New (SMEFT fits)} \\
    \toprule
    ATLAS
    & 13
    & multi-lepton
    & $\sigma_{\text{tot}}(t\bar{t}t\bar{t})$
    & $139$
    & 1
    &\cite{ATLAS:2020hpj}
    & \\
    & 
    & single-lepton
    & $\sigma_{\text{tot}}(t\bar{t}t\bar{t})$
    & $139$
    & 1
    & \cite{ATLAS:2021kqb}
    & $\checkmark$ \\
        & 
    & $\ell+$jets
    & $\sigma_{\text{tot}}(t\bar{t}b\bar{b})$
    & $36.1$
    & 1
    & \cite{ATLAS:2018fwl}
    & \\
  \midrule
      CMS
     & 13
    & multi-lepton
    & $\sigma_{\text{tot}}(t\bar{t}t\bar{t})$
    & $137$
    & 1
    & \cite{CMS:2019rvj}
    &  \\
     & 
    & single-lepton
    & $\sigma_{\text{tot}}(t\bar{t}t\bar{t})$
    & $35.8$
    & 1
    & \cite{CMS:2019jsc}
    &  \\
     & 
    & all-jet
    & $\sigma_{\text{tot}}(t\bar{t}b\bar{b})$
    & $35.9$
    & 1
    & \cite{CMS:2019eih}
    &  \\
     & 
    & dilepton
    & $\sigma_{\text{tot}}(t\bar{t}b\bar{b})$
    & $35.9$
    & 1
    & \cite{CMS:2020grm}
    & \\
     & 
    & $\ell$+jets
    & $\sigma_{\text{tot}}(t\bar{t}b\bar{b})$
    & $35.9$
    & 1
    & \cite{CMS:2020grm}
    &$\checkmark$ \\
\bottomrule
   \end{tabular}
   \vspace{0.3cm}
  \caption{\small  Same as Table~\ref{tab:input_datasets_toppair} for
    the measurements of $t\bar{t}$ production in association with a heavy quark
    pair.
    Theory predictions for these observables adopt a fixed PDF.
     \label{tab:input_datasets2c}
   }
}
\end{center}
\end{table}


Concerning the $t\bar{t}Z$ and $t\bar{t}W$ data,
from both ATLAS and CMS we use four fiducial cross section measurements at 8 TeV
and 13 TeV,
and one distribution differential in $p_T^Z$ at 13 TeV.
These measurements are particularly interesting to probe SMEFT coefficients that
modify the interactions between the top quark and the electroweak sector.
For top-quark production associated with a photon, we include the fiducial
cross-section measurements from ATLAS and CMS at 8 TeV; also available is a
differential distribution at 13 TeV from ATLAS binned in the photon transverse
momentum $p_T^\gamma$~\cite{Aad:2020axn}, but we exclude this from our analysis because of the difficulty in
producing SMEFT predictions in the fiducial phase space (in the \fitm{} analysis, its inclusion
is only approximate, and in \smefit{} this distribution is neglected entirely).
Finally, we include fiducial measurements of
$t\bar{t}b\bar{b}$ and $t\bar{t}t\bar{t}$ production at 13 TeV considering
the data with highest luminosity for each available final state.
%

\paragraph{Inclusive single-top pair production.}
The inclusive single-top production data considered here
and summarised in Table~\ref{tab:input_datasets_3}
comprises measurements of
single-top production in the $t$-channel, which have previously been included
in PDF fits~\cite{Nocera:2019wyk,NNPDF:2021njg}, as well as measurements of single-top production in the $s$-channel, which in the context of PDF studies have been implemented for the first time in this study.
For $t$-channel production, we consider the ATLAS and CMS top and anti-top fiducial cross sections
$\sqrt{s}=7,8,$ and 13 TeV, as well as normalised $y_t$ and $y_{\bar{t}}$ distributions
at 7 and 8 TeV (ATLAS) and at 13 TeV (CMS).
For  $s$-channel production, no differential measurements are available and hence
we consider fiducial cross-sections at 8 and 13 TeV from ATLAS and CMS.

\begin{table}[!ht]
  \begin{center}
{\fontsize{8pt}{8pt}\selectfont
  \centering
   \renewcommand{\arraystretch}{2}
   \setlength{\tabcolsep}{5pt}
   \begin{tabular}{lcccccc|c|c}
     \toprule \textbf{Exp.}   & $\bf{\sqrt{s}}$ \textbf{(TeV)}    &   \textbf{Channel}
    &  \textbf{Observable} & $\mathcal{L}$ (fb${}^{-1}$) & $\mathbf{n_{\rm dat}}$ & \textbf{Ref.}
     &\textbf{New (PDF fits)}
    &  \textbf{New (SMEFT fits)}\\
    \toprule
    {\bf ATLAS}
    & 7
    & $t$-channel
    & $\sigma_\text{tot}(t)$ 
    & $4.59$
    & 1
    & \cite{ATLAS:2014sxe}
      & ($\checkmark$)                                                                    
      &   $\checkmark$     \\
    & 
    & 
    & $\sigma_\text{tot}(\bar{t})$ 
    & $4.59$
    & 1
    & \cite{ATLAS:2014sxe}
          & ($\checkmark$)                                                                    
      &   $\checkmark$     \\
    & 
    &
    & $1/\sigma d\sigma(tq)/dy_t$ 
    & $4.59$
    & 3
    & \cite{ATLAS:2014sxe}
     &                                                                  
      & $\checkmark$       \\
    & 
    &
    & $1/\sigma d\sigma(\bar{t}q)/dy_{\bar{t}}$ 
    & $4.59$
    & 3
    & \cite{ATLAS:2014sxe}
     &                                                                
    &  $\checkmark$      \\
    \midrule
    & 8
    & $t$-channel
    & $\sigma_{\text{tot}}(t)$
    & $20.2$
    & 1
    & \cite{Aaboud:2017pdi}
      & ($\checkmark$)                                                                    
      &  $\checkmark$      \\
    & 
    & 
    & $\sigma_{\text{tot}}(\bar{t})$
    & $20.2$
    & 1
    & \cite{Aaboud:2017pdi}
          & ($\checkmark$)                                                                    
      &   $\checkmark$     \\
    & 
    &
    & $1/\sigma d\sigma(tq)/dy_t$ 
    & $20.2$
    & 3
    & \cite{Aaboud:2017pdi}
     &
       & ($\checkmark$)\\
    & 
    &
    & $1/\sigma d\sigma(\bar{t}q)/dy_{\bar{t}}$ 
    & $20.2$
    & 3
    & \cite{Aaboud:2017pdi}
     &
       & ($\checkmark$)\\
    & 
    & $s$-channel
    & $\sigma_{\text{tot}}(t + \bar{t})$
    & $20.3$
    & 1
    & \cite{Aad:2015upn} 
     & $\checkmark$
    & \\
    \midrule
    & 13
    & $t$-channel
    & $\sigma_{\text{tot}}(t)$
    & $3.2$
    & 1
    & \cite{Aaboud:2016ymp}  
     & ($\checkmark$)
     & \\
    & 
    &
    & $\sigma_{\text{tot}}(\bar{t})$
    & $3.2$
    & 1
    &  \cite{Aaboud:2016ymp}  
     & ($\checkmark$)
     & \\
    & 
    & $s$-channel
    & $\sigma_{\text{tot}}(t + \bar{t})$ 
    & $139$
    & 1
    & \cite{ATLAS:2022wfk}
     & $\checkmark$
     & $\checkmark$\\
    \midrule
    \midrule
      {\bf CMS}
     & 7
    & $t$-channel
    & $\sigma_{\text{tot}}(t) + \sigma_{\text{tot}}(\bar{t})$  
    & $1.17, 1.56$
    & 1
    & \cite{CMS:2012xhh}
     &
      & $\checkmark$\\
      \midrule
    & 8
    & $t$-channel
    & $\sigma_{\text{tot}}(t)$
    & $19.7$
    & 1
    & \cite{Khachatryan:2014iya}
     & ($\checkmark$)
     & \\
    & 
    &
    & $\sigma_{\text{tot}}(\bar{t})$
    & $19.7$
    & 1
    & \cite{Khachatryan:2014iya}
     & ($\checkmark$)
     & \\
    & 
    & $s$-channel
    & $\sigma_{\text{tot}}(t + \bar{t})$
    & $19.7$
    & 1
    & \cite{Khachatryan:2016ewo}  
     & $\checkmark$
      & \\
      \midrule
    & 13
    & $t$-channel
    & $\sigma_{\text{tot}}(t)$
    & $2.2$
    & 1
    & \cite{Sirunyan:2016cdg} 
        & ($\checkmark$)
     & \\
    &
    & 
    & $\sigma_{\text{tot}}(\bar{t})$
    & $2.2$
    & 1
    & \cite{Sirunyan:2016cdg} 
      & ($\checkmark$)
     & \\
    &
    & 
    & $1/\sigma d\sigma/d|y^{(t)}|$
    & $35.9$
    & 4
    & \cite{Sirunyan:2019hqb}
     & $\checkmark$
     & \\
\bottomrule
   \end{tabular}
   \vspace{0.3cm}
  \caption{\small Same as Table~\ref{tab:input_datasets_toppair} for
    the inclusive single-top production datasets.
     \label{tab:input_datasets_3}
   }
}
\end{center}
\end{table}
\paragraph{Associated single top-quark production with weak bosons.}
Finally, Table~\ref{tab:input_datasets_4}
lists the measurements of associated single-top production with vector bosons
included in  our analysis.
We consider fiducial cross-sections for $tW$ production at 8 and 13 TeV
from ATLAS and CMS in the dilepton and single-lepton final states,
as well as the $tZj$ fiducial cross-section at 13 TeV from ATLAS and CMS in
the dilepton final state.
In addition, kinematical distributions in $tZj$ production from CMS
at 13 TeV are considered
 for the first time here in an EFT fit.
 For these differential distributions, the measurement is presented binned in either $p_T^Z$ or $p_T^t$;
 here, we take the former as default for consistency with the corresponding $t\bar{t}Z$ analysis.

\begin{table}[h!]
  \begin{center}
{\fontsize{8pt}{8pt}\selectfont
  \centering
   \renewcommand{\arraystretch}{2}
   \setlength{\tabcolsep}{5pt}
   \begin{tabular}{lcccccc|c}
     \toprule \textbf{Experiment}     & $\bf{\sqrt{s}}$\textbf{(TeV)}
     &   \textbf{Channel} & \textbf{Observable} & $\mathcal{L}$ (fb${}^{-1}$) & $\mathbf{n_{\rm dat}}$ & \textbf{Ref.}   
     &  \textbf{New (SMEFT fits)} \\
    \toprule
      {\bf ATLAS}
      & 8
      & dilepton
      & $\sigma_{\text{tot}}(tW)$
      & $20.3$
      & 1
      & \cite{Aad:2015eto}
      & \\
      & 
      & single-lepton
      & $\sigma_{\text{tot}}(tW)$
      & $20.2$
      & 1
      & \cite{Aad:2020zhd}
      & \\
      \midrule
      & 13
      & dilepton
      & $\sigma_{\text{tot}}(tW)$
      & $3.2$
      & 1
      & \cite{Aaboud:2016lpj}
     &  \\
      & 
      & dilepton
      & $\sigma_{\text{fid}}(tZj)$
      & $139$
      & 1
      & \cite{Aad:2020wog}
      & \\
     \midrule
      {\bf CMS}
      & 8
      & dilepton
      & $\sigma_{\text{tot}}(tW)$
      & $12.2$
      & 1
      & \cite{Chatrchyan:2014tua}
      &   \\
      \midrule
      & 13
      & dilepton
      & $\sigma_{\text{tot}}(tW)$
      & $35.9$
      & 1
      & \cite{Sirunyan:2018lcp}
      & \\
      & 
      & dilepton
      & $\sigma_{\text{fid}}(tZj)$
      & $77.4$
      & 1
      & \cite{Sirunyan:2018zgs}
     & \\
      &
      & dilepton
      & $d\sigma_{\text{fid}}(tZj)/dp_T^t$
      & $138$
      & 3
      & \cite{CMS:2021rvz}
      & $\checkmark$\\
      & 
      & single-lepton
      & $\sigma_{\text{tot}}(tW)$
      & $36$
      & 1
      & \cite{CMS:2021vqm}
      & $\checkmark$\\
\bottomrule
   \end{tabular}
   \vspace{0.3cm}
  \caption{\small Same as Table~\ref{tab:input_datasets_toppair} for
    single-top production in association with an electroweak bosons.
}
   \label{tab:input_datasets_4}
}
\end{center}
\end{table}


\clearpage

\paragraph{Dataset selection.}
The top quark production measurements listed in
Tables~\ref{tab:input_datasets_toppair}-\ref{tab:input_datasets_4} 
summarise all datasets that have been considered for the present analysis.
In principle, however, some of these may need to be excluded from the baseline 
fit dataset to ensure that the baseline dataset is maximally consistent.
Following the dataset selection procedure adopted
in~\cite{NNPDF:2021njg}, here our baseline dataset is chosen to exclude datasets that may be either internally
inconsistent or inconsistent with other measurements of the same process type.
These inconsistencies can be of experimental origin, for instance due to
unaccounted (or underestimated) systematic errors,
or numerically unstable correlation models, as well as originating in theory,
for example whenever a given process is affected
by large missing higher-order perturbative uncertainties.
Given that the ultimate goal of a global SMEFT analysis, such as the present one,
is to unveil deviations from the SM, one should strive to deploy
objective dataset selection criteria that exclude datasets affected by such
inconsistencies, which are unrelated to BSM physics.

The first step is to run a global SM-PDF fit including all the datasets
summarised in Tables~\ref{tab:input_datasets_toppair}-\ref{tab:input_datasets_4}
(and additionally a fit with the data summarised therein, but with the CMS measurement of 
the differential $t\bar{t}$ cross-section at $13$ TeV in the $\ell+$jets channel replaced with 
the double-differential measurement) and monitor in each case the following two statistical estimators:

\begin{itemize}
\item The total $\chi^2$ per data point and the number of standard deviations $n_\sigma$ by which the value of the
  $\chi^2$ per data point differs from the median of the $\chi^2$ distribution
  for a perfectly consistent dataset,
  \begin{equation}\label{eq:nsigma}
    n_\sigma\equiv
    \frac{|\chi^2-1|}{\sigma_{\chi^2}}=\frac{|\chi^2-1|}{\sqrt{2/n_{\rm dat}}} \, ,
  \end{equation}
  where the $\chi^2$ in this case (and in the rest of the chapter unless
  specified) is the experimental $\chi^2$ per
  data point, which is defined as
  \begin{equation}\label{eq:chi2exp}
\chi^2 \equiv \chi^2_{\rm exp}/n_{\rm dat} \,=  \,\frac{1}{n_{\rm
    dat}}\,\sum_{i,j=1}^{n_{\rm dat}} (D_i-T_i^0)\,\,\left({\rm cov}_{\rm exp}^{-1}\right)_{ij}  \,(D_j-T_j^0),
  \end{equation}
  where $T_i^0$ are the theoretical predictions computed with the
  central PDF replica, which is the average over the PDF replicas, and
  the experimental covariance matrix is the one defined for example in
  Eq.~(3.1) of Ref.~\cite{Ball:2022uon}.
  
  Specifically, we single out for further examination datasets for which $n_\sigma\ge  3$
  and $\chi^2 \ge 2$ per data point, where the poor description of the data is unlikely
  to be caused by a statistical fluctuation (note that these conditions relax those given 
  in~\cite{NNPDF:2021njg}, which we hope gives the opportunity for the EFT to account for
  poor quality fits to data, rather than immediately attributing poor fits to inconsistencies).
  The question is then to ascertain whether this poor $\chi^2$ can be explained
  by non-zero EFT coefficients (and in such case it should be retained
  for the fit) or if instead there one can find other explanations,
  such as the ones mentioned above, that justify removing it from the baseline
  dataset.

\item The metric $Z$ defined in Ref.~\cite{Kassabov:2022pps} which
quantifies the stability of the $\chi^2$ with respect
to potential inaccuracies affecting the modelling of the experimental correlations.
The calculation of $Z$ relies  exclusively on the experimental covariance matrix
and is independent of the theory predictions.
A large value of the stability metric $Z$ corresponds to
datasets with an unstable covariance matrix, in the sense that small changes
in the values of the correlations between data points lead to large
increases in the corresponding $\chi^2$.
Here we single out for further inspection  datasets with $Z\ge 4$.

As also described in~\cite{Kassabov:2022pps}, it is possible to regularise
covariance matrices in a minimal manner to assess the impact of these numerical
instabilities at the PDF or SMEFT fit level, and determine how
they affect the resulting pre- and post-fit $\chi^2$.
To quantify whether datasets with large $Z$ distort the fit results in a sizable
manner, one can run fit variants applying this decorrelation procedure such that all datasets
exhibit a value of the $Z$-metric below the threshold. We do not find it necessary
to run such fits in this work.

\end{itemize}

\noindent
In  Tables~\ref{tab:datasets_selection_atlas} and~\ref{tab:datasets_selection_cms}
we list the outcome of such a global SM-PDF fit, where entries that lie above the corresponding threshold values for $\chi^2$, $n_{\sigma}$, or $Z$ are highlighted in boldface.
In the last column, we indicate whether the dataset is flagged.
For the flagged datasets, we carry out the following tests to ascertain whether
it should be retained in the fit:

\begin{itemize}

\item For datasets with  $n_\sigma> 3$ and $Z>4$,  we run a fit variant in which the covariance matrix is regularised.
  If, upon regularisation of the covariance matrix, the PDFs are
  stable and both the $\chi^2$ per data point and the $|n_\sigma|$ decrease to a value below the
  respective thresholds of 2.0 and 3.0, we retain the dataset, else we exclude it.

\item For datasets with $\chi^2>2.0$ and $n_\sigma>3$ we carry out a fit variant
  where this dataset is given a very high weight.
  If in this high-weight fit variant the $\chi^2$ and $n_\sigma$ estimators
  improve to the point that their values lie below the thresholds  without
  deteriorating the description of any of the other datasets included
  the dataset is kept, then the specific measurement is not
  inconsistent, it just does not have enough weight compared to the
  other datasets. See Ref.~\cite{NNPDF:2021njg} for a detailed
  discussion on the size of the weight depending on the size of the
  dataset. 
\end{itemize}

\begin{table}[htbp]
  \begin{center}
{\fontsize{8pt}{8pt}\selectfont
  \centering
   \renewcommand{\arraystretch}{2}
   \setlength{\tabcolsep}{5pt}
   \begin{tabularx}{\textwidth}{ccXccccl}
  \toprule
  \textbf{Experiment}
  &\textbf{$\sqrt{s}$ (TeV)}
  &\textbf{Observable, Channel}
  & \textbf{$n_{\rm dat}$}
  & $\chi^2_{\rm exp}/n_{\rm dat}$
  & \textbf{$n_\sigma$}
  & $Z$
  & \textbf{flag}\\
  \midrule
  {\bf ATLAS }
  & 7
  & $\sigma_{t\bar{t}}^{\rm tot}$, dilepton
  & 1
  & {\bf 4.63}  
  & 2.57
  & 1.00
  & no  \\
  & 
  & $\sigma^{\rm tot}(t)$, $t$-channel
  & 1
  &  0.76 
  & -0.17
  & 1.00
  & no  \\
  & 
  & $\sigma^{\rm tot}(\bar{t})$, $t$-channel
  & 1
  &  0.29
  & -0.50
  & 1.00
  & no  \\
  & 
  & $1/\sigma d(tq)/dy_t$, $t$-channel
  & 3
  & 0.97 
  & -0.04
  & 1.28
  & no  \\
  & 
  & $1/\sigma d(\bar{t}q)/dy_{\bar{t}}$, $t$-channel
  & 3
  & 0.06 
  & -1.15
  & 1.39
  & no  \\ \midrule
  & 8
    & $\sigma_{t\bar{t}}^{\rm tot}$, dilepton
  & 1
  & 0.03
  & -0.69
  & 1.00
  & no \\
    &
    & $1/\sigma d\sigma/dm_{t\bar{t}}$, dilepton
  & 5
  & 0.29
  & -1.12
  & 1.61
  & no \\
    &
    & $\sigma_{t\bar{t}}^{\rm tot}$, $\ell+$jets 
  & 1
  & 0.28 
  & -0.51
  & 1.00 
  & no \\
     &
     & $1/\sigma d\sigma/d|y_t|$, $\ell+$jets
  & 4
  & {\bf 2.86} 
  & 2.63 
  & 1.65
  & no \\
  &
     & $1/\sigma d\sigma/d|y_{t\bar{t}}|$, $\ell+$jets 
  & 4
  & {\bf 3.37}
  & {\bf 3.35}
  & 2.19 
  & {\bf yes (kept)} \\
  &
  & $A_C$, dilepton
  & 1
  & 0.67
  & -0.23
  & 1.00
  & no\\
  & 
  & $\sigma(t\bar{t}Z)$
  & 1
  & 0.23
  & -0.54
  & 1.00
  & no \\
  & 
  & $\sigma(t\bar{t}W)$
  & 1
  & {\bf 2.44}
  & 1.01
  & 1.00
  &  no \\
  & 
  & $\sigma^{\rm tot}(t + \bar{t})$, $s$-channel
  & 1
  & 0.21 
  & -0.56
  & 1.00
  & no  \\
  & 
  & $\sigma^{\rm tot}(t W)$, dilepton
  & 1
  & 0.54 
  & -0.33
  & 1.00
  & no  \\
  & 
  & $\sigma^{\rm tot}(t W)$, single-lepton
  & 1
  & 0.71  
  & -0.21
  & 1.00
  & no  \\
  \midrule
 &13
  & $\sigma_{t\bar{t}}^{\rm tot}$, dilepton
  & 1
  & 1.41
  & 0.29
  & 1.00
  & no \\
     &
     & $\sigma_{t\bar{t}}^{\rm tot}$, hadronic
  & 1
  & 0.23 
  & -0.54
  & 1.000
  & no \\
     &
     & $1/\sigma d^2\sigma/d|y_{t\bar{t}}|dm_{t\bar{t}}$, hadronic 
  & 10
  & 1.95 
  & 2.12
  & 2.33
  & no\\ 
     &
     & $\sigma_{t\bar{t}}^{\rm tot}$, $\ell+$jets 
  & 1
  & 0.50
  & -0.35
  & 1.00 
  & no\\ 
     &
     & $1/\sigma d\sigma/dm_{t\bar{t}}$, $\ell+$jets
  & 8
  & 1.83
  & 1.66
  & {\bf 7.61}
  & no\\
  &
  & $A_C$, $\ell+$jets
  & 5
  & 0.99
  & -0.02
  & 1.41
  & no\\
  & 
  & $\sigma(t\bar{t}Z)$
  & 1
  & 0.75
  & -0.18
  & 1.00
  &  no \\
  &
  & $1/\sigma d\sigma(t\bar{t}Z)/dp_T(Z)$
  & 5
  & 1.93
  & 1.47
  & 2.27 
  & no\\
  &
  & $\sigma(t\bar{t}W)$
  & 1
  & 1.43
  & 0.30
  & 1.00 
  & no\\
  & 
  & $\sigma^{\rm tot}(t)$, $t$-channel
  & 1
  & 0.72  
  & -0.20
  & 1.00
  & no  \\
  & 
  & $\sigma^{\rm tot}(\bar{t})$, $t$-channel
  & 1
  & 0.39 
  & -0.43
  & 1.00
  & no  \\
  & 
  & $\sigma^{\rm tot}(t + \bar{t})$, $s$-channel
  & 1
  & 0.70
  & -0.21
  & 1.00
  & no  \\
  & 
  & $\sigma^{\rm tot}(t W)$, dilepton
  & 1
  & 1.15
  & 0.36
  & 1.00
  & no  \\
  \bottomrule
   \end{tabularx}
   \vspace{0.3cm}
   \caption{\small For the ATLAS measurements that we consider in this work,
     we list the outcome of a global SM-PDF fit with all
     measurements listed in
     Tables~\ref{tab:input_datasets_toppair}-\ref{tab:input_datasets_4}
     included.
     We display for each dataset the  number of
data points, the $\chi^2$ per data point (Eq.~\eqref{eq:chi2exp}, the number of
standard deviations $n_\sigma$ (Eq.~\eqref{eq:nsigma}), and the
stability metric $Z$  defined in~\cite{Kassabov:2022pps}.
The entries that lie above the corresponding threshold values
are  highlighted in boldface,
In the
last column, we indicate whether the dataset is flagged and
is either kept or removed.
See text for more details.
   \label{tab:datasets_selection_atlas}
}
}
\end{center}
\end{table}

\begin{table}[htbp]
{\fontsize{8pt}{8pt}\selectfont
  \centering
   \renewcommand{\arraystretch}{2}
   \setlength{\tabcolsep}{5pt}
   \begin{tabularx}{\textwidth}{ccXccccl}
  \toprule
  \textbf{Experiment}
  &\textbf{$\sqrt{s}$ (TeV)}
  &\textbf{Observable}
  & \textbf{$n_{\rm dat}$}
  & $\chi^2_{\rm exp}/n_{\rm dat}$
  & \textbf{$n_\sigma$}
  & $Z$
  & \textbf{flag} \\
  \midrule
  CMS
  & 5
  & $\sigma_{t\bar{t}}^{\rm tot}$, combination
  & 1
  & 0.56
  & -0.31
  & 1.00
  & no \\
  & 7
  & $\sigma_{t\bar{t}}^{\rm tot}$, combination
  & 1
  & 1.08
  & 0.06
  & 1.00
  & no \\
  & 
  & $\sigma^{\rm tot}(t) + \sigma^{\rm tot}(\bar{t})$, $t$-channel
  & 1
  & 0.72
  & -0.20
  & 1.00
  & no \\
  & 8
  & $\sigma_{t\bar{t}}^{\rm tot}$, combination
  & 1
  & 0.27 
  & -0.52
  & 1.00
  & no \\
   & 
  & $1/\sigma d^2\sigma/dy_{t\bar{t}}dm_{t\bar{t}}$, dilepton 
  & 16
  & 0.98
  & -0.06
  & 2.33
  & no \\
   & 
  & $1/\sigma d\sigma/dy_{t\bar{t}}$, $\ell+$jets 
  & 9
  & 1.15
  & 0.31
  & 1.63
  & no \\
  & 
  & $A_C$, dilepton
  & 3
  & 0.05
  & -1.16
  & 1.16
  & no \\
  & 
  & $\sigma(t\bar{t}Z)$
  & 1
  & 0.47
  & -0.37
  & 1.00
  & no\\
  & 
  & $\sigma(t\bar{t}W)$
  & 1
  & {\bf 2.27}
  & 0.90
  & 1.00
  &  no\\
  & 
  & $\sigma^{\rm tot}(t)$, $t$-channel
  & 1
  & 0.01
  & -0.70
  & 1.00
  & no \\
  & 
  & $\sigma^{\rm tot}(\bar{t})$, $t$-channel
  & 1
  & 0.09
  & -0.64
  & 1.00
  & no \\
  & 
  & $\sigma^{\rm tot}(t + \bar{t})$, $s$-channel
  & 1
  & 1.11 
  & 0.08
  & 1.00
  & no \\
  & 
  & $\sigma^{\rm tot}(t W)$, dilepton
  & 1
  & 0.38
  & -0.44
  & 1.00
  & no \\
   & 13
  & $\sigma_{t\bar{t}}^{\rm tot}$, dilepton  
  & 1
  & 0.06 
  & -0.66
  & 1.00
  & no \\
   & 
  & $1/\sigma d\sigma/dm_{t\bar{t}}$, dilepton  
  & 5
  & {\bf 2.49}
  & 2.36 
  & 1.61
  & no\\
   & 
  & $\sigma_{t\bar{t}}^{\rm tot}$, $\ell+$jets channel
  & 1
  & 0.22
  & -0.55
  & 1.00
  & no \\
   & 
  & $1/\sigma d\sigma/dm_{t\bar{t}}$, $\ell+$jets
  & 14
  & 1.41 
  & 1.08
  & {\bf 4.57}
  & no \\
   & 
  & $1/\sigma d\sigma/dm_{t\bar{t}}dy_{t}$, $\ell+$jets
  & 34
  & {\bf 6.43} 
  & 22.4
  & 3.88
  & {\bf yes (excl)} \\
  & 
  & $A_C$, $\ell+$jets
  & 3
  & 0.29
  & -0.87
  & 1.00
  & no \\
  & 
  & $\sigma(t\bar{t}Z)$
  & 1
  & 1.24
  & 0.17
  & 1.00
  &  no \\
  & 
  & $1/\sigma d\sigma(t\bar{t}Z)/dp_T(Z)$
  & 3
  & 0.59
  & -0.50
  & 1.28
  &  no\\
  & 
  & $\sigma(t\bar{t}W)$
  & 1
  & 0.66
  & -0.24
  & 1.00
  &  no \\
  & 
  & $\sigma^{\rm tot}(t)$, $t$-channel
  & 1
  & 0.88
  & -0.08
  & 1.00
  &  no \\
  & 
  & $\sigma^{\rm tot}(\bar{t})$, $t$-channel
  & 1
  & 0.13 
  & -0.62
  & 1.00
  &  no \\
  & 
  & $1/\sigma d\sigma/d|y^{(t)}|$, $t$-channel
  & 4
  & 0.38
  & -0.88
  & 1.70
  &  no \\
  & 
  & $\sigma^{\rm tot}(tW)$, dilepton
  & 1
  & 0.43  
  & -0.40
  & 1.00
  &  no \\
  & 
  & $\sigma^{\rm tot}(tW)$, single-lepton
  & 1
  & {\bf 2.84} 
  & 1.30
  & 1.00
  &  no \\
  \midrule
  ATLAS-CMS combination
  & 8
  & $A_C$, $\ell$+jets
  & 6
  & 0.602
  & -0.69
  & 1.65
  & no \\
  \bottomrule
   \end{tabularx}
   \vspace{0.3cm}
  \caption{\small Same as Table \ref{tab:datasets_selection_atlas} for the
    CMS and combined ATLAS-CMS datasets. \textit{Note carefully:} the row
    corresponding to the CMS doubly-differential distribution at 13 TeV in the $\ell+$jets
    channel comes from a separate fit, where the corresponding 1D distribution is replaced
    by this dataset.
   \label{tab:datasets_selection_cms}
}
}
\end{table}


From the analysis of Tables~\ref{tab:datasets_selection_atlas} and~\ref{tab:datasets_selection_cms},
one finds that only two datasets in the inclusive
top quark pair production (lepton+jets final state) category are flagged as potentially
problematic: the ATLAS $|y_{t\bar{t}}|$ distribution at 8 TeV
and the CMS double-differential distributions in $m_{t\bar{t}}$ and $y_{t}$
at 13 TeV.
The first of these was already discussed in the \nnpdf{}
analysis~\cite{NNPDF:2021njg}. It was observed that each of the
four distributions measured by ATLAS and presented in
Ref.~\cite{Aad:2015mbv} behave somewhat differently upon being
given large weight. The $\chi^2$ of all distributions significantly improves when given large weight in the fit, as described in \cite{NNPDF:2021njg}. However, while for the top transverse momentum and top pair invariant mass distributions this improvement is accompanied by a rather significant deterioration of the global fit quality, in the case of the top and top pair rapidity distributions the global fit quality is very similar and only the description of jets deteriorates moderately. The rapidity distributions thus remain largely compatible with the rest of the dataset, hence they are kept.

    %
    Also shown in one row of Table~\ref{tab:datasets_selection_cms} is the 
    fit-quality information for the CMS double-differential distribution at 13 TeV in the $\ell+$jets 
    channel, from a separate fit wherein the CMS single differential distribution at 13 TeV in the 
    $\ell+$jets channel is replaced by this dataset. We find that the
    2D set is described very poorly, with a $\chi^2=6.43$,
    corresponding to a $22\sigma$ deviation from the median of the $\chi^2$ distribution
  for a perfectly consistent dataset. To investigate this further, we performed a weighted fit; however, we find that the $\chi^2$
    improves only moderately (from $\chi^2$=
    6.43 to $\chi^2$ = 4.56) and moreover the $\chi^2$-statistic of the other datasets deteriorates
    significantly (with total $\chi^2$ jumping from 1.20 to 1.28). The test
    indicates that the double-differential distribution is both incompatible
    with the rest of the data and also internally inconsistent given the
    standard PDF fit. Hence we exclude this dataset from our baseline and
    include instead the single-differential distribution in $m_{t\bar{t}}$,
    which is presented in the same publication~\cite{CMS:2021vhb} and is
    perfectly described in the baseline fit. Now, we proceed to review the theory predictions that we will use to compare with the experimental measurements.

\paragraph{SM cross-sections.}
Theoretical predictions for SM cross-sections are evaluated at NNLO in perturbative QCD, whenever
available, and at NLO otherwise.
Predictions accurate to NLO QCD are obtained in terms of fast interpolation grids from {\sc\small MadGraph5\_aMC@NLO}~\cite{Frederix:2018nkq,Alwall:2014hca}, 
interfaced to {\sc\small APPLgrid}~\cite{Carli:2010rw} or {\sc\small FastNLO}~\cite{Kluge:2006xs,Wobisch:2011ij,Britzger:2012bs} together 
with {\sc\small aMCfast}~\cite{Bertone:2014zva} and {\sc\small APFELcomb}~\cite{Bertone:2016lga}.
Wherever available, NNLO QCD corrections to matrix elements are  implemented  by multiplying the NLO predictions by bin-by-bin $K$-factors, see Sect.~2.3 in~\cite{Ball:2014uwa}. The potential shift of SM parameters due to SMEFT contributions is neglected for the present study.
The top mass is set to $m_t = 172.5\ \text{GeV}$ for all processes considered.

In the case of inclusive $t\bar{t}$ cross sections and charge asymmetries,
a dynamical scale choice of $\mu_R = \mu_F = H_T/4$
is adopted, where $H_T$ denotes the sum of the transverse masses of the top and anti-top,
following the recommendations of Ref.~\cite{Czakon:2016dgf}.
This scale choice ensures that
the ratio of fixed order NNLO predictions to the NNLO+NNLL ones is minimised,
allowing us to neglect theory uncertainties associated to missing higher orders beyond NNLO. 
To obtain the corresponding
NNLO $K$-factors, we use the {\sc\small HighTEA} public software~\cite{hightea}, 
an event database for distributing and analysing the results of fixed order NNLO
calculations for LHC processes.
The NNLO PDF set used in the computation of these $K$-factors is either 
{\tt NNPDF3.1} or \nnpdf{}, depending on whether a given dataset was already included
in the \nnpdf{} global fit or not, respectively.

For associated $t\bar{t}$ and $W$ or $Z$ production, dedicated fast NLO grids  have been generated.
Factorisation and renormalisation scales are fixed to $\mu_F = \mu_R = m_t + \frac{1}{2}m_V$, where $m_V = m_W, m_Z$ is the mass of the associated weak boson,
as appropriate.
This scale choice follows the recommendation of Ref.~\cite{Kulesza:2018tqz} and minimises the ratio of the NLO+NLL over the fixed-order NLO prediction.
We supplement the predictions for the total cross section for associated $W$ and $Z$-production at 13 TeV with NLO+NNLL QCD $K$-factors taken from Table 1 of~\cite{Kulesza:2018tqz}.
On the other hand, the $t\bar{t}\gamma$, $t\bar{t}t\bar{t}$ and $t\bar{t}b\bar{b}$ data are implemented as PDF independent observables, and the corresponding theory predictions are taken directly from the relevant experimental papers in each case.

The evaluation of theoretical predictions for single top production follows~\cite{Nocera:2019wyk}.
Fast NLO interpolation grids are generated for both  $s$- and $t$-channel
single top-quark and top-antiquark datasets in the 5-flavour scheme,  with fixed
factorisation and renormalisation scales set to $m_t$.
Furthermore, 
for the $t$-channel production we include the NNLO QCD corrections to
both total and differential cross sections~\cite{Berger:2017zof}.
When the top decay is calculated, it is done in the narrow-width approximation, under which the QCD corrections to the top-(anti)quark production
and the decay are factorisable and the full QCD corrections are approximated by the vertex corrections.

\paragraph{SMEFT cross-sections.}
SMEFT corrections to SM processes are computed both at LO and at NLO in QCD, 
and both at the linear and the quadratic level in the EFT expansion.
Flavour assumptions follow the LHC TOP WG prescription of~\cite{Aguilar-Saavedra:2018ksv}
which were also used in the recent {\sc\small SMEFiT} analysis~\cite{Ethier:2021bye}.
The flavour symmetry group is
given by $U(3)_l \times U(3)_e \times U(3)_d \times U(2)_u \times U(2)_q$, i.e. we single out operators that contain
top quarks (right-handed $t$ and $SU(2)$ doublet $Q$).
This also means that one works in a five-flavour scheme in which the
only massive fermion in the theory is the top.
As far as the electroweak input scheme is concerned, we work in the $m_W$-scheme, meaning that the $4$
electroweak inputs are
$\{m_W, G_F, m_h, m_Z\}$ (a more systematic assessment of the interplay between input schemes and calculations in the SMEFT can be found in Ref. \cite{Biekotter:2023xle}). Masses and wavefunctions are renormalised in the on-shell scheme, while the strong coupling and operator coefficients are renormalised in the $\overline{\text{MS}}$ scheme. More details can be found in~\cite{Degrande:2020evl}.

In the context of our study in the top sector, 25 SMEFT operators at dimension-six modify the SM Lagrangian, as described in Refs.~\cite{Aguilar-Saavedra:2018ksv,Ethier:2021bye,Ellis:2020unq}. We shall use the same notation as in~\cite{Ethier:2021bye}. In this work we neglect renormalisation group effects on the Wilson coefficients~\cite{Aoude:2022aro}. 
For hadronic data, i.e. for proton-proton collisions, which are the only data affected by the SMEFT in this study,
 the linear effect of the $n$-th SMEFT operator on a theoretical prediction can be quantified by:
\begin{equation}
  \label{eq:mult_k_fac_app1}
   R_{\rm SMEFT}^{(n)} \equiv \displaystyle \left( {\cal L}_{ ij}^{\rm NNLO} \otimes d\widehat{\sigma}_{ij,{\rm SMEFT}}^{(n)}\right)
 \big/ \left( {\cal L}_{ ij}^{\rm NNLO} \otimes d\widehat{\sigma}_{ij,{\rm SM}} \right) \, , \quad
 n=1\,\ldots, N \, ,
\end{equation}
where $i, j$ are parton indices, ${\cal L}_{ ij}^{\rm NNLO}$ is the NNLO partonic luminosity defined as 
\begin{equation}
\label{eq:lumidef}
\mathcal{L}_{ij}(\tau,M_X) = \int_{\tau}^1 \frac{d x}{x}~f_i (x,M_X) f_j (\tau/x,M_X) ~, \quad \tau=M_X^2/s,
\end{equation}
$d\widehat{\sigma}_{ij,{\rm SM}}$ the bin-by-bin partonic SM cross section, and $d\widehat{\sigma}_{ij,{\rm SMEFT}}^{(n)}$
the corresponding partonic cross section associated to the interference between 
$\mathcal{O}_n$ and the SM amplitude $\mathcal{A}_{\rm SM}$ when setting $c_n = 1$. This value of $c_n$ 
is only used to initialize the potential contributions of the
SMEFT operator; the effective values of the Wilson coefficient 
are found after the fit is performed. 
%

The computation of the SMEFT contributions is
performed numerically with the FeynRules~\cite{Alloul:2013bka} model SMEFTatNLO~\cite{Degrande:2020evl}, which allows one to include NLO QCD corrections to the observables. The obtained cross sections are then combined in so-called \textit{BSM factors} by taking the ratio with the respective SM cross sections, in order to 
produce $R_{\rm SMEFT}^{(n)}$.

With these considerations, we can account for SMEFT effects in our theoretical predictions
by mapping the SM prediction $T^{\rm SM}$ to 
\begin{equation}
\label{eq:theory_k_fac_app3}
 T = T^{\rm SM} \times  K(\{c_n\})\, , 
\end{equation}
with 
\begin{equation}
\label{eq:theory_k_fac}
    K(\{c_n\}) = 1+\sum_{n=1}^{N} c_n R_{\rm SMEFT}^{(n)}
    \, .
\end{equation}
Eq. (\ref{eq:theory_k_fac_app3}) is at the centre of the \simunet{} methodology, which we will use in the following sections. However, first we will explore
how SM PDFs are affected by the inclusion of data in the top sector.

\subsection{Global dataset and theory}
\label{subsec:data}

The dataset explored in this study builds on those already implemented in the high-mass Drell-Yan analysis 
presented in Refs.~\cite{Greljo:2021kvv,Iranipour:2022iak} and on the
top quark analysis presented in Ref.~\cite{Kassabov:2023hbm}. 
These studies extended the datasets included in the global \nnpdf{}~\cite{NNPDF:2021njg} analysis, 
by adding observables that enhance the sensitivity to the SMEFT and that constrain PDFs in the Drell-Yan and in the 
top sector respectively. 

The \nnpdf{} NNLO analysis~\cite{NNPDF:2021njg} included $n_{\rm data}\,=\,4618$ data points corresponding to a wide variety of processes in
deep-inelastic lepton-proton scattering (from the HERA $ep$ collider and from fixed-target experiments such as NMC, SLAC, 
BCDMS, CHORUS and NuTeV), fixed-target DY data from the
Fermilab E605 and E866 experiments, hadronic proton-antiproton collisions from the Tevatron, and proton-proton collisions from LHC. 
The LHC data in turn include inclusive gauge boson production data; $Z$- and $W$-boson transverse momentum production data, 
single-inclusive jet and di-jets production data, as well as gauge boson with jets and inclusive isolated photon
production data and top data. 
In Refs.~\cite{Greljo:2021kvv,Iranipour:2022iak} the high-mass Drell-Yan sector of the \nnpdf{} analysis 
was augmented by two extra measurements taken by CMS at $\sqrt{s}=8,13$ TeV, for a total of 65 extra datapoints. 
In Ref.~\cite{Kassabov:2023hbm}, besides inclusive and differential $t\bar{t}$ cross sections and $t$-channel 
single top production already implemented in \nnpdf{}, more observables were included, such as $t\bar{t}$ production asymmetries, 
$W$-helicity fractions, associated top pair production with vector bosons and heavy quarks, including 
$\bar{t}t Z$, $\bar{t}t W$, $\bar{t}t \gamma$, $\bar{t}t\bar{t}t$, $\bar{t}t\bar{b}b$, $s$-channel 
single top production, and associated single top and vector boson production, bringing the total number of datapoints to $n_{\rm data}\,=\,4710$. 

In this present study, we further extend the datasets by probing electroweak precision observables (EWPOs), 
the Higgs sector and the diboson sector. 
The datasets are described below and details such as the centre-of-mass energy, the observable, the integrated luminosity, the number 
of data points, the dataset name as implemented in \simunet \ and the source are given 
in Tables~\ref{tab:dataset_EWPO},~\ref{tab:dataset_Higgs} and~\ref{tab:dataset_diboson} respectively. 

We remind the reader that, thanks to the functionality of \simunet{} that enables users to include PDF-independent observables 
as well as to freeze PDFs for the observables in which the PDF dependence is mild, we can add observables such as the measurement of 
$\alpha_e$ that are completely independent of PDFs (but which are affected by SMEFT corrections) or Higgs signal strengths and Simplified Template Cross Section (STXS) measurements that 
are mildly dependent on PDFs but are strongly dependent on the relevant SMEFT-induced corrections. 

To summarise, in this analysis we include a total of {\bf 4985 datapoints}, of which 366 are affected by SMEFT corrections. 
Among those 366 datapoints, 210 datapoints are PDF independent. 
Hence our pure PDF analysis performed with \simunet{} includes $(4985-210) = 4775$ datapoints and is equivalent to the 
\nnpdf{} global analysis augmented by 78 datapoints from the top sector~\cite{Kassabov:2023hbm} and 
by 65 datapoints measuring the high-mass Drell-Yan tails~\cite{Greljo:2021kvv,Iranipour:2022iak}. Our pure 
SMEFT-only analysis performed with \simunet{} includes 366 datapoints, while our simultaneous PDF-SMEFT fit includes 
4985 datapoints. 

\vspace*{0.4cm}
\noindent {\bf EWPOs:}
In Table~\ref{tab:dataset_EWPO}, we list EWPOs included in this study. 
The dataset includes the pseudo-observables
measured on the $Z$ resonance by LEP~\cite{ALEPH:2005ab}, 
which include cross sections, forward-backward and polarised asymmetries. 
Branching ratios of the decay of the $W$ boson into leptons~\cite{ALEPH:2013dgf} are also included, 
along with the LEP measurement of the Bhabha scattering cross section~\cite{ALEPH:2013dgf}. 
Finally we include the measurement of the effective electromagnetic coupling constant~\cite{Workman:2022ynf}, 
for a total of {\bf 44 datapoints}.  These datasets and their SMEFT predictions
are taken from the public \smefit{} code
\cite{Giani:2023gfq,terHoeve:2023pvs}, where linear EFT corrections, LO cross sections, and flavour universality were assumed.
These datapoints are all PDF independent, hence they directly 
affect only the SMEFT part of the fits.
\begin{table}[htb!]
    \begin{center}
  {\fontsize{8pt}{8pt}\selectfont
    \centering
     \renewcommand{\arraystretch}{2}
     \setlength{\tabcolsep}{5pt}
     \begin{tabular}{lcccccc}
       \toprule \textbf{Exp.}   & $\bf{\sqrt{s}}$ \textbf{(TeV)}
      &  \textbf{Observable} & $\mathcal{L}$ (fb${}^{-1}$) & $\mathbf{n_{\rm dat}}$ & \textbf{Dataset name}
       &\textbf{Ref.}\\
      \toprule
        \bf{LEP}
        & 0.250
        & Z observables
        & 
        & 19
        & LEP\_ZDATA
        & \cite{ALEPH:2005ab}\\
      \midrule
        \bf{LEP}
        & 0.196
        & $\mathcal{B}(W \rightarrow l^{-} \bar{v}_l)$
        & 3
        & 3
        & LEP\_BRW
        & \cite{ALEPH:2013dgf}\\
      \midrule
        \bf{LEP}
        & 0.189
        & $\sigma(e^+ e^- \rightarrow e^+ e^-)$
        & 3
        & 21
        & LEP\_BHABHA
        & \cite{ALEPH:2013dgf}\\
      \midrule
        \bf{LEP}
        & 0.209
        & $\hat{\alpha}^{(5)}_{\rm}(M_Z)$
        & 3
        & 1
        & LEP\_ALPHAEW
        & \cite{Workman:2022ynf}\\
  
  \bottomrule
     \end{tabular}
     \vspace{0.3cm}
     \caption{Measurements of electroweak precision observables included in \simunet{}. The columns contain information on the 
     centre-of-mass energy, the observable, the integrated luminosity, the number 
     of data points, the dataset name as implemented in \simunet \ and the source.
      \label{tab:dataset_EWPO}
  }
  }
  \end{center}
  \end{table}

\vspace*{0.4cm}
\noindent {\bf Higgs sector:}
In Table~\ref{tab:dataset_Higgs}, we list the Higgs sector datasets included in this study. 
The Higgs dataset at the LHC includes the combination of Higgs signal strengths
by ATLAS and CMS for Run 1, and for Run 2 both signal strengths and STXS
measurements are used. ATLAS in particular provides the combination of stage 1.0 STXS bins
for the $4l$, $\gamma\gamma$, $WW$, $\tau\tau$ and $b\bar{b}$ decay channels, while for CMS we use the combination
of signal strengths of the $4l$,
$\gamma\gamma$, $WW$, $\tau^-\tau^+$, $b\bar{b}$ and $\mu^-\mu^+$ decay channels. 
We also include the $H \rightarrow Z\gamma$ 
and $H \rightarrow \mu^-\mu^+$ signal strengths from ATLAS, for a total of {\bf 73 datapoints}. 
The Run I and CMS datasets and their corresponding predictions are
taken from the \smefit{} code~\cite{Giani:2023gfq,terHoeve:2023pvs},
whereas the STXS observables and signal strength measurements of
$H \rightarrow Z\gamma$ and $H \rightarrow \mu^-\mu^+$ are taken from the \fitmaker ~\cite{Ellis:2020unq} code.
The signal strength's dependence on the PDFs is almost completely negligible, given that the 
PDF dependence cancels in the ratio, hence all the datapoints are treated as PDF independent and 
are directly affected only by the SMEFT Wilson coefficients.
\begin{table}[htb!]
    \begin{center}
  {\fontsize{8pt}{8pt}\selectfont
    \centering
     \renewcommand{\arraystretch}{2}
     \setlength{\tabcolsep}{5pt}
     \begin{tabular}{lcccccc}
       \toprule \textbf{Exp.}   & $\bf{\sqrt{s}}$ \textbf{(TeV)}
      &  \textbf{Observable} & $\mathcal{L}$ (fb${}^{-1}$) & $\mathbf{n_{\rm dat}}$ & \textbf{Dataset name}
       &\textbf{Ref.}\\
      \toprule
        \bf{ATLAS and CMS}
        & 7 and 8
        & $\mu_{H \rightarrow \mu^+ \mu^-}$
        & 5 and 20
        & 22
        & ATLAS\_CMS\_SSinc\_RunI
        & \cite{ATLAS:2016neq}\\
      \midrule
        \bf{CMS}
        & 13
        & $\mu_{H}$
        & 35.9
        & 24
        & CMS\_SSINC\_RUNII
        & \cite{CMS:2018uag}\\
      \midrule
        \bf{ATLAS}
        & 13
        & $\mu_{H}$
        & 80
        & 25
        & ATLAS\_STXS\_RUNII
        & \cite{ATLAS:2019nkf}\\
      \midrule
        \bf{ATLAS}
        & 13
        & $\mu_{H \rightarrow Z \gamma}$
        & 139
        & 1
        & ATLAS\_SSINC\_RUNII\_ZGAM
        & \cite{ATLAS:2020qcv}\\
      \midrule
        \bf{ATLAS}
        & 13
        & $\mu_{H \rightarrow \mu^+ \mu^-}$
        & 139
        & 1
        & ATLAS\_SSINC\_RUNII\_MUMU
        & \cite{ATLAS:2020fzp}\\

  \bottomrule
     \end{tabular}
     \vspace{0.3cm}
     \caption{Same as~\ref{tab:dataset_EWPO} for the measurements of the Higgs sector.
      \label{tab:dataset_Higgs}  }
  }
  \end{center}
  \end{table}

\vspace*{0.4cm}
\noindent {\bf Diboson sector:}
In Table~\ref{tab:dataset_diboson}, we list the diboson sector datasets included in this study. 
We have implemented four measurements from LEP at low energy, three from ATLAS and one from CMS both at 13 TeV. 
The LEP measurements are of the differential 
cross-section of $e^+ e^- \rightarrow W^+ W^-$ as a function of the cosine of the $W$-boson polar angle. 
The ATLAS measurements are of 
the differential cross-section of $W^+ W^-$ as a function of the invariant mass of the electron-muon pair 
and of the transverse mass of the $W$-boson. 
The third ATLAS measurement is of the differential cross-section of $Zjj$ as a function of the azimuthal angle 
between the two jets, $\Delta \phi_{jj}$.  Although this is not a diboson 
observable, it is grouped here as this observable constrains operators typically 
constrained by the diboson sector,
such as the triple gauge interaction $O_{WWW}$. The CMS dataset is a measurement of the differential cross-section of $WZ$ as 
a function of the transverse momentum 
of the $Z$-boson, for a total of {\bf 82 datapoints}. 
Almost all datasets and corresponding predictions in this sector are
taken from the \smefit{} code~\cite{Giani:2023gfq, terHoeve:2023pvs},
except for the ATLAS measurement of $Zjj$ production, taken from the \fitmaker ~\cite{Ellis:2020unq} code.
While the LEP measurements are PDF-independent by definition, 
the LHC measurements of diboson cross sections do depend on PDFs. However, their impact on the 
PDFs is extremely mild, due to much more competitive constraints in the same Bjorken-$x$ region given by other measurements
such as single boson production and single boson production in association with jets. 
Therefore we consider all diboson measurements 
to be PDF-independent, i.e. the PDFs are fixed and their parameters are not varied in the computation of these observables. 
\begin{table}[htb!]
    \begin{center}
  {\fontsize{8pt}{8pt}\selectfont
    \centering
     \renewcommand{\arraystretch}{2}
     \setlength{\tabcolsep}{5pt}
     \begin{tabular}{lcccccc}
       \toprule \textbf{Exp.}   & $\bf{\sqrt{s}}$ \textbf{(TeV)}
      &  \textbf{Observable} & $\mathcal{L}$ (fb${}^{-1}$) & $\mathbf{n_{\rm dat}}$ & \textbf{Dataset name}
       &\textbf{Ref.}\\
      \toprule
        \bf{LEP}
        & 0.182
        & $d \sigma _{WW} / d cos(\theta _W)$
        & 0.164
        & 10
        & LEP\_EEWW\_182GEV
        & \cite{ALEPH:2013dgf}\\
      \midrule
        \bf{LEP}
        & 0.189
        & $d \sigma _{WW} / d cos(\theta _W)$
        & 0.588
        & 10
        & LEP\_EEWW\_189GEV
        & \cite{ALEPH:2013dgf}\\
      \midrule
        \bf{LEP}
        & 0.198
        & $d \sigma _{WW} / d cos(\theta _W)$
        & 0.605
        & 10
        & LEP\_EEWW\_198GEV
        & \cite{ALEPH:2013dgf}\\
      \midrule
        \bf{LEP}
        & 0.206
        & $d \sigma _{WW} / d cos(\theta _W)$
        & 0.631
        & 10
        & LEP\_EEWW\_206GEV
        & \cite{ALEPH:2013dgf}\\
      \midrule
        \bf{ATLAS}
        & 13
        & $d \sigma _{W^+W^-}/d m_{e \mu}$
        & 36.1
        & 13
        & ATLAS\_WW\_13TeV\_2016\_MEMU
        & \cite{ATLAS:2019rob}\\
      \midrule
        \bf{ATLAS}
        & 13
        & $d \sigma _{WZ} / d m_{T}$
        & 36.1
        & 6
        & ATLAS\_WZ\_13TeV\_2016\_MTWZ
        & \cite{ATLAS:2019bsc}\\
      \midrule
        \bf{ATLAS}
        & 13
        & $d \sigma(Zjj)/d \Delta \phi_{jj}$
        & 139
        & 12
        & ATLAS\_Zjj\_13TeV\_2016
        & \cite{ATLAS:2020nzk}\\
      \midrule
        \bf{CMS}
        & 13
        & $d \sigma _{WZ} / d p_{T}$
        & 35.9
        & 11
        & CMS\_WZ\_13TeV\_2016\_PTZ
        & \cite{ATLAS:2020nzk}\\

  \bottomrule
     \end{tabular}
     \vspace{0.3cm}
     \caption{Same as~\ref{tab:dataset_EWPO} for the measurements of the diboson sector.
      \label{tab:dataset_diboson}
  }
  }
  \end{center}
  \end{table}

\vspace*{0.4cm}
\noindent {\bf SMEFT Wilson coefficients:}
We include a total of 40 operators from the dimension-6 SMEFT in our global analysis 
of the measurements of the
Higgs, top, diboson and electroweak precision observables outlined above.
\begin{figure}[thb!]
    \centering
    \includegraphics[width=0.7\textwidth]{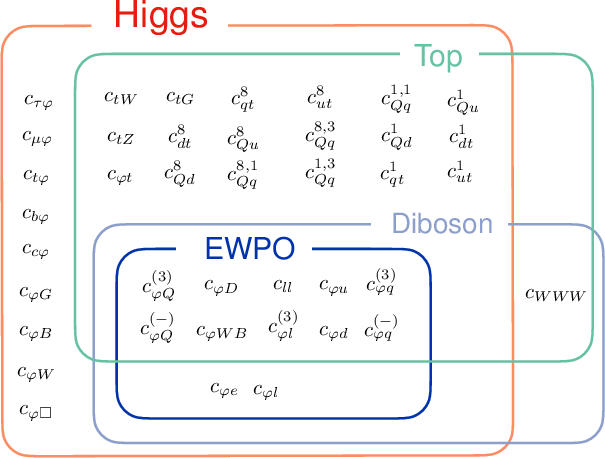}
    \caption{Schematic representation of the data categories included in this analysis and of 
    their overlapping dependences on the 40 dim-6 SMEFT operators included in our global analysis, 
    adapted from Ref.~\cite{Ellis:2020unq}.}
\label{fig:operators}
\end{figure}
In Fig.~\ref{fig:operators} we display the SMEFT Wilson coefficients included in our analysis, following
the operator conventions of Refs.~\cite{Kassabov:2023hbm,Ethier:2021bye}. 
The schematic diagram, adapted from Ref.~\cite{Ellis:2020unq}, demonstrates the sectors of data affected by each WC and highlights
the interconnected nature of the global SMEFT determination.
The overlaps between the different rectangles show explicitly how a given operator contributes to several data categories.
For example, $O_{WWW}$, contributes to both the diboson sector and the top sector (through its contribution to $tZq$ production),
while the operators $O_{\varphi e}$,$O_{\varphi l}$ contribute to the Higgs, diboson 
and electroweak sectors, but have no effect on the top sector.
\section{Impact of the top quark Run II dataset on the SM-PDFs}
\label{sec:baseline_sm_fits}

Here we present the results of a global SM-PDF determination which accounts for the constraints
of the most extensive top quark dataset considered to date
in such analyses and described in Sect.~\ref{sec:exp}.
The fitting methodology adopted follows closely the settings of the \nnpdf{}
study~\cite{NNPDF:2021njg}, see also~\cite{Ball:2022uon} for a rebuttal of some critical arguments.
This dataset includes not only the most up-to-date measurements of top quark
pair production from Run II, but it also includes all available single top
production cross-sections and distributions 
and for the first time new processes not considered in PDF studies before,
such as the $A_C$ asymmetry in $t\bar{t}$ production and the $t\bar{t}V$ and
$tV$ associated production (with $V = Z,W$).

We begin by summarising the methodological settings used for these
fits in Sect.~\ref{sec:setting}.
Then in Sect.~\ref{sec:individual} we assess the impact of adding to
a no-top baseline PDF fit various subsets of the top quark data
considered in this study. 
In particular, we assess the impact of including updated Run II $t\bar{t}$
and single-top measurements in comparison
with the subset used in the \nnpdf{} analysis, see the second-to-last
column of Tables~\ref{tab:input_datasets_toppair}--
\ref{tab:input_datasets_3}.
Furthermore, we quantify for the first time the impact of
associated vector boson and single-top ($tV$) as well as associated vector boson and
top-quark pair production ($t \bar{t} V$) data in a PDF fit.
Finally in Sect.~\ref{sec:alltop} we combine these results and present a
\nnpdf{} fit variant including all data described in Sect.~\ref{sec:exp},
which is compared to both the \nnpdfnotop baseline and to the original
\nnpdf{} set.

\subsection{Fit settings}
\label{sec:setting}

An overview of the SM-PDF fits that are discussed in this section is
presented in Table~\ref{tab:fit_list}. 
First of all, we produce a baseline fit, denoted
by \nnpdfnotop, which is based on the same dataset as \nnpdf{}
but with all top quark measurements excluded.
Then we produce fit variants A to G, which quantify the impact of
including in this baseline various subsets of
top data (indicated by a check mark in the table).
Finally, fit variant H is the main result of this section, namely the fit
in which the full set of top quark measurements described in
Sect.~\ref{sec:exp} is added to the no-top baseline.

In these fits, methodological settings such as network architecture,
learning rates, and other hyperparameters are kept the same as in
\nnpdf{}, unless otherwise specified.
One difference is the training fraction $f_{\rm tr}$ defining
the training/validation split used for the cross-validation stopping criterion.
 In \nnpdf{}, we used $f_{\rm tr}=0.75$  for all datasets.
 Here instead we adopt $f_{\rm tr}=0.75$ only for the no-top datasets
 and $f_{\rm tr}=1.0$ instead for the top datasets.
 The rationale of this choice is to ensure that the fixed-PDF 
 SMEFT analysis, where overfitting is not possible~\cite{Ethier:2021bye},
 exploits all the information contained in the top quark data considered
 in this study\footnote{The top quark dataset represents only a small fraction of the total dataset. When working with such a small dataset, the risk of overfitting becomes negligible. Consequently, it is standard practice in scenarios like this to utilise the entire dataset for training.},  and then for consistency to maintain the same settings in
 both the SM-PDF fits (this section) and
 in the joint SMEFT-PDF fits (to be discussed in Sect.~\ref{sec:joint_pdf_smeft}).
Nevertheless, we have verified that the resulting SM-PDF fits are statistically equivalent in terms of global and individual dataset fit quality, distance metrics, and the fulfilment of sum rules, among other criteria, to the fits obtained by setting the training fraction to be $0.75$ for all data, including for the top quark observables.

Fits A--G in Table~\ref{tab:fit_list}  are composed by $N_{\rm rep}=100$ Monte Carlo replicas after post-fit selection criteria, while the \nnpdfnotop baseline fit and fit H are instead composed by $N_{\rm rep}=1000$  replicas.
As customary, all fits presented in this section are iterated with respect to the $t_0$ PDF set and the pre-processing exponents.

\begin{table}[tb]
  \begin{center}
  \centering
  \tiny
   \renewcommand{\arraystretch}{1.5}
   \setlength{\tabcolsep}{5pt}
  \begin{tabular}{l|C{1.5cm}C{1.5cm}C{1.5cm}C{1.5cm}C{1.5cm}C{1.5cm}C{1.5cm}}
  \multirow{2}{*}{\textbf{Fit ID}}    &  \multicolumn{6}{c}{\textbf{Datasets included in fit}}  \\
   & \textbf{No-top baseline, Sect.~\ref{sec:baseline_data}} & \textbf{Incl. $t\bar{t}$, Table~\ref{tab:input_datasets_toppair}} & \textbf{Asymm., Table~\ref{tab:input_datasets_topasymmetries}} & \textbf{Assoc. $t\bar{t}$, Table~\ref{tab:input_datasets2}} & \textbf{Single-$t$, Table~\ref{tab:input_datasets_3}} & \textbf{Assoc. single$-t$, Table~\ref{tab:input_datasets_4}}  \\
    \midrule
    \nnpdfnotop (Baseline) & \checkmark  \\
    A (inclusive $t\bar{t}$) & \checkmark & \checkmark \\
    B (inclusive $t\bar{t}$ and charge asymmetry) & \checkmark & \checkmark & \checkmark \\
    C (single top) & \checkmark & & & & \checkmark \\
    D (all $t\bar{t}$ and single top) & \checkmark & \checkmark & \checkmark & & \checkmark \\
    E (associated $t\bar{t}$) & \checkmark & & & \checkmark & & \\
    F (associated single top) & \checkmark & & & & & \checkmark \\
    G (all associated top) & \checkmark & & & \checkmark & & \checkmark \\
    H (all top data) & \checkmark & \checkmark & \checkmark & \checkmark & \checkmark & \checkmark \\
    \bottomrule
  \end{tabular}
  \vspace{0.3cm}
  \caption{Overview of the SM-PDF fits discussed in this section.
    The baseline fit, \nnpdfnotop, is based on the same dataset as \nnpdf
    with all top quark measurements excluded.
    The fit variants A to G consider the impact of including in this baseline
    various subsets of top data, while in fit H the full set of top quark measurements described in Sect.~\ref{sec:exp} is added to the baseline.
     \label{tab:fit_list}
  }
  \end{center}
\end{table}


\subsection{Impact of individual top quark datasets}
\label{sec:individual}

First we assess the impact of specific subsets of LHC top quark data when added
to the \nnpdfnotop baseline, fits A--G in Table~\ref{tab:fit_list}. 
In the next section we discuss the outcome of fit H, which contains the full
top quark dataset considered in this work.

\begin{figure}[t]
        \centering
	\begin{subfigure}[t]{0.49\textwidth}
	\includegraphics[scale=0.5]{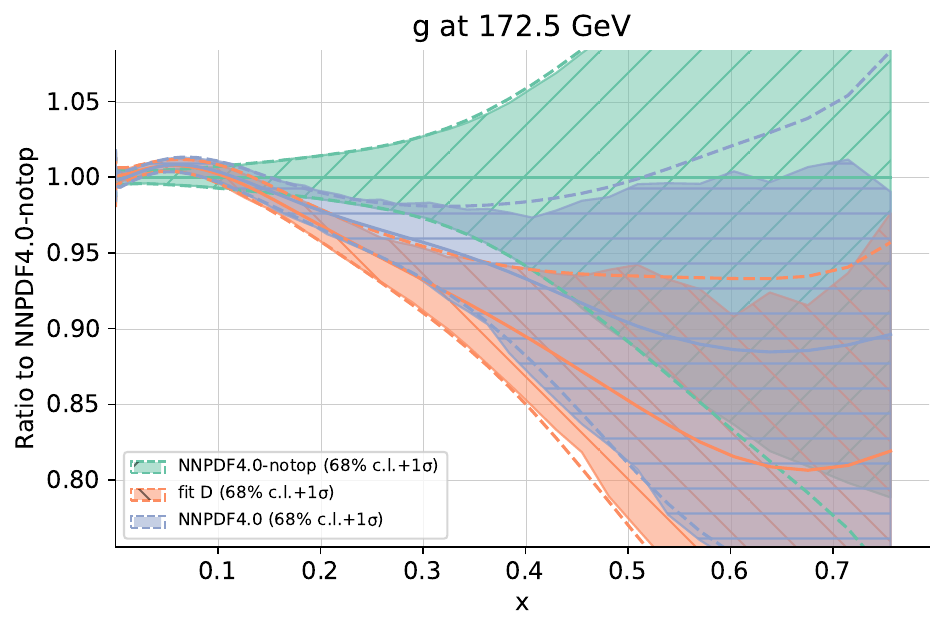}
	\end{subfigure}
	\begin{subfigure}[b]{0.49\textwidth}
	\includegraphics[scale=0.5]{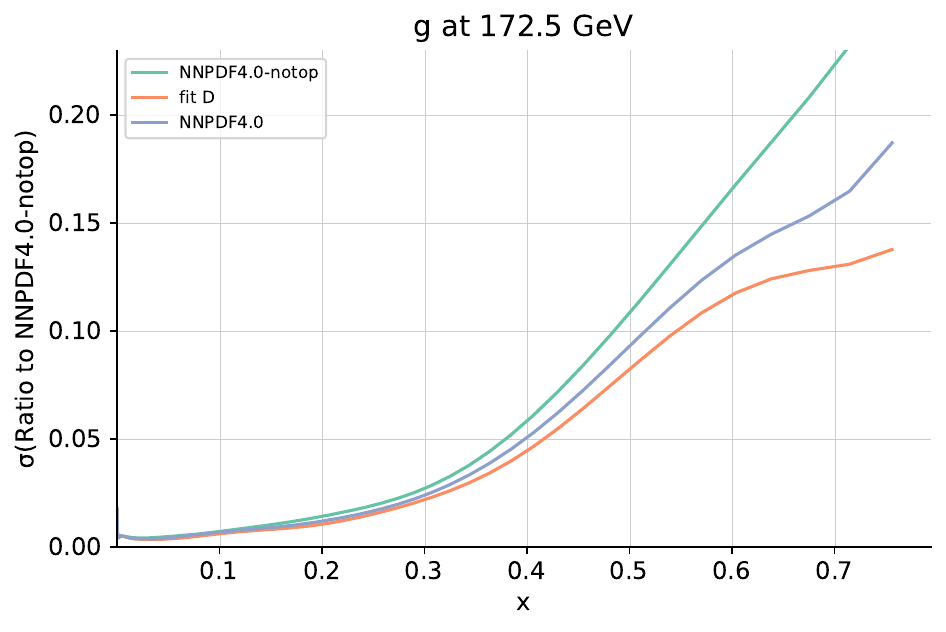}
	\end{subfigure}
        \caption{Left: comparison between the gluon PDF at $Q=m_t=172.5$ GeV
          obtained in the \nnpdf{} and \nnpdfnotop fits
          against fit D in Table~\ref{tab:fit_list}, which includes all
          top-quark
          pair (also the charge asymmetry $A_C$)
          and single-top quark production data considered in this analysis.
          Results are normalised to the central value of the \nnpdfnotop set.
          Right: same comparison now for the PDF uncertainties (all
          normalised to the central value of the \nnpdfnotop set).
}
\label{fig:fitD_gluon}
\end{figure}

Fig.~\ref{fig:fitD_gluon}
displays the comparison between the gluon PDF at $Q=m_t=172.5$ GeV obtained
in the \nnpdf{} and \nnpdfnotop fits against fit D, which includes all top-quark
pair (also the charge asymmetry $A_C$) and all single-top quark production
data considered in this analysis.
Results are normalised to the central value of the \nnpdfnotop fit,
and in the right panel we show the corresponding PDF uncertainties, all 
normalised to the central value of the \nnpdfnotop baseline. 
%
%
From Fig.~\ref{fig:fitD_gluon} one finds that the main impact of the
additional LHC Run II $t\bar{t}$ and single-top data included in fit D as compared to that
already present in \nnpdf{} is a further depletion of the large-$x$ gluon
PDF as compared to the \nnpdfnotop baseline, together with a reduction of
the PDF uncertainties in the same kinematic region.
%
While fit D and \nnpdf{} agree within uncertainties in the whole range of $x$, fit D and \nnpdfnotop exhibit some deviation in the region $x\approx [0.2,0.4]$.
These findings imply that the effect on the SM-PDFs of the new Run II top
data is consistent with, and strengthens, that of the data already part of
\nnpdf{}, and suggests a possible tension
between top quark data and other measurements in the global PDF sensitive
to the large-$x$ gluon, in particular inclusive jet and di-jet production.
The reduction of the gluon PDF uncertainties from the new measurements
can be as large as about 20\% at $x\approx 0.4$.
Differences are much reduced for the quark PDFs, and restricted
to a 5\% to 10\% uncertainty reduction in the region around $x\sim 0.2$
with central values essentially unchanged.

\begin{figure}[h!]
        \centering
	\begin{subfigure}[b]{0.49\textwidth}
	        \includegraphics[scale=0.48]{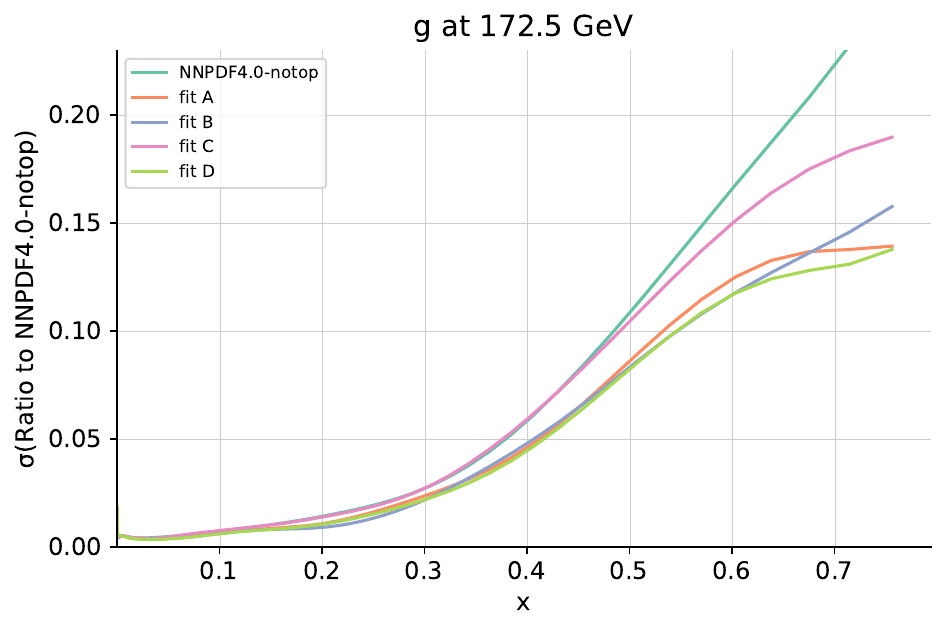}
	\end{subfigure}
	\begin{subfigure}[b]{0.49\textwidth}
	        \includegraphics[scale=0.48]{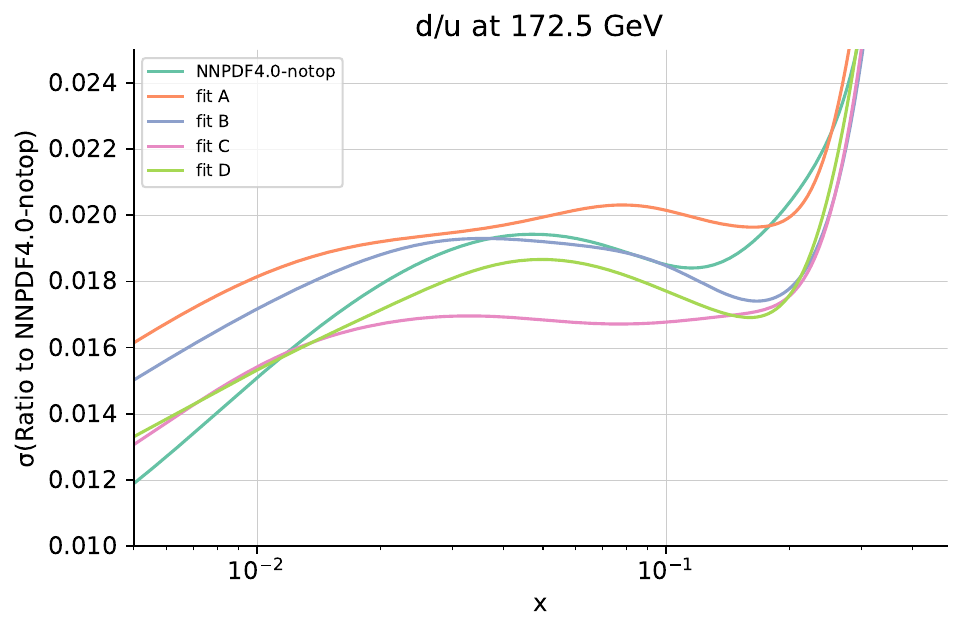}
	\end{subfigure}
        \caption{The ratio between the PDF 1$\sigma$ uncertainty and the
          central value of the \nnpdfnotop baseline in the case of the gluon
          (left panel) and in the case of the $d/u$ ratio (right panel).
          The uncertainty of  the  baseline fit at $Q=m_t=172.5$ GeV
          is compared with the uncertainty associated with fits A, B, C, and
          D in Table~\ref{tab:fit_list}.
          These fit variants include the following top datasets:  inclusive $t\bar{t}$ (A),
          inclusive $t\bar{t}$ + $A_C$ (B), single top (C), and their sum (D).}
\label{fig:fitABCD}
\end{figure}

To disentangle which of the processes
considered dominates the observed effects on the gluon and the light
quarks PDFs, Fig.~\ref{fig:fitABCD} compares the relative PDF uncertainty
on the gluon and on the $d/u$ quark ratio in the \nnpdfnotop 
baseline fit at $Q=m_t=172.5$ GeV with the results from fits A, B, C, and D.
As indicated in Table~\ref{tab:fit_list}, these fit variants include the following top datasets:  inclusive $t\bar{t}$ (A),
inclusive $t\bar{t}$ + $A_C$ (B), single top (C), and their sum (D).
The inclusion of the top charge asymmetry $A_C$ data does not have any
impact on the PDFs; indeed fits A and B are statistically equivalent.
This is not surprising, given that in  Eq.~\eqref{eq:ac} the dependence
on PDFs cancels out.
Concerning the inclusion of single top data (fit C), it does not affect
the gluon PDF but instead leads to a moderate reduction on the PDF
uncertainties on the light quark PDFs in the intermediate-$x$ region,
$x\approx[0.01,0.1]$, as shown in the right panel
displaying the relative uncertainty reduction for the $d/u$ ratio.
This observation agrees with what was pointed out by a previous
study~\cite{Nocera:2019wyk}, and the impact of  LHC single-top
measurements is more marked now
as expected since the number of data points considered here is larger.
We conclude that the inclusive $t\bar{t}$ measurements dominate
the impact on the large-$x$ gluon observed in Fig.~\ref{fig:fitD_gluon}, with
single top data moderately helping to constrain the light quark PDFs
in the intermediate-$x$ region. 

We now consider the effect of the inclusion of data that were not included before in any PDF fit, namely either $t\bar{t}$ or single-top production in association with a weak vector boson.
Although current data exhibits large experimental and theoretical uncertainties, it is interesting to
study whether  they impact PDF fits at all, in view of their increased precision expected in future measurements; in particular, it is useful to know which parton flavours are most affected.

\begin{figure}[t]
\centering
	\begin{subfigure}[b]{0.49\textwidth}
	\includegraphics[scale=0.5]{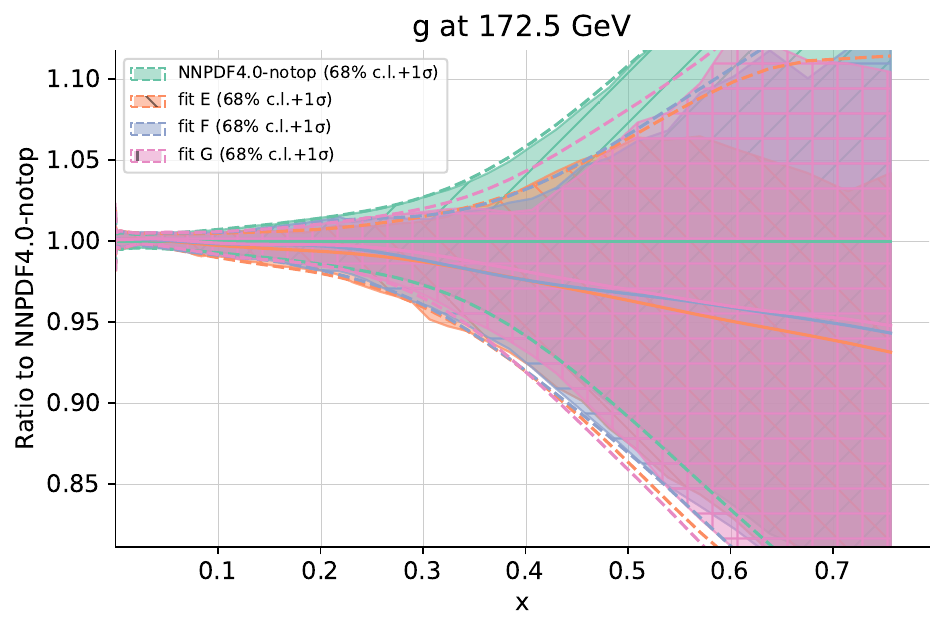}
	\end{subfigure}
	\begin{subfigure}[b]{0.49\textwidth}
	\includegraphics[scale=0.5]{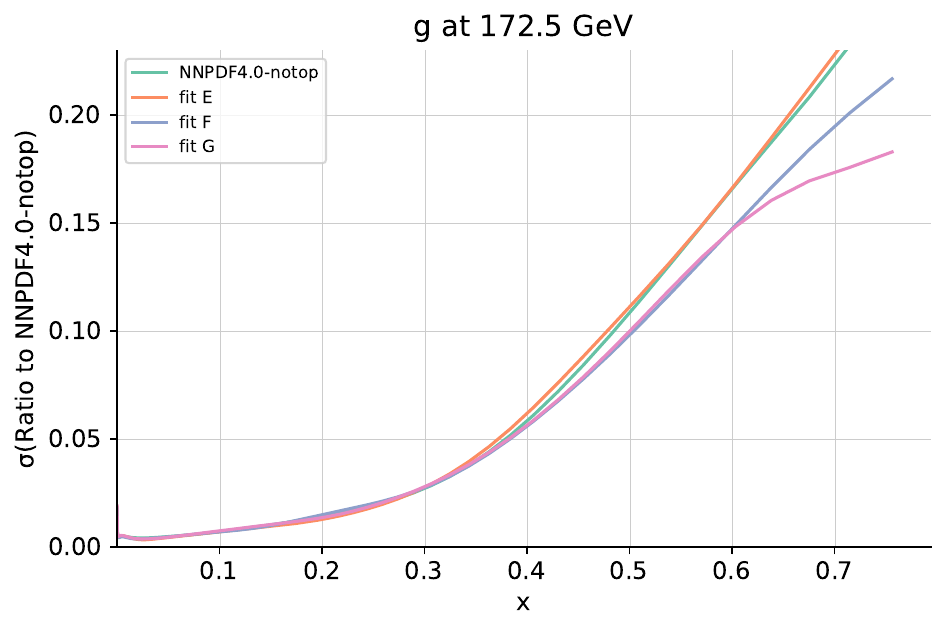}
	\end{subfigure}
        \caption{Same as Fig.~\ref{fig:fitD_gluon} comparing the \nnpdfnotop
          baseline fit with variants E, F, and G from Table~\ref{tab:fit_list}.
          These variants include associated $t\bar{t}$ and vector boson
          production data (E), associated single top and vector boson
          production data (F) and their sum (G).
        }
\label{fig:fitEFG}
\end{figure}

Fig.~\ref{fig:fitEFG} displays the same comparison as in
Fig.~\ref{fig:fitD_gluon} now for the \nnpdfnotop baseline
and the variants E, F, and G from Table~\ref{tab:fit_list}, which 
include the $t\bar{t}V$ (E) and $tV$ (F) data
as well as their sum (G).
The pull of $t\bar{t}V$ is very small, while the pull of the $tV$
measurements is in general small, but consistent
with those of the inclusive  $t\bar{t}$ measurements, namely preferring a
depletion of the large-$x$ gluon.
This result indicates that $t\bar{t}V$  and $tV$  data may be
helpful in constraining PDF once both future experimental data 
and theoretical predictions become more precise, although
the corresponding inclusive measurements are still expected to provide
the dominant constraints. 

\subsection{Combined effect of the full top quark dataset}
\label{sec:alltop}

\begin{figure}[t]
\centering
	\begin{subfigure}[b]{0.49\textwidth}
	\includegraphics[scale=0.5]{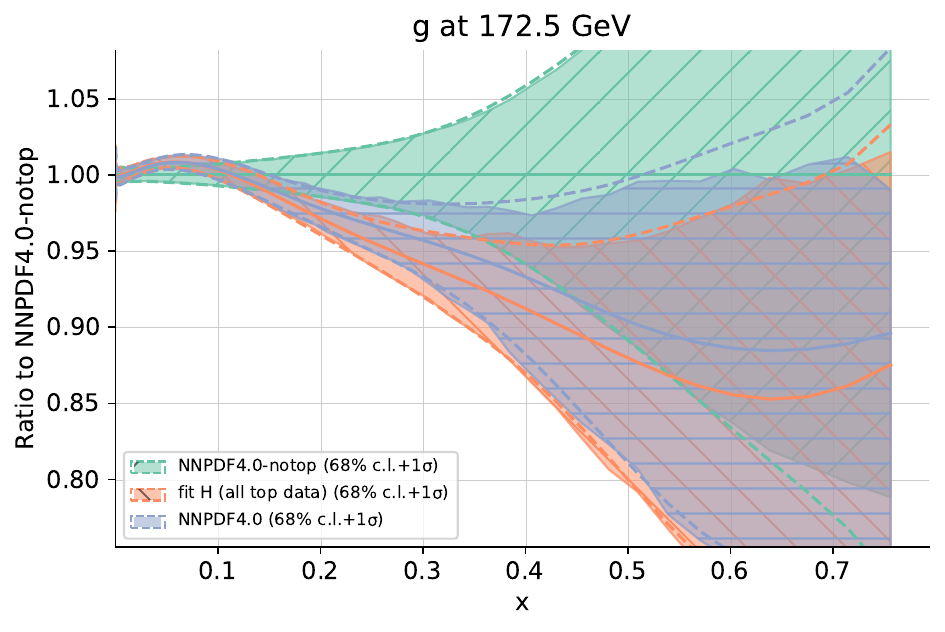}
	\end{subfigure}
	\begin{subfigure}[b]{0.49\textwidth}
	\includegraphics[scale=0.5]{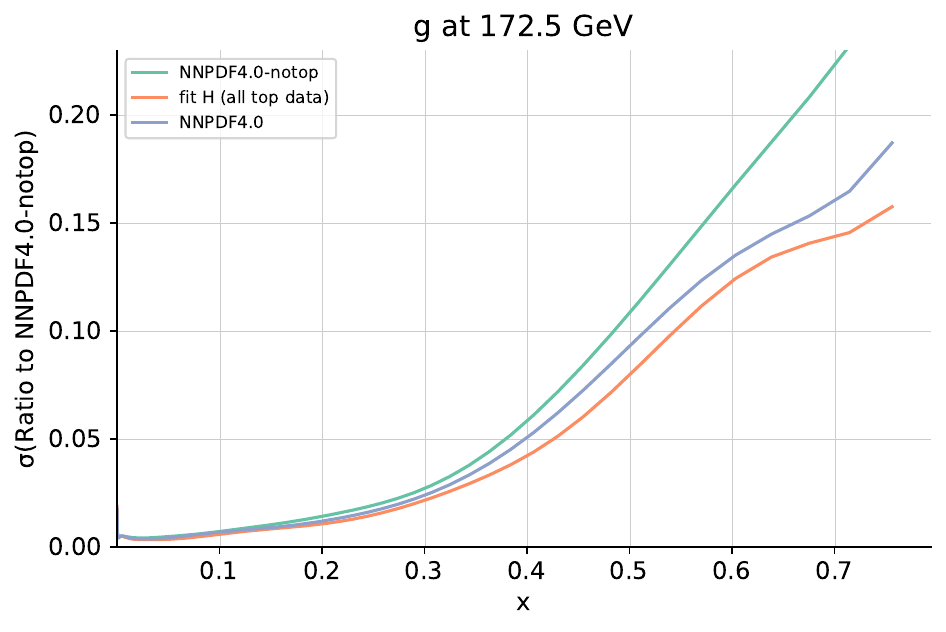}
	\end{subfigure}
        \caption{ Same as Fig.~\ref{fig:fitD_gluon} comparing NNPDF4.0
          and NNPDF4.0 (no top) with fit variant H in Table~\ref{tab:fit_list},
          which includes the full set of top quark measurements considered in this analysis.}
\label{fig:fitH_gluon}
\end{figure}

The main result of this section is displayed in Fig.~\ref{fig:fitH_gluon},
which compares the \nnpdf{} and the \nnpdfnotop fits with variant H in
Table~\ref{tab:fit_list},
namely with the fit where the full set of top quark measurements considered
in this analysis has been added to the no-top baseline.
As in the case of Fig.~\ref{fig:fitD_gluon}, we show the large-$x$ gluon
normalised to the central value of \nnpdfnotop and the associated
1$\sigma$ PDF uncertainties (all normalised to the central value of the
baseline).
The results of fit H are similar to those of fit D, although slightly
more marked. This is expected, 
since as shown above the associated production datasets $t\bar{t}V$
and $tV$ carry little weight in the fit.

From Fig.~\ref{fig:fitH_gluon} one observes
how the gluon PDF of fit H deviates from the \nnpdfnotop baseline
markedly in the data region $x \in [0.1, 0.5]$.
The shift in the gluon PDF can be up to the $2\sigma$ level, and
in particular the two PDF uncertainty bands do not overlap
in the region $x\in \lc 0.2, 0.35\rc $.
As before, we  observe that the inclusion of
the latest Run II top quark measurements enhances the effect of the top
data already
included in \nnpdf{}, by further depleting the gluon in the large-$x$ region
 and  by reducing its uncertainty by a factor up to 25\%.
Hence, one finds again that the new Run II top quark production
measurements lead to
a strong pull on the large-$x$ gluon, qualitatively
consistent but stronger as compared
with the pulls associated from the datasets already included in \nnpdf{}.

\begin{figure}[t]
\centering
\includegraphics[width=0.49\textwidth]{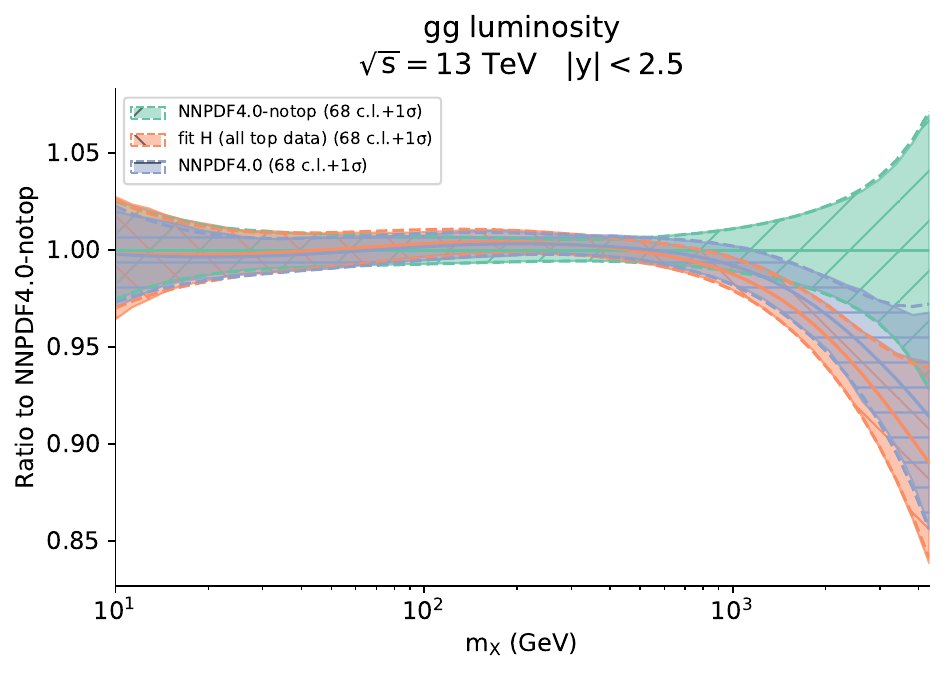}
\includegraphics[width=0.49\textwidth]{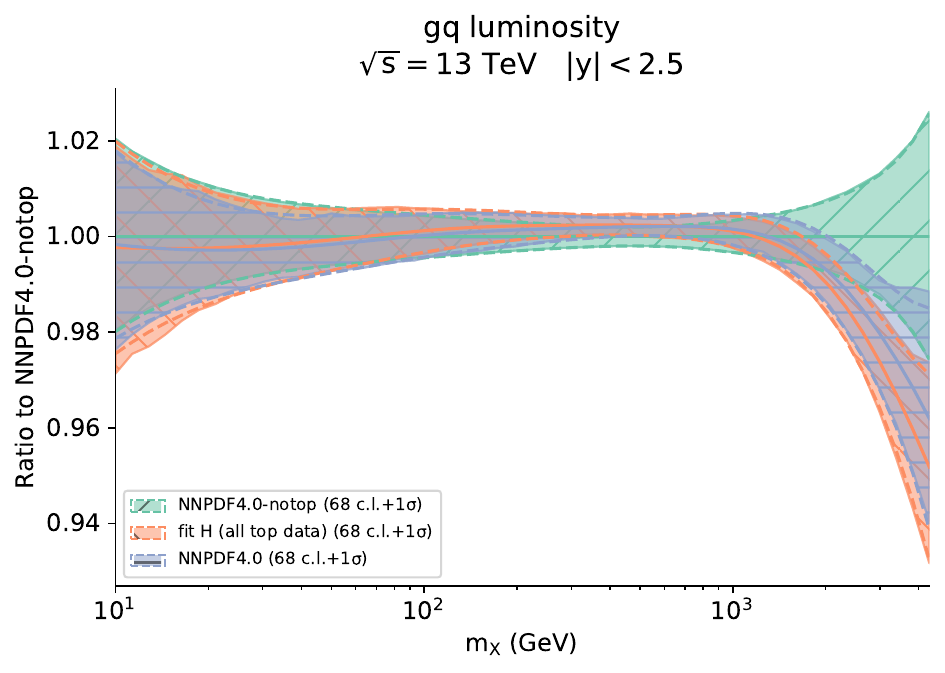}
\caption{The gluon-gluon (left) and quark-gluon (right panel)
  partonic luminosities at $\sqrt{s}=13$ TeV restricted to
  the central acceptance region with $|y|\le 2.5$.
  We compare the \nnpdf{} and \nnpdfnotop fits with the
  predictions based on fit H, which includes the full top quark dataset
  considered here.
  Results are presented as the ratio to the \nnpdfnotop baseline fit.
}
\label{fig:fith_luminosities}
\end{figure}

\begin{figure}[t]
\centering
\includegraphics[width=0.49\textwidth]{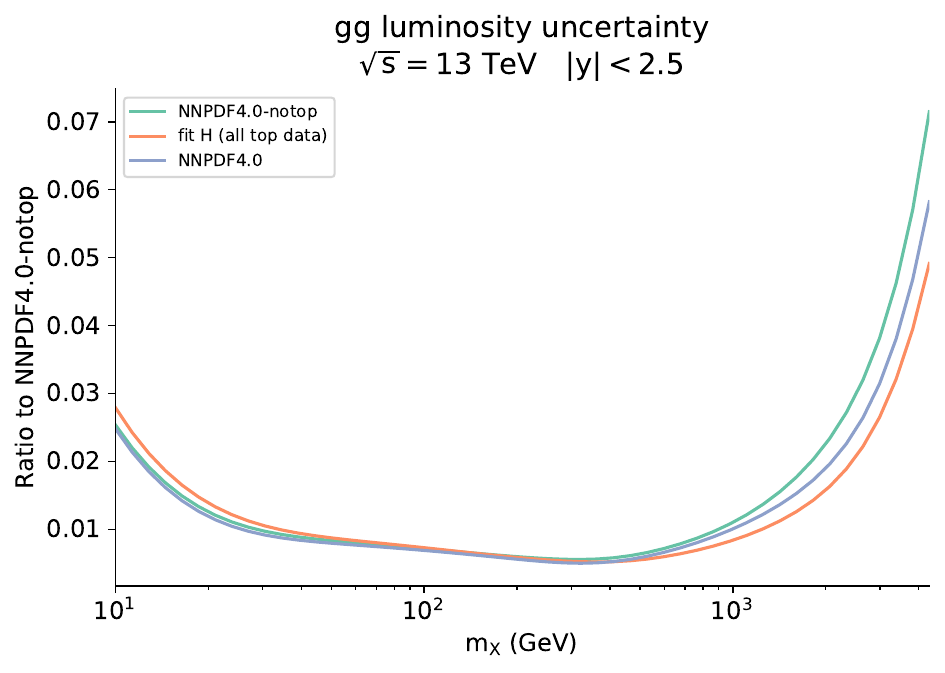}
\includegraphics[width=0.49\textwidth]{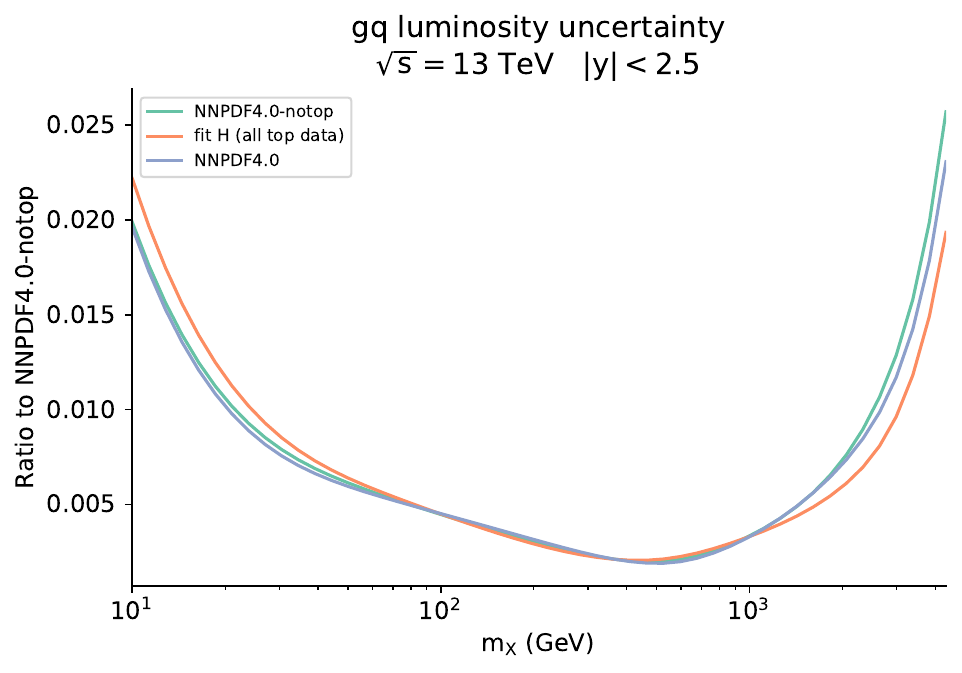}
\caption{Same as Fig.~\ref{fig:fith_luminosities}, now for the
  relative luminosity uncertainties, all normalised to the
  \nnpdfnotop baseline fit.
}
\label{fig:fith_luminosities_unc}
\end{figure}

To assess the phenomenological impact of our analysis at the level
of LHC processes, Fig.~\ref{fig:fith_luminosities}
compares the gluon-gluon and quark-gluon 
partonic luminosities at $\sqrt{s}=13$ TeV (restricted to
the central acceptance region with $|y|\le 2.5$)
between \nnpdf{}, \nnpdfnotop, and fit H including the full top quark dataset
considered here and Fig.~\ref{fig:fith_luminosities_unc} compares their
uncertainties. 
Results are presented as the ratio to the no-top baseline fit.
The $qq$ and $q\bar{q}$ luminosities of fit H are essentially
identical to those of the no-top baseline, as expected given the negligible changes in the quark PDFs observed in fit H, and hence are not discussed further. 

From Figs.~\ref{fig:fith_luminosities}-\ref{fig:fith_luminosities_unc} one
observes that both
for the quark-gluon and gluon-gluon luminosity the impact
of the LHC top quark data is concentrated on the region above $m_X \gsim 1$ TeV.
As already reported for the case of the gluon PDF, also for the luminosities
the net effect of the new LHC Run II top quark data is to further reduce
the luminosities for invariant masses in the TeV range, with a qualitatively
similar but stronger pull as compared to that obtained in \nnpdf{}.
While \nnpdf{} and its no-top variant agree at the $1\sigma$ level in the full
kinematical range considered, this is not true for fit H,  whose error bands
do not overlap with those of \nnpdfnotop for invariant masses $m_X$ between
2 and 4 TeV.
On the other hand, \nnpdf{} and fit H are fully consistent across the full
$m_X$ range, and hence we conclude that predictions for LHC observables
based on \nnpdf{} will not be significantly affected by the inclusion
of the latest LHC top quark data considered in this work.

Finally, concerning the fit quality of fit H is essentially stable, actually better relative to the \nnpdfnotop baseline. The experimental $\chi^2$ per data point on their respective datasets is $1.156$ for the no-top baseline, whilst for fit H is reduced to $1.144$. It is interesting to observe that all new top data included in fit H are already described well by using the \nnpdf{} set, although clearly the $\chi^2$ per data point improves (from $1.139$ to $1.102$) once all data are included in the fit. This confirms the overall consistency of the analysis. 

\section{Impact of the top quark Run II dataset on the SMEFT}
\label{sec:res_smeft}

We now quantify the impact of the LHC Run II legacy measurements, described in Sect.~\ref{sec:exp},
on the top quark sector of the SMEFT.
As compared to previous investigations of SMEFT operators modifying top-quark
interactions~\cite{Ellis:2020unq,Ethier:2021bye}~\cite{Aguilar-Saavedra:2018ksv,
  Buckley:2015lku, Brivio:2019ius, Bissmann:2019gfc, Hartland:2019bjb,
  Durieux:2019rbz, vanBeek:2019evb, Yates:2021udl}, the current analysis
considers a wider dataset, in particular extended
to various measurements based on the full LHC Run II luminosity.
In the last column of Tables~\ref{tab:input_datasets_toppair}--\ref{tab:input_datasets_4} we indicated
which of the datasets included here were considered
for the first time within a SMEFT interpretation.
Here we assess the constraints that the available LHC top quark measurements
provide on the SMEFT parameter space, and in particular study the
impact of the new measurements as compared to those used in~\cite{Ellis:2020unq,Ethier:2021bye}.
In this section we restrict ourselves to fixed-PDF EFT fits, where the input PDFs used in the calculation
of the SM cross-sections are kept fixed.
Subsequently, in Sect.~\ref{sec:joint_pdf_smeft},
we generalise to the case in which PDFs are extracted simultaneously
together with the EFT coefficients.

The structure of this section is as follows. We begin in Sect.~\ref{subsec:smeftfitsettings} 
by describing the methodologies used
to constrain the EFT parameter space both at linear and quadratic
order in the EFT expansion.
%
We also present results for the Fisher information matrix, which indicates which datasets
provide constraints on which operators.
In Sect.~\ref{subsec:smeftresults}, we proceed to give the results of the fixed-PDF EFT 
fits at both linear and quadratic order,
highlighting the impact of the new Run II top quark data by comparison with 
previous global SMEFT analyses.
Finally, in Sect.~\ref{subsec:pdfeftcorr} we evaluate the correlation between PDFs and EFT coefficients
to identify the kinematic region and EFT operators which are potentially
sensitive to the SMEFT-PDF interplay to be studied in Sect.~\ref{sec:joint_pdf_smeft}.

\subsection{Fit settings}
\label{subsec:smeftfitsettings}

Throughout this section, we will allow only the SMEFT coefficients to vary in the fit,
keeping the PDFs fixed to the SM-PDFs baseline obtained in the
\nnpdfnotop 
fit discussed in Sect.~\ref{sec:baseline_sm_fits}; with this choice, one removes the overlap
between the datasets entering the PDF fit and the EFT coefficients determination.
Our analysis is sensitive to the $N=25$ Wilson coefficients defined
in Ref.~\cite{Ethier:2021bye}, except at the linear level where the four-heavy
coefficients $c_{Qt}^{8}$, $c_{QQ}^{1}$, $c_{QQ}^{8}$, $c_{Qt}^{1}$ and $c_{tt}^{1}$ (which are 
constrained only by $t\bar{t}t\bar{t}$ and $t\bar{t}b\bar{b}$ data)
exhibit three flat directions~\cite{Ethier:2021bye}. In order to tackle this, we remove the five four-heavy coefficients from the linear fit\footnote{In principle one could instead rotate to the principal component analysis (PCA) basis
  and constrain the two non-flat directions in the four-heavy subspace,
  but even so, the obtained constraints
  remain much looser in comparison with those obtained in the quadratic EFT fit~\cite{Ethier:2021bye}. 
  
  In our fits, we also keep the $t\bar{t} t\bar{t}$ and $t\bar{t} b \bar{b}$ datasets after removing the five four-heavy coefficients. We 
  have verified that including or excluding these sets has no
  significant impact whatsoever on the remaining coefficients.
}
Hence, in our linear fit we have $N=20$ independent coefficients constrained
from the data, whereas in the quadratic fit, we fit all $N=25$ independent coefficients, including the four-heavy coefficients.

The linear EFT fits presented in this section are performed with the \simunet{}  methodology
in the fixed-PDF option, as described in Chapter \ref{ch:simunet}; 
we explicitly verified that the \simunet{} methodology reproduces the posterior distributions 
provided by \smefit{} (using
either the NS (Nested Sampling) or MCfit options)
for a common choice of inputs. 
However, in the case of the quadratic EFT fits we are unable to use the \simunet{} 
methodology due to a failure of the Monte-Carlo sampling method utilised in the \simunet{} and \smefit{} codes;
this is discussed in App. E of~\cite{Kassabov:2023hbm} and described in detail in~\cite{Costantini:2024wby}.
For this reason, quadratic EFT fits in this section are carried out
with the public {\sc\small SMEFiT} code using the NS
mode~\cite{Giani:2023gfq}. 
To carry out these fits, the full dataset listed in
Tables~\ref{tab:input_datasets_toppair}--\ref{tab:input_datasets_4}, together
with the corresponding SM and EFT theory calculations described earlier in the chapter have been converted into the standard \smefit{} format.

\paragraph{Fisher information.}
The sensitivity to the EFT operators of the various processes entering
the fit can be evaluated by means of the Fisher information, $F_{ij}$, which
quantifies the information carried by a given dataset on the EFT coefficients $c_i$~\cite{Brehmer:2016nyr}. 
In a linear EFT setting,
the Fisher information is given by:
\begin{equation}
\label{eq:fisherdef}
F_{ij} = L^{(i)T} ({\rm cov}_{\rm exp})^{-1} L^{(j)}
\end{equation}
where the $k$-th entry, $L^{(i)}_k$, of the vector $L^{(i)}$ is the
linear contribution multiplying $c_i$ in the SMEFT theory prediction
for the $k$-th data point, and ${\rm cov}_{\rm exp}$ is the
experimental covariance matrix. In particular, the Fisher information
is an $N \times N$ matrix, where $N$ is the number of EFT
coefficients, and it depends on the dataset. 
An important property of the Fisher information is that it is related the covariance matrix $C_{ij}$ of the maximum likelihood estimators by the \textit{Cramer-Rao bound}:
\begin{equation}
\label{eq:cramerrao}
C_{ij} \geq (F^{-1})_{ij},
\end{equation}
indicating that larger values of $F_{ij}$ will translate to tighter bounds on the
EFT coefficients.

\begin{figure}[h!]
        \centering
        \includegraphics[scale=0.42]{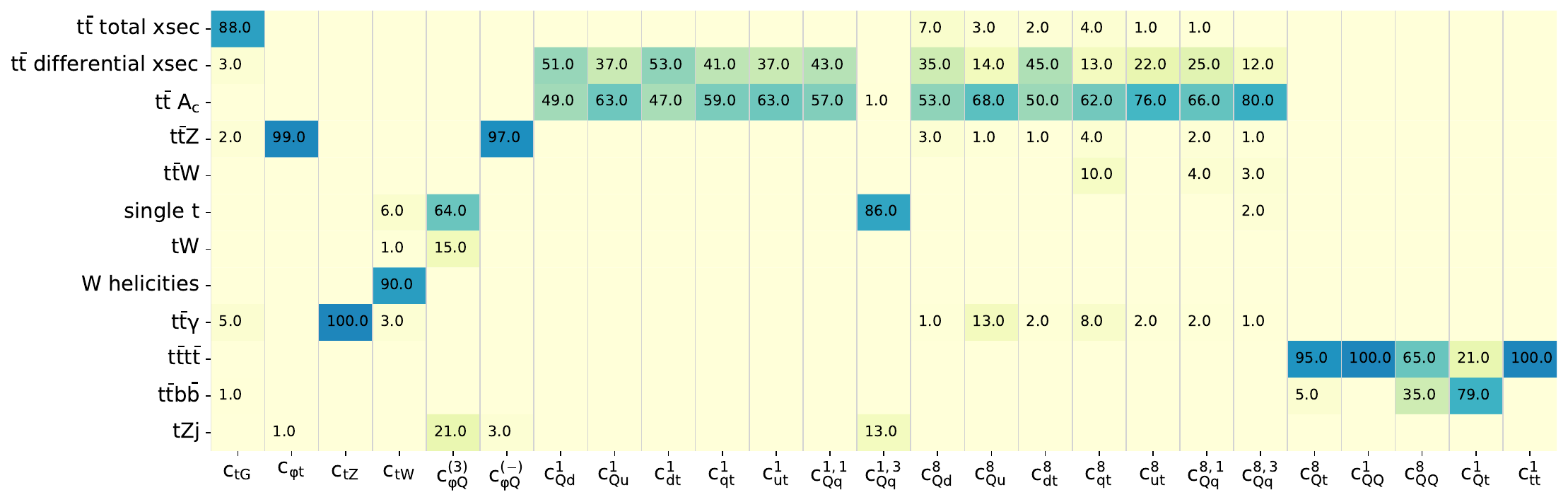}
	\caption{Relative constraining power (in \%) on each of the operators for each of the processes entering the fit, 
	as defined in Eq.~\eqref{eq:relativeconstrainingpower}.}
	\label{fig:fisher_diag}
\end{figure}

Before displaying the results of the fixed-PDF SMEFT analysis in
Sect.~\ref{subsec:smeftresults}, we use the Fisher information 
to assess the relative impact of each sector of top quark data on the
EFT parameter space; this is done in the linear analysis, including
${\cal O}(1/\Lambda^2)$ SMEFT corrections.  
In the quadratic case, once ${\cal O}(1/\Lambda^4)$ SMEFT corrections
are included, the dependence of $F_{ij}$ on the Wilson coefficients makes interpretation more difficult.
Writing $F_{ij}(D)$ for the
Fisher information matrix evaluated on the dataset $D$, we define the relative constraining power of the dataset $D$
via:
\begin{equation}
\label{eq:relativeconstrainingpower}
	\textrm{relative constraining power of }D\textrm{ on operator }c_i = F_{ii}(D) \bigg/ \displaystyle \sum_{\text{sectors }D'} F_{ii}(D').
\end{equation}
Since $F_{ii}(D)$ corresponds to the constraining power of the dataset $D$ in a one-parameter fit of the Wilson coefficient $c_i$,
this definition only quantifies how much a dataset impacts one-parameter fits of single Wilson coefficients in turn; however, this will give 
a general qualitative picture of some of the expected behaviour in the global fit too. 
We display the results of evaluating the relative constraining power of each top quark data sector on each of the
parameters in Fig.~\ref{fig:fisher_diag}, quoting the results in percent (\%).

As expected, $t \bar{t}$ total cross sections constitute the dominant source of constraints
on the coefficient $c_{tG}$.  Each of the four-fermion operators receive important constraints
from differential $t \bar{t}$ distributions and charge asymmetry measurements.
Note that this impact is magnified when we go beyond individual fits, in which
case measurements of charge asymmetries are helpful in breaking flat directions 
amongst the four-fermion operators~\cite{Brivio:2019ius,Ellis:2020unq}.
The coefficient $c_{Qq}^{1,3}$ is the exception as it 
is instead expected to be well-constrained by single top production.  
We note that the measurements of $W$ helicities are helpful in constraining the coefficient $c_{tW}$,
while $t \bar{t} Z$ measurements provide the dominant source of constraints on $c_{\varphi Q}^{(-)}$.
We observe that the neutral top coupling $c_{tZ}$ is entirely 
constrained by $t \bar{t} \gamma$, 
and that the effects of $t \bar{t} t \bar{t}$ and $t \bar{t} b \bar{b}$ are mostly restricted to the 4-heavy operators
$c_{Qt}^{8}$, $c_{QQ}^{1}$, $c_{QQ}^{8}$, $c_{Qt}^{1}$ and $c_{tt}^{1}$. 

\subsection{Fixed-PDF EFT fit results}
\label{subsec:smeftresults}

In this section, we present the results of the linear and quadratic fixed-PDF fits
with the settings described in Sect~\ref{subsec:smeftfitsettings}. 

\paragraph{Fit quality.}
We begin by discussing the fit quality of the global SMEFT determination, quantifying 
the change in the data-theory agreement relative to the SM in both the linear and quadratic SMEFT
scenarios.  Table~\ref{tab:smeft_fit_quality} provides the values of the $\chi^2$ per data point in the SM and in the case of 
the SMEFT at both linear and quadratic order in the EFT expansion for
each of the processes entering the fit\footnote{Notice that NS is designed to estimate the Bayesian evidence of a model. However, as a by-product one also obtains a sampling of the posterior that can be used in an analogous way to the MC replicas to compute summary statistics. Since \simunet{} uses the MC replica method, we use NS the posterior samples to compare using analogous metrics.}.
Here, in order to ease the comparison of our results to those of
\smefit{} and \fitm, we quote the $\chi^2$ per data point computed
by using the covariance matrix defined in Eq.~\eqref{eq:chi2exp} supplemented with the theory uncertainties coming from the PDFs, as detailed in Refs.~\cite{Hartland:2019bjb,Ethier:2021bye}. In a nutshell, the covariance matrix in this case is defined as
\begin{equation*}
        \text{cov}_{\text{exp+th}} = \text{cov}_{t_0} + \text{cov}_{\text{th}},
\end{equation*}
where 
\begin{equation*}
        \left( \text{cov}_{\text{th}} \right)_{ij} = \langle T_{i}^{(k)} T_{j}^{(k)} \rangle_{(k)} -  \langle T_{i}^{(k)} \rangle_{(k)}  \langle T_{j}^{(k)} \rangle_{(k)},
\end{equation*}
and the average is taken over the $k$ replicas.

We observe that in many sectors, the linear EFT fit improves the fit
quality compared to the SM; notably, the $\chi^2_{\rm exp+th}$ per data point for inclusive $t\bar{t}$ is vastly improved from 1.71 to 1.11. When
quadratic corrections are also considered, the fit quality is usually poorer compared to the linear fit. For example,
in inclusive $t\bar{t}$ the $\chi^2_{\rm exp+th}$ per data point deteriorates from 1.11 to 1.69. This is not unexpected, however,
since the flexibility of the quadratic fit is limited by the fact that 
for sufficiently large values of Wilson coefficients the EFT can only
make positive corrections.\footnote{This also has methodological implications.
Large quadratic corrections can negatively impact the Monte-Carlo sampling method
used by \simunet{}.}

\begin{table}[t!]
\begin{center}
\centering
\renewcommand{\arraystretch}{1.4}
\begin{tabularx}{\textwidth}{lcccc}
\toprule
Process  $\qquad $ & $n_{\rm dat}$ & {\footnotesize $\chi^2_{\rm exp+th}$ [SM]} & {\footnotesize $\chi^2_{\rm exp+th}$ [SMEFT $\mathcal{O}(\Lambda^{-2})$]}
& {\footnotesize $\chi^2_{\rm exp+th}$ [SMEFT $\mathcal{O}(\Lambda^{-4})$]}
\\ \midrule
	$t \bar{t}$ & 86 & 1.71 & 1.11 & 1.69 \\
	$t \bar{t}$ AC & 18 & 0.58 & 0.50 & 0.60 \\
	$W$ helicities & 4 & 0.71 & 0.45 & 0.47  \\
        \midrule
        $t\bar{t}Z$ & 12 & 1.19 & 1.17 & 0.94  \\ 
        $t\bar{t}W$ & 4 & 1.71 & 0.46 & 1.66 \\
        $t \bar{t} \gamma$ & 2 & 0.47 & 0.03 & 0.59 \\ 
        \midrule
        $t \bar{t} t \bar{t}$ \& $t \bar{t} b \bar{b}$ & 8 & 1.32 & 1.06 & 0.49  \\
        \midrule
        single top & 30 & 0.504& 0.33 & 0.37  \\ 
        $tW$ & 6 & 1.00 & 0.82 & 0.82 \\ 
        $tZ$ & 5 & 0.45 & 0.30 & 0.31 \\
        \midrule
        {\bf Total} & {\bf 175} & {\bf 1.24} & {\bf 0.84} & {\bf 1.14}  \\ 
\bottomrule
\end{tabularx}
\end{center}
\caption{The values of the $\chi^2$ per data point for the fixed-PDF EFT fits presented in this section,
  both for individual groups of processes, and for the total
  dataset. 
  In each case we indicate the number of data points, the $\chi^2_{\rm exp+th}$ obtained using the baseline
  SM calculations, and the results of both the linear and quadratic EFT fits.
}
\label{tab:smeft_fit_quality}
\end{table}

It is also useful to calculate the goodness of fit, quantified by the $\chi^2$ per degree of freedom, $\chi^2/n_{\rm dof} = \chi^2/(n_{\rm dat} - n_{\rm param})$, 
which additionally accounts for the complexity of the models we are
using in each fit. In our case, we find $\chi^2_{\rm exp+th}/n_{\rm dof} = 1.25$ in the SM and $\chi^2_{\rm exp+th}/n_{\rm dof} = 0.95$ 
and $1.33$ in the linear
and quadratic EFT scenarios respectively.  
We see that while the EFT at quadratic order does not provide a better fit than the SM, neglecting quadratic EFT corrections leads to
a significant improvement in the overall fit quality.

\begin{figure}[h!]
	\centering
	\includegraphics[scale=0.72]{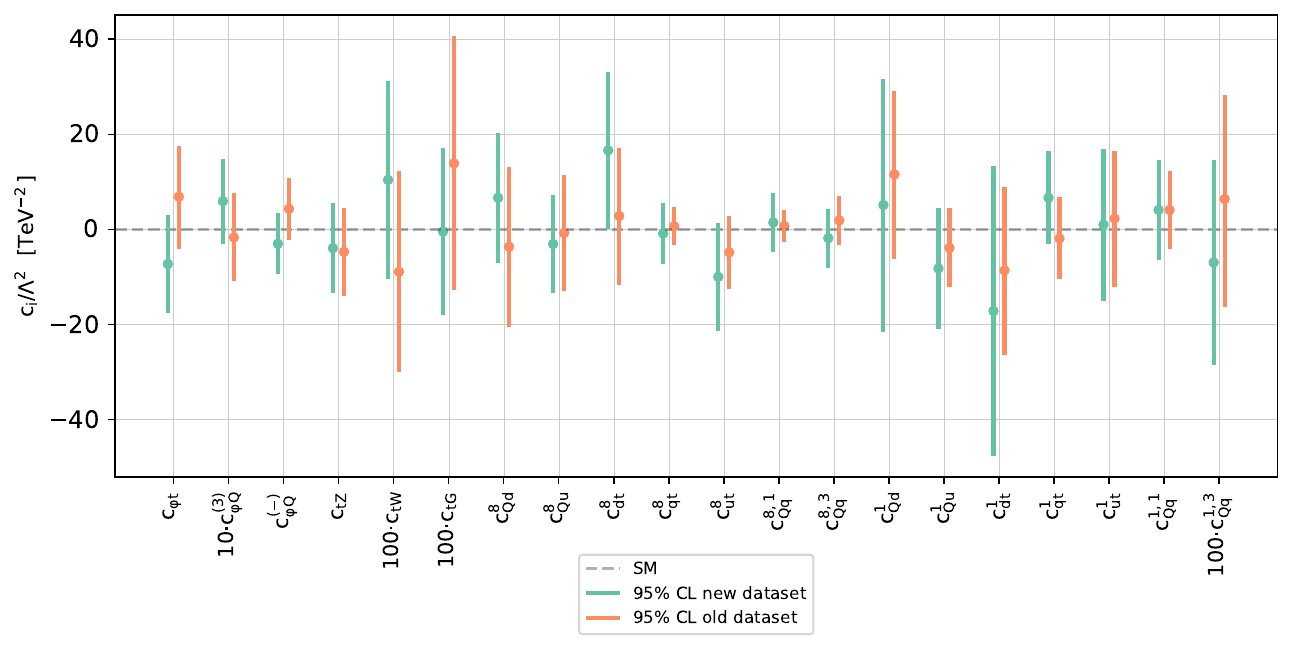}
	\caption{The 95\% CL intervals on the EFT coefficients entering the linear fit,
          evaluated with \simunet{} on the dataset considered in this work, and evaluated with \smefit{} on the
          top quark dataset entering the analysis of~\cite{Ethier:2021bye} 
          (note that at the linear EFT level, the results obtained with \simunet{}
          coincide with those provided by \smefit{} for the same dataset).
         Note also that the constraints on selected coefficients 
          are rescaled by the factors shown, for display purposes.
	  }
	\label{fig:linear_eft_constraints}
\end{figure}

\begin{figure}[tb!]
        \centering
        \includegraphics[scale=0.7]{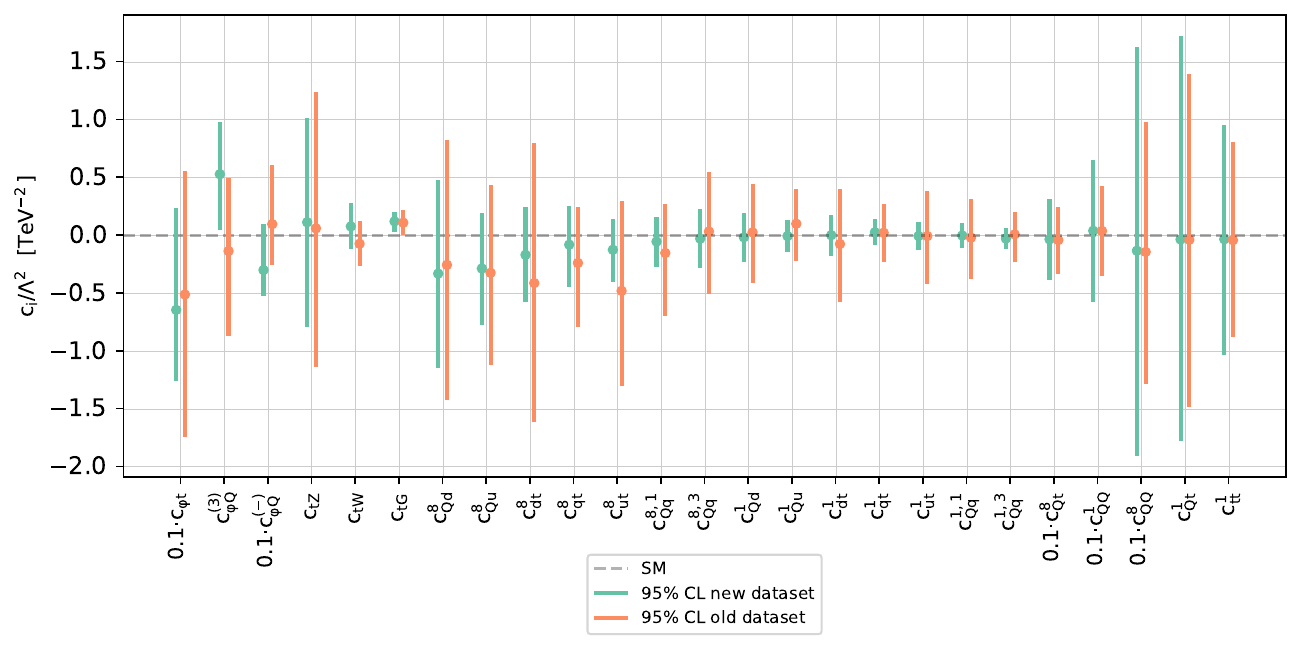}
	\caption{The 95\% CL intervals on the EFT coefficients entering the quadratic fit,
          evaluated with {\sc\small SMEFiT}.
          We compare the results based on the full top quark dataset with
          the corresponding results obtained from the subset of top quark
          measurements entering the analysis of~\cite{Ethier:2021bye}.  
         As in Fig.~\ref{fig:linear_eft_constraints}, the constraints on selected coefficients 
          are rescaled by the factors shown, for display purposes.        
        }
	\label{fig:quadratic_eft_constraints}
\end{figure}

\paragraph{Constraints on the EFT parameter space.}
Next, we present the constraints on the EFT parameter space. In
Fig.~\ref{fig:linear_eft_constraints}, we display the 95\% CL
constraints (defined as the mean of the MC replicas $\pm 2 \sigma$) on the 20 Wilson coefficients entering the linear fit. Two
sets of constraints are shown; in green, we give the intervals
obtained from a fit to the 175 data points introduced in
Sect.~\ref{sec:exp}, whilst in orange, we give in the intervals
obtained from a fit to the older top quark dataset used in the global
analysis of Ref.~\cite{Ethier:2021bye}, obtained from a fit of 150
data points. This comparison allows us to quantify the information
gained from the latest Run II datasets, relative to those available to
previous analyses. The same comparison, this time at quadratic order in the EFT expansion,  
is shown in Fig.~\ref{fig:quadratic_eft_constraints} (note that in this plot we display constraints on 
all 25 coefficients, including the 4-heavy coefficients $c_{Qt}^{8}$, $c_{QQ}^{1}$, $c_{QQ}^{8}$, $c_{Qt}^{1}$ and $c_{tt}^{1}$).

We first note that Figures~\ref{fig:linear_eft_constraints} and~\ref{fig:quadratic_eft_constraints} both
demonstrate good agreement between the fits using old and new datasets,
and consistency between the new fit SMEFT bounds and the SM.
At the linear level, the most noticeable improvement concerns $c_{tG}$; its
 95\% C.L. bounds decrease from $[-0.13, 0.41]$ to $[-0.18, 0.17]$,
 thanks to the increased amount of information in the input dataset,
 coming in particular from $t \bar{t}$ data. This results in both a 
 tightening of the constraints by about 35\% and a shift in their central value. 
For many of the other coefficients, the bounds are either stable
(e.g. $c_{tZ}$), or exhibit a shift in central value but no significant tightening (e.g. $c_{\varphi t}$, undergoes
a shift of -14.3, but a decrease in the size of the constraints 95\%
C.L. by only 1\%, and $c_{\varphi Q}^{(-)}$ undergoes a shift of
$-7.33$, but its bounds only tighten by 2\%). Finally, we note that some coefficients instead exhibit a broadening of constraints with the new dataset relative to the old
dataset (for example, some of the four-fermion operators). The increase in the size of the constraints could point to some
inconsistency within the new inclusive $t\bar{t}$ dataset; however, given that the bounds are very large anyway, at the edges of the 
intervals we are likely to approach a region where the EFT is no
longer valid in both cases, hence no definite conclusions may be
drawn. 

At the quadratic order in the EFT expansion, however, the impact of the latest Run II dataset is clear;
we see a marked improvement in many of the SMEFT constraints.
As shown in Fig.~\ref{fig:quadratic_eft_constraints}, the bounds on all 14 of the four-fermion operators become noticeably smaller as a result of 
the increase in precision in the $t \bar{t}$ sector.
The constraint on $c_{tZ}$ is improved by the inclusion of
measurements of the $t \bar{t} \gamma$ total cross sections, resulting
in a tightening of 24\%.
The addition of the $p_{T}^{\gamma}$ spectrum~\cite{Aad:2020axn} would
yield an even stronger constraint, as seen in~\cite{Ellis:2020unq}. We will make use of this observable in future work when 
unfolded measurements are made available.
Contrary to the linear fit, where we singled out $c_{\varphi t}$ and $c_{\varphi Q}^{(-)}$ as examples of coefficients which shift, but whose bounds are not improved, the constraints on $c_{\varphi t}$, $c_{\varphi Q}^{(-)}$ markedly tighten in the quadratic fit in the presence of new data; in particular, the size of the bounds on $c_{\varphi t}$, $c_{\varphi Q}^{(-)}$ decrease by 35\% and 28\%, respectively.
On the other hand, despite the addition of new $t\bar{t}t\bar{t}$,
$t\bar{t} b \bar{b}$ datasets, we find limited sensitivity to the five
four-heavy coefficients $c_{Qt}^{8}$, $c_{QQ}^{1}$, $c_{QQ}^{8}$,
$c_{Qt}^{1}$ and $c_{tt}^{1}$. In fact, with the new data we see a
broadening of the bounds. As with the linear fit, this could point to
either an inconsistency in the $t\bar{t}t\bar{t}$, $t\bar{t}b\bar{b}$ data, or
simply to the ambiguity associated to the EFT validity in that
particular region of the parameter space. 

\begin{figure}[htb!]
\centering
        \begin{subfigure}[b]{0.49\textwidth}
        \includegraphics[scale=0.33]{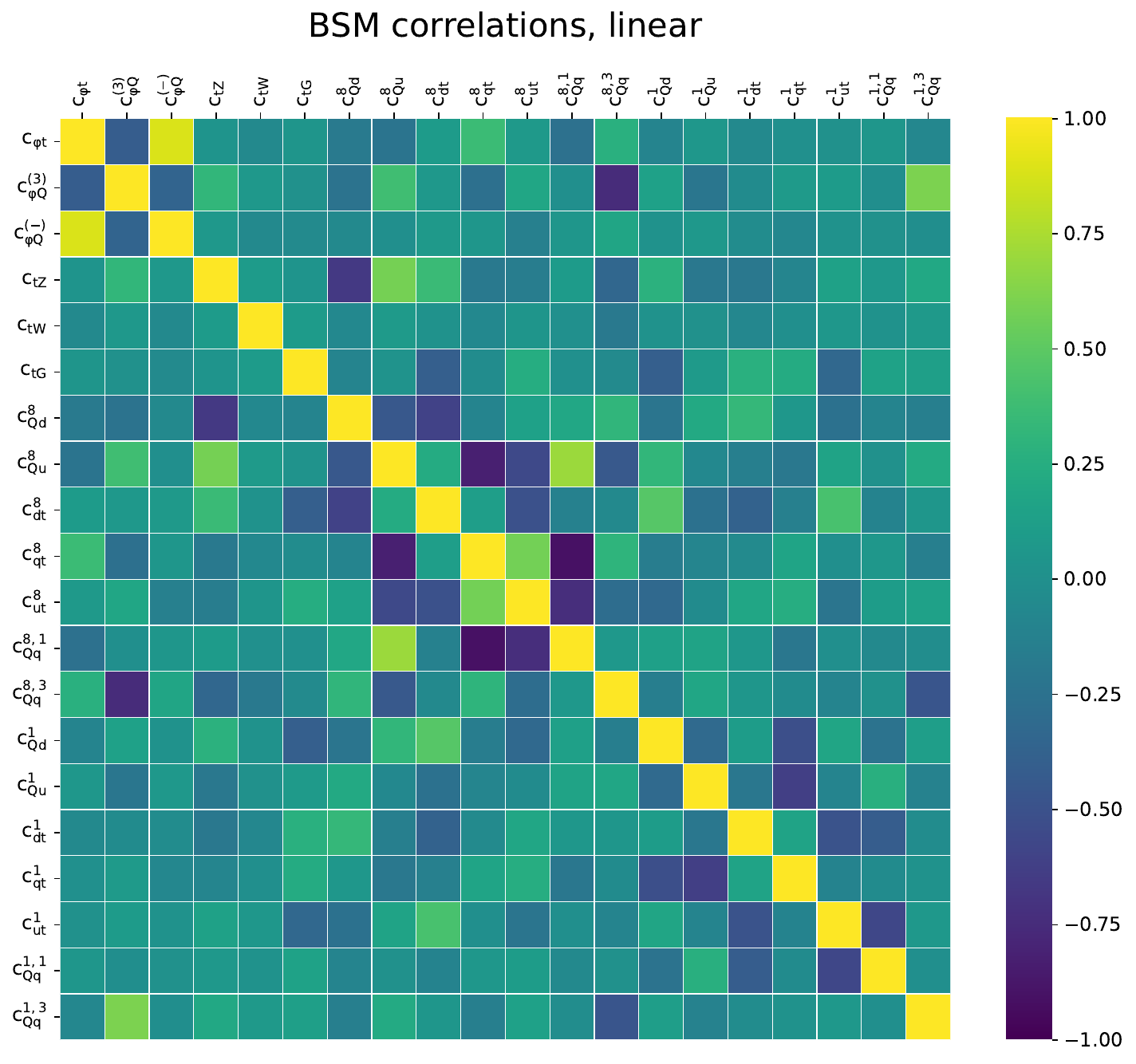}
        \end{subfigure}
        \begin{subfigure}[b]{0.49\textwidth}
        \includegraphics[scale=0.36]{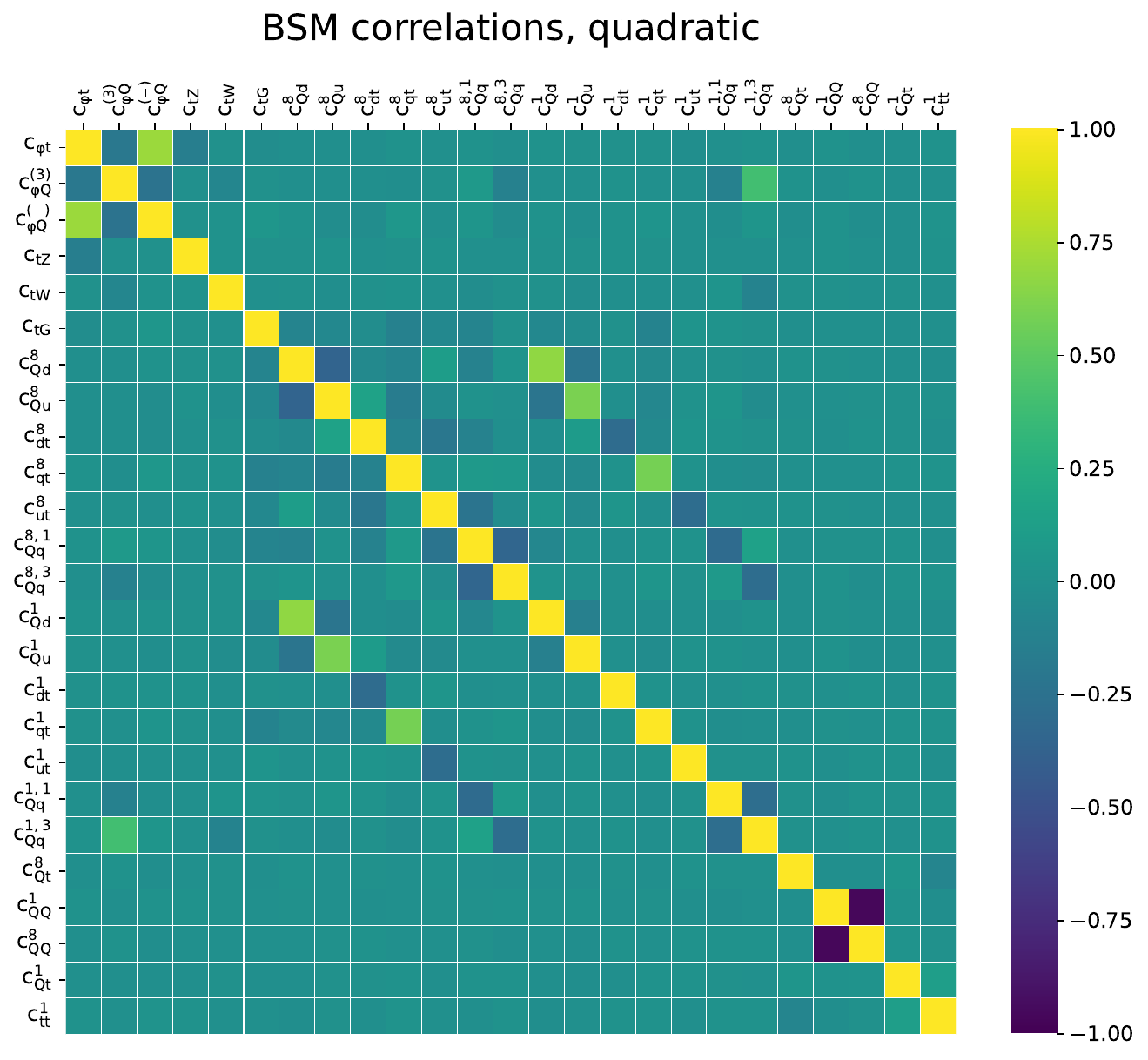}
        \end{subfigure}
	\caption{The values of the correlation coefficients evaluated
          between all pairs of Wilson coefficients entering the
          EFT fit at the  linear  (left) and quadratic  (right)  order.
          As explained in the next, the number of fitted DoFs is different
          in each case.
        }
\label{fig:bsmcorr}
\end{figure}

\paragraph{Correlations.}
Figure~\ref{fig:bsmcorr} shows the correlations between the Wilson
coefficients evaluated in this analysis both at the linear and the
quadratic order in the EFT expansion, shown on the left and right
panels respectively.
In the linear fit, we first note a number of large correlations amongst the octet four-fermion operators
which enter the $t \bar{t}$ production together.  
The singlet four-fermion operators are similarly correlated among themselves, although 
their correlations are comparatively suppressed.  
The coefficients $c_{\varphi t}$ and $c_{\varphi Q}^{(-)}$
exhibit a large positive correlation due to their
entering into $t \bar{t} Z$ production together,
while $c_{Qq}^{3,1}$ and $c_{\varphi Q}^{(3)}$ have positive
correlations through their contribution to single top production.
Further non-zero correlations are found, for example amongst the pairs $c_{Qq}^{8,3}$ \& $c_{\varphi Q}^{(3)}$,
$c_{Qd}^{8}$ \& $c_{tZ}$ and $c_{Qu}^{8}$ \& $c_{tZ}$.

At quadratic order, however, we observe that many of these correlations are suppressed,
as a result of the fact that the inclusion of quadratic corrections lifts many of the 
degeneracies in the fit.
We observe that the pairs $c_{\varphi t}$, $c_{\varphi Q}^{(-)}$
and $c_{Qq}^{1,3}$, $c_{\varphi Q}^{(3)}$ remain correlated though.
The 4-heavy operators are also included in this quadratic fit, and we find large anti-correlations between 
$c_{QQ}^{1}$ and $c_{QQ}^8$, indicating that they are poorly distinguished in the $t \bar{t} t \bar{t}$
and $t \bar{t} b \bar{b}$ processes.  Finally, note that we obtain subtle non-zero correlations
between the octet and singlet four-fermion operators constructed from the same fields, for example between $c_{Qd}^8$ and $c_{Qd}^1$.
This is a result of the fact that $t \bar{t}$ measurements provide the dominant source of constraints on these coefficients and are
very sensitive to
quadratic corrections, and at this order the contribution from these operators differs only by
a numerical factor.

\subsection{Correlations between PDFs and EFT coefficients}
\label{subsec:pdfeftcorr}

\begin{figure}[hbt!]
\centering
        \includegraphics[width=0.48\textwidth]{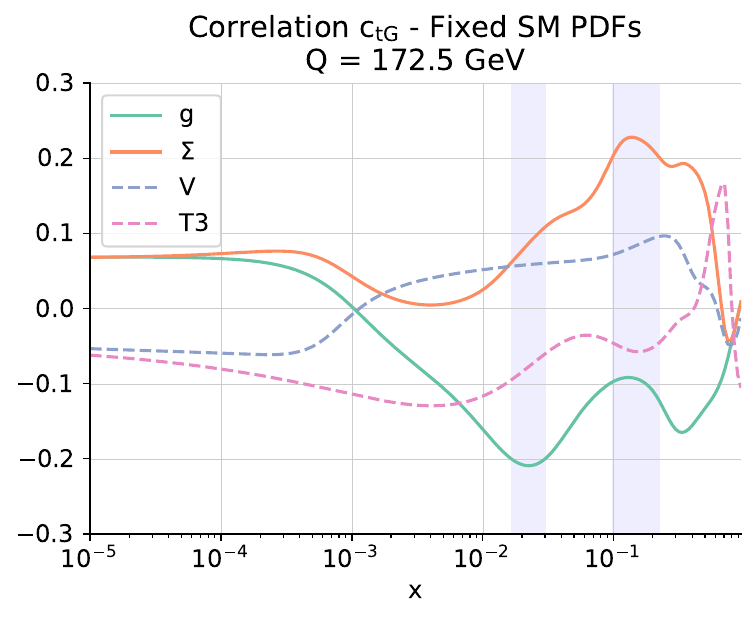}
        \includegraphics[width=0.48\textwidth]{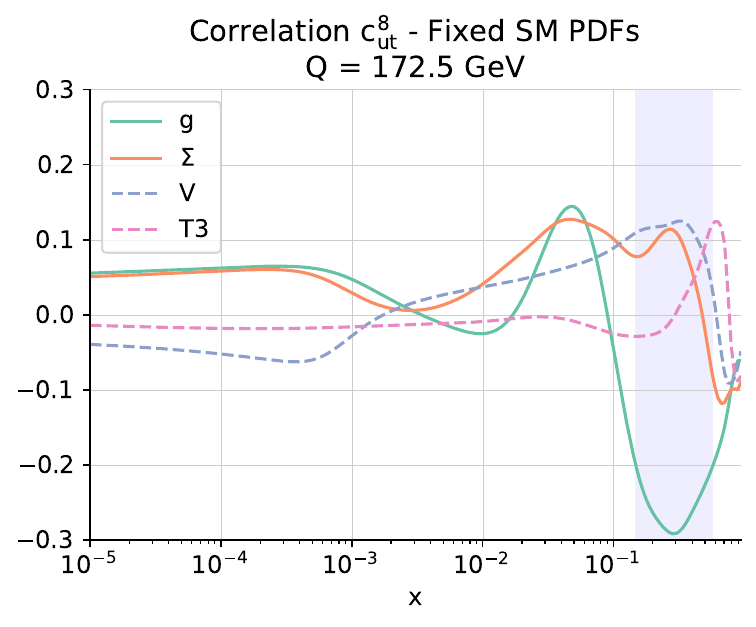}
	\caption{The value of the correlation coefficient $\rho$
          between the PDFs and selected EFT coefficients as a function of $x$
          and $Q=172.5$ GeV.
          We show the results for the gluon, the total singlet $\Sigma$, total valence $V$, and non-singlet
          triplet $T_3$ PDFs.
          We provide results for representative EFT coefficients, namely $c_{tG}$ and  $c_{ut}^{(8)}$.
        }
\label{fig:pdfbsmcorr}
\end{figure}

We conclude this section by discussing the correlations observed between the PDFs and Wilson coefficients.
The PDF-EFT correlation coefficient for a Wilson coefficient $c$ and a
PDF $f(x, Q)$ at a given $x$ and $Q^2$ is defined as
\begin{equation}
\label{eq:correlationL2CT}
\rho\lp c, f(x,Q^2)\rp=\frac{\la c^{(k)}f^{(k)}(x,Q^2)\ra_k - \la c^{(k)}\ra_k \la f^{(k)}(x,Q^2)\ra_k
}{\sqrt{\la (c^{(k)})^2\ra_k - \la c^{(k)}\ra_k^2}\sqrt{\la\left( f^{(k)}(x,Q^2)\right)^2\ra_k - \la f^{(k)}(x,Q^2)\ra_k^2}}  \, ,
\end{equation}
where $c^{(k))}$ is the best-fit value of the Wilson coefficient for
the $k$-th replica and $f^{(k)}$ is the $k$-th PDF replica computed at
a given $x$ and $Q$, and $\la \cdot \ra_k$ represents the average
over all replicas. We will compute the correlation between a SM PDF and the Wilson coefficients, both of which have been separately determined from 
the total dataset including all new top quark data.  
By doing so we hope to shed light on which Wilson coefficients,
and which PDF flavours and kinematical regions, are strongly impacted by the top quark data and therefore exhibit a potential for interplay
in a simultaneous EFT-PDF determination.
The EFT corrections will be
restricted to linear order in the EFT expansion.

Fig.~\ref{fig:pdfbsmcorr} displays a selection of the largest correlations.
We observe that the gluon PDF in the large-$x$ region is significantly correlated with the Wilson coefficients 
$c_{tG}$, $c_{ut}^{(8)}$.  On the other hand, relatively large correlations are observed between $c_{tG}$ and the total singlet $\Sigma$, while the total valence $V$, and non-singlet triplet $T_3$ PDFs show no relevant correlations
with the selected coefficients.
This is not surprising, given the impact of top quark pair production total cross sections and differential distributions
in constraining these PDFs and Wilson coefficients.  
Whilst these correlations are computed from a determination of the SMEFT in which the PDFs are fixed to SM PDFs, 
the emergence of non-zero correlations provides an indication of the potential for interplay between the PDFs and the SMEFT
coefficients; this interplay will be investigated in a simultaneous determination
in the following section.

\section{PDF-SMEFT fit from top quark data}
\label{sec:joint_pdf_smeft}

In this section we present the main results of this work, namely
the simultaneous determination of the proton PDFs and the SMEFT Wilson coefficients
from the LHC Run II top quark data described in Sect.~\ref{sec:exp},
following the \simunet{} methodology presented in Chapter \ref{ch:simunet}.
This determination of the SMEFT-PDFs from top quark data is carried
out at the linear, $\mathcal{O}(1/\Lambda^2)$, level in the EFT expansion.
We do not perform simultaneous fits at the quadratic level due to
shortcomings of the Monte Carlo replica method first described in Ref. \cite{Kassabov:2023hbm} and then formalised in Ref. \cite{Costantini:2024wby}, on which
\simunet{} is based. 

\begin{figure}[h!]
\centering
  \begin{subfigure}[b]{0.49\textwidth}
  \includegraphics[scale=0.45]{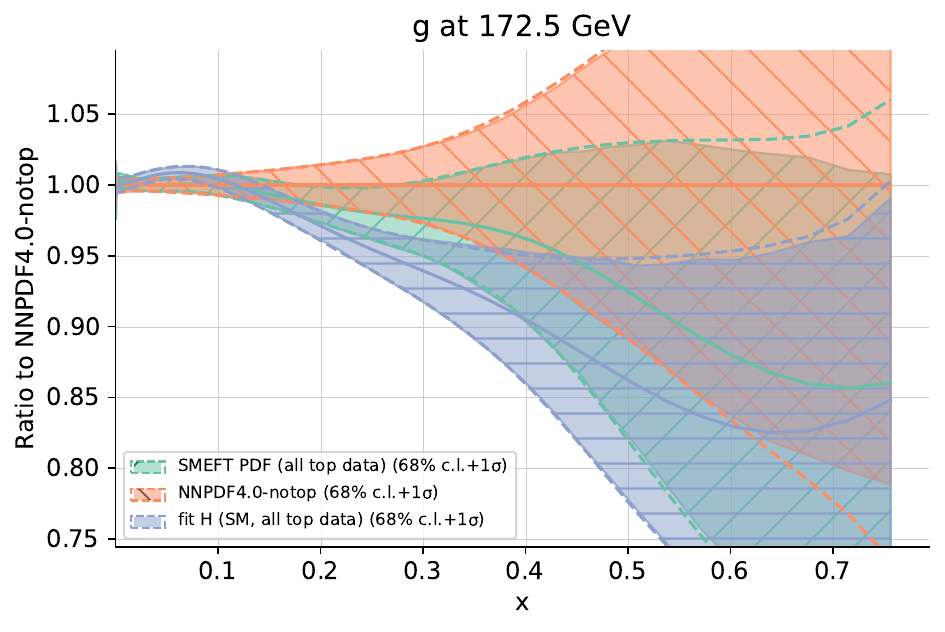}
  \end{subfigure}
  \begin{subfigure}[b]{0.49\textwidth}
    \includegraphics[scale=0.45]{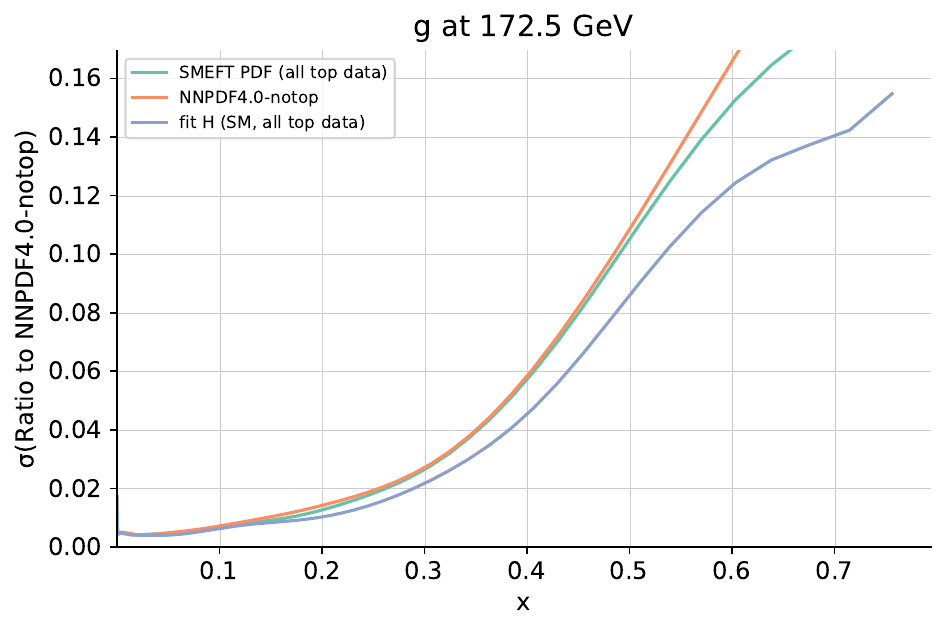}
  \end{subfigure}
   \begin{subfigure}[b]{0.49\textwidth}
  \includegraphics[scale=0.45]{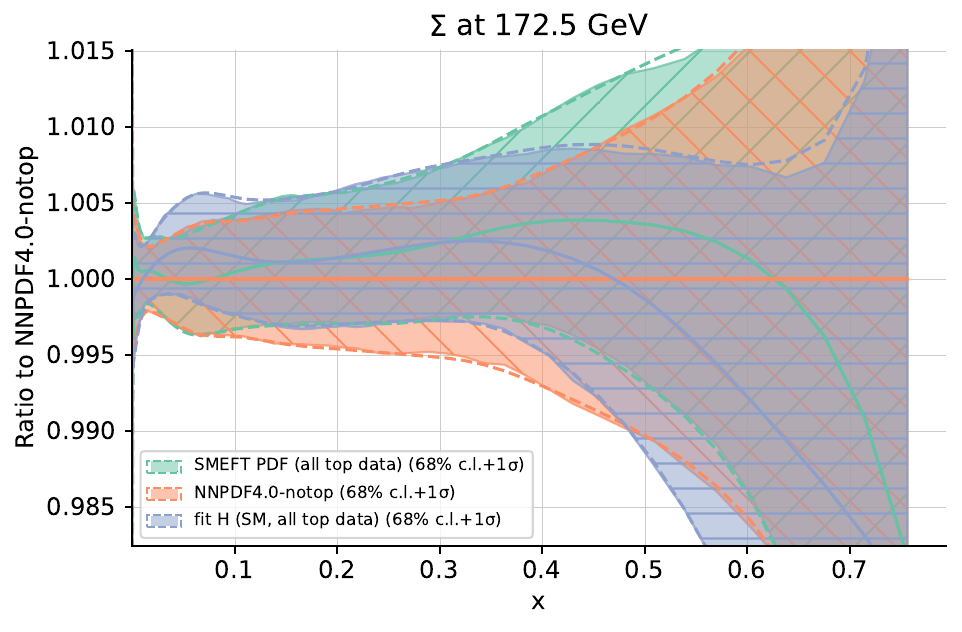}
  \end{subfigure}
  \begin{subfigure}[b]{0.49\textwidth}
  \includegraphics[scale=0.45]{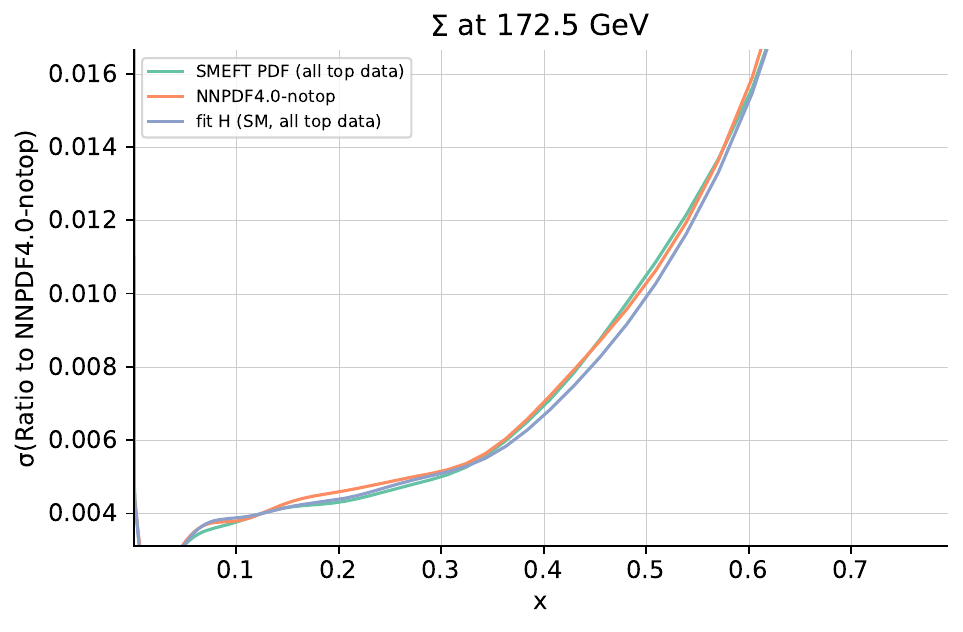}
  \end{subfigure}
        \caption{
    Left: a comparison between the gluon (upper panel)
    and quark singlet (lower panel) PDFs evaluated at $Q=m_t=172.5$ GeV
    in the large-$x$ region.
    We display the \nnpdfnotop baseline, the SM-PDFs of
    fit H in Table~\ref{tab:fit_list} which include
    the full top quark dataset, and their SMEFT-PDF counterparts.
    The results are normalised to the central value of the \nnpdfnotop fit.
    Right: the same comparison, but now for the relative $1\sigma$ PDF uncertainties.
}
\label{fig:SMEFTpdf_gluon}
\end{figure}

\paragraph{PDFs from a joint SMEFT-PDF fit.} We begin by discussing the PDFs obtained through a
joint fit of PDFs and Wilson coefficients
from the complete LHC top quark dataset considered in this work.
Simultaneously extracting the PDFs and the EFT coefficients
from top quark data has a marked impact on the former, as compared to a SM-PDF
baseline, but we shall see has much less impact on the latter, as compared to the results of the
corresponding fixed-PDF EFT analyses.
Fig.~\ref{fig:SMEFTpdf_gluon}
displays a comparison between the gluon and quark singlet PDFs, as
well as of their relative $1\sigma$ PDF uncertainties, for 
the \nnpdfnotop baseline, the SM-PDFs of
fit H in Table~\ref{tab:fit_list} which include
the full top quark dataset, and their SMEFT-PDF counterparts based
on the same dataset.
PDFs are compared in the large-$x$ region for   $Q=m_t=172.5$ GeV. In
the left panel they are normalised to the central value of the
\nnpdfnotop fit.

While differences are negligible for the quark singlet PDF, both in terms of
central values and uncertainties, they are more marked for the gluon
PDF. Two main effects are observed therein.
First, the central value of the SMEFT-PDF gluon moves upwards as compared
to the SM-PDF fit based on the same dataset, ending up halfway between the latter
and the \nnpdfnotop fit.
Second, uncertainties increase for the SMEFT-PDF determination as compared
to the SM-PDFs extracted from the same data, becoming close to the uncertainties
of the \nnpdfnotop fit except for $x\gsim 0.5$.
In both cases, differences are restricted to the large-$x$ region with $x\gsim 0.1$,
where the impact of the dominant top quark pair production measurements is
the largest.

The results of Fig.~\ref{fig:SMEFTpdf_gluon} for the gluon PDF therefore indicate that
within a simultaneous extraction of the PDFs and the EFT coefficients, the impact
of the top quark data on the PDFs is diluted, with the constraints it provides 
partially `reabsorbed' by the Wilson coefficients.
This said, there remains a pull of the top quark data as compared to
the no-top baseline fit
which is qualitatively consistent with the pull
obtained in a SM-PDF determination based on the same dataset, albeit of reduced magnitude.
Interestingly, as we show below, while the SMEFT-PDF gluon is significantly
modified in the joint fit as compared to a SM-PDF reference, much smaller
differences are observed at the level of the bounds on the EFT parameters themselves.

\begin{figure}[t]
\centering
\includegraphics[width=0.49\textwidth]{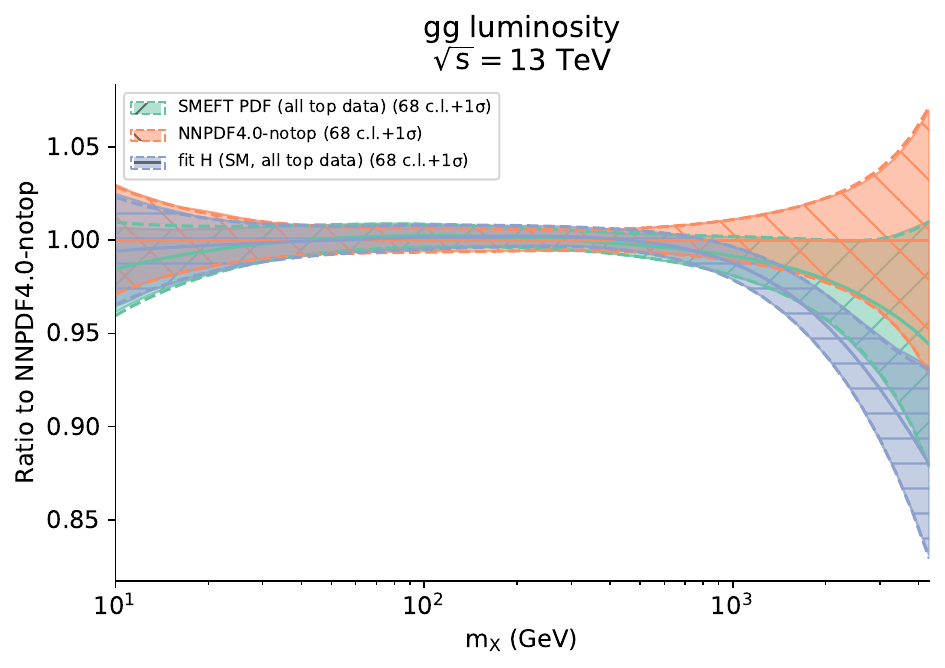}
\includegraphics[width=0.49\textwidth]{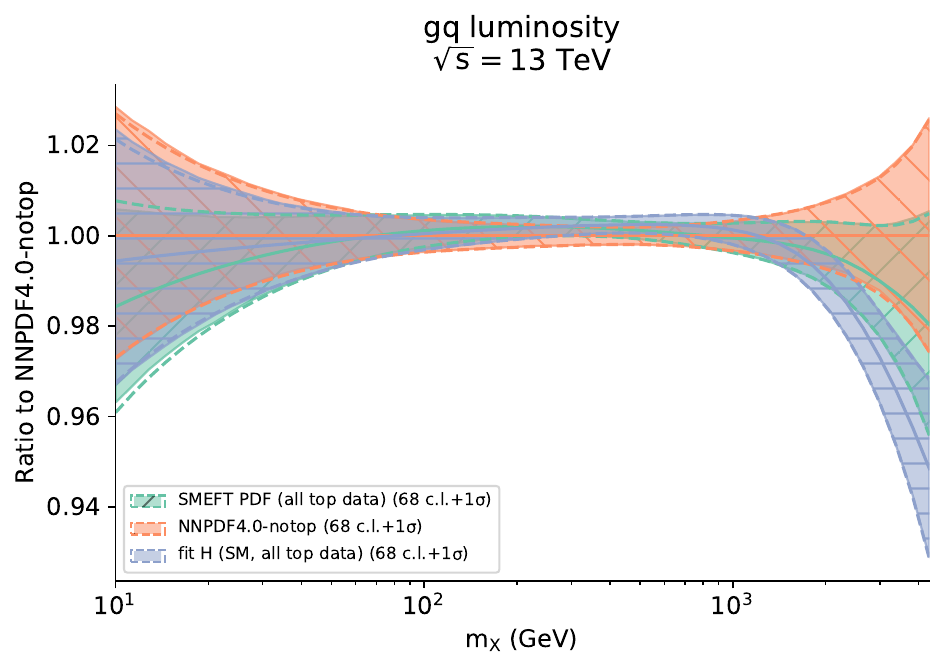}
\caption{The gluon-gluon (left panel) and quark-gluon (right panel)
  partonic luminosities at $\sqrt{s}=13$ TeV as a function of the final-state
  invariant mass $m_X$.
  We compare the\nnpdfnotop baseline fit with its SM-PDF counterpart
  including all top quark data considered (fit H in Table~\ref{tab:fit_list}) as well as with
  the SMEFT-PDF determination.
  Results are presented as the ratio to the central value of the
  \nnpdfnotop baseline.
}
\label{fig:simu_luminosities}
\end{figure}

Fig.~\ref{fig:simu_luminosities} displays the same comparison
as in Fig.~\ref{fig:SMEFTpdf_gluon} now for the case of the gluon-gluon and quark-gluon 
partonic luminosities at $\sqrt{s}=13$ TeV, as a function of the final-state
invariant mass $m_X$.
Consistently with the results obtained at the PDF level, one finds that
the three luminosities are almost identical for $m_X \le 1$ TeV, and
at higher invariant masses the SMEFT-PDF predictions are bracketed
between the \nnpdfnotop fit from above and the SM-PDF which includes
all top data (fit H in Table~\ref{tab:fit_list}) from below.
Hence, the net effect of simultaneously fitting the PDFs
and the EFT coefficients  is a dilution of constraints
provided by the top quark data on the former,
which translates into larger PDF uncertainties (which end up being rather similar to
those of \nnpdfnotop) and an increase
in the large-$m_X$ luminosity, e.g. of 5\% in the $gg$ case for $m_X\simeq 3$ TeV,
as compared to the SM-PDF luminosity.

This dilution arises because of an improved description of the top quark
data included in the fit, especially in the high $m_{t\bar{t}}$ bins.
In the SM-PDF case this can only be obtained by suppressing
the large-$x$ gluon, while in a SMEFT-PDF analysis this can also be achieved by shifting the EFT coefficients
from their SM values.
In other words, as compared to the \nnpdfnotop baseline, the gluon PDF experiences
a suppression at large-$x$ of up to 10\% when fitting the top quark data, and this
pull is reduced by approximately a factor two in the joint SMEFT-PDF determination due to the
coherent effect of the linear EFT cross-sections.

\begin{figure}[t]
\centering
\includegraphics[width=0.49\textwidth]{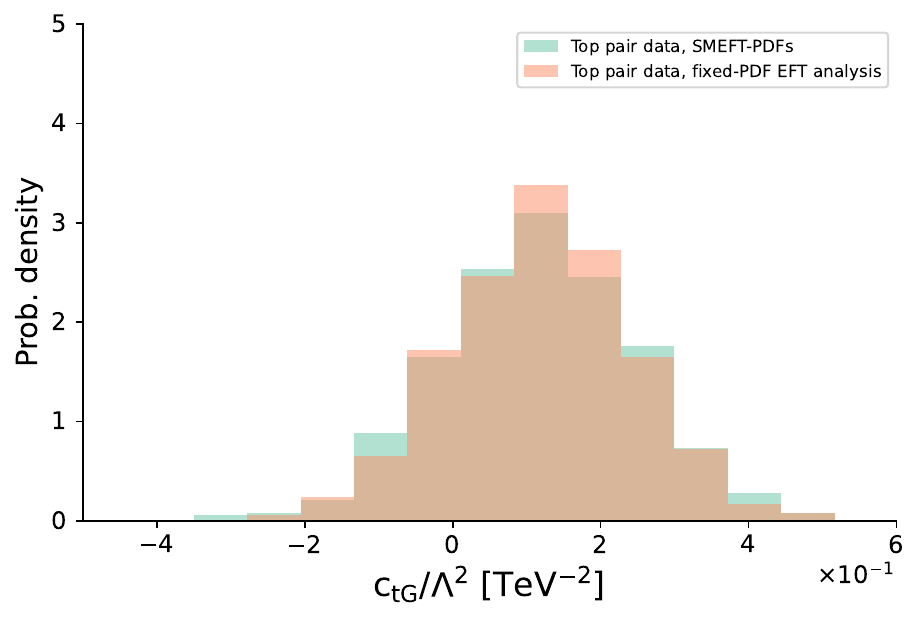}
\includegraphics[width=0.49\textwidth]{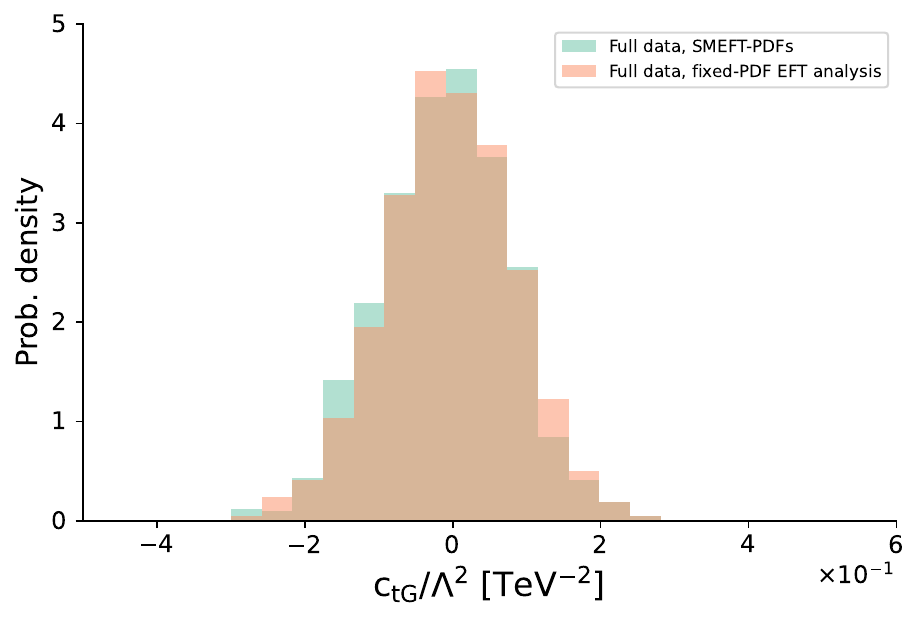}
\caption{Posterior distributions for the Wilson coefficient $c_{tG}$
  associated to the chromomagnetic operator in the joint SMEFT-PDF determination, compared
  with the corresponding results from the fixed-PDF EFT analysis whose settings are described in
  Sect.~\ref{sec:res_smeft}.
  We show results based on only top-quark pair production data (left) and
  in the whole top quark dataset considered in this work (right panel).
}
\label{fig:simultaneous_broadening}
\end{figure}

\paragraph{EFT coefficients from a joint SMEFT-PDF fit.} 
As opposed to the marked effect of the SMEFT-PDF interplay found for the large-$x$ gluon,
its impact is more moderate at the level of the bounds on the EFT coefficients,
and is restricted to mild shifts in the central values and a slight broadening
of the uncertainties.
This is illustrated by Fig.~\ref{fig:simultaneous_broadening},
showing the posterior distributions for the Wilson coefficient $c_{tG}$
  associated to the chromomagnetic operator in the joint SMEFT-PDF determination, compared
  with the corresponding results from the fixed-PDF EFT analysis whose settings are described in
  Sect.~\ref{sec:res_smeft}.
  The comparison is presented both for the fits which consider only top-quark pair production data
  and those based on the whole top quark dataset considered in this work.
  The leading effect of the chromomagnetic operator $\mathcal{O}_{tG}$ is to modify the
  total rates of 
  $t \bar{t}$ production without altering the shape of the differential distributions, and hence
  it plays an important role in a simultaneous SMEFT-PDF determination based on top quark data.

For fits based only on inclusive $t\bar{t}$ data, as shown in the left panel of
Fig.~\ref{fig:simultaneous_broadening}, the two posterior
distributions are similar; the distribution based on the SMEFT-PDF analysis
is slightly broader, approximately $10 \%$ so, as compared to the fixed-PDF EFT fit
to the same measurements.
This slight broadening is washed out in the fit to the full top quark dataset, as shown in the right panel of Fig.~\ref{fig:simultaneous_broadening}.
In both cases, the determination of $c_{tG}$ is consistent with the SM
at a 95\% CL, and the central values of the coefficient
are the same in the SMEFT-PDF and fixed-PDF EFT analyses.
Hence, in the specific case of the chromomagnetic operator, the interplay between PDFs and EFT fits
is rather moderate and restricted to a broadening of at most $10\%$ in the 95\% CL bounds.
Similar comparisons have been carried out for other EFT coefficients as well as in the context
of fits to a subset of the data and/or to a subset of the coefficients.
We find that in general the impact of the SMEFT-PDF interplay
translates to a broadening of the uncertainties
in the EFT coefficients, which at most reaches $30\%$, and
alongside which the central values remain stable\footnote{All results
  obtained with various subsets of the data and of the
  coefficients can be found at \url{https://www.pbsp.org.uk/research/topproject}}.

All in all, within  the global fit based on the best available theory predictions,
results for the EFT coefficients turn out to be very similar
in the fixed-PDF EFT and SMEFT-PDF fits.
This indicates that, provided a broad enough dataset and the most
up-to-date theory calculations are used, the PDF dependence on the cross-sections
entering an EFT interpretation of the LHC data is currently
subdominant and can be neglected (this is not the case for the
PDFs, see Fig.~\ref{fig:simu_luminosities}).
Nevertheless, this statement applies only to the dataset currently available,
and it is likely that the SMEFT-PDF interplay will become more significant in the future
once HL-LHC measurements become available~\cite{AbdulKhalek:2018rok,Azzi:2019yne},
as demonstrated in the case of high-mass Drell-Yan~\cite{Greljo:2021kvv}.

\begin{figure}[t!]
\centering
\includegraphics[width=0.99\textwidth]{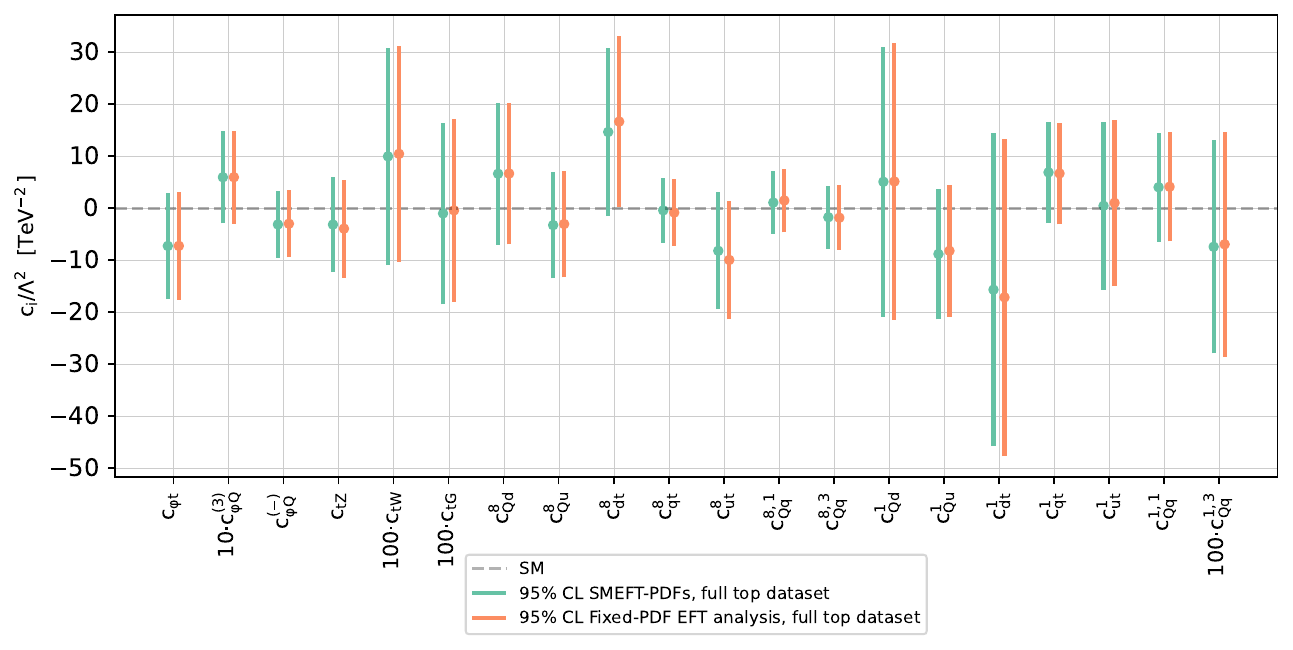}
\caption{
  Comparison of the $95 \%$ CL intervals on the 20 Wilson coefficients
  considered in this work (in the linear EFT case) between the outcome of the joint SMEFT-PDF determination
  and that of the fixed-PDF EFT analysis.
  The latter is based on SM and EFT calculations performed with
 \nnpdfnotop as input, see also  Sect.~\ref{sec:res_smeft}.
  In both cases, results are based on the full 
  top quark dataset being considered and EFT cross-sections
	are evaluated up  to linear, $\mathcal{O}\lp \Lambda^{-2}\rp$, corrections.
  The dashed horizontal line indicates the SM prediction, $c_k=0$.
  Note that some coefficients are multiplied by the indicated prefactor
  to facilitate the visualisation of the results. 
}
\label{fig:smeft_simu_bounds}
\end{figure}

The moderate impact of the SMEFT-PDF interplay on the Wilson coefficients 
for the full top quark dataset considered in this work is summarised in Fig.~\ref{fig:smeft_simu_bounds},
which 
compares the $95 \%$ CL intervals of the 20 fitted Wilson coefficients
relevant for the linear EFT fit obtained from the outcome of the joint SMEFT-PDF determination
and the fixed-PDF EFT analysis.
The latter is based on SM and EFT calculations performed with
\nnpdfnotop as input; see also the description of the settings in
Sect.~\ref{sec:res_smeft}.
The dashed horizontal line indicates the SM prediction,
and  some coefficients are multiplied by the indicated prefactor
to facilitate the visualisation of the results. 
Fig.~\ref{fig:smeft_simu_bounds} demonstrates that, other than slight
broadenings and shifts in the central values,
the results of the two analyses coincide.

\begin{figure}[t!]
\centering
\includegraphics[width=0.49\textwidth]{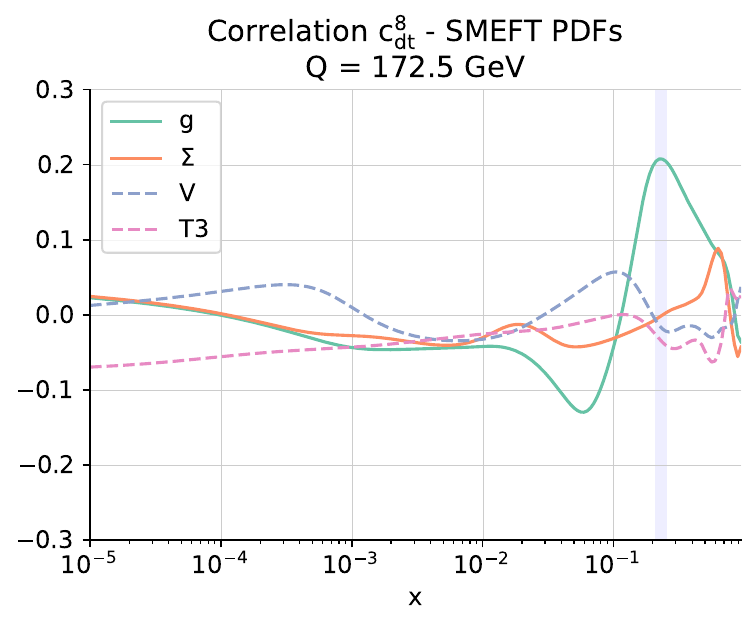}
\includegraphics[width=0.49\textwidth]{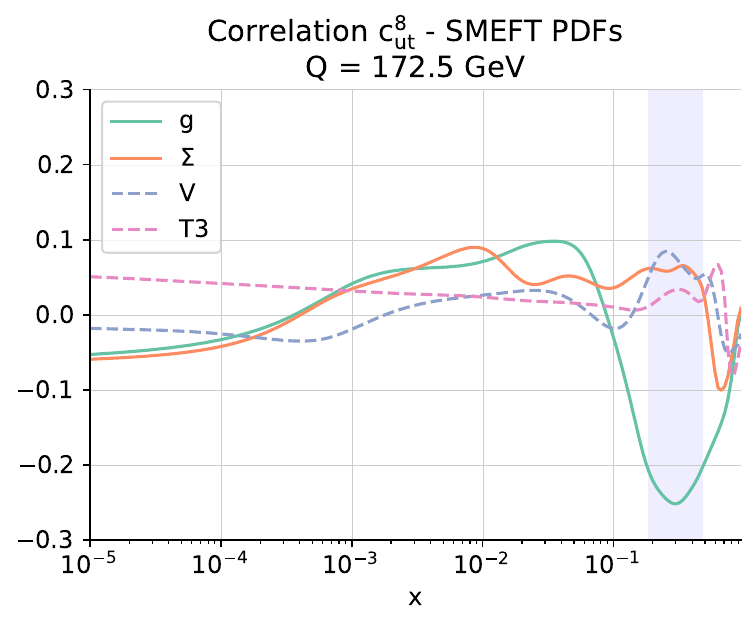}
\includegraphics[width=0.49\textwidth]{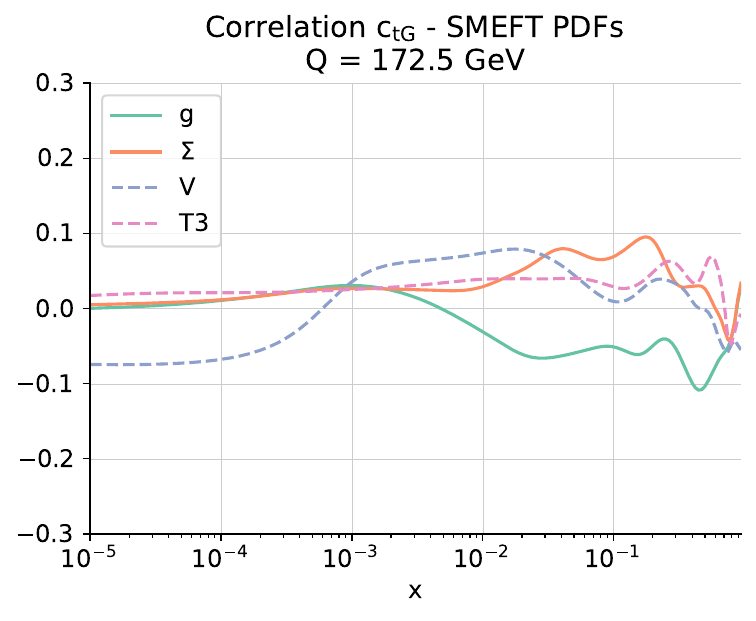}
\includegraphics[width=0.49\textwidth]{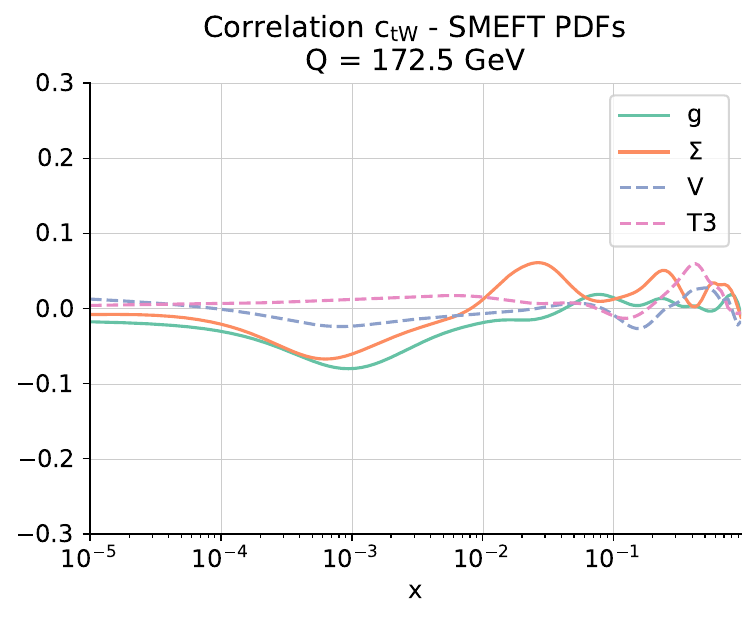}
\caption{The correlation coefficient $\rho(f_i, c_k)$
  between the SMEFT-PDFs $f_i$ and the Wilson coefficients $c_k$ evaluated at $Q=172.5$ GeV
  as a function of $x$.
  Each panel displays the correlations of the coefficient $c_k$
  with the gluon and the total singlet $\Sigma$, total valence $V$, and non-singlet
  triplet $T_3$ PDFs.
  We provide results for representative EFT coefficients, namely $c_{td}^{8}$,
  $c_{tu}^{8}$, $c_{tG}$, and $c_{tW}$.
  The largest correlations within the EFT coefficients  considered in this
  work are associated to four-fermion operators such as  $c_{td}^{8}$ and $c_{tu}^{8}$.
}
\label{fig:smeft_pdf_corr}
\end{figure}

\paragraph{Correlations.} Fig.~\ref{fig:smeft_pdf_corr} displays the
correlation coefficients~\cite{Ball:2010de}
between the SMEFT-PDFs and the Wilson coefficients evaluated at $Q=172.5$ GeV
as a function of $x$.
Each panel displays the correlations of the coefficient $c_k$
with the gluon and the total singlet $\Sigma$, total valence $V$, and non-singlet
triplet $T_3$ PDFs, and we consider four representative EFT coefficients, namely $c_{td}^{8}$,
$c_{tu}^{8}$, $c_{tG}$, and $c_{tW}$.
The largest correlations within the EFT coefficients  considered in this
work are associated to the gluon PDF and four-fermion operators such as  $c_{td}^{8}$ and $c_{tu}^{8}$
in the large-$x$ region, peaking at $x\simeq 0.3$.
Correlations for other values of $x$ and for the quark PDFs are negligible
for all operators entering the analysis.
We note that future data with an enhanced coverage of the high-$m_{t\bar{t}}$ region in top quark
pair-production
might alter this picture, given that for $m_{t\bar{t}}\gsim 3$ TeV
the  $q\bar{q}$ luminosity starts to become more relevant and eventually dominates over the $gg$
contribution.

\begin{figure}[t]
\centering
\includegraphics[width=0.98\textwidth]{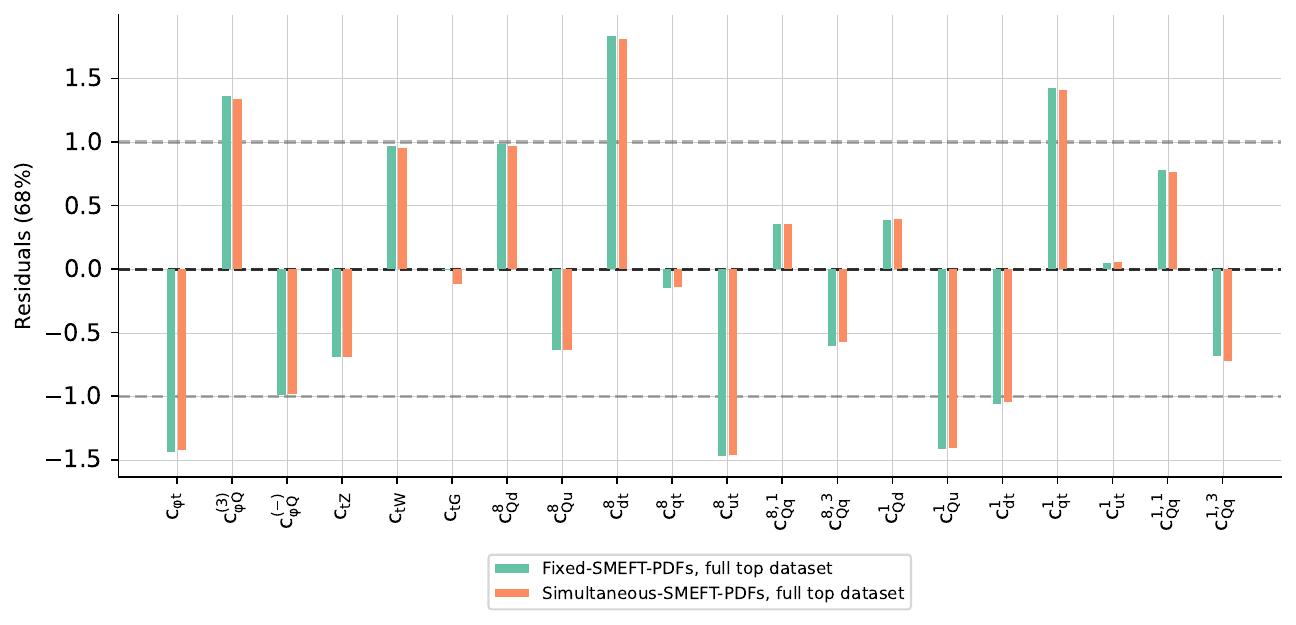}
\caption{The 68\% CL residuals, Eq.~(\ref{eq:residuals}), for the
  same Wilson coefficients displayed in Fig.~\ref{fig:smeft_simu_bounds},
  comparing the outcome of the joint SMEFT-PDF determination
  with that of a fixed-PDF EFT analysis.
  In the latter, we use as input for the theory calculations
  the SMEFT-PDFs obtained in the joint fit rather than the
  \nnpdfnotop set
  used in Sect.~\ref{sec:res_smeft}.
  The horizontal dashed lines indicate the $\pm 1\sigma$ regions.
}
\label{fig:fixedsmeftpdf_res}
\end{figure}

\paragraph{Residuals.} Finally, Fig.~\ref{fig:fixedsmeftpdf_res} displays a similar comparison as
in Fig.~\ref{fig:smeft_simu_bounds} now at the level of the 68\% CL fit residuals defined as
\begin{equation}
  \label{eq:residuals}
R_n = \frac{c_n^*}{\sigma_n} \, ,
\end{equation}
where $c_n^*$ and $\sigma_n$ are the median value and the standard deviation of the Wilson coefficient
$c_n$ respectively, with $n=1,\ldots,N$, where $N$ is the number of operators.
The outcome of the joint SMEFT-PDF determination
is compared with that of a fixed-PDF EFT analysis
where we use as input for the theory calculations
the  SMEFT-PDFs obtained in the joint fit, rather than the \nnpdfnotop set.
That is, in both cases the information provided by the top quark data on the PDFs
and Wilson coefficients is accounted for, but in one case the cross-correlations
are neglected whereas they are accounted for in the other.
The residuals are similar in the two cases; they are slightly bigger (in absolute value)
in the fixed-PDF case in which the correlations between the SMEFT-PDFs and the EFT
coefficients are ignored.
This analysis further emphasises that, for the currently available top quark data, the cross-talk
between PDFs and EFT degrees of freedom does not significantly modify the posterior
distributions in the space spanned by the Wilson coefficients.\\

\noindent In summary, on the one hand we find that from the point of view of a PDF determination,
SM-PDFs and SMEFT-PDFs extracted from top quark data differ by an amount comparable
to their respective uncertainties in the case of the large-$x$ gluon.
On the other hand, at the level of Wilson coefficients the results are unchanged irrespective
of the PDF set used as input for the theory calculations; that is, bounds based
on \nnpdfnotop or the SMEFT-PDFs are almost the same.
Hence, while EFT interpretations of top quark data can safely ignore the PDF dependence,
at least for the settings adopted in this work, a global PDF fit could be significantly
distorted if BSM physics were to be present in the large-energy top quark distributions.

\section{PDF-SMEFT fits from a global dataset}
\label{sec:global}

In this section we present the first global simultaneous fit of SMEFT Wilson coefficients and PDFs, including Deep-Inelastic-Scattering, 
Drell-Yan, top, jets, Higgs, diboson and EW precision observables.  This analysis can be performed thanks to the 
\simunet{} methodology that we have described in the previous chapters. 
In Sect.~\ref{subsec:fixedpdf} we 
present the result of the fixed-PDF analysis, 
producing a global SMEFT fit of a large number of datapoints from the Higgs, top, and EW sectors ($n_{\rm data}\,=\,366$)
to a set of $n_{\rm op}\,=\,40$ SMEFT operators. In Sect.~\ref{subsec:simu} we present the 
simultaneous PDF and SMEFT global fit ($n_{\rm data}\,=\,4985$).
We compare the resulting PDFs and SMEFT bounds to the SM PDFs and to the SMEFT bounds obtained in the fixed-PDF fit respectively, hence 
assessing the effect of the interplay between PDFs and SMEFT effects on both our description of the proton and 
on New Physics bounds. 

\subsection{Global SMEFT fit}
\label{subsec:fixedpdf}

We present the results of this new global fit in two parts, beginning with a discussion 
of the result of the fixed PDF fit, that is a global fit purely of the SMEFT Wilson 
coefficients. 
In Sect.~\ref{subsec:simu} we will present the results of the simultaneous global 
fit of SMEFT WCs and will compare them to those obtained here.

The global analysis is performed at linear order in the SMEFT operator
coefficients, accounting for the interference between the SM and the insertion of a dimension-6 SMEFT operator. 
The linear constraints can be viewed as 
provisional for operators where quadratic contributions are non-negligible. Nevertheless,
keeping those operators in the global fit typically yields conservative marginalised limits
that allows one to assess the impact on other operators and whether this is significant to a first
approximation. An analysis including quadratic contributions requires a new methodology that does not 
rely on the Monte-Carlo sampling method to propagate uncertainties~\cite{progr}, given that the latter fails at
reproducing the Bayesian confidence intervals once quadratic corrections are dominant,  
as described in App.~E of Ref.~\cite{Kassabov:2023hbm}.

Our analysis comprises  $n_{\rm op} = 40$ operators in a simultaneous combination of the constraints 
from $n_{\rm data}\,=\,366$ datapoints 
from the Higgs, EWPO, diboson and top sectors, as described in Sect. \ref{subsec:data}. Note that in the \simunet{} code there 
are more SMEFT operators implemented, in particular the set of four-fermion operators that 
affect Drell-Yan and DIS data. As demonstrated in~\cite{Carrazza:2019sec,Greljo:2021kvv}, 
the effect of the four 
fermion operators in DIS and on the current Drell-Yan data, including the high-mass Drell-Yan 
data included in \nnpdf{} is not significant enough for this data to provide strong constraints 
on the relevant set of four-fermion operators. 
On the contrary, the HL-LHC projections, which include both neutral and charged current 
Drell-Yan data have a strong potential in constraining such operators~\cite{Iranipour:2022iak}. 
As a result, Drell-Yan data are not affected by SMEFT corrections in this current analysis, 
but we leave the user the opportunity to include such effects straightforwardly, 
should the measurements by the LHC increase the constraining potential, or should the user combine 
those with flavour observables, as is done in Ref.~\cite{Bartocci:2023nvp}.

In the following, we also omit from the results the $4$ heavy quark production datasets (four tops and $t\bar{t}b\bar{b}$). 
Their effect is practically negligible on all the operators affecting the other sectors and they would simply constrain $2$ 
directions in the $5$-dimensional space of the four heavy fermion operators. These observables are therefore {\it de facto} decoupled 
from the rest of the dataset.

The input PDF set, which is kept fixed during the pure SMEFT fit, is the
\begin{center}
  \texttt{nnpdf40nnlo\_notop}
\end{center}
baseline, corresponding to the \nnpdf{} fit without the top data, to avoid possible contamination 
between PDF and EFT effects in the top sector~\cite{Kassabov:2023hbm}. 

\begin{figure}[tb!]
  \centering
  \includegraphics[width=0.9\textwidth]{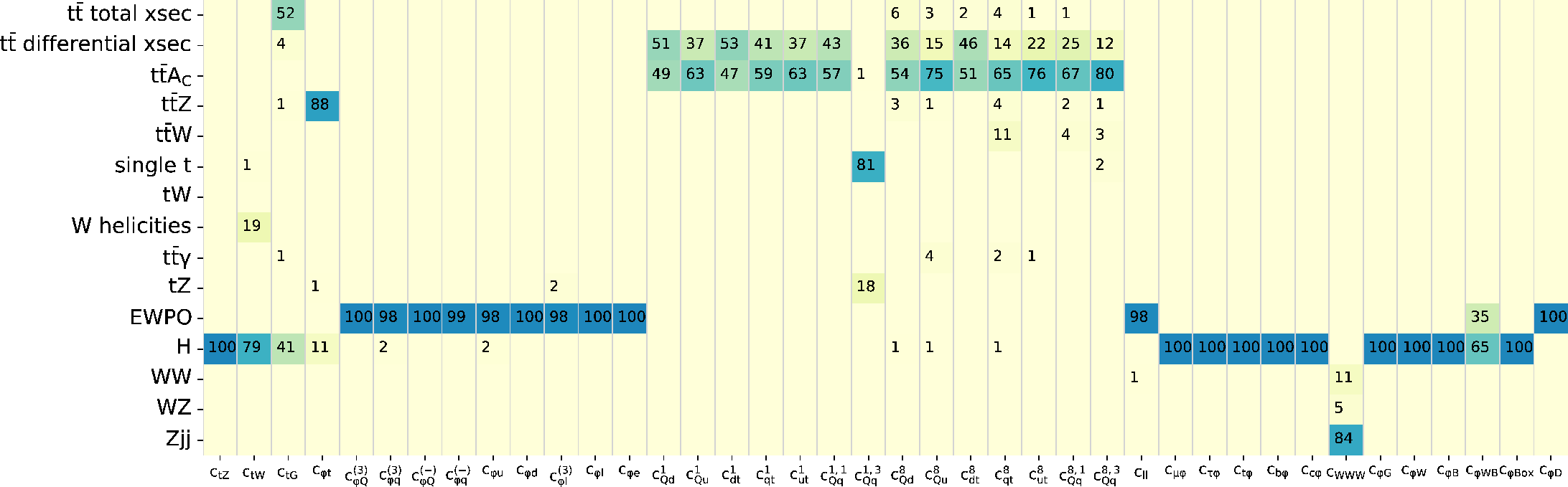}
  \caption{
Relative constraining power (in \%) on each of the operators for each of the processes entering the fit, as defined in Eq.~\eqref{eq:relativeconstrainingpower}.}
\label{fig:fisher}
\end{figure}

In Fig.~\ref{fig:fisher} we use the Fisher information, as defined in Eq. (\ref{eq:fisherdef}), to assess the relative impact of each sector of data included in our analysis on the SMEFT parameter space by plotting the relative percentage constraining power of Eq. (\ref{eq:relativeconstrainingpower}).
We observe that the coefficients $c_{tZ}$, $c_{tW}$ are dominantly constrained by the Higgs sector, while
the coefficient $c_{tG}$ is now constrained both by the $t \bar{t}$ total cross sections and by the 
Higgs measurements, as expected.
Higgs measurements also provide the dominant constraints on the bosonic coefficients
$c_{\varphi G}$, $c_{\varphi W}$, $c_{\varphi B}$, $c_{\varphi \Box}$
and on the Yukawa operators.
Coefficients coupling the Higgs to fermions,
$c_{\varphi Q}^{(3)}$,
$c_{\varphi q}^{(3)}$, $c_{\varphi Q}^{(-)}$, $c_{\varphi q}^{(-)}$, $c_{\varphi u}$, $c_{\varphi d}$,
$c_{\varphi l}^{(3)}$ and $c_{\varphi e}$, receive their dominant constraining power from the electroweak precision observables,
as does the four-fermion coefficient $c_{ll}$.
The coefficient $c_{WWW}$ is constrained by the diboson sector, with measurements of $Zjj$ production
providing the leading constraints at linear order in the SMEFT.
From the top sector of the SMEFT, the four-fermion operators are constrained by measurements of top quark pair
production total cross sections, differential distributions and charge asymmetries.  The exception to this is $c_{Qq}^{(1,3)}$,
which is constrained by both single top production and single top production in association with a $Z$ boson.
Overall we find qualitatively similar results to those shown in Refs.~\cite{Ellis:2020unq,Ethier:2021bye}.

\begin{figure}[tb!]
    \centering
    \includegraphics[width=0.8\textwidth]{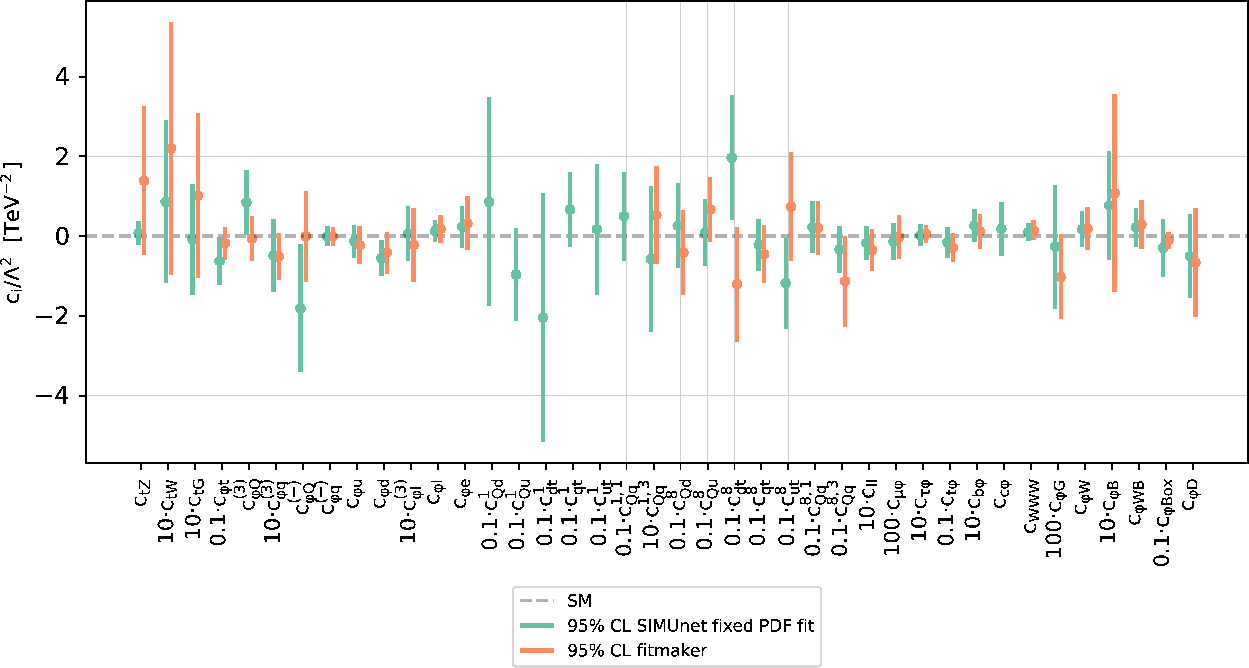}
    \caption{
  Comparison of the bounds for the SMEFT Wilson coefficients determined in the pure  
  SMEFT analysis presented in this work and the \fitmaker{} analysis~\cite{Ellis:2020unq}. 
  The dashed horizontal line is the SM prediction. In each bound, the dot represents 
  the central value and the 95\% CL interval is constructed from the envelope of 
  two standard deviations away from this central value.}
\label{fig:simunet_vs_fitmaker}
\end{figure}

To assess the strength of the bounds on the Wilson coefficients that we find in this work and 
for reference, in Fig.~\ref{fig:simunet_vs_fitmaker} we compare the bounds that we obtain here 
against the fixed-PDF SMEFT analysis presented by the \fitmaker{} 
collaboration~\cite{Ellis:2020unq}, which is the global public SMEFT analysis that currently 
includes all sectors that are considered here.
We observe that the bounds are comparable, with a few differences. First of all \fitmaker{} 
uses LO QCD predictions for the SMEFT corrections in the top sector, hence most of the singlet four 
fermion operators in the top sector are not included, while \simunet{} uses NLO QCD predictions 
and obtains bounds on those, although weak~\cite{Degrande:2020evl}. Moreover, while the 
other bounds overlap and are of comparable size, the ${\cal O}_{tZ}$ is much more constrained in 
the \simunet{} analysis, as compared to the \fitmaker{} one. 
This is due to the combination of all operators and observables, which collectively improves the constraints 
on this specific operator, thanks to the interplay between $O_{tZ}$ and the other operators that enter the same 
observables.

\begin{figure}[b!]
  \centering
  \includegraphics[width=0.8\textwidth]{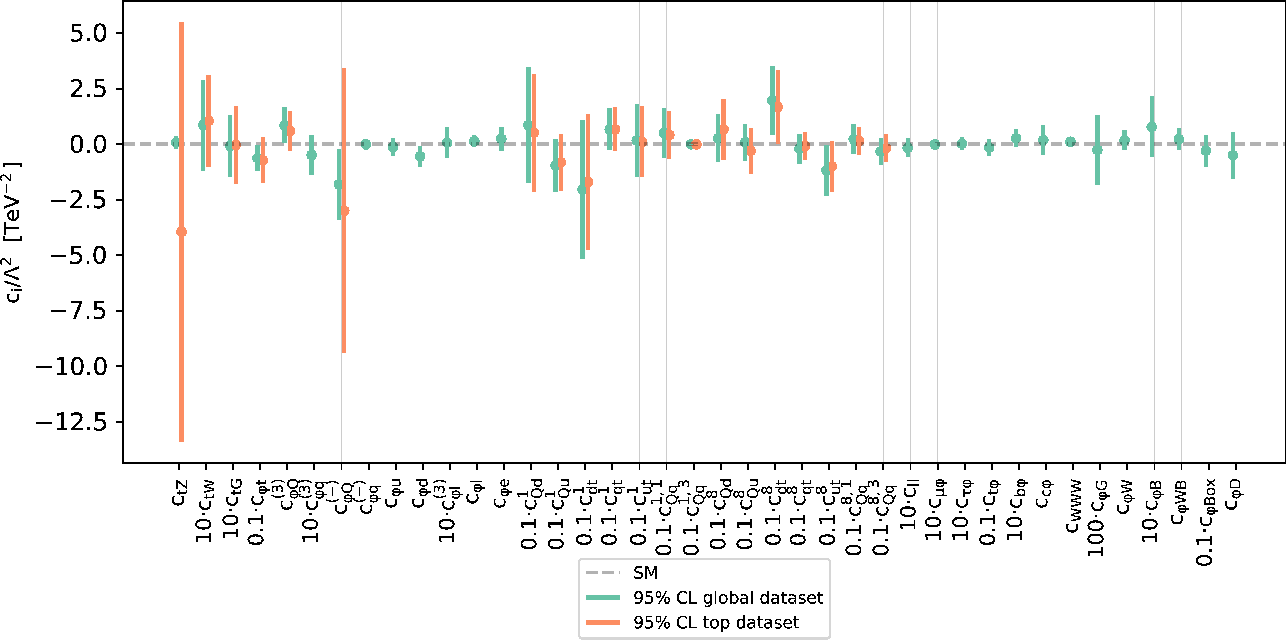}
  \caption{Same as Fig.~\ref{fig:simunet_vs_fitmaker}, now comparing the 95\% CL bounds obtained in the SMEFT top analysis 
  of Ref.~\cite{Kassabov:2023hbm} (old dataset) with the 95\% CL bounds obtained in this global SMEFT analysis.}
\label{fig:smeft_only_top_vs_global}
\end{figure}

In Fig.~\ref{fig:smeft_only_top_vs_global} we show the comparison between the results of this current global SMEFT analysis against 
the SMEFT analysis of the top sector presented in~\cite{Kassabov:2023hbm}.
In general, the results are very compatible, with bounds on most Wilson coefficients comparable between the two fits. 
However, the increased dataset size results in a marked improvement on the constraints of several Wilson coefficients, 
in particular $c_{\phi t}$, $c_{tZ}$ and $c_{\phi Q}^{(-)}$; see Table~\ref{tab:global_vs_top} for comparisons. 
\begin{table}[ht]
  \begin{center}
{\fontsize{8pt}{8pt}\selectfont
  \centering
   \renewcommand{\arraystretch}{2}
   \setlength{\tabcolsep}{5pt}
   \begin{tabular}{lcc}
   \toprule
    \textbf{Operator} & \textbf{Global fit} & \textbf{Top fit} \\
  \toprule
  $c_{\phi t}$ & (-13, -0.22)  & (-18, 3.1) \\
  \midrule
  $c_{tZ}$ & (-0.18, 0.37)  & (-13, 5.5) \\
  \midrule
  $c_{\phi Q}^{(-)}$ & (-3.6, -0.042)  & (-9.4, 3.4)\\
   \bottomrule
  \end{tabular}
  }
\end{center}
    \caption{\label{tab:global_vs_top} The 95\% confidence intervals on three key Wilson coefficients included in this analysis.}
  \end{table}
We observe that the bounds on $c_{\phi t}$ become more stringent by nearly a factor of 2, due to the extra constraints that come 
from the Higgs sector. Analogously, the Higgs constraints reduce the size of the bounds on $c_{\phi Q}^{(-)}$ by a factor of 3. 
The most remarkable effect is seen in $c_{tZ}$ which is now strongly constrained by the top loop contribution to the $H\to\gamma\gamma$ 
decay, for which we include experimental information on the signal strength from LHC Run II.

\begin{figure}[tb!]
  \centering
  \includegraphics[width=0.47\textwidth]{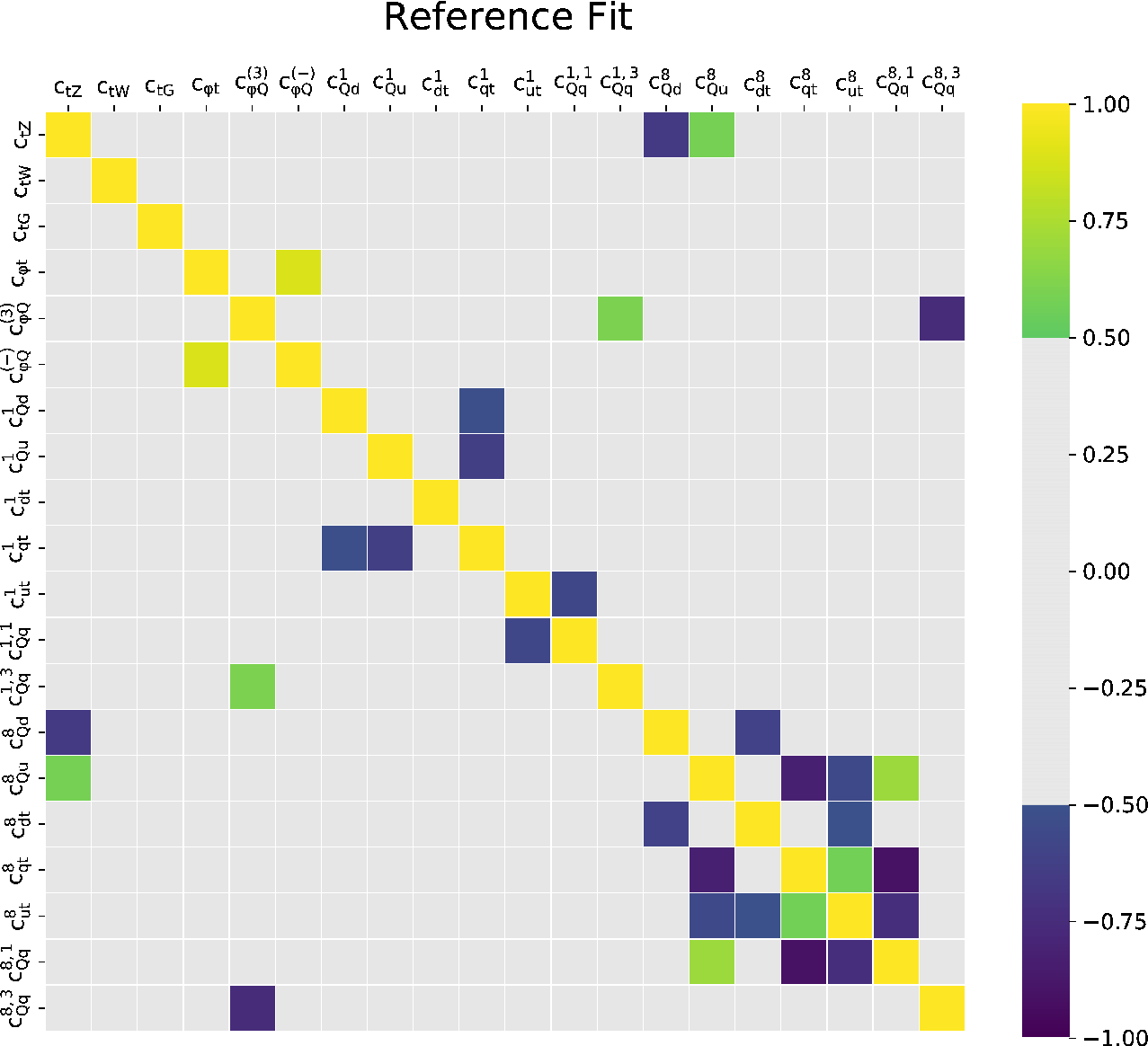}
  \includegraphics[width=0.47\textwidth]{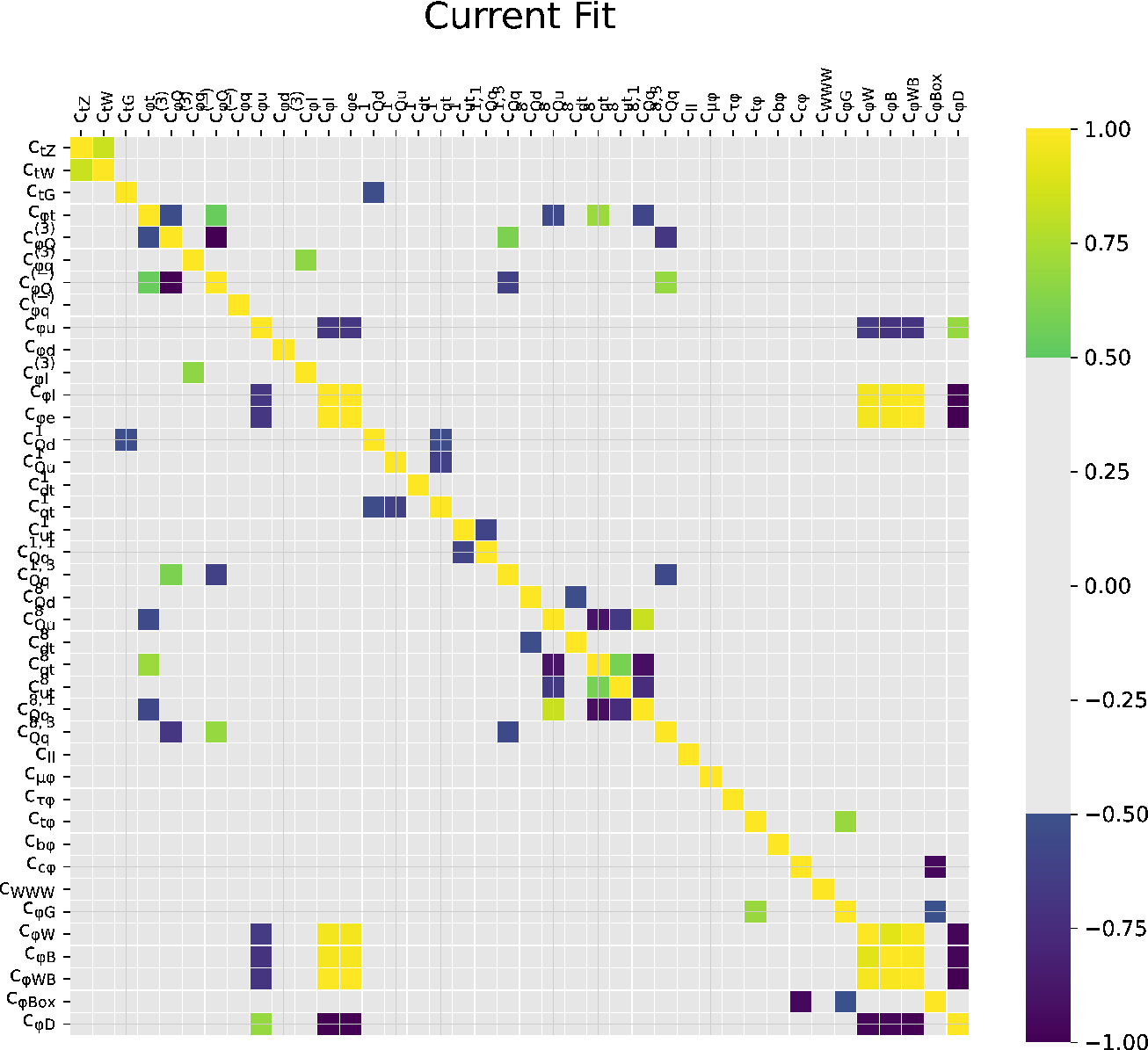}
  \caption{The values of the correlation coefficients evaluated between all pairs of Wilson coefficients entering the
  SMEFT fit of the top-only analysis (left) and the global analysis (right). }
\label{fig:fisher_info}
\end{figure}
It is interesting to compare the correlations between Wilson coefficients evaluated in this analysis
to the top-only SMEFT analysis of Ref.~\cite{Kassabov:2023hbm}. In Fig.~\ref{fig:fisher_info} we 
note that the additional sectors, particularly the Higgs one, 
introduce a degree of anti-correlation between $c_{\phi Q}^{(3)}$ and $c_{\phi t}$ and between $c_{Q q}^{(8,3)}$ and $c_{Q q}^{(1,3)}$, 
while the one between $c_{u t}^{8}$ and $c_{d t}^{8}$ goes away in the global analysis. In the EW sector, we find very strong correlations, 
in agreement with similar studies in the literature. This suggests that there is room for improvement in the sensitivity to the operators 
affecting EW observables once more optimal and targeted measurements are employed in future searches~\cite{GomezAmbrosio:2022mpm,Long:2023mrj}. 

Finally, we display the correlations observed between the PDFs and Wilson coefficients, as defined in Eq. (\ref{eq:correlationL2CT}).
\begin{figure}[hbt!]
  \centering
          \includegraphics[width=0.48\textwidth]{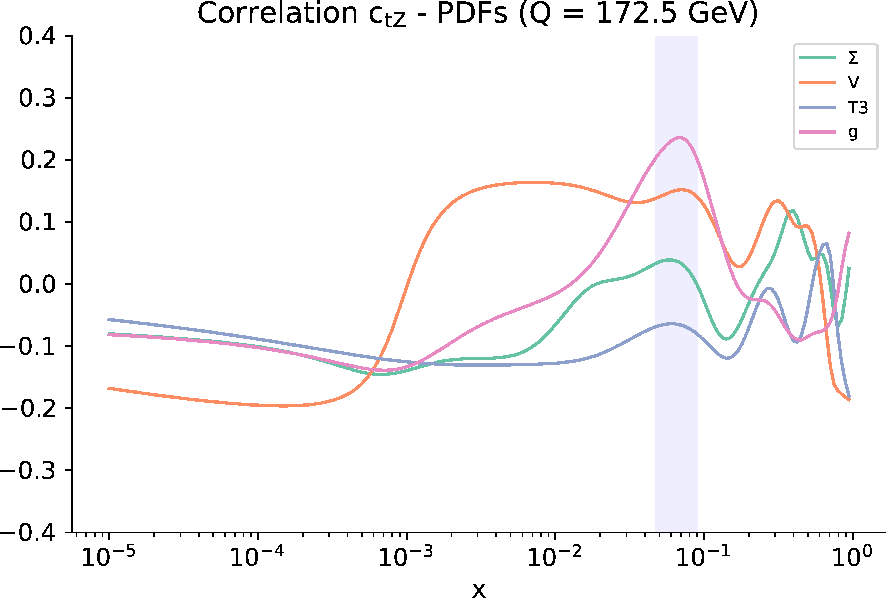}
          \includegraphics[width=0.48\textwidth]{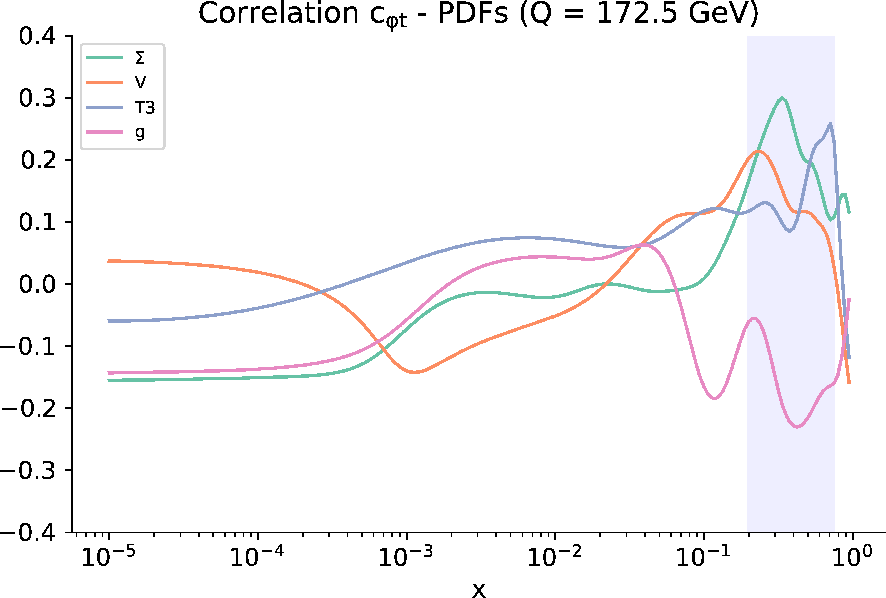}
    \caption{The value of the correlation coefficient $\rho$
            between the PDFs and selected EFT coefficients as a function of $x$
            and $Q=172.5$ GeV.
            We show the results for the gluon, the total singlet $\Sigma$, total valence $V$, and non-singlet
            triplet $T_3$ PDFs.
            We provide results for representative EFT coefficients,
            namely $c_{tZ}$ and  $c_{\phi t}$.
          }
  \label{fig:pdfbsmcorr_simu}
\end{figure}
Fig.~\ref{fig:pdfbsmcorr_simu} displays a selection of the largest correlations.
We observe that the gluon PDF in the medium to large-$x$ region is significantly correlated with the Wilson coefficients 
$c_{tZ}$, as one would expect given that $c_{tZ}$ is strongly constrained by the top loop contribution to the Higgs decay, which
in turn is predominantly correlated to the gluon-gluon parton luminosity. On the other hand $c_{\phi t}$ is anticorrelated
with the gluon and correlated to the singlet and triplet in the large-$x$ region.
This is not surprising, given the impact of top quark pair production total cross sections and differential distributions
in constraining these PDFs and Wilson coefficients.  
Whilst these correlations are computed from a determination of the SMEFT in which the PDFs are fixed to SM PDFs, 
the emergence of non-zero correlations provides an indication of the potential for interplay between the PDFs and the SMEFT
coefficients; this interplay will be investigated in a simultaneous determination
in the following section.

\subsection{Global simultaneous SMEFT and PDF fit}
\label{subsec:simu} 

In this section we present the results of the simultaneous global fit of SMEFT Wilson coefficients and PDFs. We 
compare the bounds on the coefficients obtained in the two analyses as well as the resulting PDF sets, 
to assess whether the inclusion of more PDF-independent observables modifies the interplay 
observed in the top-focussed analysis of Ref.~\cite{Kassabov:2023hbm}.

\begin{figure}[tb!]
    \centering
    \includegraphics[width=0.8\textwidth]{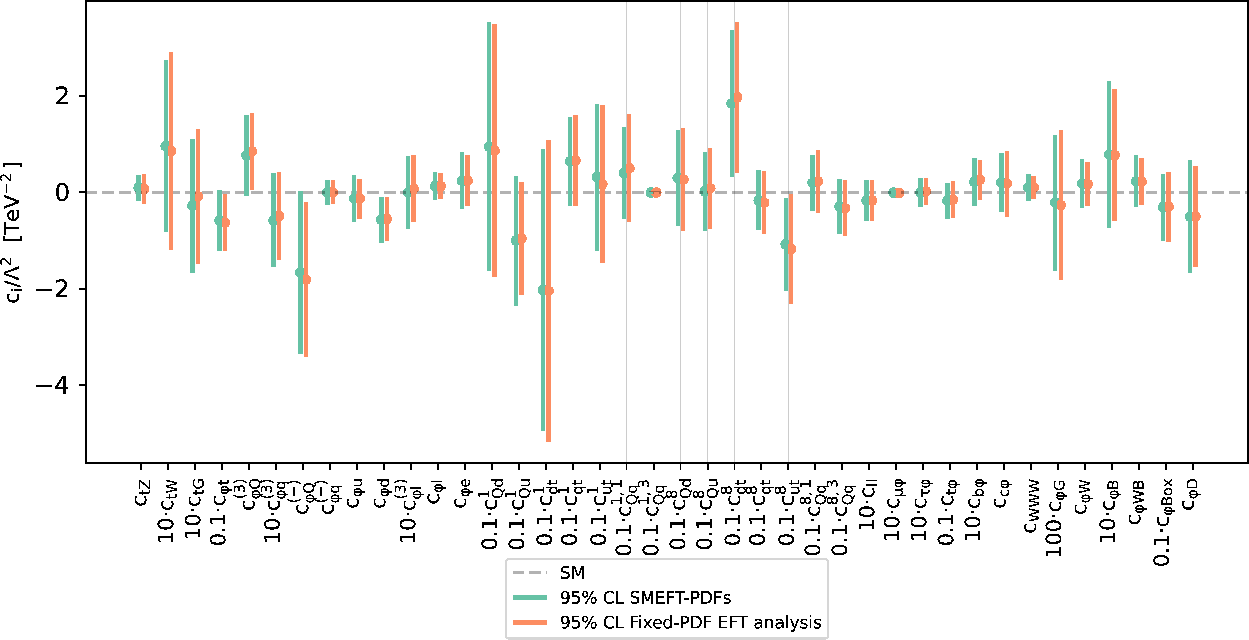}
\caption{
  Comparison of the 95\% CL intervals on the 40 Wilson coefficients considered in this work (in the linear
EFT case) between the outcome of the joint SMEFT-PDF determination and that of the fixed-PDF EFT analysis. The
latter is based on SM and EFT calculations performed with {\tt NNPDF4.0-notop} as input. In both cases,
results are based on the global dataset being considered and SMEFT cross-sections are evaluated up to linear,
corrections. The dashed horizontal line indicates the SM prediction, $c_k = 0$. Note that some coefficients are
multiplied by the indicated prefactor to facilitate the visualisation of the results.}
\label{fig:fixed_vs_simu}
\end{figure}

The first observation is that, similarly to the top-sector results observed in Ref.~\cite{Kassabov:2023hbm}, 
in the global fit presented here the interplay between PDFs and SMEFT Wilson coefficients is weak. 
Indeed in Fig.~\ref{fig:fixed_vs_simu} we observe that, including all data listed in Sect.~\ref{subsec:data}, 
the bounds on the SMEFT Wilson coefficients are essentially identical in a SMEFT-only fit and  
in a simultaneous fit of SMEFT WCs and PDFs. The only mild sign of interplay is shown by the 
operator $c_{\phi t}$, which undergoes a fair broadening in the simultaneous fit compared to the fixed-PDF 
SMEFT fit.

The PDF fit is more interesting. It was shown in Ref.~\cite{Kassabov:2023hbm} that when comparing: 
(i) a SM PDF fit excluding top data, (ii) a simultaneous PDF-SMEFT using top data, 
(iii) a SM PDF fit including top data, there is a hierarchy of shifts in the gluon-gluon luminosity 
at high invariant masses. In particular, the gluon-gluon luminosity of the simultaneous fit (ii) is reduced at high 
invariant masses compared to fit (i), whilst the gluon-gluon luminosity of the SM fit including all top data 
(iii) is even further reduced at high invariant masses relative to the result of the simultaneous fit (ii). 
This can be explained due to the additional SMEFT degrees of freedom in (ii) allowing for a better description 
of the top data with the PDF remaining compatible with the no-top PDF.

\begin{figure}[tb!]
\centering
\includegraphics[width=0.7\textwidth]{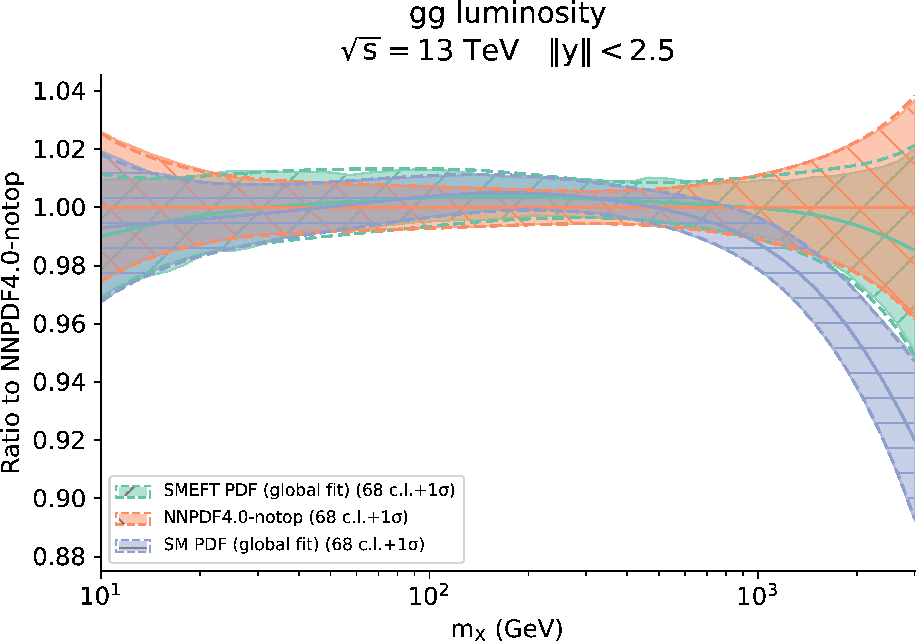}
\caption{ \label{fig:gglumi} The gluon-gluon partonic luminosities at $\sqrt{s} = 13$ TeV
as a function of the final-state invariant mass $m_X$. We compare the {\tt NNPDF4.0-notop} 
baseline fit with its SM-PDF counterpart including all new data presented in Sect.~\ref{subsec:data}, 
as well as with the simultaneous SMEFT-PDF determination. Results are presented as the ratio to the 
central value of the {\tt NNPDF4.0-notop} baseline.}
\end{figure}
As it can be observed in Fig.~\ref{fig:gglumi}, 
in the new global fit presented in this chapter, there is a shift in the gluon-gluon luminosity in the simultaneous 
PDF and SMEFT fit relative to the baseline no-top PDF set, however the shift is much smaller than the shift that one 
has by including all data in the SM, meaning that the effect of the inclusion of the top, Higgs, EWPO and diboson
data has a much smaller effect on the PDFs due to the interplay between the pull of the SMEFT coefficients and the 
pull of the new data on the PDFs.  
The shift is even smaller compared to the shift of the simultaneous top fit relative to the no-top 
PDF presented in Ref.~\cite{Kassabov:2023hbm}. 
This is due to the inclusion of new high-mass Drell-Yan data in the new fit, which were not 
included in \nnpdf{}. These data favour a softer singlet PDF combination in the high-mass region, 
and as a result they enhance the gluon in the same region, due to sum rules. 

Finally, it is also interesting to compare the fit quality of the SMEFT fit keeping PDF fixed and the
simultaneous SMEFT-PDF fit. We display the $\chi^2$ per datapoint in Tables~\ref{tab:chi2-top-pair},~\ref{tab:chi2-single-top-simu} and ~\ref{tab:chi2-other}
for each dataset appearing in both the simultaneous SMEFT-PDF fit and the fixed PDF SMEFT fit.
The fit quality for inclusive and associated top pair production is shown in Table~\ref{tab:chi2-top-pair},
for inclusive and associated single top production in Table~\ref{tab:chi2-single-top-simu} and
for Higgs, diboson and electroweak precision observables in Table~\ref{tab:chi2-other}.
We also display the total fit quality of the 366 SMEFT-affected datapoints in Table~\ref{tab:chi2-other}.

Overall, we observe that the fit quality of the total dataset remains stable between the SMEFT-PDF
fit and the SMEFT fit, with a total $\chi^2$ of 0.91 for both fits, as shown in Table~\ref{tab:chi2-other}.
This is also observed in the Higgs and electroweak precision observables, while in the diboson sector
a small improvement from $\chi^2 = 1.24$ to $\chi^2 = 1.23$
is observed.
A similar improvement in the $\chi^2$ is observed in the inclusive top quark pair production 
datasets in Tables~\ref{tab:chi2-top-pair} from $\chi^2=1.03$ to $\chi^2=1.01$, and in inclusive single top datasets 
in Tables~\ref{tab:chi2-single-top-simu} from $\chi^2=0.34$ to $\chi^2=0.33$, indicating that
in these sectors, the SMEFT-PDF fit provides a better fit to the data.  This improvement in 
the $\chi^2$ in these sectors is not as significant as the improvement observed in Ref.~\cite{Kassabov:2023hbm},
however.  As discussed below Fig.~\ref{fig:gglumi}, this is due to the inclusion of new high-mass Drell-Yan data in the new fit.

  \begin{table}[htbp]
    \begin{center}
    \renewcommand{\arraystretch}{1.4}
    \scriptsize
    \begin{tabular}{ l | c| C{1.2cm} |  C{1.2cm} }
   \toprule
   Dataset & $n_{\rm dat}$   &  \multicolumn{2}{c}{$\chi^2/n_{\rm dat}$}  \\[1ex]
    &   & SMEFT-PDF fit & SMEFT fit   \\
   \midrule
      ATLAS $\sigma(t\bar{t})$, dilepton, 7 TeV    & 1    & 1.87 &  1.84 \\
      ATLAS $\sigma(t\bar{t})$, dilepton, 8 TeV & 1 & 0.33 & 0.35 \\
      ATLAS $1/\sigma d\sigma/dm_{t\bar{t}}$, dilepton, 8 TeV & 5 & 0.30 &  0.30 \\
      ATLAS $\sigma(t\bar{t})$, $\ell+$jets, 8 TeV & 1 & 3$\cdot 10^{-5}$ &  3$\cdot 10^{-4}$ \\
      ATLAS $1/\sigma d\sigma/d|y_t|$, $\ell+$jets, 8 TeV & 4&  1.27 & 1.22 \\
      ATLAS $1/\sigma d\sigma/d|y_{t\bar{t}}|$, $\ell+$jets, 8 TeV & 4 & 2.8 & 2.95  \\
      ATLAS $\sigma(t\bar{t})$, dilepton, 13 TeV           & 1   & 2$\cdot 10^{-3}$ & 2$\cdot 10^{-4}$  \\
      ATLAS $\sigma(t\bar{t})$, hadronic, 13 TeV         & 1    & 0.09 &  0.09   \\
      ATLAS $1/\sigma d^2\sigma/d|y_{t\bar{t}}|dm_{t\bar{t}}$, hadronic, 13 TeV & 10 & 1.84 & 1.88 \\
      ATLAS $\sigma(t\bar{t})$, $\ell+$jets, 13 TeV & 1 & 0.01 & 0.01 \\
      ATLAS $1/\sigma d\sigma/dm_{t\bar{t}}$, $\ell+$jets, 13 TeV & 8 & 2.07 & 2.1  \\
      \midrule
      CMS $\sigma(t\bar{t})$, combined, 5 TeV & 1 & 0.23 &  0.22 \\
      CMS $\sigma(t\bar{t})$, combined, 7 TeV & 1 & 0.01 & 0.01  \\
      CMS $\sigma(t\bar{t})$, combined, 8 TeV & 1 & 0.12 & 0.13  \\
      CMS $1/\sigma d^2\sigma/d|y_{t\bar{t}}|dm_{t\bar{t}}$, dilepton, 8 TeV & 16 & 0.51 & 0.53  \\
      CMS $1/\sigma d\sigma/d|y_{t\bar{t}}|$, $\ell+$jets, 8 TeV & 9  & 0.93 & 0.94  \\
      CMS $\sigma(t\bar{t})$, dilepton, 13 TeV & 1 & 0.60 & 0.62 \\
      CMS $1/\sigma d\sigma/dm_{t\bar{t}}$, dilepton, 13 TeV & 5 & 2.26 & 2.24  \\
      CMS $\sigma(t\bar{t})$, $\ell+$jets, 13 TeV & 1 & 1.90 & 1.98 \\
      CMS $1/\sigma d\sigma/dm_{t\bar{t}}$, $\ell+$jets, 13 TeV & 14& 0.93 & 0.94  \\
      \midrule
      ATLAS charge asymmetry, 8 TeV & 1 & 0.63 & 0.63\\
      ATLAS charge asymmetry, 13 TeV & 5& 0.87 & 0.91 \\
      CMS charge asymmetry, 8 TeV & 3 & 0.06 & 0.06\\
      CMS charge asymmetry, 13 TeV & 3 & 0.39 & 0.36  \\
      ATLAS \& CMS combined charge asy., 8 TeV & 6 & 0.60 & 0.61\\
      \midrule
      ATLAS $W$-hel., 13 TeV & 2 & 7$\cdot 10^{-5}$ &  1$\cdot 10^{-3}$ \\
      ATLAS \& CMS combined $W$-hel., 8 TeV & 2&  0.33 & 0.34  \\
         \midrule
	  {\bf Total inclusive $t\bar{t}$}  & {\bf 108}  & {\bf 1.01}
    & {\bf 1.03}\\
       \bottomrule
    \end{tabular}
    \end{center}
    \caption{\label{tab:chi2-top-pair}
    The values of the $\chi^2$ per data point for the $t\bar{t}$
    production datasets that enter both the SMEFT-only fit and the 
    simultaneous SMEFT-PDF fit. For each dataset, we indicate the number of data points $n_{\rm dat}$ and the 
    $\chi^2$ for the SMEFT-only fit with PDF fixed to the input PDF {\tt NNPDF4.0-notop}  and for the 
    simultaneous SMEFT-PDF determination. 
  }
  \end{table}

\begin{table}[htbp]
    \begin{center}
    \renewcommand{\arraystretch}{1.60}
  \small
  \begin{tabular}{ l | c| C{1.4cm} |  C{1.4cm} }
   \toprule
   Dataset & $n_{\rm dat}$   &  \multicolumn{2}{c}{$\chi^2/n_{\rm dat}$}  \\[1.5ex]
    &   & SMEFT-PDF fit & SMEFT fit   \\
   \midrule
      ATLAS $\sigma(t\bar{t}Z)$, 8 TeV & 1 & 1.40 & 1.33 \\
      ATLAS $\sigma(t\bar{t}W)$, 8 TeV & 1 & 0.62 & 0.71   \\
      ATLAS $\sigma(t\bar{t}Z)$, 13 TeV & 1 & 2$\cdot 10^{-6}$ & 5$\cdot 10^{-3}$  \\
      ATLAS $1/\sigma d\sigma(t\bar{t}Z)/dp_T^Z$, 13 TeV & 5 & 1.85 & 1.84 \\
      ATLAS $\sigma(t\bar{t}W)$, 13 TeV & 1& 2$\cdot 10^{-2}$ & 4$\cdot 10^{-3}$\\
      \midrule
      ATLAS $\sigma(t\bar{t}\gamma)$, 8 TeV & 1 & 0.33 & 0.29  \\
      \midrule
      CMS $\sigma(t\bar{t}Z)$, 8 TeV & 1 & 2$\cdot 10^{-3}$ & 5$\cdot 10^{-3}$\\
      CMS $\sigma(t\bar{t}W)$, 8 TeV & 1 & 0.69 & 0.78  \\
      CMS $\sigma(t\bar{t}Z)$, 13 TeV & 1& 0.10 & 0.15 \\
      CMS $1/\sigma d\sigma(t\bar{t}Z)/dp_T^Z$, 13 TeV & 3 & 0.86 & 0.88 \\
      CMS $\sigma(t\bar{t}W)$, 13 TeV & 1 & 0.48 & 0.35 \\
      \midrule
      CMS $\sigma(t\bar{t}\gamma)$, 8 TeV & 1 & 0.01 & 7$\cdot 10^{-3}$ \\
      \midrule
     {\bf Total associated $t\bar{t}$}  & {\bf 18}  &  {\bf 0.86} &  {\bf 0.86 }  \\
   \bottomrule
  \end{tabular}
  \end{center}
    \caption{ \label{tab:chi2-top-pair-2}
      Same as Table~\ref{tab:chi2-top-pair} for associate $t\bar{t}$ and vector boson production.
  }
  \end{table}

\begin{table}[htbp]
    \begin{center}
    \renewcommand{\arraystretch}{1.30}
  \small
  \begin{tabular}{ l | c| C{1.4cm} |  C{1.4cm} }
   \toprule
   Dataset & $n_{\rm dat}$   &  \multicolumn{2}{c}{$\chi^2/n_{\rm dat}$}  \\[1.5ex]
    &   & SMEFT-PDF fit & SMEFT fit   \\
   \midrule 
	  ATLAS $t$-channel $\sigma(t)$, 7 TeV & 1 & 0.24 & 0.26 \\
	  ATLAS $t$-channel $\sigma(\bar{t})$, 7 TeV & 1 & 0.01 & 0.02 \\
	  ATLAS $t$-channel $1/\sigma d\sigma(tq)/dy_t$, 7 TeV & 3 & 0.9 & 0.89 \\
	  ATLAS $t$-channel $1/\sigma d\sigma(\bar{t}q)/dy_{\bar{t}}$, 7 TeV & 3 & 0.06 & 0.06 \\
	  ATLAS $t$-channel $\sigma(t)$, 8 TeV & 1 & 0.01 & 0.02 \\
	  ATLAS $t$-channel $\sigma(\bar{t})$, 8 TeV & 1 & 0.30 & 0.24 \\
	  ATLAS $t$-channel $1/\sigma d\sigma(tq)/dy_t$, 8 TeV & 3 & 0.28 & 0.28 \\
	  ATLAS $t$-channel $1/\sigma d\sigma(\bar{t}q)/dy_{\bar{t}}$, 8 TeV & 3 & 0.19 & 0.20 \\
	  ATLAS $s$-channel $\sigma(t + \bar{t})$, 8 TeV & 1 & 0.02 & 0.04 \\
	  ATLAS $t$-channel $\sigma(t)$, 13 TeV & 1 & 0.31 & 0.32 \\
	  ATLAS $t$-channel $\sigma(\bar{t})$, 13 TeV & 1 & 0.05 & 0.03 \\
	  ATLAS $s$-channel $\sigma(t + \bar{t})$, 13 TeV & 1 & 0.05 & 0.03 \\
          \midrule
	  CMS $t$-channel $\sigma(t) + \sigma(\bar{t})$, 7 TeV & 1 & 0.02 & 0.02 \\
	  CMS $t$-channel $\sigma(t)$, 8 TeV & 1 & 0.34 & 0.31 \\
	  CMS $t$-channel $\sigma(\bar{t})$, 8 TeV & 1 & 0.98 & 1.07 \\
	  CMS $s$-channel $\sigma(t + \bar{t})$, 8 TeV & 1 & 1.42 & 1.45 \\
	  CMS $t$-channel $\sigma(t)$, 13 TeV & 1 & 0.34 & 0.35 \\
	  CMS $t$-channel $\sigma(\bar{t})$, 13 TeV & 1 & 0.02 & 0.03 \\
	  CMS $t$-channel $1/\sigma d\sigma/d|y^{(t)}|$, 13 TeV & 4 & 0.40 & 0.40 \\
         \midrule
	  {\bf Total inclusive single top}  & {\bf 30}  & {\bf 0.33} & {\bf 0.34}  \\
         \midrule
         \midrule
	  ATLAS $\sigma(tW)$, dilepton, 8 TeV & 1 & 0.07 & 0.05 \\
	  ATLAS $\sigma(tW)$, single-lepton, 8 TeV & 1 & 0.35 & 0.32 \\
	  ATLAS $\sigma(tW)$, dilepton, 13 TeV & 1 & 0.74 & 0.71 \\
	  ATLAS $\sigma_{\text{fid}}(tZj)$, dilepton, 13 TeV & 1 & 0.30 & 0.24 \\
	 \midrule
	  CMS $\sigma(tW)$, dilepton, 8 TeV & 1 & 0.08 & 0.07 \\
	  CMS $\sigma(tW)$, dilepton, 13 TeV & 1 & 2.21 & 2.41 \\
	  CMS $\sigma_{\text{fid}}(tZj)$, dilepton, 13 TeV & 1 & 0.13 & 0.17 \\
	  CMS $d\sigma_{\text{fid}}(tZj)/dp_T^{t}$, dilepton, 13 TeV & 3 & 0.13 & 0.14 \\
	  CMS $\sigma(tW)$, single-lepton, 13 TeV & 1 & 1.51 & 1.42 \\
	 \midrule
         {\bf Total associated single top}  & {\bf 11}  &  {\bf 0.53} &  {\bf 0.53}  \\
	 \bottomrule
  \end{tabular}
  \end{center}
    \caption{\label{tab:chi2-single-top-simu}
	Same as Tab.~\ref{tab:chi2-top-pair}, now for inclusive and associated single top datasets.}
  \end{table}

\begin{table}[htbp]
    \begin{center}
    \renewcommand{\arraystretch}{1.60}
  \small
  \begin{tabular}{ l | c| C{1.4cm} |  C{1.4cm} }
   \toprule
   Dataset & $n_{\rm dat}$   &  \multicolumn{2}{c}{$\chi^2/n_{\rm dat}$}  \\[1.5ex]
    &   & SMEFT-PDF fit & SMEFT fit   \\
   \midrule 
	 ATLAS STXS combination, $\mu_{H}$, 13 TeV & 25 & 0.49 & 0.50 \\
	 ATLAS signal strength $H \rightarrow Z \gamma$, 13 TeV & 1 & 4$\cdot 10^{-5}$ & 3$\cdot 10^{-3}$ \\
	 ATLAS signal strength $H \rightarrow \mu^{+} \mu^{-}$, 13 TeV & 1 & 7$\cdot 10^{-3}$ & 2$\cdot 10^{-3}$ \\
	 CMS signal strength combination, $\mu_{H}$, 13 TeV & 24 & 0.66 & 0.67 \\
	 ATLAS \& CMS signal strength combination, $\mu_{H}$,  7 and 8 TeV & 22 & 0.98 & 0.96 \\
	 \midrule
	 {\bf Total Higgs}  & {\bf 73}  &  {\bf 0.68} &  {\bf 0.68}  \\
	 \midrule
	 \midrule
	 ATLAS $d \sigma _{W^+W^-}/d m_{e \mu}$, 13 TeV & 13 & 1.69 & 1.70 \\
	 ATLAS $d \sigma _{WZ} / d m_{T}$, 13 TeV & 6 & 0.77 & 0.77 \\
	 ATLAS $d \sigma(Zjj)/d \Delta \phi_{jj}$, 13 TeV & 12 & 0.79 & 0.79 \\
	 CMS $d \sigma _{WZ} / d p_{T}$, 13 TeV & 11 & 1.42 & 1.43 \\
	 LEP $d \sigma _{WW} / d cos(\theta _W)$, 0.182 TeV & 10 & 1.39 & 1.39 \\
	 LEP $d \sigma _{WW} / d cos(\theta _W)$, 0.189 TeV & 10 & 0.91 & 0.92 \\
	 LEP $d \sigma _{WW} / d cos(\theta _W)$, 0.198 TeV & 10 & 1.58 & 1.57 \\
         LEP $d \sigma _{WW} / d cos(\theta _W)$, 0.206 TeV & 10 & 1.07 & 1.08 \\
	 \midrule
	  {\bf Total Diboson}  & {\bf 82}  &  {\bf 1.23} &  {\bf 1.24 }  \\
	 \midrule 
	 \midrule
	 LEP $Z$ observables, 0.25 TeV & 19 & 0.52  & 0.52 \\
	 LEP $\mathcal{B}(W \rightarrow l^{-} \bar{v}_l)$ 0.196 TeV & 3 & 2.59 & 2.59 \\
	 LEP $\sigma(e^+ e^- \rightarrow e^+ e^-)$ 0.189 TeV & 21 & 1.02 & 1.03 \\
	 LEP $\hat{\alpha}^{(5)}_{\rm}(M_Z)$ 0.209 TeV & 1 & 0.47 & 0.25 \\
	 \midrule
	 {\bf Total EWPO}  & {\bf 44}  &  {\bf 0.90} &  {\bf 0.90 }  \\
	  \midrule
	  \midrule
	 {\bf Total} & {\bf 366} & {\bf 0.91} & {\bf 0.91} \\
	 \bottomrule
  \end{tabular}
  \end{center}
	\caption{\label{tab:chi2-other}
	Same as Tab.~\ref{tab:chi2-top-pair}, now for the Higgs, diboson and electroweak precision observables.}
  \end{table}

\section{Conclusion}
\label{sec:summary_top}

In this chapter we have carried out a comprehensive study of the PDF-BSM interplay in two cases: with the inclusion of the full LHC Run II top quark dataset, and in the context of a global dataset also including the Higgs, diboson and electroweak sectors. Let us begin by summarising what we have done in the context of the top sector.

We have integrated a global determination of PDFs with the fit of dimension-six operators modifying top quark interactions into a simultaneous determination of the SMEFT-PDFs and the EFT coefficients. The main outcome of our analysis is the assessment of the conditions under which the usual assumptions of SM-PDFs and fixed-PDF EFT analyses are valid, and establishing when the SMEFT-PDF interplay cannot be neglected without introducing a sizeable bias to the results. Furthermore, we have provided strategies to disentangle eventual BSM signals from QCD effects in the interpretation of these top quark measurements.

As a by-product of this determination of the SMEFT-PDFs, we have also presented state-of-the-art SM-PDF and fixed-PDF EFT interpretations of top quark data based on the broadest LHC experimental dataset to date, therefore fully exploiting the information contained in the legacy Run II top quark measurements. From the SM-PDF study, we have quantified the impact of the recent Run II top quark production data on the large-$x$ gluon, and assessed their compatibility with the information provided by the datasets already included in previous global fits such as \nnpdf. From the fixed-PDF SMEFT analysis, we have benchmarked the \simunet{} performance reproducing the \smefit{} results, determined the improved constraints provided by the full-luminosity Run II data, and compared our findings with the results presented from other groups. All in all, we demonstrate the unparalleled sensitivity of Run II top quark data both to probe the proton structure and to search for signatures of new physics arising as anomalous top quark properties and interactions.

Additionally, in this chapter we have shown the first ever global simultaneous fit of PDFs and Wilson coefficients. This new result shows that \simunet{} can be used to produce fits combining the Higgs, top, diboson and electroweak sectors such as those in Refs.~\cite{Ellis:2020unq,Ethier:2021bye,Giani:2023gfq}, and that more data can be added in a rather straightforward way to combine the sectors presented here alongside Drell-Yan and flavour observables, as it is done in Ref.~\cite{Bartocci:2023nvp}. We have presented the results and compared them to existing fits. We have shown that the interplay between PDFs and SMEFT coefficients, once they are fitted simultaneously, has little effect on the Wilson coefficients, while it has the effect of shifting the gluon-gluon luminosity and increasing its uncertainty in the large-$M_X$ region. 

The results presented in this chapter could be extended in a number of directions. One of the most pressing matters is the provision of a simultaneous determination of PDFs with EFT coefficients entering at quadratic order. This will involve significant modification to the Monte-Carlo replica method used by \simunet{}, and will be the subject of future work.

On a different note, top quark measurements at the LHC have also been used to determine the strong coupling $\alpha_s(m_Z)$ as well as the top quark mass $m_t$, both independently and simultaneously with the PDFs~\cite{Cooper-Sarkar:2020twv}. It could hence be interesting to extend the \simunet{} framework to account for the determination of other SM parameters in addition to the PDFs, as it is sketched in the original \simunet{} publication~\cite{Iranipour:2022iak}, and to study their interplay with eventual new physics effects.

Additionally, it would be interesting to extend future versions of the \simunet{} methodology to account for potential renormalisation group effects on the Wilson coefficients~\cite{Jenkins:2013zja,Jenkins:2013wua,Alonso:2013hga,Aoude:2022aro}.

One may also want to assess the SMEFT-PDF interplay for other types of processes beyond those considered so far, and in particular study inclusive jet and di-jet production, which are instrumental for PDF fits in the same region constrained by the top data~\cite{AbdulKhalek:2020jut} and also provide information on a large number of poorly-known SMEFT operators.

Another possible development of the  \simunet{} methodology would be to establish the interplay between PDFs and model parameters such as masses and couplings in specific UV-complete BSM scenarios by using the SMEFT as a matching bridge~\cite{Dawson:2022ewj,terHoeve:2023pvs}. 

It would also be interesting to extend the joint SMEFT-PDF determination as implemented in \simunet{} to novel types of observables with enhanced or even optimal sensitivity to either the PDFs of the EFT coefficients, such as the ML-assisted unbinned multivariate observables introduced in Ref.~\cite{GomezAmbrosio:2022mpm}.

Finally, following along the lines of the HL-LHC projections for high-mass $W$ and $Z$ production presented in~\cite{Greljo:2021kvv}, one can study the projected SMEFT-PDF synergies at future colliders, certainly at the HL-LHC but also at the Electron Ion Collider~\cite{Boughezal:2022pmb}, sensitive to directions in the SMEFT parameter space not covered by LHC data, and at the Forward Physics Facility~\cite{Feng:2022inv,Anchordoqui:2021ghd}, where proton and nuclear structure can be probed  while also obtaining information on anomalous neutrino interactions parametrised by extensions of the SMEFT.

As the LHC approaches its high-luminosity era, it becomes more and more important to develop novel analysis frameworks that make possible the full exploitation of the information contained in the legacy LHC measurements. The results presented in this chapter contribute to the full exploitation of the information contained in LHC measurements by demonstrating how to achieve a consistent simultaneous determination of the PDFs and EFT parameters globally, bypassing the need for the assumptions required for the SM-PDF and fixed-PDF EFT interpretations.

In this chapter we have discussed the PDF-BSM interplay in terms of simultaneous and individual fits. In what follows, we will assess a complementary and \textit{crucial} issue that can arise when studying this interplay: the potential absorption of BSM physics effects in PDF fits.

\chapter{Hide and seek: how PDFs can conceal new physics}
\label{ch:contamination}
  \epigraph{Sapere aude. \\ \vspace{1em} Dare to know.}{\textit{\\ from The First Book of Letters, \\ by Horace.}}

\textit{This chapter is based on Ref.~\cite{Hammou:2023heg}, and was written in collaboration with
E. Hammou, Z. Kassabov, M. Madigan, M. L. Mangano, L. Mantani, J. Moore, and M. Ubiali. 
My contributions to this work include generating the theory predictions with contaminated PDF sets in} $WZ$ \textit{production,
implementing the high rapidity luminosities in the} \texttt{validphys} \textit{module to assess the effects of contamination in the forward region, 
and producing the observable ratios in} $WW$ \textit{production to combine with Drell Yan projections, which disentangled the effects of PDF contamination in collider observables.}

\vspace{1em} 
\noindent\rule{\textwidth}{0.4pt}
\vspace{1em} 

The interpretation of LHC data, and the assessment of possible 
hints of new physics, require the precise knowledge of the proton PDFs. In this chapter, we present an application of the \simunet{} methodology, as described in Chapter \ref{ch:simunet}, to determine whether 
and how global PDF fits might inadvertently ‘fit away’ signs of 
new physics in the high-energy tails of the distributions. 
We showcase a scenario for the High-Luminosity LHC, in which 
the PDFs may completely absorb such signs of new physics, thus 
biasing theoretical predictions and interpretations. We discuss 
strategies to single out the effects in this scenario, and 
disentangle the inconsistencies that stem from them.  
Our study brings to light the synergy between the high luminosity 
programme at the LHC and future low-energy non-LHC measurements 
of large-$x$ sea quark distributions. The analysis code used in 
this work is fully integrated in the \simunet{} methodology so that any users can test the robustness 
of the signal associated to a given BSM model against absorption 
by the PDFs.

\section{Introduction}
\label{sec:intro_sd}

As previously discussed, the DGLAP equations describe the evolution of PDFs with respect to the energy scale, and have been computed, in the SM, to high precision in perturbation theory: at NNLO~\cite{Moch:1999eb,Vogt:2004mw,Moch:2004pa,Vermaseren:2005qc}, 
and partially to N$^3$LO~\cite{Davies:2016jie,Moch:2017uml,Davies:2022ofz,Henn:2019swt,Duhr:2022cob,Moch:2021qrk,Soar:2009yh,Falcioni:2023luc,Falcioni:2023vqq,Moch:2023tdj,Falcioni:2023tzp}.

In the context of BSM theories, the effect of new light particles (both coloured and colourless) on the DGLAP evolution equations has
been investigated in several publications by considering, in the early days before the start of the LHC,
light gluinos~\cite{Berger:2004mj}
and color-octet fermions~\cite{Berger:2010rj}. Then, with LHC constraints making
bounds more stringent in general, the potential effects of non-coloured partons such as leptophobic dark photons~\cite{McCullough:2022hzr}
has been considered. 

This chapter aims to complement the study of PDFs and their evolution in the context of BSM theories but, this time, by means of a SMEFT parametrisation of the new physics effects. This thesis has addressed the PDF-BSM interplay from different angles, but this is the first instance in which we explicitly consider potential heavy new physics effects in the context of PDF evolution.

The effects of the SMEFT dimension-six operators on the RGE running of SM parameters and Wilson coefficients up to one
loop was computed in Refs.~\cite{Grojean:2013kd,Jenkins:2013wua,Jenkins:2013zja,Alonso:2013hga}, and the potential effect of higher dimensional operators on $\as$ as an input parameter in the SMEFT has been studied in Ref.~\cite{Trott:2023jrw}. Additionally, the potential modification of parton shower algorithms induced by the EFT was discussed in Ref.~\cite{Englert:2018byk}. Although there are formal arguments that explain why higher dimensional operators do not modify the IR behaviour of an effective theory (and its collinear structure, in particular), as discussed in Ref.~\cite{Appelquist:1974tg}, in recent publications EFT effects in the IR were advocated, particularly hinting on the effects of EFT in the modification
of the collinear splitting functions~\cite{Martin:2023fad}. 

In this chapter we would like to clarify the discussion and look at the EFT effects on all ingredients entering  the leading order DGLAP evolution equations, which
to our knowledge have not been discussed yet in a comprehensive manner in the literature. We fill this gap and embark on a more
comprehensive exploration of the theoretical framework. We prove that such changes are only due to the effects of the SMEFT on the RGE running
of the strong coupling constants, as the splitting functions with SMEFT operators are left unchanged. 

We organise the chapter as follows. In Sect.~\ref{sec:pij} we set the formalism that allows us to 
systematically explore the effect of dimension-six operators on the leading-order 
collinear splitting functions, and prove that the SMEFT does not modify them. In Sect.~\ref{sec:as} we
calculate the SMEFT effects on the running of $\alpha_s(\mu)$. We conclude in Sect.~\ref{sec:summary_sd} with a summary of the investigation and 
future directions.

\section{New physics scenarios}
\label{sec:scenarios}
Now, following the methodology presented in Sect.~\ref{subsec:pseudodata_generation}, we
extend the SM to UV-complete models by introducing heavy new fields. 
%
We choose simple extensions of the SM corresponding to ``universal theories''~\cite{Wells:2015uba},
the effect of which can be well-described with an EFT approximation using the oblique corrections $\hat{Y}$ and $\hat{W}$~\cite{Peskin:1991sw,Altarelli:1991fk,Barbieri:2004qk}. 
It is important to bear in mind that such theories can be considered "universal" only if the RGE running of the Wilson coefficients~\cite{Jenkins:2013zja,Jenkins:2013wua,Alonso:2013hga,Aoude:2022aro} is neglected or additional prescriptions are devised~\cite{Wells:2015cre}. Neglecting the RGE running of the Wilson coefficients is an approximation that is completely adequate for our case studies, given that the running of the Wilson coefficients of the four-fermion operators and their mixing are expected to be sub-dominant for the observables considered in the study~\cite{DYpaper}.
We consider the following two scenarios:

\begin{itemize}
	\item {\bf Scenario I:} A flavour universal $Z'$, charged under a $U(1)_Y$ gauge symmetry. We give a mass to the field assuming it is generated by some higher energy physics. It corresponds to a new heavy neutral bosonic particle. At the EFT level, the effect of this model on our dataset can be  
		described by the $\hat{Y}$ parameter.
	\item {\bf Scenario II:} A flavour universal $W'$ charged under $SU(2)_L$.
		Once again, we directly add a mass term to the Lagrangian. This gives rise to two new heavy charged bosonic particles ($W'^{+}$ and $W'^{-}$) as well as a new heavy neutral boson, similar to a $Z'$ that only couples to left-handed fermions. At the EFT level, the effect of this model on our dataset can 
	be described by the $\hat{W}$ parameter.
\end{itemize}

\noindent The following subsections are devoted to describing each of these NP scenarios.
In particular, in each case we use tree-level matching to obtain a parametrisation of the model in terms of dim-6 operators of the SMEFT,
making use of the matching provided in Ref.~\cite{deBlas:2017xtg} to do so.
We identify the observables in our dataset affected by each NP scenario, and in each case we compare the UV and EFT predictions.  Finally, we identify values of 
the model parameters for which the EFT description is justified at the projected energy of the HL-LHC, and for which existing constraints on the models are avoided.

\subsection{Scenario I: A flavour-universal $Z'$ model}
\label{subsec:model-zprime}

The addition to the SM of a new spin-1 boson $Z'$ associated to a gauge symmetry $U(1)_Y$, a mass $M_{Z'}$ and a coupling coefficient $g_{Z'}$ yields the following Lagrangian:
\begin{equation} 
\label{eq:Zprime}
	\begin{split}
		\mathcal{L}^{Z'}_{\text{UV}} &= \mathcal{L}_{\text{SM}} - \frac{1}{4} Z'_{\mu \nu} {Z'}^{\mu \nu} + \frac{1}{2} M_{Z'}^{2} Z'_{\mu} {Z'}^{\mu} \\
		&\qquad - g_{Z'} Z'_{\mu} \sum_{\substack{f}} Y_{f} \bar{f} \gamma^{\mu} f - Y_{\varphi} g_{Z'} ( Z'_{\mu} \varphi^{\dagger} i  D^{\mu} \varphi + \textrm{h.c.} ) \, .
	\end{split}
\end{equation}
We sum the interactions with the fermions $f \in \{Q^i_L, u^i _R, d^i _R, \ell^i _L, e^i _R \text{ for i} \in  \{1, 2, 3\}\}$ , where $Y_f$ is the corresponding hypercharge: $( Y_{Q_L}, Y_{u_R}, Y_{d_R}, Y_{l_L}, Y_{e_R}, Y_{\varphi} ) = (\frac{1}{6}, \frac{2}{3}, -\frac{1}{3}, -\frac{1}{2}, -1, \frac{1}{2} ) $. The kinetic term is given by $Z'_{\mu \nu} = \partial_{\mu} Z'_{\nu} - \partial_{\nu} Z'_{\mu}$. We neglect the mixing between the $Z'$ and SM gauge bosons\footnote{In general, mixing terms can be written in a gauge invariant way and, therefore, are allowed in the Lagrangian. They can have phenomenological implications \cite{Dawson:2024ozw,Allanach:2024ozu}  (e.g. on $M_{Z'}$ exclusion limits) which are beyond the scope of this chapter.}. The covariant derivative is given by 
\begin{equation}
    D_{\mu} = \partial_{\mu} + \frac{1}{2} ig \sigma^{a} W_{\mu}^{a} + i g^{'} Y_{\varphi} B_{\mu} + i g_{Z'} Y_{\varphi} Z'_{\mu},
\end{equation}
where $\sigma ^a$ are the Pauli matrices for $ a \in  \{1, 2, 3\}$. Note that quark and lepton flavour indices are suppressed, and that the couplings to quark and leptons are flavour diagonal.
The new gauge interaction is anomaly free, as it has the same hypercharge-dependent couplings to fermions as the SM fields~\cite{Allanach:2018vjg}.
Models of $Z'$ bosons and their impact on LHC data have been widely studied; see for example Refs.~\cite{Salvioni:2009mt,Salvioni:2009jp,Langacker:2008yv,Panico:2021vav}.

\begin{table}[H]
\begin{center}
\begin{tabular}{ |r|l| }
 \hline
	Bosonic & $\mathcal{O}_{\varphi D}$, $\mathcal{O}_{\varphi \Box}$, $\mathcal{O}_{\varphi l}^{(1)}$, $\mathcal{O}_{\varphi q}^{(1)}$, $\mathcal{O}_{\varphi e}$, $\mathcal{O}_{\varphi u}$, $\mathcal{O}_{\varphi d}$ \\
	4-fermion $(\bar{L} L)(\bar{L} L)$ & $\mathcal{O}_{ll}$, $\mathcal{O}_{qq}^{(1)}$, $\mathcal{O}_{lq}^{(1)}$ \\
	4-fermion $(\bar{R} R)(\bar{R} R)$ & $\mathcal{O}_{ee}$, $\mathcal{O}_{uu}$, $\mathcal{O}_{dd}$, $\mathcal{O}_{ed}$, $\mathcal{O}_{eu}$, $\mathcal{O}_{ud}^{(1)}$\\
	4-fermion $(\bar{L} L)(\bar{R} R)$ & $\mathcal{O}_{l e}$, $\mathcal{O}_{l d}$, $\mathcal{O}_{l u}$, $\mathcal{O}_{q u}^{(1)}$, $\mathcal{O}_{q d}^{(1)}$\\
 \hline
\end{tabular}
\end{center}
	\caption{\label{tab:Zpops} Warsaw basis operators generated by the $Z'$ model of Eq.~\eqref{eq:Zprime}.}
\end{table}
Tree-level matching of $\mathcal{L}^{Z'}_{\text{UV}}$ to the dim-6 SMEFT produces the Warsaw basis~\cite{deBlas:2017xtg} operators in Table~\ref{tab:Zpops}. The complete SMEFT Lagrangian is given by:
\begin{equation} 
\label{eq:ZprimeEFT}
	\begin{split}
		\mathcal{L}^{Z'}_{\text{SMEFT}} &= \mathcal{L}_{\text{SM}} -\frac{g_{Z'}^{2}}{ M_{Z'}^{2}} \Big( { 2 Y_{\varphi}^{2} \mathcal{O}_{\varphi D} + \frac{1}{2} Y_{\varphi}^{2} \mathcal{O}_{\varphi \Box}}\\
		&\qquad+ Y_{\varphi} Y_{l} \mathcal{O}_{\varphi l}^{(1)} + Y_{\varphi} Y_{q}  \mathcal{O}_{\varphi q}^{(1)}  + Y_{\varphi} Y_{e}  \mathcal{O}_{\varphi e} +  Y_{\varphi} Y_{d}  \mathcal{O}_{\varphi d} + Y_{\varphi} Y_{u}  \mathcal{O}_{\varphi u} \\
		&\qquad+ \frac{1}{2} Y_{l}^{2} \mathcal{O}_{ll} + \frac{1}{2} Y_{q}^{2} \mathcal{O}_{qq}^{(1)} + Y_{q} Y_{l} \mathcal{O}_{lq}^{(1)}\\
		&\qquad+  \frac{1}{2} Y_{e}^{2} \mathcal{O}_{ee} + \frac{1}{2} Y_{u}^{2} \mathcal{O}_{uu} + \frac{1}{2} Y_{d}^{2} \mathcal{O}_{dd} + Y_{e} Y_{d} \mathcal{O}_{ed} + Y_{e} Y_{u} \mathcal{O}_{eu}  + Y_{u} Y_{d} \mathcal{O}_{ud}^{(1)}\\
		&\qquad+ Y_{e} Y_{l} \mathcal{O}_{l e} + Y_{u} Y_{l} \mathcal{O}_{l u} + Y_{d} Y_{l} \mathcal{O}_{l d} + Y_{e} Y_{q} \mathcal{O}_{q e} + Y_{u} Y_{q} \mathcal{O}_{q u}^{(1)} + Y_{d} Y_{q} \mathcal{O}_{q d}^{(1)} \Big) \, .
	\end{split}
\end{equation}
The leading effect of this model on the data entering our analysis is to modify the Drell-Yan and Deep Inelastic Scattering datasets; 
in particular the high-mass neutral current Drell-Yan 
tails~\cite{DYpaper} will be affected.  
The $Z'$ may have an additional impact on top quark and dijet data through four-quark operators such as $\mathcal{O}_{qq}^{(1)}$, however the 
effect is negligible and we do not consider it here.

The effect of the $Z'$ on high-mass Drell-Yan is dominated by the energy-growing four-fermion operators~\cite{Farina:2016rws,Torre:2020aiz,Panico:2021vav}.
By neglecting the subdominant operators involving the Higgs doublet $\varphi$,
we can add the operators of the last three lines of Eq.~\eqref{eq:ZprimeEFT} in the following way:
\begin{equation}
\label{eq:zprimeL}
	\mathcal{L}^{Z'}_{\text{SMEFT}} = \mathcal{L}_{\text{SM}} -\frac{g_{Z'}^{2}}{2 M_{Z'}^{2}} J^{\mu}_Y J_{Y, \mu}, \qquad J^{\mu}_Y = \sum_{\substack{f}} Y_{f} \bar{f} \gamma^{\mu} f \, .
\end{equation}
We can describe the new physics introduced in this type of scenario with the $\hat{Y}$ parameter:
\begin{equation}
	\mathcal{L}^{Z'}_{\text{SMEFT}} = \mathcal{L}_{\text{SM}} -\frac{{g'}^2 \hat{Y}}{2 m^2_W} J^{\mu}_Y J_{Y, \mu}, \qquad \hat{Y} = \frac{g_{Z'}^{2}}{M_{Z'}^{2}} \frac{m^2_W}{{g'}^2} \, .
\end{equation}
The $\hat{Y}$ parameter allows us to write the Lagrangian using SM parameters. We can write the relation between $\hat{Y}$ and the $Z'$ parameters $g_{Z'}$, $M_{Z'}$ as follows,
\begin{equation}
	\frac{g_{Z'}^{2}}{M_{Z'}^{2}} = 4 \sqrt{2} G_{F} \hat{Y} \left( \frac{m_{Z}^{2} - m_{W}^{2}}{m_{W}^{2}}  \right) \, ,
\end{equation}
where we make use of the \{$m_W$, $m_Z$, $G_F$\} electroweak input scheme, and take the following as input parameters:
%
%
%
%
\begin{equation*}
	G_{F} = 1.16639 \times 10^{-5} \text{ GeV}, \qquad m_{W} = 80.352 \text{ GeV}, \qquad m_{Z} = 91.1876 \text{ GeV} \, .
\end{equation*}
%

\begin{figure}[htbp]
    \centering
	\includegraphics[width=0.49\linewidth]{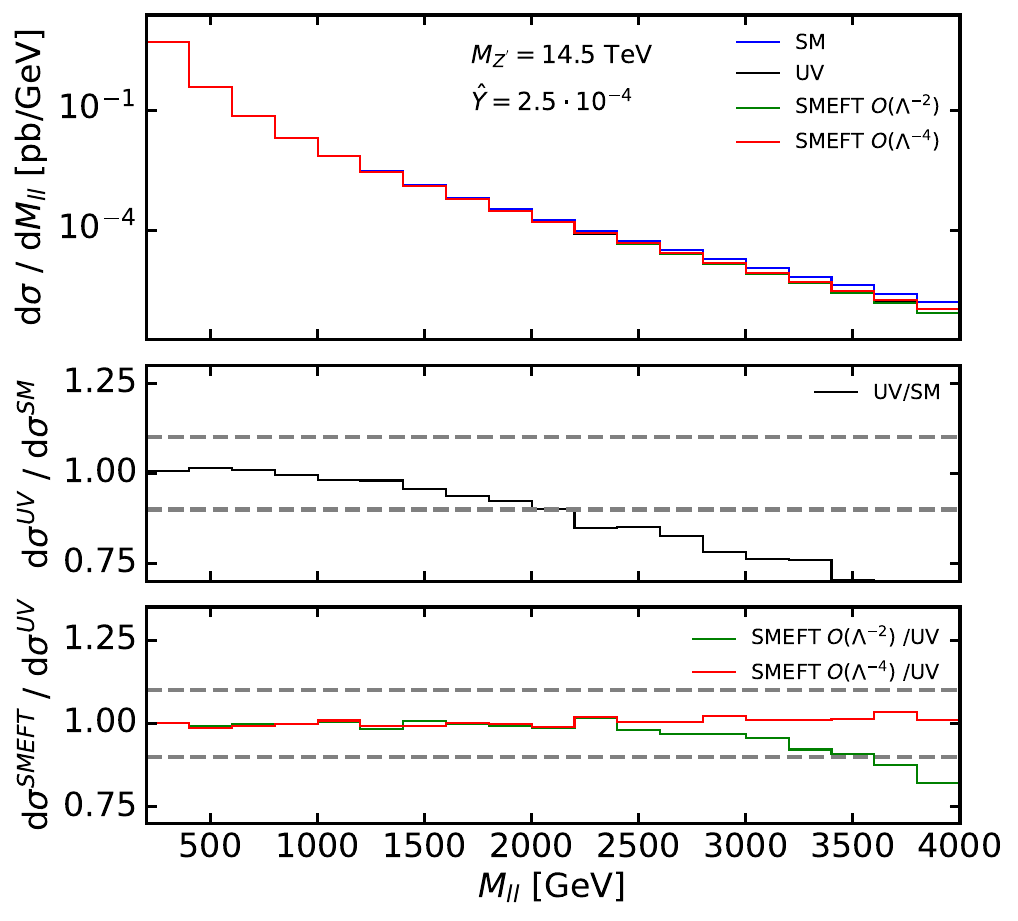}
	\includegraphics[width=0.49\linewidth]{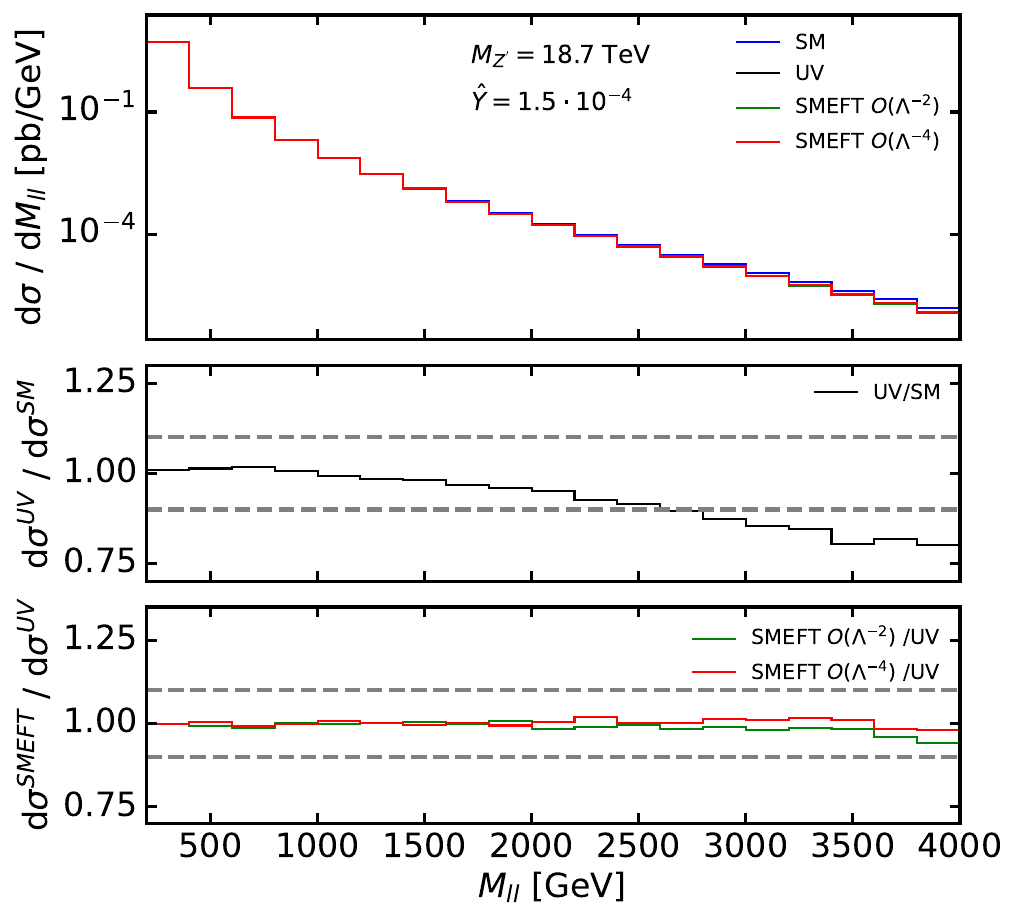}\\
	\includegraphics[width=0.49\linewidth]{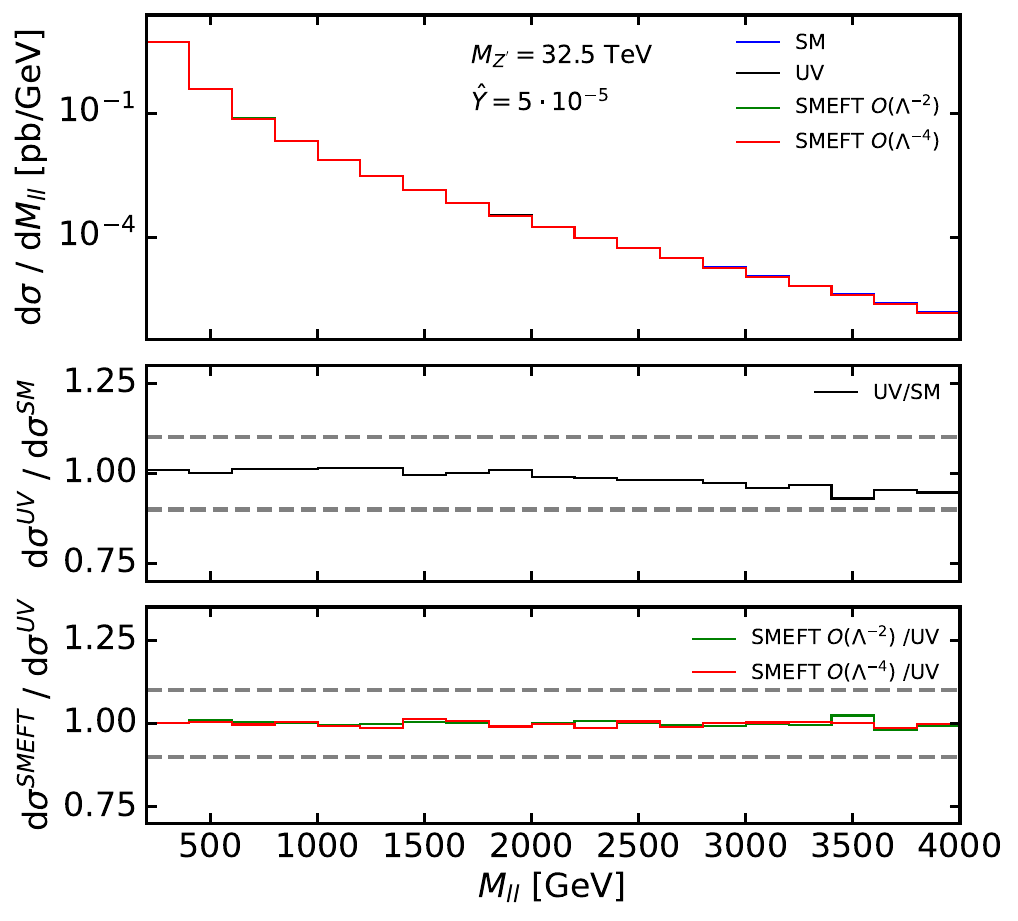}
	\caption{Predictions for neutral current Drell-Yan inclusive cross sections differential in the dilepton invariant mass $m_{\ell\ell}$, with $\ell=(e,\mu$).  
 Factorisation and renormalisation scales are set to $\mu_F=\mu_R=m_{\ell\ell}$, where $m_{\ell\ell}$ is the centre of each invariant mass bin. 
    We show the SM predictions compared to the predictions for a $Z'$ of different masses corresponding to different $\hat{Y}$ values, assuming $g_{Z'}=1$. 
    Top left: mass of $14.5$ TeV, corresponding to $\hat{Y}=25 \cdot 10^{-5}$. Top right: mass of $18.7$ TeV, corresponding to $\hat{Y}=15 \cdot 10^{-5}$. 
    Bottom: mass of $32.5$ TeV, corresponding to $\hat{Y}=5 \cdot 10^{-5}$.}
	\label{fig:Zp_DY_Comp}
\end{figure}

In Fig.~\ref{fig:Zp_DY_Comp} we compare the predictions of the UV-complete $Z'$ model and the corresponding EFT parametrisation
for the differential cross section of the Drell-Yan process $p p \rightarrow \ell^+ \ell^-$, where $\ell=e,\,\mu$.
The predictions are computed assuming $\sqrt{s}=14$ TeV, using \madgraph{} and setting $\mu_F=\mu_R=m_{\ell\ell}$, where $m_{\ell\ell}$ is the centre of each invariant mass bin.  
We compare the SM, the full UV-complete model, the linear-only $\mathcal{O}(\Lambda^{-2})$ EFT and linear-plus-quadratic $\mathcal{O}(\Lambda^{-4})$ EFT predictions 
assuming $g_{Z'} = 1$, for 
three benchmark values of the $Z'$ mass: $M_{Z'} = 14.5$ TeV, $M_{Z'} = 18.7$ TeV and $M_{Z'} = 32.5$ TeV.
Such large values of $M_{Z'}$ are clearly well beyond the possible direct reach of direct $Z'$ searches at ATLAS and CMS~\cite{Workman:2022ynf}. In the top panel we plot the differential cross section with respect to the dilepton invariant mass, 
in the middle panel we plot the ratio of the full $Z'$ model to the SM, and in the
lower panel we plot the ratio of the EFT to the full $Z'$ model predictions. 

First, we observe that the UV model predictions differ from the SM predictions by more than $20\%$ for the smaller masses of the $Z'$. 
In the lower panels, we observe the point at which the linear EFT corrections fail to describe the UV physics, and the quadratic EFT contributions begin to become non-negligible; in the same way, the quadratic dim-6 EFT description starts failing when the dim-8 SMEFT operators become important~\cite{Boughezal:2021tih}.

As displayed in the top right panel of Fig.~\ref{fig:Zp_DY_Comp}, for $M_{Z'}=14.5$ TeV the linear EFT corrections start failing to describe the UV model at higher energies. For $M_{Z'}=18.7$ TeV and heavier the linear EFT describes the UV physics faithfully for dilepton invariant masses up to 4 TeV. The deviations from new physics are over $20\%$ for $M_{Z'}=18.7$ and $M_{Z'}=14.5$ TeV.
We will implement our PDF ``contamination'' by working in the area of the parameter space where the linear EFT describes the UV physics faithfully and where the UV extension to the SM impacts significantly the observables.

Finally, it is worth noting that we have compared those SMEFT predictions involving only four-fermion operators with 
those obtained 
additionally including the SMEFT operators containing the Higgs doublet, such as $\mathcal{O}_{\varphi l}^{(1)}$ and $\mathcal{O}_{\varphi e}$. We find that these operators have no visible impact on the observables. 
They are only competitive with the four-fermion corrections at lower energies (around 500 GeV),
and at this scale the new physics has very little impact on the SM predictions. 
When the influence of the heavy new physics starts to be noticeable at higher invariant mass, the four-fermion operators, whose impact grows faster with energy, completely dominate the SMEFT corrections. Thus, we have verified that our parametrisation
in terms of the $\hat{Y}$ parameter reflects the UV physics to a very good degree.

\subsection{Scenario II: A flavour universal $W'$ model}
\label{subsec:model-wprime}

We now consider a new $SU(2)_L$ triplet field ${W'}^{a, \mu}$, where $a \in \{1,2,3\}$ denotes an $SU(2)_L$ index. We add a mass $M_{W'}$, and denote the $W'$ coupling coefficient by $g_{W'}$. Similarly to what happens with the SM $W$ field, the $W'^1$ and $W'^2$ components mix to form the $W'^+$ and $W'^-$ particles, while the $W'^3$ component gives a neutral boson similar to the $Z'$, but which only couples to left-handed fields. The 
model is described by the following Lagrangian:
\begin{equation} \label{eq:Wprime}
	\begin{split}
		\mathcal{L}^{W'}_{\text{UV}} &= \mathcal{L}_{\text{SM}} - \frac{1}{4} {W'}^{a}_{\mu \nu} {W'}^{a, \mu \nu} + \frac{1}{2} M_{W'}^{2} {W'}_{\mu}^{a} {W'}^{a, \mu}\\
		&\qquad - g_{W'} {W'}^{a,\mu} \sum_{\substack{f_L}} \bar{f}_L T^{a} \gamma^{\mu} f_L
		 - g_{W'} ({W'}^{a, \mu} \varphi^{\dagger} T^{a} i D_{\mu} \varphi + \textrm{h.c.}) \, ,
	\end{split}
\end{equation}
where we sum over the left-handed fermions: $f_L \in \{Q_L^i, \ell_L^i, \,{\rm for}\,\,\,i\in\{1,2,3\}\}$. The $SU(2)_L$ generators are given by $T^a = \frac{1}{2} \sigma ^a$ where $\sigma ^a$ are the Pauli matrices. The kinetic term is given by $W^{'a}_{\mu \nu} = \partial_{\mu} {W'}^{a}_{\nu} - \partial_{\nu} {W'}^{a}_{\mu}- ig_{W'} [{W'}^{a}_{\mu}, {W'}^{a}_{\nu} ]$. The covariant derivative is given by
\begin{equation}
    D_{\mu} = \partial_{\mu} + \frac{1}{2} ig \sigma^{a} W_{\mu}^{a} + i g^{'} Y_{\varphi} B_{\mu} + \frac{1}{2} ig_{W'} \sigma^{a} {W'}_{\mu}^{a}.
\end{equation}
As above, we neglect the mixing with the SM gauge fields.  

Tree-level matching of $\mathcal{L}^{W'}_{\text{UV}}$ to the dim-6 SMEFT produces the Warsaw basis operators in Table~\ref{tab:Wpops}, where we
have distinguished the operators
$(O_{ll})_{ij} = (l_{i} \gamma^{\mu} l_{i})(l_{j} \gamma^{\mu} l_{j})$ and $(O_{ll}')_{ij} = (l_{i} \gamma^{\mu} l_{j})(l_{j} \gamma^{\mu} l_{i})$~\cite{Brivio:2017vri}. 
\begin{table}[h!]
\begin{center}
\begin{tabular}{ |r|l| }
 \hline
	Bosonic &  $\mathcal{O}_{\varphi \Box}$, $\mathcal{O}_{\varphi}$, $\mathcal{O}_{\varphi l}^{(3)}$, $\mathcal{O}_{\varphi q}^{(3)}$\\
	Yukawa & $\mathcal{O}_{e \varphi}$, $\mathcal{O}_{d \varphi}$, $\mathcal{O}_{u \varphi}$ \\
	4-fermion $(\bar{L} L)(\bar{L} L)$ & $\mathcal{O}_{ll}$,$\mathcal{O}_{ll}^{'}$, $\mathcal{O}_{qq}^{(3)}$, $\mathcal{O}_{lq}^{(3)}$ \\
 \hline
\end{tabular}
\end{center}
        \caption{\label{tab:Wpops} Warsaw basis operators generated by the $W'$ model of Eq.~\eqref{eq:Wprime}.}
\end{table}
The complete SMEFT Lagrangian is given by~\cite{deBlas:2017xtg}:
\begin{equation} 
\label{eq:WprimeEFT}
        \begin{split}
		\mathcal{L}^{W'}_{\text{SMEFT}} &= \mathcal{L}_{\text{SM}} - \frac{g_{W}^{2}}{M_{W}^{2}} \Big(- \frac{1}{8} \mathcal{O}_{ll} + \frac{1}{4} \mathcal{O}_{ll}^{'} + \frac{1}{8} \mathcal{O}_{qq}^{(3)} + \frac{1}{4} \mathcal{O}_{lq}^{(3)}\\
		&\qquad+ \lambda_{\varphi} \mathcal{O}_{\varphi} + \frac{3}{8} \mathcal{O}_{\varphi \Box}+ \frac{1}{4} \mathcal{O}_{\varphi q}^{(3)}  + \frac{1}{4} \mathcal{O}_{\varphi l}^{(3)}     \\
		&\qquad + \frac{1}{4} (y_{e})_{ij} (\mathcal{O}_{e \varphi})_{ij} + \frac{1}{4} (y_{u})_{ij} (\mathcal{O}_{u \varphi})_{ij} + \frac{1}{4} (y_{d})_{ij} (\mathcal{O}_{d \varphi})_{ij} \Big) \, . 
	\end{split}
\end{equation}
As in the case of the $Z'$, 
the leading effect of this model on our dataset 
is to modify the Drell-Yan and Deep Inelastic Scattering datasets; however, this time both charged current and neutral current Drell-Yan will
be affected.  
This impact is dominated by the four-fermion
interactions in the first line of Eq.~\eqref{eq:WprimeEFT}, which sum to:
\begin{equation}
\label{eq:wprimeL}
	\mathcal{L}^{W'}_{\text{SMEFT}} = \mathcal{L}_{\text{SM}} -\frac{g_{W'}^{2}}{2 M^2_{W'}} J^{a, \mu}_L J^a_{L, \mu}, \qquad J^{a, \mu}_L = \sum_{\substack{f_L}} \bar{f_L} T^a \gamma^{\mu} f_L \, .
\end{equation}
We can describe the new physics introduced in this type of scenario with the $\hat{W}$ parameter:
\begin{equation}
	\mathcal{L}^{W'}_{\text{SMEFT}} = \mathcal{L}_{\text{SM}} -\frac{g^{2} \hat{W}}{2 m^2_W} J^{a, \mu}_L J^a_{L, \mu}, \qquad \hat{W} = \frac{g_{W'}^{2}}{g^{2}} \frac{m_{W}^{2}}{M_{W'}^{2}} \, .
\end{equation}
Using Fermi's constant, we can write the relation between the UV parameters and $\hat{W}$ in the following way:
\begin{equation}
	\frac{g_{W'}^{2}}{M_{W'}^{2}} = 4 \sqrt{2} G_{F} \hat{W} \, .
\end{equation}
Again, by fixing $g_{W'}=1$, each $M_{W'}$ can be associated to a value of $\hat{W}$.

\begin{figure}[htbp]
    \centering
	\includegraphics[width=0.49\linewidth]{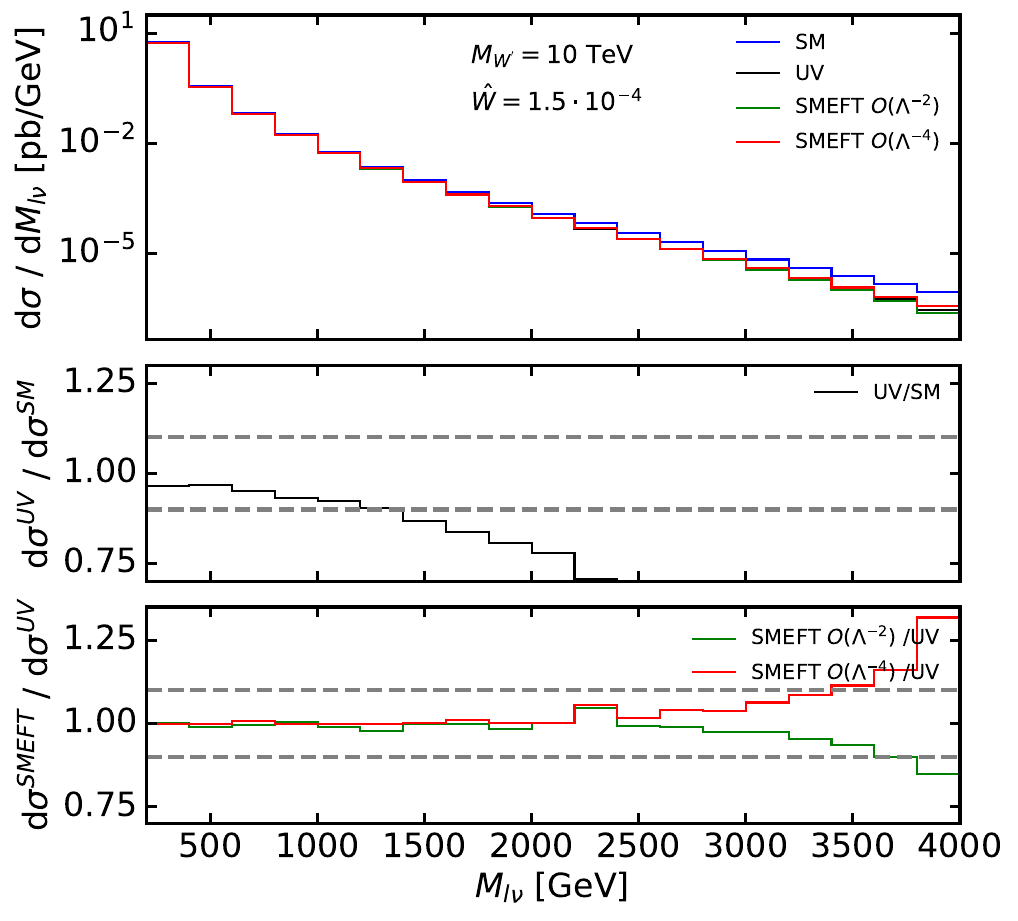}
	\includegraphics[width=0.49\linewidth]{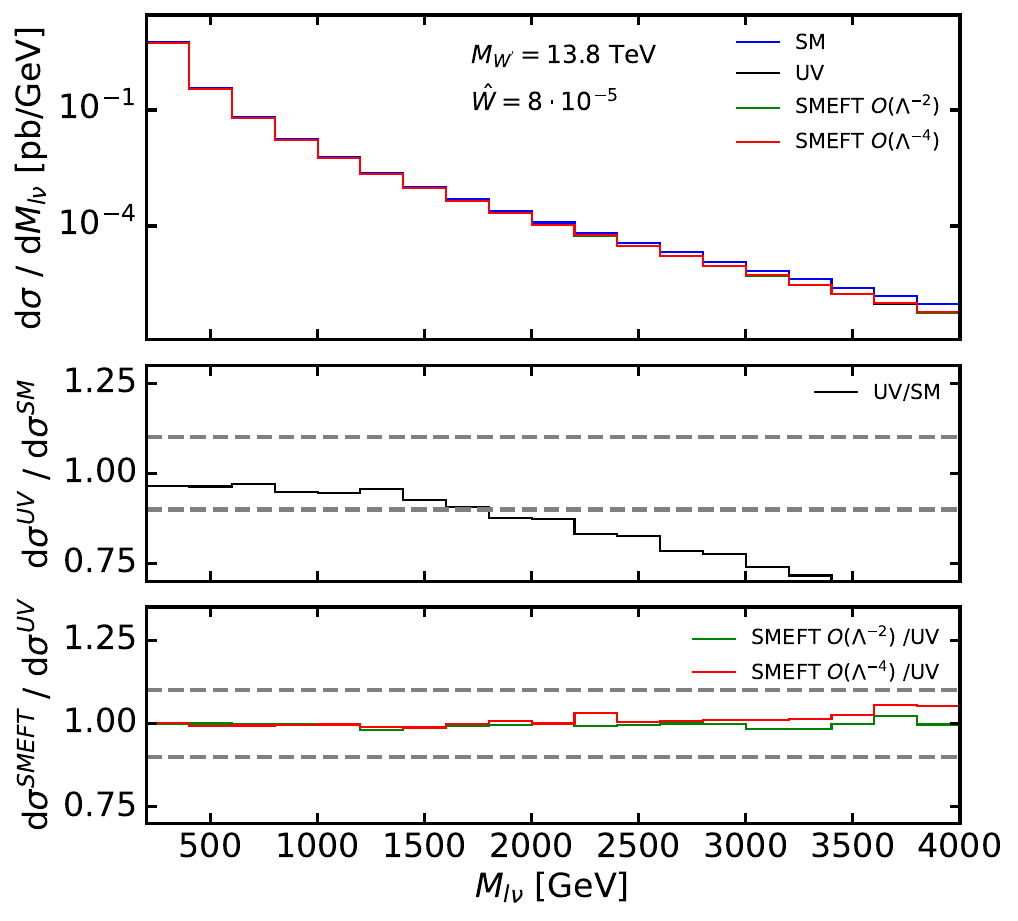}\\
	\includegraphics[width=0.49\linewidth]{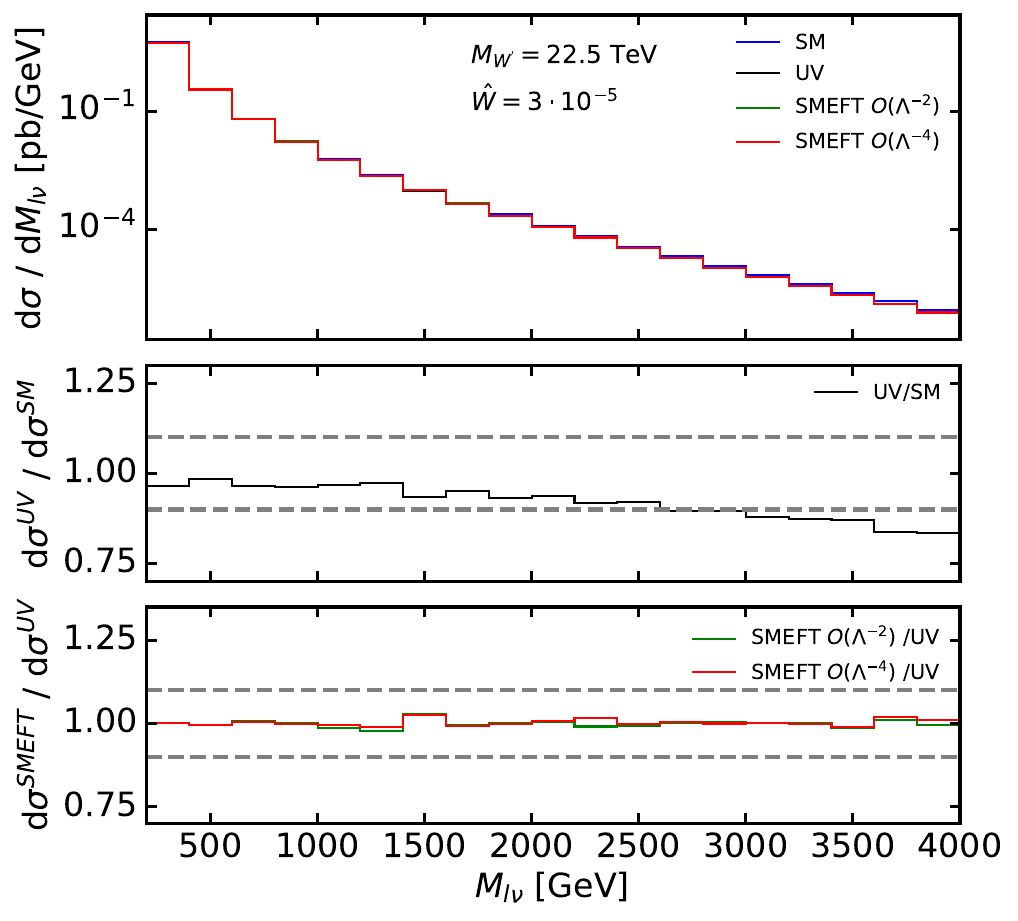}
	\caption{Predictions for charged current Drell-Yan ($pp \rightarrow l^- \bar{\nu}$ inclusive cross sections differential in $M_{\ell\nu}$, with $\ell=(e,\mu$). 
 Factorisation and renormalisation scales are set to $\mu_F=\mu_R=M_T$, where $M_T$ is the transverse mass of each bin. 
 We show the SM predictions compared to the predictions for a $W'$ of different masses corresponding to different $\hat{W}$ values, assuming $g_{W'}=1$. 
 Top left: mass of $10$ TeV, corresponding to $\hat{W}=15 \cdot 10^{-5}$. Top right: mass of $13.8$ TeV, corresponding to $\hat{W}=8 \cdot 10^{-5}$. Bottom: mass of $22.5$ TeV, corresponding to 
    $\hat{W}=3 \cdot 10^{-5}$.}
 	\label{fig:Wp_DY_Comp}
\end{figure}

In Fig.~\ref{fig:Wp_DY_Comp}
we perform a comparison of the UV-complete $W'$ model and the EFT predictions.
We assess the differences between the EFT parametrisation and the UV model description by studying the charged current Drell-Yan process, $pp \rightarrow \ell^- \bar{\nu}$,
assuming $g_{W'} = 1$, at
three benchmark values of the $W'$ mass: $M_{W'} = 10$ TeV, $M_{W'} = 13.8$ TeV and $M_{W'} = 22.5$ TeV.
A similar comparison could be made in neutral current Drell-Yan; however, we expect the dominant effect of the $W'$ to occur in charged current Drell-Yan,
and therefore this process will provide the leading sensitivity to differences between the UV model and the EFT parametrisation.
As displayed in the top right panel of Fig.~\ref{fig:Wp_DY_Comp}, for $M_{W'}=10$ TeV the linear EFT corrections start failing to describe the UV model at higher energies. For $M_{W'}=13.8$ TeV and heavier the linear EFT describes the UV physics faithfully for dilepton invariant masses up to 4 TeV. The deviations from new physics are over $20\%$ for $M_{W'}=13.8$ and $M_{W'}=10$ TeV.
We will again implement our PDF ``contamination'' by working in the area of the parameter space where the linear EFT describes the UV physics faithfully and where the UV extension to the SM impacts the observables noticeably.
Finally, our analysis also reveals that the SMEFT operators involving a Higgs doublet $\varphi$ have no impact on the predictions, for the same reason we presented in the $Z'$ case. This means that this model built with the $\hat{W}$ parameters also describes the UV physics faithfully.


\section{Contamination from Drell-Yan large invariant \\ mass distributions}
  \label{sec:dy}

In this Section, after presenting the analysis settings in terms of
theory predictions and data, we explore in detail the effects of new
heavy vector  bosons in the high-mass Drell-Yan distribution tails and how fitting
this data assuming the SM would modify the data-theory agreement and
the PDFs. We will see that in some scenarios the PDFs manage to mimic
the effects of new physics in the high tails without deteriorating the
data-theory agreement in any visible way. In this cases PDFs can
actually ``fit away'' the effects of new physics. In the following
sections we will explore the phenomenological consequences of using
such ``contaminated'' PDF sets and we will see that they might
significantly distort the interpretation of HL-LHC measurements. Subsequently, in Sect.~\ref{sec:solution},
we conclude by devising strategies to spot the contamination by
including in a PDF fit complementary observables that highlight the
incompatibility of the high-mass Drell-Yan tails with the bulk of the
data.

\subsection{Analysis settings}

For this analysis we generate a set of artificial  Monte Carlo data, which comprises 4771 data points,
spanning a broad range of processes.
The Monte Carlo data that we generate are either taken from current Run I and
Run II LHC data or from HL-LHC projections. The uncertainties in the
former category are more realistic, as they are taken from the
experimental papers (we remind the reader that the central
measurement is generated by the underlying law of Nature according to Eq.~\eqref{eq:observed_data}), 
while the uncertainties on projected HL-LHC data are generated according to specific projections. 

As far as the current data is concerned, we generate MC data that
cover all the observables included in the \nnpdf analysis~\cite{Ball:2021leu}. 
In particular, in the category of Drell-Yan, we
include the NC Drell-Yan that follows the kinematic
distributions and the errors analysed by ATLAS at \(7\) and \(8\) TeV \cite{ATLAS:2013xny, ATLAS:2016gic}, and
CMS at \(7\), \(8\), and \(13\) TeV \cite{CMS:2013zfg, CMS:2014jea,
  CMS:2018mdl}. These LHC measurements are not only used to constrain the PDFs, but are also
sufficiently sensitive to the BSM scenarios considered in Sect.~\ref{sec:scenarios}.

For the HL-LHC pseudo-data, we include the high-mass Drell-Yan projections that we produced in
Ref.~\cite{DYpaper}, inspired by the HL-LHC projections studied
in Ref.~\cite{AbdulKhalek:2018rok}. The invariant mass distribution
projections are generated at \(\sqrt{s} = 14\) TeV, assuming an
integrated luminosity of \(\mathcal{L} = 6 \text{ ab}^{-1}\) ($3 \text{ ab}^{-1}$ collected by ATLAS and  \(3
\text{ ab}^{-1}\) by CMS). 
Both in the case of NC and CC Drell-Yan cross sections, the MC data
were generated using the {\tt MadGraph5\_aMCatNLO} NLO Monte Carlo event
generator~\cite{Frederix:2018nkq} with additional $K$-factors to include the NNLO QCD and
NLO EW corrections. The MC data consist of four datasets
(associated with NC/CC distributions with muons/electrons in the final
state), each comprising 16 bins in the $m_{ll}$
invariant mass distribution or transverse mass $m_T$ distributions
with both $m_{ll}$ and $m_T$ greater than 500 GeV , with the highest energy bins reaching $m_{ll}=4$
TeV ($m_T=3.5$ TeV) for NC (CC) data.
The rationale behind the choice of number of bins and the width of each bin was outlined 
in Ref.~\cite{DYpaper}, and stemmed from the requirement that the expected number of events
per bin was big enough to ensure the applicability of Gaussian
statistics. The choice of binning for the $m_{ll}$ ($m_T$) distribution at the
HL-LHC is displayed in Fig.~5.1 of Ref.~\cite{DYpaper}.

\begin{figure}[t!]
  \centering
  \includegraphics[width=\textwidth]{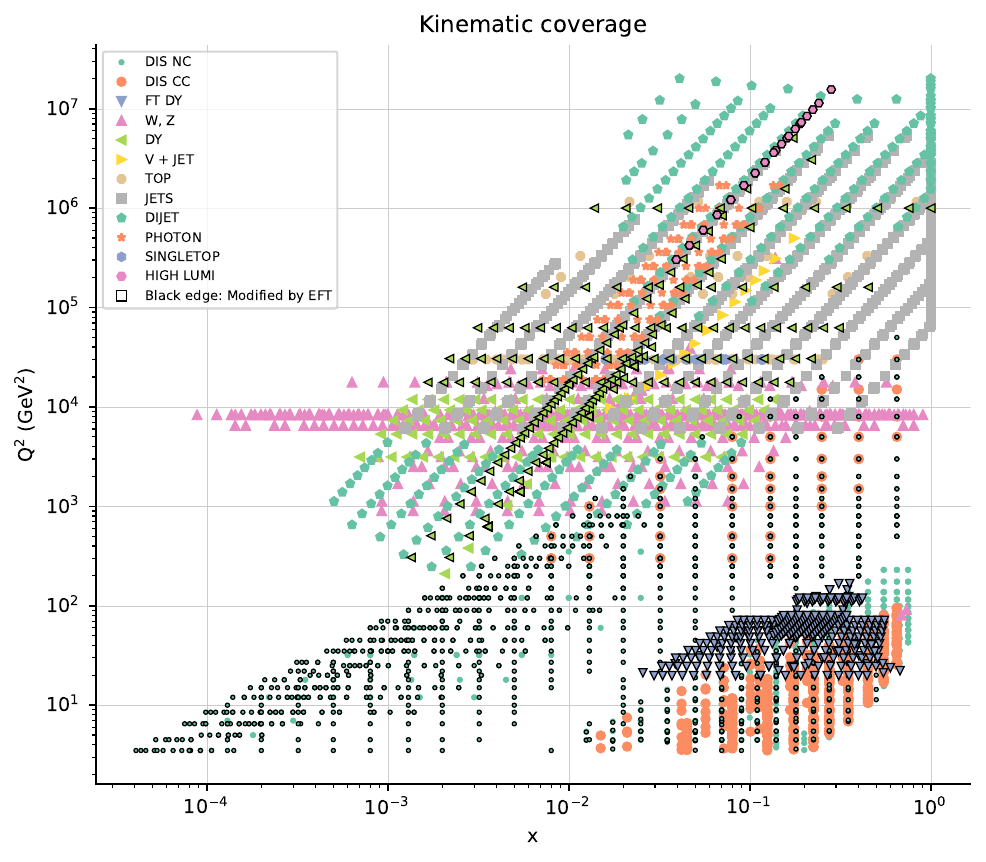}
  \caption{Kinematic coverage of data points included in the fit. The EFT corrections
  for this study have been computed for the points which are highlighted with a black
  edge. The values of $x$ have been computed using a linear order approximation. \label{fig:xq2}}
\end{figure}

The kinematic coverage of the data points used in
this study are shown in Fig.~\ref{fig:xq2}. The points are shown
in \((x,Q^2)\) space with the data points that are modified by the EFT operators
highlighted with a black border; such points thus constrain the Wilson
coefficients as well as the PDFs. We note that, while DIS theory predictions are
modified by the operators we consider in the two benchmark scenarios,
the change in the HERA DIS cross sections upon the variation of the Wilson
coefficients under consideration is minimal, as is explicitly assessed in Ref.~\cite{DYpaper}.

In what follows, we will assess the impact of the injection of NP in the data on the 
fitted PDFs by looking at the integrated luminosity for the parton pair $i,j$, which is defined as:
\begin{equation}
\label{eq:integrated_luminosity}
  {\cal L}_{ij}(m_X,\sqrt{s}) = \frac{1}{s}\int\limits_{-y}^{y}\,d\tilde{y}\,
  \left[
  f_{i}\left(\frac{m_X}{\sqrt{s}}e^{\tilde{y}},m_X\right)\, 
  f_j\left(\frac{m_X}{\sqrt{s}}e^{-\tilde{y}},m_X\right)
  +
  (i \leftrightarrow j)
  \right]
  ,
  \end{equation}
 where $f_i \equiv f_i(x,Q)$ is the PDF corresponding to the parton flavour $i$ 
 , and the integration limits are defined by:
 \begin{equation}
 \label{eq:rapidity}
  y=\ln\left(\frac{\sqrt{s}}{m_X}\right).
  \end{equation}
In particular we will focus on the luminosities that are most constrained by the Neutral Current (NC) and Charged Current (CC) Drell-Yan data respectively, namely   
  \begin{align}
  \label{eq:wzlumi}
    {\cal L}^{\rm NC}(m_X,\sqrt{s}) &= {\cal L}_{u\bar{u}}(m_X,\sqrt{s}) + {\cal L}_{d\bar{d}}(m_X,\sqrt{s}),\\[1.5ex]
      {\cal L}^{\rm CC}(m_X,\sqrt{s}) &= {\cal L}_{u\bar{d}}(m_X,\sqrt{s}) + {\cal L}_{d\bar{u}}(m_X,\sqrt{s}).
    \end{align}

\subsection{Effects of new heavy bosons in PDF fits}
\label{sec:contaminated_pdfs}

\begin{figure}[b!]
\centering
  \includegraphics[width=0.6\linewidth]{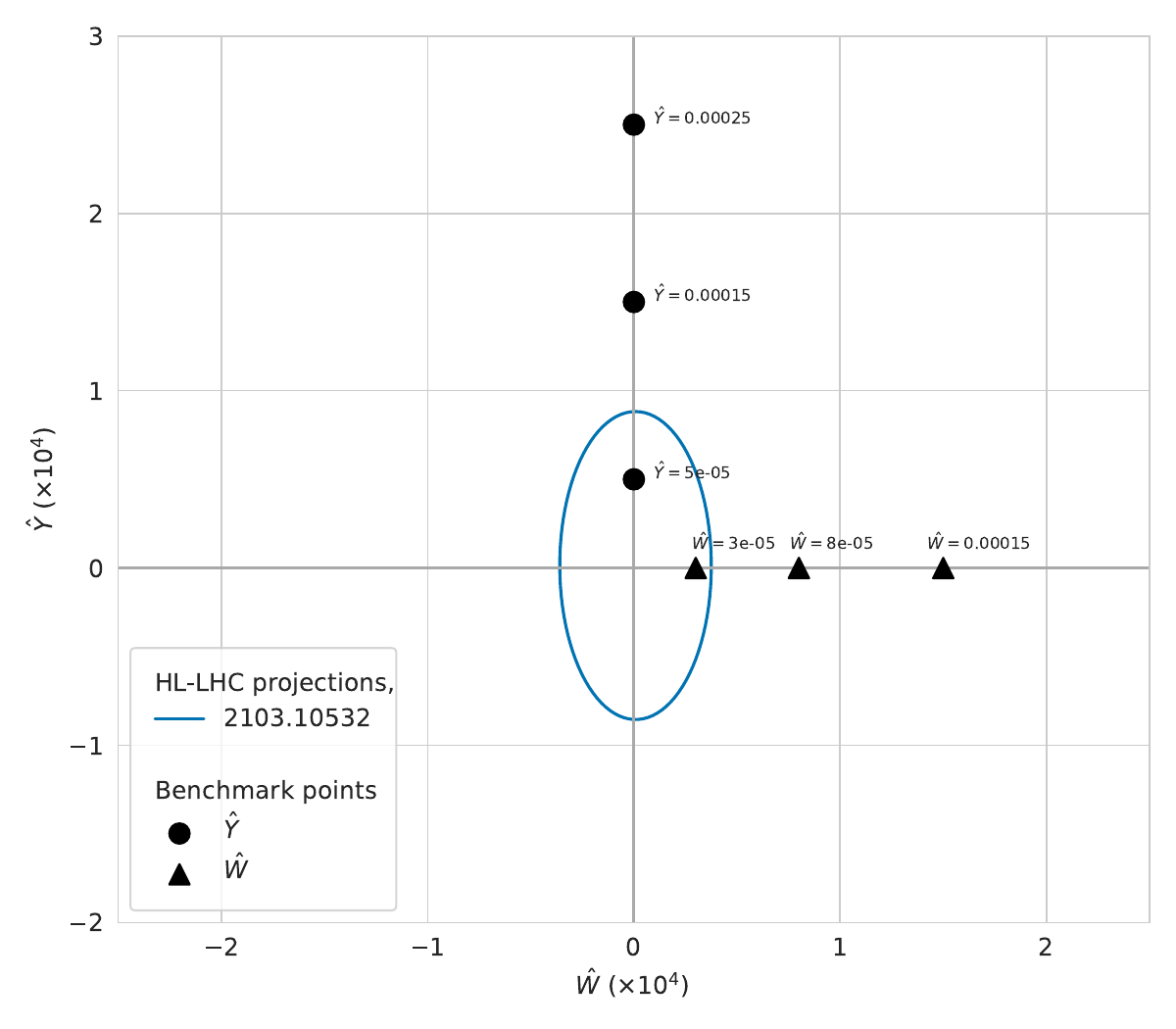}
	\caption{\label{fig:bps} Benchmark $\hat{Y}$ and $\hat{W}$ points explored in this analysis compared to 
	the constraints at 95\% CL as given by the analysis of fully-differential Drell-Yan
 projections given in Ref.~\cite{Panico:2021vav}.}
      \end{figure}

In Fig.~\ref{fig:bps} we display the benchmark points that we
consider, corresponding to the two scenarios described in
Sect.~\ref{sec:scenarios}. Namely, the points along the vertical axis
correspond to the flavour-universal $Z'$ model (Scenario I), while
those along the horizontal axis correspond to the flavour-universal
$W'$ model (Scenario II). The benchmark points are compared to projected constraints
from the HL-LHC.  In particular, we consider the most up-to-date constraints 
from the analysis of a fully-differential Drell-Yan projection in the HL-LHC regime, as 
given by Ref.~\cite{Panico:2021vav}.
%
%

In order to estimate the effect of a heavy $Z'$ ($W'$) in Nature and the ability of PDFs
to fit it away, we inject new physics in the artificial Monte Carlo data by setting $\hat{Y}\neq
0$ ($\hat{W}\neq 0$) to the values that we consider in our benchmark (see
Fig.~\ref{fig:bps}) and we measure the effect on the fit
quality and on the PDFs. To assess the fit quality, we generate
{\it L1 pseudodata}, as in Eq.~\eqref{eq:observed_data}, according to
1000 variations of the random seed $k$ and compare the distributions of the corresponding $\chi^2$-statistic per data point, 
$\chi^{2(k)}/n_{\rm dat}$, and the number of standard deviations from the expected $\chi^2$, $n^{(k)}_{\sigma}$, across the 1000
random seed variations for the
baseline and the 3 benchmark values in each of the two
scenarios.
If the distributions shift above the critical levels defined in Chapter \ref{ch:simunet}, then the PDFs have {\it not} been able to absorb
the effects of new physics and the datasets that display a bad
data-theory agreement would be excluded from a PDF fit. If instead the
distributions remain statistically compatible with those of the baseline
PDF fit, then the PDFs have been able to absorb new physics.

Note that in this exercise the distribution across random seed values is calculated by keeping the PDF fixed
to the value obtained with a given random seed, while if we were
refitting them for each random seed, we would obtain slightly
different PDFs. A comparison at the level of PDFs and parton luminosities is then performed to
assess whether the absorption of new physics shifts them significantly with
respect to the baseline PDFs. We have verified that the effect is negligible and
does not modify the results. A more detailed account of the contaminated PDF's random
seed dependence is given in App.~\ref{app:random}. 
The goal of this exercise is to estimate the maximum strength of new physics
effects beyond which PDFs are no longer able to absorb the effect, and
subsequently assess whether the effect is significant or not.\\

\subsubsection*{(i) Scenario I}

\begin{figure}[t!]
  \includegraphics[width=0.48\linewidth, page=1]{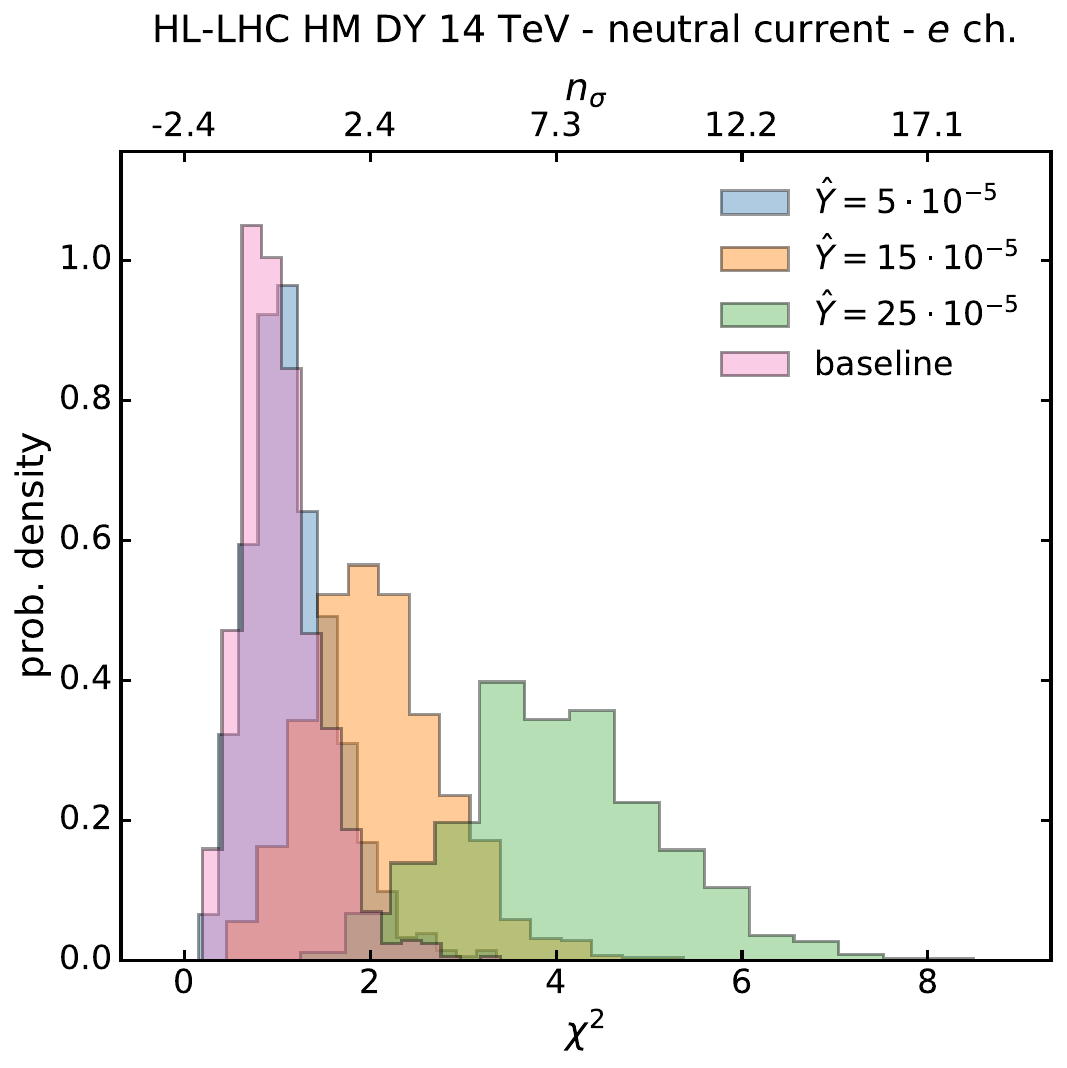}
  \includegraphics[width=0.48\linewidth, page=3]{Figures/chi2_nsigma_Y.pdf}\\
  \includegraphics[width=0.48\linewidth, page=6]{Figures/chi2_nsigma_Y.pdf}
  \includegraphics[width=0.48\linewidth, page=10]{Figures/chi2_nsigma_Y.pdf}\\
  \caption{Distribution of $\chi^2$ and $n_\sigma$ for selected datasets in the $\hat{Y}$ contamination scenarios.}
\label{fig:Y_chi2_nsigma_dist}
\end{figure}

\noindent In the case of the flavour-universal $Z'$ model, we inject three non-zero
values of $\hat{Y}=5\cdot 10^{-5},\, 15\cdot 10^{-5},\, 25\cdot
10^{-5}$. In Fig.~\ref{fig:Y_chi2_nsigma_dist} we display 
the $\chi^{2(k)}$ and $n_{\sigma}^{(k)}$ distributions
across the 1000 $k$ random seeds for a selection of the datasets
included in each of the fits. In particular, we display the datasets in
which a shift occurs either because of the direct effect of the
non-zero Wilson coefficients in the partonic cross sections (such as
the high-mass Drell-Yan in the HL-LHC projections) or
because of the indirect effect of the change of PDFs, which can alter the behaviour of other datasets
that probe the large-$x$ light quark and antiquark distributions. Full
details about the trend in the fit quality for all datasets are
given in App.~\ref{app:fit}.

As far as the quality of the fit is concerned, we observe that, for
$\hat{Y}=5\cdot 10^{-5}$, the global fit is
    equivalent to the SM baseline, while as $\hat{Y}$ is increased to
    $15\cdot 10^{-5}$ the quality of the fit deteriorates. This is due
    mostly to a worse description of the HL-LHC neutral current data (top left panel in Fig.~\ref{fig:Y_chi2_nsigma_dist}) data, while the other datasets remain
    roughly equivalent. This is an indication that there is a bulk of data points 
    in the global dataset that constrains the ${\cal L}^{\rm NC}$ luminosity behaviour at high-$x$ and 
    does not allow the PDF to shift and accommodate the HL-LHC Drell-Yan NC data.
    According to the selection criteria outlined in
    Sect.~\ref{subsec:postfit}, the deterioration of both the $\chi^2$ and the $n_{\sigma}$ indicators
    would single out the high-mass Drell-Yan data and indicate that they
    are incompatible with the rest of the data included in the PDF fit.
    As a consequence, they would be excluded from the fit and no contamination would occur. 
    Hence, in this scenario, $\hat{Y}=15\cdot10^{-5}$ falls in the interval of NP values in which the disagreement in
    the data metrics would flag the incompatibility of the high-mass Drell-Yan
    tails with the rest of the datasets.  
%

\begin{figure}[t!]
  \includegraphics[width=0.49\linewidth]{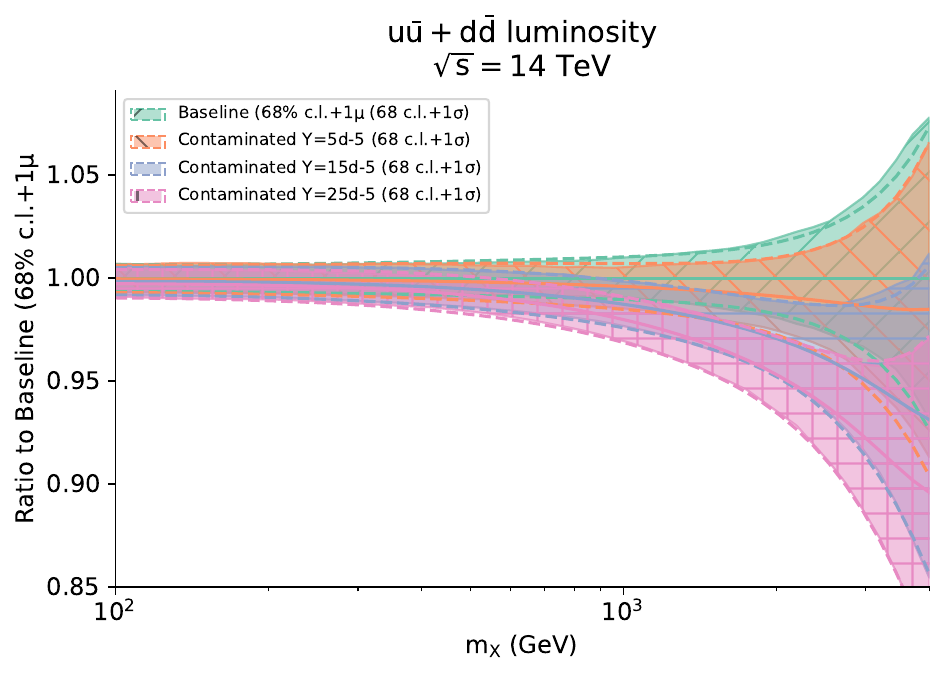}
  \includegraphics[width=0.49\linewidth]{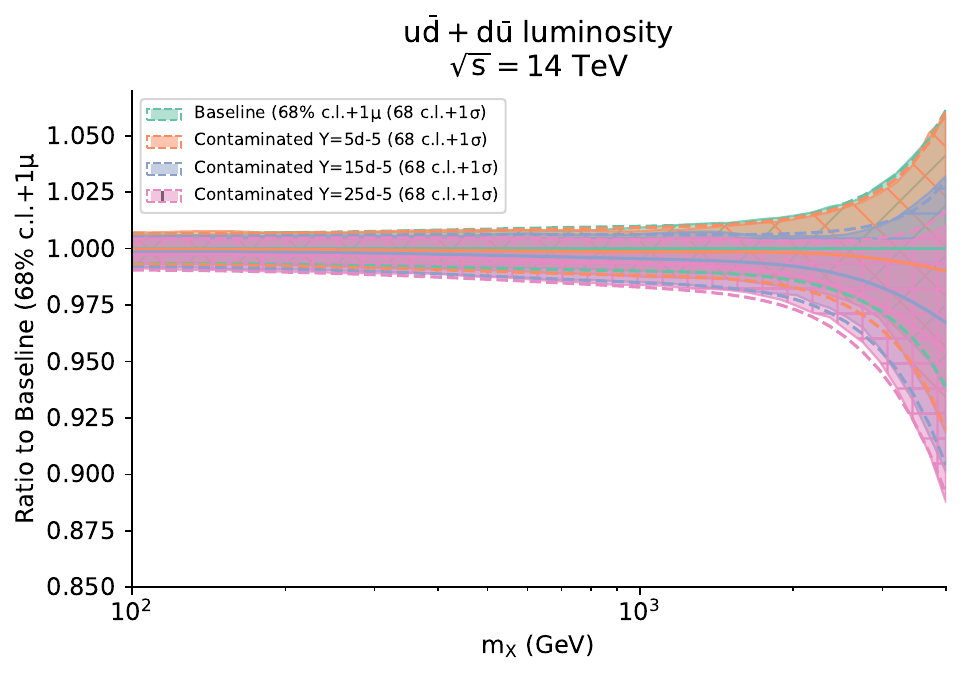}
	\caption{Contaminated versus baseline ${\cal L^{\rm NC}}$ and ${\cal L^{\rm CC}}$ (defined in Eq.~\eqref{eq:wzlumi},
        at $\sqrt{s}=$ 14 TeV in the central rapidity region. The results are normalised to the baseline SM luminosities and the 68\% C.L. band 
        is displayed. Contaminated PDFs have been obtained by fitting the MC data in which $\hat{Y}=5\cdot 10^{-5}$ (orange line),
        $\hat{Y}=15\cdot 10^{-5}$ (blue line) and $\hat{Y}=25\cdot 10^{-5}$ (pink line) has been injected.}
	\label{fig:Ylumi}
      \end{figure}

We now want to check whether, for such values, there is any significant
shift in the relevant NC and CC parton luminosities at the HL-LHC centre-of-mass energy of $\sqrt{s}=14$ TeV. They are displayed in 
Fig.~\ref{fig:Ylumi}. 
We observe that in general the PDFs do not manage to shift much to accommodate the $Z'$ induced contamination. The plots of the individual PDFs are displayed in App.~\ref{app:pdfs}. 
In general the CC luminosity remains compatible with the baseline SM one up to large values of $\hat{Y}$, while, 
as soon as the NC luminosity manages to shift beyond the 1$\sigma$ level, the fit quality of the NC high-mass data 
deteriorates. For the maximum value of new physics contamination that the PDFs can absorb in this scenario, $Y=5\cdot 10^{-5}$ (corresponding to a $Z'$ mass above 30 TeV), 
the parton luminosity shift is contained within the baseline 1$\sigma$ error bar. 
Overall, we see that there is a certain sturdiness in the fit, such that even in the presence of big $\hat{Y}$ values, the parton luminosity does not deviate much from the underlying law.

\subsubsection*{(ii) Scenario II}
\begin{figure}[t!]
  \includegraphics[width=0.48\linewidth, page=1]{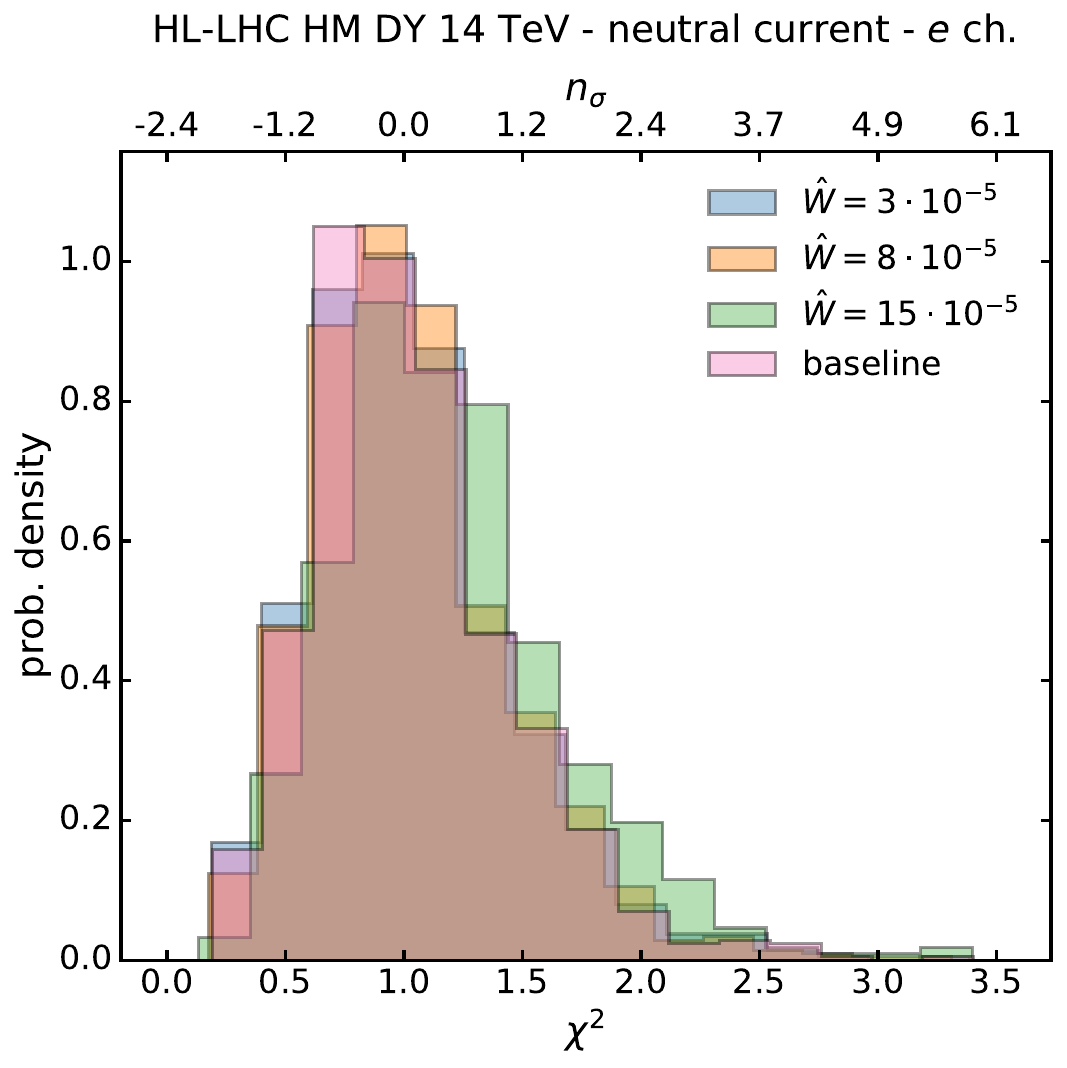}
  \includegraphics[width=0.48\linewidth, page=3]{Figures/chi2_nsigma_W.pdf}\\
  \includegraphics[width=0.48\linewidth, page=6]{Figures/chi2_nsigma_W.pdf}
  \includegraphics[width=0.48\linewidth, page=10]{Figures/chi2_nsigma_W.pdf}\\
  \caption{Distribution of $\chi^2$ and $n_\sigma$ for selected datasets in the $\hat{W}$ contamination scenarios.}
\label{fig:W_chi2_nsigma_dist}
\end{figure}

\noindent  In the flavour-universal $W'$ model we inject three non-zero
values of $\hat{W}=3\cdot 10^{-5},\, 8\cdot 10^{-5},\, 15\cdot
10^{-5}$. In Fig.~\ref{fig:W_chi2_nsigma_dist} we display 
the $\chi^{2(k)}$ and $n_{\sigma}^{(k)}$ distributions
across the 1000 random seeds $k$ for a selection of the datasets
included in each of the fits. In particular we display the datasets in
which a shift occurs either because of the direct effect of the
non-zero Wilson coefficients in the partonic cross sections (such as
the high-mass Drell-Yan in the HL-LHC projections) or
because of the indirect effect of the change of PDFs on other datasets
that probe the large-$x$ light quark and antiquark distributions. Full
details about the trend in the fit quality for all datasets is
given in App.~\ref{app:fit}.

In this case, concerning the quality of the fit, we observe that
up to $\hat{W}=8\cdot 10^{-5}$, the global fit shows
    equivalent behaviours to the SM baseline, while as $\hat{W}$ is increased to
    $15\cdot 10^{-5}$, the quality of the fit markedly deteriorates. This is due
    mostly to a worse description of the HL-LHC charged current
    $e\nu_{e}$ (top right panel in Fig.~\ref{fig:W_chi2_nsigma_dist}) as
    well as the $\mu\nu_\mu$ data. It is interesting to observe that
    also the low-mass fixed-target Drell-Yan data from the E886
    experiment experiences a deterioration in the fit quality due to
    the shift that occurs in the large-$x$ quark and antiquark PDFs. 
    For this largest value, $\hat{W} = 15 \cdot 10^{-5}$, according to the selection criteria outlined in
    Sect.~\ref{subsec:postfit}, the deterioration of both the $\chi^2$ and the $n_{\sigma}$ indicators
    would highlight the high-mass Drell-Yan data as being incompatible with the bulk of the data included in the PDF fit,
    thus excluding them from the fit; thus, no contamination would occur. 
    Hence, in this scenario, $\hat{W}=15\cdot10^{-5}$ falls in the NP parameter region 
    in which the disagreement between the data and theory predictions would unveil the presence of incompatibility of the high-mass Drell-Yan
    tails with the rest of the data; on the other hand, the contamination would go undetected for $\hat{W} = 8 \cdot 10^{-5}$.

\begin{figure}[t!]
  \includegraphics[width=0.49\linewidth]{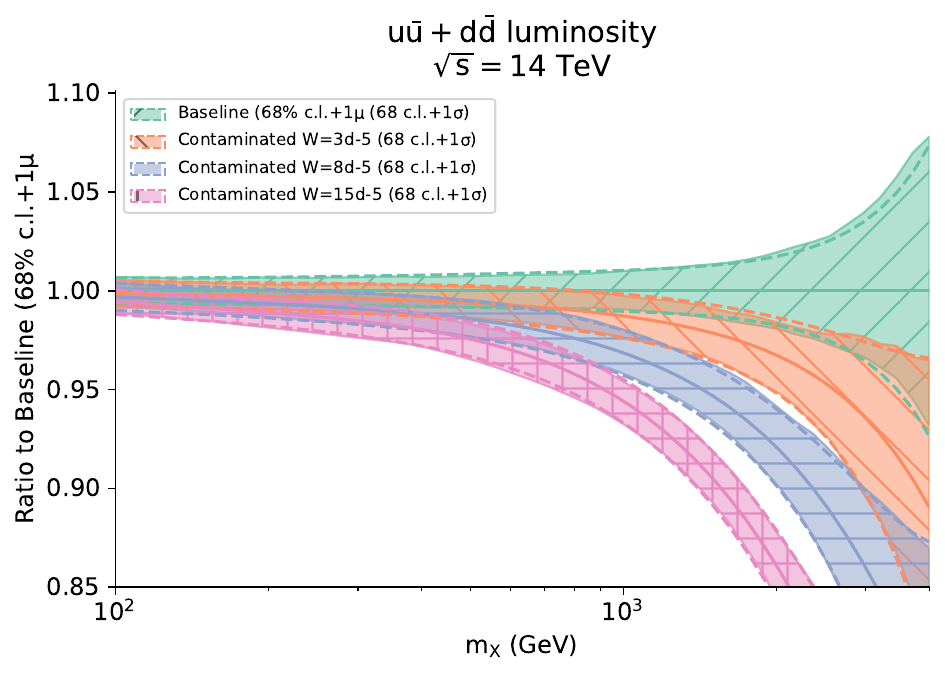}
  \includegraphics[width=0.49\linewidth]{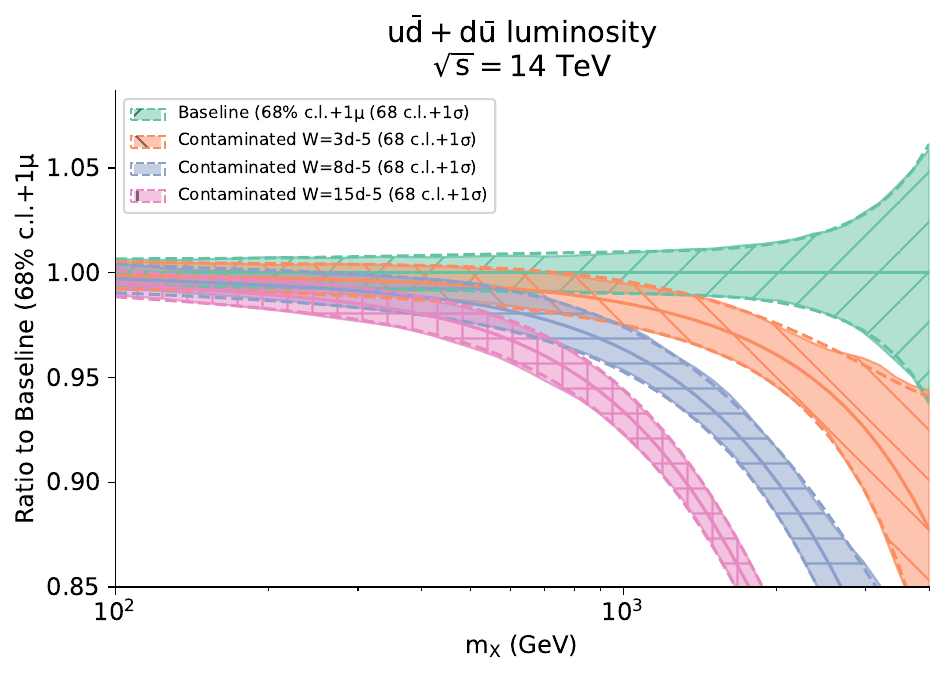}
  \caption{Same as Fig.~\ref{fig:Ylumi} for $\hat{W}=3\cdot 10^{-5}$ (orange line),
    $\hat{W}=8\cdot 10^{-5}$ (blue line) and $\hat{W}=15\cdot
    10^{-5}$ (pink line). }
\label{fig:Wlumi}
\end{figure}
We now check whether, for such $\hat{W}$ values, there is any significant
shift in the PDFs and in the parton luminosities. Individual PDFs are displayed in App.~\ref{app:pdfs}.
In Fig.~\ref{fig:Wlumi} we observe that in this scenario the
NC and CC luminosities defined in Eq.~\eqref{eq:wzlumi} can both shift significantly in the high-mass region, even for low values of $\hat{W}$ ($m_{W'}$ above 20 TeV).
Contrary to the case outlined in the $\hat{Y}$ scenario, the fit \textit{does} have enough flexibility to absorb significant deviations in the high-mass Drell-Yan without impacting 
the rest of the dataset. In particular, until the deviations become too large, the NC and CC sectors, which are both affected by the $W'$ boson, manage to compensate each other. \\

\subsubsection*{(iii) Summary}
Overall, we find that in Scenario I the presence of a new
heavy $Z'$ of about 18 TeV would affect the high-energy tails of the Drell-Yan
  distributions in such a way that they are no longer compatible with the
  bulk of the data included in a PDF analysis.
    On the other hand, in Scenario II, a model of new physics involving a $W'$ of about
  14 TeV  would affect the high-energy tails of the Drell-Yan
  distributions in a way that can be compensated by the PDFs. As a result, if there
  is such a $W'$ in Nature, then this would yield a good $\chi^2$ for
  the high-mass Drell-Yan tails that one includes in a PDF fit as well
  as for the bulk of the data included in a PDF fit, but it 
  would significantly modify PDFs.  Thus, in this case new physics
  contamination does occur.

These results are in agreement with the results of Ref.~\cite{DYpaper}, which generalises the analysis of
Ref.~\cite{Torre:2020aiz} by allowing the PDFs to vary along with the
$\hat{Y}$ and $\hat{W}$ coefficients, finding less stringent
constraints from the same HL-LHC projections.  In particular, it was found that $\hat{W}=8\cdot 10^{-5}$ would have been excluded by the HL-LHC under the assumption of SM PDFs, but that this value of $\hat{W}$ was allowed by the constraints at 95\% CL obtained by varying the PDFs along with the SMEFT.
Ref.~\cite{DYpaper} also indicated that the impact of varying the PDFs along with the
$\hat{W}$ coefficient was more significant than the impact in the $\hat{Y}$ direction, indicating 
a greater possibility to absorb the effects of new physics into the PDFs in the $\hat{W}$ direction.


%

Comparing the two scenarios considered in this section, one might wonder why 
the $Z'$ scenario does not yield any contamination, while the $W'$ does. 
%
%
Looking at the effect of the $Z'$ and $W'$ bosons on the observables included in a PDF fit (see Eqs.~\eqref{eq:zprimeL} and \eqref{eq:wprimeL} respectively), 
we see that the main difference lies in the fact that the $Z'$ scenario only affects the NC DY high-mass data, while the $W'$ scenario affects both the
NC and the CC DY high-mass data. Hence, in the former scenario, the shift required in ${\cal L}^{\rm NC}\equiv (u\bar{u}+d\bar{d})$ to accommodate the 
effect of a $Z'$ in the tail of the $m_{ll}$ distribution would cause a shift in ${\cal L}^{\rm CC}\equiv (u\bar{d}+d\bar{u})$, thus spoiling its agreement with the data, 
in particular the tails of the $m_T$ distribution -- which is unaffected by the presence of a $Z'$. 

On the other hand, in the $W'$ scenario, the shift in the $(u\bar{u}+d\bar{d})$ parton channel that accommodates the effect of a $W'$ in the tail of the 
NC DY $m_{ll}$ distribution is compensated by the shift in the $(u\bar{d}+d\bar{u})$ parton channel that accommodates the presence of a $W'$ in the tail of the 
CC DY $m_T$ distribution (as, in this scenario, they are both affected by new physics). It is as if there is a flat direction in the luminosity versus the 
matrix element space. This continues until, for sufficiently large $\hat{W}$, a critical 
point is reached in which the two effects do not manage to compensate each other as they start affecting significantly the luminosities at lower $\tau=M/\sqrt{s}$, 
hence spoiling the agreement with the other less precise datasets included in a PDF fit which are sensitive to large-$x$ antiquarks. 

\begin{figure}[tb]
  \begin{center}
    \includegraphics[width=0.49\linewidth]{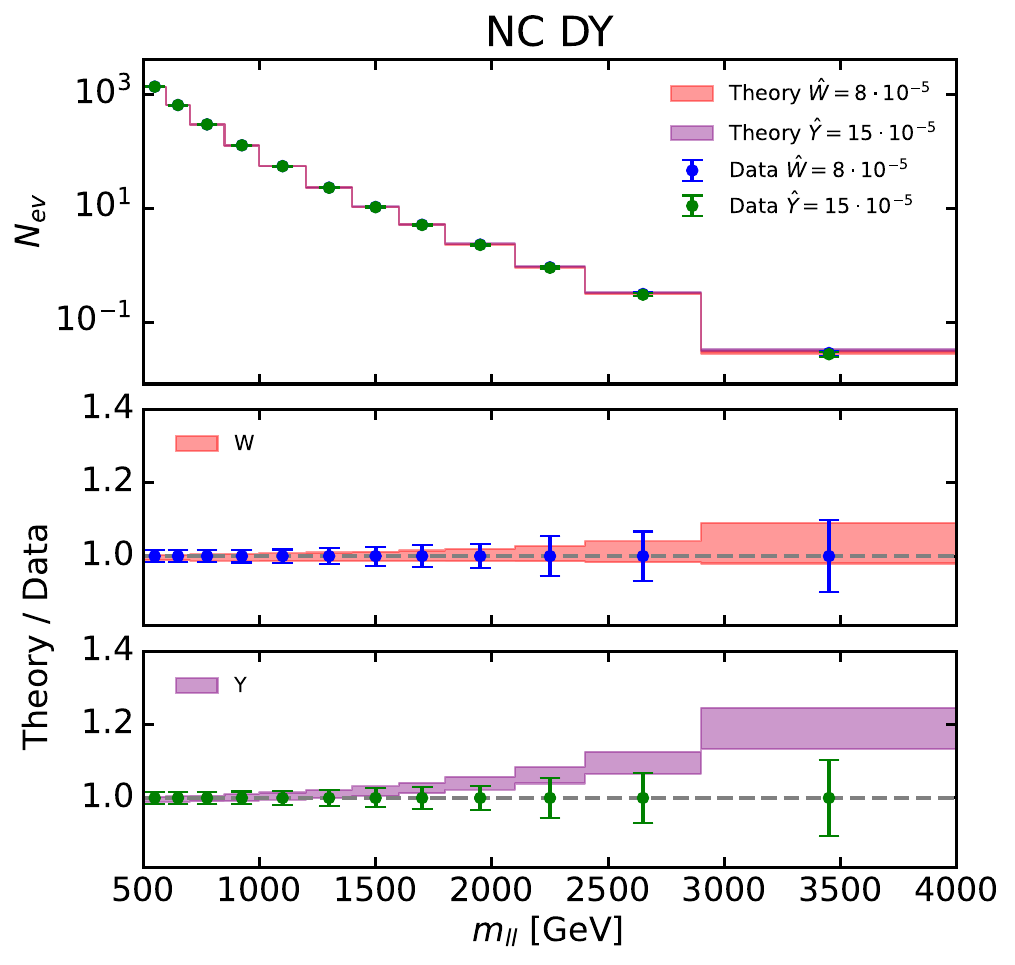}
    \includegraphics[width=0.49\linewidth]{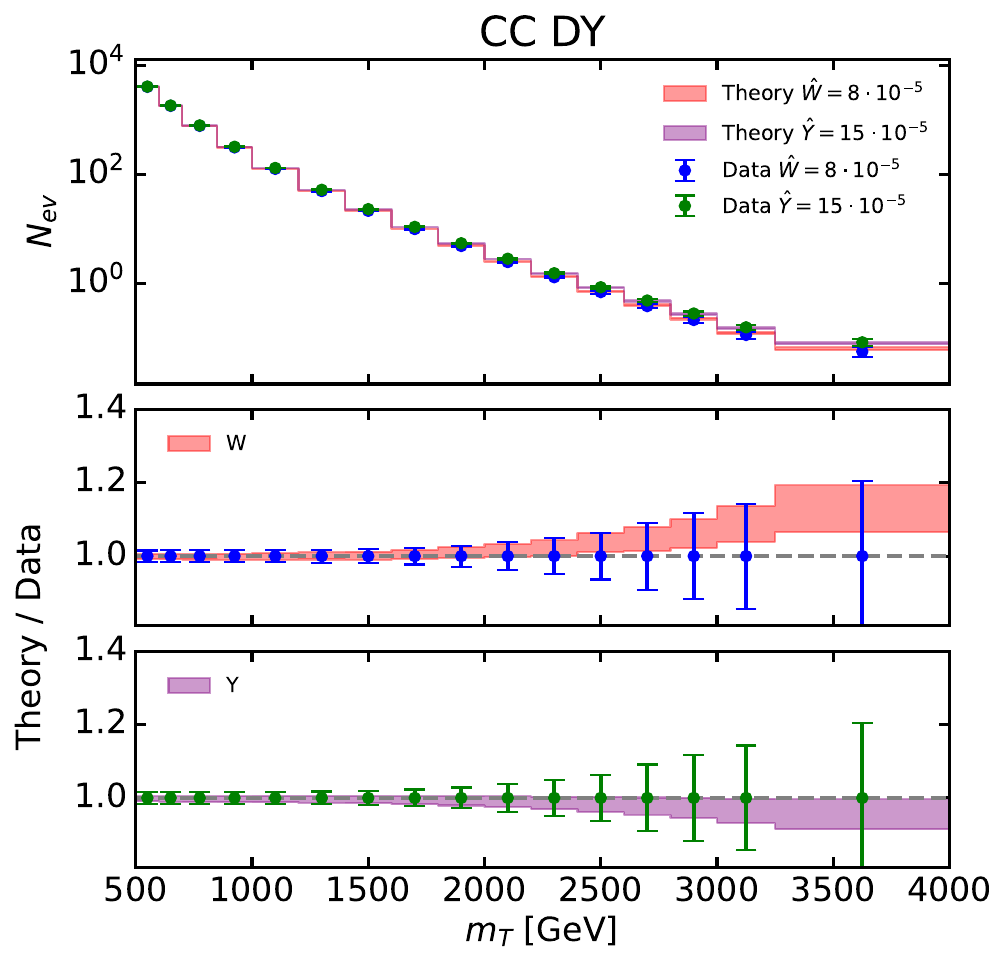}
  \end{center}
	  \caption{For two of the 
      representative scenarios that we consider, $\hat{W}=8\cdot 10^{-5}$ and $\hat{Y}=15\cdot 10^{-5}$, we show the comparison between the data expected in the presence of new physics (``Data'' points) and the SM theory predictions obtained with the potentially contaminated PDFs (``Theory'' bands). Left panel: NC Drell-Yan $m_{ ll}$ distribution. 
      Right panel: CC Drell-Yan $m_T$ distribution. 
    }
  \label{fig:wy_dy_prod}
\end{figure}

To see this more clearly, we plot in Fig.~\ref{fig:wy_dy_prod} the data-theory comparison for the HL-LHC NC and CC Drell-Yan Monte Carlo data that we include in the fit. 
%
The points labelled as ``Data'' correspond to the `truth' in the presence of the new physics, namely they are obtained by convolving the DY prediction with non-zero $\hat{Y},\hat{W}$ parameters with a non-contaminated PDF set.
The bands labelled as ``Theory'' represent the theoretical predictions for pure SM DY production, but obtained with the 
PDFs fitted with the inclusion of the DY data modified by the effect of non-zero $\hat{Y},\hat{W}$ parameters. 
We observe that the SM predictions obtained with the contaminated PDFs do fit the data well in the case of $\hat{W}=8\cdot 10^{-5}$, 
because the significant depletion of the $(u\bar{u}+d\bar{d})$ and $(u\bar{d}+d\bar{u})$ parton luminosities observed in Fig.~\ref{fig:Wlumi} compensates 
the enhancement in the partonic cross section observed in Fig.~\ref{fig:Wp_DY_Comp}. This is not the case for $\hat{Y}=15\cdot 10^{-5}$, where instead the 
much milder modification of the parton luminosities observed in Fig.~\ref{fig:Ylumi} does not manage to compensate the enhancement of the partonic cross 
section observed in Fig.~\ref{fig:Zp_DY_Comp}. We can also notice that $\hat{W}=8\cdot 10^{-5}$ is within a region in the $\hat{W}$ parameter space 
beyond which the parton luminosities do not manage to move enough to compensate the shift in the matrix elements of the $m_T$ distribution. 
To find the exact critical value of $\hat{W}$ one would need a finer scan.
Analogously, $\hat{Y}=15\cdot 10^{-5}$ is in the region of $\hat{Y}$ such that  contamination in the PDFs does not occur.
However, these values have been determined assuming a given statistical uncertainty in the distributions;
the regions in which these values fall clearly depend on the actual statistical 
uncertainty that the $m_T$ and $m_{ll}$ distributions will reach in the HL-LHC phase.

\subsection{Consequence of new physics contamination in PDF fits}
\label{sec:projections}
In the previous section, we showed that in the presence of heavy new physics effects in 
DY observables, the flexible PDF parametrisation is able to accommodate the deviations
and absorb the effects coming from the new interactions. In particular, we observe that when
data are contaminated with the presence of a $W^\prime$, we generally find good fits and are able to accommodate
even large deviations from the SM. However, it is worth reminding the reader that
the leading source of contaminated data are the HL-LHC projections, as present data would not be as susceptible to the $W^\prime$ effects.
Hence, from now on we will focus on the scenario in which data include the presence of a heavy $W^\prime$ that induces 
a modified interaction parametrised by the $\hat{W}$ parameter with value $\hat{W}=8\cdot 10^{-5}$.

\begin{figure}[htb!]
\centering
\includegraphics[width=0.9\textwidth]{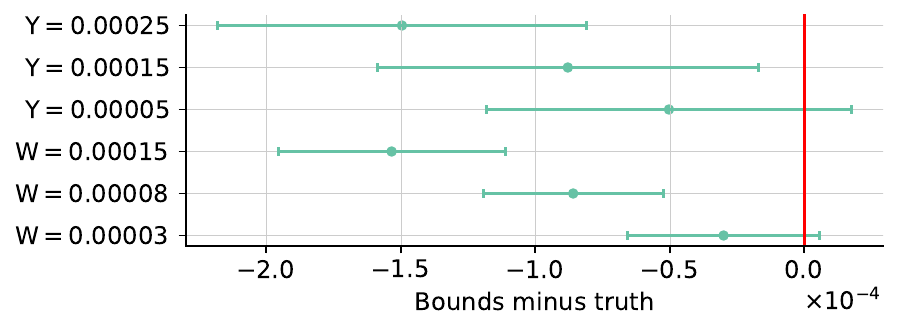}
\caption{A comparison of the 95\% C.~L.~bounds obtained using different contaminated PDFs to fit the $\hat{W}$, $\hat{Y}$ parameters 
to HL-LHC high-mass Drell-Yan projected data, relative to the true values of $\hat{W}, \hat{Y}$. In some cases, the true value is not 
contained in the 95\% C.~L.~bounds.}
\label{fig:bounds_comparison}
\end{figure}

In this section we examine the consequences of
using unknowingly contaminated PDFs, and the implications of this for possible new physics searches.
%
The first interesting consequence is that, if we use the contaminated PDF as an input set in a SMEFT study 
of HL-LHC projected data to gather knowledge on the $\hat{W}$ parameter, we find that the analysis excludes 
the ``true'' value of the SMEFT coefficients that the data should reveal.
Indeed, in Fig.~\ref{fig:bounds_comparison} we observe that, in both scenarios under consideration, and in particular for the one
corresponding to $\hat{W}=8\cdot10^{-5}$, the 95\% C.L. bounds on the Wilson coefficients that one would extract from the precise
HL-LHC data discussed in Sect.~\ref{sec:contaminated_pdfs} would agree with the SM and would not contain the true 
``values'' of the underlying law of Nature that the data should reveal.
In fact, the measured value would exclude the true value with a significance that ranges from  $\sim 1.5 \, \sigma$ to $\sim 4.5 \, \sigma$. A comparison of whether the bounds generated by the different contaminated PDFs considered in this study contain the true value is shown in Fig.~\ref{fig:bounds_comparison}.
This is all very expected, as the quark-antiquark luminosity for this specific scenario does exhibit signs of
new physics absorption in a significant amount, as can be seen in Fig.~\ref{fig:Wlumi}. 
As a matter of fact, we expect all data that entered in the PDF fit to be well described by the combination of the PDF set and the SM theory.
This simple fact is once again reminding us that it might be dangerous to perform SMEFT studies on overlapping datasets and that simultaneous studies should be preferred or, at least, a conservative approach with disjoint datasets should be undertaken. 
This was discussed in Ref.~\cite{Iranipour:2022iak}, for instance,
where it was shown that by means of a simultaneous study, one is able to recover 
both the underlying true PDFs and the presence of a new interaction. It is worth 
mentioning that the use of conservative PDF sets, while appealing given the simplicity, 
might also come with its own shortcomings, see Ref.~\cite{DYpaper} and Ref.~\cite{Kassabov:2023hbm} for detailed studies on the matter in the 
Drell-Yan sector and the top quark sector respectively. In particular, 
the extrapolated PDFs might both underestimate the error band and
have a significant bias.\\

\noindent We now turn to study the effects of the contaminated PDFs in observables and processes that did not enter the PDF fit.
We focus in particular on the EW sector, given its relevance for NP searches and the fact that the contaminated PDFs show deviations from 
the true PDFs mostly in the quark-antiquark luminosities, which are particularly relevant for theoretical predictions involving EW interactions.
The study is performed by producing projected data according to the true laws of Nature, i.e. the true PDFs of choice and 
the SM + $\hat{W}=8\cdot 10^{-5}$ in the matrix elements. 

In particular, we 
produce MC data for several diboson processes, including $H$ production in association with EW bosons. 
Given that the $\hat{W}$ operator 
induces only four-fermion interactions, $\hat{W}$ does not have an effect on these observables, and the hard scattering 
amplitudes are given by the SM ones. For each observable we build HL-LHC projections and devise bins with the objective 
of probing the high-energy tails of the distributions, scouting for new physics effects, although we know these do not exist in the ``true'' law of Nature for these observables.
We then produce predictions by convolving the contaminated PDF set obtained with a value of $\hat{W}=8\cdot 10^{-5}$ and the SM matrix elements. 
Given our knowledge of the ``true'' law of Nature, the possible deviations between theory and data are therefore
only a consequence of the shift in the PDFs coming from the contaminated Drell-Yan data.
Whenever in the presence of $W$ bosons, we decided to split the contributions of $W^+ X$ and $W^- X$ as they probe different luminosities and in particular, 
from the contaminated fits, we know that the luminosity $u \bar{d}$ deviates more severely than $d \bar{u}$ from the true luminosity.

Both SM theory and data have been produced at NLO in QCD making use of the Monte Carlo generator \madgraph{}.
In the case of $ZH$ production, the gluon fusion channel has also been taken into account.
Data are obtained by fluctuating around the central value, assuming a Gaussian distribution with total 
covariance matrix given by the sum of the statistical, luminosity and systematic covariance matrices. 
Regarding the theory predictions, we also provide an estimate 
of the PDF uncertainty.
We assume a luminosity of $3$ ab$^{-1}$ and we estimate the systematic uncertainties on each observable 
by referring to the experimental papers~\cite{ATLAS:2020osn, ATLAS:2019bsc, CMS:2019efc} .
These systematic uncertainties can be experimental or come from other additional sources such as background and signal theoretical calculations.
We also include estimates of the luminosity uncertainty by taking as a reference the CMS measurement at 13 TeV~\cite{CMS:2018mdl}. 
%
Statistical uncertainties are given by $\sqrt{N}$, where $N$ is the number of expected events in each bin. Performing a fully realistic simulation, with acceptance cuts and detector effects, is beyond the scope of the current study, and we simply simulate events at parton level and apply the branching ratios into relevant decay channels. 
Specifically, in the case of $W$ bosons we apply a branching ratio of $Br(W \to l \nu) = 0.213$ with $l=e, \, \mu$, for the $H$ boson we use $Br(H \to b \bar{b})=0.582$ and for the $Z$ boson we use $Br(Z \to l^+ l^-) = 0.066$ with $l=e, \, \mu$~\cite{Workman:2022ynf}.
Multiple sources of uncertainty are simply added in quadrature.
\begin{table}[t!]
        \small
        \centering
        \begin{tabular}{l|c|c|c|c|c|}
                \toprule
       & \multicolumn{2}{c}{HL-LHC} & \multicolumn{2}{c}{Stat. improved}\\
          \midrule
       Dataset  & $\chi^2/n_{\rm dat}$ &  $n_\sigma$ & $\chi^2/n_{\rm dat}$ & $n_\sigma$ \\
          \midrule
 $W^+ H$         & 1.17 & 0.41 & 1.77 & 1.97 \\
 $W^- H$         & 1.08 & 0.19 & 1.08 & 0.19 \\
 $W^+ Z$         & 1.08 & 0.19 & 1.49 & 1.20\\
 $W^- Z$         & 0.99 & -0.03 & 1.02 & 0.05\\
 $Z H$           & 1.19 & 0.44 & 1.67 & 1.58\\
 $W^+ W^-$       & 2.19 & 3.04 & 2.69 & 4.31\\
 VBF $\to$ H        & 0.70 & -0.74 & 0.62 & -0.90\\
                \bottomrule
        \end{tabular}
        \caption{Values of the \(\chi^2\) and $n_\sigma$ for the projected observables at HL-LHC in the EW sector. In the left column we report the values from a realistic estimate of the statistical uncertainties, while in the right columns we show what would be obtained if statistics were to improve by a factor $10$. \label{tab:chi2_hllhc}}
\end{table}

In Table~\ref{tab:chi2_hllhc}, for each process considered, we collect the computed $\chi^2$ and the corresponding value of $n_\sigma$. These numbers are obtained by performing several fluctuations of the data and then taking the average $\chi^2$ from all the replicas. As a consequence, the quoted $\chi^2$ are considered the expected $\chi^2$ and are not associated to a specific random fluctuation. The numbers are provided both in a realistic scenario, with a reasonable estimate of the statistical uncertainties, and in a scenario in which the statistics are improved by a factor $10$. 
The latter could be both the result of an increased luminosity and/or additional decay channels of the EW bosons, e.g. decays into jets. 
As it can be seen by inspection of the table, the processes that would lead to the most notable deviations between data and theory are $W^+ H$ and $W^+ W^-$, with the latter being in significant tension already in the scenario of a realistic uncertainty estimation. With improved statistics, slight tensions start to appear in $Z H$ and $W^+ Z$, both exhibiting a deviation just above 1$\sigma$.
Interestingly, the clear smoking gun process here seems to be $W^+ W^-$, which just by 
itself would point towards a significant tension with the SM, which could potentially and erroneously be interpreted in terms of new interactions.
\begin{figure}[t!]
  \includegraphics[scale=0.45]{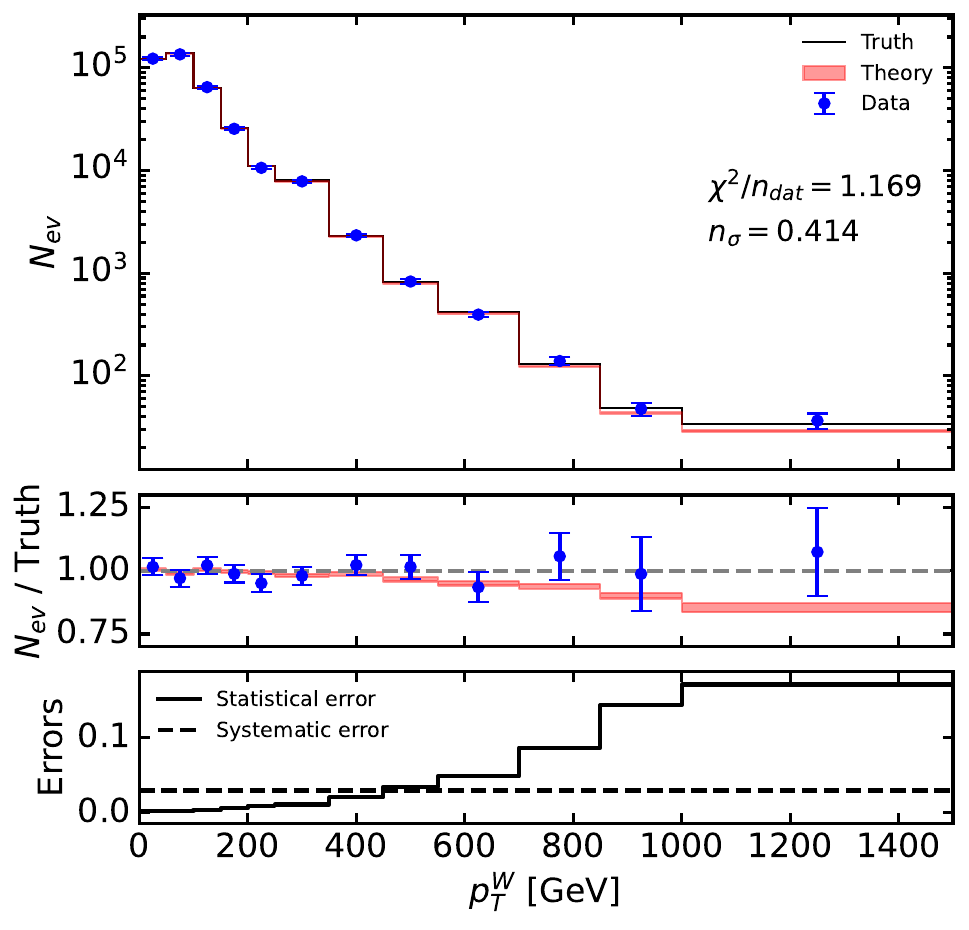}
  \includegraphics[scale=0.45]{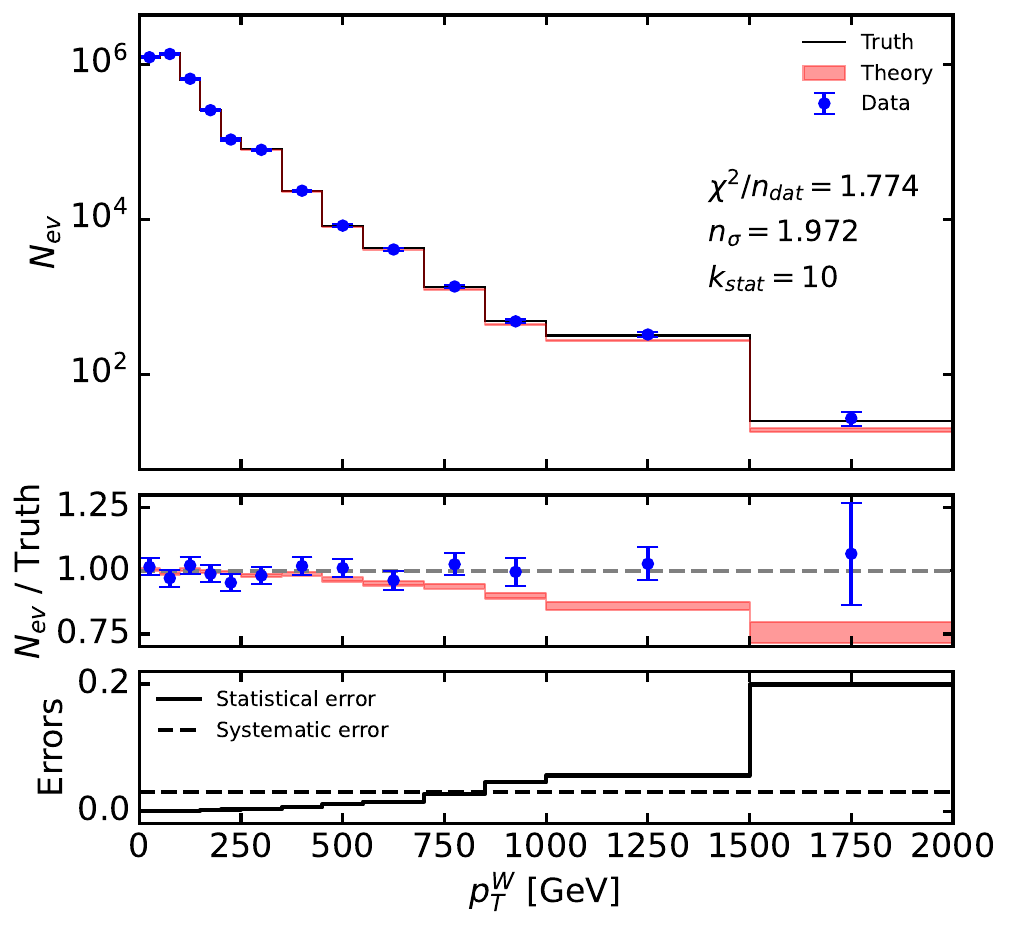}
  \caption{Predictions (with the contaminated PDF) for $W^+ H$ at the HL-LHC compared with the projected data. Left: HL-LHC projection. Right: statistics improved by a factor 10 (futuristic scenario). In the latter, an additional bin is added at high energy to take advantage of the additional expected events.}
  \label{fig:wph_hllhc}
      \end{figure}

\begin{figure}[t!]
  \includegraphics[scale=0.45]{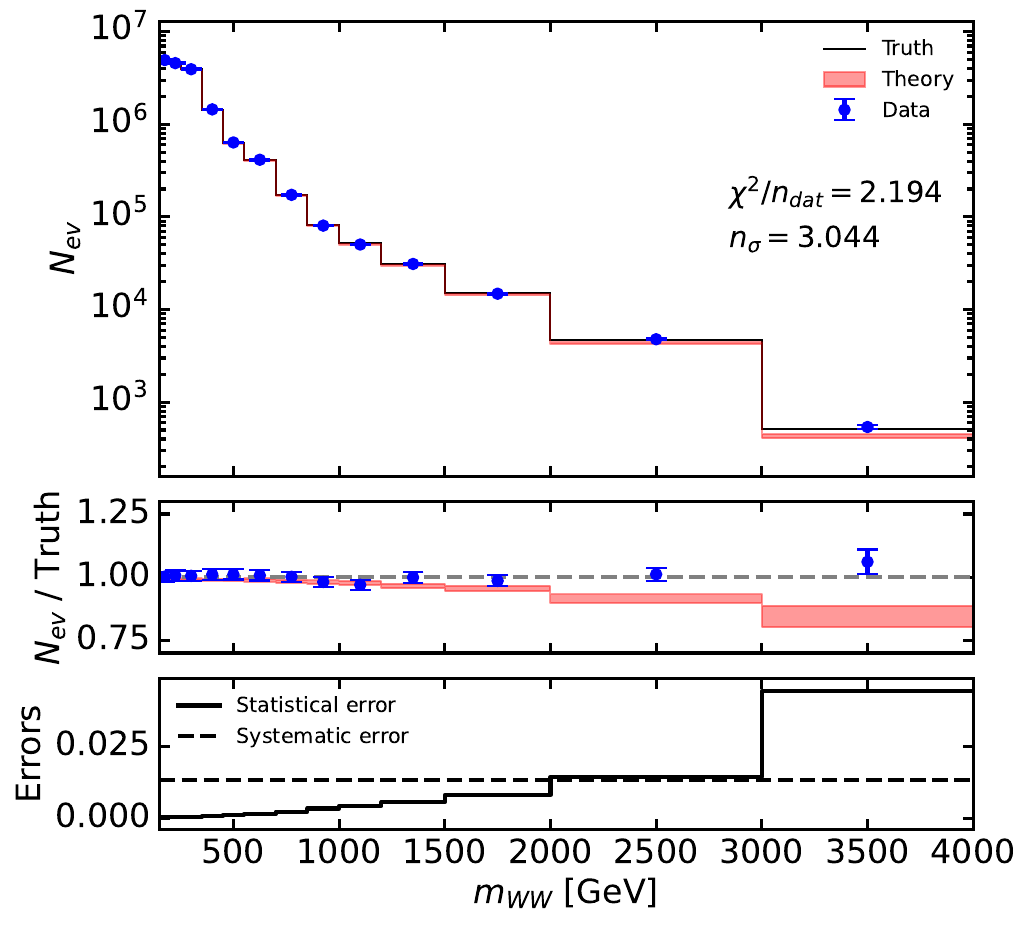}
  \includegraphics[scale=0.45]{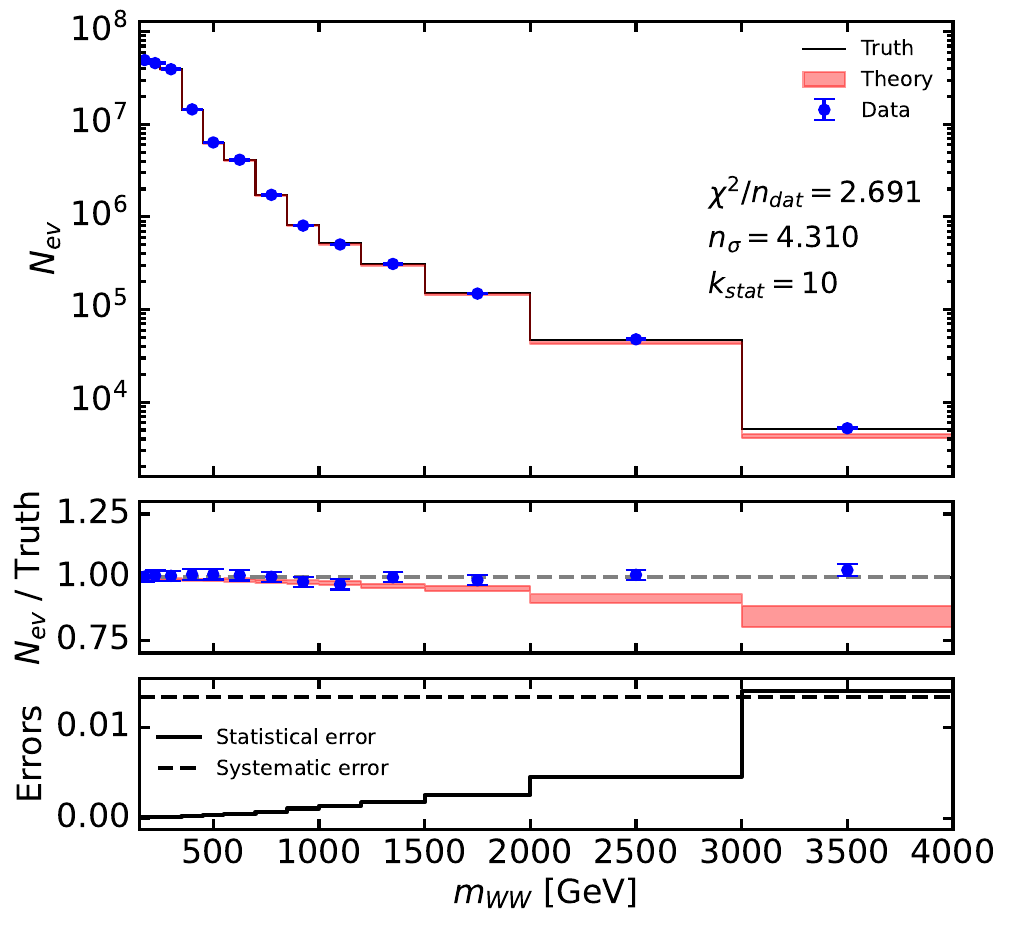}
  \caption{Predictions (with contaminated PDF) for $W^+ W^-$ at the HL-LHC compared with the projected data. Left: HL-LHC projection. Right: statistics improved by a factor 10 (futuristic scenario).}
  \label{fig:ww_hllhc}
      \end{figure}



In Figs.~\ref{fig:wph_hllhc} and ~\ref{fig:ww_hllhc}, we show plots of the two most affected observables considered in 
this section, namely $W^+ H$ and $W^+ W^-$ respectively.
While in all other processes the deviations between the true central values and the theoretical predictions obtained with the contaminated PDFs are limited, in the case of $W^+ H$ and $W^+ W^-$, they are substantial. 
It is clear that the true limiting factor is that as soon as we are in the high energy tails of the distributions and potentially sensitive to the PDF contamination, the pseudodata become statistically dominated and therefore we lose resolution. This is particularly true in the case of $W^+ H$, while $W^+ W^-$ is predicted to have a higher number of events and could potentially probe higher energies.

We also assess the ratios $W^+ Z / W^- Z$ and $W^+ H / W^- H$, and observe that in 
this case the deviations resulting from contaminated PDFs are 
no longer visible. In general the ratios cancel the effect of 
any possible contamination in the parton luminosities if they are correlated. The fact 
that the effect disappears is a proof that the $u\bar{d}$ and $d\bar{u}$ luminosities are highly correlated and 
the contamination effects are compatible.
%
%
%
%
%
%
%
%
%

In summary, the PDF contamination has the potential to generate substantial deviations in observables and processes generally considered to be good portals to new physics, which could nonetheless be unaffected by the presence of heavy states at the current probed energies, as in the scenarios considered in this work.

\section{How to disentangle new physics effects}
\label{sec:solution}

In this section we discuss several strategies that might be proposed in order to disentangle 
new physics effects in a global fit of PDFs. In Sect.~\ref{subsec:forward} we start by 
assessing the potential of precise on-shell forward vector boson production 
data in the HL-LHC phase and check whether their inclusion in a PDF fit helps to disentangle 
new physics effects in the high-mass Drell-Yan tails. 
In Sect.~\ref{subsec:cut} we scrutinise whether the data-theory agreement displays a deterioration 
that scales with the maximum energy probed by the data included in the fit. We will see that 
neither of these strategies helps to disentangle the contamination that arises in the scenario that we have 
highlighted in our study and we outline the reason for this.  We then turn to analyse the behaviour of suitable observable ratios in Sect.~\ref{subsec:ratio}; we will see that such ratios \textit{do} correctly indicate the 
presence of new physics in the observables that are affected by it, although they would not be able to distinguish between the two observables that 
enter the ratio. Finally in Sect.~\ref{subsec:lowE} we will determine the  observables in current PDF fits that are correlated to the 
large-$x$ antiquarks and we will highlight the signs of tension with the ``contaminated'' high-mass Drell-Yan data via suitably devised weighted fits. 
The result of these tests points to the need for the inclusion of independent low-energy/large-$x$ constraints in future PDF analyses, 
if one wishes to safely exploit the constraining power of high-energy data without inadvertently absorbing signs of new physics 
in the high-energy tails.

\subsection{On-shell forward boson production}
\label{subsec:forward}
The most obvious way to disentangle any possible contamination effects in the PDF is the inclusion 
of observables that probe the large-$x$ region in the PDFs at low energies, where NP-induced energy growing effects are not present.  
In this section we assess whether the inclusion of precise forward LHCb distributions measured at the $W$ and $Z$ on-shell energy 
at the HL-LHC might help spotting NP-induced inconsistencies in the high-mass distributions measured by ATLAS and CMS. 

\begin{figure}[htb]
  \centering
  \includegraphics[width=0.49\linewidth]{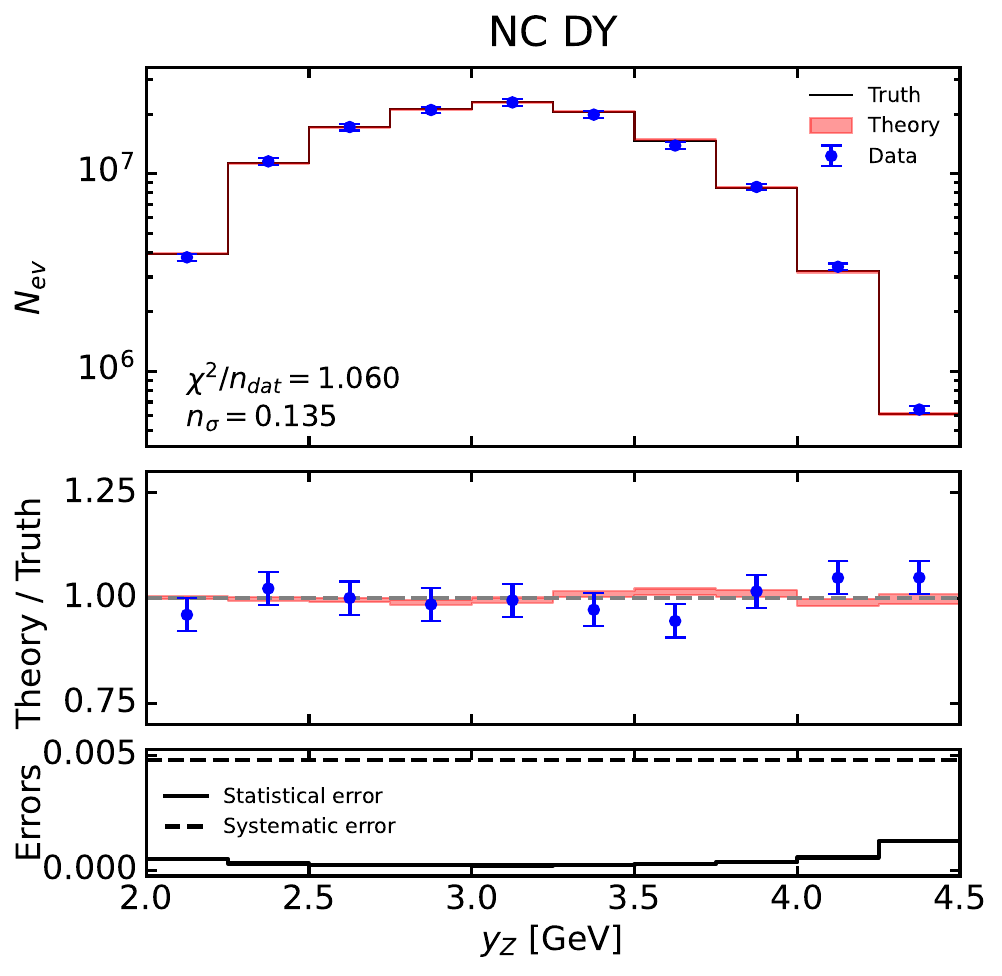}
  \includegraphics[width=0.49\linewidth]{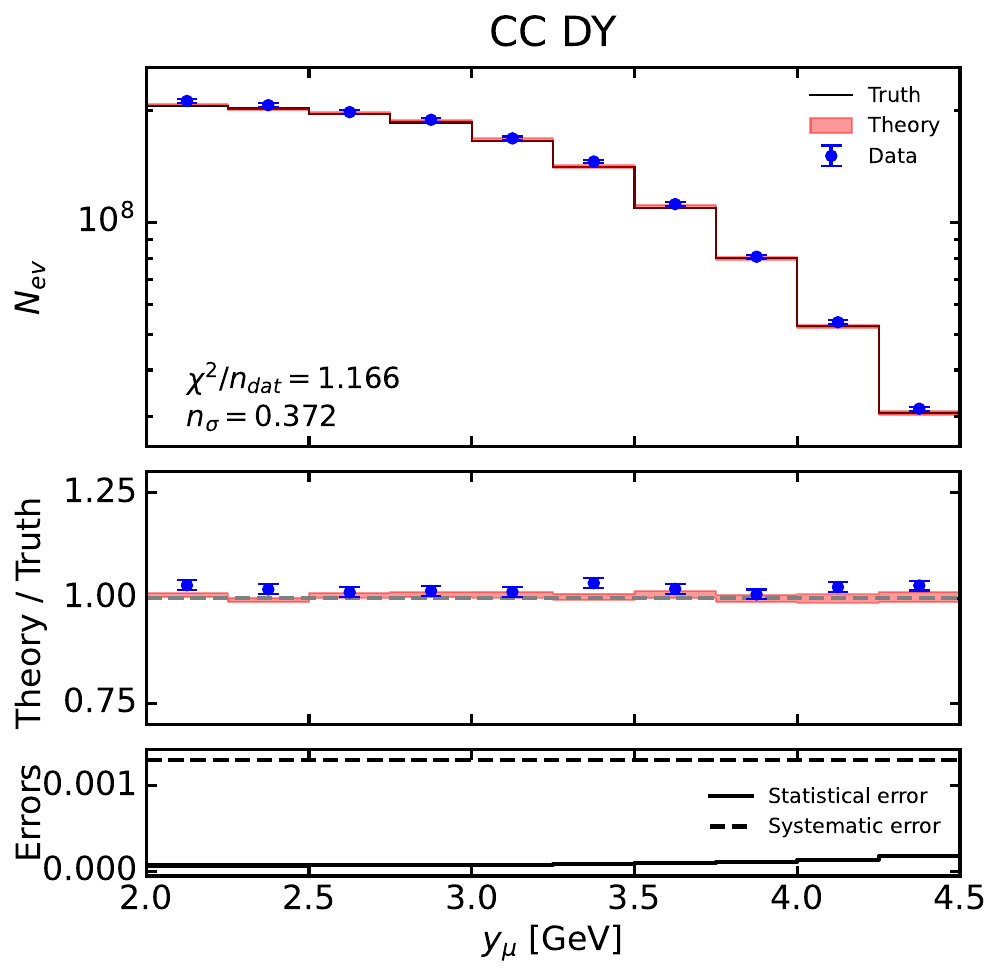}
  \caption{Predictions (with $\hat{W}=8\cdot 10^{-5}$ contaminated PDF) for forward vector bosons production 
  in the HL-LHC phase at LHCb compared with the projected data. Left panel: on-shell $Z$ production cross section as a 
  function of the $Z$ boson rapidity $y_Z$. Right panel: $W$ production cross section as a function of the final-state muon
  pseudo-rapidity $y_\mu$.}
  \label{fig:forwardZ_hllhc}
\end{figure}

In order to test this, we compute HL-LHC projections for LHCb, taking 0.3 ab$^{-1}$ as benchmark luminosity~\cite{Azzi:2019yne} 
and focusing on the forward production of $W/Z$. The $Z$ boson is produced on-shell 
($60 \text{ GeV} < m_{ll} < 120 \text{ GeV}$), while no explicit cuts are applied on  the transverse mass $m_T$ in 
the case of a produced $W$ boson decaying into a muon and a muonic neutrino, which is dominated by the mass-shell region. 
We impose the LHCb forward cuts on the lepton transverse momentum ($p^l_T > 20 \text{ GeV}$) 
and on both the $Z$ rapidity and pseudo-rapidity of the $\mu$ originated by $W$ ($2.0 < |y_{Z,\mu}| < 4.5$).
Fig.~\ref{fig:forwardZ_hllhc} shows a comparison between the pseudodata generated with the ``true'' PDFs and NP-corrected matrix elements,\footnote{Note that at the energy probed by the 
forward $W/Z$ production the NP contribution associated to the presence of a $W'$ boson is negligible} and the theory predictions 
obtained with the $\hat{W}=8\cdot 10^{-5}$ contaminated fit and the SM matrix element, for each of the two processes.
We observe that there are no significant deviations between the theory predictions obtained from a 
contaminated PDF set and the true underlying law. 
Intuitively this can be understood, as the produced leptons are in the forward region measured at LHCb, and one of the initial 
partons must have more longitudinal momentum than the other. 

\begin{figure}[htb]
  \begin{center}
    \includegraphics[width=0.8\linewidth]{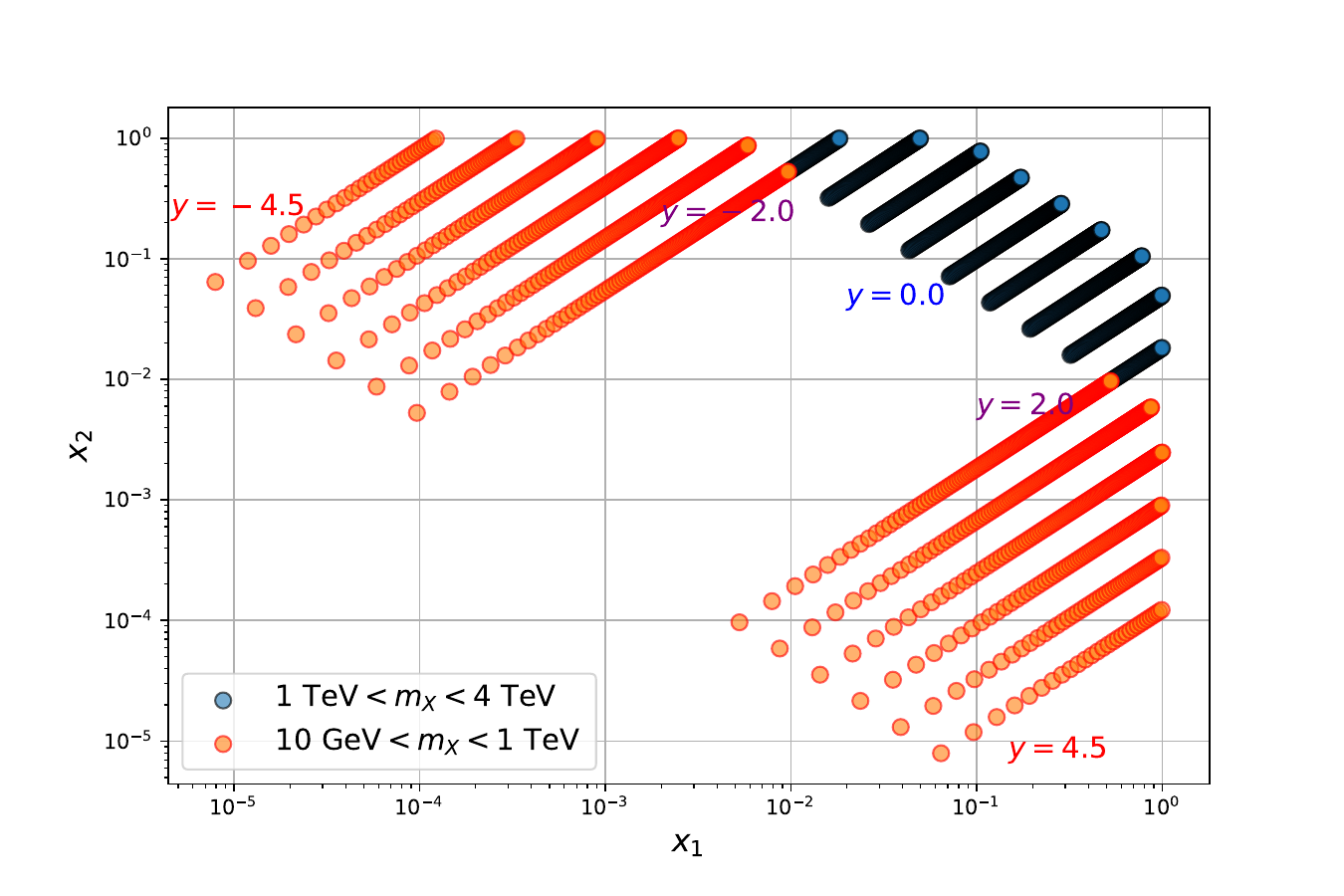}
    \end{center}
	\caption{Leading order kinematic plot of $x_{1,2} = m_ X/\sqrt{s}\,\exp(\pm y)$ in the
large-$M$ and central region, $|y|<2.0,\quad 1\,{\rm TeV}<m_X<4\,{\rm TeV}$, (black dots) and in the low-intermediate-$m_X$ and forward region, $2.0<|y|<4.5\quad 10\,{\rm GeV} < m_X < 1\,{\rm TeV}$ (red dots). Here $\sqrt{s}=14$~TeV.}
        \label{fig:kinplot}
\end{figure}

To visualise more precisely the regions in $x$ that are constrained by a measurement of a given final state at 
the energy $E\sim m_X$ and at a given rapidity $y$, we display in Figure.~\ref{fig:kinplot} the scatter plot for $x_{1,2} = m_X/\sqrt{s}\,\exp(\pm y)$ in the
large-$m_X$ and central region, namely $|y|<2.0$ and $1\,{\rm TeV}<m_X<4\,{\rm TeV}$, and compare it to the low-to-intermediate-$m_X$ 
and forward rapidity region, namely $2.0<|y|<4.5$ and $10\,{\rm GeV} < m_X < 1\,{\rm TeV}$.
We can see that, while the measurements of large-invariant mass objects in the central rapidity region constrain solely the large-$x$ region, and
where both partons carry a fraction $x$ of the proton's momentum in the region $0.01\lesssim x \lesssim 0.8$, the low-to-intermediate invariant
mass region in the forward rapidity region constrains both the small and the large $x$ region, given that at $|y|\approx 4.0$ the $x$-region
probed is around $0.1\lesssim x \lesssim 0.8$ for one parton and around $10^{-5}\lesssim x \lesssim 10^{-4}$ for the other parton. 
Given that the valence quarks are much more abundant 
at large $x$ than the sea quarks, in most collisions the up or down quarks will be the partons carrying a large fraction $x$ of the proton's momentum, 
while the antiquarks will carry a small fraction $x$. Hence, this observable will not be sensitive to the shift in the large-$x$ anti-up and 
anti-down that the global PDF fit yields in order to compensate the effect of NP in the tails. 

\subsection{Sliding ${m_{ll}}$ cut}
\label{subsec:cut}
Another way to potentially disentangle EFT effects in PDF fits was explored in Ref.~\cite{DYpaper} where, in the context of the $\hat{W}$ and $\hat{Y}$ parameters, the authors exploited the energy-growing effects of the EFT operators at high invariant dilepton mass. 
They introduced a ratio evaluation metric $R_{\chi^2}$ (in the notation of the original reference) which described how much the PDF fit quality deteriorated when data-points at high dilepton invariant mass were included in the computation of the $\chi^2$, with respect to a fit that only used low invariant mass bins.
In this way, $R_{\chi^2} \sim 1 $ indicated a fit quality similar to a purely SM scenario, as the sensitivity to energy-growing EFT effects is suppressed. 

They found that, when including higher dilepton invariant mass measurements (where the effect of EFT operators is enhanced) in the computation of the $\chi^2$, the fit quality deteriorated and $R_{\chi^2}$ growth almost monotonically away from $1$. With this metric other sources of PDF deterioration are minimised and the tension arises fundamentally because of the EFT effects.

\begin{figure}[htb]
  \centering
  \includegraphics[width=0.49\linewidth]{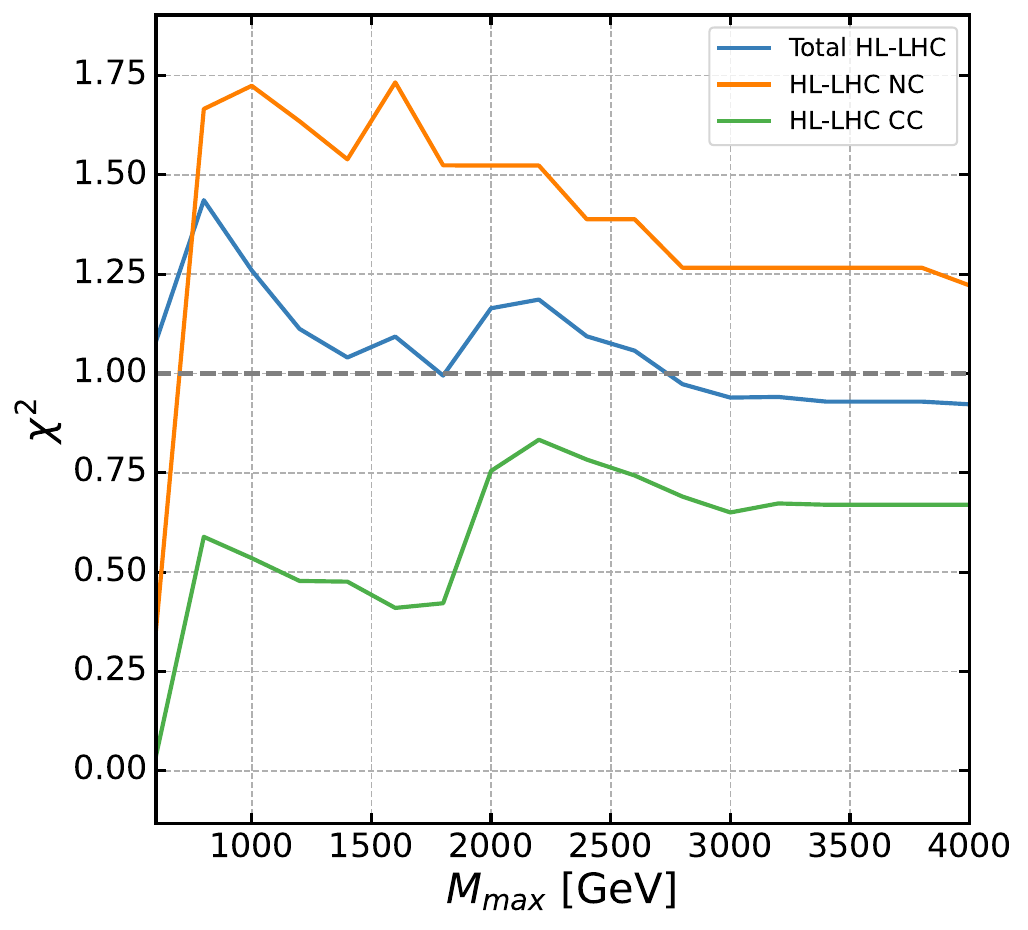}
  \includegraphics[width=0.49\linewidth]{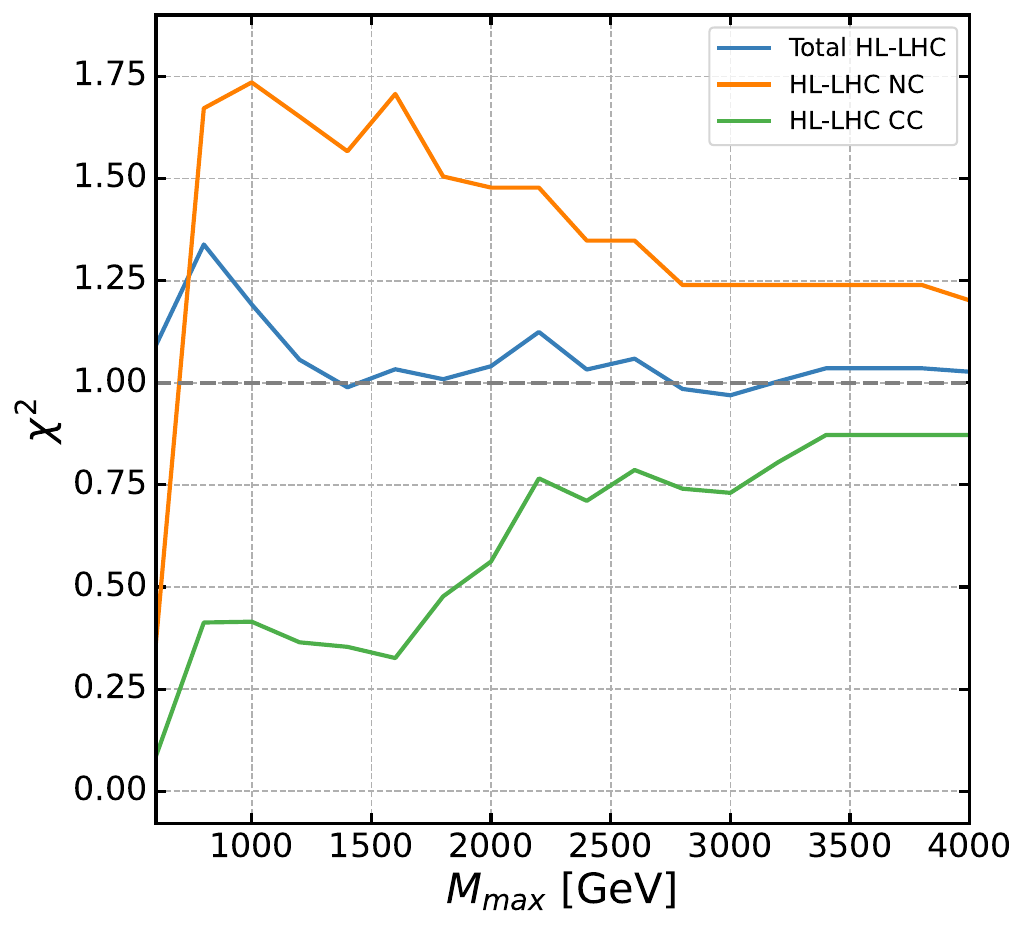}
  \caption{Value of the $\chi^2$ per HL-LHC dataset as a function of the maximum value of the invariant mass allowed in the kinematics $M_{\rm max}$. This parameter is analogous to $m_{ll}^{({\rm max})}$ of Sect. 4 of~\cite{DYpaper} . Left: baseline SM fit. Right: contaminated $\hat{W}=8\cdot 10^{-5}$.}
  \label{fig:sliding_chi2}
\end{figure}

For our study, we show in Fig. ~\ref{fig:sliding_chi2} the $\chi^2$ values for the NC and CC HL-LHC datasets, combining the two lepton channels, in the SM and $\hat{W} = 8 \cdot 10^{-5}$ contaminated cases. $M_{\rm max}$ is the maximum value of the dilepton invariant mass that we include in the computation of the $\chi^2$.
We see that after $M_{\rm max} \sim 2$ TeV, the $\chi^2$ values in both the SM and contaminated scenario stagnate and no persistent deterioration is observed. This means that it is not possible to isolate the EFT effects, which are enhanced at higher masses, in the worsening of the fit quality. In this way, it is not possible to disentangle the EFT effects on the PDF fit and other options have to be explored.

\subsection{Observable ratios}
\label{subsec:ratio}
In order to disentangle PDF contamination, another quantity worth studying is the ratio between observables whose processes have similar parton channels. Indeed, in this case the impact of the PDF is much reduced and any discrepancy between data and theory predictions can be more confidently attributed to new physics in the partonic cross-section. Practically, a deviation would mean that one of the two datasets involved in the ratio is ``contaminated'' by new physics and should therefore be excluded from the PDF fit. 

We have studied the ratio between the number of events in $WW$ production and Neutral Current Drell-Yan (NC DY), as well as between $WH$ production and Charged Current Drell-Yan (CC DY). In each pair both processes are initiated from the same parton channels. 

\begin{figure}[!h]
  \begin{center}
    \includegraphics[width=0.49\linewidth]{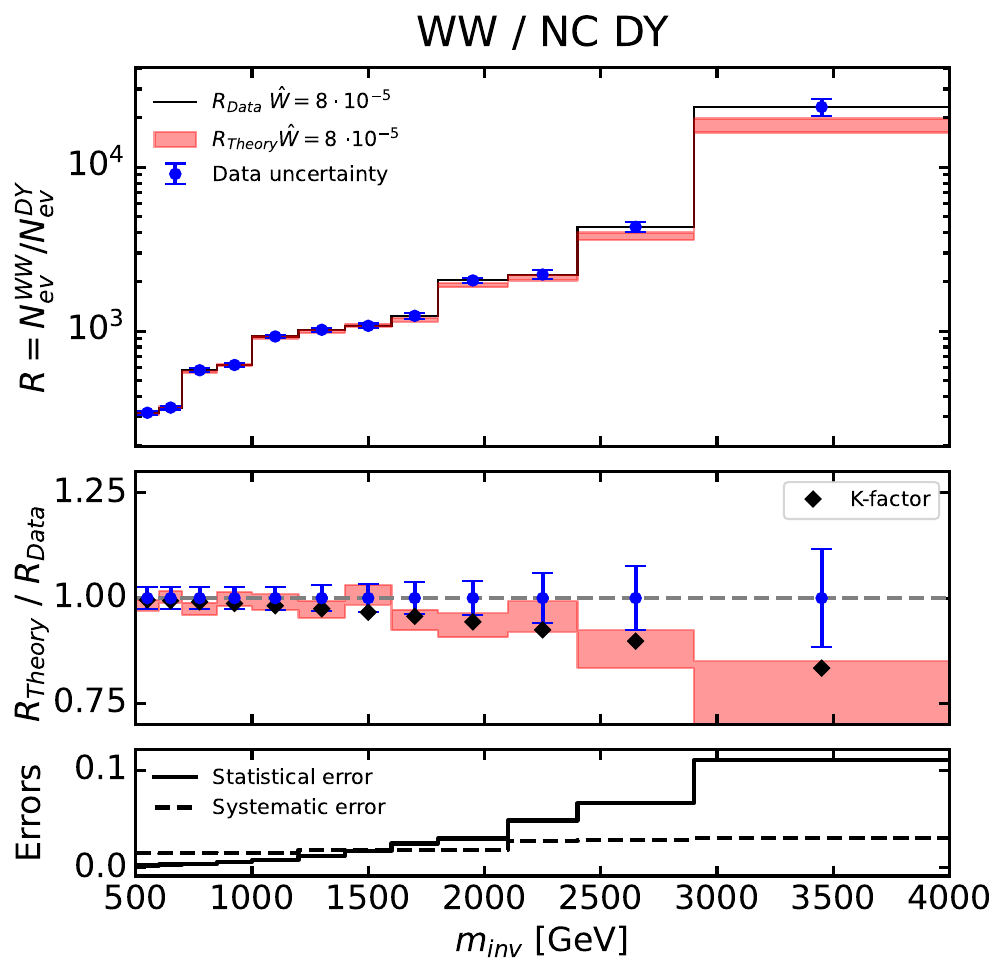}
    \includegraphics[width=0.49\linewidth]{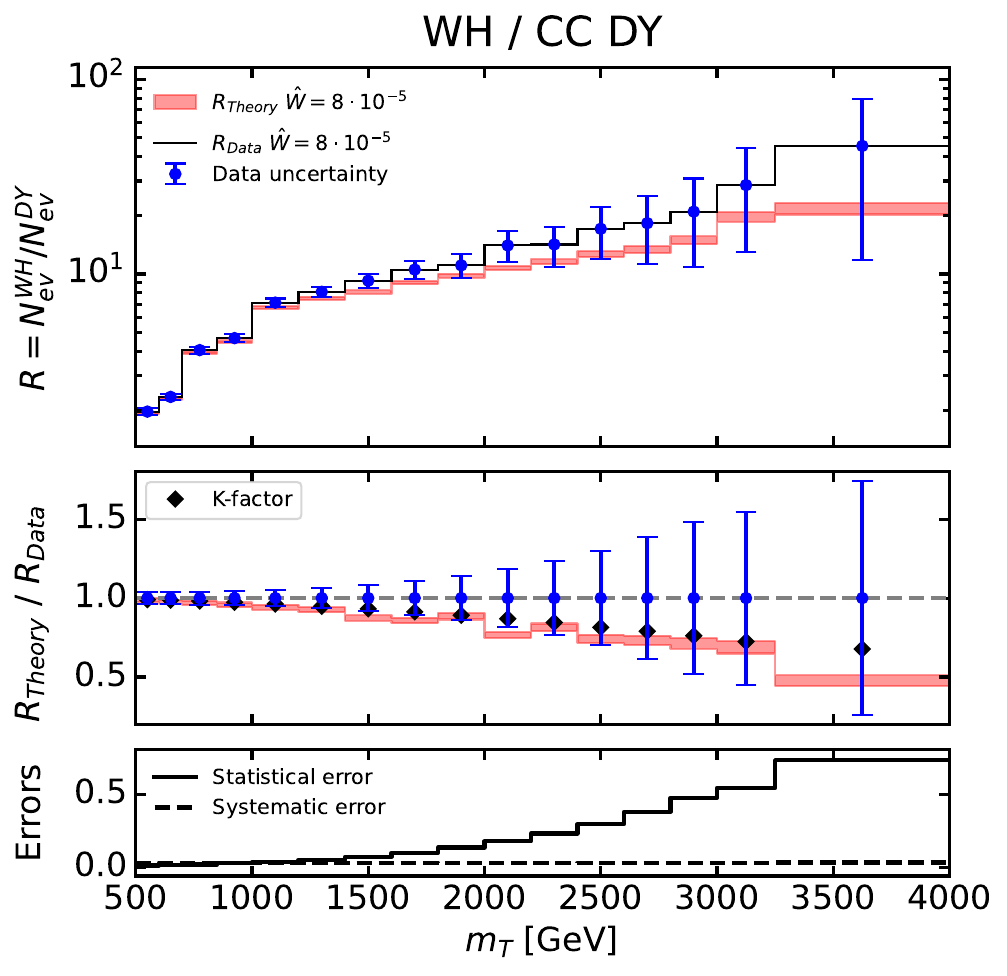}
  \end{center}
	\caption{
    Ratio between diboson production and Drell-Yan processes for the HL-LHC predictions. On the left, we show the ratio of $W^+ W^-$ to NC DY binned in invariant mass and on the right, we show the ratio of $WH$ to CC DY binned in transverse mass.
    In the top panel we plot the ratios of number of events for data and theory predictions.
    In the middle panel, we plot the ratio of those ratios (theory over data) alongside the $K$-factors. The lower panel displays the uncertainties.
  }
  \label{fig:diboson_dy_ratio}
\end{figure}

The Drell-Yan events we use are displayed in Fig. ~\ref{fig:wy_dy_prod}. The diboson events can be seen in Fig.~\ref{fig:ww_hllhc} for $WW$ and in Fig.~\ref{fig:wph_hllhc} for $W^{+}H$. However, note that we also include the $W^{-}H$ channel to measure the ratio of $WH$ and CC DY here. We plot the ratio of those quantities in Fig.~\ref{fig:diboson_dy_ratio}. We compare data and theory predictions where, as in Fig.~\ref{fig:wy_dy_prod}, data corresponds to a baseline PDF and a BSM partonic cross-section ($f_{\rm Baseline} \otimes \hat{\sigma} _{BSM}$) and theory is computed from a contaminated PDF and a SM partonic cross-section ($f_{\rm Cont} \otimes \hat{\sigma}_{SM}$). We also compare those results to $K$-factors which are obtained by taking the ratio of Drell-Yan BSM predictions over the SM ones. Practically the $K$-factors are a ratio of their respective partonic cross-section ($K = \hat{\sigma}^{DY} _{BSM} / \hat{\sigma}^{DY} _{SM}$).

We see in both cases a deviation between data and theory predictions growing with the energy. The uncertainties are smaller in the ratio $\text{WW} / \text{NC DY}$, which allows the discrepancy to be over $1 \sigma$ in the last bin. Furthermore, we also witness that the deviation follows the $K$-factors, which reinforces our initial assumption that using ratios greatly diminishes the impact of the PDFs.

As we mentioned earlier, the lesson we can get from this plot is that there is some new physics in either the DY or the diboson datasets. Unfortunately, without further information it is not possible to identify in which of those datasets the new physics is. Therefore, with just this plot in hand, the only reasonable decision would be to exclude the two datasets involved in the ratio where the deviation is observed from the fit. The downside of this disentangling method is that it might worsen the overall quality of the fit and increase the PDF uncertainties in certain regions of the parameter space. However, it proves to be an efficient solution against the sort of contamination we studied. Indeed, by excluding the DY datasets in this case, one would exclude the contamination we manually introduced there from the PDF fit.

\subsection{Alternative constraints on large-$x$ antiquarks}
\label{subsec:lowE}

In Sect.~\ref{subsec:forward}, it was shown that the inclusion of precise on-shell forward $W$ and $Z$ 
production measurements does not disentangle the contamination that new physics in the high-energy tails 
might yield. In this section, we ask ourselves whether there are any other future low-energy observables that 
might constrain large-$x$ antiquarks and show tension with the high-energy data in case the latter are 
affected by NP-induced incompatibilities.

\begin{figure}[t!]
  \begin{center}
    \includegraphics[width=0.49\linewidth]{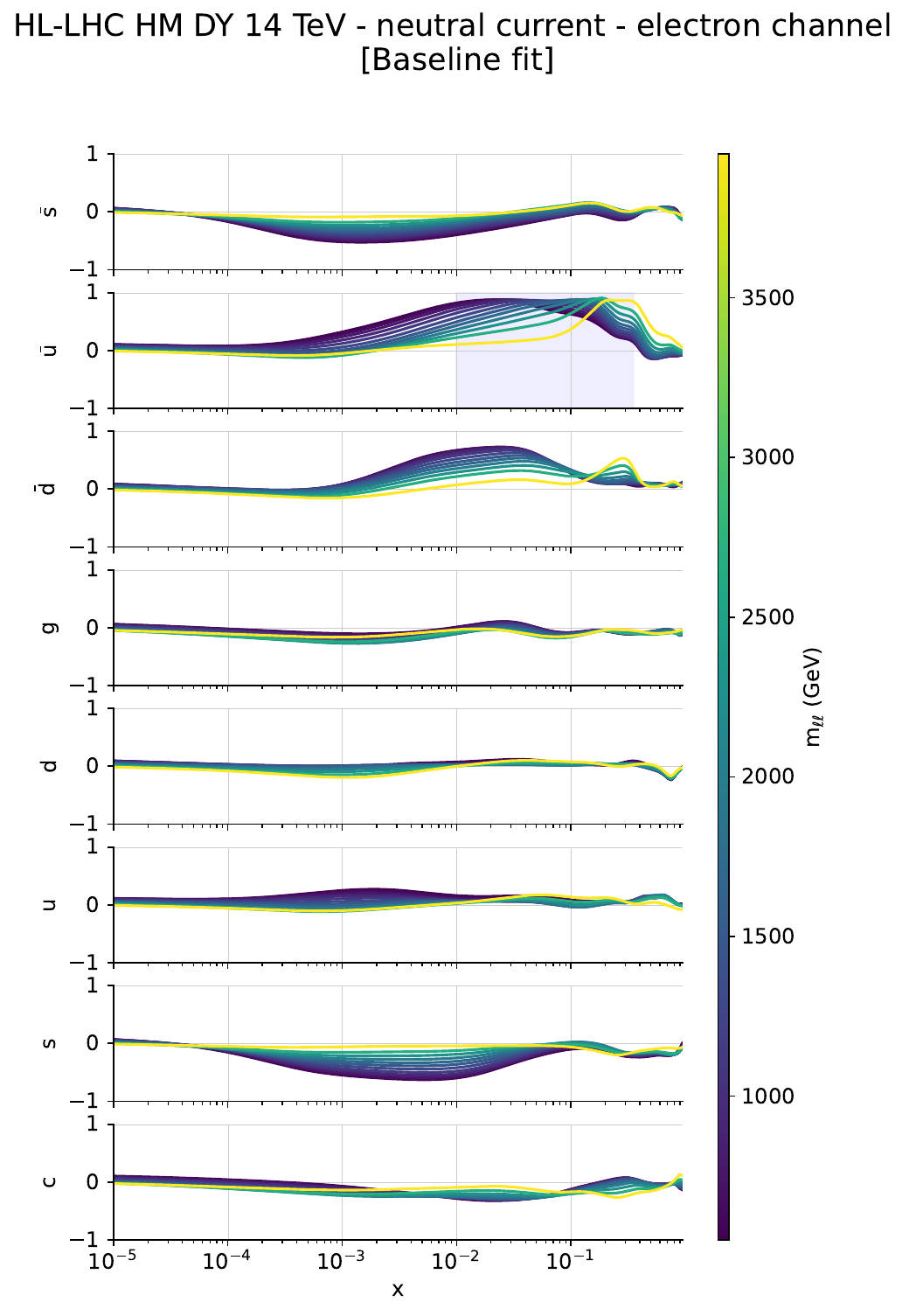}
    \includegraphics[width=0.49\linewidth]{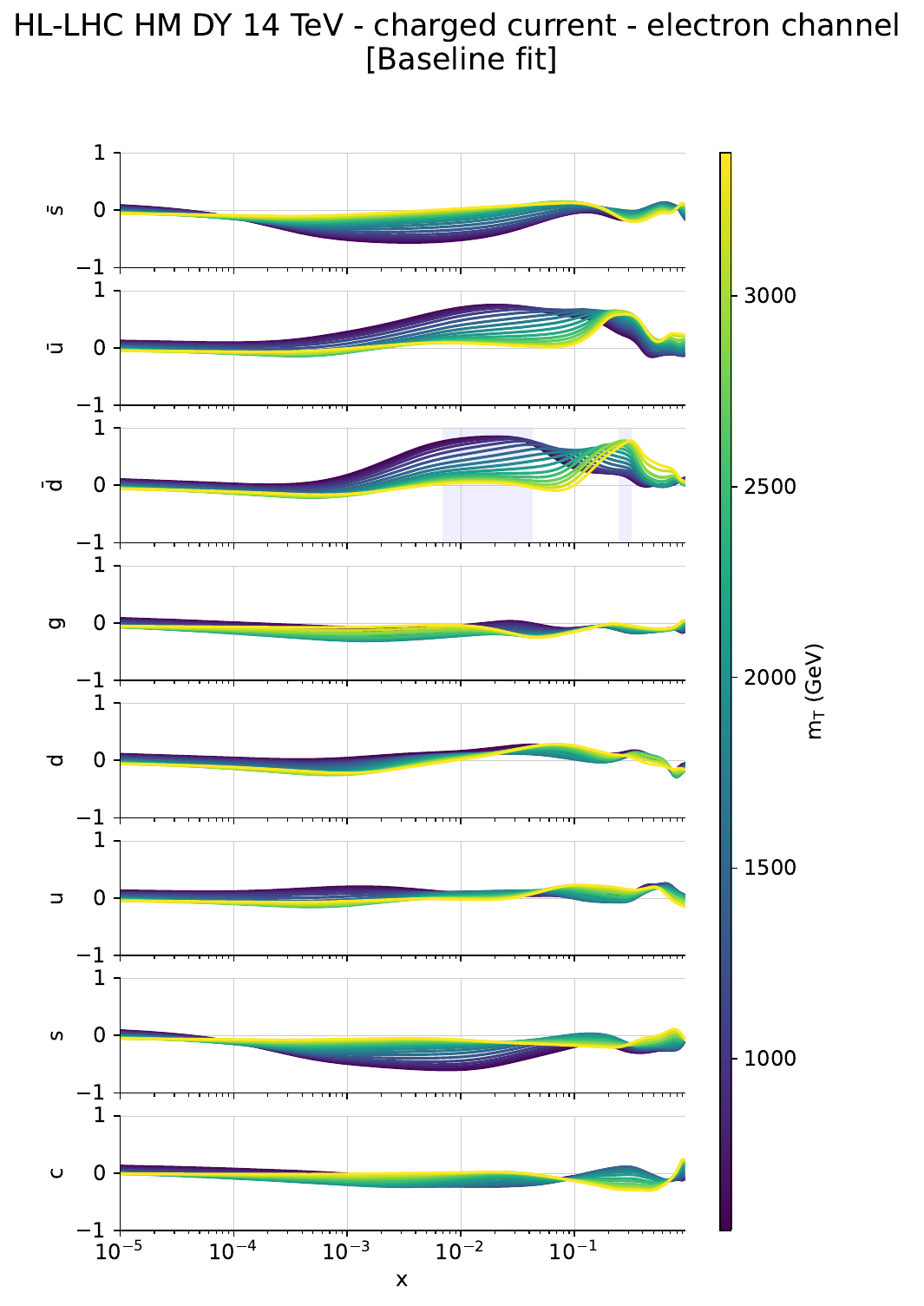}
    \end{center}
	\caption{\label{fig:smpdfplots_hl}Correlation coefficient $\rho$ defined in Eq.~\eqref{eq:correlations} between the flavour PDFs of the baseline  
        set and the HL-LHC neutral current Drell-Yan data (left panel); the HL-LHC charged current Drell-Yan data (right panel). The highlighted 
        region corresponds to $\rho>0.9\,\rho_{\rm max}$.}
\end{figure}

\begin{figure}[htb]
  \begin{center}
    \includegraphics[width=0.3\linewidth]{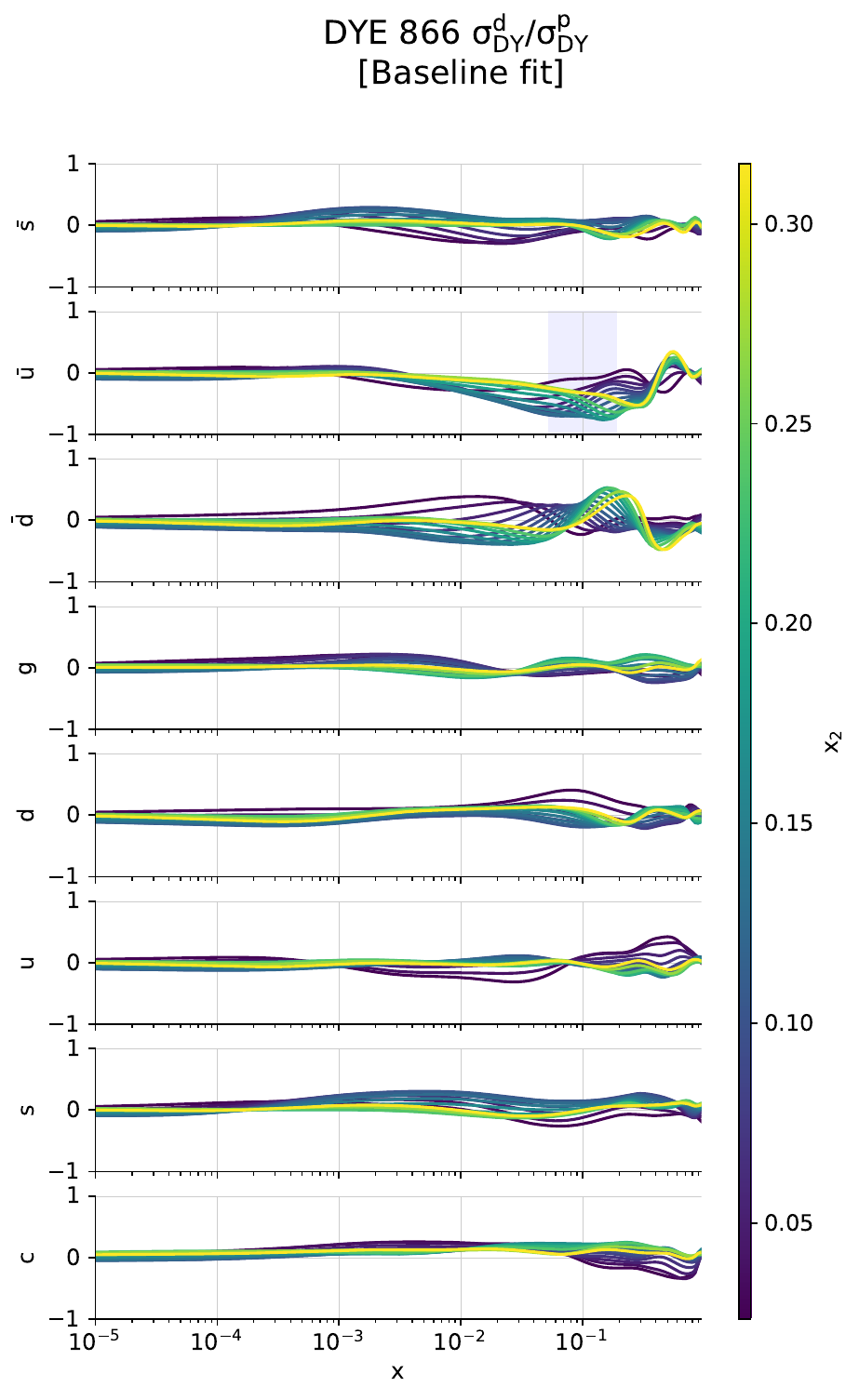}
    \includegraphics[width=0.3\linewidth]{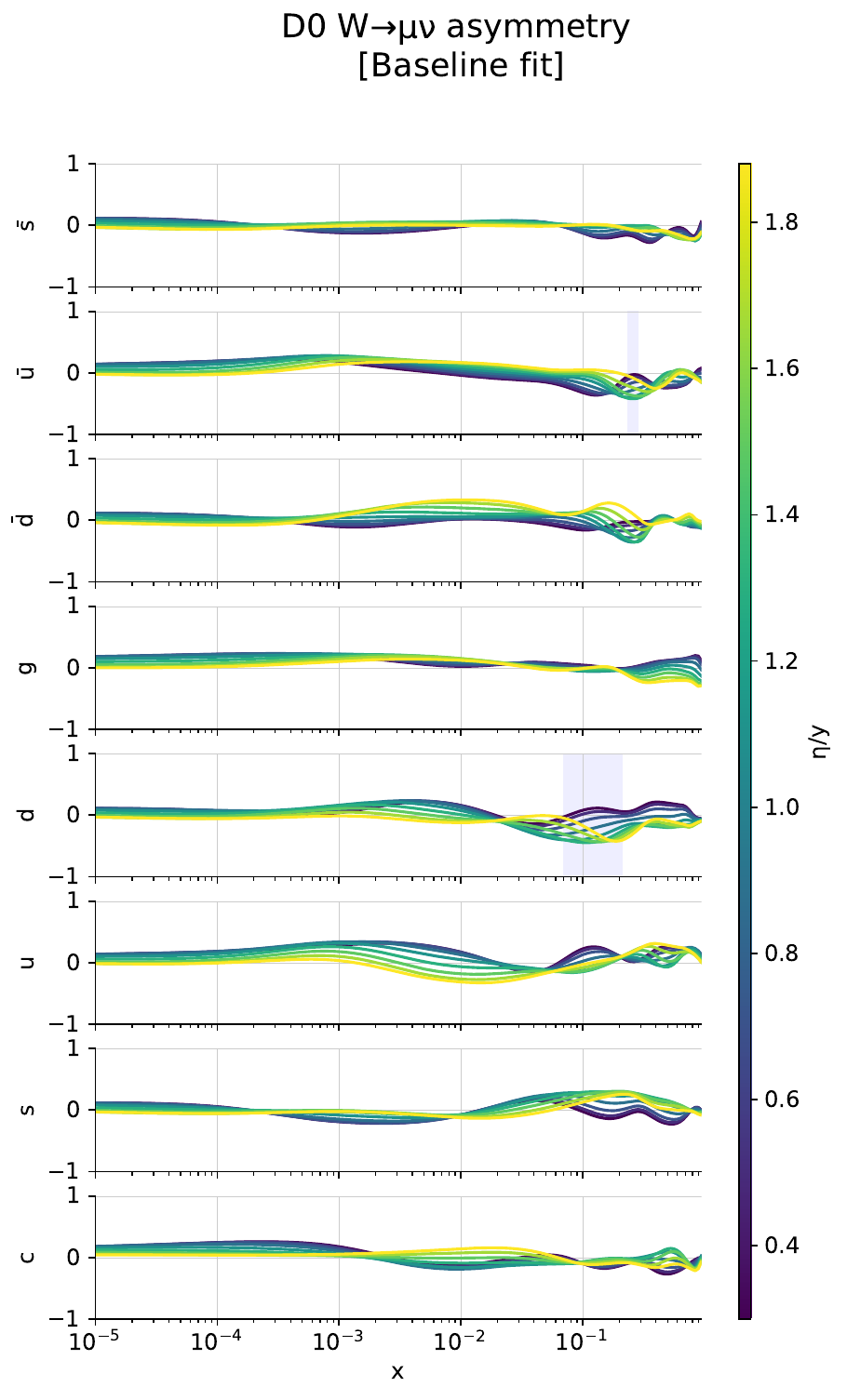}
    \includegraphics[width=0.3\linewidth]{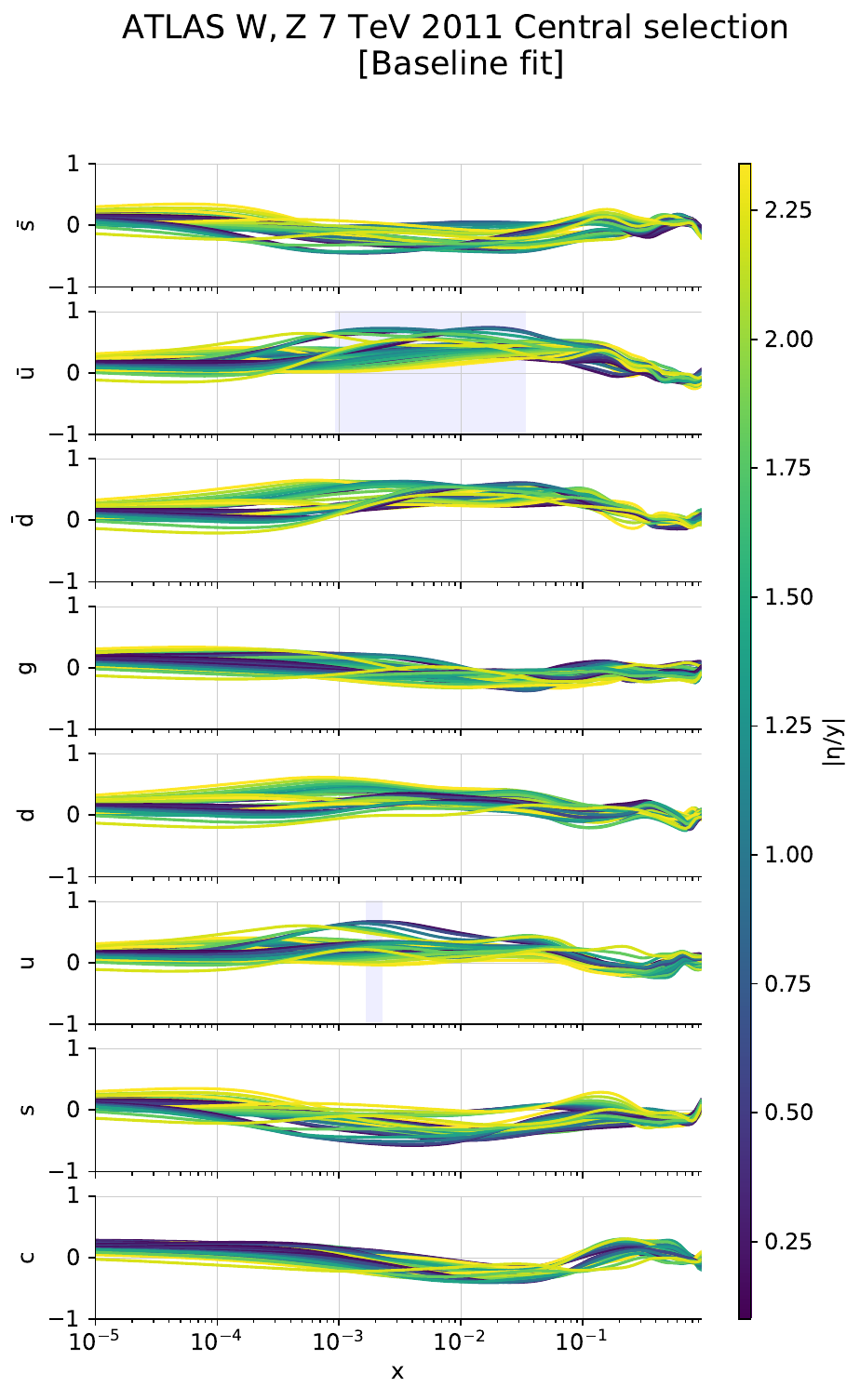}
    \end{center}
	\caption{\label{fig:smpdfplots} Same as Fig.~\ref{fig:smpdfplots_hl} for the FNAL E866 data measuring the ratio between low energy Drell-Yan muon pair production on 
 proton and deuteron targets~\cite{NuSea:2001idv} (left panel), the Tevatron D0 muon charge asymmetry~\cite{D0:2013xqc} (central panel) and the ATLAS 
      measurements of the $W$ and $Z$ differential cross-section at $\sqrt{s}=7$ TeV in the central rapidity region\cite{ATLAS:2016nqi} (right panel).}
\end{figure}

We start by looking at the correlation between the data that are currently included in our baseline PDF fit 
and the various PDF flavours. To assess the level of correlation, we plot the correlation defined in Ref.~\cite{Carrazza:2016htc}. 
The correlation function is defined as:
\begin{equation}
\label{eq:correlations}
    \rho(j,x,{\cal O})\equiv\frac{N_{\rm rep}}{N_{\rm rep}-1}
    \left(\frac{\langle f_j(x,Q)\,{\cal O}\rangle_{\rm reps}-\langle f_j(x,Q)\rangle_{\rm reps}\langle {\cal O}\rangle_{\rm reps}}{\Delta_{\rm PDF}f(x,Q)\,\Delta_{\rm PDF}{\cal O}}\right),
\end{equation}
where the PDFs are evaluated at a given scale $Q$ and the observable ${\cal O}$ is computed with the set of PDFs $f$, $j$ is the PDF 
flavour, $N_{\rm rep}$ is the number of replicas in the baseline PDF set and $\Delta_{\rm PDF}$ are the PDF uncertainties. 
In Figs.~\ref{fig:smpdfplots_hl} and~\ref{fig:smpdfplots} we show the correlation between the PDFs in the flavour basis 
and the observables which are strongly correlated with the antiquark distributions. 
The region highlighted in blue is the region in $x$ such that the correlation 
coefficient defined in Eq.~\eqref{eq:correlations} is larger than $0.9 \, \rho_{\rm max}$, where $\rho_{\rm max}$ is 
the maximum value that the correlation coefficient takes over the grid of points in $x$ and over the flavours $j$.
From Fig.~\ref{fig:smpdfplots_hl} we observe that while the largest invariant mass bins of 
the HL-LHC NC are most strongly correlated with the up antiquark distribution in the $10^{-2}\lesssim x \lesssim 3\cdot 10^{-1}$ region, 
the HL-LHC CC, particularly the lowest invariant mass bins, are most strongly correlated with the down antiquark distribution in 
the $ 7\cdot 10^{-3}\lesssim x \lesssim 5\cdot 10^{-2}$ region. This observation is quite interesting as it gives us a further insight into the 
difference between the $Z'$ and the $W'$ scenarios discussed at the end of Sect.~4.2. Indeed the $W'$ scenario affecting both the 
NC and CC distributions manages to compensate the $\bar{u}$ shift with the $\bar{d}$ shift in a slightly smaller region, hence the 
successful contamination. 

We now ask ourselves whether there are other observables that display a similar correlation pattern with the light antiquark distributions. 
In Fig.~\ref{fig:smpdfplots} we show the three most interesting showcases. In the left panel, we see that the FNAL E866/NuSea measurements 
of the Drell-Yan muon pair production cross section from an 800 \rm{GeV} proton beam incident on proton and deuterium targets~\cite{NuSea:2001idv} 
yields constraints on the ratio of anti-down to anti-up quark distributions in the proton in the large Bjorken-$x$ region, and the correlation is 
particularly strong with the anti-up in the $5\cdot 10^{-2}\lesssim x \lesssim 3\cdot 10^{-1}$ region.
The central panel shows that the Tevatron D0 muon charge asymmetry~\cite{D0:2013xqc} exhibits a strong correlation with the 
up antiquark around $x\approx 0.3$ and the down quark around $x\approx 0.1$. This is understood, as by charge conjugation the 
anti-up distribution of the proton corresponds to the up distribution of the anti-proton. Finally, on the right 
panel we see that the precise ATLAS measurements of the $W$ and $Z$ differential cross-section at $\sqrt{s}=7$ TeV~\cite{ATLAS:2016nqi} 
have a strong constraining power on the up antiquark in a slightly lower $x$ region around $3\cdot 10^{-3}\lesssim x \lesssim 2\cdot 10^{-2}$.

The results presented in Sect.~\ref{sec:dy} show that the tension with the low-energy datasets that 
constrain the same region in $x$ as the high-mass Drell-Yan HL-LHC data is not strong enough to flag the 
HL-LHC datasets. Hence, the conditions highlighted in Sect.~\ref{subsec:postfit} are necessary in order to determine a bulk of maximally
consistent datasets, but they are not sufficient, as they still allow
new physics contamination to go undetected. 
A way to emphasise the tension is to produce {\it weighted fits} which give a larger weight to the high-energy data that are affected by 
new physics effects. The rationale behind this is that, if some energy-growing
effect associated to the presence of new physics in the data shows up in the tails
of the distributions, PDFs might accommodate this effect without deteriorating the agreement with the other datasets 
only up to a point. If the datasets that are affected by new
physics are given a large weight in the fit, the tension with the datasets constraining the large-$x$ region 
that are perfectly described by the SM could in principle get worse. Hence by giving a
larger weight to a specific NP-affected dataset, the disagreement with the datasets highlighted 
above should become more apparent. 
Depending on the kind of new physics model that Nature might display in the data, 
the effect might affect either of the three
classes of the high energy processes entering PDF fits, namely: (i)
jets and dijets, (ii) top and (iii) Drell-Yan. In our example, in order to emphasise the tension with the 
low-energy Drell-Yan data and the Tevatron data, we would have to give more weight to the HL-LHC high-mass 
Drell Yan data. However, performing this exercise we observe that, although the $\chi^2$ of the HL-LHC Drell-Yan further improves and the 
ones of the highlighted data deteriorates, the level of deterioration is never strong enough to flag the tension. 
The result of this test points to the fact that one should include independent and more precise 
low-energy/large-$x$ constraints in future PDF analyses\footnote{Note that some of this data have the disadvantage of being affected by 
additional uncertainties associated with nuclear corrections, target mass corrections, higher twists and other low-energy effects.}
if one wants to safely exploit the constraining power of high-energy 
data without inadvertently absorbing signs of new physics 
in the high-energy tails. In this sense the EIC programme~\cite{Khalek:2021ulf,Abir:2023fpo}, as well as 
other low-energy data which are not exploited in the standard PDF global fits, such 
as JLAB~\cite{Accardi:2023chb,Accardi:2021ysh} or STAR and SeaQuest measurements~\cite{Accardi:2023gyr,Guzzi:2021fre,Alekhin:2023uqx}, will be a precious 
input in future PDF analyses, alongside the constraints from lattice data~\cite{Hou:2022onq}. 



\section{Conclusion}
\label{sec:summary_sd}

In this chapter we have explored the potential corrections to the DGLAP evolution equations of PDFs that can arise from higher dimensional operators in the SMEFT. We have focused on the leading order splitting functions and on the running of the strong coupling, with the aim of determining whether SMEFT effects can modify the PDF evolution in a significant manner.

Regarding the splitting functions, our primary focus was on the CP-even dimension-6 operators that can modify the quark-gluon vertex and 3-gluon vertex at tree level. Our analysis revealed that SMEFT contributions, which are kinematically enhanced by powers of transverse momentum, do not alter the collinear limit of the splitting functions. This finding is consistent with the Appelquist-Carazzone theorem, which sustains that the infrared behaviour of the theory remains unaffected by integrating out heavy degrees of freedom.

We also studied the potential effects of the SMEFT on the running of the strong coupling, determined by its $\beta$ function. To calculate the running of $\as$, we renormalised the bare theory, which involved considering quantum corrections to the propagators of gluons and quarks, as well as to the quark-gluon vertex. The divergences that emerge in loop integrals are absorbed by counterterms, which are subsequently used to calculate the running of the parameters of the theory, including $\as$. We carried out this process in the presence of the $O_{HG}$ SMEFT operator. While the quark-gluon vertex and the quark self-energy counterterms were not affected by the SMEFT operator, the gluon self-energy counterterm was modified by the presence of the $O_{HG}$. From our calculations, we derived the modified $\beta$ function with SMEFT corrections. We found that while SMEFT contributions modify the gluon self-energy, the overall impact on the $\beta$ function is minimal at current bounds for the relevant Wilson coefficient. This result suggests that SMEFT corrections to the running of $\as$ are small compared to even higher-order SM contributions, and its effect is negligible in a phenomenological context.

Under these conditions, SMEFT effects on the DGLAP evolution of PDFs can be safely ignored, as the leading order splitting functions are not modified by SMEFT operators and, while the running of $\as$ is affected by SMEFT contributions, its impact is minimal. This work provides a foundation for further investigations into the collective impact of all SMEFT operators on the running of $\as$ and other SM parameters. Several additional directions can be taken for the future publication of this work, including the study of the SMEFT on the DGLAP evolution equations at higher orders closer to the current state of the art, by including SMEFT operators at higher dimensions, higher loops in the running of the strong coupling, and an assessment of the calculation in the broken phase to consider quark mass effects, as studies for a single SMEFT operator in \cite{BuarqueFranzosi:2015jrv}.



\chapter{SMEFT effects in PDF evolution}
\label{ch:smeft_dglap}
  \epigraph{Non hominibus tantum, sed rebus persona demenda est et reddenda facies sua. \\ \vspace{1em} Not only from people, but also from things, the mask must be removed and their true face revealed.}{\textit{\\ from Moral Letters to Lucilius, \\ by Seneca.}}

\textit{This chapter is based on a forthcoming publication in collaboration with L. Mantani and M. Ubiali. My contributions
include the original idea, the calculation of the SMEFT splitting functions via on-shell methods (to be compared with L. Mantani's results from helicity amplitudes),
and the calculation of the SMEFT corrections to the running of the strong coupling constant in different setups.} 

\vspace{1em} 
\noindent\rule{\textwidth}{0.4pt}
\vspace{1em} 

This chapter focuses specifically on the potential impact of higher dimensional operators in the SMEFT on the DGLAP equations, which govern the scale dependence of the PDFs.
In this study, we discuss thoroughly for the first time whether the SMEFT affects the DGLAP evolution equations by focussing on 
the leading order splitting functions and on the running of the strong coupling constant.
We calculate the relevant matrix elements in the collinear limit and show how the kinematic enhancement
of the SMEFT operators does not alter them.
We also assess the impact of SMEFT corrections in the running of the strong coupling and
verify that it is highly suppressed.

\section{SMEFT corrections to collinear splitting functions}
\label{sec:pij}
The splitting functions are a key ingredient in the DGLAP evolution equations. In this section, we will address the leading order splitting functions at tree level, and we will study the potential effects of (CP-even) dimension-6 SMEFT operators on them. As described in Eq. (\ref{eq:split_all_ord}), the splitting functions can be expressed as an expansion in the strong coupling $\as$. In this chapter, we retain the leading contributions, and drop the superscript $(0)$ to reduce the clutter. At this order, the splitting functions at this order are just a function of $z$, and not of $\as$.

To parametrise SMEFT corrections in this chapter, we will work with the basis of operators used by the SMEFTatNLO framework \cite{Degrande:2020evl}, which is very similar to the Warsaw basis \cite{Grzadkowski:2010es}. We do this to have consistent definitions with its UFO implementation, which has been widely used in phenomenological studies. Since we are interested in SMEFT corrections to these QCD splitting functions, we need to consider only the operators that modify the 3-gluon and quark-gluon vertices, which reduces the number of operators to consider when compared to the whole basis. In this case, all the relevant SMEFT contributions must have at least one gluon field (encoded in one QCD field strength tensor, in this basis). 

Let us now discuss the relevant vertices that define the splitting functions: the 3-gluon vertex and the quark-gluon vertex. 

\textbf{3-gluon vertex.} The corrections at tree level to the 3-gluon vertex that define $P_{gg}(z)$ come from the operator
\begin{equation}
\label{eq:OG}
\mathcal{O}_{G} = g_s f^{ABC} G^{A\nu}_{\mu} G^{B\rho}_{\nu} G^{C\mu}_{\rho},
\end{equation}
where $g_s$ is the strong coupling constant, $f^{ABC}$ is the $SU(3)$ structure constant, and $G^{A}_{\mu \nu}$ is the gluon field strength tensor
\begin{equation*}
    G^{A}_{\mu \nu} = \partial_{\mu} G_{\nu}^{A} - \partial_{\nu} G_{\mu}^{A} - g_s f^{ABC} G_{\mu}^{B} G^{C}_{\nu}.
\end{equation*}
Notice the extra factor of $g_s$ in the definition of the operator in Eq. (\ref{eq:OG}) when compared to the original definition in the Warsaw basis. This is just a matter of convention, and it is done to simplify the comparison with existing phenomenological studies.

The corrections to the quark-gluon vertex that defines $P_{qq}(z)$, $P_{qg}(z)$, and $P_{gq}(z)$ come from 

\begin{equation}
    \label{eq:def_OqG}
    \begin{aligned}
        \mathcal{O}_{uG} &= g_s \overline{q} \sigma^{\mu \nu} T^A u \Tilde{H} G_{\mu \nu}^{A}, \\
        \mathcal{O}_{dG} &= g_s \overline{q} \sigma^{\mu \nu} T^A d H G_{\mu \nu}^{A},
    \end{aligned}
\end{equation}

where $q$ is the left-handed quark doublet, $\sigma^{\mu \nu} = \frac{i}{2} [ \gamma^\mu, \gamma^\nu ]$, $T^{A} = \frac{1}{2} \lambda^{A}$
are the $SU(3)$ generators, $u$ and $d$ are respectively the right-handed up-type and down-type quark singlets, and $\Tilde{H}_{i} = \epsilon_{ij} (H_{j})^*$, with $\epsilon_{12} = + 1$ being totally antisymmetric. 

We now study if the inclusion of these operators can affect the splitting functions. In what follows, it will be useful to define an expression
to account for the matrix elements and phase space factors of Eq.(\ref{eq:def_split}) before taking the collinear limit,
\begin{equation}
    \label{eq:coll_limit}
    \hat{P}_{BA} \equiv \frac{1}{2} z (1-z) \overline{\sum_{\operatorname{pols.}}} \sum_{\operatorname{cols.}}\frac{|\mathcal{M}_{A \rightarrow B + C}|^2}{p_T^2},
\end{equation}
where the parton $B$ is radiated from the parton $A$, and such that $P_{AB} = \lim_{p_T  \to  0} \hat{P}_{AB}\footnote{Note that the 'hat' above the splitting function does \textit{not} denote in this case the splitting function before dealing with the soft and collinear singularity, as it is sometimes written in the literature.}$.
We now study the potential corrections to the $P_{qq}$ splitting function. The relevant SMEFT operators are the ones of Eq. (\ref{eq:def_OqG}) and, after computing the matrix elements, we find
\begin{equation} \label{eq:P_hat_qq}
  \hat{P}_{qq}(z) = \frac{\as}{2 \pi} \left( C_F \left( \frac{1+z^2}{1-z} \right) + 4 \left( \frac{c_{uG}}{\Lambda^2} \right)^2 p_T^2 v^2 \frac{1}{(1-z)} \right),
\end{equation}
Eq. (\ref{eq:P_hat_qq}) applies for both up-type and down-type quarks. We see that the linear interference with the SM amplitude vanishes, as the
dipole operator contributes only to helicity configurations that are vanishing in the SM. This is due to the fact that the QCD vertex in the SM
preserves the fermion helicity, while the dipole operator introduces an interaction that causes an helicity flip, much like a mass term. Since we
are dealing with massless partons, only the quadratic SMEFT contribution, kinematically enhanced by $p_T^2$, survives. At this point we see that the matrix element of this 3-point function changes. However, this is not enough to modify the collinear structure of the theory. Indeed, when we take the collinear
limit of Eq. (\ref{eq:coll_limit}) we obtain
\begin{equation*}
    \lim_{p_T \to 0} \hat{P}_{\operatorname{qq}}(z) = \frac{\as}{2 \pi} C_F \frac{1+z^2}{1-z},
\end{equation*}
with no contributions from SMEFT Wilson coefficients. The expression above translates into the well known SM splitting function after regularising the soft singularity with the $+$ prescription. We have found that SMEFT contributions, which are kinematically enhanced by powers of $p_T$, vanish in the collinear limit, as expected from the Appelquist-Carazzone theorem \cite{Appelquist:1974tg}. This is a general feature of the theory that persists in the other splitting functions.

For the $P_{qg}(z)$ and $P_{gq}(z)$ splitting functions we obtain
\begin{equation}
\label{eq:P_hat_qg}
\begin{aligned}
    \hat{P}_{qg}(z) = \frac{\as}{2 \pi} \left( C_F \left( (1-z)^2 + z^2 \right) + \frac{p_T^2}{p^2 (1-z)^2}  +  2 \left( \frac{c_{uG}}{\Lambda^2} \right)^2 p_T^2 v^2  \right), \\
    \hat{P}_{gq}(z) = \frac{\as}{2 \pi} \left( C_F \left( \frac{(1-z)^2 + 1}{z} \right) + \frac{p_T^2}{3 p^2 (1-z)^2}  +  2 \left( \frac{c_{uG}}{\Lambda^2} \right)^2 \frac{1}{z} p_T^2 v^2  \right), 
\end{aligned}
\end{equation}
where we recover the SM expressions with contributions from SMEFT operators. In both cases the kinematic enhancement of the SMEFT contributions keeps the collinear limit unaltered
\begin{equation*}
    \begin{aligned}
    \lim_{p_T \to 0} \hat{P}_{\operatorname{qg}}(z) = \frac{\as}{2 \pi} C_F \left( (1-z)^2 + z^2 \right), \\
    \lim_{p_T \to 0} \hat{P}_{\operatorname{gq}}(z) = \frac{\as}{2 \pi} C_F \left( \frac{(1-z)^2 + 1}{z} \right),
    \end{aligned}
\end{equation*}
and the SM splitting functions are recovered.

Now, we turn our attention to the 3-gluon vertex, which is used to calculate the $P_{gg}(z)$ splitting function. This time, the relevant operator is
$\mathcal{O}_{G}$, defined in Eqs. (\ref{eq:OG}). We obtain 
\begin{equation} \label{eq:P_hat_gg}
  \hat{P}_{gg}(z) = 2 C_A \left( \frac{z}{(1-z)} + \frac{1-z}{z} + z(1-z) - 3 p_T^2 \frac{1-z}{z} \frac{c_G}{\Lambda^2}\right),
\end{equation}
for which we also obtain
\begin{equation*}
    \lim_{p_T \to 0} \hat{P}_{\operatorname{gg}}(z) = 2 C_A \left( \frac{z}{(1-z)} + \frac{1-z}{z} + z(1-z) \right).
\end{equation*}
In general, we find that the kinematic enhancement of the SMEFT operators by extra powers of transverse momentum $p_T$ keeps the collinear limit unaltered by
SMEFT corrections. This is consistent with the Appelquist-Carazzone theorem \cite{Appelquist:1974tg}, which states that the IR behaviour of the theory is not modified by integrating out heavy degrees of freedom.

Now, we turn our attention to the potential SMEFT modifications in the running of the strong coupling $\as$.

\section{SMEFT corrections to the running of $\alpha_{S}$}
\label{sec:as}

\begin{figure}[ht]
    \centering
    \begin{subfigure}[b]{0.3\textwidth} 
        \centering
        \includegraphics[width=\textwidth]{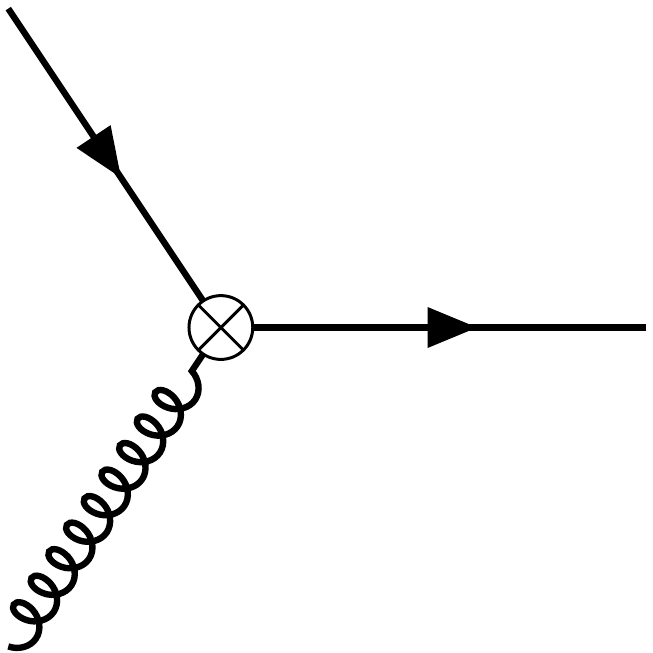} 
        \caption{Counterterm for the quark-gluon vertex.}
        \label{fig:d1_ct}
    \end{subfigure}
    \hfill 
    \begin{subfigure}[b]{0.3\textwidth} 
        \centering
        \includegraphics[width=\textwidth]{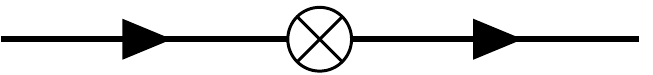} 
        \vspace{1.63cm}
        \caption{Counterterm for the self-energy of the quark.}
        \label{fig:d2_ct}
    \end{subfigure}
    \hfill 
    \begin{subfigure}[b]{0.3\textwidth} 
        \centering
        \includegraphics[width=\textwidth]{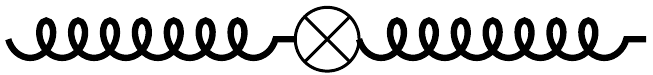} 
        \vspace{1.63cm}
        \caption{Counterterm for the self-energy of the gluon.}
        \label{fig:d3_ct}
    \end{subfigure}
    \caption{Counterterms required to obtain the running of $\as$.}
    \label{fig:cts}
\end{figure}

The DGLAP evolution equations depend on the strong coupling $\as$, which follows the running described in Eqs. (\ref{eq:beta1}), (\ref{eq:beta}), and (\ref{eq:beta3}). To obtain the RGE of $\as$, we have to renormalise the bare theory by considering quantum corrections to the propagator of the gluon and the quarks, and loop amplitudes in the theory. The divergences that appear in the loop integrals have to be absorbed by counterterms, which are then used to calculate the running of the parameters of the theory and, in particular, of $\as$. To obtain the running of $\as$ we have to calculate divergences in three types of diagrams: the quark self-energy, the gluon self-energy, and the quark-gluon vertex. Their associated counterterms are shown in Fig. \ref{fig:cts}.
In what follows, we work in the $\overline{\text{MS}}$ scheme and, following the usual conventions, we define the counterterm of the quark-gluon vertex represented in Fig. \ref{fig:cts} (a) to be
\begin{equation}
    \label{eq:ct1}
    i g_s \gamma^ \mu T^a \delta_{1}.
\end{equation}
Additionally, the self-energy wavefunction counterterm for a quark with momentum $p$ shown in Fig. \ref{fig:cts} (b) is
\begin{equation}
    \label{eq:ct2}
    i \delta_{2} \gamma^\mu p_\mu,
\end{equation}
and the gluon self-energy counterterm in Fig. \ref{fig:cts} (c), for a gluon with momentum $p$, is
\begin{equation}
    \label{eq:ct3}
    i \delta_3 \delta^{ab} (p^\mu p ^\nu - p^2 g^{\mu \nu}),
\end{equation}
where $a$ and $b$ are gluon indices. The inclusion of higher dimensional operator in the theory can change the divergences that have to be absorbed by the counterterms and, 
therefore, they can depend on Wilson coefficients. In what follows, we will study the effects of SMEFT operators in the running of $\as$ by calculating
the counterterms that appear in the diagrams of Fig. \ref{fig:cts} when higher-dimensional operators are included in the theory.

\begin{figure}[ht]
    \centering
    \includegraphics[width=0.3\textwidth]{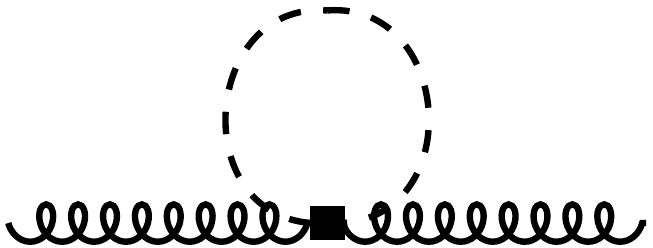}
    \caption{Example of $O_{HG}$ contribution in the self-energy of the gluon via a closed scalar loop.}
    \label{fig:OHG_diag}
\end{figure}

In this section we will work in the theory with manifest $\text{SU}(2) \times \text{U}(1)$ symmetry, namely the unbroken phase. In this approach, the SMEFT contributions to the running of $\as$, at dimension-6, are limited. The argument stems from dimensional consistency \cite{Jenkins:2013zja,Trott:2023jrw}, and the available energy scales in the theory.
The corrections from higher dimensional operators to the running of the strong coupling must all be proportional to the square mass of the Higgs, $m_h^2 \approx (126 \ \text{GeV})^2$, as this is the only mass scale of the theory in this phase. In this way, SMEFT corrections are limited, and the only relevant contributions come from operators that can form closed scalar loops. At dimension-6, these corrections come from
\begin{equation}
    \label{eq:OHG}
    O_{HG} = (H^\dagger H  - v^2 / 2) G^{A}_{\mu \nu} G^{A \mu \nu}.
\end{equation}
The $O_{HG}$ operator does not contribute to the quark self-energy diagrams as it is a purely bosonic operator and does not contain fermion fields. It also does not contribute at one-loop order to the quark-gluon vertex. However, it can contribute to the gluon self-energy via a closed Higgs loop, as shown in Fig. \ref{fig:OHG_diag}.
To obtain the modified one-loop $\beta$ function, we have to calculate the UV divergences in the self-energy diagrams and the quark-gluon vertex, and absorb them in the counterterms of Eqs. (\ref{eq:ct1}), (\ref{eq:ct2}), and (\ref{eq:ct3}). The diagrams that contribute to the quark self-energy and the quark-gluon vertex are purely SM, and they are well known. In the case of the gluon self-energy, most of the diagrams are purely SM. However, we have to consider the extra diagram with the insertion of $O_{HG}$ and the closed scalar loop, and we summarise the calculation in the following. Using dimensional regularisation in $D$ dimensions, from the diagram in Fig. \ref{fig:OHG_diag} we obtain the correction to the gluon propagator
\begin{equation}
    \label{eq:Pi}
    \Pi_{\mu \nu}^{ab} (p^2) = - \frac{i}{2} \int \frac{d^D l}{(2 \pi)^D} \frac{1}{l^2 - m_h^2 + i \epsilon} 4 \delta^{ab} \frac{c_{HG}}{\Lambda^2} (p^2 g_{\mu \nu} - p_\mu p_\nu),
\end{equation}
where $p$ is the momentum of the gluon, $l$ is the loop momentum of the closed scalar loop, and $c_{HG}$ is the Wilson coefficient of the $O_{HG}$ operator. With Eqs. (\ref{eq:ct3}) and (\ref{eq:Pi}), we have all the ingredients to obtain the counterterm for the gluon self-energy. This can be done by imposing that
\begin{equation}
    \label{eq:div_cond}
    \text{Div} \left( \Pi_{\mu \nu}^{ab} (p^2) \right) + i \delta_3 \delta^{ab} (p_\mu p_\nu - p^2 g_{\mu \nu}) = 0,
\end{equation}
where $\text{Div}( \cdots )$ denotes an action that retains the UV divergent part of the expression. The divergences arise from the loop integral in Eq. (\ref{eq:Pi}), given by
\begin{equation}
    I \equiv \int \frac{d^D l}{(2 \pi)^D} \frac{1}{l^2 - m_h^2 + i \epsilon}.
\end{equation}
Let us take a closer look at this expression. After a Wick rotation that takes $l^0 \to i l^0$, we can write it in Euclidean space as
\begin{equation}
    I = -i \frac{1}{(2 \pi)^D} \int d^D l \frac{1}{l^2 + m_h^2}.
\end{equation}
We can now perform the angular integration in $D$ dimensions, which allows us to write
\begin{equation}
    \label{eq:1}
    I = -i \frac{1}{(2 \pi)^D} S_D \int_0^\infty dl \frac{l^{D-1}}{l^2 + m_h^2},
\end{equation}
where $S_D$ is the unit sphere surface area in $D$ dimensions. It can be calculated by considering that
\begin{align*}
    (\sqrt{\pi})^D &= \int_{\mathbb{R}^D} \prod_{i=1}^D dx_i \, e^{-x_i^2} \\
    &= S_D \int_0^\infty dr \, r^{D-1} e^{-r^2} \\
    &= \frac{S_D}{2} \Gamma\left(\frac{D}{2}\right),
\end{align*}
so we have that 
\begin{equation}
    S_D = \frac{2 \pi^{D/2}}{\Gamma(D/2)}.
\end{equation}
Now we are left with a radial integral. In this context, it is useful to remember that integrals like these can be expressed in terms of
the Euler Beta function, defined as
\begin{equation}
    \label{eq:id_1}
    B(x,y) = \int_0^1 dt \, t^{x-1} (1-t)^{y-1},
\end{equation}
which are related to the Gamma function by
\begin{equation}
    \label{eq:id_2}
    B(x,y) = \frac{\Gamma(x) \Gamma(y)}{\Gamma(x+y)}.
\end{equation}
Let us work the integral in Eq. (\ref{eq:1}) to write it in a more convenient way. To make use of the identities of Eqs. (\ref{eq:id_1}) and (\ref{eq:id_2}) it would be
useful to have the integration variable in the numerator and the denominator. We can write
\begin{equation}
    I^\prime \equiv \int_0^\infty dl \frac{l^{D-1}}{l^2 + m_h^2} = \frac{1}{2} \int_0^\infty dl^2 \frac{(l^2)^{D/2 - 1}}{l^2 + m_h^2}.
\end{equation}
where $d l^2 = 2 l d l$. Using a change of variables
\begin{equation*}
    u = \frac{m^2}{k^2 + m^2}, \quad du = - \frac{m^2}{(k^2 + m^2)^2} dk^2, \ \text{and} \ 1 - u = \frac{k^2}{k^2 + m^2},
\end{equation*}
we can write
\begin{equation}
    I^\prime = \frac{m^{D-2}}{2} \int_0^1 u^{-D/2} (1-u)^{D/2 - 1} du = \frac{m^{D-2}}{2} B(1 - D/2, D/2) = \frac{m^{D-2}}{2} \frac{\Gamma(1 - D/2) \Gamma(D/2)}{\Gamma(1)}.
\end{equation}
Now, putting everything together, we have so far 
\begin{equation}
    \label{eq:hello}
    I = - \frac{i}{(2 \pi)^D} \frac{2 \pi^{D/2}}{\Gamma(D/2)} \frac{1}{2} m^{D-2} \frac{\Gamma(1 - D/2) \Gamma(D/2)}{\Gamma(1)}.
\end{equation}
Some simplifications can be performed right away. In what follows, we will take the small $\epsilon$ limit in $D = 4 - 2 \epsilon$ dimensions, we can work out the remaining Gamma function
\begin{equation*}
    \Gamma(1 - D/2) = \Gamma(1 - 2 + \epsilon) = \Gamma(-1 + \epsilon) = \frac{\Gamma(\epsilon)}{-1 + \epsilon},
\end{equation*}
where we have used the property $\Gamma(z + 1) = z \Gamma(z)$. Using the expansion of the Gamma function around $\epsilon = 0$, we have
\begin{equation}
    \Gamma(\epsilon) = \frac{1}{\epsilon} - \gamma + \mathcal{O}(\epsilon),
\end{equation}
where $\gamma \approx 0.57722$ is the Euler-Mascheroni constant. In this way, we obtain
\begin{equation}
    \label{eq:final_gamma}
    \Gamma(1 - D/2) = - \frac{1}{\epsilon} + \gamma - 1 + \cdots.
\end{equation}
where the dots represent other finite terms. Using Eq. (\ref{eq:final_gamma}) in (\ref{eq:hello}), we are left with
\begin{equation}
    I = \frac{i}{16 \pi^2} m_h^2 \left( \frac{1}{\epsilon} - \gamma + 1 + \cdots \right).
\end{equation}
With the expression above, it is straightforward to calculate the divergence in Eq. (\ref{eq:div_cond}) and obtain the relevant counterterm with the $O_{HG}$ SMEFT contribution. It is given by
\begin{equation}
    \label{eq:d3_final}
    \delta_3 = - \frac{c_{HG}}{\Lambda^2} \frac{m_h^2}{8 \pi^2 \Lambda^2} + \delta_3^{\text{SM}}.
\end{equation} 
At one-loop, this is the only counterterm that is affected by $O_{HG}$, while $\delta_1$ and $\delta_2$ do not receive any contributions
\begin{equation}
    \label{eq:d1_d2_final}
    \delta_1 = \delta_1^{\text{SM}}, \quad \delta_2 = \delta_2^{\text{SM}}.
\end{equation}
With these counterterms we can calculate the one of the strong coupling $\delta Z_{g_s}$. It can be extracted from the gauge-fermion term in the renormalised Lagrangian, and it satisfies the following expression
\begin{equation}
    \label{eq:delta_g_s}
    \delta_1 = \frac{1}{2} \delta_3 + \delta_2 + \delta Z_{g_s}.
\end{equation}
We find 
\begin{equation}
    \label{eq:delta_g_s_final}
    \delta Z_{g_s} = - \frac{g_s^2 (33 - 2 n_f)}{96 \pi^2 \epsilon} + \frac{c_{HG}}{\Lambda^2} \frac{m_h^2}{16 \pi^2 \epsilon}
\end{equation}
where we find the well known SM contribution at one-loop, the first term, and we see a new contribution from the SMEFT operator. We are left with
\begin{equation*}
    \mu \frac{d g_s}{d \mu} = - \frac{g_s^3 (33 - 2 n_f)}{48 \pi^2} + \frac{c_{HG}}{\Lambda^2} \frac{m_h^2}{8 \pi^2},
\end{equation*}
which, after defining $\as = g_s^2 / (4 \pi)$ and using
\begin{equation}
    \frac{d \as}{d t} = \frac{1}{2} \mu \frac{d \as}{d \mu} = \frac{g_s}{4 \pi} \mu \frac{d g_s}{d \mu},
\end{equation}
gives, at one loop,
\begin{equation}
    \label{eq:beta_smeft}
    \frac{d \as}{d t} = - \as^2 \left( \frac{33 - 2 n_f}{12 \pi} \right) + \as \frac{c_{HG}}{\Lambda^2} \frac{m_h^2}{8 \pi^2}.
\end{equation}

Eq. (\ref{eq:beta_smeft}) is one of the main result of this section. It shows the running of the strong coupling $\as$ in the presence of the $O_{HG}$ SMEFT operator working in the unbroken phase. The first term is the usual SM running of $\as$, while the second term is the SMEFT contribution. This result is similar to the one obtained in Ref. \cite{Jenkins:2013zja}, modulo a factor of $-1/2$. The result of Eq. (\ref{eq:beta_smeft}) is the one that we will use in the following sections to study the running of $\as$ in the presence of SMEFT operators. The complete study of the running has to account not only for the running of $\as$, but also for the running of the Wilson coefficients and the mixing among them, and the running of other SM parameters. However, here we simply aim to provide an estimate of the effects of $c_{HG}$ in a phenomenological context.

\begin{figure}[ht]
    \centering
    \includegraphics[width=0.7\textwidth]{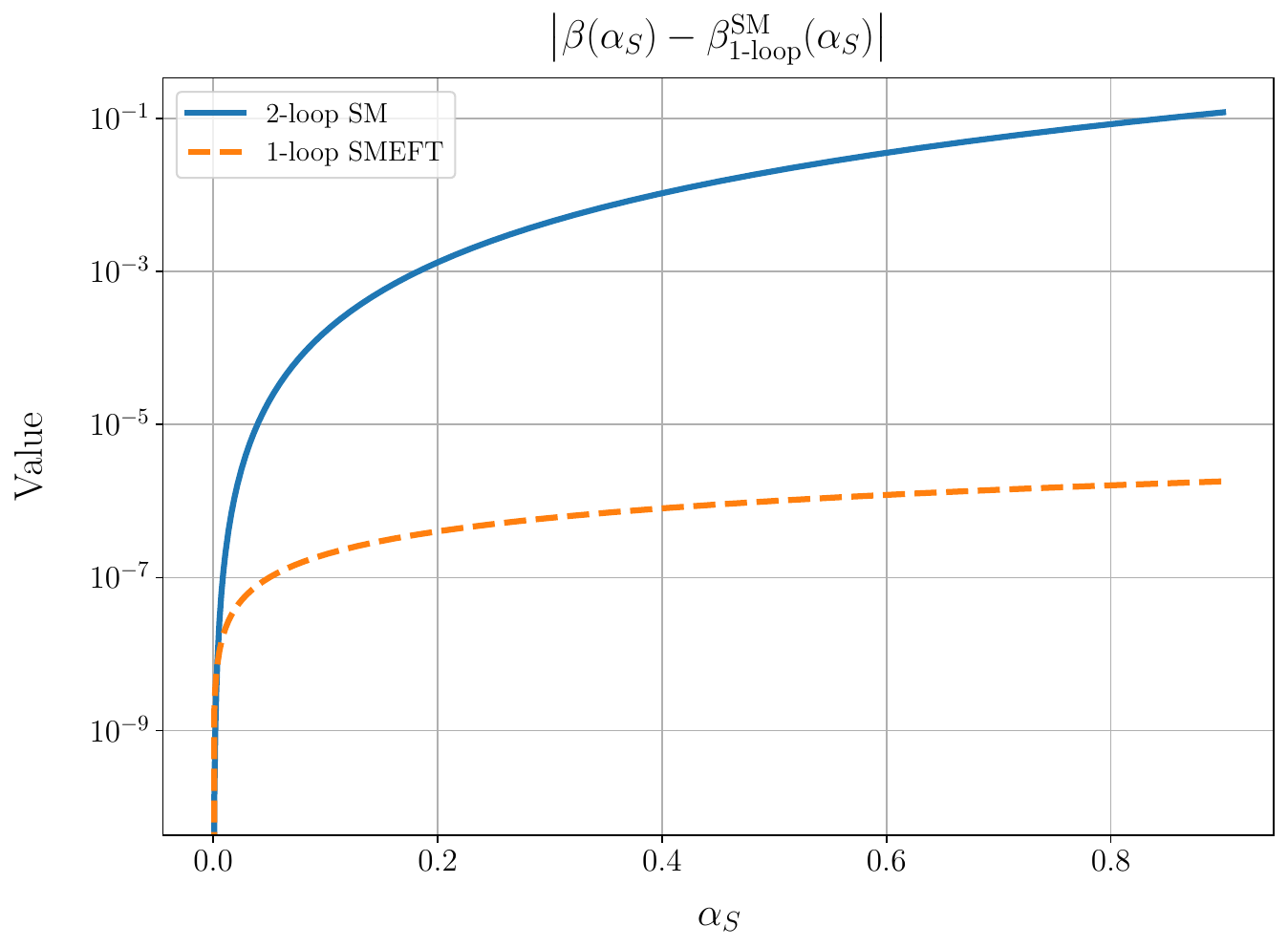}
    \caption{Absolute differences in $\beta$ functions: 2-loop SM and 1-loop SMEFT $(1/\Lambda^2)$ compared to the SM 1-loop.}
    \label{fig:beta_smeft}
\end{figure}

At $\Lambda = 1$ TeV, current bounds allow for $c_{HG} = 0.01$ \cite{Ethier:2021bye,Celada:2024mcf}. At this value, in Fig. \ref{fig:beta_smeft} we show the absolute difference between the SM $\beta$ function at 1-loop (reference), and the 2-loop SM and 1-loop with SMEFT contribution runnigs.
%
%

We can see that, at current bounds, the effect of $c_{HG}$ on the $\beta$ function is very suppressed. The shift it produces is very small compared to the 2-loop contribution in the SM. This is expected, as the Wilson coefficient is very small and the scale $\Lambda$ is high. In principle, it is unlikely that the effects of $c_{HG}$ on the running of the strong coupling will make a difference in current experiments, at the level of precision that is available. However, it is important to keep in mind that the SMEFT is a very general framework, and it is important to study the effects of all operators in the theory jointly. The study of the running of $\as$ in the presence of the $c_{HG}$ operator is a first step in this direction.

Additionally, another phenomenologically relevant scenario to consider is the potential effects of SMEFT operators on the running of $\as$ working in the broken phase of the theory, where particles have acquired mass. This increases the number of SMEFT operators to consider, and corrections proportional to the masses of the quarks, for example, can appear. Relatively recent calculations have been performed in this direction \cite{BuarqueFranzosi:2015jrv}, but only considering the effect of a single SMEFT operator at a time, namely $O_{tG}$. In the aforementioned reference, however, it is argued that the top-loop contributions decouple from the running of $\as$, in the same way as in the SM. This implies that the argument could not be extended to high energy measurements above the scale of the top quark mass.

\chapter{Symbolic regression for precision LHC physics}
\label{ch:sr}
  \epigraph{Je n’ai fait celle-ci plus longue que parce que je n’ai pas eu le loisir de la faire plus courte. \\ \vspace{1em} If I had more time, I would have written a shorter letter.}{\textit{\\ from Lettres Provinciales, \\ by B. Pascal.}}

\textit{This chapter is based on a forthcoming publication in collaboration with J. Bendavid, D. Conde, V. Sanz, and M. Ubiali. My contributions
include the generation of events and all the results displayed in this chapter. Additional results, not shown in this section, are
to be included in the final publication.} 

\vspace{1em} 
\noindent\rule{\textwidth}{0.4pt}
\vspace{1em} 

In this chapter we explore how symbolic regression can be used to infer simple, accurate, and closed form analytic expressions that can be exploited to
improve the accuracy of phenomenological analysis at the LHC. We will carry a thorough study of QED processes, whose analytical description
is well known from QFT principles, to assess the correctness of the method. This will provide a benchmark to test the robustness of the method
to then apply it to angular coefficients in $W$ and $Z$ production in the future publication.

\section{Introduction}

Symbolic Regression (SR) is a type of machine learning (ML) technique which aims to discover human-interpretable symbolic models. SR describes a supervised
learning task where the model space is spanned by analytic expressions. This is commonly solved in a multi-objective optimisation framework, 
jointly minimizing prediction error and model complexity. In this family of algorithms, instead of fitting concrete parameters
in some (potentially overparametrised) general model, one searches the space of simple analytic expressions for accurate and simple models.
SR is a way to combine the power of ML with the advantage of analytical 
intuition. In analogy to training a neural network, one can use SR to learn a general, analytic function the describes a dataset. SR gives us a way to extract
relatively simple human-readable formulas from complex datasets or simulations.

The PySR software package~\cite{Cranmer2023InterpretableML} is an open-source library for SR, based on a multi-population evolutionary algorithm.  
The population consists of symbolic expressions represented as binary expressions tree and consisting of nodes with can be features, constants, or operators. 

PySR has been deployed in a variety of domains, including collider phenomenology. It was used in Ref.~\cite{Choi:2010wa} to derive a kinematic
variable that is sensitive to the mass of the  Higgs boson in the $WW$ channel. Subsequently, in Ref.~\cite{Butter:2021rvz} it was used to
construct optimal observables for LHC processes. In Ref.~\cite{Dersy:2022bym} it was used to simplify the mathematical expression of polylogarithms. 
In Ref.~\cite{Dong:2022trn} SR was used to derive analytical formulas for the stransverse mass $M_{T,2}$ kinematical variable, which is usually defined algorithmically
through an optimisation procedure and not in terms of an analytical formula. Furthermore, an analytical approximation for the NLO kinematic distributions after
detector simulation was derived, for which no known analytical formulas currently exist.

Standard analyses of at (LHC) at CERN involve the study of distributions of
kinematic variables, which are typically defined in terms of the energies and momenta of the particles observed in
the detector. Many of these variables, such as invariant mass or missing
transverse momentum, are defined in terms of simple analytical expressions and can be readily computed from
the collections of particle 4-momenta in the event. However, there also kinematic variables which are defined algorithmically, i.e., through a well-defined
optimisation procedure which involves the minimisation of a relevant kinematic function. 
In that case, the kinematic variable is a quantity that can be computed only once the algorithm has converged, and typically there is no a priori known analytical
expression for it in the general case.

In this ongoing study our aim is to infer analytic expressions for angular coefficients in $W$ and $Z$ production. These coefficients are
very important in the context of precision physics at the LHC. The study of Z boson production and their decay into lepton pairs or lepton and neutrino
have been extensively covered in LHC experiments, as described in the previous chapters of this thesis. The
high precision of these measurements has driven significant advancements over the past decade in PDF fitting and QCD theoretical predictions in general.
Achieving sub-percent precision in absolute and normalized fiducial cross-section measurements of the Z-boson transverse momentum, $p_T$, as a function of its rapidity, $|y|$, at the Z pole, imposes
stringent constraints on state-of-the-art theoretical calculations. 

However, before being able to apply SR to the angular coefficients in $W$ and $Z$ production, we need to validate the method.
We will do this by studying QED processes, which are well known from QFT principles. We will start by considering the production of a
lepton pairs in $e^+e^-$ collisions. Then, we will stress test the method by adding Poissonian and Gaussian noise to the distributions and
assessing the stability of the results. Further results will be included at a later stage. In this text, we aim to 
demonstrate the potential of SR to derive interpretable and efficient models from complex datasets, thereby contributing
to advancements in collider phenomenology 

\section{Angular variables}

The production of vector bosons has been extensively studied at the LHC. Measurements of processes involving weak vector bosons are
important both for confirming SM EW predictions and searching for evidence of New Physics\footnote{Recently, the CMS collaboration reported the most precise measurement of the $W$ boson mass to date~\cite{CMS-PAS-SMP-23-002}. The careful determination of angular coefficients was essential to model the angular distributions in $W$ and $Z$ leptonic decays.}.

Recent LHC analyses have delivered extremely precise measurements involving W and Z bosons. In particular,
ATLAS~\cite{ATLAS:2023lsr} has released double-differential measurements 
in $p_T^Z$ and $y^Z$ of absolute and normalised cross-sections at the $Z$ pole at $\sqrt{s}=8$ TeV.
This kind of measurement can be done by using a methodology that relies on the decomposition of the lepton angular distributions in the 
Collins-Soper frame~\cite{Collins:1977iv} into nine spherical harmonic polynomials $P_i$, multiplied by 
angular coefficients $A_i$~\cite{Mirkes:1992hu,Mirkes:1994dp,Mirkes:1994eb,Mirkes:1994nr}. For both $W$- and $Z$-boson production, the full five-dimensional differential cross section 
describing the kinematics of the two Born-level leptons can be written as:
\begin{equation}
    \frac{d^5\sigma}{dp_T\, d\eta\, dm \,d\cos\theta \,d\phi} = \frac{3}{16\pi}\frac{d^3\sigma^{U+L}}{dp_T\, d\eta\, dm}\left[(1+\cos^2\theta) + \sum_{i=0}^7 P_i(\theta,\phi)\,A_i \right],
    \label{eq:full_coeff}
\end{equation}
where $p_T$ is the transverse momentum of the electroweak boson, $\eta$ its rapidity, $m$ its invariant mass, $\theta$ and $\phi$ are respectively the polar and azimuthal angle of the lepton, and
\begin{align}
    P_0(\theta,\phi)&=\frac{1}{2}(1-3\cos^2\theta),\notag\\
    P_1(\theta,\phi)&=\sin2\theta\,\cos\phi,\notag\\
    P_2(\theta,\phi)&=\frac{1}{2}\sin^2\theta\cos2\phi,\notag\\
    P_3(\theta,\phi)&=\sin\theta\cos\phi,\notag\\
    P_4(\theta,\phi)&=\cos\theta,\notag\\
    P_5(\theta,\phi)&=\sin^2\theta\sin2\phi,\notag\\
    P_6(\theta,\phi)&=\sin2\theta\sin\phi,\notag\\
    P_7(\theta,\phi)&=\sin\theta\sin\phi.\notag
\end{align}
The dependence of the differential cross section on $p_T$, $|\eta|$ and $m$ is entirely contained in the
unpolarised cross section $\sigma^{U+L}$ and in the $A_i$ angular coefficients. 
In QCD, the angular coefficients can be extracted~\cite{Bern:2011ie} by running Monte Carlo at a given perturbative order and computing
\begin{align}
    A_0&= 4-10\langle\cos^2\theta\rangle, \\
    A_1&=\langle 5\sin2\theta\,\cos\phi\rangle, \notag\\
    A_2&=\langle 10\sin^2\theta\cos2\phi\rangle, \notag\\
    A_3&=\langle 4\sin\theta\cos\phi\rangle, \notag\\
    A_4&=\langle 4\cos\theta\rangle, \notag\\
    A_5&=\langle\sin\theta\sin\phi\rangle, \notag\\
    A_6&=\langle 5\sin2\theta\sin\phi\rangle, \notag\\
    A_7&=\langle 5\sin^2\theta\sin2\phi\rangle. \notag
\end{align}

To study the angular coefficients it is useful to layout different scenarios in increasing levels of complexity. Different processes
at different orders can be considered in such a way that the method can be tested and validated, and that progressively more angular coefficients
appear in the decomposition of the cross section. One potential path is 

\begin{enumerate}
  \item Level 1: $e^+e^- \to \gamma^* \to \mu^+ \mu^-$. At this level, the differential cross section is fully known
  from QED, and the quality of the SR results can be assessed directly from first principles. There
  are no angular coefficients but just the $1 + \cos^2 \theta$ angular dependence.
  \item Level 2: $p\ p \to \gamma \to \mu^+ \mu^-$. Same angular dependence as level 1, but now adding the PDF dependence from initial hadronic states.
  \item Level 3: $p\ p \to Z \to \mu^+ \mu^-$. Same as level 2, but now $A_4 \neq 0$. 
  \item Level 4: $p \ p \to Z \ \to \mu^+ \mu^-$ at NLO QCD ($\mathcal{O}(\alpha_S)$). Same as level 3, but now also $A_{0, \dots, 3} \neq 0$.
  \item Level 5: $p \ p \to Z \ \to \mu^+ \mu^-$ at NNLO QCD ($\mathcal{O}(\alpha_S^2)$). Same as level 4, but now also $A_{5, 6, 7} \neq 0$.
\end{enumerate}

In this chapter, we will only assess level 1. Highlighting the importance of this level is crucial as it serves as a fundamental validation
of the SR method against well-understood QED processes and performing equation recovery. By establishing the method's reliability in this simplest scenario, we lay a solid foundation
for its application to more complex processes involving angular coefficients in vector boson production. The full set of results will be included in the final publication.

\section{Symbolic regression}

SR is a supervised learning task that aims to find closed analytical expressions to describe a mapping between inputs and outputs.
It presents an alternative to other regression techniques as it does not fix the functional form that the mapping has to have, as in linear regression, for example, where
the gradient and the intercept of the mapping have to be found via the optimisation of some quality metric, or in a neural network, where the weights and biases
of the network have to be found via backpropagation and gradient descent, but the form of the mapping is fixed by the architecture of the network and the activation functions.
SR aims to find a mapping that is accurate and simple (unlike a neural network which potentially has thousands of trainable parameters).
Compared to other ML techniques, SR presents the key advantage that it provides simple equations to describe relations between variables, facilitating scientific discovery,
and the interpretability of the results.

\begin{figure}[ht]
  \centering
  \begin{minipage}{0.45\linewidth}
    \centering
    \includegraphics[width=\linewidth]{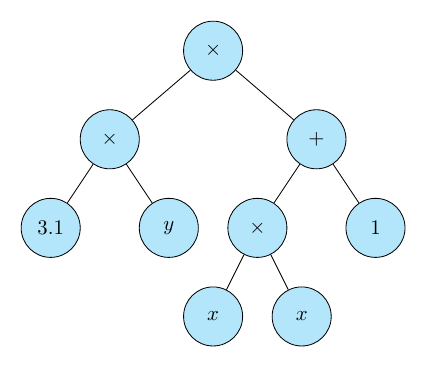} 
    \caption{Example of an expression tree.}
    \label{fig:tree}
  \end{minipage}%
  \hspace{0.05\linewidth}
  \begin{minipage}{0.45\linewidth}
    \centering
    \includegraphics[width=\linewidth]{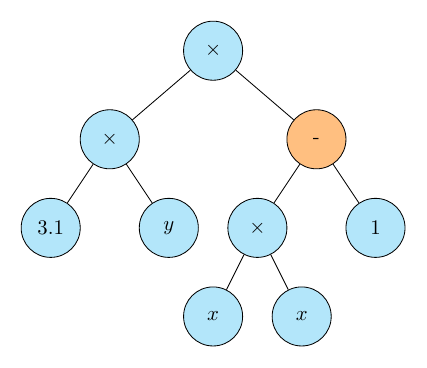} 
    \caption{Mutation of an expression tree.}
    \label{fig:tree2}
  \end{minipage}
\end{figure}

For this study, we make use of the PySR package \cite{Cranmer2023InterpretableML}, which makes use of a multi-population evolutionary algorithm that evaluates
symbolic expressions (equations) represented by expressions trees. In Fig. \ref{fig:tree} we show an example of a tree that represent the equation
$3.1 y \cdot (x^2 + 1)$. The nodes of the tree can be either features, constants, or operations. In this tree, $\times$ and $+$ represent the binary
operations of multiplication and sum, respectively. The node at the top of the tree is called the \textit{root node}, and each node can
have either 0, 1, or 2 children nodes.
The quality of the equations is evaluated by a loss function, which is minimised during the optimisation process.
The loss function can be any of the well-known metrics used in regression problems, such as the mean squared error (MSE) between the
predicted values and the actual values, the mean absolute error, or it can be a tailored loss function specific to the problem at hand. In PySR,
the loss function can even be a non-differentiable function, as fittest trees are not found via backpropagation.

During the PySR optimisation, expression trees mutate during a given number of iterations (to be defined by the user according to the specific problem to be solved)
to find better equations according to a given selection criterion. A mutation of a tree can be seen in Fig. \ref{fig:tree2}, where
where the '$+$' of the expression tree in Fig. \ref{fig:tree} has been mutated to a '$-$'. Fitter trees can also combine to produce better trees, a process known as crossover.

The selection criterion is a metric that guides the evolutionary algorithm in selecting the fittest expression trees. In PySR, there are three selection
criteria: \texttt{accuracy}, \texttt{score}, and \texttt{best}. These criteria are defined as follows:

\paragraph{\texttt{Accuracy}:} The fittest trees are the ones that minimise the loss function, whatever the definition of the loss and the complexity of the tree and the expression it represents. For example, if the MSE is chosen as the loss function, the fittest trees are the ones that minimise
\begin{equation*}
L = \frac{1}{n} \sum_{i=1}^{n} (Y_i - \hat{Y}_i)^2,
\end{equation*}
where $\hat{Y}_i$ are the values predicted by the expression tree, $Y_i$ are the actual values, and $n$ is the number of data points.

\paragraph{\texttt{Score}:} This metric maximises the decrease in the logarithm ot the loss function $L$ with respect to the complexity $c$ of the tree,
defined as the number of nodes in the tree. In this way, the score criterion maximises
\begin{equation*}
-\frac{\partial \log(L)}{\partial c}.
\end{equation*}
In this case, the fittest trees are the ones that minimise the loss function while keeping the complexity of the tree low.
A marginal decrease in the loss function is penalised if it comes with a significant increase in the complexity of the tree, so this criterion is useful to avoid overfitting.

\paragraph{\texttt{Best}:} This metric selects the tree with the highest score, as defined in the previous point, as long as its loss
$L$ is within 1.5 times the loss \(L_{\text{min}}\) of the most accurate tree, as defined in the \texttt{accuracy} criterion:
\begin{equation*}
L \leq 1.5 \times L_{\text{min}}
\end{equation*}
This criterion is a compromise between the \texttt{accuracy} and \texttt{score} criteria, as it selects the most accurate tree that is not too complex.

The selection criterion is a key aspect of the optimisation process, as it determines the balance between the accuracy of the model and its complexity.
In the presence of noisy datasets, the stability and robustness of the algorithm is a key aspect to consider. In the following sections,
we will use the three selection criteria, and their interplay with the binning in the distributions and different degrees of noise, to find
the best equations to describe collider observables.

\section{Equation recovery with symbolic regression}

In this section we will show the results of the SR algorithm applied to the process $e^+e^- \to \gamma^* \to \mu^+ \mu^-$ at leading order. We
will see if the method can recover the well-known angular distribution of the cross section (in the massless limit), and the constant if the angular dependence is factored out.
Additionally, we will assess the stability of the results in the presence of Poissonian and Gaussian noise, and in the presence of different binning in the distributions.

The differential cross section for this process is given by
\begin{equation}
    \label{eq:l1}
    \frac{d\sigma}{d\Omega} = \frac{ \alpha^2}{4s}\left(1 + \cos^2\theta\right),
\end{equation}
where $d \Omega = d (\cos \theta) \ d \phi$ is a differential solid angle measure, $\theta$ is the angle between the outgoing muon and axis of the incoming electrons,
$\phi$ is the azimuthal angle, $\alpha$ is the QED coupling, and $s$ the squared centre of mass energy. 

To train the regressor, we generate events using {\sc\small MadGraph5\_aMC@NLO}~\cite{Alwall:2014hca,Frederix:2018nkq} and
show the regressor different distributions in terms of the number of bins, and introducing different levels of Poissonian and Gaussian noise.

To introduce Poissonian noise, we understand the event count in the $i$-th bin, $N_i$, as a Poisson-distributed variable with mean equal to the actual count in the bin.
The noisy count is then given by
\begin{equation}
    N_i \sim \text{Poisson}(N_{i}^{\text{count}}),
\end{equation}
where $N_{i}^{\text{count}}$ is the actual count in the bin from the simulation.

Additionally, to stress test the methodology even further we introduce Gaussian noise to the counts in the bins, potentially already affected by Poissonian noise.
Gaussian noise is characterized by the normal distribution. The noise is added to each bin count of the histogram, neglecting correlations for the sake of simplicity.
The event count in each bin is then given by 
\begin{equation}
    N_i \sim N_i^{\text{count}} + \epsilon,
\end{equation}
where $\epsilon \sim \mathcal{N}(0, n \sigma^2)$ is a normally distributed random variable. The standard deviation is given by $\sigma = \sqrt{N_i}$,
consistent with the Poissonian distribution, and we have introduced a factor $n$ to control the level of noise in the distribution.

We begin by studying how the three selection criteria perform in the recovery of the angular distribution of the cross section for different binning.
To do this, we will use the MSE as the loss function and expect to find the $1 + \cos^2 \theta$ dependence. In Tab. \ref{tab:l1_results} we show the results for each case. 
\begin{table}[h]
\begin{center}
\resizebox{\textwidth}{!}{%
\begin{tabular}{@{}c>{\centering\arraybackslash}m{0.3\textwidth}|>{\centering\arraybackslash}m{0.3\textwidth}|>{\centering\arraybackslash}m{0.3\textwidth}@{}}
\toprule
Bins & \multicolumn{1}{c}{Accuracy} & \multicolumn{1}{c}{Best} & \multicolumn{1}{c}{Score} \\
\midrule
10 & 
$x_{0}(x_{0} + 0.00798)(0.00111 \cdot x_{0}^{3} + 0.03459) + 0.03503$ & 
$x_{0}^{2}(0.00111 \cdot x_{0} + 0.03459) + 0.03503$ & 
$0.03459 \cdot x_{0}^{2} + 0.03503$ \\
\noalign{\smallskip} \hline \noalign{\smallskip}
20 & 
$x_{0}(x_{0} + 0.01825)(-0.00155 \cdot x_{0}(x_{0} - 0.05138) + 0.03579) + 0.03485$ & 
$x_{0}(0.03447 \cdot x_{0} + 0.00064) + 0.03498$ & 
$0.03447 \cdot x_{0}^{2} + 0.03498$ \\
\noalign{\smallskip} \hline \noalign{\smallskip}
30 & 
$x_{0}^{2}(x_{0}(0.00098 - 0.00257 \cdot x_{0}) + 0.03659) + 0.03477$ & 
$0.03439 \cdot x_{0}^{2} + 0.03499$ & 
$0.03439 \cdot x_{0}^{2} + 0.03499$ \\
\noalign{\smallskip} \hline \noalign{\smallskip}
50 & 
$x_{0}(x_{0} + 0.00096)(-0.00119 \cdot x_{0}^{2} + 0.00096 \cdot x_{0} + 0.03547) + 0.03486$ & 
$0.03445 \cdot x_{0}^{2} + 0.03496$ & 
$0.03445 \cdot x_{0}^{2} + 0.03496$ \\
\noalign{\smallskip} \hline \noalign{\smallskip}
100 & 
$x_{0}^{2}(-0.00125 \cdot x_{0}(x_{0}^{2} + x_{0} - 1.60285) + 0.03553) + 0.03485$ & 
$0.03446 \cdot x_{0}^{2} + 0.03496$ & 
$0.03446 \cdot x_{0}^{2} + 0.03496$ \\
\noalign{\smallskip} \hline \noalign{\smallskip}
200 & 
$x_{0}^{2}(-0.64647 \cdot x_{0}(0.00119 \cdot x_{0} - 0.00151) + 0.03495) + 0.03495$ & 
$0.03447 \cdot x_{0}^{2} + 0.03495$ & 
$0.03447 \cdot x_{0}^{2} + 0.03495$ \\
\noalign{\smallskip} \hline \noalign{\smallskip}
500 & 
$x_{0}(x_{0} + 0.00604)(x_{0}(0.00069 - 0.00118 \cdot x_{0}) + 0.03548) + 0.03485$ & 
$0.03447 \cdot x_{0}^{2} + 0.03495$ & 
$0.03447 \cdot x_{0}^{2} + 0.03495$ \\
\noalign{\smallskip} \hline \noalign{\smallskip}
1000 & 
$0.03447 \cdot x_{0}(0.01821 \cdot x_{0}^{2} + x_{0} + 0.00723) + 0.03495$ & 
$0.03447 \cdot x_{0}^{2} + 0.03495$ & 
$0.03447 \cdot x_{0}^{2} + 0.03495$ \\
\bottomrule
\end{tabular}
}
\caption{Equations according to the three selection criteria for different bin sizes with $x_{0} \equiv \cos \theta$. The numbers that appear in these expressions have been approximated to the 5th decimal place.}
\label{tab:l1_results}
\end{center}
\end{table}
We see that, in all cases, the \texttt{accuracy} criterion misses the correct functional form by overfitting. Odd powers of the cosine account for
a measure of asymmetry in the distribution which does not come from the underlying theory but from the binning and the Monte Carlo integration error of the cross section.
The expressions are overcorrected by even higher powers of the cosine that effectively are fitting the noise in the distribution.

The \texttt{best} criterion misses the correct functional form in the case of 10 and 20 bins, but it recovers the correct form for 30 bins and above. This could happen
because of the fact that, with fewer bins, the amount of datapoints to train over is smaller, and more complex (and potentially overfitting) solutions can cause more damage.
Since the \texttt{best} criterion has a hard constraint on the accuracy of the expression, and the \texttt{accuracy} criterion is overfitting, the \texttt{best} criterion can
miss the correct form in the case of fewer bins. For a higher number of datapoints, the \texttt{best} criterion can recover the correct form. Still, in the case of
unknown functional forms, it is useful to consider the \texttt{best} criterion to ensure that accuracy is not sacrificed for simplicity.

The \texttt{score} criterion is able to recover the correct functional form for all bin sizes. The correct form for the distribution that is well-known and simple,
and so it makes sense that the \texttt{score} criterion is able to recover it. In the case of unknown functional forms, the \texttt{score} criterion can be a good choice
to look for the simplest expression that describes the data but it is important that accuracy does not deteriorate significantly.

\begin{figure}[ht!]
  \centering
  \includegraphics[width=0.8\linewidth]{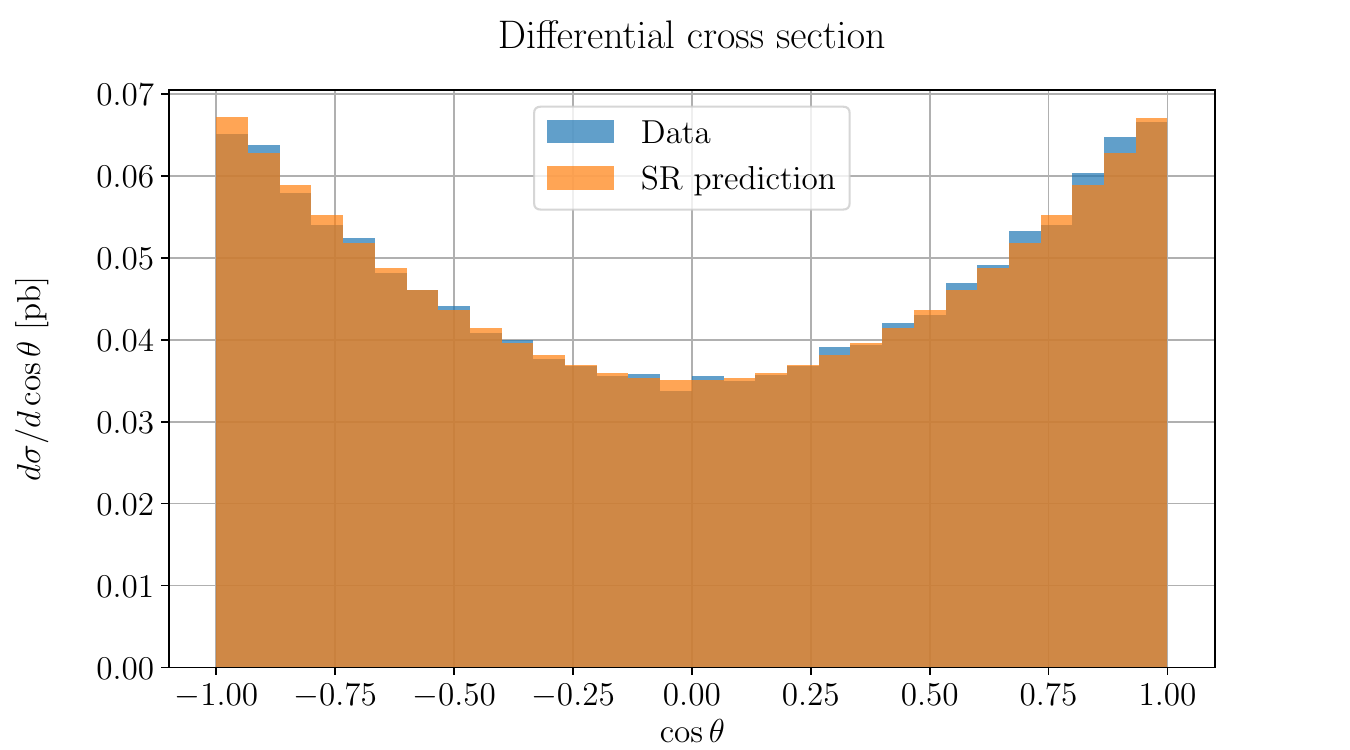}
  \caption{Distribution in $\cos \theta$.}
  \label{fig:sr_pred_l1_1}
\end{figure}

In Fig. \ref{fig:sr_pred_l1_1} we show a graphical comparison between the data and the predictions of the SR algorithm for the case of 30 bins, showing very good agreement.

\begin{figure}[ht!]
  \centering
  \includegraphics[width=0.8\linewidth]{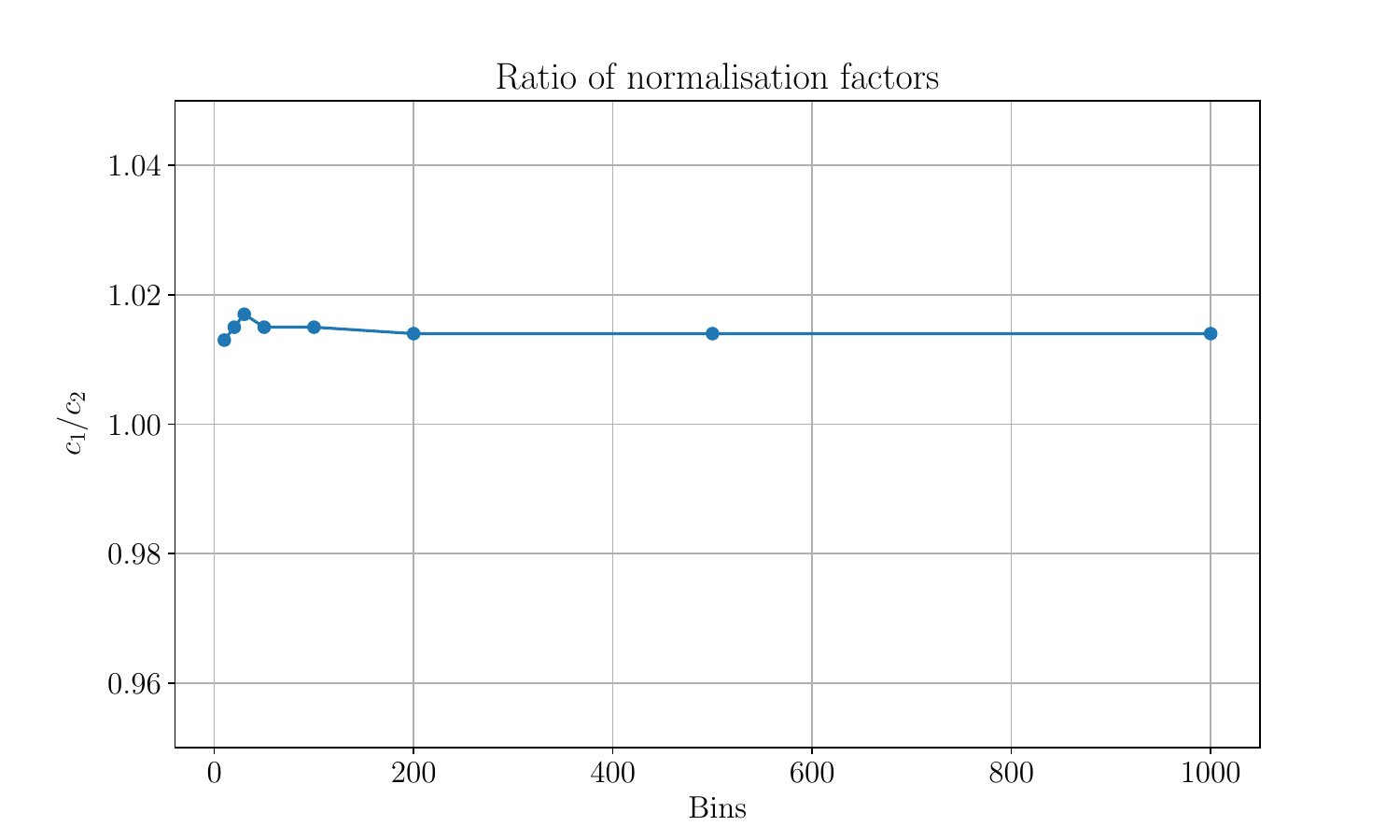}
  \caption{Normalisation coefficients.}
  \label{fig:sr_l1_norm}
\end{figure}

\begin{figure}[ht!]
  \centering
  \begin{minipage}{0.49\linewidth}
    \centering
    \includegraphics[width=\linewidth]{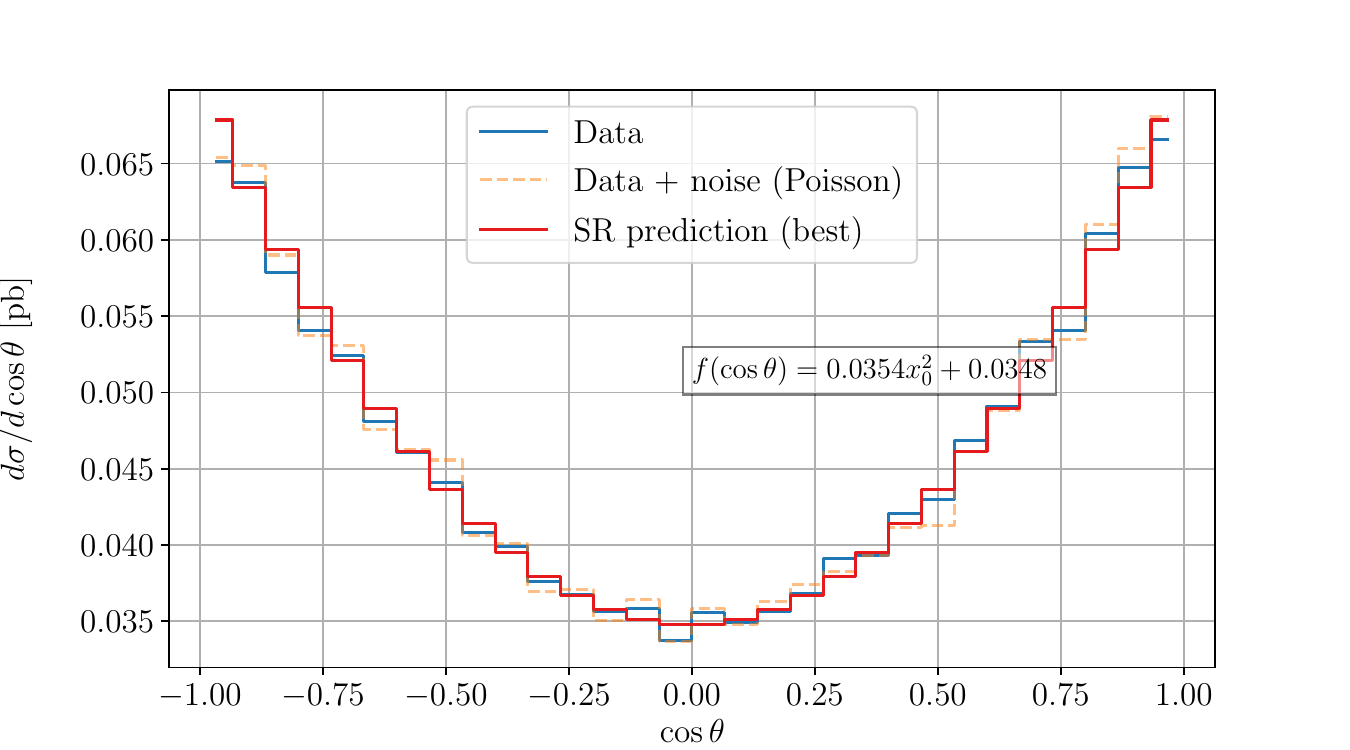}
    \caption*{(a)}
  \end{minipage}%
  \hspace{0.01\linewidth}
  \begin{minipage}{0.49\linewidth}
    \centering
    \includegraphics[width=\linewidth]{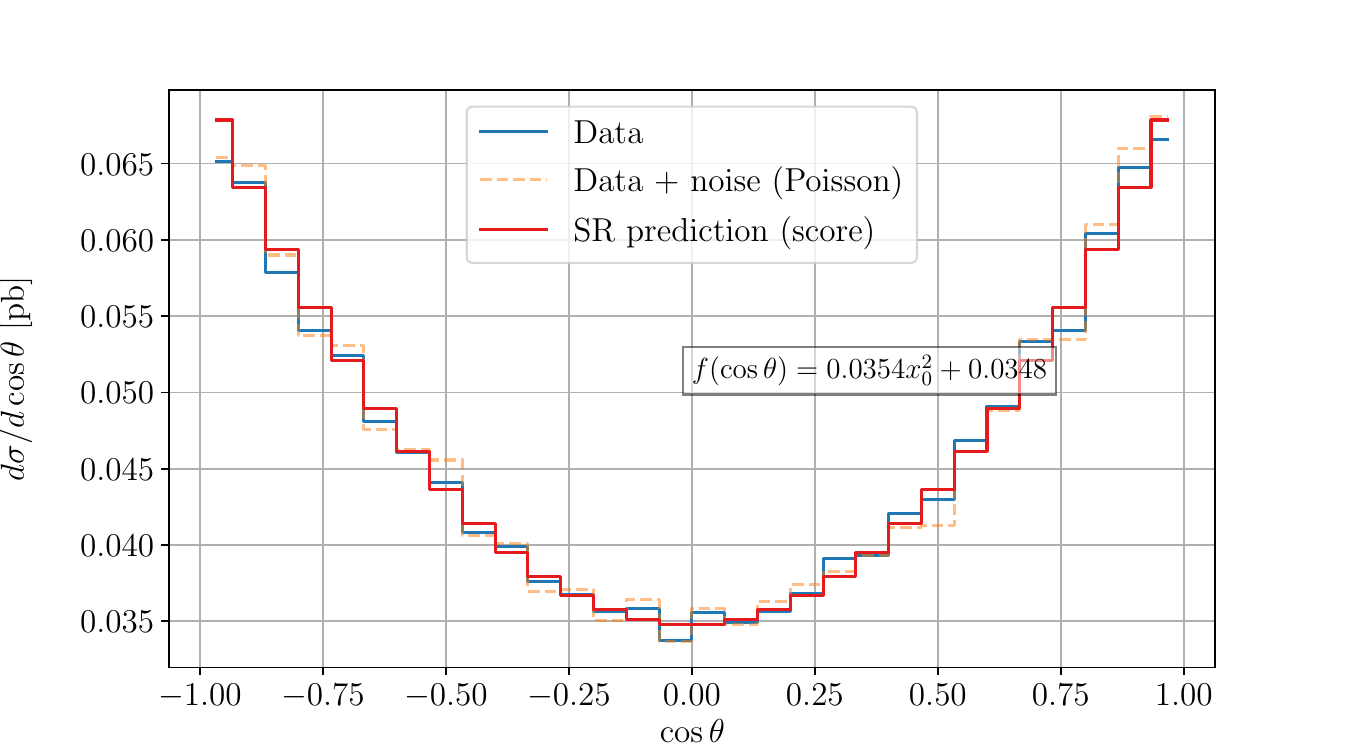}
    \caption*{(b)}
  \end{minipage}

  \vspace{0.2cm} 

  \begin{minipage}{0.49\linewidth}
    \centering
    \includegraphics[width=\linewidth]{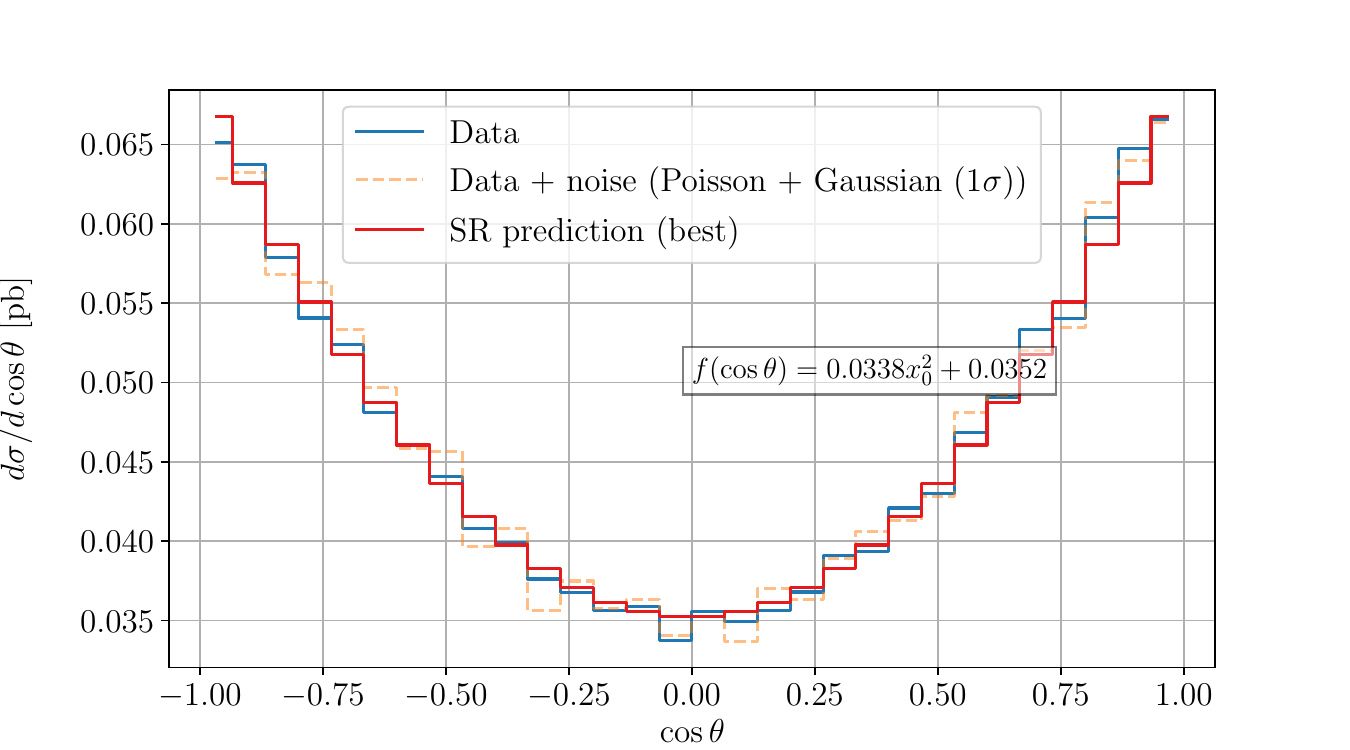}
    \caption*{(c)}
  \end{minipage}%
  \hspace{0.01\linewidth}
  \begin{minipage}{0.49\linewidth}
    \centering
    \includegraphics[width=\linewidth]{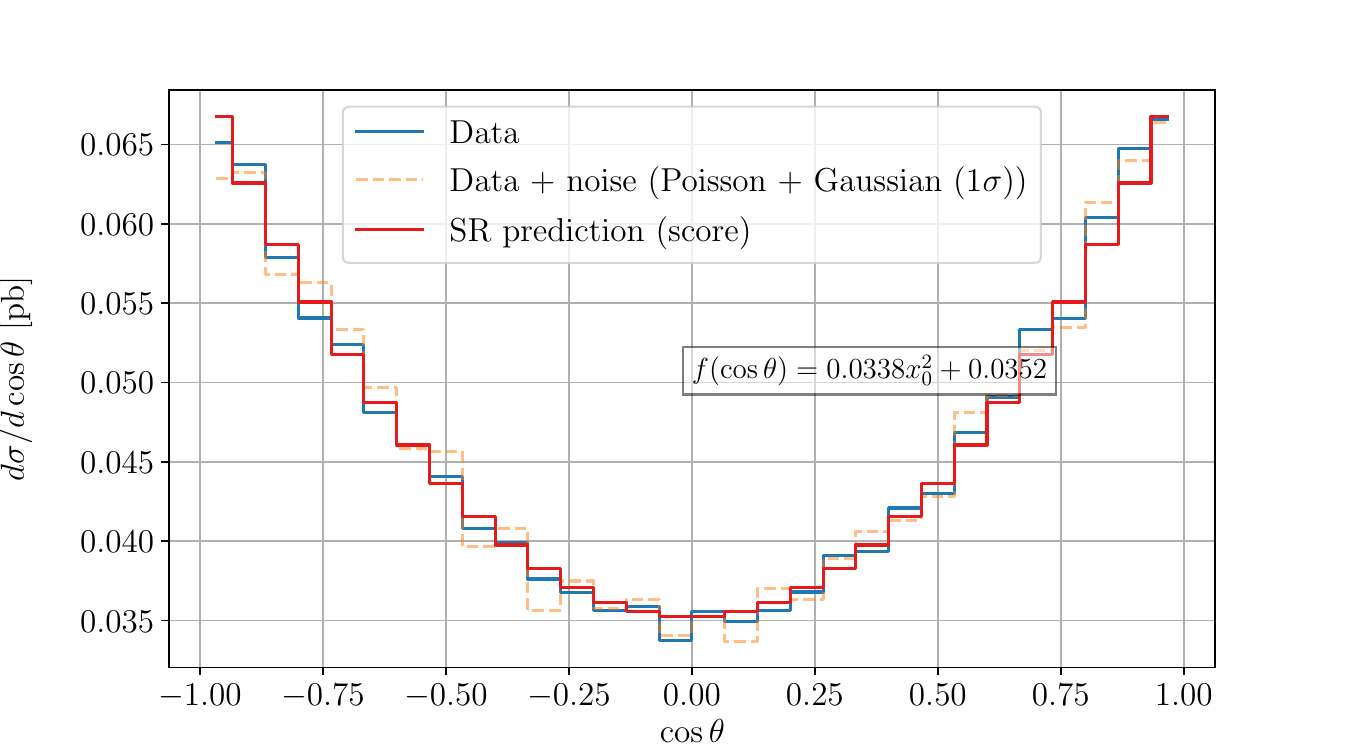}
    \caption*{(d)}
  \end{minipage}

  \vspace{0.2cm} 

  \begin{minipage}{0.49\linewidth}
    \centering
    \includegraphics[width=\linewidth]{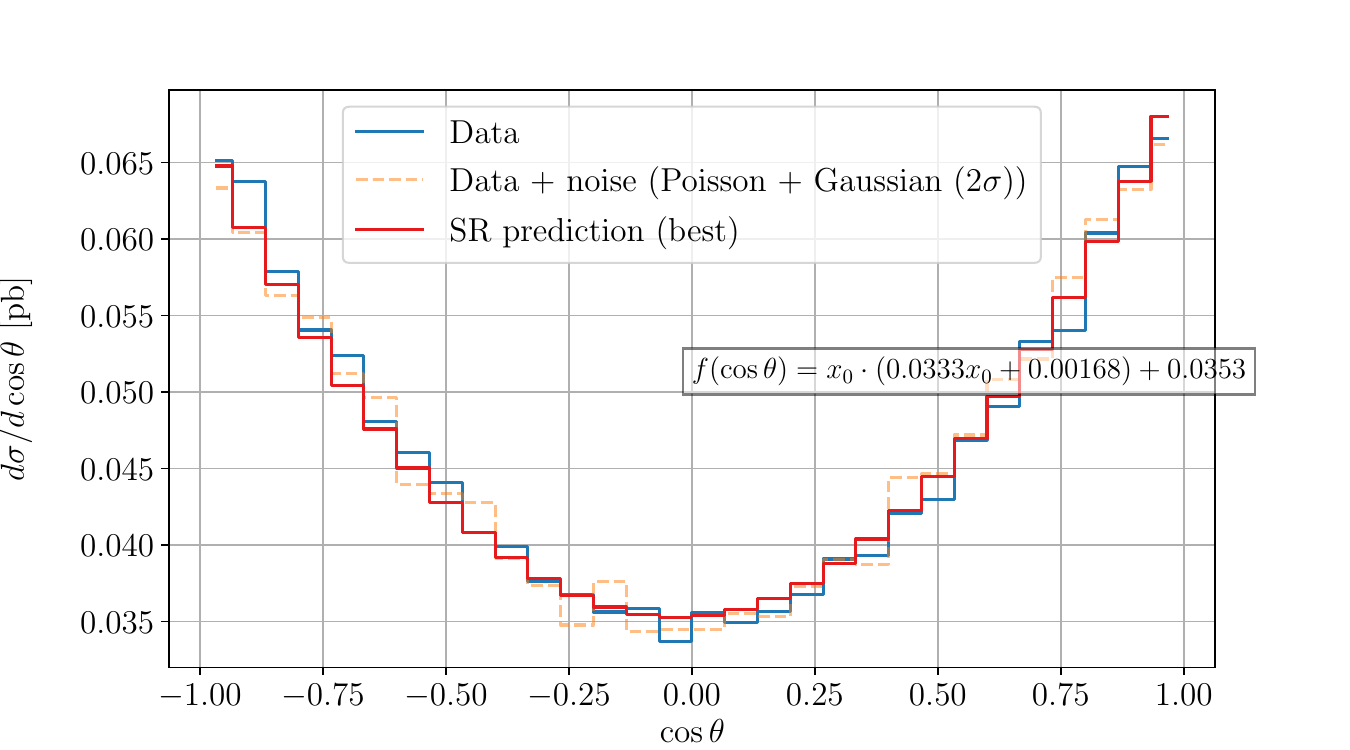}
    \caption*{(e)}
  \end{minipage}%
  \hspace{0.01\linewidth}
  \begin{minipage}{0.49\linewidth}
    \centering
    \includegraphics[width=\linewidth]{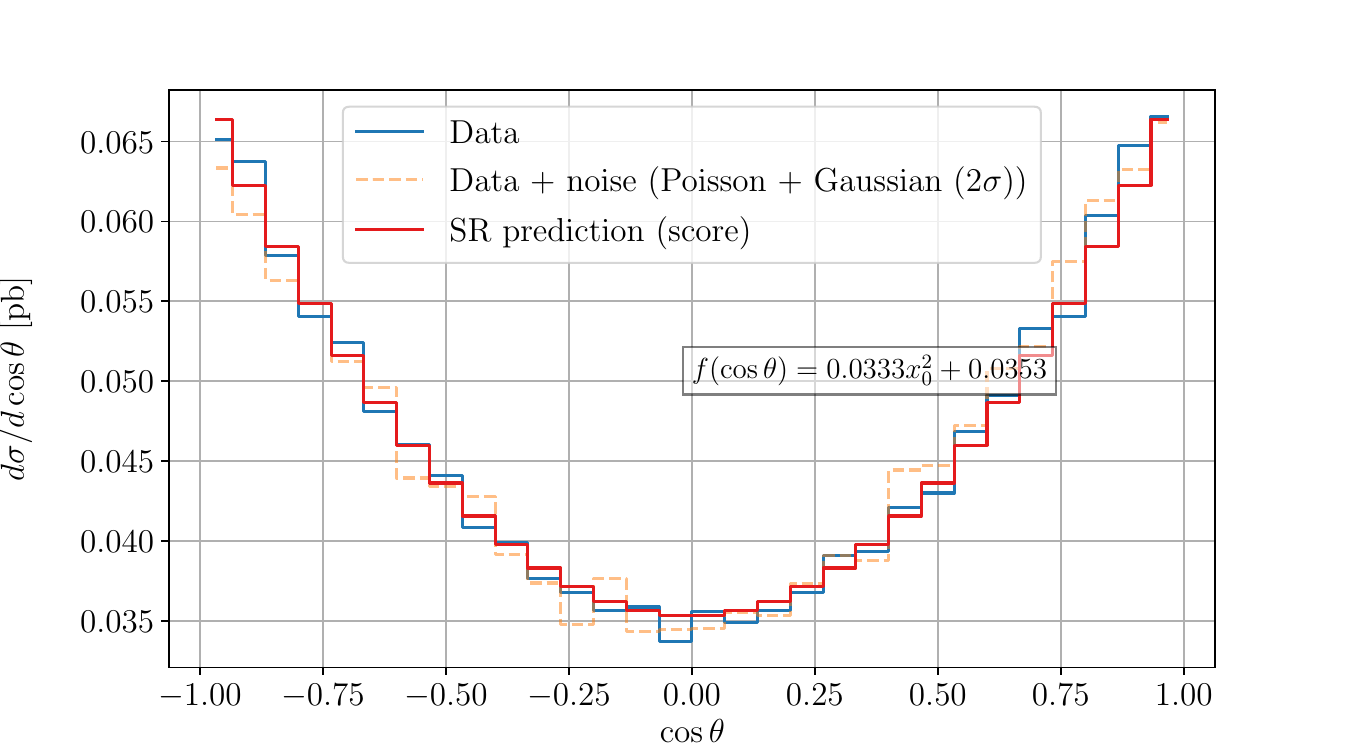}
    \caption*{(f)}
  \end{minipage}

  \vspace{0.2cm} 

  \begin{minipage}{0.49\linewidth}
    \centering
    \includegraphics[width=\linewidth]{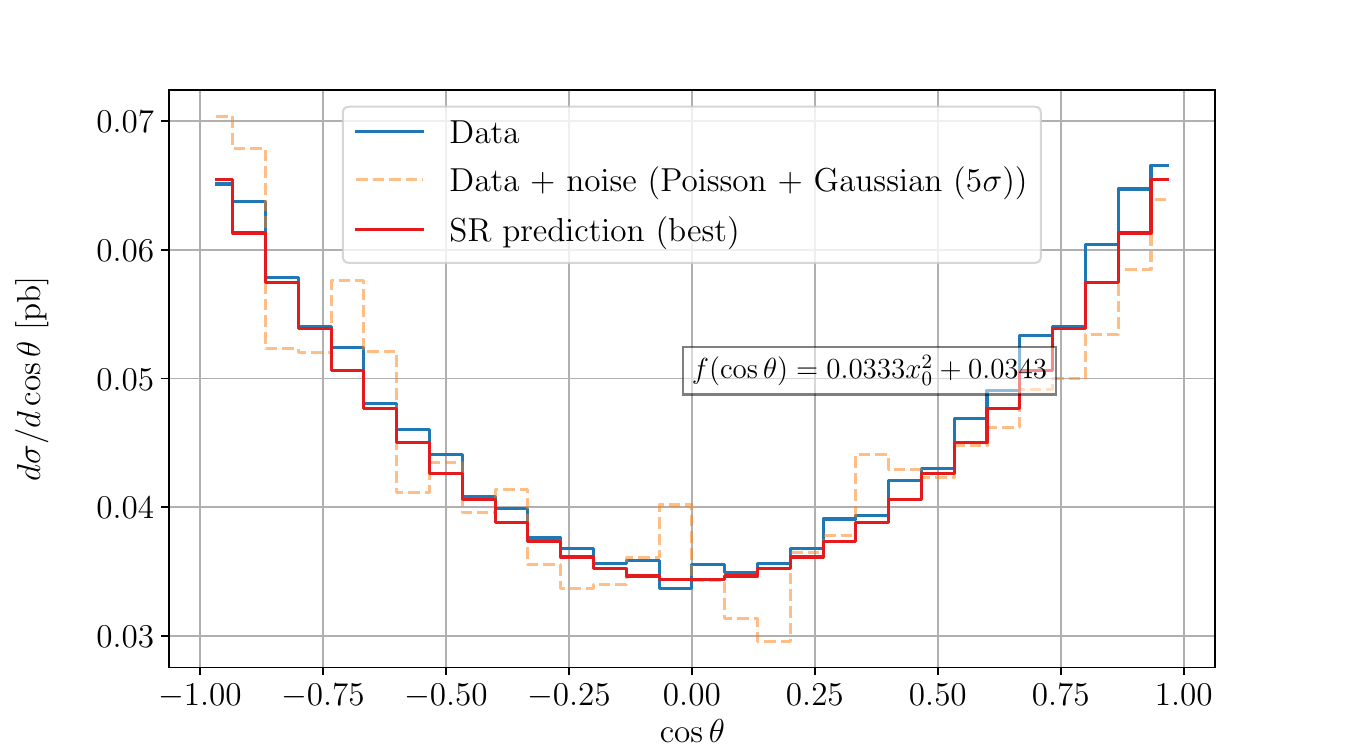}
    \caption*{(g)}
  \end{minipage}%
  \hspace{0.01\linewidth}
  \begin{minipage}{0.49\linewidth}
    \centering
    \includegraphics[width=\linewidth]{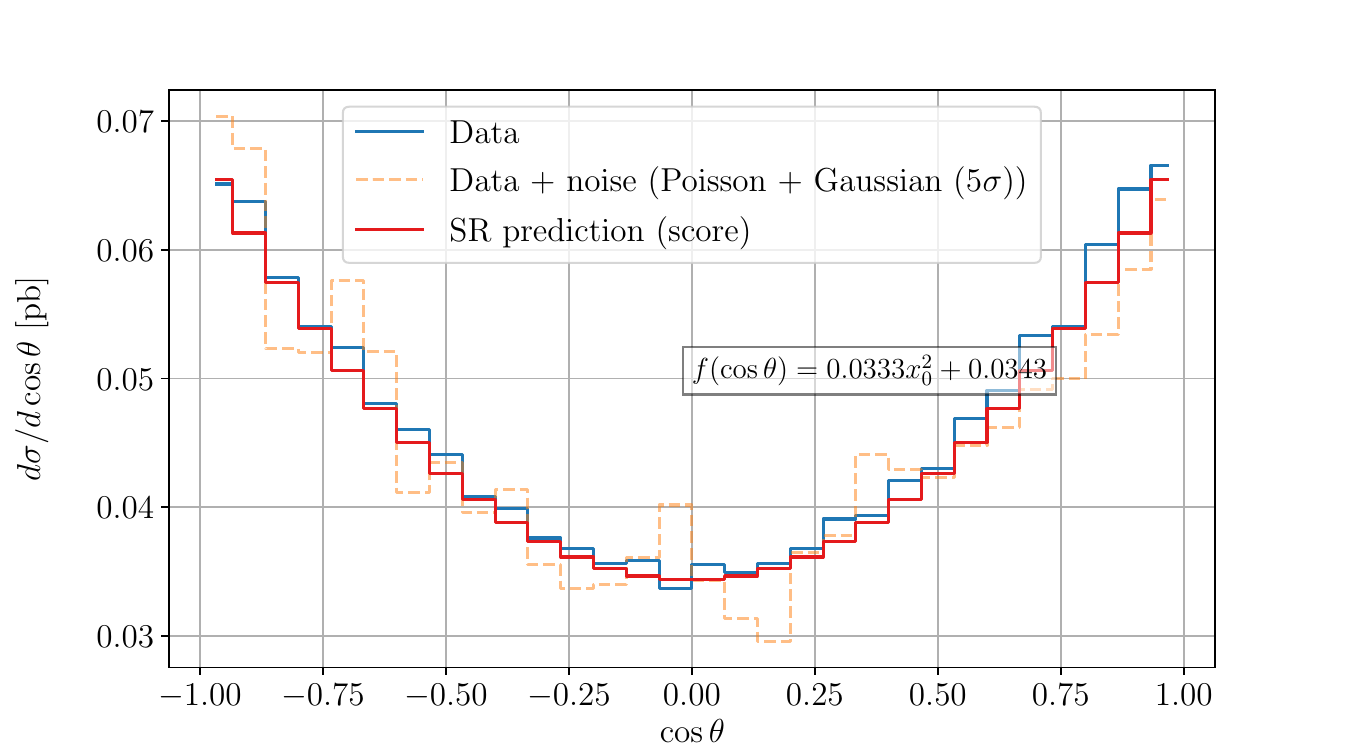}
    \caption*{(h)}
  \end{minipage}
  
  \caption{Distributions in the presence of low to moderate noise.}
  \label{fig:noise_images_1}
\end{figure}

\begin{figure}[ht!]
  \centering
  \begin{minipage}{0.49\linewidth}
    \centering
    \includegraphics[width=\linewidth]{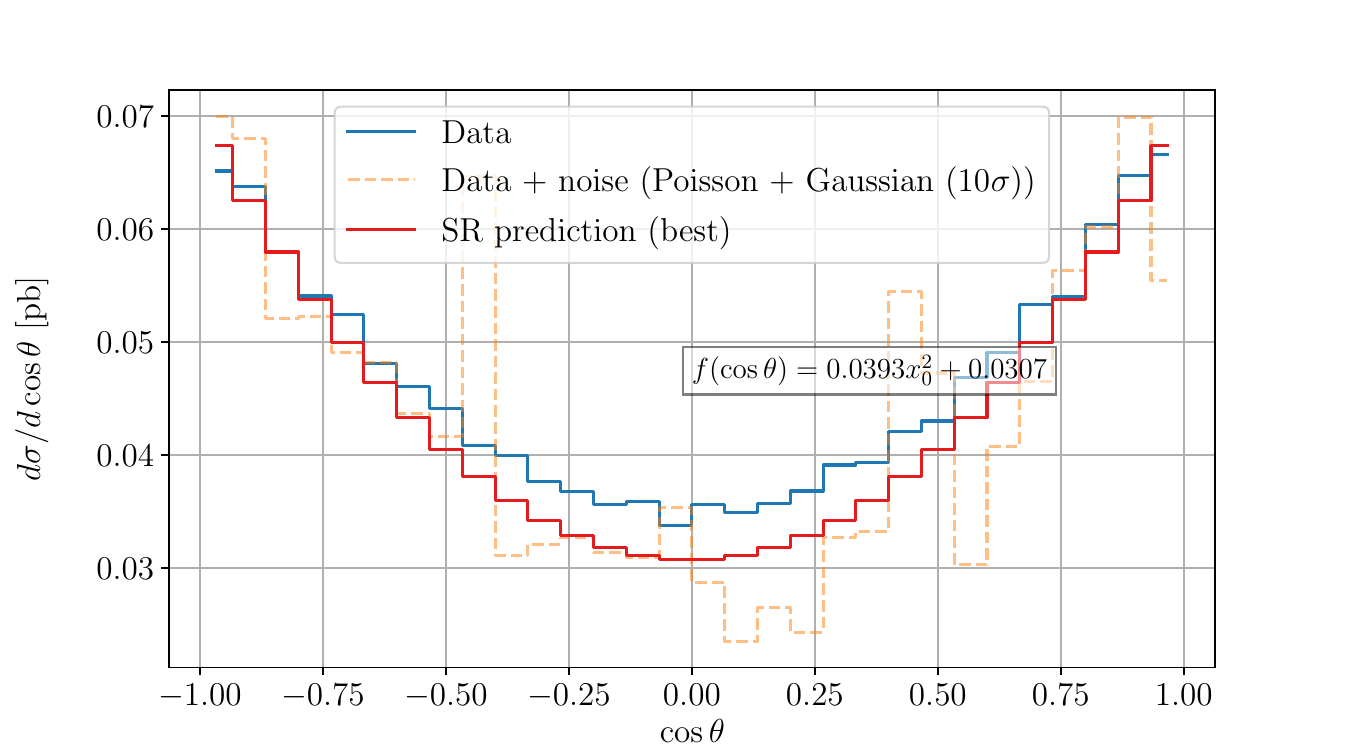}
    \caption*{(a)}
  \end{minipage}%
  \hspace{0.01\linewidth}
  \begin{minipage}{0.49\linewidth}
    \centering
    \includegraphics[width=\linewidth]{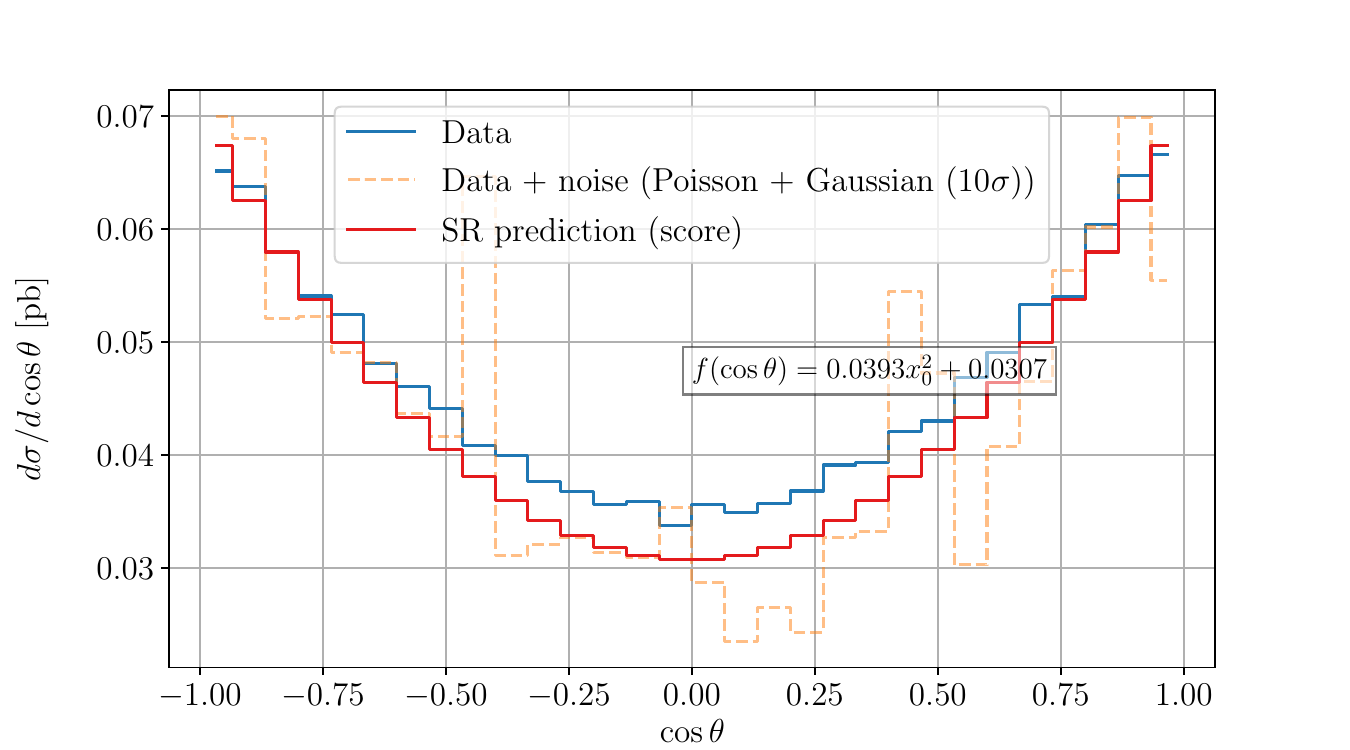}
    \caption*{(b)}
  \end{minipage}

  \vspace{0.2cm} 

  \begin{minipage}{0.49\linewidth}
    \centering
    \includegraphics[width=\linewidth]{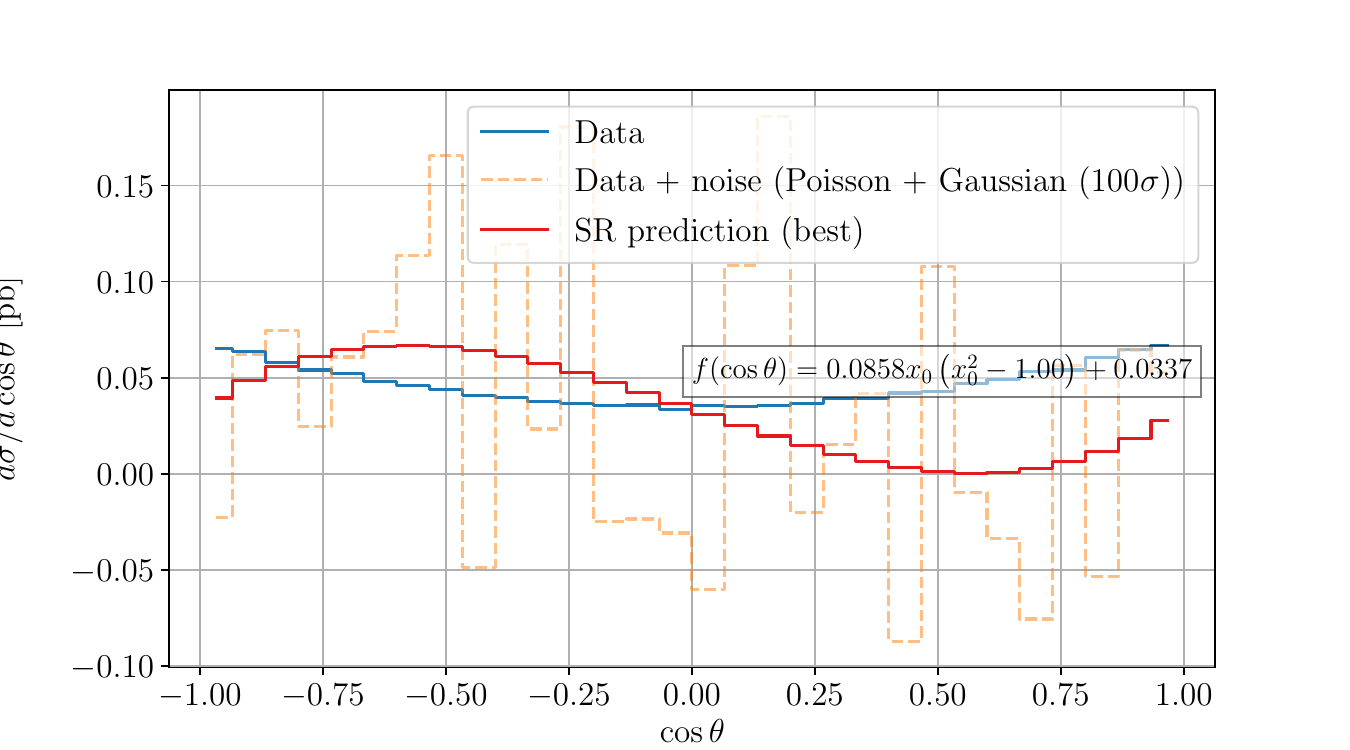}
    \caption*{(c)}
  \end{minipage}%
  \hspace{0.01\linewidth}
  \begin{minipage}{0.49\linewidth}
    \centering
    \includegraphics[width=\linewidth]{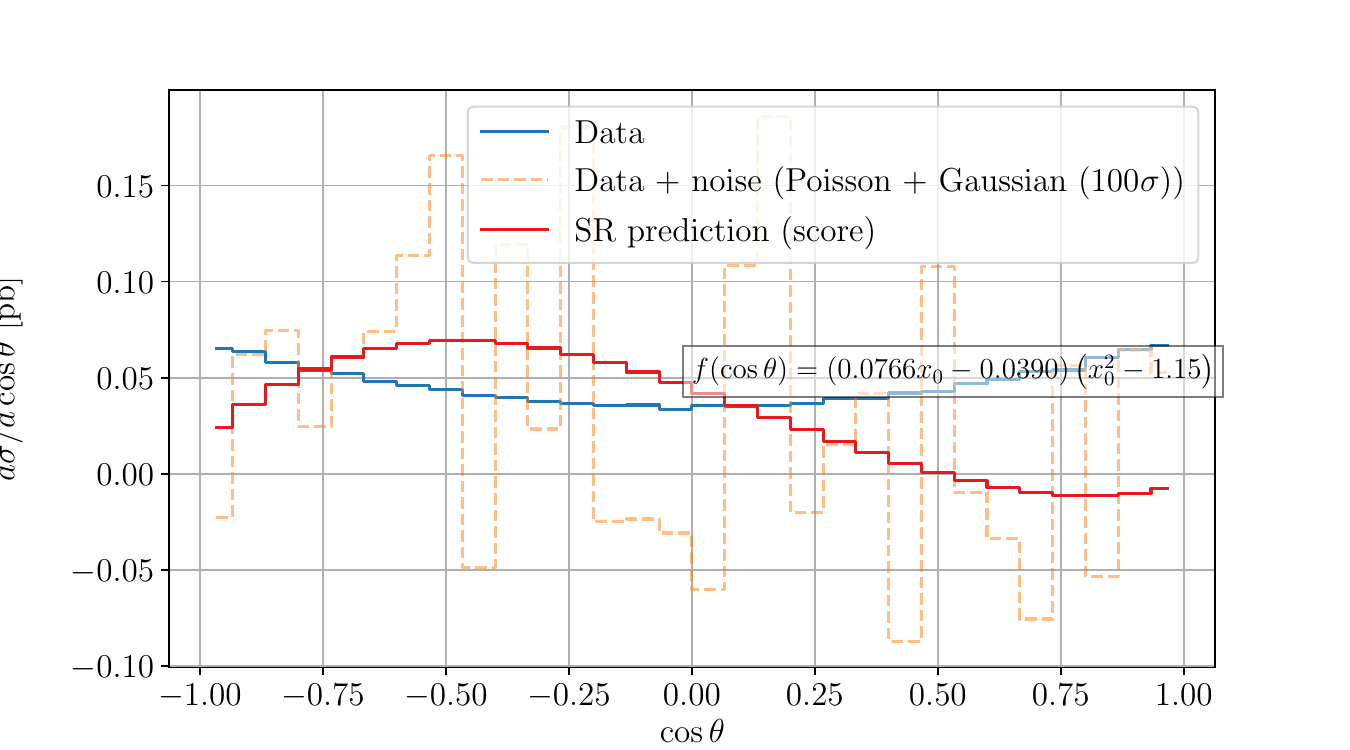}
    \caption*{(d)}
  \end{minipage}
  
  \caption{Distributions in the presence of high to extreme noise.}
  \label{fig:noise_images_2}
\end{figure}

Not only the correct functional dependence in $\cos^2 \theta$ is recovered, but also the constants to a very good accuracy. We can
assess this by parametrising the SR prediction as
\begin{equation}
  \text{SR} (\cos \theta) = c_1 + c_2 \cdot \cos^2 \theta,
\end{equation}
wth $c_1$ and $c_2$ being constants of the prediction. In Fig. \ref{fig:sr_l1_norm} we show their relative size when training
the regressor on distributions with different numbers of bins. We see that the constants are recovered within a $2\%$ precision,
even in the case of very granular binnings.
Now, we carry out a detailed analyses of the SR predictions in the presence of noisy data
to different levels. We fix the number of bins to 30, and we inject noise in the simulation using the prescription
described earlier. In Figs. \ref{fig:noise_images_1} we show the measured cross section, the noisy data that is used to
train the regressor, and the SR prediction, for different levels of noise ranging from purely Poissonian noise to 
$n=1, 2, 5$ Gaussian noise. The correct analytical formula from QED is recovered in all cases by the regressor.
In Figs. \ref{fig:noise_images_2} we see the results in the presence of high and extreme noisy. In the case of $n=10$ Gaussian noise
the SR show still a good description of the underlying distribution, although the values of the constants has slightly changed.
In the presence of extreme noise $n=100$, the signal stops being descriptive and, naturally, the SR results is not correct any more.
These results show that, at least in the case of relatively simple analytical formulas, SR is able to produce good results even
in the case of underlying noise. Robustness to noise is a key feature in any regression method that uses real data and, in this way,
it can be used to contribute to scientific discovery.
SR can be used to study distributions whose analytical description from first principles is just partially known.
If we look at Eq. (\ref{eq:full_coeff}) and consider the level 2 setup $p \ p \to \gamma \to \mu^+ \mu^-$, where
there are no angular coefficients but just the $1 + \cos^2 \theta$ angular dependence, we can reweight the distribution
by a $(1 + \cos^2 \theta)^{-1}$ factor and train the regressor on a reweighted distribution which is simply described by PDFs
and structure functions, whose analytical form is not known. In this way, we use our knowledge of the underlying theory to
isolate the data that cannot be described by known formulas, and train a regressor on it. 

This procedure, which implies functional constraints on the equation that has to be worked out by the regressor,
is the one that we are currently developing to study the full angular coefficients in $W$ and $Z$ production, to be
completed in the near future.




\section{Conclusion}

In this chapter we have discussed the potential of SR to obtain simple and accurate formulas from simulated and potentially complex datasets, in particular
in the context of phenomenological analyses of collider data. By focusing on well-understood QED processes, we established a robust benchmark to validate the
SR framework. Our study demonstrated that SR could effectively recover known analytic distributions from simulated data, even in the presence
of varying levels of noise, and subject to different binning conditions.

We implemented SR using the PySR software package, which uses a multi-population evolutionary algorithm. This approach allowed
us to optimise for both accuracy and simplicity, crucial for ensuring the interpretability of the derived models. Our results
confirmed that SR could recover the expected angular distributions at very good accuracy, even with fine binning and high noise levels.
The stability of the SR results under different noise conditions was thoroughly assessed, underscoring the robustness of this method in handling data uncertainties
and fluctuations.

We also outlined a strategic pathway to progressively apply SR to more complex processes involving angular coefficients in $W$ and $Z$ boson production. This involves initial validation using simpler processes and gradually introducing complexity by considering the effects of PDFs, EW corrections, and higher order QCD corrections. This is mapped into more complex functional forms to be fitted, and higher number of angular coefficients to account for the decomposition of the cross section. By reweighting distributions to isolate unknown components, we can effectively use SR to gain new insights into these coefficients.

The successful recovery of known analytic expressions from simulated data and the stability of SR under noise conditions highlight its potential to contribute significantly to collider phenomenology and to enhance scientific discovery. The ongoing development aims to apply this validated SR approach to infer analytic expressions for angular coefficients in vector boson production, which is critical for precision measurements at the LHC.

There are a number of ways in which this work can be extended aiming towards interpretable ML for scientific discovery in theoretical physics. One reason for this is that, while in experimental particle physics ML is used as a data science tool, in theoretical particle physics the problems tend to be of symbolic nature: looking for patterns, working in approximations, generalising from examples, proving theorems, and so on, as it has been discussed in Refs. \cite{Moura2021,palmskog}. ML frameworks for symbolic problems have also been able to find new solutions to long standing problems in mathematics \cite{RomeraParedes2023}, simplifying polylogarithms \cite{Dersy:2022bym} and spinor-helicity amplitudes.

SR presents a powerful tool for combining ML with analytical intuition, providing a means to derive interpretable and efficient models from complex datasets. This chapter lays the groundwork for future applications of SR in collider phenomenology, promising advancements in our understanding and analysis of high energy physics processes. Further results and detailed studies will be included in the final publication, contributing to the ongoing efforts in precision physics at the LHC and, more in general, to ML-assisted scientific discovery in theoretical physics.

\chapter{Conclusions}
\label{ch:conclusions}
  \epigraph{Gracias a la vida que me ha dado tanto \\ Me dio el corazón que agita su marco \\ Cuando miro el fruto del cerebro humano \\ \vspace{1em} Thanks to life, which has given me so much \\ It gave me a heart that shakes in its frame \\ When I see the fruit of the human brain}{\textit{\\ from Gracias a la vida, \\ by Violeta Parra (trad. M. M. A.).}}

In this thesis we have explored the PDF-BSM interplay. We have addressed it in the context of simultaneous fits and their comparison with PDF-only and BSM-only fits, and by assessing the potential absorption of BSM effects by the PDFs. This is an important discussion that has been largely overlooked in the literature since the PDF and BSM communities do not always interact. A more integrated approach is \textit{crucial} in the era of precision physics, where the LHC is probing the high energy frontier with unprecedented accuracy. 

The development of \simunet{}, an open-source deep learning methodology, stands out as a significant contribution to the high energy physics toolkit by enabling, among other possibilities, simultaneous global fits of EFT Wilson coefficients and PDFs, and the assessment of the potential absorption of BSM physics by the PDFs using heavy new physics models defined by the user. 

Our application of the \simunet{} methodology in the top quark sector and in a global dataset has shown the importance of integrated PDF-BSM fits. Naturally, our results in SM PDF fits, and fixed-PDF fits recover the literature results, but from a simultaneous determination we have obtained improved constraints that take into account the cross talk between the structure of the proton and BSM physics. In this way, more accurate and reliable PDF and BSM fits can be achieved.

We also showed the potential of the \simunet{} methodology in assessing the potential absorption of BSM effects by the PDFs. We have shown that PDFs, when fitted under the assumption of the SM, can absorb BSM physics. This can render the interpretation of experimental data ambiguous by missing the presence of new physics. Additionally, we have shown how the use of BSM `contaminated' PDF sets can lead to apparent but spurious tensions between theoretical predictions and experimental measurements, detecting the presence of new physics where there is none. We have also devised a methodology to minimise the impact of this absorption in collider observables. This is an important point to consider in the interpretation of experimental data in searches of BSM physics.

We have also explored how potential BSM effects, parametrised in terms of the SMEFT, can affect the evolution of the PDFs with the energy scale, as described by the DGLAP equations. We have shown that the kinematic enhancement of SMEFT operators keeps the collinear limit safe from BSM effects, leaving the splitting functions unchanged. We have also discussed how SMEFT operators can change the running of the strong coupling, a crucial parameter in PDF evolution. We showed that, although SMEFT corrections enter the running, their impact is numerically small.

Neural networks are at the core of the \simunet{} methodology. In this thesis, however, we have also explored the potential of other machine learning techniques. We have shown how symbolic regression can be used to provide simple and highly accurate models for collider physics. We have done this by recovering first principle calculations in QFT from experimental distributions in the presence of even severe noise. Our findings pave the way for future research to integrate novel machine learning tools which could enhance the precision and interpretability of theoretical models.

The results obtained in this thesis are simply a step in the direction of a more comprehensive and integrated approach to study the interplay between PDFs and BSM physics and the study of fundamental physics through the lens of new machine learning techniques, but much work remains to be done.

The \simunet{} methodology is tailored to perform fits with an arbitrary number of operators and using data from different experiments, current and future, and processes. It will be interesting to see how new experimental measurements improve the constraints on BSM coefficients, and how an increasing number of operators can account for the presence of new physics in a global footing. In the future, it would be important that the methodology is adapted to perform simultaneous fits along SM parameters, that takes fully into account the potential effects of quadratic EFT corrections, that is able to include the action of higher dimensional EFT operators, and that considers the running of SM and BSM parameters. It is also relevant that, in the future, the constraints coming from simultaneous PDF-BSM fits can be mapped to explicit UV-completions. 

To carefully assess the uncertainties in PDF and BSM fits, and to incorporate prior knowledge on future analyses, it could be relevant to perform a full Bayesian analysis of the PDF-BSM interplay. Naturally, as the computational complexity of a Bayesian analysis is high, it would be important to develop new methodologies to perform such analysis in a computationally efficient way. Additionally, methods that allow for a more flexible treatment of the likelihood, such as simulation based inference, could be explored.

More in general, the rapid development in machine learning techniques to solve problems in science and technology presents a good opportunity to explore fundamental physics through novel approaches. As we discussed in this thesis, symbolic regression can be used to provide simple and highly accurate models that describe experimental data, and it would be interesting see if PDFs can be parametrised in a similar way: this would provide the flexibility that fixed PDF parametrisations lack, and the simplicity and interpretability that a neural network parametrisation lacks. Machine learning is already widely used in experimental particle physics in the context of data analysis, but it is still underused in theoretical particle physics, where problems tend to be of symbolic nature. Still, big scale machine learning systems such as foundation models or large language models have seen impressive applications across different fields, and it would be interesting to see how they can be applied to fundamental physics. 

As physicists, we get the immense privilege of exploring nature at the most fundamental level. With open and curious minds we can push the boundaries of human knowledge for the pleasure of understanding and, hopefully, for the benefit of society. The search for new physics is a long and arduous journey, but it can also be a journey full of surprises and opportunities. Let us be ready to seize them!

%

\renewcommand{\bibname}{References}
\cleardoublepage
\phantomsection
\addcontentsline{toc}{chapter}{References}
\bibliographystyle{unsrt}
\bibliography{thesis}

\begin{thebibliography}{100}

\bibitem{laertius2020lives}
Diogenes Laertius.
\newblock {\em Lives of the eminent philosophers}.
\newblock Oxford University Press, 2020.

\bibitem{Bjorken:1965zz}
James~D. Bjorken and Sidney~D. Drell.
\newblock {Relativistic quantum fields}.
\newblock 1965.

\bibitem{Peskin:257493}
Michael~E Peskin and Daniel~V Schroeder.
\newblock {\em {An introduction to quantum field theory}}.
\newblock Westview, Boulder, CO, 1995.
\newblock Includes exercises.

\bibitem{Caswell:1974gg}
William~E. Caswell.
\newblock {Asymptotic Behavior of Nonabelian Gauge Theories to Two Loop Order}.
\newblock {\em Phys. Rev. Lett.}, 33:244, 1974.

\bibitem{Jones:1974mm}
D.~R.~T. Jones.
\newblock {Two Loop Diagrams in Yang-Mills Theory}.
\newblock {\em Nucl. Phys. B}, 75:531, 1974.

\bibitem{Tarasov:1980au}
O.~V. Tarasov, A.~A. Vladimirov, and A.~Yu. Zharkov.
\newblock {The Gell-Mann-Low Function of QCD in the Three Loop Approximation}.
\newblock {\em Phys. Lett. B}, 93:429--432, 1980.

\bibitem{Larin:1993tp}
S.~A. Larin and J.~A.~M. Vermaseren.
\newblock {The Three loop QCD Beta function and anomalous dimensions}.
\newblock {\em Phys. Lett. B}, 303:334--336, 1993.

\bibitem{vanRitbergen:1997va}
T.~van Ritbergen, J.~A.~M. Vermaseren, and S.~A. Larin.
\newblock {The Four loop beta function in quantum chromodynamics}.
\newblock {\em Phys. Lett. B}, 400:379--384, 1997.

\bibitem{Baikov:2016tgj}
P.~A. Baikov, K.~G. Chetyrkin, and J.~H. K\"uhn.
\newblock {Five-Loop Running of the QCD coupling constant}.
\newblock {\em Phys. Rev. Lett.}, 118(8):082002, 2017.

\bibitem{Bjorken:1969ja}
J.~D. Bjorken and Emmanuel~A. Paschos.
\newblock {Inelastic Electron Proton and gamma Proton Scattering, and the
  Structure of the Nucleon}.
\newblock {\em Phys. Rev.}, 185:1975--1982, 1969.

\bibitem{feynmanparton}
Richard~P. Feynman.
\newblock Very high-energy collisions of hadrons.
\newblock {\em Phys. Rev. Lett.}, 23:1415--1417, Dec 1969.

\bibitem{callangross.22.156}
C.~G. Callan and David~J. Gross.
\newblock High-energy electroproduction and the constitution of the electric
  current.
\newblock {\em Phys. Rev. Lett.}, 22:156--159, Jan 1969.

\bibitem{Perkins:1982xb}
D.~H. Perkins.
\newblock {\em {Introduction to high energy physics}}.
\newblock 1982.

\bibitem{Dokshitzer:1977sg}
Yuri~L. Dokshitzer.
\newblock {Calculation of the Structure Functions for Deep Inelastic Scattering
  and e+ e- Annihilation by Perturbation Theory in Quantum Chromodynamics.}
\newblock {\em Sov. Phys. JETP}, 46:641--653, 1977.

\bibitem{Gribov:1972ri}
V.~N. Gribov and L.~N. Lipatov.
\newblock {Deep inelastic e p scattering in perturbation theory}.
\newblock {\em Sov. J. Nucl. Phys.}, 15:438--450, 1972.

\bibitem{Altarelli:1977zs}
Guido Altarelli and G.~Parisi.
\newblock {Asymptotic Freedom in Parton Language}.
\newblock {\em Nucl. Phys. B}, 126:298--318, 1977.

\bibitem{Moch:1999eb}
S.~Moch and J.~A.~M. Vermaseren.
\newblock {Deep inelastic structure functions at two loops}.
\newblock {\em Nucl. Phys. B}, 573:853--907, 2000.

\bibitem{Vogt:2004mw}
A.~Vogt, S.~Moch, and J.~A.~M. Vermaseren.
\newblock {The Three-loop splitting functions in QCD: The Singlet case}.
\newblock {\em Nucl. Phys. B}, 691:129--181, 2004.

\bibitem{Moch:2004pa}
S.~Moch, J.~A.~M. Vermaseren, and A.~Vogt.
\newblock {The Three loop splitting functions in QCD: The Nonsinglet case}.
\newblock {\em Nucl. Phys. B}, 688:101--134, 2004.

\bibitem{Vermaseren:2005qc}
J.~A.~M. Vermaseren, A.~Vogt, and S.~Moch.
\newblock {The third-order QCD corrections to deep-inelastic scattering by
  photon exchange}.
\newblock {\em Nucl. Phys.}, B724:3, 2005.

\bibitem{Davies:2016jie}
J.~Davies, A.~Vogt, B.~Ruijl, T.~Ueda, and J.~A.~M. Vermaseren.
\newblock {Large-nf contributions to the four-loop splitting functions in QCD}.
\newblock {\em Nucl. Phys. B}, 915:335--362, 2017.

\bibitem{Moch:2017uml}
S.~Moch, B.~Ruijl, T.~Ueda, J.~A.~M. Vermaseren, and A.~Vogt.
\newblock {Four-Loop Non-Singlet Splitting Functions in the Planar Limit and
  Beyond}.
\newblock {\em JHEP}, 10:041, 2017.

\bibitem{Davies:2022ofz}
J.~Davies, C.~H. Kom, S.~Moch, and A.~Vogt.
\newblock {Resummation of small-x double logarithms in QCD: inclusive
  deep-inelastic scattering}.
\newblock {\em JHEP}, 08:135, 2022.

\bibitem{Henn:2019swt}
Johannes~M. Henn, Gregory~P. Korchemsky, and Bernhard Mistlberger.
\newblock {The full four-loop cusp anomalous dimension in $\mathcal{N}=4$ super
  Yang-Mills and QCD}.
\newblock {\em JHEP}, 04:018, 2020.

\bibitem{Duhr:2022cob}
Claude Duhr, Bernhard Mistlberger, and Gherardo Vita.
\newblock {Soft integrals and soft anomalous dimensions at N$^{3}$LO and
  beyond}.
\newblock {\em JHEP}, 09:155, 2022.

\bibitem{Moch:2021qrk}
S.~Moch, B.~Ruijl, T.~Ueda, J.~A.~M. Vermaseren, and A.~Vogt.
\newblock {Low moments of the four-loop splitting functions in QCD}.
\newblock {\em Phys. Lett. B}, 825:136853, 2022.

\bibitem{Soar:2009yh}
G.~Soar, S.~Moch, J.~A.~M. Vermaseren, and A.~Vogt.
\newblock {On Higgs-exchange DIS, physical evolution kernels and fourth-order
  splitting functions at large x}.
\newblock {\em Nucl. Phys. B}, 832:152--227, 2010.

\bibitem{Falcioni:2023luc}
G.~Falcioni, F.~Herzog, S.~Moch, and A.~Vogt.
\newblock {Four-loop splitting functions in QCD \textendash{} The quark-quark
  case}.
\newblock {\em Phys. Lett. B}, 842:137944, 2023.

\bibitem{Falcioni:2023vqq}
G.~Falcioni, F.~Herzog, S.~Moch, and A.~Vogt.
\newblock {Four-loop splitting functions in QCD \textendash{} The
  gluon-to-quark case}.
\newblock {\em Phys. Lett. B}, 846:138215, 2023.

\bibitem{Moch:2023tdj}
S.~Moch, B.~Ruijl, T.~Ueda, J.~Vermaseren, and A.~Vogt.
\newblock {Additional moments and x-space approximations of four-loop splitting
  functions in QCD}.
\newblock 10 2023.

\bibitem{Falcioni:2023tzp}
G.~Falcioni, F.~Herzog, S.~Moch, J.~Vermaseren, and A.~Vogt.
\newblock {The double fermionic contribution to the four-loop quark-to-gluon
  splitting function}.
\newblock {\em Phys. Lett. B}, 848:138351, 2024.

\bibitem{Appelquist:1974tg}
Thomas Appelquist and J.~Carazzone.
\newblock {Infrared Singularities and Massive Fields}.
\newblock {\em Phys. Rev. D}, 11:2856, 1975.

\bibitem{Aebischer:2023nnv}
Lukas Allwicher et~al.
\newblock {Computing tools for effective field theories: SMEFT-Tools 2022
  Workshop Report, 14\textendash{}16th September 2022, Z\"urich}.
\newblock {\em Eur. Phys. J. C}, 84(2):170, 2024.

\bibitem{Jenkins:2013zja}
Elizabeth~E. Jenkins, Aneesh~V. Manohar, and Michael Trott.
\newblock {Renormalization Group Evolution of the Standard Model Dimension Six
  Operators I: Formalism and lambda Dependence}.
\newblock {\em JHEP}, 10:087, 2013.

\bibitem{Jenkins:2013wua}
Elizabeth~E. Jenkins, Aneesh~V. Manohar, and Michael Trott.
\newblock {Renormalization Group Evolution of the Standard Model Dimension Six
  Operators II: Yukawa Dependence}.
\newblock {\em JHEP}, 01:035, 2014.

\bibitem{Alonso:2013hga}
Rodrigo Alonso, Elizabeth~E. Jenkins, Aneesh~V. Manohar, and Michael Trott.
\newblock {Renormalization Group Evolution of the Standard Model Dimension Six
  Operators III: Gauge Coupling Dependence and Phenomenology}.
\newblock {\em JHEP}, 04:159, 2014.

\bibitem{Alonso:2014zka}
Rodrigo Alonso, Hsi-Ming Chang, Elizabeth~E. Jenkins, Aneesh~V. Manohar, and
  Brian Shotwell.
\newblock {Renormalization group evolution of dimension-six baryon number
  violating operators}.
\newblock {\em Phys. Lett. B}, 734:302--307, 2014.

\bibitem{Elias-Miro:2013mua}
J.~Elias-Miro, J.~R. Espinosa, E.~Masso, and A.~Pomarol.
\newblock {Higgs windows to new physics through d=6 operators: constraints and
  one-loop anomalous dimensions}.
\newblock {\em JHEP}, 11:066, 2013.

\bibitem{Elias-Miro:2013eta}
Joan Elias-Mir\'o, Christophe Grojean, Rick~S. Gupta, and David Marzocca.
\newblock {Scaling and tuning of EW and Higgs observables}.
\newblock {\em JHEP}, 05:019, 2014.

\bibitem{murayama}
Brian Henning, Xiaochuan Lu, and Hitoshi Murayama.
\newblock How to use the standard model effective field theory.
\newblock {\em Journal of High Energy Physics}, 2016(1), January 2016.

\bibitem{Buchmuller:1985jz}
W.~Buchmuller and D.~Wyler.
\newblock {Effective Lagrangian Analysis of New Interactions and Flavor
  Conservation}.
\newblock {\em Nucl. Phys. B}, 268:621--653, 1986.

\bibitem{Grzadkowski:2010es}
B.~Grzadkowski, M.~Iskrzynski, M.~Misiak, and J.~Rosiek.
\newblock {Dimension-Six Terms in the Standard Model Lagrangian}.
\newblock {\em JHEP}, 10:085, 2010.

\bibitem{Brivio:2017vri}
Ilaria Brivio and Michael Trott.
\newblock {The Standard Model as an Effective Field Theory}.
\newblock {\em Phys. Rept.}, 793:1--98, 2019.

\bibitem{Carrazza:2019sec}
Stefano Carrazza, Celine Degrande, Shayan Iranipour, Juan Rojo, and Maria
  Ubiali.
\newblock {Can New Physics hide inside the proton?}
\newblock {\em Phys. Rev. Lett.}, 123(13):132001, 2019.

\bibitem{Greljo:2021kvv}
Admir Greljo, Shayan Iranipour, Zahari Kassabov, Maeve Madigan, James Moore,
  Juan Rojo, Maria Ubiali, and Cameron Voisey.
\newblock {Parton distributions in the SMEFT from high-energy Drell-Yan tails}.
\newblock 4 2021.

\bibitem{Gao:2022srd}
Jun Gao, MeiSen Gao, T.~J. Hobbs, DianYu Liu, and XiaoMin Shen.
\newblock {Simultaneous CTEQ-TEA extraction of PDFs and SMEFT parameters from
  jet and $t{\bar t}$ data}.
\newblock 11 2022.

\bibitem{Iranipour:2022iak}
Shayan Iranipour and Maria Ubiali.
\newblock {A new generation of simultaneous fits to LHC data using deep
  learning}.
\newblock {\em JHEP}, 05:032, 2022.

\bibitem{Kassabov:2023hbm}
Zahari Kassabov, Maeve Madigan, Luca Mantani, James Moore, Manuel~Morales
  Alvarado, Juan Rojo, and Maria Ubiali.
\newblock {The top quark legacy of the LHC Run II for PDF and SMEFT analyses}.
\newblock 3 2023.

\bibitem{Hammou:2023heg}
Elie Hammou, Zahari Kassabov, Maeve Madigan, Michelangelo~L. Mangano, Luca
  Mantani, James Moore, Manuel~Morales Alvarado, and Maria Ubiali.
\newblock {Hide and seek: how PDFs can conceal new physics}.
\newblock {\em JHEP}, 11:090, 2023.

\bibitem{Madigan:2021uho}
Maeve Madigan and James Moore.
\newblock {Parton Distributions in the SMEFT from high-energy Drell-Yan tails}.
\newblock {\em PoS}, EPS-HEP2021:424, 2022.

\bibitem{Costantini:2024xae}
Mark~N. Costantini, Elie Hammou, Zahari Kassabov, Maeve Madigan, Luca Mantani,
  Manuel Morales~Alvarado, James~M. Moore, and Maria Ubiali.
\newblock {SIMUnet: an open-source tool for simultaneous global fits of EFT
  Wilson coefficients and PDFs}.
\newblock 2 2024.

\bibitem{NNPDF:2021njg}
Richard~D. Ball et~al.
\newblock {The path to proton structure at 1\% accuracy}.
\newblock {\em Eur. Phys. J. C}, 82(5):428, 2022.

\bibitem{NNPDF:2021uiq}
Richard~D. Ball et~al.
\newblock {An open-source machine learning framework for global analyses of
  parton distributions}.
\newblock {\em Eur. Phys. J. C}, 81(10):958, 2021.

\bibitem{nnpdfcode}
Richard~D. Ball et~al.
\newblock Nnpdf/nnpdf: nnpdf v4.0.3.
\newblock \url{https://doi.org/10.5281/zenodo.5362228}, September 2021.

\bibitem{NNPDF:2024dpb}
Richard~D. Ball et~al.
\newblock {Determination of the theory uncertainties from missing higher orders
  on NNLO parton distributions with percent accuracy}.
\newblock 1 2024.

\bibitem{NNPDF:2024djq}
Richard~D. Ball et~al.
\newblock {Photons in the proton: implications for the LHC}.
\newblock 1 2024.

\bibitem{zahari_kassabov_2019_2571601}
Zahari Kassabov.
\newblock {Reportengine: A framework for declarative data analysis}.
\newblock \url{https://doi.org/10.5281/zenodo.2571601}, February 2019.

\bibitem{Bailey:2020ooq}
S.~Bailey, T.~Cridge, L.A. Harland-Lang, A.D. Martin, and R.S. Thorne.
\newblock {Parton distributions from LHC, HERA, Tevatron and fixed target data:
  MSHT20 PDFs}.
\newblock 12 2020.

\bibitem{Hou:2019efy}
Tie-Jiun Hou et~al.
\newblock {New CTEQ global analysis of quantum chromodynamics with
  high-precision data from the LHC}.
\newblock {\em Phys. Rev. D}, 103(1):014013, 2021.

\bibitem{Carrazza:2021yrg}
Stefano Carrazza, Juan~M. Cruz-Martinez, and Roy Stegeman.
\newblock {A data-based parametrization of parton distribution functions}.
\newblock {\em Eur. Phys. J. C}, 82(2):163, 2022.

\bibitem{Ball:2018twp}
Richard~D. Ball, Emanuele~R. Nocera, and Rosalyn~L. Pearson.
\newblock {Nuclear Uncertainties in the Determination of Proton PDFs}.
\newblock {\em Eur. Phys. J.}, C79(3):282, 2019.

\bibitem{Ball:2020xqw}
Richard~D. Ball, Emanuele~R. Nocera, and Rosalyn~L. Pearson.
\newblock {Deuteron Uncertainties in the Determination of Proton PDFs}.
\newblock {\em Eur. Phys. J. C}, 81(1):37, 2021.

\bibitem{AbdulKhalek:2019bux}
Rabah Abdul~Khalek et~al.
\newblock {A First Determination of Parton Distributions with Theoretical
  Uncertainties}.
\newblock 2019.

\bibitem{AbdulKhalek:2019ihb}
Rabah Abdul~Khalek et~al.
\newblock {Parton Distributions with Theory Uncertainties: General Formalism
  and First Phenomenological Studies}.
\newblock {\em Eur. Phys. J. C}, 79(11):931, 2019.

\bibitem{Kassabov:2022orn}
Zahari Kassabov, Maria Ubiali, and Cameron Voisey.
\newblock {Parton distributions with scale uncertainties: a Monte Carlo
  sampling approach}.
\newblock {\em JHEP}, 03:148, 2023.

\bibitem{DAgostini:1993arp}
G.~D'Agostini.
\newblock {On the use of the covariance matrix to fit correlated data}.
\newblock {\em Nucl. Instrum. Meth. A}, 346:306--311, 1994.

\bibitem{Ball:2009qv}
Richard~D. Ball et~al.
\newblock {Fitting Parton Distribution Data with Multiplicative Normalization
  Uncertainties}.
\newblock {\em JHEP}, 05:075, 2010.

\bibitem{Maguire:2017ypu}
Eamonn Maguire, Lukas Heinrich, and Graeme Watt.
\newblock {HEPData: a repository for high energy physics data}.
\newblock {\em J. Phys. Conf. Ser.}, 898(10):102006, 2017.

\bibitem{Bertone:2016lga}
Valerio Bertone, Stefano Carrazza, and Nathan~P. Hartland.
\newblock {APFELgrid: a high performance tool for parton density
  determinations}.
\newblock {\em Comput. Phys. Commun.}, 212:205--209, 2017.

\bibitem{Carli:2010rw}
Tancredi Carli et~al.
\newblock {A posteriori inclusion of parton density functions in NLO QCD
  final-state calculations at hadron colliders: The APPLGRID Project}.
\newblock {\em Eur.Phys.J.}, C66:503, 2010.

\bibitem{Wobisch:2011ij}
M.~Wobisch, D.~Britzger, T.~Kluge, K.~Rabbertz, and F.~Stober.
\newblock {Theory-Data Comparisons for Jet Measurements in Hadron-Induced
  Processes}.
\newblock 2011.

\bibitem{Bertone:2013vaa}
Valerio Bertone, Stefano Carrazza, and Juan Rojo.
\newblock {APFEL: A PDF Evolution Library with QED corrections}.
\newblock {\em Comput.Phys.Commun.}, 185:1647, 2014.

\bibitem{abadi2016tensorflow}
Mart{\'\i}n Abadi, Paul Barham, Jianmin Chen, Zhifeng Chen, Andy Davis, Jeffrey
  Dean, Matthieu Devin, Sanjay Ghemawat, Geoffrey Irving, Michael Isard, et~al.
\newblock Tensorflow: A system for large-scale machine learning.
\newblock In {\em 12th $\{$USENIX$\}$ Symposium on Operating Systems Design and
  Implementation ($\{$OSDI$\}$ 16)}, pages 265--283, 2016.

\bibitem{2015CS&D....8a4008B}
James {Bergstra}, Brent {Komer}, Chris {Eliasmith}, Dan {Yamins}, and David~D.
  {Cox}.
\newblock {Hyperopt: a Python library for model selection and hyperparameter
  optimization}.
\newblock {\em Computational Science and Discovery}, 8(1):014008, January 2015.

\bibitem{Candido:2020yat}
Alessandro Candido, Stefano Forte, and Felix Hekhorn.
\newblock {Can $ \overline{\mathrm{MS}} $ parton distributions be negative?}
\newblock {\em JHEP}, 11:129, 2020.

\bibitem{NNPDF:2014otw}
Richard~D. Ball et~al.
\newblock {Parton distributions for the LHC Run II}.
\newblock {\em JHEP}, 04:040, 2015.

\bibitem{DelDebbio:2021whr}
Luigi Del~Debbio, Tommaso Giani, and Michael Wilson.
\newblock {Bayesian approach to inverse problems: an application to NNPDF
  closure testing}.
\newblock {\em Eur. Phys. J. C}, 82(4):330, 2022.

\bibitem{Costantini:2024wby}
Mark~N. Costantini, Maeve Madigan, Luca Mantani, and James~M. Moore.
\newblock {A critical study of the Monte Carlo replica method}.
\newblock 4 2024.

\bibitem{NewMuon:1996fwh}
M.~Arneodo et~al.
\newblock {Measurement of the proton and deuteron structure functions, F2(p)
  and F2(d), and of the ratio sigma-L / sigma-T}.
\newblock {\em Nucl. Phys. B}, 483:3--43, 1997.

\bibitem{NewMuon:1996uwk}
M.~Arneodo et~al.
\newblock {Accurate measurement of F2(d) / F2(p) and R**d - R**p}.
\newblock {\em Nucl. Phys. B}, 487:3--26, 1997.

\bibitem{ATLAS:2014nxi}
Georges Aad et~al.
\newblock {Measurement of the $t\bar{t}$ production cross-section using $e\mu $
  events with b-tagged jets in pp collisions at $\sqrt{s}$ = 7 and 8
  $\,\mathrm{TeV}$ with the ATLAS detector}.
\newblock {\em Eur. Phys. J. C}, 74(10):3109, 2014.
\newblock [Addendum: Eur.Phys.J.C 76, 642 (2016)].

\bibitem{CMS:2014fut}
Serguei Chatrchyan et~al.
\newblock {Observation of the associated production of a single top quark and a
  $W$ boson in $pp$ collisions at $\sqrt s = $8 TeV}.
\newblock {\em Phys. Rev. Lett.}, 112(23):231802, 2014.

\bibitem{Ball:2021leu}
Richard~D. Ball et~al.
\newblock {The path to proton structure at 1\% accuracy}.
\newblock {\em Eur. Phys. J. C}, 82(5):428, 2022.

\bibitem{Forte:2020pyp}
Stefano Forte and Zahari Kassabov.
\newblock {Why $\alpha _s$ cannot be determined from hadronic processes without
  simultaneously determining the parton distributions}.
\newblock {\em Eur. Phys. J. C}, 80(3):182, 2020.

\bibitem{Isidori:2001bm}
Gino Isidori, Giovanni Ridolfi, and Alessandro Strumia.
\newblock {On the metastability of the standard model vacuum}.
\newblock {\em Nucl. Phys. B}, 609:387--409, 2001.

\bibitem{Buttazzo:2013uya}
Dario Buttazzo, Giuseppe Degrassi, Pier~Paolo Giardino, Gian~F. Giudice,
  Filippo Sala, Alberto Salvio, and Alessandro Strumia.
\newblock {Investigating the near-criticality of the Higgs boson}.
\newblock {\em JHEP}, 12:089, 2013.

\bibitem{DiLuzio:2015iua}
Luca Di~Luzio, Gino Isidori, and Giovanni Ridolfi.
\newblock {Stability of the electroweak ground state in the Standard Model and
  its extensions}.
\newblock {\em Phys. Lett. B}, 753:150--160, 2016.

\bibitem{Schwienhorst:2022yqu}
K.~Agashe et~al.
\newblock {Report of the Topical Group on Top quark physics and heavy flavor
  production for Snowmass 2021}.
\newblock 9 2022.

\bibitem{CDF:1995wbb}
F.~Abe et~al.
\newblock {Observation of top quark production in $\bar{p}p$ collisions}.
\newblock {\em Phys. Rev. Lett.}, 74:2626--2631, 1995.

\bibitem{D0:1995jca}
S.~Abachi et~al.
\newblock {Observation of the top quark}.
\newblock {\em Phys. Rev. Lett.}, 74:2632--2637, 1995.

\bibitem{CMS:2014rml}
Serguei Chatrchyan et~al.
\newblock {Determination of the Top-Quark Pole Mass and Strong Coupling
  Constant from the $t \bar{t}$ Production Cross Section in $pp$ Collisions at
  $\sqrt{s}$ = 7 TeV}.
\newblock {\em Phys. Lett. B}, 728:496--517, 2014.
\newblock [Erratum: Phys.Lett.B 738, 526--528 (2014)].

\bibitem{Cooper-Sarkar:2020twv}
Amanda~M. Cooper-Sarkar, Michal Czakon, Matthew~A. Lim, Alexander Mitov, and
  Andrew~S. Papanastasiou.
\newblock {Simultaneous extraction of $\alpha_s$ and $m_t$ from LHC $t\bar{t}$
  differential distributions}.
\newblock 10 2020.

\bibitem{CMS:2020vac}
Albert~M Sirunyan et~al.
\newblock {Measurement of CKM matrix elements in single top quark $t$-channel
  production in proton-proton collisions at $\sqrt{s} = $ 13 TeV}.
\newblock {\em Phys. Lett. B}, 808:135609, 2020.

\bibitem{ATLAS:2022jbw}
{Measurement of the top-quark mass using a leptonic invariant mass in $pp$
  collisions at $\sqrt{s}=13~\textrm{TeV}$ with the ATLAS detector}.
\newblock 9 2022.

\bibitem{CMS:2023ebf}
{Measurement of the top quark mass using a profile likelihood approach with the
  lepton+jets final states in proton-proton collisions at $\sqrt{s}$ = 13 TeV}.
\newblock 2 2023.

\bibitem{Gao:2017yyd}
Jun Gao, Lucian Harland-Lang, and Juan Rojo.
\newblock {The Structure of the Proton in the LHC Precision Era}.
\newblock {\em Phys. Rept.}, 742:1--121, 2018.

\bibitem{Amoroso:2022eow}
S.~Amoroso et~al.
\newblock {Snowmass 2021 whitepaper: Proton structure at the precision
  frontier}.
\newblock {\em Acta Phys. Polon. B}, 53(12):A1, 2022.

\bibitem{Czakon:2013tha}
Michal Czakon, Michelangelo~L. Mangano, Alexander Mitov, and Juan Rojo.
\newblock {Constraints on the gluon PDF from top quark pair production at
  hadron colliders}.
\newblock {\em JHEP}, 1307:167, 2013.

\bibitem{Czakon:2016olj}
Michał Czakon, Nathan~P. Hartland, Alexander Mitov, Emanuele~R. Nocera, and
  Juan Rojo.
\newblock {Pinning down the large-x gluon with NNLO top-quark pair differential
  distributions}.
\newblock {\em JHEP}, 04:044, 2017.

\bibitem{Czakon:2019yrx}
Micha\l{} Czakon, Sayipjamal Dulat, Tie-Jiun Hou, Joey Huston, Alexander Mitov,
  Andrew~S. Papanastasiou, Ibrahim Sitiwaldi, Zhite Yu, and C.~P. Yuan.
\newblock {An exploratory study of the impact of CMS double-differential top
  distributions on the gluon parton distribution function}.
\newblock {\em J. Phys. G}, 48(1):015003, 2020.

\bibitem{Nocera:2019wyk}
Emanuele~R. Nocera, Maria Ubiali, and Cameron Voisey.
\newblock {Single Top Production in PDF fits}.
\newblock {\em JHEP}, 05:067, 2020.

\bibitem{Campbell:2021qgd}
John Campbell, Tobias Neumann, and Zack Sullivan.
\newblock {Testing parton distribution functions with t-channel
  single-top-quark production}.
\newblock {\em Phys. Rev. D}, 104(9):094042, 2021.

\bibitem{Czakon:2016dgf}
Michal Czakon, David Heymes, and Alexander Mitov.
\newblock {Dynamical scales for multi-TeV top-pair production at the LHC}.
\newblock {\em JHEP}, 04:071, 2017.

\bibitem{Czakon:2020qbd}
Michal Czakon, Alexander Mitov, and Rene Poncelet.
\newblock {NNLO QCD corrections to leptonic observables in top-quark pair
  production and decay}.
\newblock {\em JHEP}, 05:212, 2021.

\bibitem{Catani:2019iny}
Stefano Catani, Simone Devoto, Massimiliano Grazzini, Stefan Kallweit, Javier
  Mazzitelli, and Hayk Sargsyan.
\newblock {Top-quark pair hadroproduction at next-to-next-to-leading order in
  QCD}.
\newblock {\em Phys. Rev. D}, 99(5):051501, 2019.

\bibitem{Catani:2019hip}
Stefano Catani, Simone Devoto, Massimiliano Grazzini, Stefan Kallweit, and
  Javier Mazzitelli.
\newblock {Top-quark pair production at the LHC: Fully differential QCD
  predictions at NNLO}.
\newblock {\em JHEP}, 07:100, 2019.

\bibitem{Grazzini:2017mhc}
Massimiliano Grazzini, Stefan Kallweit, and Marius Wiesemann.
\newblock {Fully differential NNLO computations with MATRIX}.
\newblock {\em Eur. Phys. J.}, C78(7):537, 2018.

\bibitem{Czakon:2017wor}
Michal Czakon, David Heymes, Alexander Mitov, Davide Pagani, Ioannis Tsinikos,
  and Marco Zaro.
\newblock {Top-pair production at the LHC through NNLO QCD and NLO EW}.
\newblock 2017.

\bibitem{Czakon:2018nun}
Michal Czakon, Andrea Ferroglia, David Heymes, Alexander Mitov, Ben~D. Pecjak,
  Darren~J. Scott, Xing Wang, and Li~Lin Yang.
\newblock {Resummation for (boosted) top-quark pair production at NNLO+NNLL' in
  QCD}.
\newblock {\em JHEP}, 05:149, 2018.

\bibitem{Mazzitelli:2021mmm}
Javier Mazzitelli, Pier~Francesco Monni, Paolo Nason, Emanuele Re, Marius
  Wiesemann, and Giulia Zanderighi.
\newblock {Top-pair production at the LHC with MINNLO$_{PS}$}.
\newblock {\em JHEP}, 04:079, 2022.

\bibitem{Berger:2017zof}
Edmond~L. Berger, Jun Gao, and Hua~Xing Zhu.
\newblock {Differential Distributions for t-channel Single Top-Quark Production
  and Decay at Next-to-Next-to-Leading Order in QCD}.
\newblock {\em JHEP}, 11:158, 2017.

\bibitem{Berger:2016oht}
Edmond~L. Berger, Jun Gao, C.~P. Yuan, and Hua~Xing Zhu.
\newblock {NNLO QCD Corrections to t-channel Single Top-Quark Production and
  Decay}.
\newblock {\em Phys. Rev. D}, 94(7):071501, 2016.

\bibitem{Liu:2018gxa}
Ze~Long Liu and Jun Gao.
\newblock {s -channel single top quark production and decay at
  next-to-next-to-leading-order in QCD}.
\newblock {\em Phys. Rev. D}, 98(7):071501, 2018.

\bibitem{Aguilar-Saavedra:2018ksv}
D.~Barducci et~al.
\newblock {Interpreting top-quark LHC measurements in the standard-model
  effective field theory}.
\newblock 2 2018.

\bibitem{Buckley:2015lku}
Andy Buckley, Christoph Englert, James Ferrando, David~J. Miller, Liam Moore,
  Michael Russell, and Chris~D. White.
\newblock {Constraining top quark effective theory in the LHC Run II era}.
\newblock {\em JHEP}, 04:015, 2016.

\bibitem{Brivio:2019ius}
Ilaria Brivio, Sebastian Bruggisser, Fabio Maltoni, Rhea Moutafis, Tilman
  Plehn, Eleni Vryonidou, Susanne Westhoff, and C.~Zhang.
\newblock {O new physics, where art thou? A global search in the top sector}.
\newblock {\em JHEP}, 02:131, 2020.

\bibitem{Bissmann:2019gfc}
Stefan Bi\ss{}mann, Johannes Erdmann, Cornelius Grunwald, Gudrun Hiller, and
  Kevin Kr\"oninger.
\newblock {Constraining top-quark couplings combining top-quark and
  $\boldsymbol{B}$ decay observables}.
\newblock {\em Eur. Phys. J. C}, 80(2):136, 2020.

\bibitem{Hartland:2019bjb}
Nathan~P. Hartland, Fabio Maltoni, Emanuele~R. Nocera, Juan Rojo, Emma Slade,
  Eleni Vryonidou, and Cen Zhang.
\newblock {A Monte Carlo global analysis of the Standard Model Effective Field
  Theory: the top quark sector}.
\newblock {\em JHEP}, 04:100, 2019.

\bibitem{Durieux:2019rbz}
Gauthier Durieux, Adrian Irles, V\'\i{}ctor Miralles, Ana Pe\~nuelas, Roman
  P\"oschl, Mart\'\i{}n Perell\'o, and Marcel Vos.
\newblock {The electro-weak couplings of the top and bottom quarks
  \textemdash{} Global fit and future prospects}.
\newblock {\em JHEP}, 12:98, 2019.
\newblock [Erratum: JHEP 01, 195 (2021)].

\bibitem{vanBeek:2019evb}
Samuel van Beek, Emanuele~R. Nocera, Juan Rojo, and Emma Slade.
\newblock {Constraining the SMEFT with Bayesian reweighting}.
\newblock {\em SciPost Phys.}, 7(5):070, 2019.

\bibitem{Yates:2021udl}
Brent~R. Yates.
\newblock {Using associated top quark production to probe for new physics
  within the framework of effective field theory}.
\newblock In {\em {13th International Workshop on Top Quark Physics}}, 1 2021.

\bibitem{Aoude:2022imd}
Rafael Aoude, Eric Madge, Fabio Maltoni, and Luca Mantani.
\newblock {Quantum SMEFT tomography: Top quark pair production at the LHC}.
\newblock {\em Phys. Rev. D}, 106(5):055007, 2022.

\bibitem{Maltoni:2019aot}
Fabio Maltoni, Luca Mantani, and Ken Mimasu.
\newblock {Top-quark electroweak interactions at high energy}.
\newblock {\em JHEP}, 10:004, 2019.

\bibitem{Aoude:2022deh}
Rafael Aoude, Hesham El~Faham, Fabio Maltoni, and Eleni Vryonidou.
\newblock {Complete SMEFT predictions for four top quark production at hadron
  colliders}.
\newblock {\em JHEP}, 10:163, 2022.

\bibitem{Degrande:2018fog}
Celine Degrande, Fabio Maltoni, Ken Mimasu, Eleni Vryonidou, and Cen Zhang.
\newblock {Single-top associated production with a $Z$ or $H$ boson at the LHC:
  the SMEFT interpretation}.
\newblock {\em JHEP}, 10:005, 2018.

\bibitem{Faham:2021zet}
Hesham~El Faham, Fabio Maltoni, Ken Mimasu, and Marco Zaro.
\newblock {Single top production in association with a WZ pair at the LHC in
  the SMEFT}.
\newblock {\em JHEP}, 01:100, 2022.

\bibitem{Severi:2022qjy}
Claudio Severi and Eleni Vryonidou.
\newblock {Quantum entanglement and top spin correlations in SMEFT at higher
  orders}.
\newblock {\em JHEP}, 01:148, 2023.

\bibitem{Cao:2020npb}
Qing-Hong Cao, Bin Yan, C.~P. Yuan, and Ya~Zhang.
\newblock {Probing $Zt\bar{t}$ couplings using $Z$ boson polarization in $ZZ$
  production at hadron colliders}.
\newblock {\em Phys. Rev. D}, 102(5):055010, 2020.

\bibitem{Cao:2015doa}
Qing-Hong Cao, Bin Yan, Jiang-Hao Yu, and Chen Zhang.
\newblock {A General Analysis of Wtb anomalous Couplings}.
\newblock {\em Chin. Phys. C}, 41(6):063101, 2017.

\bibitem{Han:2004az}
Zhenyu Han and Witold Skiba.
\newblock {Effective theory analysis of precision electroweak data}.
\newblock {\em Phys. Rev. D}, 71:075009, 2005.

\bibitem{daSilvaAlmeida:2018iqo}
Eduardo da~Silva~Almeida, Alexandre Alves, N.~Rosa~Agostinho, Oscar J.~P.
  \'Eboli, and M.~C. Gonzalez-Garcia.
\newblock {Electroweak Sector Under Scrutiny: A Combined Analysis of LHC and
  Electroweak Precision Data}.
\newblock {\em Phys. Rev. D}, 99(3):033001, 2019.

\bibitem{Ellis:2014jta}
John Ellis, Veronica Sanz, and Tevong You.
\newblock {The Effective Standard Model after LHC Run I}.
\newblock {\em JHEP}, 03:157, 2015.

\bibitem{Almeida:2021asy}
Eduardo da~Silva Almeida, Alexandre Alves, Oscar J.~P. \'Eboli, and M.~C.
  Gonzalez-Garcia.
\newblock {Electroweak legacy of the LHC run II}.
\newblock {\em Phys. Rev. D}, 105(1):013006, 2022.

\bibitem{Biekotter:2018ohn}
Anke Biek\"otter, Tyler Corbett, and Tilman Plehn.
\newblock {The Gauge-Higgs Legacy of the LHC Run II}.
\newblock {\em SciPost Phys.}, 6(6):064, 2019.

\bibitem{Kraml:2019sis}
Sabine Kraml, Tran~Quang Loc, Dao~Thi Nhung, and Le~Duc Ninh.
\newblock {Constraining new physics from Higgs measurements with Lilith: update
  to LHC Run 2 results}.
\newblock {\em SciPost Phys.}, 7(4):052, 2019.

\bibitem{Ellis:2018gqa}
John Ellis, Christopher~W. Murphy, Ver\'onica Sanz, and Tevong You.
\newblock {Updated Global SMEFT Fit to Higgs, Diboson and Electroweak Data}.
\newblock {\em JHEP}, 06:146, 2018.

\bibitem{Corbett:2012ja}
Tyler Corbett, O.~J.~P. Eboli, J.~Gonzalez-Fraile, and M.~C. Gonzalez-Garcia.
\newblock {Robust Determination of the Higgs Couplings: Power to the Data}.
\newblock {\em Phys. Rev. D}, 87:015022, 2013.

\bibitem{Ethier:2021ydt}
Jacob~J. Ethier, Raquel Gomez-Ambrosio, Giacomo Magni, and Juan Rojo.
\newblock {SMEFT analysis of vector boson scattering and diboson data from the
  LHC Run II}.
\newblock {\em Eur. Phys. J. C}, 81(6):560, 2021.

\bibitem{Elmer:2023wtr}
Nina Elmer, Maeve Madigan, Tilman Plehn, and Nikita Schmal.
\newblock {Staying on Top of SMEFT-Likelihood Analyses}.
\newblock 12 2023.

\bibitem{Allwicher:2022gkm}
Lukas Allwicher, Darius~A. Faroughy, Florentin Jaffredo, Olcyr Sumensari, and
  Felix Wilsch.
\newblock {Drell-Yan tails beyond the Standard Model}.
\newblock {\em JHEP}, 03:064, 2023.

\bibitem{Boughezal:2022nof}
Radja Boughezal, Yingsheng Huang, and Frank Petriello.
\newblock {Exploring the SMEFT at dimension eight with Drell-Yan transverse
  momentum measurements}.
\newblock {\em Phys. Rev. D}, 106(3):036020, 2022.

\bibitem{Ellis:2020unq}
John Ellis, Maeve Madigan, Ken Mimasu, Veronica Sanz, and Tevong You.
\newblock {Top, Higgs, Diboson and Electroweak Fit to the Standard Model
  Effective Field Theory}.
\newblock {\em JHEP}, 04:279, 2021.

\bibitem{Ethier:2021bye}
Jacob~J. Ethier, Giacomo Magni, Fabio Maltoni, Luca Mantani, Emanuele~R.
  Nocera, Juan Rojo, Emma Slade, Eleni Vryonidou, and Cen Zhang.
\newblock {Combined SMEFT interpretation of Higgs, diboson, and top quark data
  from the LHC}.
\newblock {\em JHEP}, 11:089, 2021.

\bibitem{Giani:2023gfq}
Tommaso Giani, Giacomo Magni, and Juan Rojo.
\newblock {SMEFiT: a flexible toolbox for global interpretations of particle
  physics data with effective field theories}.
\newblock 2 2023.

\bibitem{Cirigliano:2016nyn}
V.~Cirigliano, W.~Dekens, J.~de~Vries, and E.~Mereghetti.
\newblock {Constraining the top-Higgs sector of the Standard Model Effective
  Field Theory}.
\newblock {\em Phys. Rev. D}, 94(3):034031, 2016.

\bibitem{Cirigliano:2022qdm}
Vincenzo Cirigliano, Wouter Dekens, Jordy de~Vries, Emanuele Mereghetti, and
  Tom Tong.
\newblock {Beta-decay implications for the W-boson mass anomaly}.
\newblock {\em Phys. Rev. D}, 106(7):075001, 2022.

\bibitem{Cirigliano:2023nol}
Vincenzo Cirigliano, Wouter Dekens, Jordy de~Vries, Emanuele Mereghetti, and
  Tom Tong.
\newblock {Anomalies in global SMEFT analyses. A case study of first-row CKM
  unitarity}.
\newblock {\em JHEP}, 03:033, 2024.

\bibitem{Bartocci:2023nvp}
Riccardo Bartocci, Anke Biek\"otter, and Tobias Hurth.
\newblock {A global analysis of the SMEFT under the minimal MFV assumption}.
\newblock 11 2023.

\bibitem{terHoeve:2023pvs}
Jaco ter Hoeve, Giacomo Magni, Juan Rojo, Alejo~N. Rossia, and Eleni Vryonidou.
\newblock {The automation of SMEFT-assisted constraints on UV-complete models}.
\newblock {\em JHEP}, 01:179, 2024.

\bibitem{Celada:2024mcf}
Eugenia Celada, Tommaso Giani, Jaco ter Hoeve, Luca Mantani, Juan Rojo,
  Alejo~N. Rossia, Marion O.~A. Thomas, and Eleni Vryonidou.
\newblock {Mapping the SMEFT at High-Energy Colliders: from LEP and the
  (HL-)LHC to the FCC-ee}.
\newblock 4 2024.

\bibitem{Arneodo:1996kd}
M.~Arneodo et~al.
\newblock {Accurate measurement of $F_2^d/F_2^p$ and $R_d-R_p$}.
\newblock {\em Nucl. Phys.}, B487:3--26, 1997.

\bibitem{Arneodo:1996qe}
M.~Arneodo et~al.
\newblock {Measurement of the proton and deuteron structure functions, $F_2^p$
  and $F_2^d$, and of the ratio $\sigma_L/\sigma_T$}.
\newblock {\em Nucl. Phys.}, B483:3--43, 1997.

\bibitem{Whitlow:1991uw}
L.~W. Whitlow, E.~M. Riordan, S.~Dasu, Stephen Rock, and A.~Bodek.
\newblock {Precise measurements of the proton and deuteron structure functions
  from a global analysis of the SLAC deep inelastic electron scattering
  cross-sections}.
\newblock {\em Phys. Lett.}, B282:475--482, 1992.

\bibitem{Benvenuti:1989rh}
A.~C. Benvenuti et~al.
\newblock {A High Statistics Measurement of the Proton Structure Functions
  $F_2(x, Q^2)$ and $R$ from Deep Inelastic Muon Scattering at High $Q^2$}.
\newblock {\em Phys. Lett.}, B223:485, 1989.

\bibitem{Onengut:2005kv}
G.~Onengut et~al.
\newblock {Measurement of nucleon structure functions in neutrino scattering}.
\newblock {\em Phys. Lett.}, B632:65--75, 2006.

\bibitem{Goncharov:2001qe}
M.~Goncharov et~al.
\newblock {Precise measurement of dimuon production cross-sections in
  $\nu_{\mu}$Fe and $\bar{\nu}_{\mu}$Fe deep inelastic scattering at the
  Tevatron}.
\newblock {\em Phys. Rev.}, D64:112006, 2001.

\bibitem{MasonPhD}
David~Alexander Mason.
\newblock {Measurement of the strange - antistrange asymmetry at NLO in QCD
  from NuTeV dimuon data}.
\newblock FERMILAB-THESIS-2006-01.

\bibitem{Abramowicz:2015mha}
H.~Abramowicz et~al.
\newblock {Combination of measurements of inclusive deep inelastic ${e^{\pm
  }p}$ scattering cross sections and QCD analysis of HERA data}.
\newblock {\em Eur. Phys. J.}, C75(12):580, 2015.

\bibitem{H1:2018flt}
H.~Abramowicz et~al.
\newblock {Combination and QCD analysis of charm and beauty production
  cross-section measurements in deep inelastic $ep$ scattering at HERA}.
\newblock {\em Eur. Phys. J.}, C78(6):473, 2018.

\bibitem{Moreno:1990sf}
G.~Moreno et~al.
\newblock {Dimuon production in proton - copper collisions at $\sqrt{s}$ =
  38.8-GeV}.
\newblock {\em Phys. Rev.}, D43:2815--2836, 1991.

\bibitem{Webb:2003ps}
J.~C. Webb et~al.
\newblock {Absolute Drell-Yan dimuon cross sections in 800-GeV/c p p and p d
  collisions}.
\newblock 2003.

\bibitem{Towell:2001nh}
R.~S. Towell et~al.
\newblock {Improved measurement of the anti-d/anti-u asymmetry in the nucleon
  sea}.
\newblock {\em Phys. Rev.}, D64:052002, 2001.

\bibitem{Aaltonen:2010zza}
Timo~Antero Aaltonen et~al.
\newblock {Measurement of $d\sigma/dy$ of Drell-Yan $e^+e^-$ pairs in the $Z$
  Mass Region from $p\bar{p}$ Collisions at $\sqrt{s}=1.96$ TeV}.
\newblock {\em Phys. Lett.}, B692:232--239, 2010.

\bibitem{Abazov:2007jy}
V.~M. Abazov et~al.
\newblock {Measurement of the shape of the boson rapidity distribution for $p
  \bar{p} \to Z/\gamma^* \to e^{+} e^{-}$ + $X$ events produced at
  $\sqrt{s}$=1.96-TeV}.
\newblock {\em Phys. Rev.}, D76:012003, 2007.

\bibitem{Abazov:2013rja}
Victor~Mukhamedovich Abazov et~al.
\newblock {Measurement of the muon charge asymmetry in $p\bar{p}$ $\to$ W+X
  $\to$ $\mu$$\nu$ + X events at $\sqrt{s}$=1.96 TeV}.
\newblock {\em Phys.Rev.}, D88:091102, 2013.

\bibitem{D0:2014kma}
Victor~Mukhamedovich Abazov et~al.
\newblock {Measurement of the electron charge asymmetry in
  $\boldsymbol{p\bar{p}\rightarrow W+X \rightarrow e\nu +X}$ decays in
  $\boldsymbol{p\bar{p}}$ collisions at $\boldsymbol{\sqrt{s}=1.96}$ TeV}.
\newblock {\em Phys. Rev.}, D91(3):032007, 2015.
\newblock [Erratum: Phys. Rev.D91,no.7,079901(2015)].

\bibitem{Abulencia:2007ez}
A.~Abulencia et~al.
\newblock {Measurement of the Inclusive Jet Cross Section using the $k_{\rm T}$
  algorithm in $p\overline{p}$ Collisions at $\sqrt{s}$=1.96 TeV with the CDF
  II Detector}.
\newblock {\em Phys. Rev.}, D75:092006, 2007.

\bibitem{Aad:2011dm}
Georges Aad et~al.
\newblock {Measurement of the inclusive $W^{\pm}$ and $Z/\gamma^*$ cross
  sections in the electron and muon decay channels in pp collisions at
  $\sqrt{s}$= 7 TeV with the ATLAS detector}.
\newblock {\em Phys.Rev.}, D85:072004, 2012.

\bibitem{Aaboud:2016btc}
Morad Aaboud et~al.
\newblock {Precision measurement and interpretation of inclusive $W^+$ , $W^-$
  and $Z/\gamma ^*$ production cross sections with the ATLAS detector}.
\newblock {\em Eur. Phys. J.}, C77(6):367, 2017.

\bibitem{Aad:2014qja}
Georges Aad et~al.
\newblock {Measurement of the low-mass Drell-Yan differential cross section at
  $\sqrt{s}$ = 7 TeV using the ATLAS detector}.
\newblock {\em JHEP}, 06:112, 2014.

\bibitem{Aad:2013iua}
Georges Aad et~al.
\newblock {Measurement of the high-mass Drell--Yan differential cross-section
  in pp collisions at $\sqrt{s}$=7 TeV with the ATLAS detector}.
\newblock {\em Phys.Lett.}, B725:223, 2013.

\bibitem{Chatrchyan:2012xt}
Serguei Chatrchyan et~al.
\newblock {Measurement of the electron charge asymmetry in inclusive W
  production in pp collisions at $\sqrt{s}$ = 7 TeV}.
\newblock {\em Phys.Rev.Lett.}, 109:111806, 2012.

\bibitem{Chatrchyan:2013mza}
Serguei Chatrchyan et~al.
\newblock {Measurement of the muon charge asymmetry in inclusive pp to WX
  production at $\sqrt{s}$ = 7 TeV and an improved determination of light
  parton distribution functions}.
\newblock {\em Phys.Rev.}, D90:032004, 2014.

\bibitem{Chatrchyan:2013tia}
Sergei Chatrchyan et~al.
\newblock {Measurement of the differential and double-differential Drell-Yan
  cross sections in proton-proton collisions at $\sqrt{s} =$ 7 TeV}.
\newblock {\em JHEP}, 1312:030, 2013.

\bibitem{Khachatryan:2016pev}
Vardan Khachatryan et~al.
\newblock {Measurement of the differential cross section and charge asymmetry
  for inclusive $\mathrm {p}\mathrm {p}\rightarrow \mathrm {W}^{\pm }+X$
  production at ${\sqrt{s}} = 8$ TeV}.
\newblock {\em Eur. Phys. J.}, C76(8):469, 2016.

\bibitem{Aaij:2012mda}
R~Aaij et~al.
\newblock {Measurement of the cross-section for $Z \to e^+e^-$ production in
  $pp$ collisions at $\sqrt{s}=7$ TeV}.
\newblock {\em JHEP}, 1302:106, 2013.

\bibitem{Aaij:2015gna}
Roel Aaij et~al.
\newblock {Measurement of the forward $Z$ boson production cross-section in
  $pp$ collisions at $\sqrt{s}=7$ TeV}.
\newblock {\em JHEP}, 08:039, 2015.

\bibitem{Aaij:2015vua}
Roel Aaij et~al.
\newblock {Measurement of forward $\rm Z\rightarrow e^+e^-$ production at
  $\sqrt{s}=8$ TeV}.
\newblock {\em JHEP}, 05:109, 2015.

\bibitem{Aaij:2015zlq}
Roel Aaij et~al.
\newblock {Measurement of forward W and Z boson production in $pp$ collisions
  at $ \sqrt{s}=8 $ TeV}.
\newblock {\em JHEP}, 01:155, 2016.

\bibitem{Aad:2016zzw}
Georges Aad et~al.
\newblock {Measurement of the double-differential high-mass Drell-Yan cross
  section in pp collisions at $ \sqrt{s}=8 $ TeV with the ATLAS detector}.
\newblock {\em JHEP}, 08:009, 2016.

\bibitem{Aaboud:2017ffb}
M.~Aaboud et~al.
\newblock {Measurement of the Drell-Yan triple-differential cross section in
  $pp$ collisions at $\sqrt{s} = 8$ TeV}.
\newblock {\em JHEP}, 12:059, 2017.

\bibitem{Aad:2019rou}
Georges Aad et~al.
\newblock {Measurement of the cross-section and charge asymmetry of $W$ bosons
  produced in proton\textendash{}proton collisions at $\sqrt{s}=8~\text {TeV}$
  with the ATLAS detector}.
\newblock {\em Eur. Phys. J. C}, 79(9):760, 2019.

\bibitem{Aaij:2016qqz}
Roel Aaij et~al.
\newblock {Measurement of forward $W\to e\nu$ production in $pp$ collisions at
  $\sqrt{s}=8\,$TeV}.
\newblock {\em JHEP}, 10:030, 2016.

\bibitem{Aad:2016naf}
Georges Aad et~al.
\newblock {Measurement of $W^{\pm}$ and $Z$-boson production cross sections in
  $pp$ collisions at $\sqrt{s}=13$ TeV with the ATLAS detector}.
\newblock {\em Phys. Lett.}, B759:601--621, 2016.

\bibitem{Aaij:2016mgv}
Roel Aaij et~al.
\newblock {Measurement of the forward Z boson production cross-section in pp
  collisions at $\sqrt{s} = 13$ TeV}.
\newblock {\em JHEP}, 09:136, 2016.

\bibitem{Aad:2015auj}
Georges Aad et~al.
\newblock {Measurement of the transverse momentum and $\phi ^*_{\eta }$
  distributions of Drell–Yan lepton pairs in proton–proton collisions at
  $\sqrt{s}=8$ TeV with the ATLAS detector}.
\newblock {\em Eur. Phys. J.}, C76(5):291, 2016.

\bibitem{Khachatryan:2015oaa}
Vardan Khachatryan et~al.
\newblock {Measurement of the Z boson differential cross section in transverse
  momentum and rapidity in proton–proton collisions at 8 TeV}.
\newblock {\em Phys. Lett.}, B749:187--209, 2015.

\bibitem{Aaboud:2017soa}
Morad Aaboud et~al.
\newblock {Measurement of differential cross sections and $W^+/W^-$
  cross-section ratios for $W$ boson production in association with jets at
  $\sqrt{s}=8$ TeV with the ATLAS detector}.
\newblock {\em JHEP}, 05:077, 2018.
\newblock [Erratum: JHEP 10, 048 (2020)].

\bibitem{Sirunyan:2017wgx}
Albert~M Sirunyan et~al.
\newblock {Measurement of the differential cross sections for the associated
  production of a $W$ boson and jets in proton-proton collisions at
  $\sqrt{s}=13$ TeV}.
\newblock {\em Phys. Rev. D}, 96(7):072005, 2017.

\bibitem{Aad:2011fc}
Georges Aad et~al.
\newblock {Measurement of inclusive jet and dijet production in pp collisions
  at $\sqrt{s}$ = 7 TeV using the ATLAS detector}.
\newblock {\em Phys. Rev.}, D86:014022, 2012.

\bibitem{Aad:2013lpa}
Georges Aad et~al.
\newblock {Measurement of the inclusive jet cross section in pp collisions at
  $\sqrt{s}$=2.76 TeV and comparison to the inclusive jet cross section at
  $\sqrt{s}$=7 TeV using the ATLAS detector}.
\newblock {\em Eur.Phys.J.}, C73:2509, 2013.

\bibitem{Aad:2014vwa}
Georges Aad et~al.
\newblock {Measurement of the inclusive jet cross-section in proton-proton
  collisions at $ \sqrt{s}=7 $ TeV using 4.5 fb$^{-1}$ of data with the ATLAS
  detector}.
\newblock {\em JHEP}, 02:153, 2015.
\newblock [Erratum: JHEP09,141(2015)].

\bibitem{Chatrchyan:2012bja}
Serguei Chatrchyan et~al.
\newblock {Measurements of differential jet cross sections in proton-proton
  collisions at $\sqrt{s}=7$ TeV with the CMS detector}.
\newblock {\em Phys.Rev.}, D87:112002, 2013.

\bibitem{Khachatryan:2015luy}
Vardan Khachatryan et~al.
\newblock {Measurement of the inclusive jet cross section in pp collisions at
  $\sqrt{s} = 2.76\,\text {TeV}$}.
\newblock {\em Eur. Phys. J.}, C76(5):265, 2016.

\bibitem{Aaboud:2017dvo}
Morad Aaboud et~al.
\newblock {Measurement of the inclusive jet cross-sections in proton-proton
  collisions at $ \sqrt{s}=8 $ TeV with the ATLAS detector}.
\newblock {\em JHEP}, 09:020, 2017.

\bibitem{Khachatryan:2016mlc}
Vardan Khachatryan et~al.
\newblock {Measurement and QCD analysis of double-differential inclusive jet
  cross sections in pp collisions at $ \sqrt{s}=8 $ TeV and cross section
  ratios to 2.76 and 7 TeV}.
\newblock {\em JHEP}, 03:156, 2017.

\bibitem{Aad:2016xcr}
Georges Aad et~al.
\newblock {Measurement of the inclusive isolated prompt photon cross section in
  pp collisions at $ \sqrt{s}=8 $ TeV with the ATLAS detector}.
\newblock {\em JHEP}, 08:005, 2016.

\bibitem{ATLAS:2017nah}
Morad Aaboud et~al.
\newblock {Measurement of the cross section for inclusive isolated-photon
  production in $pp$ collisions at $\sqrt s=13$ TeV using the ATLAS detector}.
\newblock {\em Phys. Lett. B}, 770:473--493, 2017.

\bibitem{Aad:2015mbv}
Georges Aad et~al.
\newblock {Measurements of top-quark pair differential cross-sections in the
  lepton+jets channel in $pp$ collisions at $\sqrt{s}=8$ TeV using the ATLAS
  detector}.
\newblock {\em Eur. Phys. J.}, C76(10):538, 2016.

\bibitem{Aaboud:2016iot}
Morad Aaboud et~al.
\newblock {Measurement of top quark pair differential cross-sections in the
  dilepton channel in $pp$ collisions at $\sqrt{s}$ = 7 and 8 TeV with ATLAS}.
\newblock {\em Phys. Rev. D}, 94(9):092003, 2016.
\newblock [Addendum: Phys.Rev.D 101, 119901 (2020)].

\bibitem{ATLAS:2017wvi}
Morad Aaboud et~al.
\newblock {Measurement of the inclusive and fiducial $t\bar{t}$ production
  cross-sections in the lepton+jets channel in $pp$ collisions at $\sqrt{s} =
  8$ TeV with the ATLAS detector}.
\newblock {\em Eur. Phys. J. C}, 78:487, 2018.

\bibitem{ATLAS:2019hau}
Georges Aad et~al.
\newblock {Measurement of the $t\bar{t}$ production cross-section and lepton
  differential distributions in $e\mu $ dilepton events from $pp$ collisions at
  $\sqrt{s}=13\,\text {TeV}$ with the ATLAS detector}.
\newblock {\em Eur. Phys. J. C}, 80(6):528, 2020.

\bibitem{ATLAS:2020ccu}
Georges Aad et~al.
\newblock {Measurements of top-quark pair single- and double-differential
  cross-sections in the all-hadronic channel in $pp$ collisions at
  $\sqrt{s}=13~\textrm{TeV}$ using the ATLAS detector}.
\newblock {\em JHEP}, 01:033, 2021.

\bibitem{ATLAS:2020aln}
Georges Aad et~al.
\newblock {Measurement of the $t\bar{t}$ production cross-section in the
  lepton+jets channel at $\sqrt{s}=13$ TeV with the ATLAS experiment}.
\newblock {\em Phys. Lett. B}, 810:135797, 2020.

\bibitem{Aad:2019ntk}
Georges Aad et~al.
\newblock {Measurements of top-quark pair differential and double-differential
  cross-sections in the $\ell$+jets channel with $pp$ collisions at
  $\sqrt{s}=13$ TeV using the ATLAS detector}.
\newblock {\em Eur. Phys. J. C}, 79(12):1028, 2019.
\newblock [Erratum: Eur.Phys.J.C 80, 1092 (2020)].

\bibitem{CMS:2017zpm}
A.~M. Sirunyan et~al.
\newblock {Measurement of the inclusive $ \mathrm{t}\overline{\mathrm{t}} $
  cross section in pp collisions at $ \sqrt{s}=5.02 $ TeV using final states
  with at least one charged lepton}.
\newblock {\em JHEP}, 03:115, 2018.

\bibitem{Spannagel:2016cqt}
Simon Spannagel.
\newblock {Top quark mass measurements with the CMS experiment at the LHC}.
\newblock {\em PoS}, DIS2016:150, 2016.

\bibitem{Sirunyan:2017azo}
Albert~M Sirunyan et~al.
\newblock {Measurement of double-differential cross sections for top quark pair
  production in pp collisions at $\sqrt{s} = 8$ $\,\text {TeV}$ and impact on
  parton distribution functions}.
\newblock {\em Eur. Phys. J. C}, 77(7):459, 2017.

\bibitem{Khachatryan:2015oqa}
Vardan Khachatryan et~al.
\newblock {Measurement of the differential cross section for top quark pair
  production in pp collisions at $\sqrt{s} = 8\,\text {TeV} $}.
\newblock {\em Eur. Phys. J.}, C75(11):542, 2015.

\bibitem{CMS:2015yky}
Vardan Khachatryan et~al.
\newblock {Measurement of the top quark pair production cross section in
  proton-proton collisions at $\sqrt{s} =$ 13 TeV}.
\newblock {\em Phys. Rev. Lett.}, 116(5):052002, 2016.

\bibitem{Sirunyan:2018ucr}
Albert~M Sirunyan et~al.
\newblock {Measurements of $\mathrm{t\overline{t}}$ differential cross sections
  in proton-proton collisions at $\sqrt{s}=$ 13 TeV using events containing two
  leptons}.
\newblock {\em JHEP}, 02:149, 2019.

\bibitem{CMS:2021vhb}
Armen Tumasyan et~al.
\newblock {Measurement of differential $\text{t}\overline{\text{t}}$ production
  cross sections in the full kinematic range using lepton+jets events from
  proton-proton collisions at $\sqrt{s} = $ 13 TeV}.
\newblock 8 2021.

\bibitem{Aad:2016ove}
Georges Aad et~al.
\newblock {Measurements of the charge asymmetry in top-quark pair production in
  the dilepton final state at $\sqrt{s}=8$ TeV with the ATLAS detector}.
\newblock {\em Phys. Rev. D}, 94(3):032006, 2016.

\bibitem{ATLAS:2022waa}
{Evidence for the charge asymmetry in $pp \rightarrow t\bar{t}$ production at
  $\sqrt{s}= 13$ TeV with the ATLAS detector}.
\newblock 8 2022.

\bibitem{Khachatryan:2016ysn}
Vardan Khachatryan et~al.
\newblock {Measurements of $t \bar t$ charge asymmetry using dilepton final
  states in pp collisions at $\sqrt s=8$ TeV}.
\newblock {\em Phys. Lett. B}, 760:365--386, 2016.

\bibitem{CMS-PAS-TOP-21-014}
{Measurement of the ttbar charge asymmetry in highly boosted events in the
  single-lepton channel at 13 TeV}.
\newblock Technical report, CERN, Geneva, 2022.

\bibitem{Sirunyan:2017lvd}
Morad Aaboud et~al.
\newblock {Combination of inclusive and differential $
  \mathrm{t}\overline{\mathrm{t}} $ charge asymmetry measurements using ATLAS
  and CMS data at $ \sqrt{s}=7 $ and 8 TeV}.
\newblock {\em JHEP}, 04:033, 2018.

\bibitem{Aad:2020jvx}
Georges Aad et~al.
\newblock {Combination of the W boson polarization measurements in top quark
  decays using ATLAS and CMS data at $\sqrt{s} =$ 8 TeV}.
\newblock {\em JHEP}, 08(08):051, 2020.

\bibitem{ATLAS:2022bdg}
{Measurement of the polarisation of $W$ bosons produced in top-quark decays
  using di-lepton events at $\sqrt{s} = 13$ TeV with the ATLAS experiment}.
\newblock 2022.

\bibitem{Aad:2015eua}
Georges Aad et~al.
\newblock {Measurement of the $ t\overline{t}W $ and $ t\overline{t}Z $
  production cross sections in pp collisions at $ \sqrt{s}=8 $ TeV with the
  ATLAS detector}.
\newblock {\em JHEP}, 11:172, 2015.

\bibitem{Aaboud:2019njj}
Morad Aaboud et~al.
\newblock {Measurement of the $t\bar{t}Z$ and $t\bar{t}W$ cross sections in
  proton-proton collisions at $\sqrt{s}=13$ TeV with the ATLAS detector}.
\newblock {\em Phys. Rev. D}, 99(7):072009, 2019.

\bibitem{ATLAS:2021fzm}
Georges Aad et~al.
\newblock {Measurements of the inclusive and differential production cross
  sections of a top-quark\textendash{}antiquark pair in association with a
  Z~boson at $\sqrt{s} = 13$~TeV with the ATLAS detector}.
\newblock {\em Eur. Phys. J. C}, 81(8):737, 2021.

\bibitem{Khachatryan:2015sha}
Vardan Khachatryan et~al.
\newblock {Observation of top quark pairs produced in association with a vector
  boson in pp collisions at $ \sqrt{s}=8 $ TeV}.
\newblock {\em JHEP}, 01:096, 2016.

\bibitem{Sirunyan:2017uzs}
Albert~M Sirunyan et~al.
\newblock {Measurement of the cross section for top quark pair production in
  association with a W or Z boson in proton-proton collisions at $\sqrt{s} =$
  13 TeV}.
\newblock {\em JHEP}, 08:011, 2018.

\bibitem{CMS:2019too}
Albert~M Sirunyan et~al.
\newblock {Measurement of top quark pair production in association with a Z
  boson in proton-proton collisions at $\sqrt{s}=$ 13 TeV}.
\newblock {\em JHEP}, 03:056, 2020.

\bibitem{Aaboud:2017era}
Morad Aaboud et~al.
\newblock {Measurement of the $ t\overline{t}\gamma $ production cross section
  in proton-proton collisions at $ \sqrt{s}=8 $ TeV with the ATLAS detector}.
\newblock {\em JHEP}, 11:086, 2017.

\bibitem{Sirunyan:2017iyh}
Albert~M Sirunyan et~al.
\newblock {Measurement of the semileptonic $ \mathrm{t}\overline{\mathrm{t}} $
  + \ensuremath{\gamma} production cross section in pp collisions at $
  \sqrt{s}=8 $ TeV}.
\newblock {\em JHEP}, 10:006, 2017.

\bibitem{ATLAS:2020hpj}
Georges Aad et~al.
\newblock {Evidence for $t\bar{t}t\bar{t}$ production in the multilepton final
  state in proton\textendash{}proton collisions at $\sqrt{s}=13$ $\text {TeV}$
  with the ATLAS detector}.
\newblock {\em Eur. Phys. J. C}, 80(11):1085, 2020.

\bibitem{ATLAS:2021kqb}
Georges Aad et~al.
\newblock {Measurement of the t$ \overline{t} $t$ \overline{t} $ production
  cross section in $pp$ collisions at $ \sqrt{s} $ = 13 TeV with the ATLAS
  detector}.
\newblock {\em JHEP}, 11:118, 2021.

\bibitem{ATLAS:2018fwl}
Morad Aaboud et~al.
\newblock {Measurements of inclusive and differential fiducial cross-sections
  of $ t\overline{t} $ production with additional heavy-flavour jets in
  proton-proton collisions at $ \sqrt{s} $ = 13 TeV with the ATLAS detector}.
\newblock {\em JHEP}, 04:046, 2019.

\bibitem{CMS:2019rvj}
Albert~M Sirunyan et~al.
\newblock {Search for production of four top quarks in final states with
  same-sign or multiple leptons in proton-proton collisions at $\sqrt{s}=$ 13
  TeV}.
\newblock {\em Eur. Phys. J. C}, 80(2):75, 2020.

\bibitem{CMS:2019jsc}
Albert~M Sirunyan et~al.
\newblock {Search for the production of four top quarks in the single-lepton
  and opposite-sign dilepton final states in proton-proton collisions at $
  \sqrt{s} $ = 13 TeV}.
\newblock {\em JHEP}, 11:082, 2019.

\bibitem{CMS:2019eih}
Albert~M Sirunyan et~al.
\newblock {Measurement of the $\mathrm{t\bar{t}}\mathrm{b\bar{b}}$ production
  cross section in the all-jet final state in pp collisions at $\sqrt{s} =$ 13
  TeV}.
\newblock {\em Phys. Lett. B}, 803:135285, 2020.

\bibitem{CMS:2020grm}
Albert~M Sirunyan et~al.
\newblock {Measurement of the cross section for $\text{t}\bar{\text{t}}$
  production with additional jets and b jets in pp collisions at $\sqrt{s}=$ 13
  TeV}.
\newblock {\em JHEP}, 07:125, 2020.

\bibitem{Aad:2020axn}
Georges Aad et~al.
\newblock {Measurements of inclusive and differential cross-sections of
  combined $ t\overline{t}\gamma $ and $tW\gamma$ production in the e$\mu$
  channel at 13 TeV with the ATLAS detector}.
\newblock {\em JHEP}, 09:049, 2020.

\bibitem{ATLAS:2014sxe}
Georges Aad et~al.
\newblock {Comprehensive measurements of $t$-channel single top-quark
  production cross sections at $\sqrt{s} = 7$ TeV with the ATLAS detector}.
\newblock {\em Phys. Rev. D}, 90(11):112006, 2014.

\bibitem{Aaboud:2017pdi}
Morad Aaboud et~al.
\newblock {Fiducial, total and differential cross-section measurements of
  $t$-channel single top-quark production in $pp$ collisions at 8 TeV using
  data collected by the ATLAS detector}.
\newblock {\em Eur. Phys. J. C}, 77(8):531, 2017.

\bibitem{Aad:2015upn}
Georges Aad et~al.
\newblock {Evidence for single top-quark production in the $s$-channel in
  proton-proton collisions at $\sqrt{s}=$8 TeV with the ATLAS detector using
  the Matrix Element Method}.
\newblock {\em Phys. Lett. B}, 756:228--246, 2016.

\bibitem{Aaboud:2016ymp}
Morad Aaboud et~al.
\newblock {Measurement of the inclusive cross-sections of single top-quark and
  top-antiquark $t$-channel production in $pp$ collisions at $\sqrt{s}$ = 13
  TeV with the ATLAS detector}.
\newblock {\em JHEP}, 04:086, 2017.

\bibitem{ATLAS:2022wfk}
{Measurement of single top-quark production in the s-channel in proton$-$proton
  collisions at $\mathrm{\sqrt{s}=13}$ TeV with the ATLAS detector}.
\newblock 9 2022.

\bibitem{CMS:2012xhh}
Serguei Chatrchyan et~al.
\newblock {Measurement of the Single-Top-Quark $t$-Channel Cross Section in
  $pp$ Collisions at $\sqrt{s}=7$ TeV}.
\newblock {\em JHEP}, 12:035, 2012.

\bibitem{Khachatryan:2014iya}
Vardan Khachatryan et~al.
\newblock {Measurement of the t-channel single-top-quark production cross
  section and of the $\mid V_{tb} \mid$ CKM matrix element in pp collisions at
  $\sqrt{s}$= 8 TeV}.
\newblock {\em JHEP}, 06:090, 2014.

\bibitem{Khachatryan:2016ewo}
Vardan Khachatryan et~al.
\newblock {Search for s channel single top quark production in pp collisions at
  $ \sqrt{s}=7 $ and 8 TeV}.
\newblock {\em JHEP}, 09:027, 2016.

\bibitem{Sirunyan:2016cdg}
Albert~M Sirunyan et~al.
\newblock {Cross section measurement of $t$-channel single top quark production
  in pp collisions at $\sqrt s =$ 13 TeV}.
\newblock {\em Phys. Lett. B}, 772:752--776, 2017.

\bibitem{Sirunyan:2019hqb}
Albert~M Sirunyan et~al.
\newblock {Measurement of differential cross sections and charge ratios for
  t-channel single top quark production in proton\textendash{}proton collisions
  at $\sqrt{s}=13\,\text {Te}\text {V}$}.
\newblock {\em Eur. Phys. J. C}, 80(5):370, 2020.

\bibitem{Aad:2015eto}
Georges Aad et~al.
\newblock {Measurement of the production cross-section of a single top quark in
  association with a $W$ boson at 8 TeV with the ATLAS experiment}.
\newblock {\em JHEP}, 01:064, 2016.

\bibitem{Aad:2020zhd}
Georges Aad et~al.
\newblock {Measurement of single top-quark production in association with a $W$
  boson in the single-lepton channel at $\sqrt{s} = 8\,\text {TeV}$ with the
  ATLAS detector}.
\newblock {\em Eur. Phys. J. C}, 81(8):720, 2021.

\bibitem{Aaboud:2016lpj}
Morad Aaboud et~al.
\newblock {Measurement of the cross-section for producing a W boson in
  association with a single top quark in pp collisions at $ \sqrt{s}=13 $ TeV
  with ATLAS}.
\newblock {\em JHEP}, 01:063, 2018.

\bibitem{Aad:2020wog}
Georges Aad et~al.
\newblock {Observation of the associated production of a top quark and a $Z$
  boson in $pp$ collisions at $\sqrt{s} = 13$ TeV with the ATLAS detector}.
\newblock {\em JHEP}, 07:124, 2020.

\bibitem{Chatrchyan:2014tua}
Serguei Chatrchyan et~al.
\newblock {Observation of the associated production of a single top quark and a
  $W$ boson in $pp$ collisions at $\sqrt s = $8 TeV}.
\newblock {\em Phys. Rev. Lett.}, 112(23):231802, 2014.

\bibitem{Sirunyan:2018lcp}
Albert~M Sirunyan et~al.
\newblock {Measurement of the production cross section for single top quarks in
  association with W bosons in proton-proton collisions at $ \sqrt{s}=13 $
  TeV}.
\newblock {\em JHEP}, 10:117, 2018.

\bibitem{Sirunyan:2018zgs}
Albert~M Sirunyan et~al.
\newblock {Observation of Single Top Quark Production in Association with a $Z$
  Boson in Proton-Proton Collisions at $\sqrt {s}$ =13 TeV}.
\newblock {\em Phys. Rev. Lett.}, 122(13):132003, 2019.

\bibitem{CMS:2021rvz}
{Inclusive and differential cross section measurements of single top quark
  production in association with a Z boson in proton-proton collisions at
  $\sqrt{s} = 13~\mathrm{TeV}$}.
\newblock 2021.

\bibitem{CMS:2021vqm}
Armen Tumasyan et~al.
\newblock {Observation of tW production in the single-lepton channel in pp
  collisions at $ \sqrt{s} $ = 13 TeV}.
\newblock {\em JHEP}, 11:111, 2021.

\bibitem{Ball:2022uon}
Richard~D. Ball, Juan Cruz-Martinez, Luigi Del~Debbio, Stefano Forte, Zahari
  Kassabov, Emanuele~R. Nocera, Juan Rojo, Roy Stegeman, and Maria Ubiali.
\newblock {Response to ''Parton distributions need representative sampling''}.
\newblock 11 2022.

\bibitem{Kassabov:2022pps}
Zahari Kassabov, Emanuele~R. Nocera, and Michael Wilson.
\newblock {Regularising experimental correlations in LHC data: theory and
  application to a global analysis of parton distributions}.
\newblock 7 2022.

\bibitem{Frederix:2018nkq}
R.~Frederix, S.~Frixione, V.~Hirschi, D.~Pagani, H.~S. Shao, and M.~Zaro.
\newblock {The automation of next-to-leading order electroweak calculations}.
\newblock {\em JHEP}, 07:185, 2018.

\bibitem{Alwall:2014hca}
J.~Alwall, R.~Frederix, S.~Frixione, V.~Hirschi, F.~Maltoni, et~al.
\newblock {The automated computation of tree-level and next-to-leading order
  differential cross sections, and their matching to parton shower
  simulations}.
\newblock {\em JHEP}, 1407:079, 2014.

\bibitem{Kluge:2006xs}
T.~Kluge, K.~Rabbertz, and M.~Wobisch.
\newblock {Fast pQCD calculations for PDF fits}.
\newblock 2006.

\bibitem{Britzger:2012bs}
Daniel Britzger, Klaus Rabbertz, Fred Stober, and Markus Wobisch.
\newblock {New features in version 2 of the fastNLO project}.
\newblock In {\em {20th International Workshop on Deep-Inelastic Scattering and
  Related Subjects}}, 8 2012.

\bibitem{Bertone:2014zva}
Valerio Bertone, Rikkert Frederix, Stefano Frixione, Juan Rojo, and Mark
  Sutton.
\newblock {aMCfast: automation of fast NLO computations for PDF fits}.
\newblock {\em JHEP}, 08:166, 2014.

\bibitem{Ball:2014uwa}
Richard~D. Ball et~al.
\newblock {Parton distributions for the LHC Run II}.
\newblock {\em JHEP}, 04:040, 2015.

\bibitem{hightea}


\bibitem{Kulesza:2018tqz}
Anna Kulesza, Leszek Motyka, Daniel Schwartl\"ander, Tomasz Stebel, and Vincent
  Theeuwes.
\newblock {Associated production of a top quark pair with a heavy electroweak
  gauge boson at NLO$+$NNLL accuracy}.
\newblock {\em Eur. Phys. J. C}, 79(3):249, 2019.

\bibitem{Biekotter:2023xle}
Anke Biek\"otter, Benjamin~D. Pecjak, Darren~J. Scott, and Tommy Smith.
\newblock {Electroweak input schemes and universal corrections in SMEFT}.
\newblock {\em JHEP}, 07:115, 2023.

\bibitem{Degrande:2020evl}
C\'eline Degrande, Gauthier Durieux, Fabio Maltoni, Ken Mimasu, Eleni
  Vryonidou, and Cen Zhang.
\newblock {Automated one-loop computations in the standard model effective
  field theory}.
\newblock {\em Phys. Rev. D}, 103(9):096024, 2021.

\bibitem{Aoude:2022aro}
Rafael Aoude, Fabio Maltoni, Olivier Mattelaer, Claudio Severi, and Eleni
  Vryonidou.
\newblock {Renormalisation group effects on SMEFT interpretations of LHC data}.
\newblock 12 2022.

\bibitem{Alloul:2013bka}
Adam Alloul, Neil~D. Christensen, C\'eline Degrande, Claude Duhr, and Benjamin
  Fuks.
\newblock {FeynRules 2.0 - A complete toolbox for tree-level phenomenology}.
\newblock {\em Comput. Phys. Commun.}, 185:2250--2300, 2014.

\bibitem{ALEPH:2005ab}
S.~Schael et~al.
\newblock {Precision electroweak measurements on the $Z$ resonance}.
\newblock {\em Phys. Rept.}, 427:257--454, 2006.

\bibitem{ALEPH:2013dgf}
S.~Schael et~al.
\newblock {Electroweak Measurements in Electron-Positron Collisions at
  W-Boson-Pair Energies at LEP}.
\newblock {\em Phys. Rept.}, 532:119--244, 2013.

\bibitem{Workman:2022ynf}
R.~L. Workman and Others.
\newblock {Review of Particle Physics}.
\newblock {\em PTEP}, 2022:083C01, 2022.

\bibitem{ATLAS:2016neq}
Georges Aad et~al.
\newblock {Measurements of the Higgs boson production and decay rates and
  constraints on its couplings from a combined ATLAS and CMS analysis of the
  LHC pp collision data at $ \sqrt{s}=7 $ and 8 TeV}.
\newblock {\em JHEP}, 08:045, 2016.

\bibitem{CMS:2018uag}
Albert~M Sirunyan et~al.
\newblock {Combined measurements of Higgs boson couplings in
  proton\textendash{}proton collisions at $\sqrt{s}=13\,\text {Te}\text {V} $}.
\newblock {\em Eur. Phys. J. C}, 79(5):421, 2019.

\bibitem{ATLAS:2019nkf}
Georges Aad et~al.
\newblock {Combined measurements of Higgs boson production and decay using up
  to $80$ fb$^{-1}$ of proton-proton collision data at $\sqrt{s}=$ 13 TeV
  collected with the ATLAS experiment}.
\newblock {\em Phys. Rev. D}, 101(1):012002, 2020.

\bibitem{ATLAS:2020qcv}
Georges Aad et~al.
\newblock {A search for the $Z\gamma$ decay mode of the Higgs boson in $pp$
  collisions at $\sqrt{s}$ = 13 TeV with the ATLAS detector}.
\newblock {\em Phys. Lett. B}, 809:135754, 2020.

\bibitem{ATLAS:2020fzp}
Georges Aad et~al.
\newblock {A search for the dimuon decay of the Standard Model Higgs boson with
  the ATLAS detector}.
\newblock {\em Phys. Lett. B}, 812:135980, 2021.

\bibitem{ATLAS:2019rob}
Morad Aaboud et~al.
\newblock {Measurement of fiducial and differential $W^+W^-$ production
  cross-sections at $\sqrt{s}=13$ TeV with the ATLAS detector}.
\newblock {\em Eur. Phys. J. C}, 79(10):884, 2019.

\bibitem{ATLAS:2019bsc}
Morad Aaboud et~al.
\newblock {Measurement of $W^{\pm}Z$ production cross sections and gauge boson
  polarisation in $pp$ collisions at $\sqrt{s} = 13$ TeV with the ATLAS
  detector}.
\newblock {\em Eur. Phys. J. C}, 79(6):535, 2019.

\bibitem{ATLAS:2020nzk}
Georges Aad et~al.
\newblock {Differential cross-section measurements for the electroweak
  production of dijets in association with a $Z$ boson in
  proton\textendash{}proton collisions at ATLAS}.
\newblock {\em Eur. Phys. J. C}, 81(2):163, 2021.

\bibitem{Brehmer:2016nyr}
Johann Brehmer, Kyle Cranmer, Felix Kling, and Tilman Plehn.
\newblock {Better Higgs boson measurements through information geometry}.
\newblock {\em Phys. Rev. D}, 95(7):073002, 2017.

\bibitem{AbdulKhalek:2018rok}
Rabah Abdul~Khalek, Shaun Bailey, Jun Gao, Lucian Harland-Lang, and Juan Rojo.
\newblock {Towards Ultimate Parton Distributions at the High-Luminosity LHC}.
\newblock {\em Eur. Phys. J. C}, 78(11):962, 2018.

\bibitem{Azzi:2019yne}
P.~Azzi et~al.
\newblock {Report from Working Group 1}: {Standard Model Physics at the HL-LHC
  and HE-LHC}.
\newblock {\em CERN Yellow Rep. Monogr.}, 7:1--220, 2019.

\bibitem{Ball:2010de}
Richard~D. Ball et~al.
\newblock {A first unbiased global NLO determination of parton distributions
  and their uncertainties}.
\newblock {\em Nucl. Phys.}, B838:136, 2010.

\bibitem{progr}
Mark Costantini, Maeve Madigan, Luca Mantani, and James Moore.
\newblock {In progress}.

\bibitem{GomezAmbrosio:2022mpm}
Raquel Gomez~Ambrosio, Jaco ter Hoeve, Maeve Madigan, Juan Rojo, and Veronica
  Sanz.
\newblock {Unbinned multivariate observables for global SMEFT analyses from
  machine learning}.
\newblock 11 2022.

\bibitem{Long:2023mrj}
Owen Long and Benjamin Nachman.
\newblock {Designing Observables for Measurements with Deep Learning}.
\newblock 10 2023.

\bibitem{AbdulKhalek:2020jut}
Rabah Abdul~Khalek et~al.
\newblock {Phenomenology of NNLO jet production at the LHC and its impact on
  parton distributions}.
\newblock {\em Eur. Phys. J. C}, 80(8):797, 2020.

\bibitem{Dawson:2022ewj}
Sally Dawson et~al.
\newblock {LHC EFT WG Note: Precision matching of microscopic physics to the
  Standard Model Effective Field Theory (SMEFT)}.
\newblock 12 2022.

\bibitem{Boughezal:2022pmb}
Radja Boughezal, Alexander Emmert, Tyler Kutz, Sonny Mantry, Michael Nycz,
  Frank Petriello, Ka\u{g}an \c{S}im\c{s}ek, Daniel Wiegand, and Xiaochao
  Zheng.
\newblock {Neutral-current electroweak physics and SMEFT studies at the EIC}.
\newblock {\em Phys. Rev. D}, 106(1):016006, 2022.

\bibitem{Feng:2022inv}
Jonathan~L. Feng et~al.
\newblock {The Forward Physics Facility at the High-Luminosity LHC}.
\newblock 3 2022.

\bibitem{Anchordoqui:2021ghd}
Luis~A. Anchordoqui et~al.
\newblock {The Forward Physics Facility: Sites, experiments, and physics
  potential}.
\newblock {\em Phys. Rept.}, 968:1--50, 2022.

\bibitem{Jing:2023isu}
Xiaoxian Jing et~al.
\newblock {Quantifying the interplay of experimental constraints in analyses of
  parton distributions}.
\newblock 6 2023.

\bibitem{Nocera:2017zge}
Emanuele~Roberto Nocera and Maria Ubiali.
\newblock {Constraining the gluon PDF at large x with LHC data}.
\newblock {\em PoS}, DIS2017:008, 2018.

\bibitem{Accomando:2019vqt}
Elena Accomando et~al.
\newblock {PDF Profiling Using the Forward-Backward Asymmetry in Neutral
  Current Drell-Yan Production}.
\newblock {\em JHEP}, 10:176, 2019.

\bibitem{Fiaschi:2021okg}
Juri Fiaschi, Francesco Giuli, Francesco Hautmann, and Stefano Moretti.
\newblock {Lepton-Charge and Forward-Backward Asymmetries in Drell-Yan
  Processes for Precision Electroweak Measurements and New Physics Searches}.
\newblock {\em Nucl. Phys. B}, 968:115444, 2021.

\bibitem{Fiaschi:2022wgl}
J.~Fiaschi, F.~Giuli, F.~Hautmann, S.~Moch, and S.~Moretti.
\newblock {Z'-boson dilepton searches and the high-x quark density}.
\newblock {\em Phys. Lett. B}, 841:137915, 2023.

\bibitem{Ball:2022qtp}
Richard~D. Ball, Alessandro Candido, Stefano Forte, Felix Hekhorn, Emanuele~R.
  Nocera, Juan Rojo, and Christopher Schwan.
\newblock {Parton distributions and new physics searches: the
  Drell\textendash{}Yan forward\textendash{}backward asymmetry as a case
  study}.
\newblock {\em Eur. Phys. J. C}, 82(12):1160, 2022.

\bibitem{Fu:2023rrs}
Yao Fu, Raymond Brock, Daniel Hayden, and Chien-Peng Yuan.
\newblock {Probing Parton distribution functions at large x via Drell-Yan
  Forward-Backward Asymmetry}.
\newblock 7 2023.

\bibitem{Wells:2015uba}
James~D. Wells and Zhengkang Zhang.
\newblock {Effective theories of universal theories}.
\newblock {\em JHEP}, 01:123, 2016.

\bibitem{Peskin:1991sw}
Michael~E. Peskin and Tatsu Takeuchi.
\newblock {Estimation of oblique electroweak corrections}.
\newblock {\em Phys. Rev.}, D46:381--409, 1992.

\bibitem{Altarelli:1991fk}
Guido Altarelli, Riccardo Barbieri, and S.~Jadach.
\newblock {Toward a model independent analysis of electroweak data}.
\newblock {\em Nucl. Phys.}, B369:3--32, 1992.
\newblock [Erratum: Nucl. Phys.B376,444(1992)].

\bibitem{Barbieri:2004qk}
Riccardo Barbieri, Alex Pomarol, Riccardo Rattazzi, and Alessandro Strumia.
\newblock {Electroweak symmetry breaking after LEP-1 and LEP-2}.
\newblock {\em Nucl. Phys.}, B703:127--146, 2004.

\bibitem{Wells:2015cre}
James~D. Wells and Zhengkang Zhang.
\newblock {Renormalization group evolution of the universal theories EFT}.
\newblock {\em JHEP}, 06:122, 2016.

\bibitem{DYpaper}
Admir Greljo, Shayan Iranipour, Zahari Kassabov, Maeve Madigan, James Moore,
  Juan Rojo, Maria Ubiali, and Cameron Voisey.
\newblock {Parton distributions in the SMEFT from high-energy Drell-Yan tails}.
\newblock {\em JHEP}, 07:122, 2021.

\bibitem{deBlas:2017xtg}
J.~de~Blas, J.~C. Criado, M.~Perez-Victoria, and J.~Santiago.
\newblock {Effective description of general extensions of the Standard Model:
  the complete tree-level dictionary}.
\newblock {\em JHEP}, 03:109, 2018.

\bibitem{Dawson:2024ozw}
Sally Dawson, Matthew Forslund, and Marvin Schnubel.
\newblock {SMEFT matching to Z' models at dimension eight}.
\newblock {\em Phys. Rev. D}, 110(1):015002, 2024.

\bibitem{Allanach:2024ozu}
Ben Allanach and Nico Gubernari.
\newblock {Malaphoric $Z'$ models for $b \rightarrow s \ell^+ \ell^-$
  anomalies}.
\newblock 9 2024.

\bibitem{Allanach:2018vjg}
B.~C. Allanach, Joe Davighi, and Scott Melville.
\newblock {An Anomaly-free Atlas: charting the space of flavour-dependent
  gauged $U(1)$ extensions of the Standard Model}.
\newblock {\em JHEP}, 02:082, 2019.
\newblock [Erratum: JHEP 08, 064 (2019)].

\bibitem{Salvioni:2009mt}
Ennio Salvioni, Giovanni Villadoro, and Fabio Zwirner.
\newblock {Minimal Z-prime models: Present bounds and early LHC reach}.
\newblock {\em JHEP}, 11:068, 2009.

\bibitem{Salvioni:2009jp}
Ennio Salvioni, Alessandro Strumia, Giovanni Villadoro, and Fabio Zwirner.
\newblock {Non-universal minimal Z' models: present bounds and early LHC
  reach}.
\newblock {\em JHEP}, 03:010, 2010.

\bibitem{Langacker:2008yv}
Paul Langacker.
\newblock {The Physics of Heavy $Z^\prime$ Gauge Bosons}.
\newblock {\em Rev. Mod. Phys.}, 81:1199--1228, 2009.

\bibitem{Panico:2021vav}
Giuliano Panico, Lorenzo Ricci, and Andrea Wulzer.
\newblock {High-energy EFT probes with fully differential Drell-Yan
  measurements}.
\newblock {\em JHEP}, 07:086, 2021.

\bibitem{Farina:2016rws}
Marco Farina, Giuliano Panico, Duccio Pappadopulo, Joshua~T. Ruderman, Riccardo
  Torre, and Andrea Wulzer.
\newblock {Energy helps accuracy: electroweak precision tests at hadron
  colliders}.
\newblock {\em Phys. Lett.}, B772:210--215, 2017.

\bibitem{Torre:2020aiz}
Riccardo Torre, Lorenzo Ricci, and Andrea Wulzer.
\newblock {On the W\&Y interpretation of high-energy Drell-Yan measurements}.
\newblock {\em JHEP}, 02:144, 2021.

\bibitem{Boughezal:2021tih}
Radja Boughezal, Emanuele Mereghetti, and Frank Petriello.
\newblock {Dilepton production in the SMEFT at O(1/\ensuremath{\Lambda}4)}.
\newblock {\em Phys. Rev. D}, 104(9):095022, 2021.

\bibitem{ATLAS:2013xny}
Georges Aad et~al.
\newblock {Measurement of the high-mass Drell--Yan differential cross-section
  in pp collisions at sqrt(s)=7 TeV with the ATLAS detector}.
\newblock {\em Phys. Lett. B}, 725:223--242, 2013.

\bibitem{ATLAS:2016gic}
Georges Aad et~al.
\newblock {Measurement of the double-differential high-mass Drell-Yan cross
  section in pp collisions at $ \sqrt{s}=8 $ TeV with the ATLAS detector}.
\newblock {\em JHEP}, 08:009, 2016.

\bibitem{CMS:2013zfg}
Serguei Chatrchyan et~al.
\newblock {Measurement of the Differential and Double-Differential Drell-Yan
  Cross Sections in Proton-Proton Collisions at $\sqrt{s} =$ 7 TeV}.
\newblock {\em JHEP}, 12:030, 2013.

\bibitem{CMS:2014jea}
Vardan Khachatryan et~al.
\newblock {Measurements of differential and double-differential Drell-Yan cross
  sections in proton-proton collisions at 8 TeV}.
\newblock {\em Eur. Phys. J.}, C75(4):147, 2015.

\bibitem{CMS:2018mdl}
Albert~M Sirunyan et~al.
\newblock {Measurement of the differential Drell-Yan cross section in
  proton-proton collisions at $ \sqrt{\mathrm{s}} $ = 13 TeV}.
\newblock {\em JHEP}, 12:059, 2019.

\bibitem{ATLAS:2020osn}
{Measurements of $WH$ and $ZH$ production in the $H\rightarrow b\bar{b}$ decay
  channel in $pp$ collisions at $13\,$TeV with the ATLAS detector}.
\newblock 4 2020.

\bibitem{CMS:2019efc}
Albert~M Sirunyan et~al.
\newblock {Measurements of the pp $\to$ WZ inclusive and differential
  production cross section and constraints on charged anomalous triple gauge
  couplings at $\sqrt{s} =$ 13 TeV}.
\newblock {\em JHEP}, 04:122, 2019.

\bibitem{Carrazza:2016htc}
Stefano Carrazza, Stefano Forte, Zahari Kassabov, and Juan Rojo.
\newblock {Specialized minimal PDFs for optimized LHC calculations}.
\newblock {\em Eur. Phys. J.}, C76(4):205, 2016.

\bibitem{NuSea:2001idv}
R.~S. Towell et~al.
\newblock {Improved measurement of the anti-d / anti-u asymmetry in the nucleon
  sea}.
\newblock {\em Phys. Rev. D}, 64:052002, 2001.

\bibitem{D0:2013xqc}
Victor~Mukhamedovich Abazov et~al.
\newblock {Measurement of the Muon Charge Asymmetry in $p\bar{p}$ $\to$ W+X
  $\to$ $\mu\nu$ + X Events at $\sqrt{s}$=1.96 TeV}.
\newblock {\em Phys. Rev. D}, 88:091102, 2013.

\bibitem{ATLAS:2016nqi}
Morad Aaboud et~al.
\newblock {Precision measurement and interpretation of inclusive $W^+$ , $W^-$
  and $Z/\gamma ^*$ production cross sections with the ATLAS detector}.
\newblock {\em Eur. Phys. J. C}, 77(6):367, 2017.

\bibitem{Khalek:2021ulf}
Rabah~Abdul Khalek, Jacob~J. Ethier, Emanuele~R. Nocera, and Juan Rojo.
\newblock {Self-consistent determination of proton and nuclear PDFs at the
  Electron Ion Collider}.
\newblock {\em Phys. Rev. D}, 103(9):096005, 2021.

\bibitem{Abir:2023fpo}
Raktim Abir et~al.
\newblock {The case for an EIC Theory Alliance: Theoretical Challenges of the
  EIC}.
\newblock 5 2023.

\bibitem{Accardi:2023chb}
A.~Accardi et~al.
\newblock {Strong Interaction Physics at the Luminosity Frontier with 22 GeV
  Electrons at Jefferson Lab}.
\newblock 6 2023.

\bibitem{Accardi:2021ysh}
A.~Accardi, T.~J. Hobbs, X.~Jing, and P.~M. Nadolsky.
\newblock {Deuterium scattering experiments in CTEQ global QCD analyses: a
  comparative investigation}.
\newblock {\em Eur. Phys. J. C}, 81(7):603, 2021.

\bibitem{Accardi:2023gyr}
Alberto Accardi, Xiaoxian Jing, Joseph~Francis Owens, and Sanghwa Park.
\newblock {Light quark and antiquark constraints from new electroweak data}.
\newblock {\em Phys. Rev. D}, 107(11):113005, 2023.

\bibitem{Guzzi:2021fre}
Marco Guzzi et~al.
\newblock {NNLO constraints on proton PDFs from the SeaQuest and STAR
  experiments and other developments in the CTEQ-TEA global analysis}.
\newblock {\em SciPost Phys. Proc.}, 8:005, 2022.

\bibitem{Alekhin:2023uqx}
S.~Alekhin, M.~V. Garzelli, S.~Kulagin, and S.~O. Moch.
\newblock {Impact of SeaQuest data on PDF fits at large $x$}.
\newblock 6 2023.

\bibitem{Hou:2022onq}
Tie-Jiun Hou, Huey-Wen Lin, Mengshi Yan, and C.~P. Yuan.
\newblock {Impact of lattice strangeness asymmetry data in the CTEQ-TEA global
  analysis}.
\newblock {\em Phys. Rev. D}, 107(7):076018, 2023.

\bibitem{Berger:2004mj}
Edmond~L. Berger, Pavel~M. Nadolsky, Fredrick~I. Olness, and Jon Pumplin.
\newblock {Light gluino constituents of hadrons and a global analysis of hadron
  scattering data}.
\newblock {\em Phys. Rev. D}, 71:014007, 2005.

\bibitem{Berger:2010rj}
Edmond~L. Berger, Marco Guzzi, Hung-Liang Lai, Pavel~M. Nadolsky, and
  Fredrick~I. Olness.
\newblock {Constraints on color-octet fermions from a global parton
  distribution analysis}.
\newblock {\em Phys. Rev. D}, 82:114023, 2010.

\bibitem{McCullough:2022hzr}
Matthew McCullough, James Moore, and Maria Ubiali.
\newblock {The dark side of the proton}.
\newblock {\em JHEP}, 08:019, 2022.

\bibitem{Grojean:2013kd}
Christophe Grojean, Elizabeth~E. Jenkins, Aneesh~V. Manohar, and Michael Trott.
\newblock {Renormalization Group Scaling of Higgs Operators and
  \textbackslash{}Gamma(h -\ensuremath{>} \textbackslash{}gamma
  \textbackslash{}gamma)}.
\newblock {\em JHEP}, 04:016, 2013.

\bibitem{Trott:2023jrw}
Michael Trott.
\newblock {$\alpha_s$ as an input parameter in the SMEFT}.
\newblock 6 2023.

\bibitem{Englert:2018byk}
Christoph Englert, Michael Russell, and Chris~D. White.
\newblock {Effective Field Theory in the top sector: do multijets help?}
\newblock {\em Phys. Rev. D}, 99(3):035019, 2019.

\bibitem{Martin:2023fad}
Adam Martin and Michael Trott.
\newblock {More accurate $ \sigma \left(\mathcal{GG}\to h\right),\Gamma
  \left(h\to \mathcal{GG},\mathcal{AA},\overline{\Psi}\Psi \right) $ and Higgs
  width results via the geoSMEFT}.
\newblock {\em JHEP}, 01:170, 2024.

\bibitem{BuarqueFranzosi:2015jrv}
Diogo Buarque~Franzosi and Cen Zhang.
\newblock {Probing the top-quark chromomagnetic dipole moment at
  next-to-leading order in QCD}.
\newblock {\em Phys. Rev. D}, 91(11):114010, 2015.

\bibitem{Cranmer2023InterpretableML}
M.~Cranmer.
\newblock Interpretable machine learning for science with pysr and
  symbolicregression.jl.
\newblock {\em ArXiv}, abs/2305.01582, 2023.

\bibitem{Choi:2010wa}
Suyong Choi.
\newblock {Construction of a Kinematic Variable Sensitive to the Mass of the
  Standard Model Higgs Boson in $H -> WW^* -> l^+ \nu l^- \bar{\nu}$ using
  Symbolic Regression}.
\newblock {\em JHEP}, 08:110, 2011.

\bibitem{Butter:2021rvz}
Anja Butter, Tilman Plehn, Nathalie Soybelman, and Johann Brehmer.
\newblock {Back to the Formula -- LHC Edition}.
\newblock 9 2021.

\bibitem{Dersy:2022bym}
Aur\'elien Dersy, Matthew~D. Schwartz, and Xiaoyuan Zhang.
\newblock {Simplifying Polylogarithms with Machine Learning}.
\newblock 6 2022.

\bibitem{Dong:2022trn}
Zhongtian Dong, Kyoungchul Kong, Konstantin~T. Matchev, and Katia Matcheva.
\newblock {Is the machine smarter than the theorist: Deriving formulas for
  particle kinematics with symbolic regression}.
\newblock {\em Phys. Rev. D}, 107(5):055018, 2023.

\bibitem{CMS-PAS-SMP-23-002}
{Measurement of the W boson mass in proton-proton collisions at sqrts = 13
  TeV}.
\newblock Technical report, CERN, Geneva, 2024.

\bibitem{ATLAS:2023lsr}
Georges Aad et~al.
\newblock {A precise measurement of the Z-boson double-differential transverse
  momentum and rapidity distributions in the full phase space of the decay
  leptons with the ATLAS experiment at $\sqrt s$ = 8 TeV}.
\newblock 9 2023.

\bibitem{Collins:1977iv}
John~C. Collins and Davison~E. Soper.
\newblock {Angular Distribution of Dileptons in High-Energy Hadron Collisions}.
\newblock {\em Phys. Rev. D}, 16:2219, 1977.

\bibitem{Mirkes:1992hu}
E.~Mirkes.
\newblock {Angular decay distribution of leptons from W bosons at NLO in
  hadronic collisions}.
\newblock {\em Nucl. Phys. B}, 387:3--85, 1992.

\bibitem{Mirkes:1994dp}
E.~Mirkes and J.~Ohnemus.
\newblock {Angular distributions of Drell-Yan lepton pairs at the Tevatron:
  Order $\alpha-s^{2}$ corrections and Monte Carlo studies}.
\newblock {\em Phys. Rev. D}, 51:4891--4904, 1995.

\bibitem{Mirkes:1994eb}
E.~Mirkes and J.~Ohnemus.
\newblock {$W$ and $Z$ polarization effects in hadronic collisions}.
\newblock {\em Phys. Rev. D}, 50:5692--5703, 1994.

\bibitem{Mirkes:1994nr}
E.~Mirkes and J.~Ohnemus.
\newblock {Polarization effects in Drell-Yan type processes h1 + h2
  ---\ensuremath{>} (W, Z, gamma*, J / psi) + x}.
\newblock In {\em {1994 Meeting of the American Physical Society, Division of
  Particles and Fields (DPF 94)}}, pages 1721--1723, 8 1994.

\bibitem{Bern:2011ie}
Z.~Bern et~al.
\newblock {Left-Handed W Bosons at the LHC}.
\newblock {\em Phys. Rev. D}, 84:034008, 2011.

\bibitem{Moura2021}
Leonardo~de Moura and Sebastian Ullrich.
\newblock {\em The Lean 4 Theorem Prover and Programming Language}, page
  625–635.
\newblock Springer International Publishing, 2021.

\bibitem{palmskog}
Karl Palmskog, Enrico Tassi, and Théo Zimmermann.
\newblock Reliably reproducing machine-checked proofs with the coq platform,
  2022.

\bibitem{RomeraParedes2023}
Bernardino Romera-Paredes, Mohammadamin Barekatain, Alexander Novikov, Matej
  Balog, M.~Pawan Kumar, Emilien Dupont, Francisco J.~R. Ruiz, Jordan~S.
  Ellenberg, Pengming Wang, Omar Fawzi, Pushmeet Kohli, and Alhussein Fawzi.
\newblock Mathematical discoveries from program search with large language
  models.
\newblock {\em Nature}, 625(7995):468–475, December 2023.

\end{thebibliography}


\appendix

\chapter{Group generators and algebra}

In this appendix we provide the explicit group generators and algebra discussed in Chapter \ref{ch:introduction} of this thesis.

\label{app:appendix_int}

The Pauli matrices $\tau^{I}$ are defined as
\begin{equation}
    \label{eq:tau_matrices}
    \begin{aligned}
    \tau^1 &= \begin{pmatrix}
    0 & 1 \\
    1 & 0
    \end{pmatrix}, & \quad
    \tau^2 &= \begin{pmatrix}
    0 & -i \\
    i & 0
    \end{pmatrix}, & \quad
    \tau^3 &= \begin{pmatrix}
    1 & 0 \\
    0 & -1
    \end{pmatrix}.
    \end{aligned}
\end{equation}

The Gell-Mann matrices $\lambda^A$ are given by
\begin{equation}
    \label{eq:gell_mann_matrices}
    \begin{aligned}
    \lambda^1 &= \begin{pmatrix}
    0 & 1 & 0 \\
    1 & 0 & 0 \\
    0 & 0 & 0
    \end{pmatrix}, & \quad
    \lambda^2 &= \begin{pmatrix}
    0 & -i & 0 \\
    i & 0 & 0 \\
    0 & 0 & 0
    \end{pmatrix}, & \quad
    \lambda^3 &= \begin{pmatrix}
    1 & 0 & 0 \\
    0 & -1 & 0 \\
    0 & 0 & 0
    \end{pmatrix}, \\
    \lambda^4 &= \begin{pmatrix}
    0 & 0 & 1 \\
    0 & 0 & 0 \\
    1 & 0 & 0
    \end{pmatrix}, & \quad
    \lambda^5 &= \begin{pmatrix}
    0 & 0 & -i \\
    0 & 0 & 0 \\
    i & 0 & 0
    \end{pmatrix}, & \quad
    \lambda^6 &= \begin{pmatrix}
    0 & 0 & 0 \\
    0 & 0 & 1 \\
    0 & 1 & 0
    \end{pmatrix}, \\
    \lambda^7 &= \begin{pmatrix}
    0 & 0 & 0 \\
    0 & 0 & -i \\
    0 & i & 0
    \end{pmatrix}, & \quad
    \lambda^8 &= \frac{1}{\sqrt{3}} \begin{pmatrix}
    1 & 0 & 0 \\
    0 & 1 & 0 \\
    0 & 0 & -2
    \end{pmatrix}.
    \end{aligned}
\end{equation}

The commutation relations for the group generators of $\text{SU}(2)_{L}$ and $\text{SU}(3)_{C}$ are given by
\begin{equation}
    \label{eq:commutation_relations}
    \begin{aligned}
        [T^A, T^B] &= i f^{ABC} T^C, \\
        [S^I, S^J] &= i \epsilon^{IJK} S^K.
    \end{aligned}
\end{equation}
%


\chapter{Additional resources for the study of possible contamination in PDF fits from new physics}

In this appendix we provide additional material that complements Chapter \ref{ch:contamination} of this thesis. 

\section{Fit quality}
\label{app:fit}

\begin{figure}[htb]

  \includegraphics[width=0.98\linewidth]{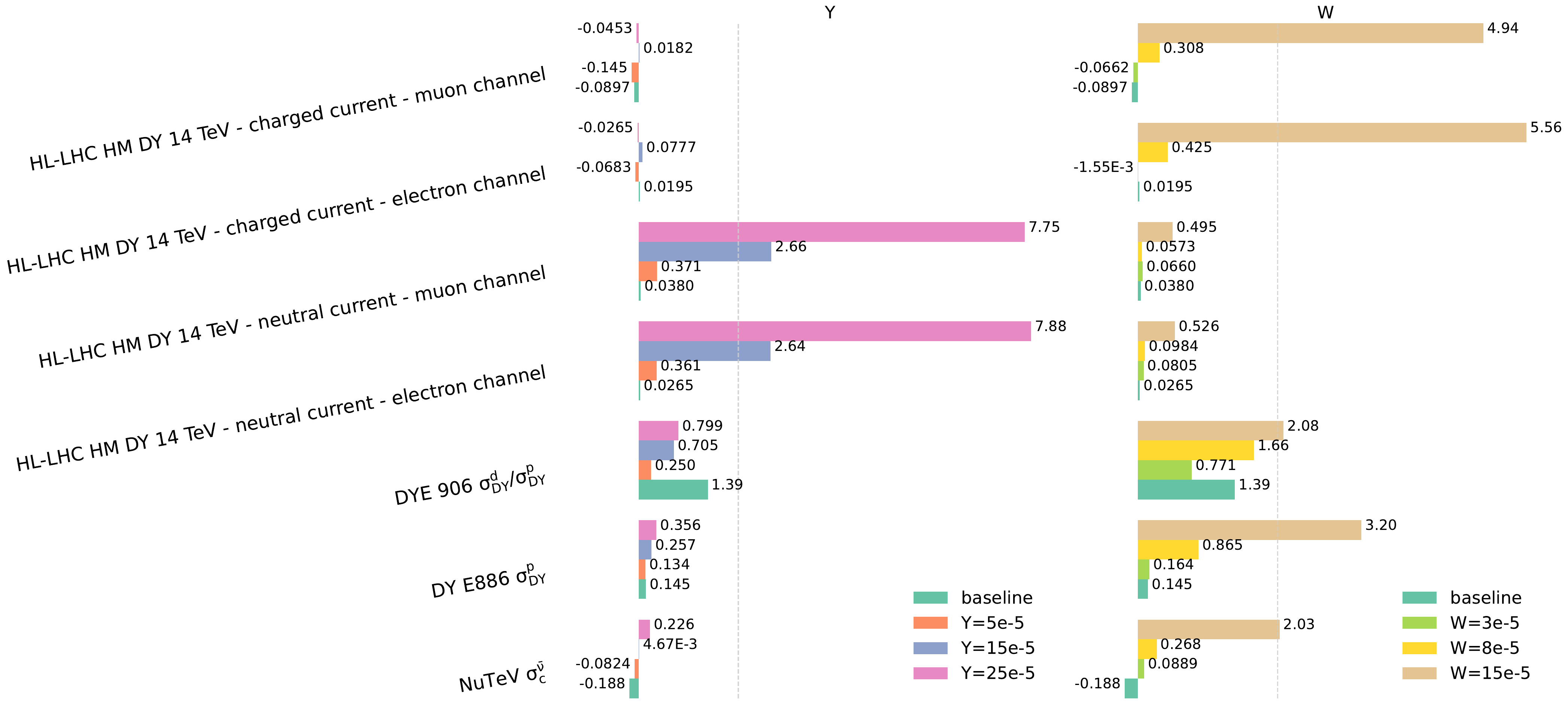}
  \caption{
    \label{fig:nsigma_multi}
  Value of $n_{\sigma}$, defined in Eq.~\eqref{eq:nsigma}
    for all datasets that pass the threshold criterion of
    $n_{\sigma}>2$ discussed in Sect.~\ref{subsec:postfit} in each of the
    three fits performed by injecting various degrees of new physics.
    The figure on the left, new physics signals in the data are added
    according to Scenario I (flavour-universal $Z^{'}$ model), namely the baseline $\hat{Y}=0$ (green bars),
    $\hat{Y}=5\cdot 10^{-5}$ (orange bars), $\hat{Y}=15\cdot 10^{-5}$
    (blue bars) and $\hat{Y}=25\cdot 10^{-5}$ (pink bars).
    In the figure on the right, signals are added according to Scenario II
    (flavour-universal $W^{'}$ model), namely the baseline $\hat{W}=0$ (again, green
    bars), $\hat{W}=3\cdot 10^{-5}$ (light green bars), $\hat{W}=8\cdot 10^{-5}$
    (yellow bars) and $\hat{W}=15\cdot 10^{-5}$ (brown bars)}
\label{fig:Ynsigma}
\end{figure}

\begin{figure}[htb]
  
  \includegraphics[width=0.90\linewidth]{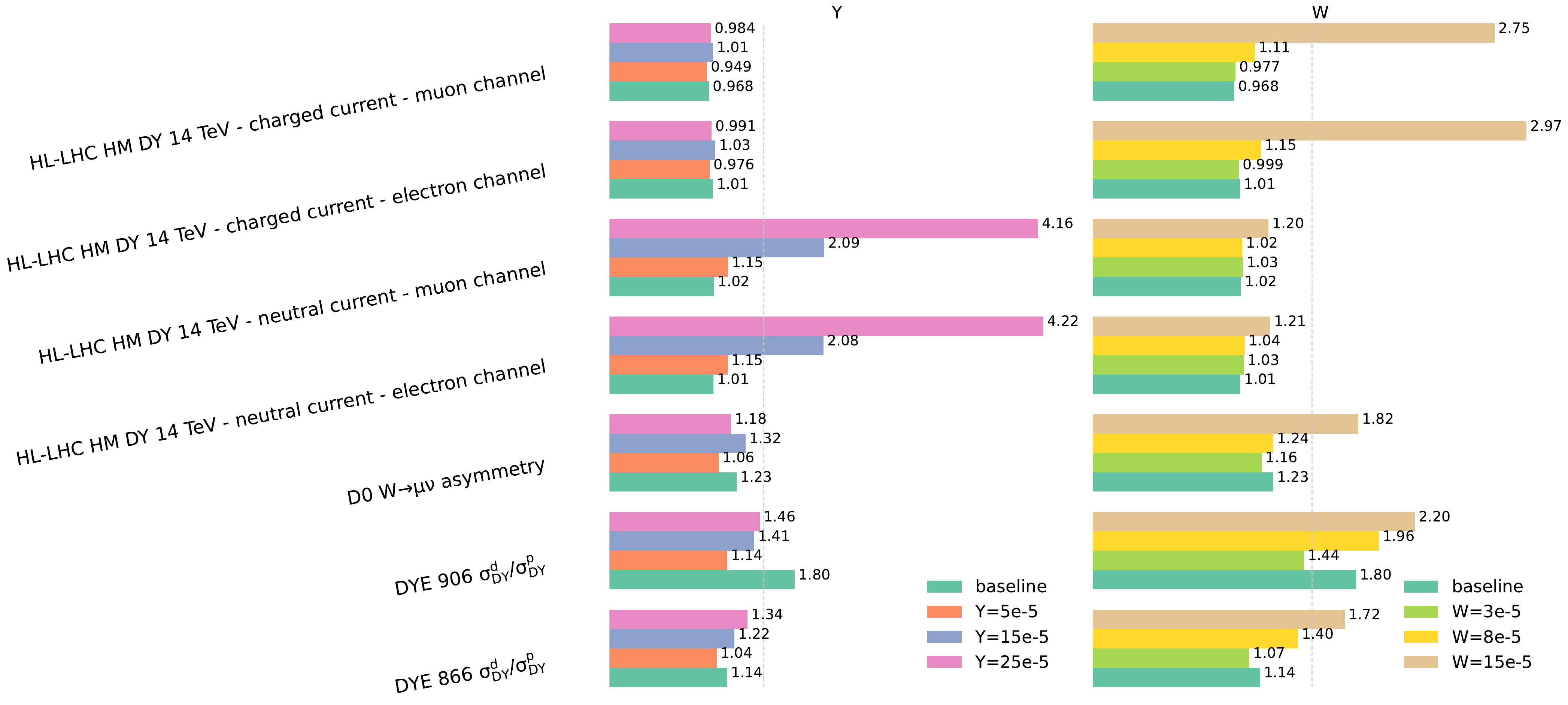}
  \caption{
  \label{fig:chi2_multi}
  $\chi^2/n_{\rm dat}$ distribution  for all datasets that pass the threshold criterion of
    $\chi^2/n_{\rm dat}>1.5$ discussed in Sect.~\ref{subsec:postfit} in each of the
    three fits performed by injecting various degrees of new physics
    signals in the data according to Scenario I (left panel) and
    Scenario II (right panel)  }
\end{figure}

In this appendix we give details about the fit-quality of the closure tests presented in this work.
In Table~\ref{tab:chi2_y} and~\ref{tab:chi2_w}, we list the value of the reduced $\chi^2/n_{\rm dat}$ 
as well as of the $n_\sigma$ estimator (see Sect.~\ref{subsec:postfit} for details) 
for each dataset included in the fit, under all the contamination scenarios we have tested. 
We have highlighted the datasets whose fit quality deteriorates the most in Figs.~\ref{fig:nsigma_multi} and~\ref{fig:chi2_multi}. 
In particular, the two figures showcase the tension between the fixed-target datasets and the HL-LHC projected data as the value of the $\hat{W}$ increases.

\begin{table}[t]
        \tiny
        \centering
\label{tab:chi2_y}
\begin{tabular}{lrrrrrrrr}
\toprule
 & \multicolumn{2}{r}{baseline} & \multicolumn{2}{r}{Y=5e-5} & \multicolumn{2}{r}{Y=15e-5} & \multicolumn{2}{r}{Y=25e-5} \\
 & $\chi^2$ & $n_\sigma$ & $\chi^2$ & $n_\sigma$ & $\chi^2$ & $n_\sigma$ & $\chi^2$ & $n_\sigma$ \\
\midrule
NMC $d/p$ & 1.02 & 0.14 & 1.02 & 0.12 & 1.03 & 0.24 & 1.06 & 0.45 \\
NMC $p$ & 1.03 & 0.26 & 1.02 & 0.23 & 1.02 & 0.18 & 1.02 & 0.18 \\
SLAC $p$ & 1.02 & 0.06 & 1.01 & 0.05 & 1.01 & 0.03 & 1.02 & 0.07 \\
SLAC $d$ & 1.00 & -0.01 & 0.98 & -0.07 & 0.99 & -0.02 & 0.99 & -0.04 \\
BCDMS $p$ & 1.02 & 0.20 & 1.00 & 0.06 & 1.02 & 0.21 & 1.01 & 0.11 \\
BCDMS $d$ & 1.01 & 0.07 & 1.00 & 0.01 & 1.01 & 0.08 & 1.00 & 0.03 \\
CHORUS $\sigma_{CC}^{\nu}$ & 1.00 & 0.02 & 1.00 & -0.06 & 1.00 & 0.06 & 1.00 & 0.01 \\
CHORUS $\sigma_{CC}^{\bar{\nu}}$ & 0.99 & -0.13 & 0.99 & -0.13 & 1.00 & -0.03 & 0.99 & -0.08 \\
NuTeV $\sigma_{c}^{\nu}$ & 0.99 & -0.06 & 0.99 & -0.05 & 0.99 & -0.02 & 1.00 & 0.02 \\
NuTeV $\sigma_{c}^{\bar{\nu}}$ & 0.96 & -0.19 & 0.98 & -0.08 & 1.00 & 0.00 & 1.05 & 0.23 \\
HERA I+II inclusive NC $e^-p$ & 1.00 & -0.02 & 1.01 & 0.12 & 1.00 & 0.02 & 1.02 & 0.17 \\
HERA I+II inclusive NC $e^+p$ 460 GeV & 1.01 & 0.08 & 1.01 & 0.12 & 1.01 & 0.13 & 1.02 & 0.18 \\
HERA I+II inclusive NC $e^+p$ 575 GeV & 0.98 & -0.21 & 1.00 & 0.01 & 0.99 & -0.17 & 1.01 & 0.07 \\
HERA I+II inclusive NC $e^+p$ 820 GeV & 1.00 & -0.00 & 1.01 & 0.07 & 1.00 & -0.01 & 1.01 & 0.09 \\
HERA I+II inclusive NC $e^+p$ 920 GeV & 1.02 & 0.29 & 1.05 & 0.63 & 1.03 & 0.35 & 1.05 & 0.67 \\
HERA I+II inclusive CC $e^-p$ & 0.99 & -0.05 & 1.03 & 0.13 & 0.99 & -0.03 & 1.03 & 0.15 \\
HERA I+II inclusive CC $e^+p$ & 1.02 & 0.08 & 1.02 & 0.07 & 1.03 & 0.11 & 1.04 & 0.18 \\
HERA comb. $\sigma_{c\bar c}^{\rm red}$ & 1.00 & 0.02 & 1.02 & 0.09 & 1.01 & 0.05 & 1.03 & 0.11 \\
HERA comb. $\sigma_{b\bar b}^{\rm red}$ & 1.12 & 0.43 & 1.13 & 0.45 & 1.14 & 0.49 & 1.13 & 0.48 \\
DYE 866 $\sigma^d_{\rm DY}/\sigma^p_{\rm DY}$ & 1.14 & 0.40 & 1.04 & 0.12 & 1.22 & 0.59 & 1.34 & 0.94 \\
DY E886 $\sigma^p_{\rm DY}$ & 1.02 & 0.14 & 1.02 & 0.13 & 1.04 & 0.26 & 1.05 & 0.36 \\
DY E605 $\sigma^p_{\rm DY}$ & 1.08 & 0.53 & 1.07 & 0.43 & 1.06 & 0.42 & 1.06 & 0.39 \\
DYE 906 $\sigma^d_{\rm DY}/\sigma^p_{\rm DY}$ & 1.80 & 1.39 & 1.14 & 0.25 & 1.41 & 0.70 & 1.46 & 0.80 \\
CDF $Z$ rapidity (new) & 1.06 & 0.21 & 1.03 & 0.12 & 1.06 & 0.21 & 1.01 & 0.06 \\
D0 $Z$ rapidity & 1.03 & 0.10 & 1.02 & 0.08 & 1.04 & 0.17 & 1.03 & 0.11 \\
D0 $W\to \mu\nu$ asymmetry & 1.23 & 0.50 & 1.06 & 0.13 & 1.32 & 0.69 & 1.18 & 0.38 \\
ATLAS $W,Z$ 7 TeV 2010 & 1.05 & 0.20 & 1.04 & 0.17 & 1.05 & 0.20 & 1.05 & 0.20 \\
ATLAS HM DY 7 TeV & 1.02 & 0.04 & 1.05 & 0.12 & 1.01 & 0.02 & 1.03 & 0.09 \\
ATLAS low-mass DY 2011 & 0.90 & -0.17 & 1.04 & 0.07 & 0.87 & -0.22 & 1.01 & 0.01 \\
ATLAS $W,Z$ 7 TeV 2011 Central selection & 1.06 & 0.28 & 1.07 & 0.36 & 1.07 & 0.31 & 1.07 & 0.35 \\
ATLAS $W,Z$ 7 TeV 2011 Forward selection & 0.91 & -0.25 & 1.33 & 0.90 & 0.90 & -0.29 & 1.32 & 0.87 \\
ATLAS DY 2D 8 TeV high mass & 1.02 & 0.11 & 1.03 & 0.13 & 1.02 & 0.08 & 1.03 & 0.12 \\
ATLAS DY 2D 8 TeV low mass & 1.03 & 0.16 & 1.00 & 0.00 & 1.02 & 0.13 & 0.99 & -0.04 \\
ATLAS $W,Z$ inclusive 13 TeV & 1.07 & 0.09 & 1.08 & 0.10 & 1.09 & 0.11 & 1.13 & 0.16 \\
ATLAS $W^+$+jet 8 TeV & 1.17 & 0.46 & 0.96 & -0.11 & 1.17 & 0.48 & 0.95 & -0.13 \\
ATLAS $W^-$+jet 8 TeV & 1.19 & 0.51 & 0.97 & -0.09 & 1.20 & 0.56 & 0.97 & -0.08 \\
ATLAS $Z$ $p_T$ 8 TeV $(p_T^{ll},M_{ll})$ & 1.01 & 0.03 & 0.98 & -0.07 & 1.01 & 0.04 & 0.99 & -0.05 \\
ATLAS $Z$ $p_T$ 8 TeV $(p_T^{ll},y_{ll})$ & 0.98 & -0.10 & 0.94 & -0.31 & 0.98 & -0.10 & 0.93 & -0.36 \\
ATLAS $\sigma_{tt}^{\rm tot}$ & 1.03 & 0.02 & 1.14 & 0.10 & 1.05 & 0.03 & 1.19 & 0.14 \\
ATLAS $\sigma_{tt}^{\rm tot}$ 8 TeV & 1.31 & 0.22 & 1.12 & 0.08 & 1.27 & 0.19 & 1.08 & 0.06 \\
ATLAS $\sigma_{tt}^{\rm tot}$ 13 TeV Run II full lumi & 0.92 & -0.06 & 0.93 & -0.05 & 0.97 & -0.02 & 0.98 & -0.01 \\
ATLAS $t\bar{t}$ $y_t$ & 1.03 & 0.05 & 1.06 & 0.08 & 1.04 & 0.06 & 1.04 & 0.05 \\
ATLAS $t\bar{t}$ $y_{t\bar{t}}$ & 1.04 & 0.05 & 1.04 & 0.05 & 1.06 & 0.08 & 1.05 & 0.07 \\
ATLAS $t\bar{t}$ normalised $|y_t|$ & 1.13 & 0.21 & 1.13 & 0.20 & 1.15 & 0.24 & 1.17 & 0.26 \\
ATLAS jets 8 TeV, R=0.6 & 0.83 & -1.53 & 0.94 & -0.58 & 0.83 & -1.55 & 0.94 & -0.60 \\
ATLAS dijets 7 TeV, R=0.6 & 1.03 & 0.19 & 1.00 & -0.00 & 1.04 & 0.24 & 1.01 & 0.09 \\
ATLAS direct photon production 13 TeV & 0.97 & -0.16 & 1.03 & 0.14 & 0.98 & -0.11 & 1.04 & 0.21 \\
ATLAS single top $R_{t}$ 7 TeV & 1.14 & 0.10 & 1.26 & 0.18 & 1.05 & 0.03 & 1.15 & 0.11 \\
ATLAS single top $R_{t}$ 13 TeV & 0.91 & -0.07 & 1.01 & 0.01 & 0.93 & -0.05 & 1.04 & 0.03 \\
ATLAS single top $y_t$ (normalised) & 0.94 & -0.07 & 1.07 & 0.09 & 0.93 & -0.09 & 1.04 & 0.04 \\
ATLAS single antitop $y$ (normalised) & 0.92 & -0.10 & 0.91 & -0.11 & 0.97 & -0.04 & 0.98 & -0.03 \\
CMS $W$ asymmetry 840 pb & 0.99 & -0.03 & 0.98 & -0.04 & 0.98 & -0.04 & 1.00 & -0.01 \\
CMS $W$ asymmetry 4.7 fb & 0.97 & -0.07 & 0.97 & -0.06 & 0.97 & -0.08 & 1.00 & 0.00 \\
CMS Drell-Yan 2D 7 TeV 2011 & 1.01 & 0.05 & 1.01 & 0.07 & 1.00 & 0.04 & 1.01 & 0.10 \\
CMS $W$ rapidity 8 TeV & 1.06 & 0.21 & 1.12 & 0.39 & 1.07 & 0.22 & 1.13 & 0.42 \\
CMS $Z$ $p_T$ 8 TeV $(p_T^{ll},y_{ll})$ & 1.03 & 0.12 & 1.03 & 0.11 & 1.03 & 0.12 & 1.04 & 0.14 \\
CMS dijets 7 TeV & 0.97 & -0.15 & 1.05 & 0.24 & 0.97 & -0.14 & 1.05 & 0.25 \\
CMS jets 8 TeV & 0.99 & -0.11 & 1.00 & -0.03 & 0.99 & -0.09 & 1.00 & -0.01 \\
CMS $\sigma_{tt}^{\rm tot}$ 7 TeV & 0.86 & -0.10 & 0.95 & -0.03 & 0.86 & -0.10 & 1.00 & 0.00 \\
CMS $\sigma_{tt}^{\rm tot}$ 8 TeV & 1.18 & 0.13 & 1.09 & 0.06 & 1.21 & 0.15 & 1.07 & 0.05 \\
CMS $\sigma_{tt}^{\rm tot}$ 13 TeV & 0.98 & -0.01 & 1.11 & 0.08 & 0.99 & -0.00 & 1.12 & 0.09 \\
CMS $t\bar{t}$ rapidity $y_{t\bar{t}}$ & 1.06 & 0.12 & 1.04 & 0.08 & 1.04 & 0.09 & 1.01 & 0.02 \\
CMS $\sigma_{tt}^{\rm tot}$ 5 TeV & 0.86 & -0.10 & 0.77 & -0.17 & 0.82 & -0.13 & 0.75 & -0.18 \\
CMS $t\bar{t}$ double differential $(m_{t\bar{t}},y_{t\bar{t}})$ & 0.99 & -0.04 & 1.00 & 0.01 & 1.02 & 0.04 & 1.03 & 0.07 \\
CMS $t\bar{t}$ absolute $y_t$ & 1.01 & 0.03 & 1.02 & 0.04 & 1.02 & 0.05 & 1.03 & 0.06 \\
CMS $t\bar{t}$ absolute $|y_t|$ & 0.98 & -0.05 & 0.99 & -0.03 & 0.97 & -0.06 & 0.96 & -0.09 \\
CMS single top $\sigma_{t}+\sigma_{\bar{t}}$ 7 TeV & 0.93 & -0.05 & 0.91 & -0.06 & 0.89 & -0.08 & 0.86 & -0.10 \\
CMS single top $R_{t}$ 8 TeV & 0.64 & -0.26 & 1.21 & 0.15 & 0.63 & -0.26 & 1.14 & 0.10 \\
CMS single top $R_{t}$ 13 TeV & 1.50 & 0.35 & 1.44 & 0.31 & 1.46 & 0.33 & 1.42 & 0.30 \\
LHCb $Z$ 940 pb & 1.08 & 0.17 & 0.96 & -0.09 & 1.11 & 0.24 & 0.97 & -0.06 \\
LHCb $Z\to ee$ 2 fb & 1.03 & 0.08 & 1.04 & 0.13 & 1.01 & 0.02 & 1.04 & 0.11 \\
LHCb $W,Z \to \mu$ 7 TeV & 0.98 & -0.07 & 0.97 & -0.10 & 1.02 & 0.06 & 1.01 & 0.05 \\
LHCb $W,Z \to \mu$ 8 TeV & 1.08 & 0.32 & 1.13 & 0.51 & 1.09 & 0.35 & 1.18 & 0.68 \\
LHCb $Z\to \mu\mu$ & 1.09 & 0.26 & 1.05 & 0.15 & 1.10 & 0.27 & 1.05 & 0.13 \\
LHCb $Z\to ee$ & 1.07 & 0.19 & 1.03 & 0.08 & 1.07 & 0.19 & 1.04 & 0.10 \\
CMS HM DY 8 TeV & 0.98 & -0.09 & 0.98 & -0.08 & 0.97 & -0.13 & 0.98 & -0.10 \\
CMS HM DY 13 TeV - combined channel & 0.97 & -0.14 & 0.97 & -0.16 & 0.97 & -0.14 & 0.97 & -0.15 \\
HL-LHC HM DY 14 TeV - neutral current - electron channel & 1.01 & 0.03 & 1.15 & 0.36 & 2.08 & 2.64 & 4.22 & 7.88 \\
HL-LHC HM DY 14 TeV - neutral current - muon channel & 1.02 & 0.04 & 1.15 & 0.37 & 2.09 & 2.66 & 4.16 & 7.75 \\
HL-LHC HM DY 14 TeV - charged current - electron channel & 1.01 & 0.02 & 0.98 & -0.07 & 1.03 & 0.08 & 0.99 & -0.03 \\
HL-LHC HM DY 14 TeV - charged current - muon channel & 0.97 & -0.09 & 0.95 & -0.14 & 1.01 & 0.02 & 0.98 & -0.05 \\
\bottomrule
\end{tabular}
\caption{\label{tab:chi2_y} Fit quality in fits contaminated with the $\hat{Y}$ parameter.}
\end{table}

\begin{table}[t]
        \tiny
        \centering
\begin{tabular}{lrrrrrrrr}
\toprule
 & \multicolumn{2}{r}{baseline} & \multicolumn{2}{r}{W=3e-5} & \multicolumn{2}{r}{W=8e-5} & \multicolumn{2}{r}{W=15e-5} \\
 & $\chi^2$ & $n_\sigma$ & $\chi^2$ & $n_\sigma$ & $\chi^2$ & $n_\sigma$ & $\chi^2$ & $n_\sigma$ \\
\midrule
NMC $d/p$ & 1.02 & 0.14 & 1.01 & 0.04 & 1.04 & 0.31 & 1.05 & 0.42 \\
NMC $p$ & 1.03 & 0.26 & 1.03 & 0.27 & 1.02 & 0.22 & 1.03 & 0.28 \\
SLAC $p$ & 1.02 & 0.06 & 1.02 & 0.07 & 1.01 & 0.03 & 1.02 & 0.06 \\
SLAC $d$ & 1.00 & -0.01 & 0.98 & -0.07 & 0.99 & -0.05 & 1.00 & 0.02 \\
BCDMS $p$ & 1.02 & 0.20 & 1.01 & 0.07 & 1.02 & 0.24 & 1.01 & 0.11 \\
BCDMS $d$ & 1.01 & 0.07 & 1.00 & 0.01 & 1.01 & 0.10 & 1.00 & 0.02 \\
CHORUS $\sigma_{CC}^{\nu}$ & 1.00 & 0.02 & 1.00 & -0.07 & 1.00 & 0.06 & 1.00 & 0.04 \\
CHORUS $\sigma_{CC}^{\bar{\nu}}$ & 0.99 & -0.13 & 0.99 & -0.13 & 1.00 & -0.00 & 1.00 & -0.02 \\
NuTeV $\sigma_{c}^{\nu}$ & 0.99 & -0.06 & 0.99 & -0.05 & 1.01 & 0.05 & 1.00 & 0.01 \\
NuTeV $\sigma_{c}^{\bar{\nu}}$ & 0.96 & -0.19 & 1.02 & 0.09 & 1.06 & 0.27 & 1.47 & 2.03 \\
HERA I+II inclusive NC $e^-p$ & 1.00 & -0.02 & 1.01 & 0.13 & 1.00 & 0.03 & 1.02 & 0.19 \\
HERA I+II inclusive NC $e^+p$ 460 GeV & 1.01 & 0.08 & 1.01 & 0.12 & 1.01 & 0.12 & 1.02 & 0.20 \\
HERA I+II inclusive NC $e^+p$ 575 GeV & 0.98 & -0.21 & 1.00 & 0.01 & 0.98 & -0.18 & 1.01 & 0.10 \\
HERA I+II inclusive NC $e^+p$ 820 GeV & 1.00 & -0.00 & 1.01 & 0.07 & 1.00 & -0.02 & 1.02 & 0.10 \\
HERA I+II inclusive NC $e^+p$ 920 GeV & 1.02 & 0.29 & 1.06 & 0.76 & 1.04 & 0.54 & 1.09 & 1.23 \\
HERA I+II inclusive CC $e^-p$ & 0.99 & -0.05 & 1.03 & 0.13 & 1.00 & -0.00 & 1.03 & 0.15 \\
HERA I+II inclusive CC $e^+p$ & 1.02 & 0.08 & 1.02 & 0.08 & 1.04 & 0.19 & 1.10 & 0.45 \\
HERA comb. $\sigma_{c\bar c}^{\rm red}$ & 1.00 & 0.02 & 1.02 & 0.08 & 1.01 & 0.02 & 1.01 & 0.04 \\
HERA comb. $\sigma_{b\bar b}^{\rm red}$ & 1.12 & 0.43 & 1.13 & 0.45 & 1.13 & 0.48 & 1.13 & 0.47 \\
DYE 866 $\sigma^d_{\rm DY}/\sigma^p_{\rm DY}$ & 1.14 & 0.40 & 1.07 & 0.20 & 1.40 & 1.11 & 1.72 & 1.98 \\
DY E886 $\sigma^p_{\rm DY}$ & 1.02 & 0.14 & 1.02 & 0.16 & 1.13 & 0.87 & 1.48 & 3.20 \\
DY E605 $\sigma^p_{\rm DY}$ & 1.08 & 0.53 & 1.07 & 0.44 & 1.07 & 0.47 & 1.08 & 0.50 \\
DYE 906 $\sigma^d_{\rm DY}/\sigma^p_{\rm DY}$ & 1.80 & 1.39 & 1.44 & 0.77 & 1.96 & 1.66 & 2.20 & 2.08 \\
CDF $Z$ rapidity (new) & 1.06 & 0.21 & 1.03 & 0.12 & 1.06 & 0.22 & 1.02 & 0.07 \\
D0 $Z$ rapidity & 1.03 & 0.10 & 1.02 & 0.07 & 1.04 & 0.16 & 1.02 & 0.07 \\
D0 $W\to \mu\nu$ asymmetry & 1.23 & 0.50 & 1.16 & 0.33 & 1.24 & 0.50 & 1.82 & 1.73 \\
ATLAS $W,Z$ 7 TeV 2010 & 1.05 & 0.20 & 1.04 & 0.17 & 1.06 & 0.22 & 1.05 & 0.18 \\
ATLAS HM DY 7 TeV & 1.02 & 0.04 & 1.05 & 0.12 & 1.01 & 0.04 & 1.03 & 0.06 \\
ATLAS low-mass DY 2011 & 0.90 & -0.17 & 1.04 & 0.07 & 0.87 & -0.23 & 0.99 & -0.02 \\
ATLAS $W,Z$ 7 TeV 2011 Central selection & 1.06 & 0.28 & 1.07 & 0.35 & 1.06 & 0.28 & 1.08 & 0.37 \\
ATLAS $W,Z$ 7 TeV 2011 Forward selection & 0.91 & -0.25 & 1.33 & 0.90 & 0.90 & -0.29 & 1.31 & 0.84 \\
ATLAS DY 2D 8 TeV high mass & 1.02 & 0.11 & 1.03 & 0.14 & 1.02 & 0.10 & 1.04 & 0.20 \\
ATLAS DY 2D 8 TeV low mass & 1.03 & 0.16 & 1.00 & 0.00 & 1.03 & 0.16 & 0.99 & -0.04 \\
ATLAS $W,Z$ inclusive 13 TeV & 1.07 & 0.09 & 1.07 & 0.09 & 1.09 & 0.11 & 1.08 & 0.10 \\
ATLAS $W^+$+jet 8 TeV & 1.17 & 0.46 & 0.96 & -0.10 & 1.17 & 0.48 & 0.96 & -0.12 \\
ATLAS $W^-$+jet 8 TeV & 1.19 & 0.51 & 0.97 & -0.10 & 1.21 & 0.58 & 0.98 & -0.06 \\
ATLAS $Z$ $p_T$ 8 TeV $(p_T^{ll},M_{ll})$ & 1.01 & 0.03 & 0.98 & -0.07 & 1.01 & 0.03 & 0.99 & -0.05 \\
ATLAS $Z$ $p_T$ 8 TeV $(p_T^{ll},y_{ll})$ & 0.98 & -0.10 & 0.94 & -0.29 & 0.99 & -0.06 & 0.96 & -0.21 \\
ATLAS $\sigma_{tt}^{\rm tot}$ & 1.03 & 0.02 & 1.14 & 0.10 & 1.04 & 0.03 & 1.17 & 0.12 \\
ATLAS $\sigma_{tt}^{\rm tot}$ 8 TeV & 1.31 & 0.22 & 1.12 & 0.09 & 1.30 & 0.21 & 1.13 & 0.09 \\
ATLAS $\sigma_{tt}^{\rm tot}$ 13 TeV Run II full lumi & 0.92 & -0.06 & 0.93 & -0.05 & 0.93 & -0.05 & 0.97 & -0.02 \\
ATLAS $t\bar{t}$ $y_t$ & 1.03 & 0.05 & 1.06 & 0.09 & 1.03 & 0.04 & 1.06 & 0.08 \\
ATLAS $t\bar{t}$ $y_{t\bar{t}}$ & 1.04 & 0.05 & 1.04 & 0.06 & 1.05 & 0.08 & 1.09 & 0.12 \\
ATLAS $t\bar{t}$ normalised $|y_t|$ & 1.13 & 0.21 & 1.13 & 0.21 & 1.14 & 0.22 & 1.18 & 0.28 \\
ATLAS jets 8 TeV, R=0.6 & 0.83 & -1.53 & 0.94 & -0.57 & 0.83 & -1.53 & 0.94 & -0.54 \\
ATLAS dijets 7 TeV, R=0.6 & 1.03 & 0.19 & 1.00 & 0.00 & 1.03 & 0.18 & 1.01 & 0.10 \\
ATLAS direct photon production 13 TeV & 0.97 & -0.16 & 1.03 & 0.14 & 0.98 & -0.13 & 1.03 & 0.16 \\
ATLAS single top $R_{t}$ 7 TeV & 1.14 & 0.10 & 1.25 & 0.18 & 1.06 & 0.04 & 1.16 & 0.11 \\
ATLAS single top $R_{t}$ 13 TeV & 0.91 & -0.07 & 1.01 & 0.01 & 0.94 & -0.04 & 1.05 & 0.03 \\
ATLAS single top $y_t$ (normalised) & 0.94 & -0.07 & 1.07 & 0.09 & 0.94 & -0.08 & 1.04 & 0.04 \\
ATLAS single antitop $y$ (normalised) & 0.92 & -0.10 & 0.91 & -0.11 & 0.94 & -0.07 & 0.98 & -0.03 \\
CMS $W$ asymmetry 840 pb & 0.99 & -0.03 & 0.99 & -0.02 & 0.97 & -0.08 & 1.05 & 0.12 \\
CMS $W$ asymmetry 4.7 fb & 0.97 & -0.07 & 0.97 & -0.06 & 0.97 & -0.06 & 0.97 & -0.06 \\
CMS Drell-Yan 2D 7 TeV 2011 & 1.01 & 0.05 & 1.01 & 0.07 & 1.01 & 0.04 & 1.01 & 0.08 \\
CMS $W$ rapidity 8 TeV & 1.06 & 0.21 & 1.11 & 0.38 & 1.07 & 0.25 & 1.11 & 0.38 \\
CMS $Z$ $p_T$ 8 TeV $(p_T^{ll},y_{ll})$ & 1.03 & 0.12 & 1.03 & 0.12 & 1.04 & 0.13 & 1.06 & 0.21 \\
CMS dijets 7 TeV & 0.97 & -0.15 & 1.05 & 0.24 & 0.97 & -0.13 & 1.05 & 0.28 \\
CMS jets 8 TeV & 0.99 & -0.11 & 1.00 & -0.02 & 0.99 & -0.05 & 1.01 & 0.06 \\
CMS $\sigma_{tt}^{\rm tot}$ 7 TeV & 0.86 & -0.10 & 0.95 & -0.03 & 0.86 & -0.10 & 0.99 & -0.00 \\
CMS $\sigma_{tt}^{\rm tot}$ 8 TeV & 1.18 & 0.13 & 1.08 & 0.06 & 1.22 & 0.16 & 1.07 & 0.05 \\
CMS $\sigma_{tt}^{\rm tot}$ 13 TeV & 0.98 & -0.01 & 1.11 & 0.08 & 0.99 & -0.01 & 1.13 & 0.09 \\
CMS $t\bar{t}$ rapidity $y_{t\bar{t}}$ & 1.06 & 0.12 & 1.04 & 0.08 & 1.03 & 0.07 & 1.02 & 0.05 \\
CMS $\sigma_{tt}^{\rm tot}$ 5 TeV & 0.86 & -0.10 & 0.77 & -0.16 & 0.81 & -0.13 & 0.73 & -0.19 \\
CMS $t\bar{t}$ double differential $(m_{t\bar{t}},y_{t\bar{t}})$ & 0.99 & -0.04 & 1.01 & 0.02 & 1.01 & 0.04 & 1.03 & 0.08 \\
CMS $t\bar{t}$ absolute $y_t$ & 1.01 & 0.03 & 1.02 & 0.04 & 1.02 & 0.04 & 1.05 & 0.11 \\
CMS $t\bar{t}$ absolute $|y_t|$ & 0.98 & -0.05 & 0.99 & -0.03 & 0.98 & -0.05 & 0.96 & -0.10 \\
CMS single top $\sigma_{t}+\sigma_{\bar{t}}$ 7 TeV & 0.93 & -0.05 & 0.91 & -0.06 & 0.88 & -0.09 & 0.86 & -0.10 \\
CMS single top $R_{t}$ 8 TeV & 0.64 & -0.26 & 1.21 & 0.15 & 0.65 & -0.25 & 1.15 & 0.10 \\
CMS single top $R_{t}$ 13 TeV & 1.50 & 0.35 & 1.44 & 0.31 & 1.46 & 0.32 & 1.40 & 0.28 \\
LHCb $Z$ 940 pb & 1.08 & 0.17 & 0.95 & -0.10 & 1.12 & 0.25 & 0.96 & -0.08 \\
LHCb $Z\to ee$ 2 fb & 1.03 & 0.08 & 1.04 & 0.12 & 1.01 & 0.02 & 1.01 & 0.03 \\
LHCb $W,Z \to \mu$ 7 TeV & 0.98 & -0.07 & 0.96 & -0.17 & 1.07 & 0.26 & 1.13 & 0.48 \\
LHCb $W,Z \to \mu$ 8 TeV & 1.08 & 0.32 & 1.12 & 0.45 & 1.17 & 0.65 & 1.32 & 1.22 \\
LHCb $Z\to \mu\mu$ & 1.09 & 0.26 & 1.05 & 0.14 & 1.10 & 0.28 & 1.05 & 0.15 \\
LHCb $Z\to ee$ & 1.07 & 0.19 & 1.03 & 0.08 & 1.08 & 0.22 & 1.04 & 0.11 \\
CMS HM DY 8 TeV & 0.98 & -0.09 & 0.98 & -0.08 & 0.99 & -0.05 & 1.00 & -0.02 \\
CMS HM DY 13 TeV - combined channel & 0.97 & -0.14 & 0.97 & -0.16 & 0.97 & -0.14 & 0.97 & -0.12 \\
HL-LHC HM DY 14 TeV - neutral current - electron channel & 1.01 & 0.03 & 1.03 & 0.08 & 1.04 & 0.10 & 1.21 & 0.53 \\
HL-LHC HM DY 14 TeV - neutral current - muon channel & 1.02 & 0.04 & 1.03 & 0.07 & 1.02 & 0.06 & 1.20 & 0.49 \\
HL-LHC HM DY 14 TeV - charged current - electron channel & 1.01 & 0.02 & 1.00 & -0.00 & 1.15 & 0.42 & 2.97 & 5.56 \\
HL-LHC HM DY 14 TeV - charged current - muon channel & 0.97 & -0.09 & 0.98 & -0.07 & 1.11 & 0.31 & 2.75 & 4.94 \\
\bottomrule
\end{tabular}
\caption{\label{tab:chi2_w} Fit quality in fits contaminated with the $\hat{W}$ parameter.}
\end{table}

\section{PDF comparison}
\label{app:pdfs}
\begin{figure}[t!]
  \includegraphics[width=0.49\linewidth]{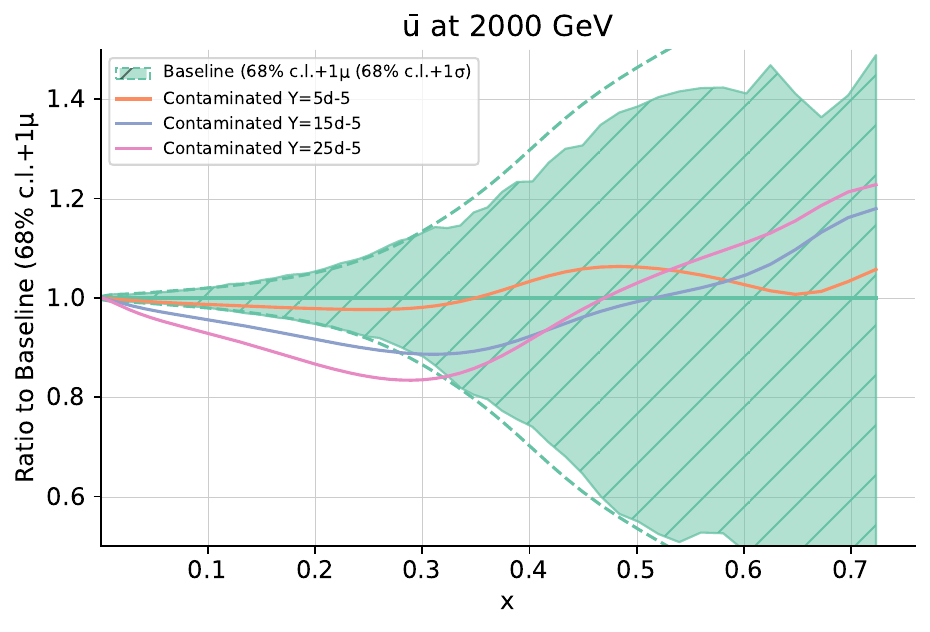}
  \includegraphics[width=0.49\linewidth]{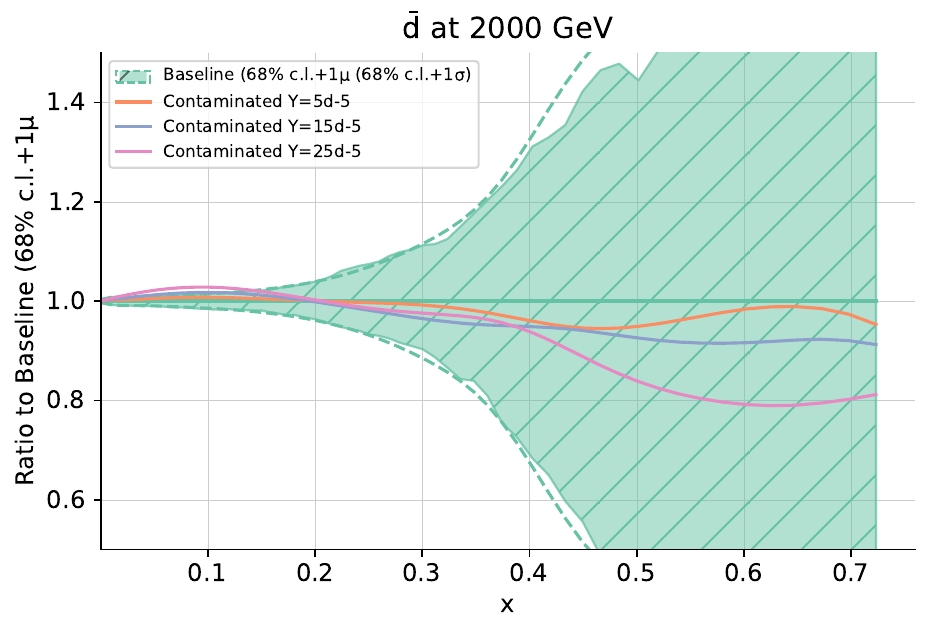}
  \includegraphics[width=0.49\linewidth]{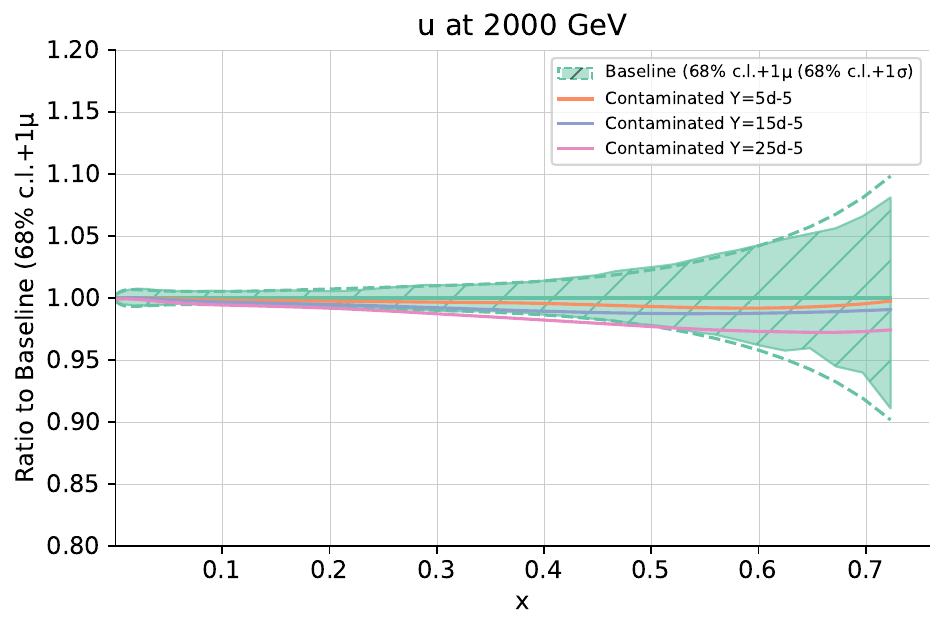}
  \includegraphics[width=0.49\linewidth]{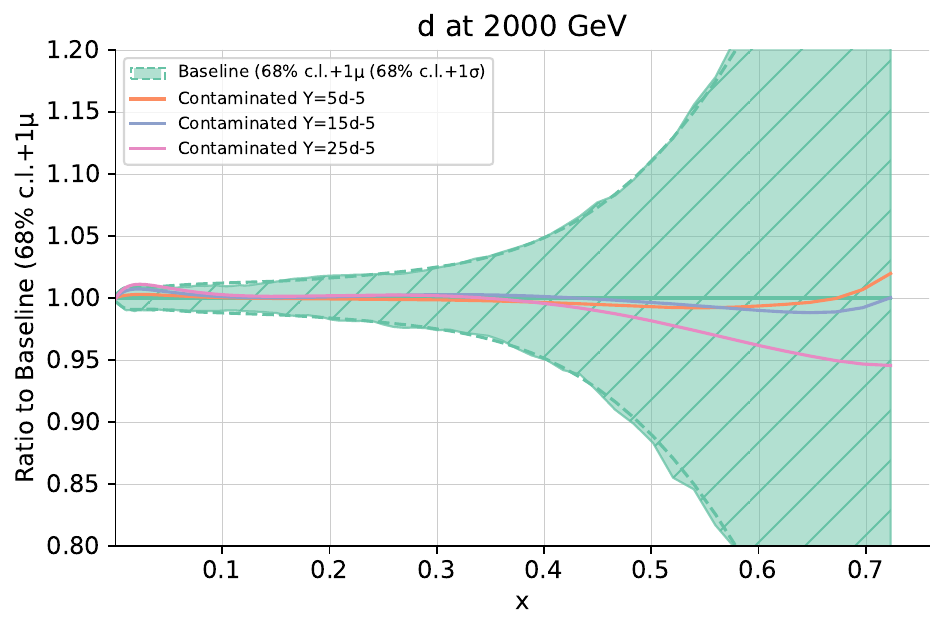}
  \caption{Contaminated versus baseline anti-up (top-left panel), anti-down (top-right panel), up (bottom-left panel) and down (bottom-right panel) PDFs at
    $Q=2$ TeV. The results are normalised to the baseline SM PDFs and
    the 68\% C.L. band is displayed. Contaminated PDFs have been
    obtained by fitting the Monte Carlo pseudodata
    produced with $\hat{Y}=5\cdot 10^{-5}$ (orange line),
    $\hat{Y}=15\cdot 10^{-5}$ (blue line) and $\hat{Y}=25\cdot
    10^{-5}$ (pink line) assuming the SM in the theory predictions. }
	\label{fig:Ypdfs}
      \end{figure}

In Fig.~\ref{fig:Ypdfs}, we display the PDFs that are mostly affected
by the new physics contamination in Scenario I, namely the anti-up and anti-down
distributions at $Q=$ 2 TeV in the large-$x$ region. We see that for
$\hat{Y} = 5\cdot 10^{-5}$, PDFs are statistically equivalent to the baseline ones.

      \begin{figure}[t!]
    \includegraphics[width=0.49\linewidth]{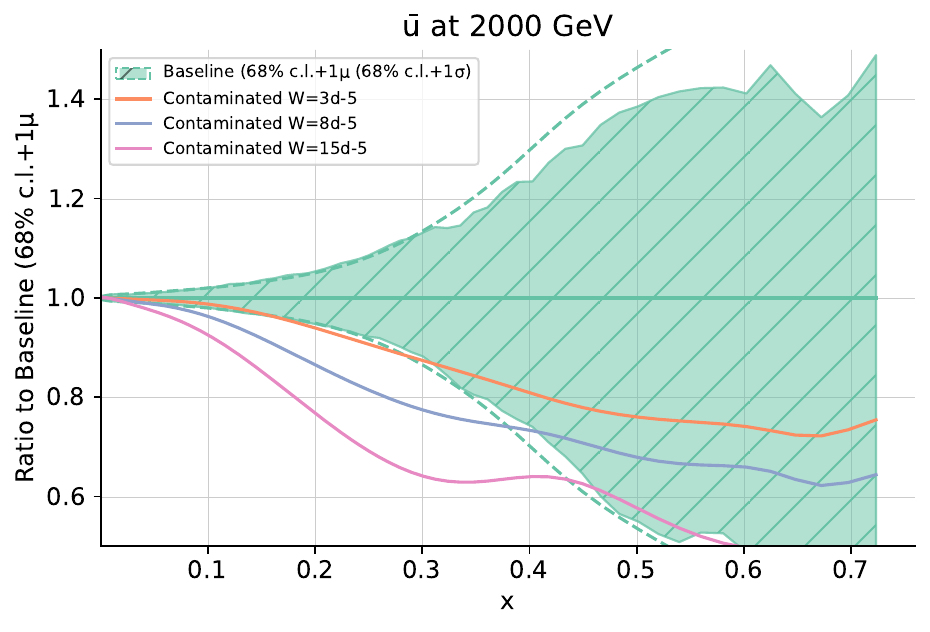}
  \includegraphics[width=0.49\linewidth]{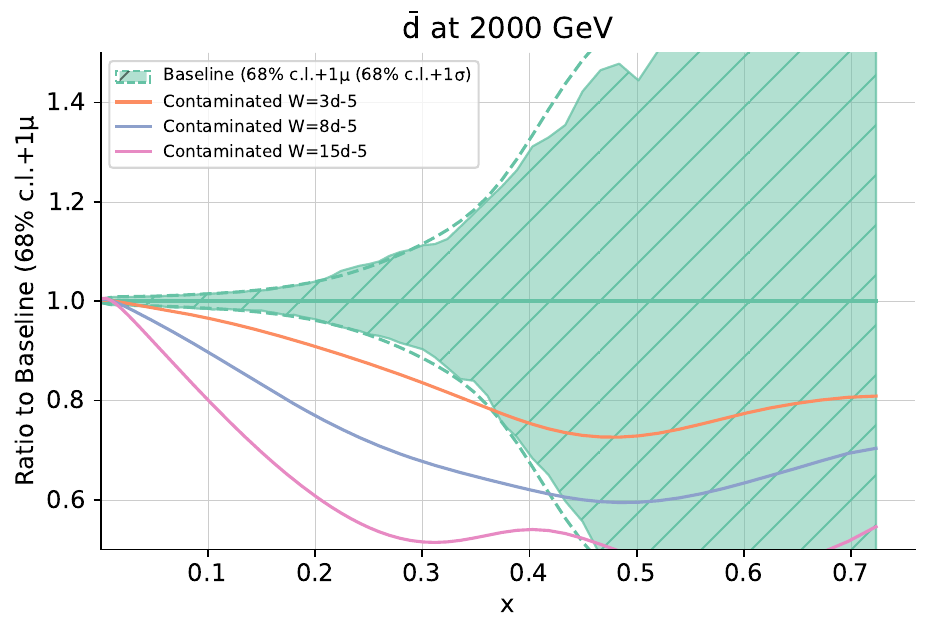}\\
  \includegraphics[width=0.49\linewidth]{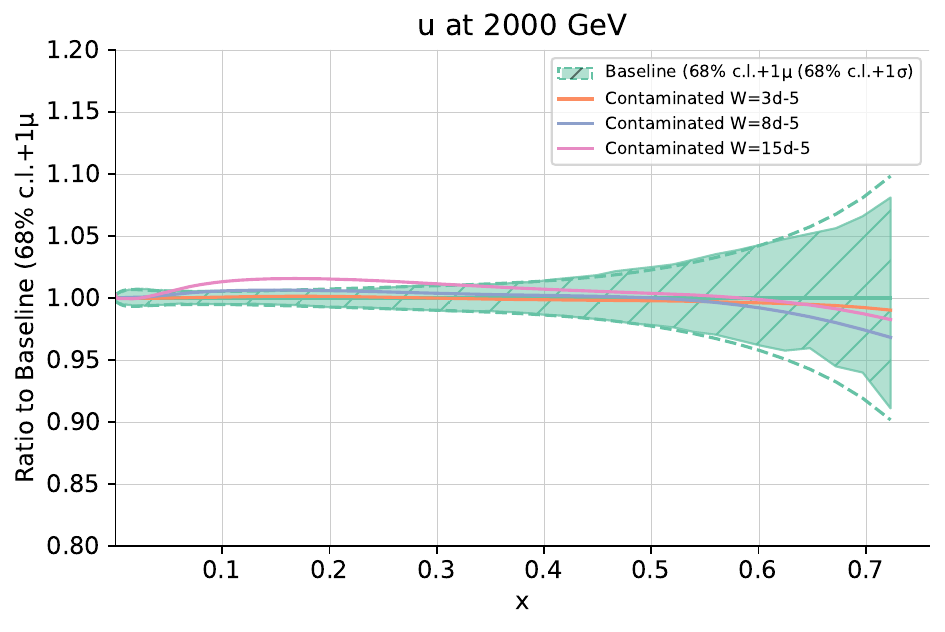}
  \includegraphics[width=0.49\linewidth]{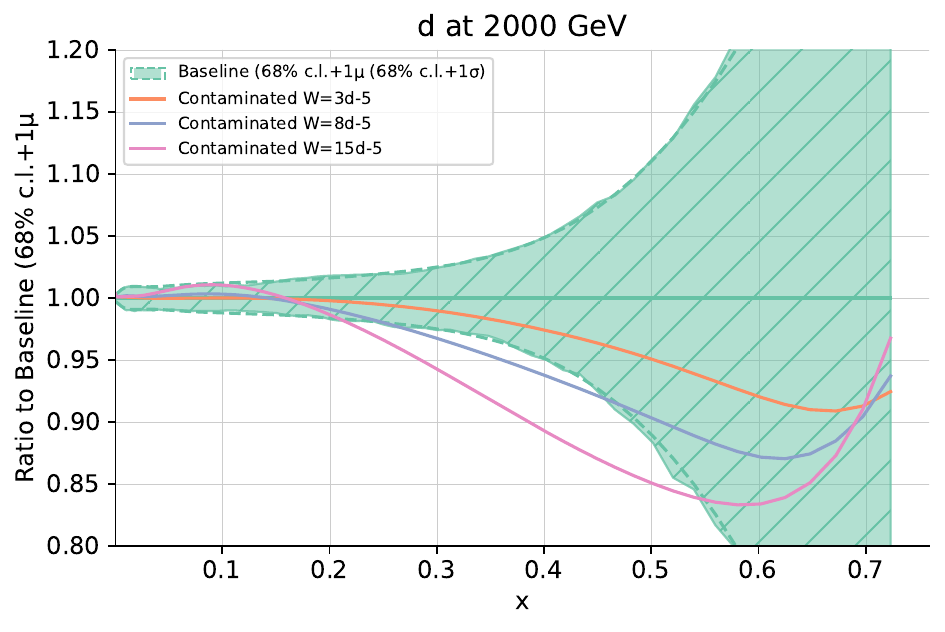}
  \caption{Same as Fig.~\ref{fig:Ypdfs} for $\hat{W}=3\cdot 10^{-5}$ (orange line),
    $\hat{W}=8\cdot 10^{-5}$ (blue line) and $\hat{W}=15\cdot
    10^{-5}$ (pink line).}
  \label{fig:Wpdfs}
      \end{figure}
      In Fig.~\ref{fig:Wpdfs}, we display the PDFs that are mostly affected
by the Scenario II new physics contamination, namely the up, down, anti-up and anti-down
distributions at $Q=$ 2 TeV in the large-$x$ region. We see that for
the critical value $\hat{W} = 8\cdot 10^{-5}$ the shift in the
antiquark PDFs is above the 2$\sigma$ level for all of the distributions
from $x\gtrsim 0.2$, apart from the up-quark PDF in which the shift is visible but below the 2$\sigma$ level.
\section{Random seed dependence}
\label{app:random}
As described in Eq.~\eqref{eq:observed_data}, the pseudodata used in this study is stochastic, fluctuated around the supposed law of Nature in order to simulate random experimental noise. This noise is generated in a reproducible manner using the NNPDF closure test code by selecting a particular \textit{seed} for the generation algorithm; different choices of seed lead to different choices of noise.

This has consequences for the resulting contaminated PDF fits, which in principle can depend on the seed used for the random noise. In certain parts of this work, in particular in the production of Figs.~\ref{fig:Y_chi2_nsigma_dist} and \ref{fig:W_chi2_nsigma_dist}, we have made the approximation that the contaminated PDFs do not depend significantly on the choice of random seed; rather, we hope that their behaviour is most importantly affected by whether or not new physics is present in the pseudodata or not. This is a useful approximation to make, since it avoids the requirement of running a large quantity of PDF fits, which is computationally expensive.

We justify this approximation in this brief appendix by comparing the PDF luminosities in various contaminated fits produced using different seeds for the random pseudodata. The luminosities are the relevant quantity to compare, since these are the quantities which enter the theoretical predictions for the hadronic data, in particular the Drell-Yan data, the focus of this study.

\begin{figure}[h!]
  \includegraphics[width=0.49\linewidth]{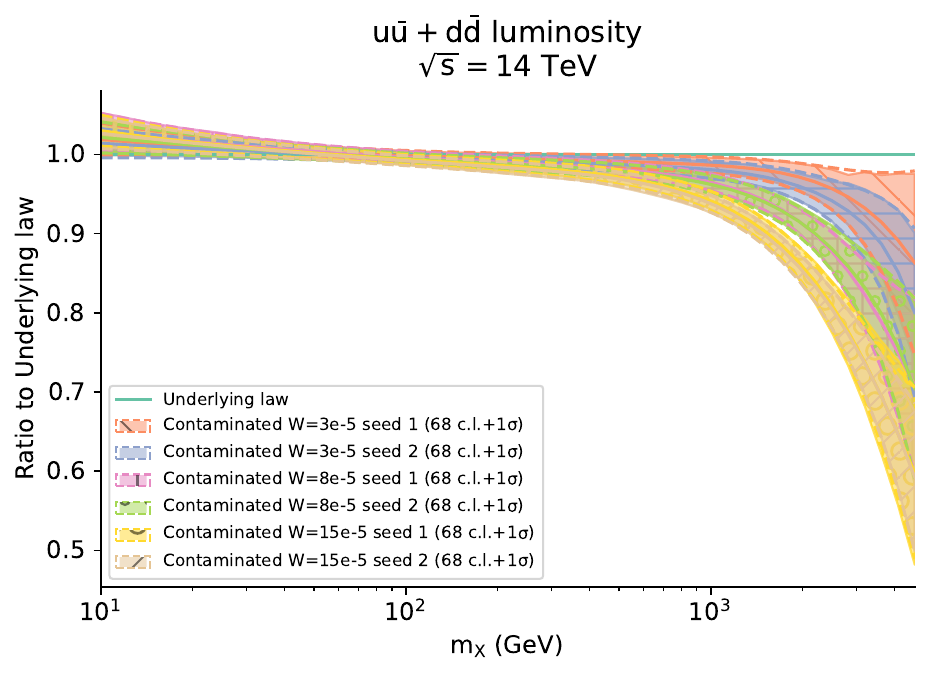}
  \includegraphics[width=0.49\linewidth]{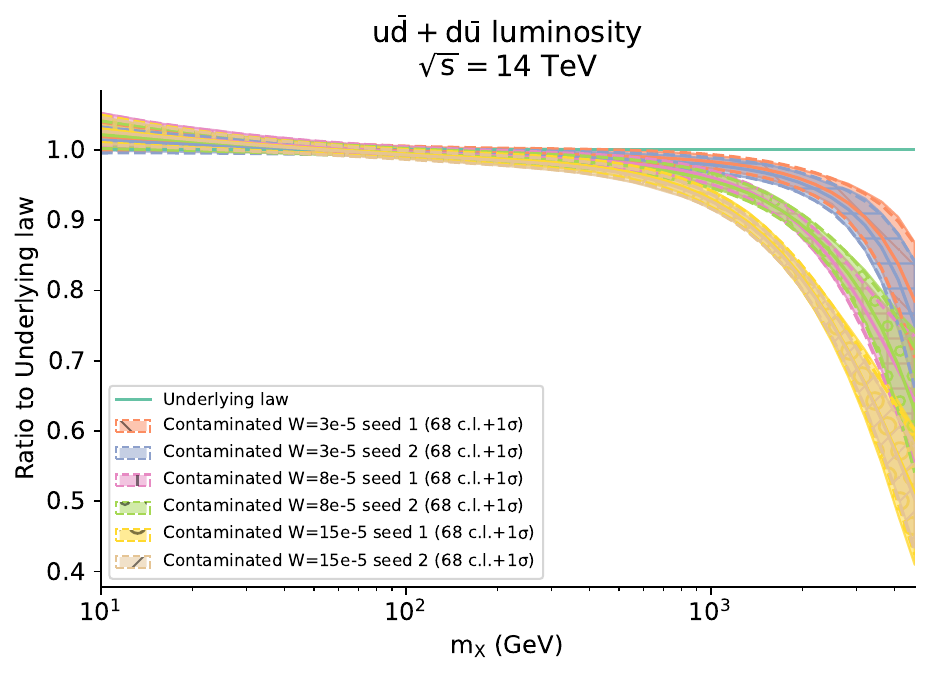}
  \caption{Comparison between luminosities obtained in contaminated fits using two different random seeds in the generation of pseudodata. In each case, we display six contaminated fits: two fits for each of the benchmark values $\hat{W} = 3 \times 10^{-5}, 8 \times 10^{-5}, 15 \times 10^{-5}$, trained on pseudodata generated with random seed $1$ and random seed $2$ respectively.}
\label{fig:lumi_random}
\end{figure}

In Fig.~\ref{fig:lumi_random}, we plot the luminosities obtained
from contaminated fits resulting from setting the $\hat{W}$ parameter to the benchmark values $\hat{W} = 3 \times 10^{-5}$, $\hat{W} = 8 \times 10^{-5}$ and $\hat{W} = 15 \times 10^{-5}$. We display the results for two separate contaminated fits for each of the benchmark values; in each case, one of the fits results from the use of a particular random seed (called \textit{seed $1$} in the plots), whilst the other results from the use of another random seed (called \textit{seed $2$} in the plots). 
We observe that the luminosities are completely statistically equivalent between the two seeds, but that across different benchmark values of $\hat{W}$, there is indeed a statistical difference between the luminosities. This justifies that the leading effect on the contaminated fits is the injection of new physics into the pseudodata, rather than the random noise added to the pseudodata. In particular, the approximation in Sect.~\ref{sec:dy} is fully justified. Similar conclusions hold for the $\hat{Y}$ parameter.








\printthesisindex

\end{document}